\documentclass[krantz1]{krantz}
\usepackage{fixltx2e,fix-cm}
\usepackage{amssymb, bbm, bm, mathtools}
\usepackage{amsmath}
\usepackage[pdftex]{graphicx}
\usepackage{color, rotating}
\usepackage{subcaption}
\usepackage{makeidx}
\usepackage{multicol}
\usepackage[all,import]{xy}
\usepackage{natbib} 
\usepackage{hyperref}

\definecolor{mygray}{gray}{0.6}
\usepackage{listings}

\lstnewenvironment{rc}[1][]{\lstset{language=R}}{}
\newcommand{\ri}[1]{\lstinline{#1}}   

\renewcommand{\theparagraph}{\arabic{chapter}.\arabic{paragraph}}
\newenvironment{myproof}[2] {\noindent {\bfseries Proof of {#1} {#2}:}}{\hfill$\square$}

\newtheorem{lemma}{Lemma}
\newtheorem{proposition}{Proposition}
\newtheorem{assumption}{Assumption}

\newtheorem{condition}{Condition}
\newtheorem{corollary}{Corollary}

\def\ind{\begin{picture}(9,8)
         \put(0,0){\line(1,0){9}}
         \put(3,0){\line(0,1){8}}
         \put(6,0){\line(0,1){8}}
         \end{picture}
        }
\def\nind{\begin{picture}(9,8)
         \put(0,0){\line(1,0){9}}
         \put(3,0){\line(0,1){8}}
         \put(6,0){\line(0,1){8}}
         \put(1,0){{\it /}}
         \end{picture}
    }

\def\var{\textup{var}}
\def\cov{\textup{cov}}
\def\sumn{\sum_{i=1}^n}

\def\iidsim{\stackrel{\textup{IID}}{\sim}}

\def\pr{\textup{pr}}
\def\RD{\textsc{rd}}
\def\RR{\textsc{rr}}
\def\OR{\textsc{or}}
\def\true{\textup{true}}
\def\obs{\textup{obs}}

\def\d{\textnormal{d}}
\def\N01{\textsc{N}(0,1)}
\def\HF{H_{0\textsc{f}}}
\def\HN{H_{0\textsc{n}}}
\newcommand*{\tran }{^{\mkern-1.5mu\mathsf{T}}}
\def\asim{\stackrel{\cdot}{\sim}}
\def\diff{\textup{d}}
\def\at{\textup{a}}
\def\nt{\textup{n}}
\def\cp{\textup{c}}
\def\df{\textup{d}}
\def\logit{\textup{logit}}
\def\expit{\textup{expit}}

\def\CDE{\textsc{cde}}
\def\NDE{\textsc{nde}}
\def\NIE{\textsc{nie}}  
\def\se{\textup{se}}  
\def\frt{\textsc{frt}}

\makeindex

\newtheorem{theorem}{Theorem}

\newtheorem{example}{Example}
\newtheorem{definition}{Definition}

\begin{document}

\frontmatter

\title{A First Course in Causal Inference}
\author{Peng Ding}

\maketitle

\cleardoublepage
\thispagestyle{empty}
\vspace*{\stretch{1}}
\begin{center}
\Large\itshape

To students and readers who are interested in causal inference

\end{center}
\vspace{\stretch{2}}
\tableofcontents
\chapter*{Preface}

\subsection*{Causal inference research and education in the past decade}

The past decade has witnessed an explosion of interest in research and education in causal inference, due to its wide applications in biomedical research, social sciences, tech companies, etc. It was quite different even ten years ago when I was a Ph.D. student in statistics. At that time, causal inference was not a mainstream research topic in statistics and very few undergraduate and graduate programs offered courses in causal inference.  In the academic world of statistics, many people were still very skeptical about the foundation of causal inference.  Many leading statisticians were reluctant to accept causal inference because of the fundamental conceptual difficulties, which differ from the traditional training of mathematical statistics.

The applications of causal inference in empirical research have changed the field of statistics in both research and education. In the end, statistics is not only about abstract theory but also about solving real-world problems.  Many talented researchers have joined the effort to advance our knowledge of causal inference.  Many students are eager to learn state-of-the-art theory and methods in causal inference so that they are better equipped to solve problems from various fields.

Due to the needs of the students, my colleagues encouraged me to develop a course in causal inference. Initially, I taught a graduate-level course cross-listed under Political Science and Statistics, which was taught by my former colleague Jas Sekhon for many years at UC Berkeley. Later, I developed this course for both undergraduate and graduate students. At UC Berkeley, the course numbers for ``Causal Inference'' are Stat 156 and Stat 256, with undergraduate students in Stat 156 and graduate students in Stat 256.  Students in both sessions used the same lecture notes and attended the same lectures given by me and my teaching assistants, although they needed to finish different homework problems, reading assignments, and final projects.

Given the mixed levels of technical preparations of my students, the most challenging part of my teaching was to balance the interests of both undergraduate and graduate students. On the one hand, I wanted to present the materials in an intuitive way and only required the undergraduate students to have the basic knowledge of probability, statistics, linear regression, and logistic regression. On the other hand, I also wanted to introduce recent research topics and results to the graduate students. This book is a product of my efforts in the past seven years.

\subsection*{Recommendations for instructors}

This book contains 29 chapters in the main text and 3 chapters in the appendix.  UC Berkeley is on the semester system and each semester has 14 weeks of lectures. I could not finish all 32 chapters in one semester.  Here are some recommendations based on my own teaching experience.

\paragraph*{Appendix}
I started with the chapters in the main text but asked my teaching assistants to review the basics in Chapters \ref{chapter::basic-prob-append} and \ref{appendix::basic-linear-regression}.  To encourage the students to review Chapters \ref{chapter::basic-prob-append}--\ref{chapter::SRS-appendix} before reading the main text, I also assigned several homework problems from Chapters \ref{chapter::basic-prob-append}--\ref{chapter::SRS-appendix} at the beginning of the semester.

\paragraph*{Part \ref{part::introduction}}
The key topic in Chapter \ref{chapter::correlationassociation} is the Yule--Simpson Paradox.  Chapter \ref{chapter::potential-outcomes} introduces the notion of potential outcomes, which is the foundation for the whole book.

\paragraph*{Part \ref{part::rcts}}
Different researchers and instructors may have quite different views on the materials in Part \ref{part::rcts} on randomized experiments.  I have talked to many friends about the pros and cons of the current presentation in Part \ref{part::rcts}.  Causal inference in randomized experiments is relatively straightforward because randomization eliminates unmeasured confounding. So some friends feel that Chapters \ref{ch::frt-cre}--\ref{chapter::bridging} are too long at least for beginners of causal inference.  This also disappointed some of my students when I spent a month on randomized experiments. 
On the other hand,  I was trained from the book of \citet{imbens2015causal} and believed that to understand observational studies, it is better to understand randomized experiments first. Moreover, I am a big fan of the canonical research of \citet{Neyman:1923} and \citet{Fisher:1935}.  Therefore, Part \ref{part::rcts} deeply reflects my own intellectual history and personal taste in statistics. Other instructors may not want to spend a month on randomized experiments and can cover Chapters \ref{chapter::stratification-poststratification}, \ref{chapter::mpe}, \ref{chapter::unification-fisher-neyman} and \ref{chapter::bridging} quickly.

\paragraph*{Part \ref{part::observational-studies}}
Part \ref{part::observational-studies} covers the key ideas in observational studies without unmeasured confounding.  Four pillars of observational studies are outcome regression, inverse propensity score weighting, doubly robust, and matching estimators, which are covered in Chapters \ref{chapter::observational-studies}, \ref{chapter::pscore-key}, \ref{chapter::doubly-robust} and \ref{chapter::matching-obs}, respectively. 
Chapters \ref{chapter::ATT-and-other} and \ref{chapter::pscore-unification} are optional in teaching. But the results in Chapters \ref{chapter::ATT-and-other} and \ref{chapter::pscore-unification} are not uninteresting, so I sometimes covered one or two results there, asked the teaching assistants to cover more in the lab sessions, and encouraged students to read them by assigning some homework problems from those chapters.

\paragraph*{Part \ref{part::challenges-os}}
Part \ref{part::challenges-os} is a novel treatment of the fundamental difficulties of observational studies including unmeasured confounding and overlap. However,  this part is far from perfect due to the complexities and subtleties of the issues.  Chapters \ref{chapter::evalue}, \ref{chapter::sensitivity-ACE} and \ref{chapter::overlap} are central, whereas Chapters \ref{chapter::difficulties-observational-studies} and \ref{sec::rosenbaum-sensitivity-analysis} are optional.

\paragraph*{Part \ref{part::instrumentalvariables}}
Part \ref{part::instrumentalvariables} discusses the idea of the instrumental variable.  Chapters \ref{chapter::iv-experiment}, \ref{chapter::iv-econometric} and \ref{chapter::iv-frdd} are key, whereas Chapters \ref{chapter::iv-inequalities} and \ref{chapter::iv-mr} are optional.

\paragraph*{Part \ref{part::post-treatmentvariable}}
Part \ref{part::post-treatmentvariable} are some special topics.  They are all optional in some sense. Probably it is worth teaching Chapter \ref{chapter::mediation} given the popularity of the Baron--Kenny method in mediation analysis.

\paragraph*{Omitted topics}
This book does not cover some popular econometric methods including the difference in differences, panel data, and synthetic controls. Instructors can use \citet{angrist2008mostly} as a reference for those topics. 
This book assumes minimal preparation for the background knowledge in probability and statistics.  Since most introductory statistics courses use the frequentists' view that assumes the unknown parameters are fixed, I adopt this view in this book and omit the Bayesian view for causal inference. In fact, many fundamental ideas of causal inference are from the Bayesian view, starting from \citet{rubin1978bayesian}. If readers and students are interested in Bayesian causal inference, please read the review paper by \citet{li2023bayesian}.

\paragraph*{Help from teaching assistants}
My teaching assistants offered invaluable help for my courses at UC Berkeley. Since I could not cover everything in my notes, I consistently relied on them to cover some technical details or \ri{R} program sessions in their labs.

\paragraph*{Solution to some homework problems}
I have also prepared the solutions to most theory problems. If you are an instructor for a causal inference course, please contact me for the solutions with detailed information about your course.

\subsection*{Additional recommendations for readers and students}

Readers and students can first read my recommendations for instructors above.  In addition, I have two other recommendations.

\paragraph*{Homework problems}
Each chapter of this book contains homework problems. To deepen the understanding,  it is important to try some homework problems. Moreover, some homework problems contain useful theoretical results. Even if you do not have time to figure out the details for those problems,  it is helpful to at least read the statements of the problems. 

\paragraph*{Recommended reading}
Each chapter of this book contains recommended reading.  If you want to do research in causal inference, those recommended papers can be useful background knowledge of the literature.  When I taught the graduate-level causal inference course at UC Berkeley, I assigned the following papers to the students as weekly reading from week one to the end of the semester:  
\begin{itemize}
\item 
\citet{bickel1975sex}; 
\item 
\citet{holland1986statistics};
\item
\citet{miratrix:2013};
\item
\citet{lin2013};
\item 
\citet{li2018asymptotic};
\item 
\citet{rosenbaum1983central};
\item 
\citet{lunceford2004stratification};
\item
\citet{ding2016sensitivity};
\item
\citet{pearl1995causal};
\item
\citet{angrist1996identification};
\item
\citet{imbens2014instrumental};
\item
\citet{frangakis2002principal}. 
\end{itemize}
Many students gave me positive feedback about their experience of reading the papers above. 
I recommend reading the above papers even if you do not read this book.

\subsection*{Features of the book}

There are already many excellent causal inference books published in the last decade.  Some of them have profound influences on me. 
When I was in college, I read some draft chapters of \citet{imbens2015causal} from the internet. They completely challenged my way of thinking about statistics and helped to build my research interest in causal inference.  
I read \citet{angrist2008mostly} many times and have gained new insights each time I re-read it. 
\citet{rosenbaum2002design}, \citet{morgan2015counterfactuals}, and \citet{hernan2020causal} are another three excellent books from leading researchers in causal inference.
When I was preparing for the book,  \citet{cunningham2021causal}, \citet{huntington2022effect},  \citet{brumback2022fundamentals} and \citet{huber2023causal} appeared as four recent excellent books on causal inference.

Thanks to my teaching experience at UC Berkeley, this book has the following features that instructors, students, and readers may find attractive. 
\begin{itemize}
\item
This book assumes minimal preparation for causal inference and reviews the basic probability and statistics knowledge in the appendix.

\item
This book covers causal inference from the statistics,  biostatistics, and econometrics perspectives, and draws applications from various fields. 

\item
This book uses \ri{R} code and data analysis to illustrate the ideas of causal inference. All the \ri{R} code and datasets are publicly available at Harvard Dataverse: \url{https://doi.org/10.7910/DVN/ZX3VEV}

\item
This book contains homework problems and can be used as a textbook for both undergraduate and graduate students. Instructors can also ask me for solutions to some homework problems. 
\end{itemize}

\subsection*{Acknowledgments}

Professor Zhi Geng at Peking University introduced me to the area of causal inference when I was studying in college. Professors Luke Miratrix, Tirthankar Dasgupta, and Don Rubin served on my Ph.D. thesis committee at the Harvard Statistics Department. Professor Tyler VanderWeele supervised me as a postdoctoral researcher in Epidemiology at the Harvard T. H. Chan School of Public Health.

My colleagues at the Berkeley Statistics Department have created a critical and productive research environment.  Bin Yu and Jas Sekhon have been very supportive since I was a junior faculty.  My department chairs, Deb Nolan, Sandrine Dudoit, and Haiyan Huang, encouraged me to develop the ``Causal Inference'' course.  It has been a rewarding experience for me.

I have been lucky to work with many collaborators,  in particular, Avi Feller,  Laura Forastiere, Zhichao Jiang, Fang Han, Fan Li, Xinran Li, Alessandra Mattei, Fabrizia Mealli, Shu Yang, and Anqi Zhao.  I will report in this book what I have learned from them.

Many students at UC Berkeley made critical and constructive comments on early versions of my lecture notes.  As teaching assistants for my ``Causal Inference'' course,  Emily Flanagan and Sizhu Lu read early versions of my book carefully and helped me to improve the book a lot.

Professor Joe Blitzstein read an early version of the book carefully and made very detailed comments. Addressing his comments leads to significant improvement in the book.  Professors Hongyuan Cao and Zhichao Jiang taught ``Causal Inference'' courses based on an early version of the book. They made very valuable suggestions.

I am also very grateful for the suggestions from Young Woong Min, Fangzhou Su, Chaoran Yu, and Lo-Hua Yuan.

If you identify any errors, please feel free to email me.

\chapter*{Acronyms}

To simplify the writing, I will use the many acronyms in this book. The following table gives the acronyms, their full names, and the first chapters in which they  appear.

\begin{tabular}{lll}
\hline 
acronym & full name & first chapter \\
\hline 
ACE& average causal effect&  \ref{chapter::potential-outcomes} \\
AI & Abadie and Imbens (for matching estimators) & \ref{chapter::matching-obs} \\
ANCOVA & analysis of covariance & \ref{chapter:rerandomization-regression}\\
BMI & body mass index & \ref{chapter::potential-outcomes} \\
BRE & Bernoulli randomized experiment & \ref{ch::frt-cre}\\
CACE & complier average causal effect & \ref{chapter::iv-experiment} \\
CATE & conditional average causal effect & \ref{chapter::observational-studies}\\
CDE & controlled direct effect & \ref{chapter::cde} \\
CLT & central limit theorem & \ref{ch::frt-cre} and \ref{chapter::basic-prob-append} \\
CRE & completely randomized experiment & \ref{ch::frt-cre} \\
EHW& Eicker--Huber--White (robust standard error) & \ref{chapter::neyman-cr} and \ref{appendix::basic-linear-regression}\\
FAR & Fieller--Anderson--Rubin (confidence set) & \ref{chapter::iv-experiment}\\
FRT & Fisher randomization test & \ref{ch::frt-cre} \\
FWL & Frisch--Waugh--Lovell (theorem) & \ref{appendix::basic-linear-regression} \\
HT& Horvitz--Thompson (estimator) & \ref{chapter::pscore-key}\\
  IID & independent and identically distributed &  \ref{ch::frt-cre} and \ref{chapter::basic-prob-append} \\
  ILS & indirect least squares & \ref{chapter::iv-econometric} \\
   IPW& inverse propensity score weighting & \ref{chapter::pscore-key}\\
ITT & intention-to-treat (analysis) &  \ref{chapter::iv-experiment} \\
 IV & instrumental variable &  \ref{chapter::iv-experiment} \\
 LASSO & least absolute shrinkage and selection operator & \ref{chapter:rerandomization-regression}\\
   LATE & local average treatment effect & \ref{chapter::iv-experiment} \\
    MLE & maximum likelihood estimate & \ref{appendix::basic-linear-regression} \\
 MPE & matched-pairs experiment & \ref{chapter::mpe}\\
  MR & Mendelian randomization & \ref{chapter::iv-mr} \\
 MSM & marginal structural model & \ref{chapter::cde} \\
 NDE& natural direct effect & \ref{chapter::mediation} \\
 NHANES & National Health and Nutrition Examination Survey & \ref{chapter::observational-studies}\\
 NIE & natural indirect effect & \ref{chapter::mediation} \\
    OLS & ordinary least squares & \ref{chapter::neyman-cr} and \ref{appendix::basic-linear-regression}\\
OR & odds ratio & \ref{chapter::correlationassociation}\\
 RCT & randomized controlled trial & \ref{chapter::correlationassociation}\\
 RD & risk difference & \ref{chapter::correlationassociation} \\
 ReM & rerandomization using the Mahalanobis distance & \ref{chapter:rerandomization-regression}\\
RR & risk ratio or relative risk & \ref{chapter::correlationassociation}\\
SNP & single nucleotide polymorphism & \ref{chapter::iv-mr} \\
  SRE& stratified randomized experiment & \ref{chapter::stratification-poststratification}\\
 SUTVA & stable unit treatment value assumption & \ref{chapter::potential-outcomes} \\
 TSLS & two-stage least squares & \ref{chapter::iv-econometric} \\
   WLS & weighted least squares & \ref{chapter::pscore-unification} and \ref{appendix::basic-linear-regression}\\
\hline 
\end{tabular}

\chapter*{Notation}

I use the following conventional notation in this book.

\subsection*{Math}

\begin{tabular}{ll}
\hline 
$\binom{n}{m}$ & ``$n$ choose $m$'' which equals $\frac{n!}{m!(n-m)!}$ \\
$\sum$ & summation, e.g., $\sumn a_i = a_1 + \cdots + a_n$ \\
$I(\cdot)$ & indicator function, i.e.,  $I(A)=1$ if $A$ happens and 0 otherwise\\
$\#$ & counting the number of units in a set\\
$\approx$ & approximately equal \\
$\propto$ & proportional to (by dropping some unimportant constant) \\
$\text{logit}$ &  $\text{logit}(x) = \log \frac{x}{1-x}$ \\
$\expit$ & $\expit(x) =   \frac{e^x}{1+e^x} =  (1 + e^{-x})^{-1}$ \\
\hline 
\end{tabular}

\subsection*{Basic probability and statistics}

\begin{tabular}{ll}
\hline 
$\pr(\cdot)$ & probability \\
$E(\cdot)$ & expectation of a random variable\\
$\var(\cdot)$ & variance of a random variable \\
$\cov(\cdot)$ & covariance between random variables\\
$\asim$ & ``$A \asim B$'' means that $A$ and $B$ have the same asymptotic distribution\\
$\rho_{YX}$ & Pearson correlation coefficient between $Y$ and $X$ \\
$R^2_{YX}$ & squared multiple correlation coefficient between $Y$ and $X$ \\
\hline 
\end{tabular}

\subsection*{Random variables}

\begin{tabular}{ll}
\hline 
Bernoulli$(p)$ & Bernoulli distribution with probability $p$\\
Binomial$(n,p)$& Binomial distribution with $n$ trials and probability $p$\\
$\textsc{N}(\mu, \sigma^2)$ & Normal distribution with mean $\mu$ and variance $\sigma^2$\\
$t_\nu$ & $t$ distribution with degrees of freedom $\nu$ \\
$\chi^2_\nu$ & chi-squared distribution with degrees of freedom $\nu$\\
$z_{1-\alpha/2}$ & the $1-\alpha/2$ upper quantile of $\N01$, e.g., $z_{0.975}=1.96$\\
\hline 
\end{tabular}

\subsection*{Causal inference}

\begin{tabular}{ll}
\hline 
 $Y_i(1), Y_i(0)$ & potential outcomes of unit $i$ under treatment and control\\
$\tau$ & finite-population average causal effect $\tau = n^{-1}\sumn \{Y_i(1)-Y_i(0)\}$ \\
$\tau$ & super-population average causal effect $\tau = E \{Y(1)-Y(0)\}$\\
$\mu_1(X)$ & outcome model $\mu_1(X) = E(Y\mid Z=1, X)$\\
$\mu_0(X)$ & outcome model $\mu_0(X) = E(Y\mid Z=0, X)$\\
$e(X)$ & propensity score $e(X) = \pr(Z=1\mid X)$ \\
$\tau_\cp$ & complier average causal effect $ \tau_\cp = E\{ Y(1) - Y(0)\mid U=\cp \} $ \\
$U$ & unmeasured confounder \\
$U$ & latent compliance status $U=(D(1), D(0))$ \\
$Y(z,M_{z'})$ & nested potential outcome (for mediation analysis)\\
 \hline 
\end{tabular}

\mainmatter

\makeatletter
\@addtoreset{assumption}{chapter}
\@addtoreset{example}{chapter}
\@addtoreset{lemma}{chapter}
\@addtoreset{remark}{chapter}
\@addtoreset{theorem}{chapter}
\@addtoreset{corollary}{chapter}
\@addtoreset{proposition}{chapter}
\@addtoreset{definition}{chapter}
\makeatother

\renewcommand {\theexample} {\thechapter.\arabic{example}}
\renewcommand {\theassumption} {\thechapter.\arabic{assumption}}
\renewcommand {\thelemma} {\thechapter.\arabic{lemma}}
\renewcommand {\thetheorem} {\thechapter.\arabic{theorem}}
\renewcommand {\theremark} {\thechapter.\arabic{remark}}
\renewcommand {\thecondition} {\thechapter.\arabic{condition}}
\renewcommand {\thecorollary} {\thechapter.\arabic{corollary}}
\renewcommand {\theproposition} {\thechapter.\arabic{proposition}}
\renewcommand {\thedefinition} {\thechapter.\arabic{definition}}

\part{Introduction}\label{part::introduction}
 
\chapter{Correlation, Association, and the Yule--Simpson Paradox}
 \label{chapter::correlationassociation}

Causality is central to human knowledge. Two famous quotes from ancient Greeks are below.

\begin{quote}
``I would rather discover one causal law than be King of Persia.''

---  Democritus 
\end{quote}

\begin{quote}
``We do not have knowledge of a thing until we grasped its cause.''

--- Aristotle 
\end{quote}

However, the major part of classic statistics is about association rather than causation.  This chapter will review some basic association measures and point out their fundamental limitations.

\section{Traditional view of statistics}

A traditional view of statistics is to infer correlation or association among variables. Based on this view, there is no role for causal inference in statistics. Two famous aphorisms associated with this view are as follows:

\begin{itemize}
\item
``Correlation does not imply causation.''

\item
``You can not prove causality with statistics.'' 
\end{itemize}

This book has a very different view: statistics is crucial for understanding causality. The main focus of this book is to introduce the formal language for causal inference and develop statistical methods to estimate causal effects in randomized experiments and observational studies.

\section{Some commonly-used measures of association}

\subsection{Correlation and regression}
\label{section:correlatioandregressionintro}

The Pearson correlation coefficient between two random variables $Z$ and $Y$ is
$$
\rho_{ZY} = \frac{  \cov(Z, Y)   }{   \sqrt{   \var(Z) \var(Y)   }   } ,
$$
which measures the linear dependence of $Z$ and $Y.$

The linear regression of $Y$ on $Z$ is the model 
\begin{equation}
Y = \alpha + \beta Z + \varepsilon ,
\label{eq::short-regression}
\end{equation} 
 where $E(   \varepsilon  ) = 0$ and $E(   \varepsilon  Z ) = 0 $. We can show that the regression coefficient $\beta$ equals  
 $$
 \beta = \frac{  \cov(Z, Y)   }{   \var(Z)    }  = \rho_{ZY}   \sqrt{     \frac{  \var(Y)     }{ \var(Z)  }    }  . 
 $$
 So $\beta$ and $\rho_{ZY} $ always have the same sign.

We can also define multiple regression of $Y$ on $Z$ and $X$: 
\begin{equation}
 Y = \alpha  + \beta Z + \gamma  X + \varepsilon ,
 \label{eq::long-regression}
\end{equation}
 where $E( \varepsilon) = 0, E( \varepsilon Z) = 0$ and $ E( \varepsilon X) = 0$. 
We usually interpret $\beta$ as the ``effect'' of $Z$ on $Y$, {\it holding $X$ constant} or {\it conditioning on $X$} or {\it controlling for $X$}. Chapter \ref{appendix::basic-linear-regression} reviews the basics of linear regression.

More interestingly, the $\beta$'s in the above two regressions \eqref{eq::short-regression} and \eqref{eq::long-regression} can be different; they can even have different signs. The following \ri{R} code reanalyzed the LaLonde observational data used by \citet{hainmueller2012entropy}. The main question of interest is the ``causal effect'' of a job training program on earnings. The regression controlling for all covariates gives a  coefficient \ri{1067.5461} for \ri{treat}, whereas the regression not controlling for any covariates gives a coefficient \ri{-8506.4954} for \ri{treat}.

\begin{lstlisting}
> dat <- read.table("cps1re74.csv", header = TRUE)
> dat$u74 <- as.numeric(dat$re74==0)
> dat$u75 <- as.numeric(dat$re75==0)
> 
> ## linear regression on the outcome
> ## . means regression on all other variable in dat
> lmoutcome = lm(re78 ~ ., data = dat)
> round(summary(lmoutcome)$coef[2, ], 3)
  Estimate Std. Error    t value   Pr(>|t|) 
  1067.546    554.060      1.927      0.054 
> 
> lmoutcome = lm(re78 ~ treat, data = dat)
> round(summary(lmoutcome)$coef[2, ], 3)
  Estimate Std. Error    t value   Pr(>|t|) 
 -8506.495    712.766    -11.934      0.000 
\end{lstlisting}

 \subsection{Contingency tables}
 \label{sec::twobytwotables-withresume}
 
 We can represent the joint distribution of two binary variables $Z$ and $Y$ by a two-by-two contingency table. With $p_{zy} = \pr( Z=z, Y=y )$, we can summarize the joint distribution in the following table:

 \begin{center}
\begin{tabular}{lcc}
\hline 
        & $Y =1$ & $Y =0$\\\hline
$Z =1$ & $p_{11}$ & $p_{10}$\\ 
$Z =0$ & $p_{01}$ & $p_{00}$\\ 
\hline 
\end{tabular}
\end{center}

 Viewing $Z$ as the treatment or exposure and $Y$ as the outcome, we can define the risk difference as
\begin{eqnarray*}
\RD  &=& \pr(Y=1\mid Z=1) - \pr(Y=1\mid Z=0)  \\
&=& \frac{p_{11}}{ p_{11} + p_{10} } -  \frac{p_{01}}{ p_{01} + p_{00} } ,
\end{eqnarray*}
 the risk ratio as
\begin{eqnarray*}
 \RR &=&  \frac{ \pr(Y=1\mid Z=1)  }{    \pr(Y=1\mid Z=0)    }  \\
& =&   \frac{p_{11}}{ p_{11} + p_{10} }  \Big /  \frac{p_{01}}{ p_{01} + p_{00} } ,
\end{eqnarray*}
 and the odds ratio\footnote{In probability theory, the odds of an event is defined as the ratio of the probability that the event happens over the probability that the event does not happen.} as
\begin{eqnarray*}
 \OR &=& \frac{ \pr(Y=1\mid Z=1) /  \pr(Y=0\mid Z=1)  }{    \pr(Y=1\mid Z=0) /\pr(Y=0\mid Z=0)   }  \\
& =& \frac{      \frac{p_{11}}{ p_{11} + p_{10} } /  \frac{p_{10}}{ p_{11} + p_{10} }  }{  \frac{p_{01}}{ p_{01} + p_{00} } / \frac{p_{00}}{ p_{01} + p_{00} }     }  \\
 &=& \frac{ p_{11} p_{00}  }{  p_{10} p_{01} }.
\end{eqnarray*}

The terminology ``risk difference'', ``risk ratio'', and ``odds ratio'' comes from epidemiology. Because the outcomes in epidemiology are often diseases, it is natural to use the name ``risk'' for the probability of having diseases.

We have the following simple facts about these measures.
\begin{proposition}\label{prop::2x2-independence}
(1) The following statements are all equivalent\footnote{This book uses the notation $\ind$ to denote independence or conditional independence of random variables. The notation is due to \citet{dawid1979conditional}.}: $Z\ind Y $, $\RD = 0$, $\RR = 1$, and $\OR = 1$. 
(2) If $p_{zy}$'s are all positive, then $\RD >0$ is equivalent to $\RR >1$ and is also equivalent to $\OR > 1$. 
(3)
$\OR \approx \RR$ if  $\pr(Y=1\mid Z=1)$ and $ \pr(Y=1\mid Z=0)$ are small. 
\end{proposition}

I leave the proofs of statements (1) and (2) as Problem \ref{hw::2x2-equivalence}. Statement (3) is informal. The approximation holds because the odds $p/(1-p)$ is close to the probability $p$ for rare diseases with $p\approx 0$: by Taylor expansion $p/(1-p) = p + p^2 + \cdots\approx p $. In epidemiology, if the outcome represents the occurrence of a rare disease, then it is reasonable to assume that $\pr(Y=1\mid X=1)$ and $ \pr(Y=1\mid X=0)$ are small.

We can also define conditional versions of the \textsc{rd}, \textsc{rr}, and \textsc{or}  if the probabilities are replaced by the conditional probabilities given another variable $X$, i.e., $\pr(Y=1\mid Z=1, X=x) $ and $ \pr(Y=1\mid Z=0, X=x)  .$

With counts $n_{zy} = \#\{i:  Z_i=z, Y_i=y \}$, we can summarize the observed data in the following two-by-two table:
 \begin{center}
\begin{tabular}{lcc}
\hline 
        & $Y =1$ & $Y =0$\\\hline
$Z =1$ & $n_{11}$ & $n_{10}$\\ 
$Z =0$ & $n_{01}$ & $n_{00}$\\ 
\hline 
\end{tabular}
\end{center}
We can estimate  \textsc{rd}, \textsc{rr}, and \textsc{or}  by replacing the true probabilities by the sample proportions $\hat p_{zy} = n_{zy}  / n$, where $n$ is the total sample size (see Chapter \ref{section::asymptotic-inference-2x2}). In \ri{R}, functions \ri{fisher.test} performs an exact test and \ri{chisq.test} performs an asymptotic test for   $Z \ind Y$ based on a two-by-two table of observed data.

 \begin{example}\label{eg::bertrand2004emily-chapter1}
 \citet{bertrand2004emily} conducted a randomized experiment on resumes to study the effect of perceived race on callbacks for interviews. They randomly assigned Black-sounding or White-sounding names on fictitious resumes to help-wanted ads in Boston and Chicago newspapers. The following two-by-two table summarizes perceived race and callback:
 
 \begin{lstlisting} 
> resume = read.csv("resume.csv")
> Alltable = table(resume$race, resume$call)
> Alltable
       
           0    1
  black 2278  157
  white 2200  235
    \end{lstlisting}

The two rows have the same total count, so it is apparent that White-sounding names received more callbacks. Fisher's exact test below shows that this difference is statistically significant.    
     \begin{lstlisting} 
> fisher.test(Alltable)

	Fisher's Exact Test for Count Data

data:  Alltable
p-value = 4.759e-05
alternative hypothesis: true odds ratio is not equal to 1
95 percent confidence interval:
 1.249828 1.925573
sample estimates:
odds ratio 
  1.549732 
  \end{lstlisting} 
 
 \end{example}
 
\section{An example of the Yule--Simpson Paradox} 

\subsection{Data}
The classic kidney stone example is from \citet{charig1986comparison}, where $Z$ is the treatment with $1$ for an open surgical procedure and $0$ for a small puncture procedure, and $Y$ is the outcome with $1$ for success and $0$ for failure. The treatment and outcome data can be summarized in the following two-by-two table: 

 \begin{center}
\begin{tabular}{lcc}
\hline 
        & $Y =1$ & $Y =0$\\\hline
$Z =1$ & $273$ & $77$\\ 
$Z =0$ & $289$ & $61$\\ 
\hline 
\end{tabular}
\end{center}

The estimated \textsc{rd} is 
$$
\widehat{\RD} = \frac{273}{273 + 77} - \frac{289}{289 + 61} = 78\% - 83\% = -5\% <0. 
$$
Treatment 0 seems better, that is, the small puncture leads to a higher success rate compared to the open surgical procedure.

However, the data were not from a randomized controlled trial (RCT)\footnote{In an RCT, patients are randomly assigned to the treatment arms. Part \ref{part::rcts} of this book will focus on RCTs.}. Patients receiving treatment $1$ can be very different from patients receiving treatment $0$. A ``lurking variable'' in this study is the severity of the case: some patients have smaller stones and some patients have larger stones. We can split the data according to the size of the stones.

For patients with smaller stones, the treatment and outcome data can be summarized in the following two-by-two table: 
 \begin{center}
\begin{tabular}{lcc}
\hline 
        & $Y =1$ & $Y =0$\\\hline
$Z =1$ & $81$ & $6$\\ 
$Z =0$ & $234$ & $36$\\
\hline 
\end{tabular}
\end{center}
For patients with larger stones, the treatment and outcome data can be summarized in the following two-by-two  table:  
 \begin{center}
\begin{tabular}{lcc}
\hline 
        & $Y =1$ & $Y =0$\\\hline
$Z =1$ & $192$ & $71$\\ 
$Z =0$ & $55$ & $25$\\
\hline 
\end{tabular}
\end{center}

The latter two tables must add up to the first table:
$$
81 + 192 = 273,\quad 6 + 71 =77,\quad   234 + 55 = 289, \quad 36 + 25 = 61 .  
$$
From the table for patients with smaller stones, the estimated \textsc{rd} is
$$
\widehat{\RD}_\text{smaller} = \frac{81}{81 + 6} - \frac{234}{234 + 36} = 93\% - 87\% = 6\% > 0,
$$
suggesting that treatment 1 is better. From the table for patients with larger stones, the estimated \textsc{rd}  is 
$$
\widehat{\RD}_\text{larger} = \frac{192}{192 + 71} - \frac{55}{55+25} = 73\% - 69\% = 4\% > 0 ,
$$
also suggesting that treatment 1 is better.

The above data analysis leads to
$$
\widehat{\RD} < 0,\quad
\widehat{\RD}_\text{smaller} >0, \quad
\widehat{\RD}_\text{larger}  > 0 .
$$
Informally, treatment 1 is better for both patients with smaller and larger stones, but treatment 1 is worse for the whole population. This interpretation is quite confusing if the goal is to infer the treatment effect. In statistics, this is called the Yule--Simpson Paradox or Simpson's Paradox in which the marginal association has the opposite sign to the conditional associations at all levels.

\subsection{Explanation}

Let $X$ be the binary indicator with $X=1$ for smaller stones and $X=0$ for larger stones. 
Let us first take a look at the $X$--$Z$ relationship by comparing the probabilities of receiving treatment 1 among patients with smaller and larger stones:
\begin{eqnarray*}
&& \widehat{\pr}(Z=1 \mid  X=1) - \widehat{\pr}(Z=1 \mid X=0) \\
&=& \frac{81 + 6}{81 + 6 + 234 + 36} - \frac{192+71}{192 + 71 + 55 + 25} \\
&=&  24\% - 77\% \\
&=& - 53\% < 0 .
\end{eqnarray*}
So patients with larger stones tend to take treatment 1 more frequently. Statistically, $X$ and $Z$ have negative association.

Let us then take a look at the $X$--$Y$ relationship by comparing the probabilities of success among patients with smaller and larger stones: under treatment 1,
\begin{eqnarray*}
&& \widehat{\pr}(Y=1 \mid Z=1,  X=1) - \widehat{\pr}(Y=1 \mid Z=1,  X=0) \\
&=& \frac{81}{81 + 6} - \frac{192}{192 + 71} \\
&=& 93\% - 73\% \\
&=& 20\%
> 0;
\end{eqnarray*}
under treatment 0, 
\begin{eqnarray*}
&& \widehat{\pr}(Y=1 \mid Z=0, X=1) - \widehat{\pr}(Y=1 \mid Z=0,   X=0) \\
&=& \frac{234}{234 + 36} - \frac{55}{55 + 25} \\
&=& 87\% - 69\% \\
&=& 18\% 
> 0.
\end{eqnarray*}
So under both treatment levels, patients with smaller stones have higher success probabilities. Statistically, $X$ and $Y$ have positive association conditional on both treatment levels.

\begin{figure}
$$
\xymatrix{
 & X\ar_{-}[ld]\ar^{+}[rd] \\ 
Z \ar_{+}[rr] &&Y 
}
$$
\caption{A diagram for the kidney stone example. The signs indicate the associations of two variables, conditioning on other variables pointing to the downstream variable.}\label{fig::kidney-stone}
\end{figure}

We can summarize the qualitative associations in the diagram in Figure \ref{fig::kidney-stone}. In technical terms, the treatment has a positive path $(Z\rightarrow Y)$ and a more negative path $(Z\leftarrow X \rightarrow Y)$ to the outcome, so the overall association is negative between the treatment and outcome. 
In plain English, when less effective treatment 0 is applied more frequently to the less severe cases, it can appear to be a more effective treatment.

In general, the association between $Z$ and $Y$ can differ qualitatively from the conditional association between $Z$ and $Y$ given $X$ due to the association between $X$ and $Z$ and the association between $X$ and $Y$.  When the Yule--Simpson Paradox happens,  we say that $X$ is a {\it confounding variable} or {\it confounder} for the relationship between $Z$ and $Y$, or $Z$ and $Y$ are {\it confounded} by $X$.  See Chapters \ref{chapter::observational-studies} and \ref{chapter::evalue} for more in-depth discussion.

\subsection{Geometry of the Yule--Simpson Paradox}

Assume that the two-by-two table based on the aggregated data has counts

 \begin{center}
\begin{tabular}{lcc}
\hline 
 whole population       & $Y =1$ & $Y =0$\\\hline
$Z =1$ & $n_{11}$ & $n_{10}$\\ 
$Z =0$ & $n_{01}$ & $n_{00}$\\
\hline 
\end{tabular}
\end{center}

The two two-by-two tables based on subgroups have counts
 \begin{center}
\begin{tabular}{lcc}
\hline 
subpopulation $X=1$        & $Y =1$ & $Y =0$\\\hline
$Z=1$ & $n_{11|1}$ & $n_{10|1}$\\ 
$Z=0$ & $n_{01|1}$ & $n_{00|1}$\\
\hline 
\end{tabular}
\end{center}
for the subgroup with $X=1$ and
 \begin{center}
\begin{tabular}{lcc}
\hline 
subpopulation    $X=0$    & $Y =1$ & $Y =0$\\\hline
$Z =1$ & $n_{11|0}$ & $n_{10|0}$\\ 
$Z =0$ & $n_{01|0}$ & $n_{00|0}$\\
\hline 
\end{tabular}
\end{center}
for the subgroup with $X=0.$

\begin{figure}
\includegraphics[width=\textwidth]{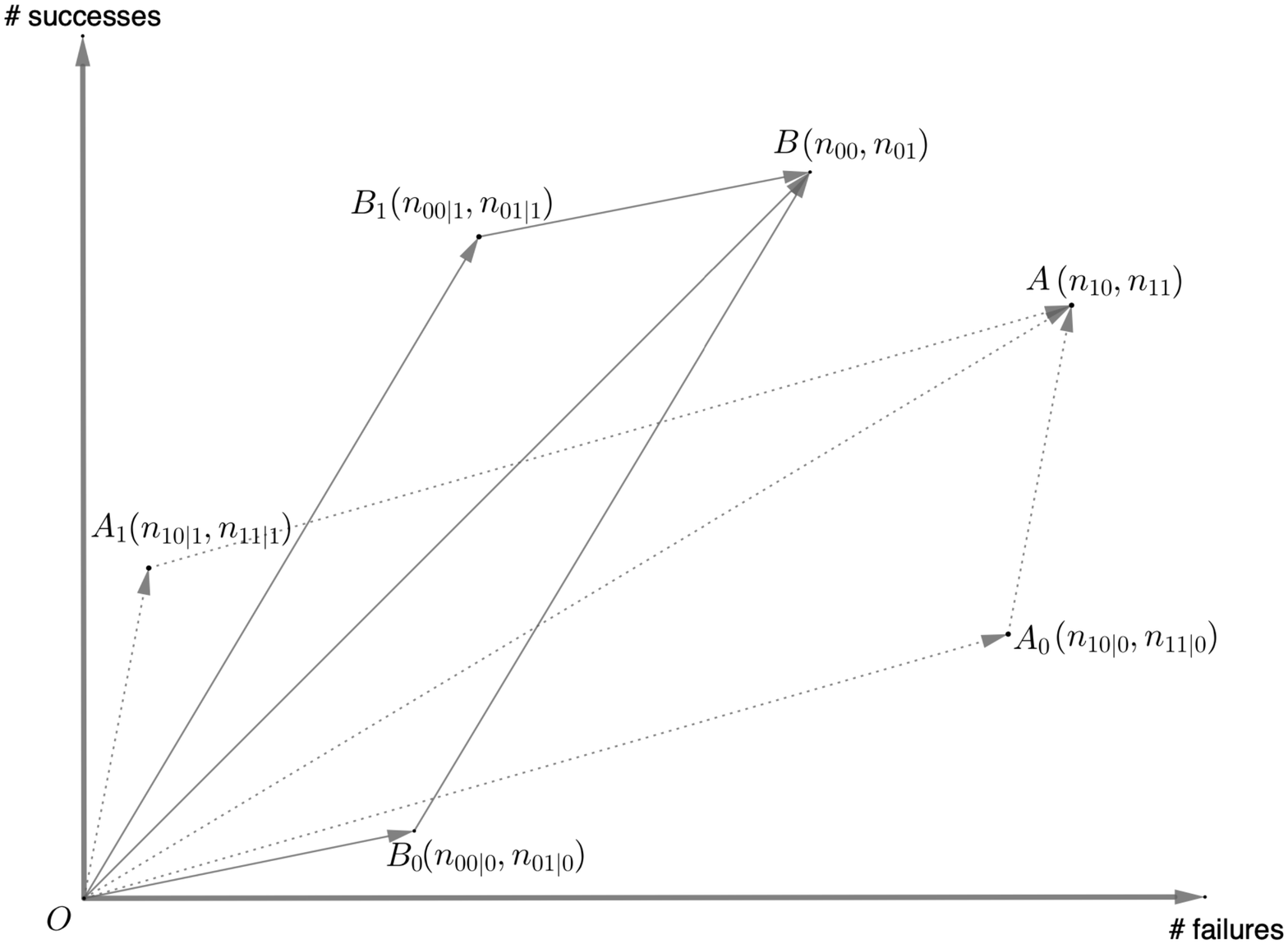}
\caption{Geometry of the Yule--Simpson Paradox}\label{fig::geomtry-simpson}
\end{figure}

Figure \ref{fig::geomtry-simpson} shows the geometry of the Yule--Simpson Paradox. The y-axis shows the count of successes with $Y=1$ and the x-axis shows the count of failures with $Y=0$. The two parallelograms correspond to aggregating the counts of successes and failures under two treatment levels. The slope of  $OA_1$ is larger than that of $OB_1$, and the slope of  $OA_0$ is larger than that of $OB_0$. So the treatment seems beneficial to the outcome within both levels of $X$. However, the slope of $OA$ is smaller than that of $OB$. So the treatment seems harmful to the outcome for the whole population. The Yule--Simpson Paradox arises.

\section{The Berkeley graduate school admission data}
\label{sec::berkeleydata}

\citet{bickel1975sex} investigated the admission rates of male and female students into the graduate school of Berkeley. The \ri{R} package \ri{datasets} contains the original data \ri{UCBAdmissions}. The raw data by the six largest departments are shown below: 
\begin{lstlisting}
> library(datasets)
> UCBAdmissions = aperm(UCBAdmissions, c(2, 1, 3))
> UCBAdmissions
, , Dept = A

        Admit
Gender   Admitted Rejected
  Male        512      313
  Female       89       19

, , Dept = B

        Admit
Gender   Admitted Rejected
  Male        353      207
  Female       17        8

, , Dept = C

        Admit
Gender   Admitted Rejected
  Male        120      205
  Female      202      391

, , Dept = D

        Admit
Gender   Admitted Rejected
  Male        138      279
  Female      131      244

, , Dept = E

        Admit
Gender   Admitted Rejected
  Male         53      138
  Female       94      299

, , Dept = F

        Admit
Gender   Admitted Rejected
  Male         22      351
  Female       24      317
\end{lstlisting}

Aggregating the data over departments, we have a simple two-by-two table:
\begin{lstlisting} 
> UCBAdmissions.sum = apply(UCBAdmissions, c(1, 2), sum)
> UCBAdmissions.sum
        Admit
Gender   Admitted Rejected
  Male       1198     1493
  Female      557     1278
\end{lstlisting}

The following function, building upon \ri{chisq.test}, have a two-by-two table as the input and the estimated \textsc{rd} and $p$-value as output:
\begin{lstlisting}
> risk.difference = function(tb2)
+ {
+   p1      = tb2[1, 1]/(tb2[1, 1] + tb2[1, 2])
+   p2      = tb2[2, 1]/(tb2[2, 1] + tb2[2, 2])
+   testp   = chisq.test(tb2)
+   
+   return(list(p.diff = p1 - p2,  
+               pv = testp$p.value))	    
+ }
\end{lstlisting}
With this function, we find a large, significant difference between the admission rates of male and female students:
\begin{lstlisting}
> risk.difference(UCBAdmissions.sum)
$p.diff
[1] 0.1416454

$pv
[1] 1.055797e-21
\end{lstlisting}

Stratifying on the departments, we find smaller and insignificant differences between the admission rates of male and female students. In department A, the difference is significant but negative. 
\begin{lstlisting}
> P.diff = rep(0, 6)
> PV     = rep(0, 6)
> for(dd in 1:6)
+ {
+ 	 department = risk.difference(UCBAdmissions[, , dd])
+ 	 P.diff[dd] = department$p.diff
+ 	 PV[dd]     = department$pv 
+ }
> 
> round(P.diff, 2)
[1] -0.20 -0.05  0.03 -0.02  0.04 -0.01
> round(PV, 2)
[1] 0.00 0.77 0.43 0.64 0.37 0.64
\end{lstlisting}

\section{Homework Problems}

\paragraph{Independence in two-by-two tables}\label{hw::2x2-equivalence}

Prove (1) and (2) in Proposition \ref{prop::2x2-independence}.

\paragraph{More examples of the Yule--Simpson Paradox}

Give a numeric example of a two-by-two-by-two table in which the Yule--Simpson Paradox arises. 

Find a real-life example in which the Yule--Simpson Paradox arises.

\paragraph{Correlation and partial correlation}\label{hw::partial-correlation}

Consider a three-dimensional Normal random vector:
$$
\begin{pmatrix}
X\\
Y\\
Z
\end{pmatrix}
\sim 
\textsc{N}
\left(
\begin{pmatrix}
0\\
0\\
0
\end{pmatrix},
\begin{pmatrix}
1 & \rho_{XY} &\rho_{XZ} \\
  \rho_{XY}  & 1 & \rho_{YZ} \\
  \rho_{XZ} &  \rho_{YZ} & 1
\end{pmatrix}
\right) . 
$$
The correlation coefficient between $Y$ and $Z$ is $\rho_{YZ}$. 
There are many equivalent definitions of the partial correlation coefficient. 
For a multivariate Normal vector, let $\rho_{YZ|X}$ denote the partial correlation coefficient between $Y$ and $Z$ given $X$, which is defined as their correlation coefficient in the conditional distribution $(Y, Z)\mid X$. Show that
$$
\rho_{YZ\mid X} = \frac{   \rho_{YZ} - \rho_{YX} \rho_{ZX}  }{  \sqrt{  1-\rho_{YX}^2 }  \sqrt{1- \rho_{ZX}^2  }  }  . 
$$

Give a numerical example with $\rho_{YZ} > 0$ and $\rho_{YZ|X} < 0$. 

Remark: This is the Yule--Simpson Paradox for a Normal random vector. 
You can use the results in Chapter \ref{sec::multivariate-normal} to prove the formula for the partial correlation coefficient.

\paragraph{Specification searches}\label{hw::lalonde-obs-reg}

Section \ref{section:correlatioandregressionintro} re-analyzes the data used by \citet{hainmueller2012entropy}. We used the outcome named \ri{re78} and the treatment named \ri{treat} in the analysis. Moreover, the data also contain $10$ covariates and therefore $2^{10} = 1024$ possible subsets of covariates in the linear regression. Run $1024$ linear regressions with all possible subsets of covariates, and report the regression coefficients of the treatment. How many coefficients of the treatment are positively significant, how many are negatively significant, and how many are not significant? You can also report other interesting findings from these regressions.

\paragraph{More on racial discrimination}

Section \ref{sec::twobytwotables-withresume} re-analyzes the data collected by \citet{bertrand2004emily}. Conduct analyses separately for males and females. What do you find from these subgroup analyses?

\paragraph{Recommended reading}

\citet{bickel1975sex} is the original paper for the paradox reported in Section \ref{sec::berkeleydata}. 
\citet{pearl2018book} recently revisited the study from the causal inference perspective.

  \chapter{Potential Outcomes}
 \label{chapter::potential-outcomes}

\section{Experimentalists' view of causal inference}

\citet{rubin1975bayesian} and \citet{holland1986statistics} made up the aphorism:
\begin{quote}
``no causation without manipulation.''
\end{quote}
Not everybody agrees with this point of view. However, it is quite helpful to clarify ambiguity in thinking about causal relationships. This book follows this view and defines causal effects using the potential outcomes framework \citep{Neyman:1923, Rubin:1974}. In this framework, an experiment, or at least a thought experiment, has a treatment, and we are interested in its effect on an outcome or multiple outcomes. Sometimes, the {\it treatment} is also called an {\it intervention} or a {\it manipulation}.

\begin{example}\label{eg::aspirin}
If we are interested in the effect of taking aspirin or not on the relief of headaches, the intervention is taking aspirin or not. 
\end{example}

\begin{example}\label{eg::jobtraining}
If we are interested in the effect of participating in a job training program or not on employment and wage, the intervention is participating in the job training program or not. 
\end{example}

\begin{example}\label{eg::classroom}
If we are interested in the effect of studying in a small classroom or a large classroom on standardized test scores, the intervention is studying in a small classroom or not. 
\end{example} 

\begin{example}\label{eg::gotv}
\citet{gerber2008social} were interested in the effect of different get-out-to-vote messages on the voting behavior. The intervention is the different get-out-to-vote messages. 
\end{example}

\begin{example}\label{eg::bmi}
\citet{pearl2018does} claimed that we could infer the effect of obesity on life span. 
A popular measure of obesity is the body mass index (BMI), defined as the body mass divided by the square of the body height in units of kg/m$^2$. So the intervention can be different levels of BMI. 
\end{example}

However, there are different levels of ambiguity in the interventions above. The meanings of interventions in Examples \ref{eg::aspirin}--\ref{eg::gotv} are relatively clear, but the meaning of intervention on BMI in Example \ref{eg::bmi} is less clear. In particular, we can imagine different versions of BMI reduction: healthier diet, more physical exercise, bariatric surgery, etc. These different versions of the intervention can have quite different effects on the outcome. In this book, we will view the intervention in Example \ref{eg::bmi} as ill-defined without further clarifications.

Another ill-defined intervention is race. Racial discrimination is an important issue in the labor market, but it is not easy to imagine an experiment to change the race of any experimental unit. \citet{bertrand2004emily} give an interesting experiment that partially answers the question. 

\begin{example}\label{eg::race}
Recall Example \ref{eg::bertrand2004emily-chapter1}. 
\citet{bertrand2004emily} randomly change the names on the resumes, and compare the callback rates of resumes with African-American- or White-sounding names. For each resume, the intervention is the binary indicator of a Black-sounding or White-sounding name, and the outcome is the binary indicator of callback. I analyzed the following two-by-two table in Section \ref{sec::twobytwotables-withresume}: 

\begin{center} 
\begin{tabular}{ccc}
\hline 
 &   callback & no  callback \\
 \hline 
Black  & 157 & 2278\\
White   & 235 & 2200\\
\hline 
\end{tabular}
\end{center}

From the above, we can compare the probabilities of being called back among  Black-sounding and White-sounding names:
$$
\frac{157}{2278 + 157} - \frac{235}{  2200 + 235 } = 6.45\% - 9.65\% = - 3.20\% < 0
$$
with $p$-value from the Fisher exact test being much smaller than $0.001$. 
\end{example}

In \citet{bertrand2004emily}'s experiment, the treatment is the {\it perceived race} which can be manipulated by experimenters. They design an experiment to answer a well-defined causal question.  However, critics may raise the concern that the causal effect of the perceived race may differ from the causal effect of the actual race.

\section{Formal notation of potential outcomes}
\label{sec::po-sciencetable}

Consider a study with $n$ experimental units indexed by $i= 1, \ldots, n$. As a starting point, we focus on a treatment with two levels: $1$ for the treatment and $0$ for the control. For each unit $i$, the outcome of interest $Y$ has two versions: 
$$
Y_i(1)  \text{ and } Y_i(0),
$$ 
which are potential outcomes under the hypothetical interventions 1 and 0. 
\citet{Neyman:1923} first used this notation. It seems intuitive but has some hidden assumptions. \citet{Rubin:1980} made the following clarifications on the hidden assumptions.

\begin{assumption}
[no interference]
\label{sutva1}
Unit $i$'s potential outcomes do not depend on other units' treatments. This is sometimes called the {\it no-interference} assumption. 
\end{assumption}

\begin{assumption}
[consistency]
\label{sutva2}
There are no other versions of the treatment. Equivalently, we require that the treatment levels be well-defined, or have no ambiguity at least for the outcome of interest. This is sometimes called the {\it consistency} assumption.\footnote{This notion of consistency is totally different from the one in Definition \ref{def::consistency} in Chapter \ref{chapter::basic-prob-append}.} 
\end{assumption}

%
%

Assumption \ref{sutva1} can be violated in infectious diseases or network experiments. For instance, if some of my friends receive flu shots, my chance of getting the flu decreases even if I do not receive the flu shot; if my friends see an advertisement on Facebook, my chance of buying that product increases even if I do not see the advertisement directly. It is an active research area to study situations with interfering units in modern causal inference literature \citep[e.g.,][]{hudgens2008toward}.

Assumption \ref{sutva2} can be violated for treatments with complex components. For instance, when studying the effect of cigarette smoking on lung cancer, the type of cigarettes may matter; when studying the effect of college education on income, the type and major of college education may matter.

\citet{Rubin:1980} called the Assumptions \ref{sutva1} and \ref{sutva2} above together the Stable Unit Treatment Value Assumption (SUTVA).

\begin{assumption}
[SUTVA]
\label{sutva-complete}
Both Assumptions \ref{sutva1} and \ref{sutva2} hold.
\end{assumption}

Under SUTVA, \citet{Rubin:2005} called the $n\times 2$ matrix of potential outcomes the Science Table:
\begin{center}
\begin{tabular}{ccc}
\hline 
$i$ & $Y_i(1)$ &  $Y_i(0)$ \\
 \hline 
$1$ & $Y_1(1)$ &  $Y_1(0)$ \\
$2$ & $Y_2(1)$ &  $Y_2(0)$ \\
\vdots  &\vdots & \vdots   \\
$n$ & $Y_n(1)$ &  $Y_n(0)$ \\
\hline 
\end{tabular}
\end{center}

Due to the fundamental contributions of Neyman and Rubin to statistical causal inference, the potential outcomes framework is sometimes referred to as the Neyman Model, the Neyman--Rubin Model, or the Rubin Causal Model. I prefer using  ``the potential outcomes framework'' in this book and my other scientific writings.

Causal effects are functions of the Science Table. Inferring individual causal effects 
$$
\tau_i =  Y_i(1) - Y_i(0) ,\quad (i=1, \ldots, n)
$$ 
is fundamentally challenging because we can only observe either $Y_i(1)$ or $Y_i(0)$ for each unit $i$, that is, we can observe only half of the Science Table. As a starting point, most parts of the book focus on the average causal effect (ACE): 
\begin{eqnarray*}
\tau  &=&  n^{-1} \sumn \{  Y_i(1) - Y_i(0) \}  \\
&=&  n^{-1} \sumn  Y_i(1) -  n^{-1} \sumn  Y_i(0).
\end{eqnarray*} 
But we can extend our discussion to many other parameters (also called {\it estimands}).  Problem \ref{hw::nonlinear-causal-estimands} gives some examples.

\subsection{Causal effects, subgroups, and the non-existence of Yule--Simpson Paradox}

If we have two subgroups defined by a binary variable $X_i $, we can define the subgroup causal effects as
$$
\tau_x = \frac{  \sumn  I(X_i = x) \{  Y_i(1) - Y_i(0) \}    }{   \sumn  I(X_i = x)    },\quad (x=0,1) 
$$
where $I(\cdot)$ is the indicator function. 
A simple identity is that
$$
\tau =  \pi_1 \tau_1  +  \pi_0 \tau_0
$$
where $\pi_x =  \sumn  I(X_i = x)  / n$ is the proportion of units with $X_i = x$ $(x=0,1)$. Therefore, if $\tau_1  > 0$ and $ \tau_0 > 0$, we must have $\tau > 0$. The Yule--Simpson Paradox thus cannot happen to causal effects.

\subsection{Subtlety of the definition of the experimental unit}

I now discuss a subtlety related to the definition of the experimental unit. Simply speaking, the experimental unit can be different from the physical unit. For example, if I did not take aspirin before and my headache did not go away, but I take aspirin now and my headache goes away, then you might think that we can observe my potential outcomes under both control and treatment. Let $i$ index myself, and let $Y=1$ denote the indicator of no headache. Then, the above heuristic suggests that $Y_i(0) = 0$ and $Y_i(1) = 1$, so it seems that aspirin kills my headache. But this logic is very wrong because of the misunderstanding of the definition of the experimental unit. At different time points, I, the same physical person, become two distinct experimental units, indexed by ``$i,\textup{before}$'' and ``$i,\textup{after}$''. Therefore, we have four potential outcomes
$$
Y_{i,\textup{before}}(0)=0, \quad  Y_{i,\textup{before}}(1)=?,\quad
Y_{i,\textup{after}}(0)=?, \quad  Y_{i,\textup{after}}(1)=1,
$$ 
with two of them observed and two of them missing. The individual causal effects 
$$
Y_{i,\textup{before}}(1) -  Y_{i,\textup{before}}(0) = ? - 0 \text{ and } Y_{i,\textup{after}}(1)  - Y_{i,\textup{after}}(0) = 1 - ?
$$ 
are unknown. It is possible that my headache goes away even if I do not take aspirin:
$$
Y_{i,\textup{after}}(0)=1, \quad  Y_{i,\textup{after}}(1)=1
$$
which implies zero effect; it is also possible that my headache does not go away if I do not take aspirin:
$$
Y_{i,\textup{after}}(0)=0, \quad  Y_{i,\textup{after}}(1)=1
$$
which implies a positive effect of aspirin. 

The wrong heuristic argument might get the right answer if the control potential outcomes are stable at the before and after periods: 
$$
Y_{i,\textup{before}}(0) = Y_{i,\textup{after}}(0) = 0 .
$$ 
But this assumption is rather strong and fundamentally untestable. \citet{rubin2001comment} offered a related discussion in the context of self-experimentation for causal effects.

\section{Treatment assignment mechanism}\label{sec::Treatment-ScienceTable}

Let $Z_i$ be the binary treatment indicator for unit $i$, vectorized as $\bm{Z} = (Z_1, \ldots, Z_n)$. The observed outcome of unit $i$ is a function of the potential outcomes and the treatment indicator: 
\begin{eqnarray} 
Y_i &=& \begin{cases}
   Y_i(1),& \text{if } Z_i = 1\\
   Y_i(0), & \text{if } Z_i = 0
\end{cases} \label{eq::yobsdef} \\
&=& Z_i Y_i(1) + (1-Z_i)Y_i(0) \label{eq::fundamentalbridge} \\
&=& Y_i(0)  + Z_i\{  Y_i(1) - Y_i(0)  \} \label{eq::teh} \\
&=&Y_i(0)  + Z_i\tau_i . \label{eq::teh2} 
\end{eqnarray}  
Equation \eqref{eq::yobsdef} is the definition of the observed outcome. Equation \eqref{eq::fundamentalbridge} is equivalent to \eqref{eq::yobsdef}. It is a trivial fact, but \citet{pearl2010brief} viewed it as the fundamental bridge between the potential outcomes and the observed outcome. Equations \eqref{eq::teh} and  \eqref{eq::teh2}  highlight the fact that the individual causal effect $\tau_i =  Y_i(1) - Y_i(0)  $ can be heterogeneous across units.

The experiment reveals only one of unit $i$'s potential outcomes with the other one missing:
\begin{eqnarray*} 
Y_i^\textup{mis} &=& \begin{cases}
   Y_i(0),& \text{if } Z_i = 1\\
   Y_i(1), & \text{if } Z_i = 0
\end{cases}  \\
&=& Z_i Y_i(0) + (1-Z_i)Y_i(1) .
\end{eqnarray*}  
The missing potential outcome corresponds to the opposite treatment level of unit $i$. For this reason, the potential outcomes framework is also called the {\it counterfactual} framework.  \footnote{The terminology ``counterfactual'' is more popular in philosophy, probably due to \citet{lewis1973causation}. }
This name can be confusing because before the experiment, both the potential outcomes can be observed. Only after the experiment, one potential outcome is observed whereas the other potential outcome is counterfactual.

The treatment assignment mechanism, i.e., the probability distribution of $\bm{Z} $, plays an important role in inferring causal effects. The following simple numerical examples illustrate this point. We first generate potential outcomes from Normal distributions with the average causal effect close to $-0.5$. 
\begin{lstlisting}
> n  = 500
> Y0 = rnorm(n)
> tau = - 0.5 + Y0
> Y1 = Y0 + tau
\end{lstlisting} 
A perfect doctor assigns the treatment to the patient if s/he knows that the individual causal effect is non-negative. This results in a positive difference in means of the observed outcomes: 
\begin{lstlisting}
> Z = (tau >= 0)
> Y = Z*Y1 + (1 - Z)*Y0
> mean(Y[Z==1]) - mean(Y[Z==0])
[1] 2.166509
\end{lstlisting} 
A clueless doctor does not know any information about the individual causal effects and assigns the treatment to patients by flipping a fair coin. This results in a difference in means of the observed outcomes close to the true average causal effect: 
\begin{lstlisting}
> Z = rbinom(n, 1, 0.5)
> Y = Z*Y1 + (1 - Z)*Y0
> mean(Y[Z==1]) - mean(Y[Z==0])
[1] -0.552064 
  \end{lstlisting} 
  
 The above examples are hypothetical since no doctors perfectly know the individual causal effects. However, the examples do demonstrate the crucial role of the treatment assignment mechanism. This book will organize the topics based on the treatment assignment mechanism.

\section{Homework Problems}

\paragraph{A perfect doctor}
\label{hw::perfect-doctor}

Following the first perfect doctor example in Section \ref{sec::Treatment-ScienceTable}, assume the potential outcomes are random variables generated from 
$$
Y(0) \sim \textsc{N}(0,1), \quad \tau = -0.5 + Y(0),\quad Y(1)  = Y(0) + \tau .
$$
The binary treatment is determined by the treatment effect as $Z = 1(\tau \geq 0)$, and the observed outcome is determined by the potential outcomes and the treatment by $Y = ZY(1) + (1-Z) Y(0)$. Calculate the difference in means
$$
E(Y\mid Z=1) - E(Y\mid Z=0).
$$

Remark: The mean of a truncated Normal random variable equals 
$$
E(X\mid   a < X < b ) = \mu - \sigma \frac{   \phi \left(  \frac{b-\mu}{\sigma} \right )   -  \phi \left(  \frac{a-\mu}{\sigma} \right )   }
{   \Phi \left(  \frac{b-\mu}{\sigma} \right )   -  \Phi \left(  \frac{a-\mu}{\sigma} \right )   } , 
$$
where $X \sim \textsc{N}(\mu, \sigma^2)$, and $\phi(\cdot)$ and $\Phi(\cdot)$ are the probability density and cumulative distribution functions of a standard Normal random variable $\textsc{N}(0,1)$.

\paragraph{Nonlinear causal estimands}
\label{hw::nonlinear-causal-estimands}

With potential outcomes $\{  (Y_i(1), Y_i(0) \}_{i=1}^n$ for $n$ units under the treatment and control, the difference in means equals the mean of the individual treatment effects:
$$
\bar{Y}(1) - \bar{Y}(0) = n^{-1} \sum_{i=1}^n  \{ Y_i(1) - Y_i(0) \}.
$$
Therefore, the average treatment effect is a {\it linear} causal estimand in the sense that the difference in average potential outcomes equals the average of the differences in individual potential outcomes.

Other estimands may not be linear. For instance, we can define the median treatment effect as
$$
\delta_1 = \text{median}\{  (Y_i(1) \}_{i=1}^n - \text{median}\{  (Y_i(0) \}_{i=1}^n,
$$
which is, in general, different from the median of the individual treatment effect
$$
\delta_2 = \text{median}\{  (Y_i(1) - Y_i(0) \}_{i=1}^n .
$$

\begin{enumerate}
\item
Give numerical examples that have $\delta_1 = \delta_2$, $\delta_1 > \delta_2$, and $\delta_1 < \delta_2$, respectively.

\item
Which estimand makes more sense, $\delta_1$ or $\delta_2$? Why? Use examples to justify your conclusion. If you feel that both $\delta_1$ and $\delta_2$ can make sense in different applications, you can also give examples to justify both estimands. 
\end{enumerate}

\paragraph{Average and individual effects}

Give a numerical example in which $\tau = n^{-1} \sum_{i=1}^n  \{ Y_i(1) - Y_i(0) \}  >0$ but the proportion of units with $Y_i(1) > Y_i(0)$ is smaller than $0.5$. That is, the average causal effect is positive, but the treatment benefits less than half of the units.

\paragraph{Recommended reading}

\citet{holland1986statistics} is a classic review article on statistical causal inference. It popularized the name ``Rubin Causal Model'' for the potential outcomes framework. At the University of California Berkeley, we call it the ``Neyman Model'' for obvious reasons.

\part{Randomized experiments}\label{part::rcts}
  
\chapter{The Completely Randomized Experiment and the Fisher Randomization Test}\label{ch::frt-cre}
 \chaptermark{Completely Randomized Experiment and FRT}

The potential outcomes framework has intrinsic connections with randomized experiments. Understanding causal inference in various randomized experiments is fundamental and helpful for understanding causal inference in more complicated non-experimental studies.

Part \ref{part::rcts} of this book focuses on randomized experiments. This chapter focuses on the simplest experiment, the completely randomized experiment (CRE). 

\section{CRE}

Consider an experiment with $n$ units, with $n_1$ receiving the treatment and $n_0$ receiving the control. We can define the CRE based on its treatment assignment mechanism\footnote{Readers may think that a CRE has $Z_i$'s as independent and identically distributed (IID) Bernoulli random variables with probability $\pi$, in which $n_1$ is a Binomial$(n,\pi)$ random variable. This is called the Bernoulli randomized experiment (BRE), which reduces to the CRE if we condition on $(n_1, n_0)$.
I will give more details for the BRE in Problem \ref{hw::bre-inference} in Chapter \ref{chapter::neyman-cr}. 
}.

\begin{definition}
[CRE]\label{def::CRE}
Fix $n_1$ and $n_0$ with $n = n_1 + n_0$.
A CRE has the treatment assignment mechanism:
$$
\pr(  \bm{Z} = \bm{z} ) = 1 \Big / \binom{n}{n_1},
$$
where $\bm{z} = (z_1,\ldots, z_n)$ satisfies $\sumn z_i  = n_1$ and $\sumn (1-z_i)  = n_0 . $
\end{definition}

In Definition \ref{def::CRE}, we assume that the potential outcome vector under treatment $\bm{Y}(1) = (Y_1(1), \ldots, Y_n(1) )$ and the potential outcome vector under control $\bm{Y}(0) = (Y_1(0), \ldots, Y_n(0) )$ are both fixed. Even if we view them as random, we can condition on them and the treatment assignment mechanism becomes 
$$
\pr\{    \bm{Z} = \bm{z} \mid \bm{Y}(1), \bm{Y}(0)  \} = 1 \Big / \binom{n}{n_1} 
$$
because $ \bm{Z} \ind \{\bm{Y}(1), \bm{Y}(0)\}$ in a CRE. In a CRE, the treatment vector $\bm{Z}$ is from a random permutation of $n_1$ 1's and $n_0$ 0's.

In his seminal book {\it Design of Experiments}, \citet{Fisher:1935} pointed out the following advantages of randomization:
\begin{enumerate}
[(1)]
\item\label{eq::comparable}
It creates comparable treatment and control groups on average.

\item\label{eq::reasonedbasis}
It serves as a ``reasoned basis'' for statistical inference. 
 \end{enumerate}

Point \ref{eq::comparable} is intuitive because the random treatment assignment does not bias toward the treatment or the control. Most people understand point 1 well. Point \ref{eq::reasonedbasis} is more subtle. What Fisher meant is that randomization justifies a statistical test, which is now called the Fisher Randomization Test (FRT). This chapter illustrates the basic idea of the FRT  under a CRE.

\section{FRT}
\label{sec::frt-basics}

\citet{Fisher:1935} was interested in testing the following null hypothesis:\footnote{Actually, \citet{Fisher:1935} did not use this form of $\HF$ since he did not use the notation of potential outcomes. This form was due to \citet{Rubin:1980}.}
$$
\HF: Y_i(1) = Y_i(0) \text{ for all units } i=1,\ldots, n.
$$
\citet{Rubin:1980} called it the {\it sharp null hypothesis} in the sense that it can determine all the potential outcomes based on the observed data: $\bm{Y}(1) = \bm{Y}(0) =  \bm{Y} = (Y_1, \ldots, Y_n)$, the vector of the observed outcomes. It is also called the {\it strong null hypothesis} \citep[e.g.,][]{wu2018randomization}.

Conceptually, under $\HF$, the FRT works for any test statistic 
\begin{eqnarray}\label{eq::teststatistic}
T = T( \bm{Z}, \bm{Y})  ,
\end{eqnarray}
which is a function of the observed data. The observed outcome vector $\bm{Y}$ is fixed under $\HF$, so the only random component in the test statistic $T$ is the treatment vector $\bm{Z}$. The experimenter determines the distribution of $\bm{Z}$, which in turn determines the distribution of $T$ under $\HF$. This is the basis for calculating the $p$-value. I will give more details below.

In a CRE, $\bm{Z}$ is uniform over the set
$$
\{  \bm{z}^1, \ldots, \bm{z}^M  \}
$$
where $M = \binom{n}{n_1}$, and the $ \bm{z}^m$'s are all possible vectors with $n_1$ $1$'s and $n_0$ $0$'s. 
For instance, with $n=5$ and $n_1=3$, we can enumerate $M = \binom{5}{3} = 10$ vectors as follows:
\begin{lstlisting}
> permutation10 = function(n, n1){
+  M  = choose(n, n1)
+  treat.index = combn(n, n1)
+  Z = matrix(0, n, M)
+  for(m in 1:M){
+    treat  = treat.index[, m]
+    Z[treat, m] = 1
+  }
+  Z
+ }
> 
> permutation10(5, 3)
     [,1] [,2] [,3] [,4] [,5] [,6] [,7] [,8] [,9] [,10]
[1,]    1    1    1    1    1    1    0    0    0     0
[2,]    1    1    1    0    0    0    1    1    1     0
[3,]    1    0    0    1    1    0    1    1    0     1
[4,]    0    1    0    1    0    1    1    0    1     1
[5,]    0    0    1    0    1    1    0    1    1     1
\end{lstlisting}

As a consequence, $T$ is uniform over the set (with possible duplications)
$$
\{   T(\bm{z}^1,\bm{Y}) , \ldots, T( \bm{z}^M, \bm{Y})  \}.
$$
That is, the distribution of $T$ is known due to the design of the CRE. We will call this distribution of $T$ the {\it randomization distribution}.

\begin{figure}[ht]
\centering 
\includegraphics[width =  \textwidth]{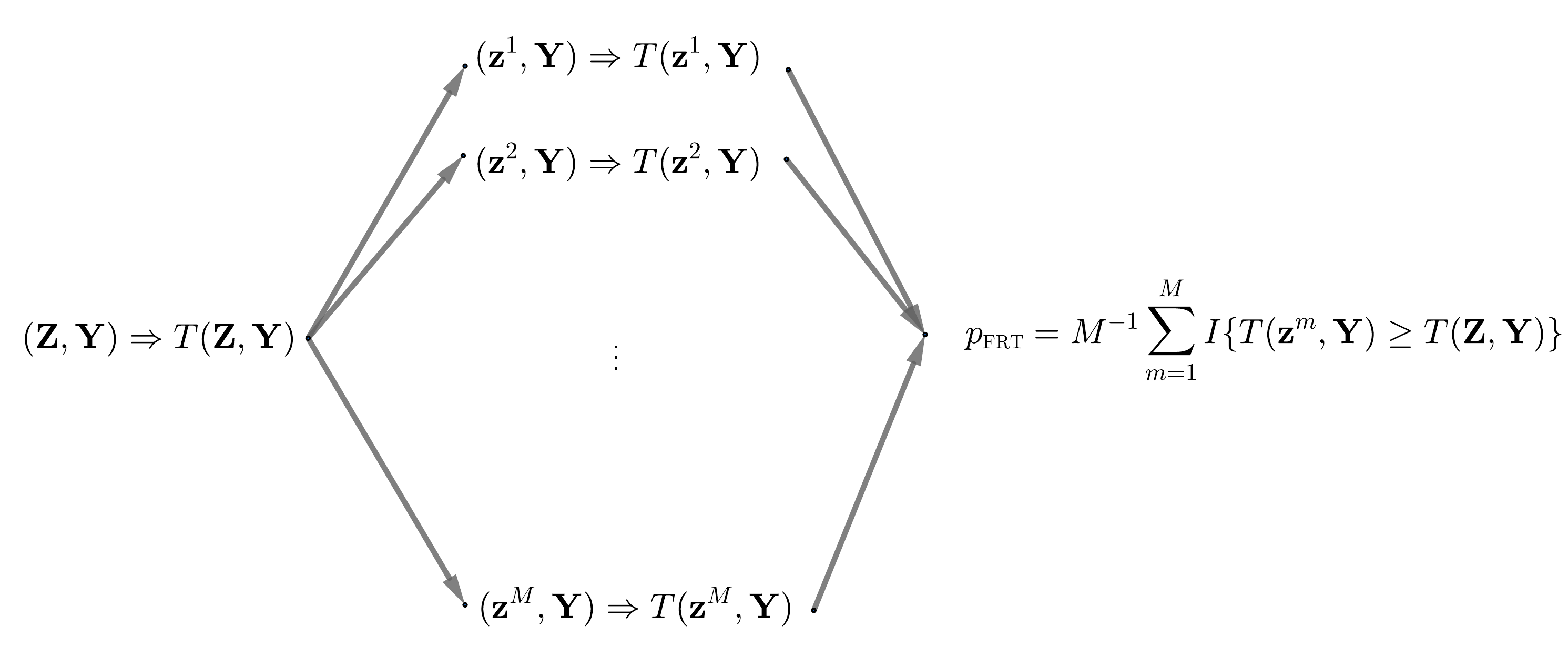}
\caption{Illustration of the FRT}\label{fig::frt}
\end{figure}

If larger values are more extreme for $T$, we can use the following tail probability to measure the extremeness of the test statistic with respect to its randomization distribution:\footnote{Because of this, the $p_\frt$ in \eqref{eq::exactpvalue} is one-sided. Sometimes, we may want to define two-sided $p$-values. One possibility is to use the absolute value of $T$ if it is distributed around $0$. }
\begin{equation}\label{eq::exactpvalue}
p_\frt = M^{-1} \sum_{m=1}^M   I\{ T(\bm{z}^m,\bm{Y})   \geq T(\bm{Z},\bm{Y})   \},
\end{equation}
which is called the $p$-value by Fisher. Figure \ref{fig::frt} illustrates the computational process of $p_\frt $.

The $p$-value, $p_\frt$,  in \eqref{eq::exactpvalue} works for any choice of test statistic and any outcome-generating process. It also extends naturally to any experiment, which will be a topic repeatedly discussed in the following chapters. Importantly, it is finite-sample exact in the sense\footnote{This is the standard definition of the $p$-value in mathematical statistics. The inequality is often due to the discreteness of the test statistic, and when the equality holds, the $p$-value is Uniform$(0,1)$ under the null hypothesis. Let $F(\cdot)$ be the distribution function of $T(\bm{Z},\bm{Y})$. Even though it is a step function, we assume that it is continuous and strictly increasing as if it is the distribution function of a continuous random variable taking values on the whole real line.  So $p_\frt = 1- F(T)$, and  
$$
\pr(p_\frt \leq u) = \pr\{ 1-F(T) \leq u\} = \pr\{  T \geq F^{-1}(1-u)  \} = 1- F( F^{-1}(1-u)) = u.
$$
The discreteness of $T$ does cause some technical issues in the proof, yielding an inequality instead of an equality. I leave the technical details to Problem \ref{hw::exact-frt}.}  that under $\HF$,
\begin{eqnarray}\label{eq::p-value-valid}
\pr(p_\frt \leq u) \leq u \quad  \text{ for all } \quad 0 \leq u \leq 1.
\end{eqnarray}

In practice, $M$ is often too large (e.g., with $n=100, n_1=50$, we have $M > 10^{29}$), and it is computationally infeasible to enumerate all possible values of the treatment vector. We often approximate $p_\frt$ by Monte Carlo (see Section \ref{sec::monte-carlo-method} for a review of the basic Monte Carlo method). To be more specific, we take independent random draws from all possible values of the treatment vector, or, equivalently, we randomly permute $\bm{Z}$, and approximate $p_\frt$ by
\begin{equation}\label{eq::mcpvalue}
\hat p_\frt  =  R^{-1} \sum_{r=1}^R   I\{ T(\bm{z}^r,\bm{Y})   \geq T(\bm{Z},\bm{Y})   \},
\end{equation}
where the $\bm{z}^r$'s are the $R$ random permutations of $\bm{Z}$.
The $p$-value in \eqref{eq::mcpvalue} has Monte Carlo error decreasing fast with an increasing $R$; see Problem \ref{para::monte-carlo-error-pfrt}. Because the calculation of the $p$-value in \eqref{eq::mcpvalue} involves permutations of $\bm{Z}$, the FRT is sometimes called the {\it permutation test} in the context of the CRE. However, the idea of FRT is more general than the permutation test in more complex experiments.

\section{Canonical choices of the test statistic}
\label{sec::frt-choiceofstatistic}

From the above discussion, the FRT generates finite-sample exact $p$-value for any choice of the test statistic. This is a feature of the FRT. However, this feature should not encourage arbitrary choice of the test statistic. Intuitively, we must choose test statistics that give information for possible violations of $\HF$.  Below I will review some canonical choices.

\begin{example}[difference-in-means]\label{eq::frt-difference-in-means}
The difference-in-means statistic is
$$
\hat{\tau}   = \hat{\bar{Y}}(1) - \hat{\bar{Y}}(0)
$$
where
$$
 \hat{\bar{Y}}(1)  =  n_1^{-1} \sum_{Z_i=1} Y_i  = n_1^{-1} \sumn Z_iY_i 
$$
is the sample mean of the outcomes under the treatment and
$$
\hat{\bar{Y}}(0) = n_0^{-1}  \sum_{Z_i=0} Y_i   = n_0^{-1}  \sumn (1-Z_i) Y_i 
$$
is the sample mean of the outcomes under the control, respectively.
%
Under $\HF$, it has mean 
$$
E(  \hat{\tau}    ) = n_1^{-1}\sumn E(Z_i)Y_i  - n_0^{-1}  \sumn E(1-Z_i) Y_i  = 0
$$
and variance
\begin{eqnarray*}
\var( \hat{\tau}   ) &=& \var\left\{  n_1^{-1} \sumn Z_iY_i - n_0^{-1}  \sumn (1-Z_i) Y_i  \right\} \\
&=&\var\left(  \frac{n}{n_0}    \frac{1}{n_1} \sumn  Z_i Y_i         \right) \\
&=_*& \frac{n^2}{n_0^2} \left( 1 - \frac{n_1}{n} \right) \frac{s^2}{n_1} \\ 
&=& \frac{n}{n_1 n_0} s^2,
\end{eqnarray*}
where $=_*$ follows from Lemma  \ref{lemma::samplemean} for simple random sampling with
$$
\bar{Y} = n^{-1}  \sumn Y_i,\quad 
s^2 = (n-1)^{-1} \sumn (Y_i - \bar{Y})^2 .  
$$ 
Furthermore, the randomization distribution of $\hat{\tau}$ is approximately Normal due to the finite population central limit theorem (CLT) in Lemma \ref{lemma::finite-population-clt}: 
\begin{equation}\label{eq::frt-tau-hat}
\frac{  \hat{\tau}      }{  \sqrt{\frac{n}{n_1 n_0} s^2}    }  
\rightarrow  \N01 
\end{equation}
in distribution. Since $s^2$ is fixed under $\HF$, it is equivalent to use 
$$
\frac{  \hat{\tau}      }{  \sqrt{\frac{n}{n_1 n_0} s^2}    }  
$$
as the test statistic in the FRT, which is asymptotically $\textsc{N}(0,1)$ as shown in \eqref{eq::frt-tau-hat}. Then we can calculate an approximate $p$-value. 
\end{example}

The observed data are $\{  Y_i: Z_i = 1 \}$ and $\{ Y_i : Z_i = 0 \}$, so the problem is essentially a two-sample problem. 
Under the assumption of independent and identically distributed (IID) Normal outcomes (see Section \ref{sec::bf-problem}), the classic two-sample $t$-test assuming equal variance is based on
\begin{equation}\label{eq::two-sample-t-equalv}
\frac{  \hat{\tau}      }{  \sqrt{\frac{n}{n_1 n_0(n-2)}  \left[  \sum_{Z_i=1}  \{Y_i - \hat{\bar{Y}}(1)\}^2
+ \sum_{Z_i=0}  \{Y_i - \hat{\bar{Y}}(0)\}^2
    \right]  }    }   \sim t_{n-2}.
\end{equation}
Based on some algebra (see Problem \ref{hw::frt-algebraic-detail}), we have the expansion 
\begin{equation}\label{eq::frt-algebra-difference-in-means}
(n-1)s^2 = \sum_{Z_i=1}  \{Y_i - \hat{\bar{Y}}(1)\}^2 + \sum_{Z_i=0}  \{Y_i - \hat{\bar{Y}}(0)\}^2
+ \frac{n_1 n_0}{n} \hat{\tau}^2 .
\end{equation}
With a large sample size $n$, we can ignore the difference between $\N01$ and $t_{n-2}$ and the difference between $n-1$ and $n-2$.
Moreover, under $\HF,$ $\hat{\tau}$ converges to zero in probability, so $ n_1 n_0 / n \hat{\tau}^2$ can be ignored asymptotically. Therefore, under $\HF,$ the approximate $p$-value in Example \ref{eq::frt-difference-in-means} is close to the $p$-value from the classic two-sample $t$-test assuming equal variance, which can be calculated by \ri{t.test} with \ri{var.equal = TRUE}. Under alternative hypotheses with nonzero $\tau$, the additional term $\frac{n_1 n_0}{n} \hat{\tau}^2$ in the expansion \eqref{eq::frt-algebra-difference-in-means} can make the FRT less powerful than the usual $t$-test; \citet{Ding:2014} made this point.

Based on the above discussion, the FRT with $\hat{\tau}$ effectively uses a pooled variance ignoring the heteroskedasticity between these two groups. In classical statistics, the two-sample problem with heteroskedastic Normal outcomes is called the Behrens--Fisher problem (see Section \ref{sec::bf-problem}). In the Behrens--Fisher problem, a standard choice of the test statistic is the studentized statistic below.

\begin{example}[studentized statistic]
\label{eg::studentized-statistic}
The studentized statistic\footnote{The $t$ notation is intentional because it is related to the $t$ distribution. But it should not be confused with the notation $t_\nu$, which denotes the $t$ distribution with degrees of freedom $\nu$.} is 
$$
t = \frac{    \hat{\bar{Y}}(1)  -   \hat{\bar{Y}}(0)    }{  \sqrt{  \frac{  \hat{S}^2(1)  }{n_1}  + \frac{\hat{S}^2(0)}{n_0}  }   },
$$
where 
$$
\hat{S}^2(1) = (n_1 - 1)^{-1} \sum_{Z_i =1} \{ Y_i  -\hat{\bar{Y}}(1)   \}^2,\quad
\hat{S}^2(0) = (n_0 - 1)^{-1} \sum_{Z_i =0} \{ Y_i  -\hat{\bar{Y}}(0)   \}^2 
$$
are the sample variances of the observed outcomes under the treatment and control, respectively. Under $\HF$, the finite population CLT again implies that $t$ is asymptotically Normal: 
$$
t \rightarrow \N01 
$$
in distribution. 
Then we can calculate an approximate $p$-value which is close to the $p$-value from \ri{t.test} with \ri{var.equal = FALSE}.
\end{example}

An extremely important point is that the FRT justifies the traditional $t$-tests using \ri{t.test} with either \ri{var.equal = TRUE} or \ri{var.equal = FALSE}, even if the underlying distributions are not Normal. Standard statistics textbooks motivate the $t$-tests based on the Normality assumption, but the assumption can be too strong in practice. Fortunately, the $t$-test procedures can still be used as long as the finite population CLTs hold. Even if we do not believe the CLTs, we can still use $\hat{\tau}$ and $t$ as test statistics in the FRT to obtain finite-sample exact $p$-values.

We will motivate this studentized statistic from another perspective in Chapter \ref{chapter::unification-fisher-neyman}. The theory there shows that using $t$ in FRT is more robust to heteroskedasticity across the two groups.

The following test statistic is robust to outliers resulting from heavy-tailed outcome data.

\begin{example}[Wilcoxon rank sum]\label{example::wilcoxon-cre}
The difference-in-means statistic $ \hat{\tau} $ uses the sample means of the original outcomes, and its sampling distribution depends on the variances of the outcomes. The studentized statistic $t$ uses the sample means and variances of the original outcomes, and its sampling distribution depends on higher moments of the outcomes. 
This makes $ \hat{\tau} $ and $t$ sensitive to outliers. 

Another popular test statistic is based on the ranks of the pooled observed outcomes. Let $R_i$ denote the rank of $Y_i$ in the pooled samples $\bm{Y}$: 
$$
R_i = \#\{j : Y_j \leq Y_i  \}.
$$ 
The Wilcoxon rank sum statistic is the sum of the ranks under treatment:
$$
W = \sumn Z_i R_i.
$$
For algebraic simplicity, we assume that there are no ties in the outcomes.\footnote{ 
The FRT can be applied regardless of the existence of ties. With ties, we can either add small random noises to the original outcomes or use the average ranks for the tied outcomes.
For the case with ties, see \citet[][Chapter 1 Section 4]{lehmann2006nonparametrics}.
}
Because the sum of the ranks of the pooled samples is fixed at $1+2+\cdots + n = n(n+1)/2$, the Wilcoxon statistic is equivalent to the difference in the means of the ranks under the treatment and control. 
Under $\HF$, the $R_i$'s are fixed, so
$W$ has mean
$$
E(W)  =  \sumn E(Z_i) R_i  =\frac{n_1}{n}  \sumn  i   = \frac{n_1}{n} \times \frac{n(n+1)}{2} = \frac{n_1(n+1)}{2} 
$$
and variance
\begin{eqnarray*}
\var(W) &=& \var\left( n_1 \frac{1}{n_1} \sumn Z_i R_i  \right) \\
&=_*&  n_1^2  \left( 1-\frac{n_1}{n} \right)  \frac{1}{n_1}  \frac{1}{n-1} \sumn \left(R_i - \frac{n+1}{2}  \right)^2\\
&=&  \frac{n_1 n_0}{n(n-1)} \sumn \left(i - \frac{n+1}{2}  \right)^2\\
&=&   \frac{n_1 n_0}{n(n-1)}   \left\{ \sumn i^2 - n \left(  \frac{n+1}{2}\right)^2     \right\}\\
&=&  \frac{n_1 n_0}{n(n-1)}  \left\{     \frac{n(n+1)(2n+1)}{6} -    n \left( \frac{n+1}{2}   \right)^2 \right\}  \\
&=& \frac{n_1 n_0 (n+1)}{12},
\end{eqnarray*}
where $=_*$ follows from Lemma \ref{lemma::samplemean}. 
Furthermore, under $\HF$, the finite population CLT ensures that the randomization distribution of $\widehat{\tau}$ is approximately Normal:
\begin{eqnarray}
\label{eq::wilcox-normal}
\frac{  \sumn Z_i R_i - \frac{n_1(n+1)}{2}   }{  \sqrt{  \frac{n_1 n_0 (n+1)}{12}   }}   \rightarrow \N01 
\end{eqnarray}
in distribution. 
Based on \eqref{eq::wilcox-normal}, we can conduct an asymptotic test. In \ri{R}, the function \ri{wilcox.test} can compute both exact and asymptotic $p$-values based on the statistic $W - n_1(n_1+1)/2$. Based on some asymptotic analyses, \citet{lehmann2006nonparametrics} showed that the FRT using $W$ has reasonable power over a wide range of data-generating processes.
\end{example}

\begin{example}
 [Kolmogorov--Smirnov statistic]
\label{eg::ks-CRE}
The treatment may affect the outcome in different ways. It seems natural to summarize the treatment outcomes and control outcomes based on the empirical distributions:
$$
\hat{F}_1( y) = n_1^{-1} \sumn Z_i I(Y_i \leq y),\quad
\hat{F}_0( y ) = n_0^{-1} \sumn (1-Z_i) I(Y_i \leq y) . 
$$ 
Comparing these two empirical distributions yields the famous Kolmogorov--Smirnov statistic
$$
D = \max_y  \Big |
\hat{F}_1( y) 
- \hat{F}_0( y ) 
\Big |.
$$
It is a challenging mathematics problem to derive the distribution of $D$. With large sample sizes $n_1\rightarrow \infty$ and $n_0\rightarrow \infty$, its distribution function converges to
$$
\pr\left(  \frac{n_1 n_0}{n}  D \leq  x   \right  ) \rightarrow   \frac{\sqrt{2\pi}}{x}   \sum_{j=1}^\infty  e^{     - (2j-1)^2 \pi^2 / (8x^2) }     ,
$$
based on which we calculate an asymptotic $p$-value \citep{van2000asymptotic}. In \ri{R}, the function \ri{ks.test} can compute both exact and asymptotic $p$-values.
\end{example}

\section{A case study of the LaLonde experimental data}
\label{sec::frt-lalonde-experiment}

\begin{figure}[ht]
\centering 
\includegraphics[width =  \textwidth]{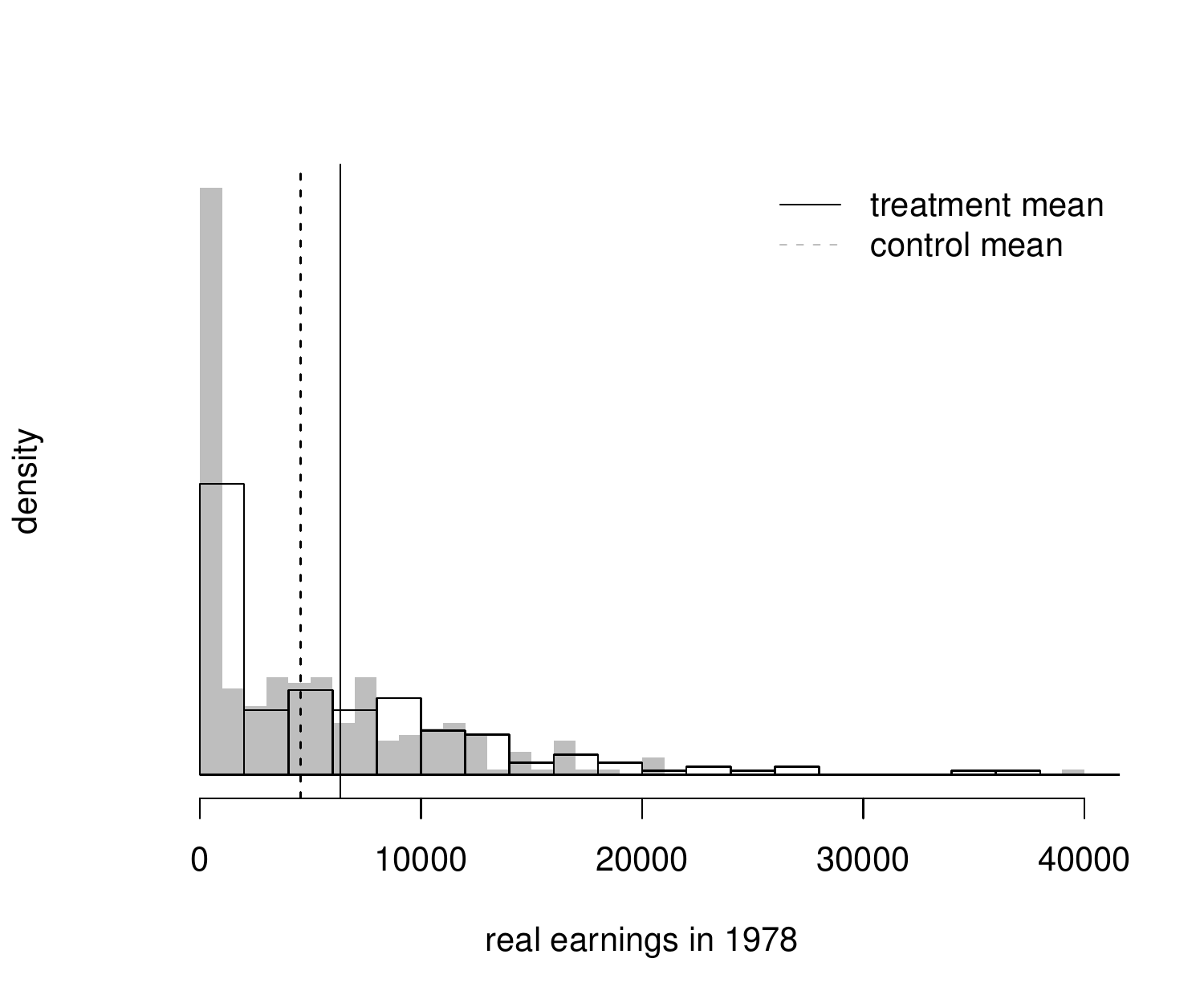}
\caption{Histograms of the outcomes in the LaLonde experimental data: the treatment in white and the control in grey. The vertical lines are the sample means of the outcomes: the solid line for the treatment and the dashed line for the control.}\label{fig::lalonde-histograms}
\end{figure}

I use \citet{lalonde1986evaluating}'s experimental data to illustrate the FRT. The data are available in the \ri{Matching} package \citep{sekhon2011multivariate}: 

\begin{lstlisting}
> library(Matching)
> data(lalonde)
> z = lalonde$treat
> y = lalonde$re78
\end{lstlisting}

In the above, \ri{z} denotes the binary treatment vector indicating whether a unit was randomly assigned to the job training program or not, and \ri{y} is the outcome vector representing a unit's real earnings in 1978. 
Figure \ref{fig::lalonde-histograms} shows the histograms of the outcomes under the treatment and control.

The following code computes the observed values of the test statistics using existing functions:
\begin{lstlisting}
> tauhat = t.test(y[z == 1], y[z == 0], 
+                 var.equal = TRUE)$statistic
> tauhat
       t 
2.835321 
> student = t.test(y[z == 1], y[z == 0],
+                 var.equal = FALSE)$statistic
> student
       t 
2.674146 
> W = wilcox.test(y[z == 1], y[z == 0])$statistic
> W
      W 
27402.5 
> D = ks.test(y[z == 1], y[z == 0])$statistic
> D
        D 
0.1321206 
\end{lstlisting}

By randomly permuting the treatment vector, we can obtain the Monte Carlo approximation of the randomization distributions of the test statistics, stored in four vectors \ri{Tauhat}, \ri{Student}, \ri{Wilcox}, and \ri{Ks}. 
\begin{lstlisting}
> MC = 10^4
> Tauhat   = rep(0, MC)
> Student  = rep(0, MC)
> Wilcox   = rep(0, MC)
> Ks       = rep(0, MC)
> for(mc in 1:MC)
+ {
+   zperm       = sample(z)
+   Tauhat[mc]  = t.test(y[zperm == 1], y[zperm == 0], 
+                        var.equal = TRUE)$statistic 
+   Student[mc] = t.test(y[zperm == 1], y[zperm == 0], 
+                        var.equal = FALSE)$statistic
+   Wilcox[mc]  = wilcox.test(y[zperm == 1], 
+                             y[zperm == 0])$statistic 
+   Ks[mc]      = ks.test(y[zperm == 1], 
+                         y[zperm == 0])$statistic
+ }
\end{lstlisting}

The one-sided $p$-values based on the FRT are all smaller than 0.05: 
\begin{lstlisting}
> exact.pv = c(mean(Tauhat >= tauhat),
+              mean(Student >= student),
+              mean(Wilcox >= W),
+              mean(Ks >= D))
> round(exact.pv, 3)
[1] 0.002 0.002 0.006 0.040
\end{lstlisting}
Without using Monte Carlo, we can also compute the asymptotic $p$-values which are all smaller than 0.05: 
\begin{lstlisting}
> asym.pv = c(t.test(y[z == 1], y[z == 0], 
+                    var.equal = TRUE)$p.value, 
+             t.test(y[z == 1], y[z == 0], 
+                    var.equal = FALSE)$p.value,
+             wilcox.test(y[z == 1], y[z == 0])$p.value,
+             ks.test(y[z == 1], y[z == 0])$p.value)
> round(asym.pv, 3)
[1] 0.005 0.008 0.011 0.046
\end{lstlisting}

The differences between the $p$-values are due to the asymptotic approximations as well as the fact that the default choices for \ri{t.test} and \ri{wilcox.test} are two-sided tests. To make fair comparisons, we need to multiply the first three $p_\frt$'s by a factor of $2$.

\begin{figure}[ht]
\centering 
\includegraphics[width =  \textwidth]{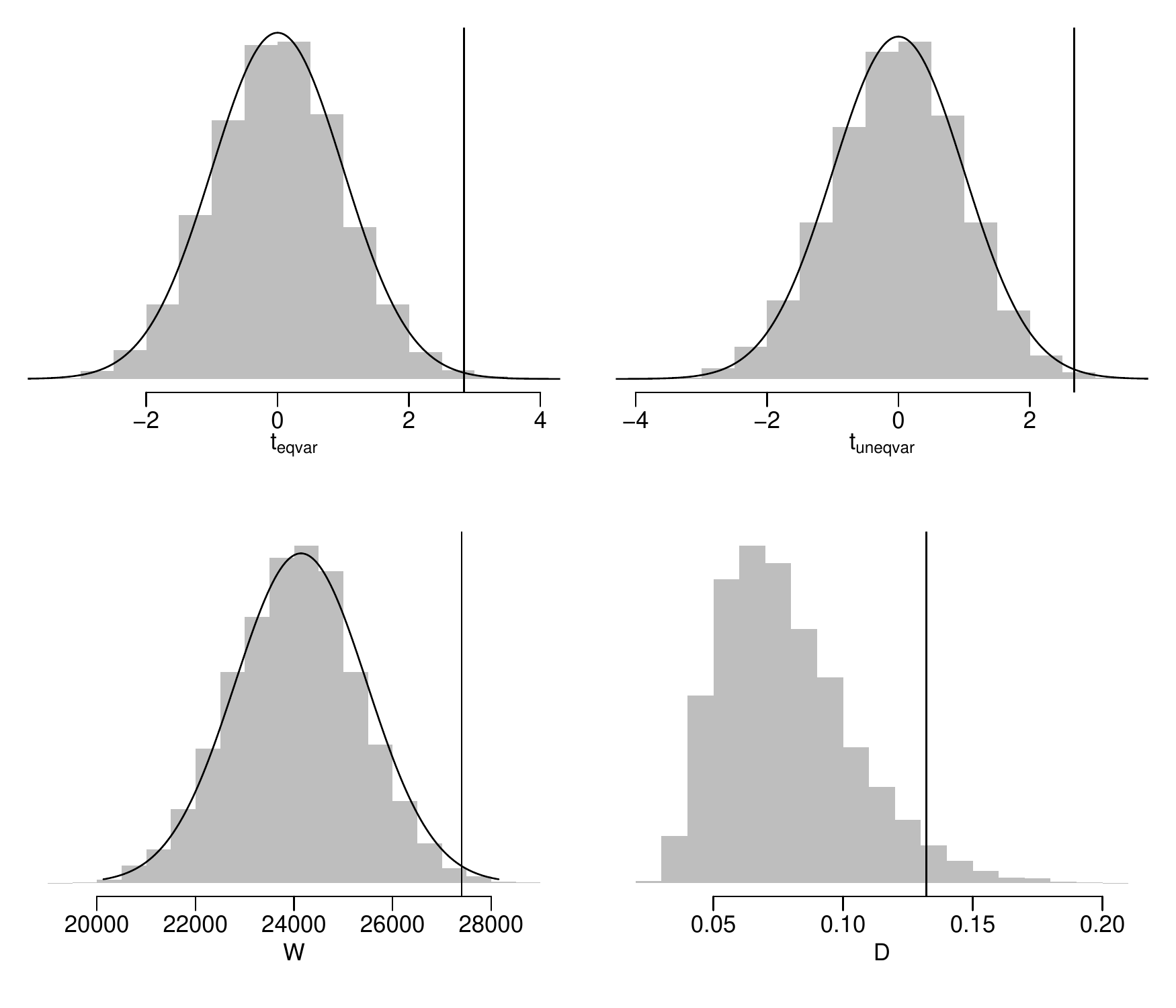}
\caption{The randomization distributions of four test statistics based on the LaLonde experimental data} \label{fig::randomizationdistribution-lalonde}
\end{figure}

Figure \ref{fig::randomizationdistribution-lalonde} shows the histograms of the randomization distributions of four test statistics, as well as their corresponding observed values. For the first three test statistics, the Normal approximations work quite well even though the underlying outcome data distribution is far from Normal as shown in Figure \ref{fig::lalonde-histograms}. In general, a figure like Figure \ref{fig::randomizationdistribution-lalonde} can give clearer information for testing the sharp null hypothesis. Recently, \citet{bind2020possible} proposes, in the title of their paper, that ``when possible, report a Fisher-exact  $p$-value and display its underlying null randomization distribution.'' I agree. 

\section{Some history of randomized experiments and FRT}

\subsection{James Lind's experiment}\label{section::lind}

James Lind (1716---1794) was a Scottish doctor and a pioneer of naval hygiene in the Royal Navy. At his time,  scurvy was a major cause of death among sailors. He conducted one of the earliest randomized experiments with clear documentation of the details and concluded that citrus fruits cured scurvy before the discovery of Vitamin C.

In \citet{lind1753treatise}, he described the following randomized experiment with $12$ patients of scurvy assigned to $6$ groups. With some simplifications, the $6$ groups are:
\begin{enumerate}
[(1)]
\item
two received a quart of cider every day; 

\item
two received twenty-five drops of sulfuric acid three times every day;

\item
two received two spoonfuls of vinegar three times every day;

\item
two received half a pint of seawater every day;

\item
two received two oranges and one lemon every day;

\item 
two received a spicy paste plus a drink of barley water every day.
\end{enumerate}

After six days, patients in the fifth group recovered, but patients in other groups did not. If we simplify the treatment as
$$
Z_i  = 1(\text{unit } i \text{ received citrus fruits})
$$
and the outcome as
$$
Y_i = 1(\text{unit } i \text{ recovered after six days}),
$$
then we have a two-by-two table

\begin{center}
\begin{tabular}{lcc}
\hline 
      & $Y =1$ & $Y =0$\\\hline
$Z =1$ & $2$ & $0$\\ 
$Z =0$ & $0$ & $10$\\
\hline 
\end{tabular}
\end{center}

This is the extremest possible two-by-two table we can observe under this experiment, and the data contain strong evidence for the positive effect of citrus fruits for curing scurvy. Statistically, how do we measure the strength of the evidence?

Following the logic of the FRT, if the treatment has no effect at all (under $\HF$), this extreme two-by-two table will occur with probability 
$$
 \frac{1}{\binom{12}{2}} = \frac{1}{66} = 0.015
$$
which is the $p_\frt$. 
This seems a surprise under $\HF$: we can easily reject $\HF$ at the level of $0.05.$

\subsection{Lady tasting tea}\label{section::lady-tasting-tea}

\citet{Fisher:1935} described the following famous experiment of {\it Lady Tasting Tea}.\footnote{This later became the title of a book on the modern history of statistics by \citet{salsburg2001lady}.}  A lady claimed that she could tell the difference between the two ways of making milk tea: one with milk added first, and the other with tea added first. This might sound odd to most people. As a statistician, Fisher designed an experiment to test whether the lady could actually tell the difference between the two ways of making milk tea. 

He made $8$ cups of tea, $4$ with milk added first and the other $4$ with tea added first. Then he presented these $8$ cups of tea in a random order to the lady and asked the lady to pick up the $4$ with milk added first. The final experiment result can be summarized in the following two-by-two table: 

\begin{tabular}{lccc}
\hline 
      &    milk first (lady)&    tea first (lady) & column sum \\ \hline
      milk first (Fisher) & $X$ & $4-X$ & 4\\ 
   tea first (Fisher) & $4-X$ & $X$ & 4\\
   row sum & 4 & 4 &8 \\
   \hline 
\end{tabular}

The $X$ can be $0, 1, 2, 3, 4$. In the real experiment, $X = 4$, which is the most extreme data, strongly suggesting that the lady could tell the difference between the two ways of making milk tea. Again, how do we measure the strength of the evidence?

Under the null hypothesis that the lady could not tell the difference, only one of the $\binom{8}{4} = 70$ possible orders yields the two-by-two table with $X=4$. So the $p$-value is
$$
p_\frt = \frac{1}{70} = 0.014.
$$
Given the significance level of $0.05$, we reject the null hypothesis.

\subsection{Two Fisherian principles for experiments}\label{sec::fisherian-principles-experiments}

In the above two examples in Sections \ref{section::lind} and \ref{section::lady-tasting-tea}, the $p_\frt$'s are justified by the randomization of the experiments. This highlights the first Fisherian principle of experiments: {\it randomization}.

Moreover, the above two experiments are in some sense the smallest possible experiments that can yield statistically meaningful results. For instance, if Lind only assigned one patient to each of the six groups, then the smallest $p$-value is
$$
\frac{1}{\binom{6}{1}}  =\frac{1}{6} = 0.167;
$$
if Fisher only made $6$ cups of tea, $3$ with milk added first and the other $3$ with tea added first, then the smallest $p$-value is
$$
\frac{1}{\binom{6}{3}} = \frac{1}{20} = 0.05.
$$
We can never reject the null hypotheses at the level of $0.05$. This highlights the second Fisherian principle of experiments: {\it replications}. This is, to ensure the FRT to have power, we must have enough units in experiments.

Chapter \ref{chapter::stratification-poststratification} will discuss the third Fisherian principle of experiments: {\it blocking}.

\section{Discussion}

\subsection{Other sharp null hypotheses and confidence intervals}\label{sec::frt-invert-interval}

I focus on the sharp null hypothesis $\HF$ above. In fact, the logic of the FRT also works for other sharp null hypotheses. For instance, we can test
$$
H_0(\bm{\tau}): Y_i(1) - Y_i(0) = \tau_i \text{ for all } i=1,\ldots, n
$$
for a known vector $\bm{\tau} = (\tau_1, \ldots, \tau_n)$. Because the individual causal effects are all known under $H_0(\bm{\tau})$, we can impute all missing potential outcomes based on the observed data. With known potential outcomes, the distribution of any test statistic $T = T(\bm{Z}, \bm{Y}(1), \bm{Y}(0))$ is completely determined by the treatment assignment mechanism, and therefore, we can compute the corresponding $p_\frt$ as a function of $\bm{\tau} $, denoted by $p_\frt(\bm{\tau} )$. If we can specify all possible $\bm{\tau}$'s, then we can compute a series of $p_\frt(\bm{\tau} )$'s. By duality of hypothesis testing and confidence set (see Section \ref{section::duality-ci-testing}), we can obtain a $(1-\alpha)$-level confidence set for the average causal effect:
$$
\left \{ \tau = n^{-1}\sum_{i=1}^n \tau_i :  p_\frt(\bm{\tau} ) \geq \alpha \right \} .
$$
Although this strategy is conceptually straightforward, it has practical complexities due to the large number of all possible $\bm{\tau}$'s. In the special case of a binary outcome, \citet{Rigdon:2015} and \citet{li2016exact} proposed some computationally feasible approaches to constructing confidence intervals for $\tau$ based on the FRT. For general unbounded outcomes, this strategy is often computationally infeasible.

A canonical simplification \citep{rosenbaum2002design, Rosenbaum:2010} is to consider a subclass of the sharp null hypotheses with constant individual causal effects:
$$
H_0(c): Y_i(1) - Y_i(0) =  c\text{ for all } i=1,\ldots, n
$$
for a known constant $c$. Given $c$, we can compute $p_\frt(c)$. By duality, we can obtain 
a $(1-\alpha)$-level confidence set for the average causal effect:
$$
\{ c :  p_\frt(c ) \geq \alpha \} .
$$
Because this procedure only involves a one-dimensional search, it is computationally feasible. 
However, the constant individual causal effect assumption is too strong. In particular, it does not hold for binary outcomes unless all units have effects $0,$ $-1$, or $1$. In general, the constant individual causal effect assumption has testable implications that can be rejected by the observed data; \citet{ding2015randomization} proposed a formal statistical test.

\subsection{Other test statistics}\label{section::other-statistic}

The FRT is a general strategy because it is applicable in any randomized experiment with any test statistic. 
I have given several examples of test statistics in Section \ref{sec::frt-choiceofstatistic}. In fact, the definition of a test statistic can be much more general. For instance, with pre-treatment covariate matrix $\bm{X}$ with the $i$th row being $X_i$ for unit $i$ $(i = 1, \ldots, n)$ \footnote{In causal inference, we call $X_i$ a covariate if it is not affected by the treatment. That is, if the covariate has two potential outcomes $X_i(1)$ and $X_i(0)$, then they must satisfy $X_i(1) = X_i(0)$. Standard statistics books often do not distinguish the treatment and covariates because they often appear on the right-hand side of a regression model for the outcome. They are both called covariates in those statistical models. This book distinguishes the treatment and covariates because they play different roles in causal inference. }, we can allow the test statistic $T(\bm{Z}, \bm{Y}, \bm{X})$ to be a function of the treatment vector, outcome vector, and the covariate matrix.  Problem \ref{hw::frt-covariates} gives an example.

\subsection{Final remarks}
For a general experiment, the probability distribution of $\bm{Z}$ is not uniform over all possible permutations of $n_1$ 1's and $n_0$ 0's. However, its distribution is completely known by the experimenter. Therefore, we can always simulate its distribution which in turn implies the distribution of any test statistic under the sharp null hypothesis. A finite-sample exact $p$-value equals 
$$
p_\textsc{frt} = \pr'\{  T(\bm{Z}', \bm{Y}) \geq T(\bm{Z}, \bm{Y}) \}
$$ 
where $\pr'$ is average over the distribution of $\bm{Z}'$ conditional on data. I will discuss other experiments in the subsequent chapters and I want to emphasize that the FRT works beyond the specific experiments discussed in this book.

The FRT works with any test statistic. However, this does not answer the practical question of how to choose a test statistic in the data analysis. If the goal is to find surprises with respect to the sharp null hypothesis, it is desirable to choose a test statistic that yields high power under alternative hypotheses. In general, no test statistic can dominate others in terms of power because power depends on the alternative hypothesis. The four test statistics in Section \ref{sec::frt-choiceofstatistic} are motivated by different alternative hypotheses. For instance, $\hat{\tau}$ and $t$ are motivated by an alternative hypothesis with a nonzero average treatment effect; $W$ is motivated by an alternative hypothesis with a constant causal effect with heavy-tailed outcomes. Specifying a working alternative hypothesis is often helpful for constructing a test statistic although it does not have to be precise to guarantee the validity of the FRT. Problems \ref{hw::frt-covariates} and \ref{hw::frt-glm} illustrate the idea of using a working alternative hypothesis or statistical model to construct test statistics.

\section{Homework Problems}

\paragraph{Exactness of $p_\frt$}
\label{hw::exact-frt}

Prove \eqref{eq::p-value-valid}.

\paragraph{Monte Carlo error of $\hat p_\frt$ }\label{para::monte-carlo-error-pfrt}

Given data, $p_\frt$ is a fixed number while its Monte Carlo estimator $\hat p_\frt$ in \eqref{eq::mcpvalue} is random. 
Show that
$$
E_{\text{mc}} (\hat p_\frt) =   p_\frt
$$
and
$$
\var_{\text{mc}} (\hat p_\frt)  \leq \frac{1}{4R},
$$
where the subscript ``mc'' signifies the randomness due to Monte Carlo, that is, $\hat p_\frt$ is random because $\bm{z}^r$'s are $R$ independent random draws from all possible values of $\bm{Z}$.

Remark: $p_\frt$ is random because $\bm{Z}$ is random. But in this problem, we condition on data, so $p_\frt$ becomes a fixed number. $\hat p_\frt$ is random because the $\bm{z}^r$'s are random permutations of $\bm{Z}$.
Problem \ref{para::monte-carlo-error-pfrt} shows that $\hat p_\frt$ is unbiased for $p_\frt$ over the Monte Carlo randomness and gives an upper bound on the variance of $\hat p_\frt$. \citet[][Theorem 2]{luo2021leveraging} gives a more delicate bound on the Monte Carlo error.

\paragraph{A finite-sample valid Monte Carlo approximation of $p_\frt$ }\label{para::finite-sample-exact-monte-carlo-frt}

Although $\hat p_\frt$ is unbiased for $p_\frt$ by the result in Problem \ref{para::monte-carlo-error-pfrt}, it may not be a valid $p$-value in the sense that $\pr(\hat p_\frt \leq u) \leq u$ for all $u\in (0,1)$ due to Monte Carlo error with a finite $R$. The following modified Monte Carlo approximation is always a finite-sample valid $p$-value. \citet{phipson2010permutation} pointed out this trick in the permutation test.

Define 
$$
\tilde p_\frt  =  \frac{  1 + \sum_{r=1}^R   I\{ T(\bm{z}^r,\bm{Y})   \geq T(\bm{Z},\bm{Y})   \}  }{1+R}
$$ 
where the $\bm{z}^r$'s the $R$ random permutations of $\bm{Z}$. Show that with an arbitrary $R$, the Monte Carlo approximation $\tilde p_\frt $ is always a finite-sample valid $p$-value in the sense that $\pr(\tilde p_\frt \leq u) \leq u$ for all $u\in (0,1)$.

Remark: You can use the following two basic probability results to prove the claim in Problem \ref{para::finite-sample-exact-monte-carlo-frt}.

\begin{lemma}
\label{lemma::binomial-order}
For two Binomial random variables $X_1\sim \text{Binomial}(R, p_1)$ and $X_2\sim \text{Binomial}(R, p_2)$ with $p_1\geq p_2$, we have  $\pr(X_1\leq x) \leq  \pr(X_2\leq x) $ for all $x.$
\end{lemma}

\begin{lemma}
\label{lemma::uniform}
If $p\sim \text{Uniform}(0,1)$ and $X\mid p \sim \text{Binomial}(R, p)$, then, marginally, $X$ is a uniform random variable over $\{ 0,1,\ldots, R  \}$. 
\end{lemma}

\paragraph{Fisher's exact test}
\label{hw::frt-fisherexacttest}

Consider a CRE with a binary outcome, with data summarized in the following two-by-two table:
 \begin{center}
\begin{tabular}{lccc}
\hline 
        & $Y =1$ & $Y =0$& total\\\hline
$Z =1$ & $n_{11}$ & $n_{10}$ & $n_1$\\ 
$Z =0$ & $n_{01}$ & $n_{00}$ & $n_0$\\ 
\hline 
\end{tabular}
\end{center}
Under $\HF$, show that any test statistic $T(n_{11}, n_{10}, n_{01}, n_{00})$ is a function of $n_{11}$ and other non-random fixed constants, and the exact distribution of $n_{11}$ is Hypergeometric. Specify the parameters for the Hypergeometric distribution. 

Remark: 
\citet{barnard1947significance} and \citet{Ding:2015} pointed out the equivalence of Fisher's exact test (reviewed in Section \ref{sec::fisher-exact-test}) and the FRT under a CRE  with a binary outcome.

\paragraph{More details for lady tasting tea}
\label{hw::more-tea}

Recall the example in Section \ref{section::lady-tasting-tea}. Calculate $\pr(X=k)$ for $k=0,1,2,3,4$.

\paragraph{Covariate-adjusted FRT}
\label{hw::frt-covariates}

This problem gives more details for Section \ref{section::other-statistic}.

Section \ref{sec::frt-lalonde-experiment} re-analyzed the LaLonde experimental data using the FRT with four test statistics. With additional covariates, the FRT can be more general with at least the following two additional strategies. Assume all potential outcomes and covariates are fixed numbers. 

First, we can use test statistics based on residuals from the linear regression. Run a linear regression of the outcomes on the covariates, and obtain the residuals (i.e., view the residuals as the ``pseudo outcomes''). Then define the four test statistics based on the residuals. Conduct the FRT using these four new test statistics. Report the corresponding $p$-values.

Second, we can define the test statistic as the coefficient in the linear regression of the outcomes on the treatment and covariates. Conduct the FRT using this test statistic. Report the corresponding $p$-value.

Why are the five $p$-values from the above two strategies finite-sample exact? Justify them.

\paragraph{FRT with a generalized linear model}
\label{hw::frt-glm}

Use the same dataset as Problem \ref{hw::frt-covariates} but change the outcome to a binary indicator for whether \ri{re78} is positive or not. Run logistic regression of the outcome on the treatment and covariates. Is the coefficient of the treatment significant and what is the $p$-value? Calculate the $p$-value from the FRT using the coefficient of the treatment as the test statistic.

\paragraph{An algebraic detail}
\label{hw::frt-algebraic-detail}

Verify \eqref{eq::frt-algebra-difference-in-means}.

\paragraph{Recommended reading}

\citet{bind2020possible} is a recent paper advocating the use of $p$-values as well as the display of the corresponding randomization distributions in analyzing complex experiments.

\chapter{Neymanian Repeated Sampling Inference in Completely Randomized Experiments}\label{chapter::neyman-cr}
 \chaptermark{Completely Randomized Experiment and Neymanian Inference}

In his seminal paper,  \citet{Neyman:1923} not only proposed the notation of potential outcomes but also derived rigorous mathematical results for making inference for the average causal effect under a CRE. In contrast to Fisher's idea of calculating the $p$-value under the sharp null hypothesis, \citet{Neyman:1923} proposed an unbiased point estimator and a conservative confidence interval based on the sampling distribution of the point estimator. This chapter will introduce \citet{Neyman:1923}'s fundamental results, which are very important for understanding later chapters in Part \ref{part::rcts} of this book.

\section{Finite population quantities}

Consider a CRE with $n$ units, where $n_1$ of them receive the treatment and $n_0$ of them receive the control. For unit $i = 1, \ldots, n$, we have potential outcomes $Y_i(1)$ and $Y_i(0)$, and individual effect $\tau_i = Y_i(1) - Y_i(0).$ The potential outcomes have finite population means  
$$
\bar{Y}(1) = n^{-1} \sumn Y_i(1),\quad
\bar{Y}(0) = n^{-1} \sumn Y_i(0),
$$
variances\footnote{Here the divisor $n-1$ makes the main theorems in this chapter more elegant. Changing the divisor to $n$ complicates the formulas but does not change the results fundamentally. With large $n$, the difference is minor.}
$$
S^2(1) = (n-1)^{-1} \sumn \{ Y_i(1) -\bar{Y}(1)  \}^2,\quad
S^2(0) = (n-1)^{-1} \sumn \{ Y_i(0) -\bar{Y}(0)  \}^2 ,
$$
and covariance
$$
S(1,0) = (n-1)^{-1} \sumn \{ Y_i(1) -\bar{Y}(1)  \}\{ Y_i(0) -\bar{Y}(0)  \} . 
$$
The individual effects have mean
$$
\tau = n^{-1} \sumn \tau_i = \bar{Y}(1) - \bar{Y}(0) .
$$
and variance
$$
S^2(\tau) = (n-1)^{-1} \sumn (\tau_i - \tau)^2.
$$
We have the following relationship between the variances and covariance.
\begin{lemma}\label{lemma::variancecovariancetau}
$2S(1,0) = S^2(1)+S^2(0) - S^2(\tau).$
\end{lemma}

Lemma \ref{lemma::variancecovariancetau}  is a basic result and will be useful for later discussion. 
The proof of Lemma \ref{lemma::variancecovariancetau} follows from elementary algebra. I leave it as Problem \ref{hw::proof-covariance-basic}. 

These fixed quantities are functions of the Science Table $\{ Y_i(1), Y_i(0)  \}_{i=1}^n$. We are interested in estimating the average causal effect $\tau$ based on the data $ (Z_i, Y_i)_{i=1}^n$ from a CRE.

\section{\citet{Neyman:1923}'s theorem}

Based on the observed outcomes, we can calculate the sample means
$$
\hat{\bar{Y}}(1) = n_1^{-1} \sumn Z_i Y_i,\quad
\hat{\bar{Y}}(0) = n_0^{-1} \sumn (1-Z_i) Y_i,
$$
the sample variances
$$
\hat{S}^2(1) = (n_1 - 1)^{-1} \sumn Z_i  \{  Y_i - \hat{\bar{Y}}(1) \}^2,\quad
\hat{S}^2(0) = (n_0 - 1)^{-1} \sumn (1-Z_i)  \{  Y_i - \hat{\bar{Y}}(0) \}^2 . 
$$
But there are no sample versions of $S(1,0)$ and $S^2(\tau)$ because the potential outcomes $Y_i(1)$ and $Y_i(0)$ are never jointly observed for each unit $i$. \citet{Neyman:1923} proved the following theorem.

\begin{theorem}
\label{thm::neyman1923}
Under a CRE, 
\begin{enumerate}
\item
the difference-in-means estimator $\hat{\tau} =\hat{\bar{Y}}(1) -  \hat{\bar{Y}}(0)$ is unbiased for $\tau$:
$$
E(\hat{\tau}) =\tau ;
$$ 
\item
$\hat{\tau} $ has variance 
\begin{eqnarray}
\var(\hat{\tau})  &=&  \frac{S^2(1)}{n_1} + \frac{S^2(0)}{n_0} - \frac{S^2(\tau)}{n} \label{eq::neyman23var} \\
&=& \frac{n_0}{n_1 n} S^2(1) + \frac{n_1}{n_0 n} S^2(0) + \frac{2}{n} S(1,0) ;  \label{eq::neyman23var-equivalent}
\end{eqnarray}
\item\label{neyman-3}
the variance estimator 
$$
\hat{V} = \frac{\hat{S}^2(1)}{n_1} + \frac{\hat{S}^2(0)}{n_0}
$$
is conservative for estimating $\var(\hat{\tau})$ in the sense that 
$$
E(\hat{V} ) - \var(\hat{\tau})  = \frac{S^2(\tau)}{n}  \geq 0
$$
with equality holding if and only if $\tau_i = \tau$ for all units. 
\end{enumerate}
\end{theorem}

\begin{figure}[ht]
\centering 
\includegraphics[width =  \textwidth]{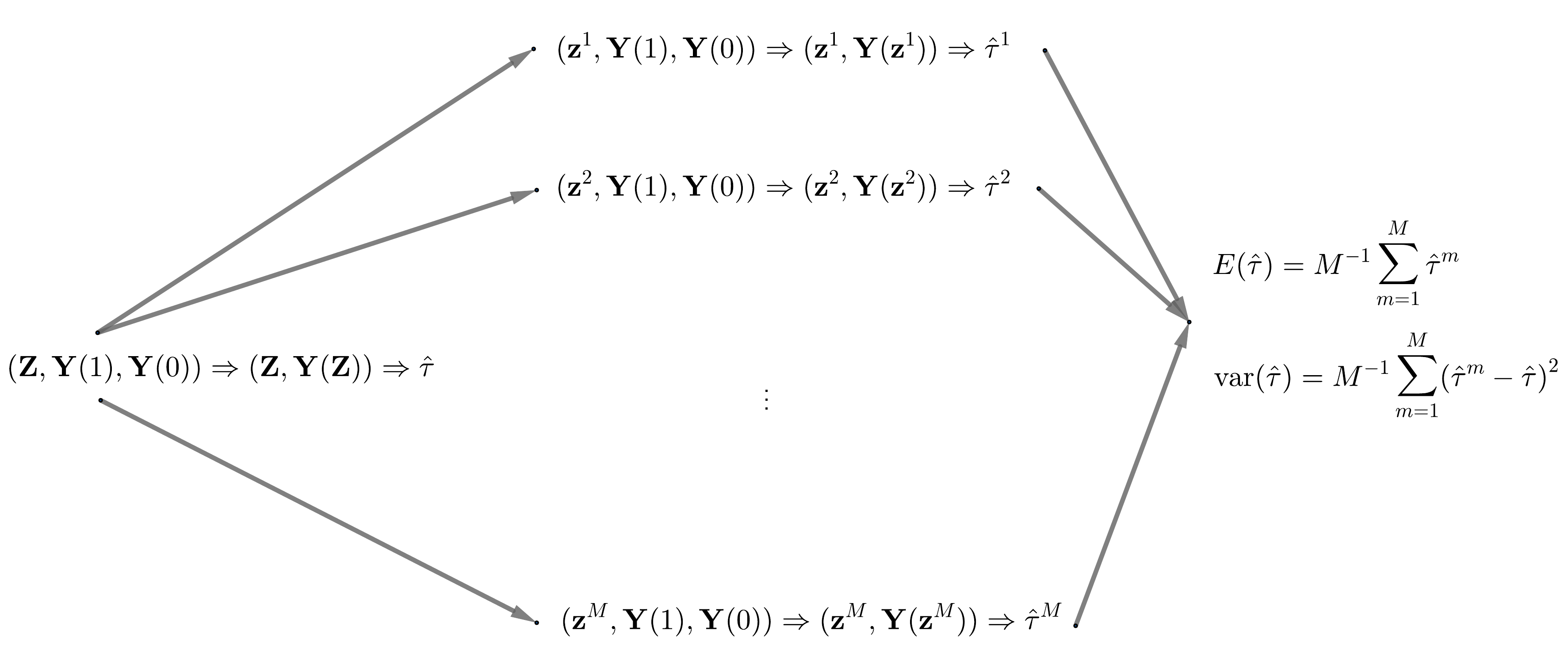}
\caption{Illustration of \citet{Neyman:1923}'s theorem, where $\bm{Y}(\bm{z}^m)$ is the observed outcome vector under the treatment vector $\bm{z}^m$.}\label{fig::neyman1923}
\end{figure}

I will present the proof of Theorem \ref{thm::neyman1923} in Section \ref{sec::prove-neyman-1923}. Before proving Theorem \ref{thm::neyman1923},  it is important to clarify the meanings of $E(\cdot)$ and $\var(\cdot)$ in Theorem \ref{thm::neyman1923}. The potential outcomes are all fixed numbers, and only the treatment indicators $Z_i$'s are random. Therefore, the expectations and variances are all over the randomness of the $Z_i$'s, which are random permutations of $n_1$ 1's and $n_0$ 0's. Figure \ref{fig::neyman1923} illustrates the randomness of $\hat\tau$, which is a discrete uniform distribution over $\{ \hat\tau^1 , \ldots, \hat\tau^M\}$ induced by  $M=\binom{n}{n_1}$ possible treatment allocations. 
Compare Figure \ref{fig::neyman1923} with Figure \ref{fig::frt} to see the key differences between the FRT and \citet{Neyman:1923}'s theorem:
\begin{enumerate}
\item
The FRT works for any test statistic but \citet{Neyman:1923}'s theorem is only about the difference in means. Although we could derive the properties of other estimators similar to \citet{Neyman:1923}'s theorem,  this mathematical exercise is often quite challenging for general estimators. 

\item
In Figure \ref{fig::frt}, the observed outcome vector $\bm{Y}$ is fixed but in Figure \ref{fig::neyman1923}, the observed outcome vector $\bm{Y}(\bm{z}^m)$ changes as $\bm{z}^m$ changes. 

\item
The $T(\bm{z}^m,\bm{Y}) $'s In Figure \ref{fig::frt} are all computable based on the observed data, but the $\hat\tau^m$'s in Figure \ref{fig::neyman1923} are hypothetical values because not all potential outcomes are known. 
\end{enumerate}

The point estimator $\hat\tau$ is standard but it has a non-trivial variance under the potential outcomes framework with a CRE. The variance formula \eqref{eq::neyman23var} differs from the classic variance formula for the difference in means\footnote{In the classic two-sample problem, the outcomes under the treatment are IID draws from a distribution with mean $\mu_1$ and variance $\sigma_1^2$, and the outcomes under the control are IID draws from a distribution with mean $\mu_0$ and variance $\sigma_0^2$. Under this assumption, we have 
$$
\var(\hat{\tau} ) = \frac{\sigma_1^2}{n_1} +  \frac{\sigma_0^2}{n_0}.
$$
Here, $\var(\cdot)$ is over the randomness of the outcomes. This variance formula does not involve a third term that depends on the variance of the individual causal effects. 
} because it not only depends on the finite population variances of the potential outcomes but also depends on the finite population variance of the individual effects, or, equivalently, the finite population covariance of the potential outcomes. Unfortunately, $S^2(\tau)$ and $S(1,0)$ are not identifiable from the data because  $Y_i(1)$ and $Y_i(0)$ are never jointly observed.

The formula \eqref{eq::neyman23var} is a little puzzling in that the more heterogeneous the individual effects are the smaller the variability of $\hat{\tau}$ is. Section \ref{sec::simulation-neyman-formula} will use numerical examples to verify \eqref{eq::neyman23var}. What is the intuition here? I give an explanation based on the equivalent form \eqref{eq::neyman23var-equivalent}. Compare the case with positively correlated potential outcomes and the case with negatively correlated potential outcomes. 
Although the treatment group is a simple random sample from the finite population of $n$ units, it is possible to observe relatively large treatment potential outcomes in a realized experiment. Assume this happens. 
\begin{enumerate}
\item
If $S(1,0) > 0$, then those treated units also have relatively large control potential outcomes. Consequently, the observed outcomes under control are relatively small, which results in large $\hat\tau$. 
\item
If $S(1,0) < 0$, then those treated units have relatively small control potential outcomes. Consequently, the observed outcomes under control are relatively large, which results in small $\hat\tau$. 
\end{enumerate}
The reverse can also happen if we observe relatively small treatment potential outcomes in a realized experiment. Overall, although the unbiasedness of $\hat\tau$ does not depend on the correlation between the potential outcomes,  it is more likely to observe more extreme $\hat\tau$ under $S(1,0) > 0$ than under $S(1,0) < 0$. So the variance of $\hat{\tau}$ is larger when the potential outcomes are positively correlated.

\citet[][Theorem 5 and Proposition 3]{li2017general} further proved the following asymptotic Normality of $\hat{\tau}$ based on the finite population CLT.
\begin{theorem}
\label{thm::clt-difference-in-means}
Let $n\rightarrow \infty$ and $n_1 \rightarrow \infty$. 
If $n_1/n$ has a limiting value in $(0,1)$,  $\{S^2(1), S^2(0), S(1,0)\}$ have limiting values, and 
$$  
\max_{1\leq i \leq n}\{ Y_i(1) - \bar{Y}(1) \}^2/n \rightarrow 0, \quad 
\max_{1\leq i \leq n}\{ Y_i(0) - \bar{Y}(0) \}^2/n \rightarrow 0,
$$
then 
$$
\frac{   \hat{\tau} - \tau }{  \sqrt{ \var(\hat{\tau}) } } \rightarrow \N01
$$
in distribution, and
$$
\hat{S}^2(1) \rightarrow S^2(1),\quad 
\hat{S}^2(0) \rightarrow S^2(0)
$$
in probability. 
\end{theorem}

The proof of  Theorem \ref{thm::clt-difference-in-means} is technical and beyond the scope of this book. It ensures that the sampling distribution of $\hat{\tau} $ can be approximated by a Normal distribution with a large sample size and some regularity conditions. Moreover, it ensures that the sample variances of the outcomes are consistent for the population variances, which further ensures that the probability limit of \cite{Neyman:1923}'s variance estimator is larger than the true variance of $\hat{\tau} $. This justifies a conservative large-sample confidence interval for $\tau$: 
$$
\hat{\tau} \pm z_{1-\alpha/2} \sqrt{  \hat{V}  },
$$
where $z_{1-\alpha/2}$ is the $1-\alpha/2$ upper quantile of the standard Normal random variable. It is the same as the confidence interval for the standard two-sample problem asymptotically (see Chapter \ref{sec::bf-problem}). This confidence interval covers $\tau$ with probability at least as large as $1-\alpha$ when the sample size is large enough. By duality, the confidence interval implies a test for 
$$
\HN: \tau = 0,
$$
which is called the weak null hypothesis.

Due to the fundamental problem of missing one potential outcome, we can at most obtain a conservative variance estimator. In statistics, the definition of the confidence interval allows for over coverage and thus conservativeness in variance estimation (see Chapter \ref{sec::statistical-inference}).  Conservativeness is not a big problem if underreporting the treatment effect is not a big problem in practice. Sometimes, it can be problematic if the outcomes measure the side effects of a treatment. In medical experiments, underreporting the side effects of a new drug can have severe consequences on patients' health.

\section{Proofs}\label{sec::prove-neyman-1923}

In this section, I will prove  Theorem \ref{thm::neyman1923}.

First, the unbiasedness of $\hat{\tau}$ follows from the representation
\begin{eqnarray*}
\hat{\tau} &=& n_1^{-1} \sumn Z_i Y_i - n_0^{-1} \sumn (1-Z_i) Y_i \\
&=& n_1^{-1} \sumn Z_i Y_i(1) - n_0^{-1} \sumn (1-Z_i) Y_i(0) 
\end{eqnarray*}
and the linearity of the expectation:
\begin{eqnarray*}
E( \hat{\tau}  ) &=& E\left\{    n_1^{-1} \sumn Z_i Y_i(1) - n_0^{-1} \sumn (1-Z_i) Y_i(0)   \right\} \\
&=& n_1^{-1} \sumn E(Z_i) Y_i(1) - n_0^{-1} \sumn E(1-Z_i) Y_i(0) \\
&=& n_1^{-1} \sumn \frac{n_1}{n} Y_i(1) - n_0^{-1} \sumn \frac{n_0}{n} Y_i(0) \\
&=&  n^{-1} \sumn Y_i(1) - n^{-1} \sumn Y_i(0) \\
&=& \tau .
\end{eqnarray*}

Second, we can further write $\hat{\tau}$ as
\begin{eqnarray*}
\hat{\tau} &=&  \sumn Z_i  \left\{    \frac{  Y_i(1) }{ n_1 } + \frac{  Y_i(0) }{ n_0 }   \right\}
- n_0^{-1} \sumn Y_i(0).
\end{eqnarray*}
The variance of $\hat{\tau}$ follows from Lemma \ref{lemma::samplemean} of simple random sampling:
\begin{eqnarray*}
\var(\hat{\tau}) &=&  \frac{  n_1 n_0  }{  n(n-1)  }  \sumn  
\left\{    \frac{  Y_i(1) }{ n_1 } + \frac{  Y_i(0) }{ n_0 }  -  \frac{  \bar{Y}(1) }{ n_1 } - \frac{  \bar{Y}(0) }{ n_0 }    \right\}^2 \\
&=&  \frac{  n_1 n_0  }{  n(n-1)  }  \left[
\frac{1}{n_1^{2}}   \sumn  \{  Y_i(1) -  \bar{Y}(1)\}^2 + \frac{1}{n_0^{2}}  \sumn  \{  Y_i(0) -  \bar{Y}(0)\}^2 \right. \\
&& \left. \qquad \qquad 
+ \frac{2}{n_1 n_0}  \sumn  \{  Y_i(1) -  \bar{Y}(1)\}  \{  Y_i(0) -  \bar{Y}(0)\} 
\right]\\
&=& \frac{n_0}{n_1n} S^2(1) + \frac{n_1}{n_0n}S^2(0) + \frac{2}{n} S(1,0). 
\end{eqnarray*}
From Lemma \ref{lemma::variancecovariancetau}, we can also write the variance as
\begin{eqnarray*}
\var(\hat{\tau})
&=& \frac{n_0}{n_1n} S^2(1) + \frac{n_1}{n_0n}S^2(0) + \frac{1}{n} \{ S^2(1)+S^2(0) - S^2(\tau) \}\\
&=&  \frac{S^2(1)}{n_1} + \frac{S^2(0)}{n_0} - \frac{S^2(\tau)}{n}.
\end{eqnarray*}

Third, because the treatment group is a simple random sample of size $n_1$ from the $n$ units, Lemma \ref{lemma::variance-unbiased} ensures that the sample variance of $Y_i(1)$'s is unbiased for its population variance: 
$$
E\{  \hat{S}^2(1) \} = S^2(1).
$$
Similarly, $E\{  \hat{S}^2(0) \} = S^2(0).$ Therefore, $\hat{V} $ is unbiased for the first two terms in \eqref{eq::neyman23var}.

 \section{Regression analysis of the CRE}

Practitioners often use regression-based inference for the average causal effect $\tau$. A standard approach is to run the ordinary least squares (OLS) of the outcomes on the treatment indicators with an intercept
$$
(\hat\alpha, \hat\beta) = \arg\min_{(a,b)} \sum_{i=1}^n (Y_i - a - bZ_i)^2,
$$
and use the coefficient of the treatment $ \hat\beta$ as the estimator for the average causal effect. 
We can show the coefficient $ \hat\beta$  equals the difference in means:
\begin{equation}
\label{eq::ols-difference-in-means}
\hat\beta = \hat\tau .
\end{equation}

However, the usual variance estimator from the OLS (see \eqref{eq::v-hat-ols} in Chapter \ref{appendix::basic-linear-regression}), e.g., the output from the \ri{lm} function of \ri{R}, equals
\begin{eqnarray}
\label{eq::ols-se}
\hat V_\textsc{ols} &=& \frac{n\left(n_1 -1\right)}{( n-2 ) n_1  n_0 } \hat S^2(1) +\frac{n\left(n_0 -1\right)}{( n-2 ) n_1  n_0 } \hat S^2(0)\\
\nonumber &\approx& \frac{\hat S^2(1)}{n_0 }+\frac{ \hat S^2(0) }{n_1 },
\end{eqnarray}
where the approximation holds with large $ n_1 $ and $ n_0 $. It differs from $\hat{V}$ even with large $ n_1 $ and $ n_0 $.

Fortunately, the Eicker--Huber--White (EHW) robust variance estimator (see \eqref{eq::v-hat-ehw} in Chapter \ref{appendix::basic-linear-regression}) is close to $\hat{V}$:
 \begin{eqnarray}
\label{eq::ols-ehw}
\hat V_\textsc{ehw} &=& \frac{\hat S^2(1)}{n_1 } \frac{n_1 -1}{n_1 }+\frac{\hat S^2(0)}{n_0 } \frac{n_0 -1}{n_0 } \\
\nonumber &\approx& \frac{ \hat S^2(1)  }{n_1 }+\frac{ \hat S^2(0) }{n_0 }
\end{eqnarray}
where the approximation holds with large $ n_1 $ and $ n_0 $. It is almost identical to $\hat{V}$. Moreover, the so-called HC2 variant of the EHW robust variance estimator is identical to  $\hat{V}$. The \ri{hccm} function in the \ri{car} package returns the EHW robust variance estimator as well as its HC2 variant. 

Problem \ref{hw::neyman-ols-nox} provides more technical details for \eqref{eq::ols-difference-in-means}--\eqref{eq::ols-ehw}.

\section{Examples}

\subsection{Simulation}
\label{sec::simulation-neyman-formula}

\begin{figure}
 \centering 
 \includegraphics[width = \textwidth]{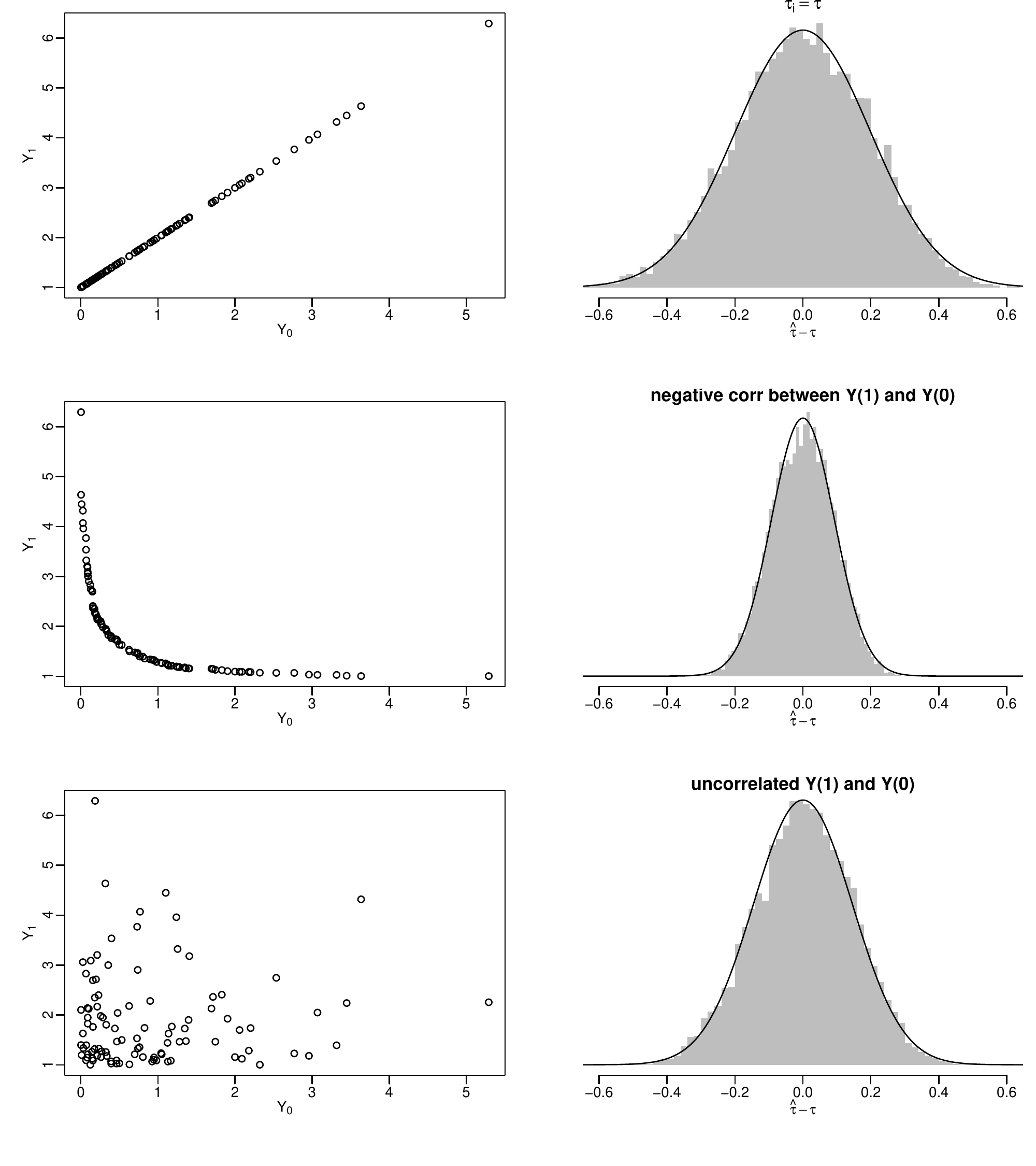}
 \caption{
Left: joint distributions of the potential outcomes with the same marginal distributions. 
Right: sampling distribution of $\hat{\tau} - \tau$ over $10^4$ simulations with different joint distributions of the potential outcomes.}\label{fig::simulation-neyman}
 \end{figure}

I first choose the sample size as $n=100$ with $60$ treated and $40$ control units, and generate the potential outcomes with constant individual causal effects. 
\begin{lstlisting}
n  = 100
n1 = 60
n0 = 40
y0 = rexp(n)
y0 = sort(y0, decreasing = TRUE)
y1 = y0 + 1
 \end{lstlisting}
With the Science Table fixed, I repeatedly generate CREs and apply Theorem \ref{thm::neyman1923} to obtain the point estimator, the conservative variance estimator, and the confidence interval based on the Normal approximation. 
The $(1,1)$th panel of Figure \ref{fig::simulation-neyman} shows the scatter plot of the potential outcomes,
and the $(1,2)$th panel of Figure \ref{fig::simulation-neyman}  shows the histogram of $\hat{\tau} - \tau$. 

I then change the potential outcomes by sorting the control potential outcomes in reverse order
\begin{lstlisting}
y0 = sort(y0, decreasing = FALSE)
 \end{lstlisting}
and repeat the above simulation. 
The $(2,1)$th panel of Figure \ref{fig::simulation-neyman} shows the scatter plot of the potential outcomes,
and the $(2,2)$th panel of Figure \ref{fig::simulation-neyman}  shows the histogram of $\hat{\tau} - \tau$.

I finally permute the control potential outcomes randomly 
 \begin{lstlisting}
y0 = sample(y0)
 \end{lstlisting}
 and repeat the above simulation. 
 The $(3,1)$th panel of Figure \ref{fig::simulation-neyman} shows the scatter plot of the potential outcomes,
and the $(2,2)$th panel of Figure \ref{fig::simulation-neyman}  shows the histogram of $\hat{\tau} - \tau$.

 Importantly, in the above three sets of simulations, the correlations between potential outcomes are different but the marginal distributions are the same. The following table compares the true variances, the average estimated variances, and the coverage rates of the $95\%$ confidence intervals. 
 \begin{lstlisting}
              constant negative independent
var              0.036    0.007       0.020
estimated var    0.036    0.036       0.036
coverege rate    0.947    1.000       0.989
 \end{lstlisting}
The true variance depends on the correlation between the potential outcomes, with positively correlated potential outcomes corresponding to a larger sampling variance. This verifies \eqref{eq::neyman23var-equivalent}. The estimated variances are almost identical because the formula of $\hat{V}$ depends only on the marginal distributions of the potential outcomes. Due to the discrepancy between the true and estimated variances, the coverage rates differ across the three sets of simulations. Only with constant causal effects, the estimated variance is identical to the true variance, verifying point \ref{neyman-3} of Theorem \ref{thm::neyman1923}. 
 
Figure \ref{fig::simulation-neyman} also shows the Normal density curves based on the CLT for $\hat{\tau} $. They are very close to the histogram over simulations, verifying Theorem \ref{thm::clt-difference-in-means}.

 \subsection{Heavy-tailed outcome and failure of Normal approximations}

The  CLT of $\hat{\tau}  $ in Theorem \ref{thm::clt-difference-in-means} holds under some regularity conditions. Those conditions will be violated with heavy-tailed potential outcomes. We can modify the above simulation studies to illustrate this point. Assume the individual causal effects are constant but the control potential outcomes are contaminated by a Cauchy component with probability $0.1, 0.3$, or $0.5$. The following code generates the potential outcomes with the probability of contamination being $0.1$. 

\begin{lstlisting}
eps = rbinom(n, 1, 0.1)
y0 = (1 - eps)*rexp(n) + eps*rcauchy(n)
y1 = y0 + 1
\end{lstlisting}

\begin{figure}
 \centering 
 \includegraphics[width = \textwidth]{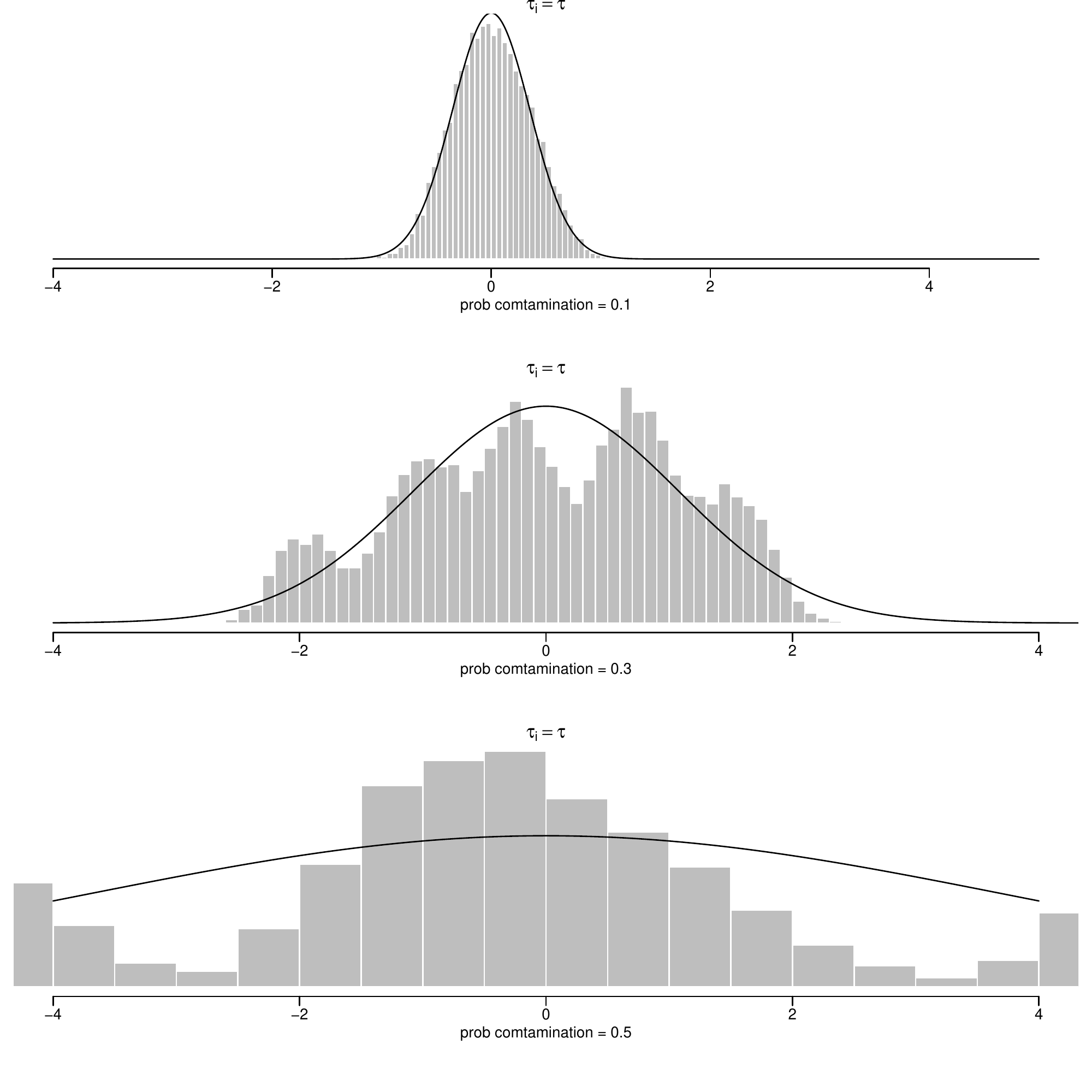}
 \caption{Sampling distribution of $\hat{\tau} - \tau$ with contaminated potential outcomes (with different contamination probabilities): realization one }\label{fig::simulation-neyman-heavytail-1}
 \end{figure}

\begin{figure}
 \centering 
 \includegraphics[width = \textwidth]{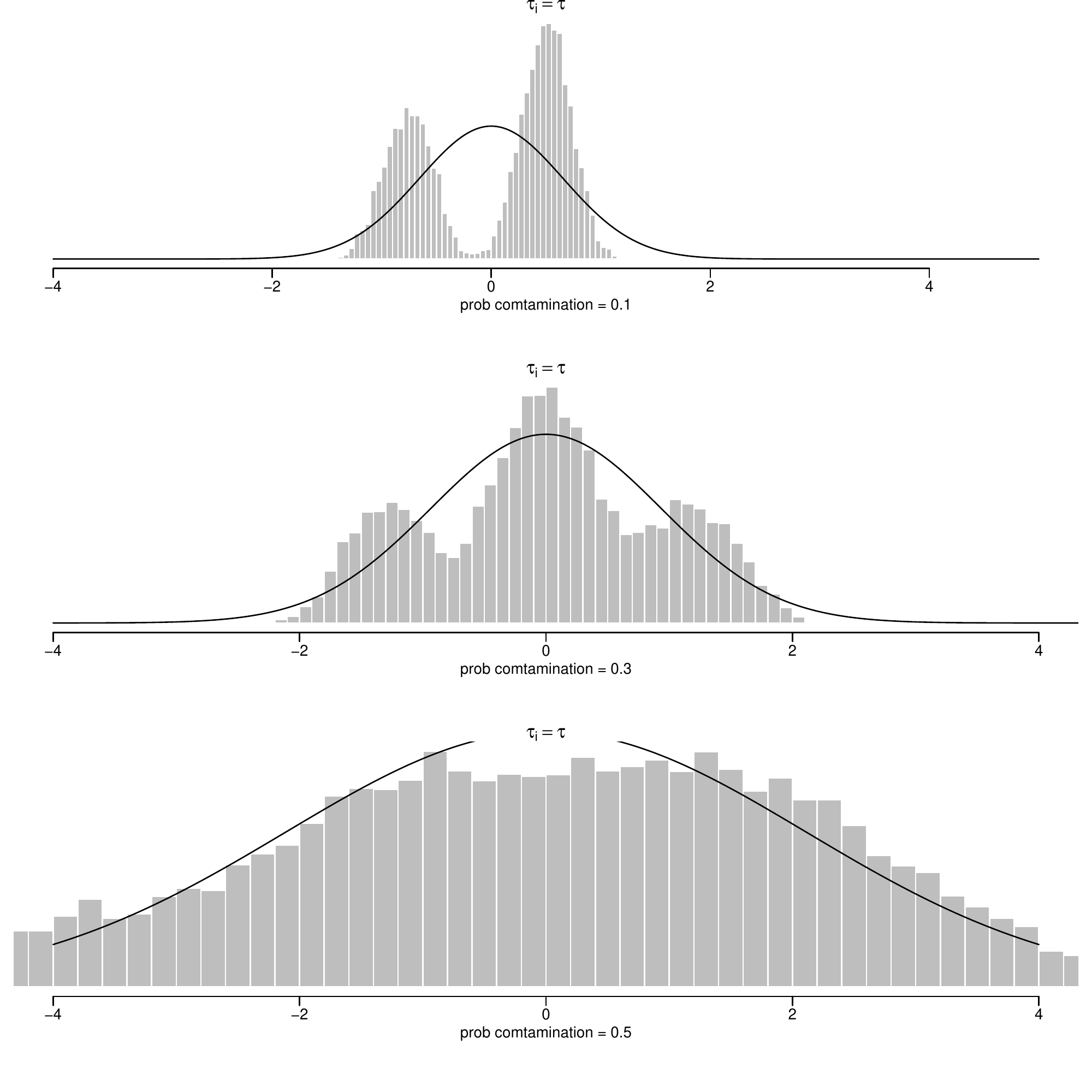}
 \caption{Sampling distribution of $\hat{\tau} - \tau$ with contaminated potential outcomes (with different contamination probabilities): realization two }\label{fig::simulation-neyman-heavytail-2}
 \end{figure}

Figures \ref{fig::simulation-neyman-heavytail-1} and \ref{fig::simulation-neyman-heavytail-2} show two realizations of the histograms of $\hat{\tau} - \tau$ with the corresponding Normal approximations. With heavy-tailed potential outcomes, the Normal approximations are quite poor. Moreover, unlike Figure \ref{fig::simulation-neyman}, the histograms are quite sensitive to the random seed of the simulation.

\subsection{Application}
\label{sec::example-neyman-formula}

I analyzed the \ri{lalonde} data in Chapter \ref{sec::frt-lalonde-experiment} to illustrate the idea of the FRT. Now I use the data to illustrate the theory in this chapter. 
\begin{lstlisting}
> library(Matching)
> data(lalonde)
> z = lalonde$treat
> y = lalonde$re78
\end{lstlisting} 

We can easily calculate the point estimator and standard error based on the formulas in Theorem \ref{thm::neyman1923}:
\begin{lstlisting}
> n1= sum(z)
> n0= length(z) - n1
> tauhat = mean(y[z==1]) - mean(y[z==0])
> vhat   = var(y[z==1])/n1 + var(y[z==0])/n0
> sehat  = sqrt(vhat)
> tauhat
[1] 1794.343
> sehat
[1] 670.9967
\end{lstlisting} 

Practitioners often use OLS to estimate the average causal effect which also gives a standard error. 
\begin{lstlisting}
> olsfit = lm(y ~ z)
> summary(olsfit)$coef[2, 1: 2]
  Estimate Std. Error 
 1794.3431   632.8536 
 \end{lstlisting} 
However, the above standard error seems too small compared to the one based on  Theorem \ref{thm::neyman1923}.  This can be solved by using the EHW robust standard error. 
\begin{lstlisting}
> library(car)
> sqrt(hccm(olsfit)[2, 2])
[1] 672.6823
> sqrt(hccm(olsfit, type = "hc0")[2, 2])
[1] 669.3155
> sqrt(hccm(olsfit, type = "hc2")[2, 2])
[1] 670.9967
 \end{lstlisting}

\section{Homework Problems}

\paragraph{Proof of Lemma \ref{lemma::variancecovariancetau}}\label{hw::proof-covariance-basic}

Prove Lemma \ref{lemma::variancecovariancetau}.

\paragraph{Alternative proof of Theorem \ref{thm::neyman1923}}\label{hw::neyman-alternative-proof}

Under a CRE, calculate 
$$
\var\{  \hat{\bar{Y}}(1)  \} ,\quad
\var\{  \hat{\bar{Y}}(0)  \} ,\quad
\cov\{  \hat{\bar{Y}}(1) ,  \hat{\bar{Y}}(0)  \} 
$$
and use these formulas to calculate $\var(\hat{\tau} )$. 

Remark: Use the results in Chapter \ref{chapter::SRS-appendix}.

\paragraph{Neymanian inference and OLS}
\label{hw::neyman-ols-nox}

Prove \eqref{eq::ols-difference-in-means}--\eqref{eq::ols-ehw}. Moreover, prove that the HC2 variant of the EHW robust variance estimator recovers $\hat{V}$ exactly. 

Remark: Chapter \ref{appendix::basic-linear-regression} reviews some important technical results about OLS.

\paragraph{Treatment effect heterogeneity}\label{hw::treatment-effect-heterogeneity-neyman}

Show that $S^2(\tau) = 0$ implies that $S^2(1) = S^2(0)$. Given a counterexample with $S^2(1) = S^2(0)$ but $S^2(\tau) \neq  0$.

Show that $S^2(1) < S^2(0)$ implies that 
$$
S(Y(0),\tau)  = (n-1)^{-1} \sumn \{ Y_i(0)  - \bar{Y}(0) \} (\tau_i - \tau) <0.
$$
Give a counterexample with $S^2(1) > S^2(0)$ but $S(Y(0),\tau)  < 0 .$

Remark: The first result states that no treatment effect heterogeneity implies equal variances in the treated and control potential outcomes. But the converse is not true. The second result states that if the treated potential outcome has a smaller variance than the control potential outcome, then the individual treatment effect is negatively correlated with the control potential outcome. But the converse is not true. \citet[][page 293]{gerber2012field} and \citet[][Appendix B.3]{ding2019decomposing} gave related discussions.

\paragraph{A better bound of the variance formula}
\label{hw::better-bound-neyman}

\cite{Neyman:1923}'s conservative variance estimator essentially uses the following upper bound on the true variance:
$$
\var(\widehat{\tau}) = \frac{S^2(1)}{n_1} + \frac{S^2(0)}{n_0} - \frac{S^2(\tau)}{n} \leq \frac{S^2(1)}{n_1} + \frac{S^2(0)}{n_0},
$$
which uses the trivial fact that $S^2(\tau) \geq 0$. 
Show the following upper bound
\begin{eqnarray}\label{hw::cs-bound-neyman}
\var(\widehat{\tau})  \leq   \frac{1}{n} \left\{   \sqrt{  \frac{n_0}{n_1}  } S(1) +  \sqrt{  \frac{n_1}{n_0}  } S(0)   \right\}^2.
\end{eqnarray}
When does the equality in \eqref{hw::cs-bound-neyman} hold? 

The upper bound \eqref{hw::cs-bound-neyman} motivates another conservative variance estimator 
$$
\hat{V}' =  \frac{1}{n} \left\{   \sqrt{  \frac{n_0}{n_1}  } \hat S(1) +  \sqrt{  \frac{n_1}{n_0}  } \hat S(0)   \right\}^2.
$$
Section \ref{sec::simulation-neyman-formula} used $\hat{V}$ in the simulation. Repeat the simulation with an additional comparison with the variance estimator $\hat{V}' $ and the associated confidence interval.

Remark: 
You may find Section \ref{sec::cauchy-schwarz-inequality} useful for the proof. 
The upper bound \eqref{hw::cs-bound-neyman} can be further improved. 
\citet{aronow2014sharp} derived the sharp upper bound for $\var(\widehat{\tau}) $ using the Frechet--Hoeffding inequality. Those improvements are rarely used in practice mainly for two reasons. First, they are more complicated than $\hat{V}$ which can be conveniently implemented by OLS. Second, the confidence interval based on  $\hat{V}$ also works under other formulations, for example, under a true linear model of the outcome on the treatment, but those improvements do not. Although they are theoretically interesting, those improvements have little practical impact.

\paragraph{Vector version of \citet{Neyman:1923}} \label{hw::vector-neyman1923}

The classic result of \citet{Neyman:1923} is about a scalar outcome. It is common to have multiple outcomes in practice. Therefore, we now extend the potential outcomes to vectors. We consider the average causal effect on a vector outcome $\bm{V}  \in \mathbb{R}^K$: 
$$
\tau_{\bm{V}} = \frac{1}{n}  \sumn \left\{  \bm{V}_i(1) - \bm{V}_i(0) \right\},
$$
where $\bm{V}_i(1)$ and $\bm{V}_i(0)$ are the potential outcomes of $\bm{V}$ for unit $i$. 
The Neyman-type estimator for $\tau_{\bm{V}}$ is the difference between the sample mean vectors of the observed outcomes under treatment and control:
$$
\widehat{ \tau}_{\bm{V}} = \bar{\bm{V}}_1  - \bar{\bm{V}}_0 
= \frac{1}{n_1}   \sumn Z_i  \bm{V}_i   - \frac{1}{n_0}  \sumn (1-Z_i) \bm{V}_i . 
$$

Consider a CRE. Show that $\widehat{ \tau}_{\bm{V}} $ is unbiased for $\tau_{\bm{V}} $. Find the covariance matrix of $\widehat{ \tau}_{\bm{V}} $. Find a (possibly conservative) estimator for the variance.

\paragraph{Inference in the BRE}\label{hw::bre-inference}

In this book, I use the following definition for the BRE. 

\begin{definition}
[BRE]\label{def::BRE}
The treatment indicators $Z_i$'s are IID Bernoulli$(\pi)$ with $n_1 = \sumn Z_i$ receiving the treatment and $n_0 = \sumn (1-Z_i)$ receiving the control, respectively. 
\end{definition}

First, we can use the FRT to analyze the BRE. How do we test $\HF$ in the BRE? Can we use the same FRT procedure as in the CRE if the actual experiment is the BRE? If yes, give a justification; if no, explain why.

Second, we can obtain point estimators for $\tau$ and find the associated variance estimators, as \citet{Neyman:1923} did for the CRE. 
\begin{enumerate}
\item
Is $\hat{\tau}$ unbiased for $\tau$? Is it consistent?
\item
Find an unbiased estimator for $\tau$.
\item
Compare the variance of the above unbiased estimator and the asymptotic variance of $\hat{\tau}$. 
\end{enumerate}

Remark: 
Under the BRE, the estimator $\hat{\tau}$ does not have finite variance but the variance of its asymptotic distribution is finite.

\paragraph{Recommended reading}

 \citet{Ding:2014} compared the Fisherian approach and Neymanian approach to analyzing the CRE.

\chapter{Stratification and Post-Stratification in Randomized Experiments}
 \chaptermark{Stratification and Post-Stratification}
\label{chapter::stratification-poststratification}

\begin{quotation}
Block what you can and randomize what you cannot.  

--- \citet[][page 103]{box1978statistics}
\end{quotation}

This is the second most famous quote from George Box\footnote{His most famous quote is ``all models are wrong but some are useful'' \citep[][page 202]{box1979robustness}.}. This chapter will explain its meaning.

\section{Stratification}

A CRE may generate an undesired treatment allocation by chance. Let us start with a CRE with a discrete covariate $X_i   \in \{ 1, \ldots, K \}$, and define $n_{[k]} = \#\{ i: X_i    =k  \}$ and $\pi_{[k]} = n_{[k]} / n$ as the number and proportion of units  in stratum $k$ $(k=1, \ldots, K)$. A CRE assigns $n_1$ units to the treatment group and $n_0$ units to the control group, which results in 
$$
n_{[k]1} = \#\{ i: X_i   = k, Z_i = 1 \},\quad
n_{[k]0} = \#\{ i: X_i   = k, Z_i = 0 \}
$$
units in the treatment and control groups, respectively, within stratum $k$. With positive probability, $n_{[k]1}$ or $n_{[k]0}$ is zero for some $k$, that is, it is possible that some strata only have treated or control units. Even if none of the $n_{[k]1}$'s or $n_{[k]0}$'s are zero, with high probability 
\begin{equation}
\frac{  n_{[k]1} }{n_1}  -  \frac{  n_{[k]0} }{n_0} \neq 0 ,
\label{eq::covarianceimbK}
\end{equation} 
and the magnitude can be quite large. So  the proportions of units in stratum $k$ are different across the treatment and control groups although on average their difference is zero (see Problem \ref{hw::balance-discrete-CRE}):
\begin{equation}\label{eq::balance-discrete-CRE}
E\left(  \frac{  n_{[k]1} }{n_1}  -  \frac{  n_{[k]0} }{n_0} \right) = 0.
\end{equation}
When $n_{[k]1}  / n_1  -  n_{[k]0} / n_0$ is large for some strata $k$'s, the treatment and control groups have undesirable covariate imbalance. Such covariate imbalance deteriorates the quality of the experiment, making it difficult to interpret the results of the experiment since the difference in the outcomes may be attributed to the treatment or the covariate imbalance.

How can we actively avoid covariate imbalance in the experiment? We can conduct stratified randomized experiments (SRE).

\begin{definition}[SRE]
\label{def::SRE}
Fix the $n_{[k]1}$'s or $n_{[k]0}$'s.
We conduct $K$ independent  CREs within the $K$ strata of a discrete covariate $X$.
\end{definition}

In agricultural experiments, the SRE is also called the {\it randomized block design}, with the strata called the blocks. Analogously, {\it stratified randomization} is also called {\it block randomization}.\footnote{
Chapter \ref{sec::fisherian-principles-experiments} mentioned two Fisherian principles for experiments: {\it randomization} and {\it replication}. {\it Blocking} is the third Fisherian principle. 
} The total number of randomizations in an SRE equals 
$$
\prod_{k=1}^K  \binom{  n_{[k]} }{  n_{[k]1} } ,
$$
and each feasible randomization has equal probability. Within stratum $k$, the proportion of units receiving the treatment is
$$
e_{[k]} = \frac{n_{[k]1}}{n_{[k]}},
$$
which is also called the {\it propensity score}, a concept that will play a central role in Part \ref{part::observational-studies} of this book (see Definition \ref{def::pscore}). 
An SRE is different from a CRE: first, all feasible randomizations in an SRE form a subset of all feasible randomizations in a CRE, so 
$$
\prod_{k=1}^K  \binom{  n_{[k]} }{  n_{[k]1} } < 
\binom{n}{n_1} ;
$$
second, $e_{[k]} $ is fixed in an SRE but random in a CRE.

For every unit $i$, we have potential outcomes $Y_i(1)$ and $Y_i(0)$, and individual causal effect $\tau_i = Y_i(1) - Y_i(0) . $
For stratum $k$, we have the stratum-specific average causal effect
$$
\tau_{[k]} = n_{[k]}^{-1} \sum_{X_i   = k} \tau_i .
$$
The average causal effect is 
$$
\tau = n^{-1} \sumn \tau_i = n^{-1} \sum_{k=1}^K \sum_{X_i   = k} \tau_i =  \sum_{k=1}^K  \pi_{[k]}  \tau_{[k]},
$$
which is a weighted average of the stratum-specific average causal effects.

If we are interested in $\tau_{[k]} $, then we can use the methods in Chapters \ref{ch::frt-cre} and \ref{chapter::neyman-cr} for the CRE within stratum $k$. Below I will discuss statistical inference for $\tau$.

\section{FRT}
\label{sec::sre-frt}

\subsection{Theory}
In parallel with the discussion of a CRE, I will start with the FRT in an SRE. The sharp null hypothesis is still 
$$
\HF: Y_i(1) = Y_i(0) \text{ for all units } i=1, \ldots, n.
$$
The fundamental idea of the FRT applies to any randomized experiment: we can use any test statistic $T = T(\bm{Z}, \bm{Y}, \bm{X})$, where $\bm{Z}, \bm{Y}$ and $\bm{X}$ are the treatment vector, observed outcome vector and the covariate matrix, respectively. Under the SRE and $\HF$,  the test statistic $T$ has a known distribution because $\bm{Z}$ has a known distribution under the SRE. However, we must be careful with two subtle issues. First, when we simulate the treatment vector, we must permute the treatment indicators within strata of $X$ according to Definition \ref{def::SRE}. The resulting FRT is sometimes called the {\it conditional randomization test} or {\it  conditional permutation test}. Second, we should choose test statistics that can reflect the nature of the SRE. Below I give some canonical choices of the test statistic.

\begin{example}[Stratified estimator]
\label{eg::stratifiedestimator}
Motivated by estimating $\tau$ (see Chapter \ref{sec::sre-neyman} below), we can use the following stratified estimator in the FRT:
$$
\hat{\tau}_\textsc{S} = \sum_{k=1}^K  \pi_{[k]} \hat{\tau}_{[k]},
$$
where 
$$
\hat{\tau}_{[k]} = n_{[k]1}^{-1} \sumn I(X_i  =k, Z_i=1) Y_i - n_{[k]0}^{-1} \sumn I(X_i  =k, Z_i=0) Y_i 
$$
is the stratum-specific difference-in-means within stratum $k$. 
\end{example}

\begin{example}[Studentized stratified estimator]
\label{eg::student-stratifiedestimator}
Motivated by the studentized statistic in the simple two-sample problem, we can use the following studentized statistic for the stratified estimator in the FRT:
$$
t_\textsc{S} = 
\frac{ \hat{\tau}_\textsc{S}  }{  \sqrt{\hat{V}_\textsc{S}  }},
$$
with 
$$
\hat{V}_\textsc{S} = \sum_{k=1}^K  \pi_{[k]}^2 \left(  \frac{\hat{S}_{[k]}^2(1)}{ n_{[k]1} } +  \frac{ \hat{S}_{[k]}^2(0)}{ n_{[k]0} } \right)
$$
where $\hat{S}_{[k]}^2(1)$ and $\hat{S}_{[k]}^2(0)$ are the stratum-specific sample variances of the outcomes under the treatment and control, respectively. The exact form of this statistic is motivated by the Neymanian perspective discussed in Section \ref{sec::sre-neyman} below.
\end{example}

\begin{example}[Combining Wilcoxon rank-sum statistics]
\label{eg::vanelteren}
We first compute the Wilcoxon rank sum statistic $W_{[k]}$ within stratum $k$ (recall Example \ref{example::wilcoxon-cre}) and then combine them as
$$
W_\textsc{S} = \sum_{k=1}^K   c_{[k]}  W_{[k]}   . 
$$
Based on different asymptotic schemes and optimality criteria, \citet{van1960combination} proposed two weighting methods, one with
$$
c_{[k]}  = \frac{1}{  n_{[k]1}  n_{[k]0} },
$$
and the other with 
$$ 
c_{[k]}  = \frac{1}{  n_{[k]}  +1 } .
$$
The motivations for these weights are quite technical, and other choices of weights may also be reasonable. 
\end{example}

\begin{example}[\citet{hodges1962rank}'s aligned rank statistic]
\label{eg::hodgeslehmann}
\citet{van1960combination}'s statistic works well with a few large strata. However, it does not work well with many small strata since it does not make enough comparisons, potentially losing information in the data.  \citet{hodges1962rank} proposed a test statistic that makes more comparisons across strata after standardizing the outcomes. They suggested first centering the outcomes as
$$
\tilde{Y}_i = Y_i  - \bar{Y}_{[k]}
$$
with the stratum-specific mean
$
\bar{Y}_{[k]} = n_{[k]}^{-1} \sum_{X_i   = k} Y_i
$
if $X_i = k$, 
then obtaining the ranks $(\tilde{R}_1, \ldots, \tilde{R}_n)$ of the pooled outcomes $(\tilde{Y}_1, \ldots, \tilde{Y}_n)$, and finally constructing the test statistic 
$$
\tilde{W} = \sumn Z_i \tilde{R}_i.
$$
\end{example}

We can simulate the exact distributions of the above test statistics under the SRE. We can also calculate their means and variances and obtain the $p$-values based on Normal approximations.

After searching for a while,  I failed to find a detailed discussion of the Kolmogorov--Smirnov statistic for the SRE. Below is my proposal.

 \begin{example}[Kolmogorov--Smirnov statistic]
\label{eg::ks-stratified}
We compute $D_{[k]}$, the maximum difference between the empirical distributions of the outcomes under treatment and control within stratum $k$. The final test statistic can be 
$$
D_\textsc{S}  = \sum_{k=1}^K    c_{[k]}    D_{[k]}
$$
or 
$$
D_{\max} = \max_{1\leq k \leq K}  c_{[k]}   D_{[k]},
$$ 
where $ c_{[k]}   = \sqrt{   n_{[k]1}  n_{[k]0} /  n_{[k]}  }$ is motivated by the limiting distribution of $D_{[k]}$ with $n_{[k]1} $ and $ n_{[k]0} $ approaching infinity (see Example \ref{eg::ks-CRE}). The statistics $D_\textsc{S} $ and $D_{\max} $ are more appropriate when all strata have large sample sizes. 
Another reasonable choice is
$$
D =  \max_y \Big |  \sum_{k=1}^K  \pi_{[k]}  \{  \hat{F}_{[k]1}(y)  - \hat{F}_{[k]0}(y) \}   \Big | ,
$$
where $\hat{F}_{[k]1}(y)  $ and $ \hat{F}_{[k]0}(y)$ are the stratum-specific empirical distribution functions of the outcomes under the treatment and control, respectively. The statistic $D$ is appropriate in both the cases with large strata and the cases with many small strata. 
\end{example}

 \subsection{An application}
 \label{sec::frt-pennbonus}

The Penn Bonus experiment is an example to illustrate the FRT in the SRE. The dataset used by  \citet{koenker2002inference}  is from a job training program stratified on \ri{quarter}, with the outcome being the duration before employment. 

\begin{lstlisting}
> penndata = read.table("Penn46_ascii.txt")
> z = penndata$treatment
> y = log(penndata$duration)
> block = penndata$quarter
> table(penndata$treatment, penndata$quarter)
   
      0   1   2   3   4   5
  0 234  41 687 794 738 860
  1  87  48 757 866 811 461
\end{lstlisting}

I will focus on $\hat{\tau}_\textsc{S}$ and $W_\textsc{S}$, and leave the FRT with other statistics as Problem \ref{hw::FRT-SRE-more}. 
The following function computes $\hat{\tau}_\textsc{S}$ and $W_\textsc{S}$:

\begin{lstlisting}
stat_SRE = function(z, y, x)
{
       xlevels = unique(x)
       K       = length(xlevels)
       PiK     = rep(0, K)
       TauK    = rep(0, K)
       WK      = rep(0, K)
       for(k in 1:K)
       {
             xk         = xlevels[k]
             zk         = z[x == xk]
             yk         = y[x == xk]
             PiK[k]     = length(zk)/length(z)
             TauK[k]    = mean(yk[zk==1]) - mean(yk[zk==0])
             WK[k]      = wilcox.test(yk[zk==1], yk[zk==0])$statistic
       }
       
       return(c(sum(PiK*TauK), sum(WK/PiK)))
}
\end{lstlisting}

The following function generates a random treatment assignment in the SRE based on the observed data:

\begin{lstlisting}
zRandomSRE = function(z, x)
{
         xlevels = unique(x)
         K       = length(xlevels)
         zrandom = z
         for(k in 1:K)
         {
              xk = xlevels[k]   
              zrandom[x == xk] = sample(z[x == xk])
         }
         
         return(zrandom)
}
\end{lstlisting}

\begin{figure}
\centering
\includegraphics[width = \textwidth]{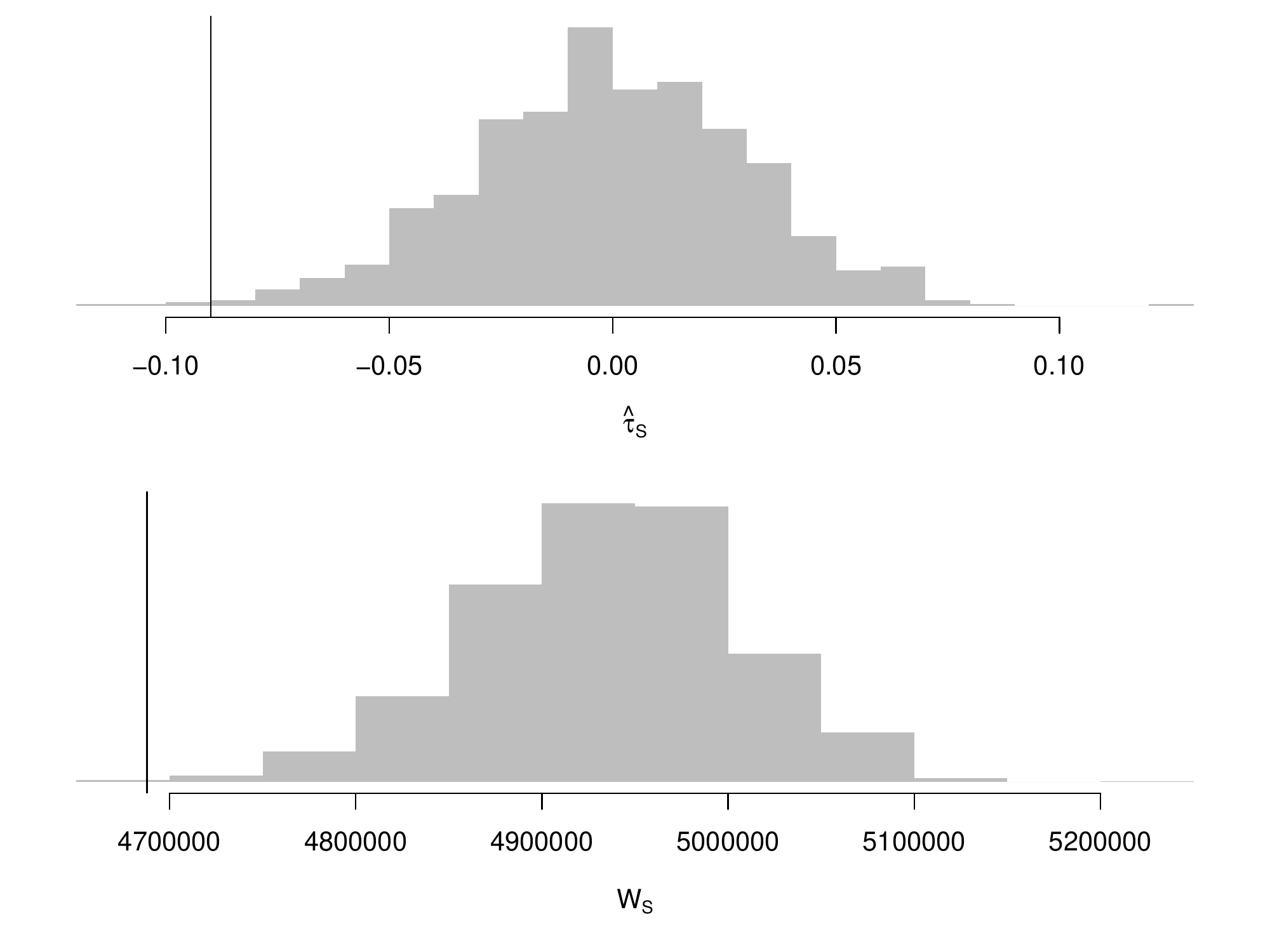}
\caption{The randomization distributions of $\hat{\tau}_\textsc{S}$ and $W_\textsc{S}$ based on the data from the Penn Bonus experiment, with $10^4$ Monte Carlo draws.}\label{fig::frt_sre_penn}
\end{figure}

Based on the above data and functions, we can simulate the randomization distributions of the test statistics and compute the $p$-values. 
\begin{lstlisting}
> stat.obs = stat_SRE(z, y, block)
> MC = 10^3
> statSREMC = matrix(0, MC, 2)
> for(mc in 1:MC)
+ {
+     zrandom         = zRandomSRE(z, block) 
+     statSREMC[mc, ] = stat_SRE(zrandom, y, block)     
+ }
> mean(statSREMC[, 1] <= stat.obs[1])
[1] 0.002
> mean(statSREMC[, 2] <= stat.obs[2])
[1] 0.001
\end{lstlisting} 
In the above, I calculate the $p$-values based on left-tail probabilities because the treatment has a negative effect on the outcome. 
See Figure \ref{fig::frt_sre_penn} for more details about the observed test statistics and randomization distributions.

\section{Neymanian inference}
\label{sec::sre-neyman}
\subsection{Point and interval estimation}

Statistical inference for an SRE builds on the fact that it essentially consists of $K$ independent CREs. Based on this, we can extend \citet{Neyman:1923}'s results to the SRE.  Within stratum $k$, the difference-in-means $\hat{\tau}_{[k]}$ is unbiased for $\tau_{[k]}$ with variance
$$
\var ( \hat{\tau}_{[k]}) =  \frac{S_{[k]}^2(1)}{ n_{[k]1} } +  \frac{S_{[k]}^2(0)}{ n_{[k]0} }
- \frac{  S_{[k]}^2(\tau) }{ n_{[k]} },
$$
where $ S_{[k]}^2(1), S_{[k]}^2(0)$ and $S_{[k]}^2(\tau)$ are the stratum-specific variances of potential outcomes and the individual causal effects, respectively. 
Therefore, the stratified estimator $\hat{\tau}_\textsc{S} = \sum_{k=1}^K  \pi_{[k]} \hat{\tau}_{[k]}$ is unbiased for $\tau =\sum_{k=1}^K  \pi_{[k]}  \tau_{[k]} $ with variance
$$
\var( \hat{\tau}_\textsc{S}   ) = \sum_{k=1}^K  \pi_{[k]}^2   \var ( \hat{\tau}_{[k]}).
$$
If $n_{[k]1} \geq 2$ and $n_{[k]0} \geq 2$, then we can obtain the sample variances $\hat{S}_{[k]}^2(1)$ and $\hat{S}_{[k]}^2(0)$ of the outcomes within stratum $k$ and construct a conservative variance estimator
$$
\hat{V}_\textsc{S}  = \sum_{k=1}^K  \pi_{[k]}^2 \left(  \frac{\hat{S}_{[k]}^2(1)}{ n_{[k]1} } +  \frac{ \hat{S}_{[k]}^2(0)}{ n_{[k]0} } \right), 
$$  
where $\hat{S}_{[k]}^2(1)$ and $\hat{S}_{[k]}^2(0)$ are the stratum-specific sample variances of the outcomes under treatment and control, respectively. Based on a Normal approximation of $\hat{\tau}_\textsc{S} $, we can construct a Wald-type $1-\alpha$ confidence interval for $\tau$:
$$
\hat{\tau}_\textsc{S} \pm z_{1-\alpha/2} \sqrt{  \hat{V}_\textsc{S}  } . 
$$
From a hypothesis testing perspective, under $\HN: \tau=0$, we can compare $t_\textsc{S} =   \hat{\tau}_\textsc{S} / \sqrt{  \hat{V}_\textsc{S}  }$ with the standard Normal quantiles to obtain asymptotic $p$-values. The statistic $t_\textsc{S} $ appears in Example \ref{eg::student-stratifiedestimator} for the FRT. Chapter \ref{chapter::unification-fisher-neyman} later will show that using  $t_\textsc{S} $ in the FRT yields finite-sample exact $p$-value under $\HF$  and asymptotically valid $p$-value under $\HN$.

Here I omit the technical details for the CLT of $\hat{\tau}_\textsc{S} $. See \citet{liu2020regression} for a proof,  which includes the two regimes with a few large strata and many small strata. I will illustrate this theoretical issue using numerical examples in Chapter \ref{sec::sre-neyman-example} below.

\subsection{Numerical examples}
\label{sec::sre-neyman-example}

The following function computes the Neymanian point and variance estimators under the SRE:
\begin{lstlisting}
Neyman_SRE = function(z, y, x)
{
       xlevels = unique(x)
       K       = length(xlevels)
       PiK     = rep(0, K)
       TauK    = rep(0, K)
       varK    = rep(0, K)
       for(k in 1:K)
       {
             xk         = xlevels[k]
             zk         = z[x == xk]
             yk         = y[x == xk]
             PiK[k]     = length(zk)/length(z)
             TauK[k]    = mean(yk[zk==1]) - mean(yk[zk==0])
             varK[k]    = var(yk[zk==1])/sum(zk) + 
                               var(yk[zk==0])/sum(1 - zk)
       }
       
       return(c(sum(PiK*TauK), sum(PiK^2*varK)))
}
\end{lstlisting}

\begin{figure}
\centering
\includegraphics[width = \textwidth]{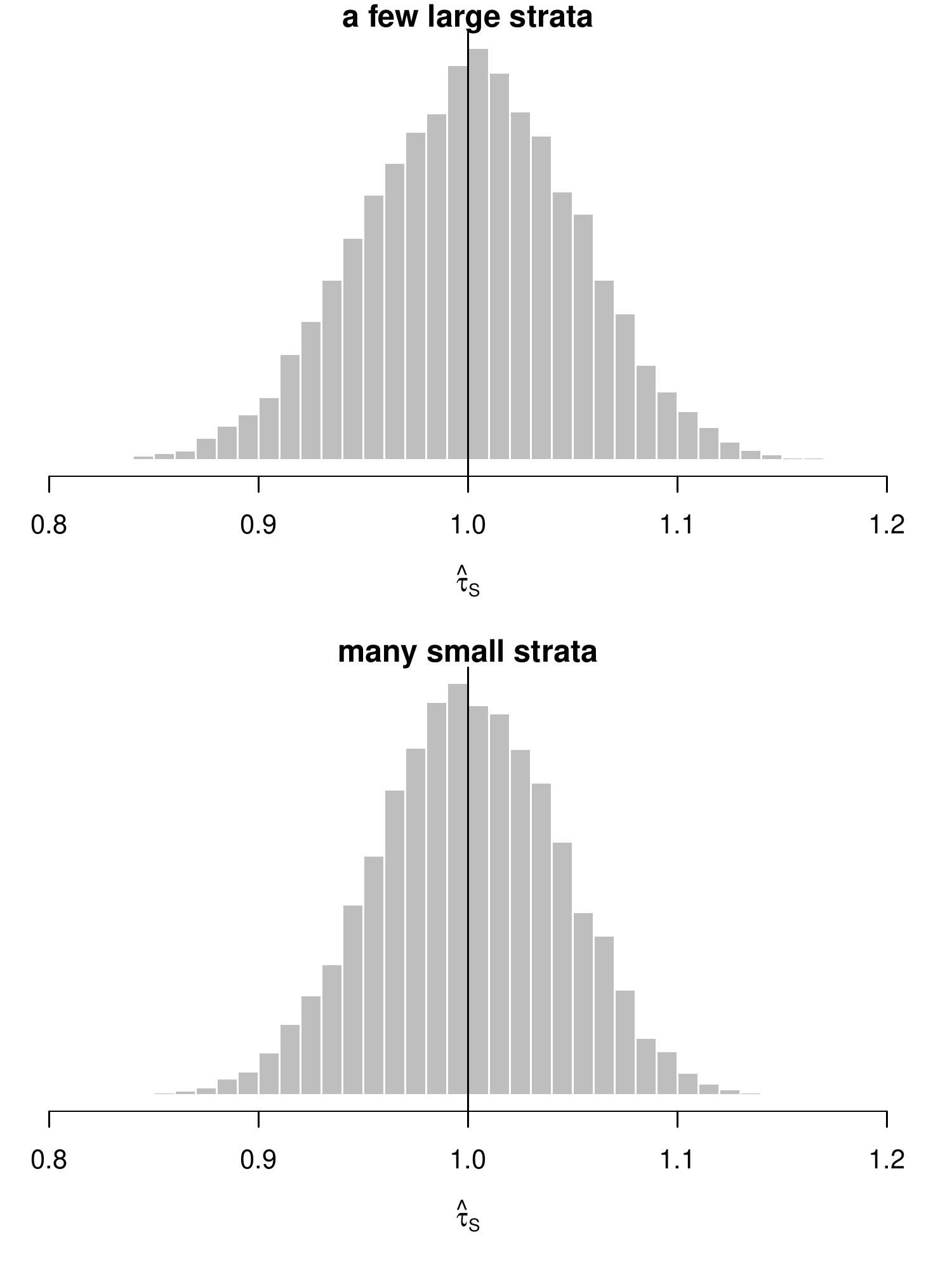}
\caption{Normal approximations under two regimes}\label{fig::2CLTs}
\end{figure}

The first simulation setting has $K=5$ and each stratum has $80$ units. \ri{TauHat} and \ri{VarHat} are the point and variance estimators over $10^4$ simulations. 
\begin{lstlisting}
> K  = 5
> n  = 80
> n1 = 50
> n0 = 30
> x  = rep(1:K, each = n)
> y0 = rexp(n*K, rate = x)
> y1 = y0 + 1
> zb = c(rep(1, n1), rep(0, n0))
> MC = 10^4
> TauHat = rep(0, MC)
> VarHat = rep(0, MC)
> for(mc in 1:MC)
+ {
+   z  = replicate(K, sample(zb))
+   z  = as.vector(z)
+   y  = z*y1 + (1-z)*y0
+   est = Neyman_SRE(z, y, x) 
+   TauHat[mc] = est[1]
+   VarHat[mc] = est[2]
+ }
> var(TauHat)
[1] 0.002248925
> mean(VarHat)
[1] 0.002266396
\end{lstlisting} 
The upper panel of Figure \ref{fig::2CLTs} shows the histogram of the point estimator, which is symmetric and bell-shaped around the true parameter. From the above, the average value of the variance estimator is almost identical to the variance of the estimators because the individual causal effects are constant.

The second simulation setting has $K=50$ and each stratum has $8$ units. 
\begin{lstlisting}
> K  = 50
> n  = 8
> n1 = 5
> n0 = 3
> x  = rep(1:K, each = n)
> y0 = rexp(n*K, rate = log(x + 1))
> y1 = y0 + 1
> zb = c(rep(1, n1), rep(0, n0))
> MC = 10^4
> TauHat = rep(0, MC)
> VarHat = rep(0, MC)
> for(mc in 1:MC)
+ {
+   z  = replicate(K, sample(zb))
+   z  = as.vector(z)
+   y  = z*y1 + (1-z)*y0
+   est = Neyman_SRE(z, y, x) 
+   TauHat[mc] = est[1]
+   VarHat[mc] = est[2]
+ }
> 
> hist(TauHat, xlab = expression(hat(tau)[S]),
+      ylab = "", main = "many small strata", 
+      border = FALSE, col = "grey",
+      breaks = 30, yaxt = 'n',
+      xlim = c(0.8, 1.2))
> abline(v = 1)
> 
> var(TauHat)
[1] 0.001443111
> mean(VarHat)
[1] 0.001473616
\end{lstlisting} 
The lower panel of Figure \ref{fig::2CLTs} shows the histogram of the point estimator, which is symmetric and bell-shaped around the true parameter.  Again, the average value of the variance estimator is almost identical to the variance of the estimators because the individual causal effects are constant. 

We finally use the Penn Bonus Experiment to illustrate the Neymanian inference in an SRE. 
Applying the function \ri{Neyman_SRE} to the dataset, we obtain:
\begin{lstlisting}
> penndata = read.table("Penn46_ascii.txt")
> z = penndata$treatment
> y = log(penndata$duration)
> block = penndata$quarter
> est = Neyman_SRE(z, y, block)
> est[1]
[1] -0.08990646
> sqrt(est[2])
[1] 0.03079775
\end{lstlisting} 
So the job training program significantly shortens the log of the duration time before employment.

\subsection{Comparing the SRE and the CRE}

What are the benefits of the SRE compared with the CRE?  I have motivated the SRE from the covariate balance perspective. In addition, I will show that better covariate balance in turn results in better estimation precision of the average causal effect. To make a fair comparison, I assume that $e_{[k]} = e$ for all $k$ which ensures that the difference in means equals the stratified estimator: 
\begin{equation}
\label{eq::dim=stratified}
\hat{\tau} = \hat{\tau}_\textsc{S}.
\end{equation}
I leave the proof of \eqref{eq::dim=stratified} as Problem \ref{hw::sre-additional-neyman}.

We now compare the sampling variances. The  classic analysis of variance technique motivates the decomposition of the total variance into the summation of the within-strata and between-strata variances, yielding
\begin{eqnarray*}
S^2(1) &=&  (n-1)^{-1} \sumn \{  Y_i(1) - \bar{Y}(1) \}^2 \\
&=& (n-1)^{-1} \sum_{k=1}^K \sum_{X_i   = k} \{ Y_i(1) - \bar{Y}_{[k]}(1)  + \bar{Y}_{[k]}(1)  -   \bar{Y}(1)  \}^2 \\
&=& (n-1)^{-1} \sum_{k=1}^K \sum_{X_i   = k} \left[  \{ Y_i(1) - \bar{Y}_{[k]}(1) \}^2 +  \{ \bar{Y}_{[k]}(1)  -   \bar{Y}(1)  \}^2 \right] \\
&=& \sum_{k=1}^K \left[   \frac{n_{[k]} - 1}{n-1}   S_{[k]}^2(1)  +  \frac{ n_{[k]}}{ n-1} \{ \bar{Y}_{[k]}(1) - \bar{Y}(1) \}^2  \right],
\end{eqnarray*} 
and similarly,
\begin{eqnarray*}
S^2(0) &=& \sum_{k=1}^K \left[   \frac{n_{[k]} - 1}{n-1}   S_{[k]}^2(0)  +  \frac{ n_{[k]}}{ n-1} \{ \bar{Y}_{[k]}(0) - \bar{Y}(0) \}^2  \right],\\
S^2(\tau) &=& \sum_{k=1}^K \left[   \frac{n_{[k]} - 1}{n-1}   S_{[k]}^2(\tau)  +  \frac{ n_{[k]}}{ n-1} \{ \tau_{[k]} - \tau \}^2  \right] .
\end{eqnarray*}
The variance of the difference-in-means estimator under the CRE decomposes into 
\begin{eqnarray*}
&&\var_{\text{CRE}}(  \hat{\tau} )  \\
&=& \frac{S^2(1)}{ n_{1} } +  \frac{S^2(0)}{ n_{0} }
- \frac{  S^2(\tau) }{ n }  \\ 
&=& \sum_{k=1}^K  \left[   \frac{ n_{[k]} - 1}{(n-1)n_1}  S_{[k]}^2(1) + \frac{ n_{[k]} - 1}{ (n-1)n_0}  S_{[k]}^2(0) -  \frac{ n_{[k]} - 1}{(n-1)n}    S_{[k]}^2(\tau) \right] \\
&&  + \sum_{k=1}^K  \left[    \frac{ n_{[k]} - 1}{ (n-1)n_1}  \{ \bar{Y}_{[k]}(1) - \bar{Y}(1) \}^2 +  \frac{ n_{[k]} - 1}{ (n-1)n_0}  \{ \bar{Y}_{[k]}(0) - \bar{Y}(0) \}^2   \right. \\
&& \left.  - \frac{ n_{[k]} - 1  }{ (n-1)n} \{ \tau_{[k]} - \tau \}^2  \right] .
 \end{eqnarray*}
 With large $n_{[k]}$'s, it is approximately
 \begin{eqnarray*}
 &&\var_{\text{CRE}}(  \hat{\tau} )  \\
&\approx & 
\sum_{k=1}^K  \left[   \frac{ \pi_{[k]}}{n_1}  S_{[k]}^2(1) + \frac{ \pi_{[k]}}{n_0}  S_{[k]}^2(0) -  \frac{ \pi_{[k]}}{n}    S_{[k]}^2(\tau) \right] \\
&&  + \sum_{k=1}^K  \left[    \frac{ \pi_{[k]}}{n_1}  \{ \bar{Y}_{[k]}(1) - \bar{Y}(1) \}^2 +  \frac{ \pi_{[k]}}{n_0}  \{ \bar{Y}_{[k]}(0) - \bar{Y}(0) \}^2  
 - \frac{ \pi_{[k]}  }{n} \{ \tau_{[k]} - \tau \}^2  \right] .
\end{eqnarray*}
The constant propensity scores assumption ensures 
$$ 
\pi_{[k]} / n_{[k]1}  = 1/(ne), \quad   \pi_{[k]} / n_{[k]0} = 1/\{n(1-e) \},\quad \pi_{[k]} / n_{[k]} = 1/n,
$$
which allow us to  rewrite the variance of $\hat{\tau}_\textsc{S}$ under the SRE as
\begin{eqnarray*}
\var_\textsc{SRE}(\hat{\tau}_\textsc{S}) &= & \sum_{k=1}^K  \pi_{[k]}^2 \left[ \frac{S_{[k]}^2(1)}{ n_{[k]1} } +  \frac{S_{[k]}^2(0)}{ n_{[k]0} }
- \frac{  S_{[k]}^2(\tau) }{ n_{[k]} } \right]\\
&=&\sum_{k=1}^K  \left[   \frac{ \pi_{[k]}}{n_1}  S_{[k]}^2(1) + \frac{ \pi_{[k]}}{n_0}  S_{[k]}^2(0) -  \frac{ \pi_{[k]}}{n}    S_{[k]}^2(\tau) \right] .
\end{eqnarray*}
 With large $n_{[k]}$'s, approximately, the difference between $\var_{\text{CRE}}(  \hat{\tau} )$ and $\var_\textsc{SRE}(\hat{\tau}_\textsc{S})$ is 
$$
\sum_{k=1}^K  \left[    \frac{ \pi_{[k]}}{n_1}  \{ \bar{Y}_{[k]}(1) - \bar{Y}(1) \}^2 +  \frac{ \pi_{[k]}}{n_0}  \{ \bar{Y}_{[k]}(0) - \bar{Y}(0) \}^2  
 - \frac{  \pi_{[k]}  }{n}  ( \tau_{[k]} - \tau  ) ^2  \right] 
 $$
 which is nonnegative because it equals (see Problem \ref{hw::compare-CRE-SRE})
 \begin{eqnarray}
\label{eq::difference}
\sum_{k=1}^K\frac{ \pi_{[k]}}{n} 
 \left\{
\sqrt{ \frac{n_0}{n_1} } \{ \bar{Y}_{[k]}(1) - \bar{Y}(1) \} +  \sqrt{ \frac{n_1}{n_0} } \{ \bar{Y}_{[k]}(0) - \bar{Y}(0) \}
 \right\}^2 \geq 0,
\end{eqnarray} 
The difference in \eqref{eq::difference} is zero only in the extreme case that
$$
\sqrt{ \frac{n_0}{n_1} } \{ \bar{Y}_{[k]}(1) - \bar{Y}(1) \} +  \sqrt{ \frac{n_1}{n_0} } \{ \bar{Y}_{[k]}(0) - \bar{Y}(0) \}
= 0 
$$
for $k=1,\ldots, K$. When the covariate is predictive of the potential outcomes, the above quantities are usually not all zeros, which ensures the efficiency gain of the SRE compared with the CRE. Only in the extreme cases that the covariate is not predictive at all, the large-sample efficiency gain is zero. In those cases, the SRE can even result in less efficient estimators in finite samples. The above discussion corroborates the quote from George Box at the beginning of this chapter.

I will end this section with several remarks. First, the above comparison is based on the sampling variance, and we can also compare the estimated variances under the SRE and the CRE. The results are similar. Second,  increasing $K$ improves efficiency, but this argument depends on the large strata assumption. So we face a trade-off in practice.  We cannot arbitrarily increase $K$, and the most extreme case is $n_{[k]1} = n_{[k]0} = 1$, which is called the matched-pairs experiment and will be discussed in Chapter \ref{chapter::mpe}.

\section{Post-stratification in a CRE}\label{sec::poststra-cre}

In a CRE with a discrete covariate $X$, the numbers of units receiving the treatment and control are random within stratum $k$. In a SRE, these numbers are fixed. But if we conduct conditional inference given $\bm{n} = \{  n_{[k]1} , n_{[k]0} \}_{k=1}^K$, then a CRE becomes a SRE. Mathematically, if none of the components of $\bm{n}$ are zero, then 
\begin{equation}\label{eq::condition-to-SRE}
\pr_\textsc{CRE}(\bm{Z} = \bm{z} \mid\bm{n} )
= \frac{ \pr_\textsc{CRE}( \bm{Z} = \bm{z},  \bm{n})   }{ \pr_\textsc{CRE}(   \bm{n})   }
=  \frac{1}{  \prod_{k=1}^K  \binom{  n_{[k]} }{  n_{[k]1} } }  ,
\end{equation}
that is, the conditional distribution of $\bm{Z}$ from a CRE given $\bm{n}$ is identical to the distribution of $\bm{Z}$ from an SRE. 
I leave the proof of \eqref{eq::condition-to-SRE} as Problem \ref{hw::CRE-SRE}. 
So conditional on $\bm{n}$, we can analyze a CRE with a discrete covariate $X$ in the same way as in a SRE. 
In particular,  the FRT becomes a {\it conditional FRT}, and the Neymanian analysis becomes {\it post-stratification}:
$$
\hat{\tau}_\textsc{PS} = \sum_{k=1}^K \pi_{[k]} \hat{\tau}_{[k]},
$$
which has an identical form as $\hat{\tau}_\textsc{S}$. The variance of $\hat{\tau}_\textsc{PS} $ conditioning on $\bm{n}$ is identical to the variance of $\hat{\tau}_\textsc{S}$ under the SRE.

\citet{hennessy2016conditional} used simulation to show that the conditional FRT is often more powerful than the unconditional one. \citet{miratrix:2013} used theory to show that in many cases, post-stratification improves efficiency compared with $\hat{\tau}$. Both results hold if $X$ is predictive of the outcome.  However, the simulation is based on a limited number of data-generating processes, and the theory assumes all strata are large enough. We can not go too extreme in the conditional FRT or post-stratification because with a larger $K$ it is more likely that some $n_{[k]1} $ or $ n_{[k]0}$ become zero. Small or zero values of $n_{[k]1} $ or $ n_{[k]0}$ greatly reduce the number of randomizations in the FRT, possibly reducing the power dramatically. The problem for the Neymanian counterpart is more salient because we cannot even define $\hat{\tau}_\textsc{PS}$ and the corresponding variance estimator.

Stratification uses $X$ in the design stage and post-stratification uses $X$ in the analysis stage.  They are duals for using $X$. 
Asymptotically, their difference is small with large strata \citep{miratrix:2013}.

\subsection{\citet{meinert1970study}'s Example}\label{sec::ps-merinert}

We use the data from a CRE reported in \citet{meinert1970study}, which were also used by \citet{rothman2008modern}. The treatment is tolbutamide and the control is a placebo. 
\begin{center} 
\begin{tabular}{ccc}
\multicolumn{3}{c}{ Age $< 55$} \\
\hline 
 & Surviving & Dead \\
$Z=1$ &    98 & 8 \\
$Z=0$  & 115 & 5 \\
\hline 
\end{tabular}
\begin{tabular}{ccc}
\multicolumn{3}{c}{ Age $\geq  55$} \\
\hline 
 & Surviving & Dead \\
$Z=1$ &    76 & 22 \\
$Z=0$  &  69 & 16 \\
\hline 
\end{tabular}
\begin{tabular}{ccc}
\multicolumn{3}{c}{Total } \\
\hline 
 & Surviving & Dead \\
$Z=1$ &    174 & 30 \\
$Z=0$  & 184 & 21 \\
\hline 
\end{tabular}
\end{center}

The following table shows the estimates for two strata separately, the post-stratified estimator, and the crude estimator ignoring the binary covariate, as well as the corresponding standard errors.  
\begin{center}
\begin{tabular}{crrrr}
\hline 
    &stratum 1& stratum 2 &post-stratification &  crude \\
est &    $-0.034$ &   $-0.036$ &  $-0.035$ & $-0.045$ \\
se   &   $0.031$    & $0.060$   & $0.032$   & $0.033$\\
\hline 
\end{tabular}
\end{center}

The crude estimator and the post-stratification estimator do not lead to fundamentally different results.
However, the crude estimator is larger than both of the stratum-specific estimators, whereas the post-stratification estimator is within the range.

\subsection{\citet{chong2016iron}'s Example}\label{sec::ps-chong}

\citet{chong2016iron} conducted an SRE on   219 students of a rural secondary school in the Cajamarca district of Peru during the 2009 school year. They first provided the village clinic with iron supplements and trained the local staff to distribute one free iron pill to any adolescent who requested one in person. They then randomly assigned students to three arms with three different types of videos: in the first video, a popular soccer player was encouraging the use of iron supplements to maximize energy (``soccer'' arm); in the second video, a physician was encouraging the use of iron supplements to improve overall health (``physician'' arm); the third video did not mention iron at all (``control'' arm). The experiment was stratified on the class level (1--5). The treatment and control group sizes within classes are shown below:

\begin{lstlisting}
> library("foreign")
> dat_chong = read.dta("chong.dta")
> table(dat_chong$treatment, dat_chong$class_level)
               
                 1  2  3  4  5
  Soccer Player 16 19 15 10 10
  Physician     17 20 15 11 10
  Placebo       15 19 16 12 10
\end{lstlisting}

One outcome of interest is the average grades in the third and fourth quarters of 2009, and an important background covariate was the anemia status at baseline. 
I will only use a subset of the original data in this chapter. 

\begin{lstlisting}
> use.vars = c("treatment", 
+              "gradesq34", 
+              "class_level", 
+              "anemic_base_re")
> dat_physician = subset(dat_chong,
+                        treatment != "Soccer Player",
+                        select = use.vars)
> dat_physician$z = (dat_physician$treatment=="Physician")
> dat_physician$y = dat_physician$gradesq34
> table(dat_physician$z, 
+       dat_physician$class_level)
       
         1  2  3  4  5
  FALSE 15 19 16 12 10
  TRUE  17 20 15 11 10
> table(dat_physician$z, 
+       dat_physician$class_level,
+       dat_physician$anemic_base_re)
, ,  = No

       
         1  2  3  4  5
  FALSE  6 14 12  7  4
  TRUE   8 12  9  5  6

, ,  = Yes

       
         1  2  3  4  5
  FALSE  9  5  4  5  6
  TRUE   9  8  6  6  4
\end{lstlisting}

We can use the \ri{Neyman_SRE} function defined before to compute the stratified estimator and its estimated variance. 
\begin{lstlisting}
tauS = with(dat_physician,
            Neyman_SRE(z, gradesq34, class_level))
\end{lstlisting}

An important additional covariate is the baseline anemic indicator which is quite important for predicting the outcome. Further conditioning the baseline anemic indicator, we have an experiment with $5\times 2 =10$ strata, with the treatment and control group sizes shown above. 
 Again we can use the \ri{Neyman_SRE} function defined before to compute the post-stratified estimator and its estimated variance. 
 \begin{lstlisting}
> tauSPS = with(dat_physician, {
+   sps = interaction(class_level, anemic_base_re)
+   Neyman_SRE(z, gradesq34, sps)
+ })
\end{lstlisting}

The following table compares these two estimators. The post-stratified estimator yields a much smaller $p$-value. 
\begin{center}
\begin{tabular}{crrrr}
                          & est &   se &t.stat& p.value \\
stratify                & 0.406 &0.202 & 2.005  & 0.045 \\
stratify and post-stratify & 0.463 &0.190 & 2.434 &  0.015
\end{tabular}
\end{center} 

This example illustrates that post-stratification can be used not only in the CRE but also in the SRE with additional discrete covariates.

\section{Practical questions}

How do we choose $X$ to construct an SRE? Theoretically, $X$ should be predictive of the potential outcomes. In some cases, the experimenter has enough background knowledge about the predictive covariates based on, for example, some pilot studies. 
Then the choice of $X$ should be straightforward. In some other cases, this background knowledge may not be clear enough. Experimenters instead choose $X$ based on logistical convenience, for example, $X$ can be the indicator for the study areas or the cohort of students.

The choice of $K$ is a related problem. Theoretically, more stratification increases the estimation efficiency if all strata are large enough. However, extremely large $K$ may even decrease the estimation efficiency. In simulation studies, we observe diminishing marginal returns of increasing $K$. Anecdotally, $K = 5$ often suffices for efficiency gain (the magic number $5$ will appear again in Chapter \ref{chapter::pscore-key}). Some experimenter prefers the most extreme version of the SRE with $K=n/2$. This results in the matched-pairs experiment, which will be discussed in Chapter \ref{chapter::mpe} later.

Some experiments have multidimensional continuous covariates. Can the SRE still be used? If we have a pilot study, we can build a model for the potential outcome $Y(0)$ given those covariates, and then we can choose $X$ as a discretized version of the predictor $\hat{Y}(0)$. In general, if we do not have such a pilot study or we do not want to make ad hoc discretizations, we can use a more general strategy called rerandomization, which will be the topic for Chapter \ref{chapter:rerandomization-regression}.

 \section{Homework Problems}

  \paragraph{Covariate balance in the CRE}\label{hw::balance-discrete-CRE}
  
 Under the CRE, prove \eqref{eq::balance-discrete-CRE}. 
 
 \paragraph{Consequence of the constant propensity score}
\label{hw::sre-additional-neyman}

Prove \eqref{eq::dim=stratified}.

\paragraph{Consquence of constant individual causal effects}

Assume that the individual causal effects are constant $\tau_i = \tau$ for all $i=1,\ldots, n$. 
Consider the following class of weighted estimator for $\tau$:
$$
\hat{\tau}_w = \sum_{k=1}^K  w_{[k]}  \hat{\tau}_{[k]} ,
$$
where the weights $w_{[k]}  $'s are non-negative for all $k$.

Find the condition on the $w_{[k]}$'s such that $\hat{\tau}_w $ is unbiased for $\tau$. Among all unbiased estimators, find the weights that give the $\hat{\tau}_w$ with the minimum variance.

\paragraph{Compare the CRE and SRE}\label{hw::compare-CRE-SRE}

Prove \eqref{eq::difference}.

\paragraph{From the CRE to the SRE}\label{hw::CRE-SRE}

Prove \eqref{eq::condition-to-SRE}.

 \paragraph{More FRTs for Section \ref{sec::frt-pennbonus}}\label{hw::FRT-SRE-more}
 
 Extend the analysis in Section \ref{sec::frt-pennbonus} using FRTs with other test statistics.

 \paragraph{FRT for an SRE in \citet{imbens2015causal}}

\citet{imbens2015causal} discussed an SRE from the Student/Teacher Achievement Ratio experiment conducted in 1985--1986 in Tennessee. 
The kindergarten data are below:
 \begin{lstlisting}
treatment = list(c(1,1,0,0), 
                 c(1,1,0,0), 
                 c(1,1,1,0,0), 
                 c(1,1,0,0),
                 c(1,1,0,0),
                 c(1,1,0,0),
                 c(1,1,0,0), 
                 c(1,1,1,1,0,0),
                 c(1,1,0,0),
                 c(1,1,0,0),
                 c(1,1,0,0),
                 c(1,1,1,0,0),
                 c(1,1,0,0),
                 c(1,1,0,0),
                 c(1,1,0,0),
                 c(1,1,0,0))
outcome = list(c(0.165,0.321,-0.197,0.236), 
               c(0.918,-0.202,1.19,0.117),
               c(0.341,0.561,-0.059,-0.496,0.225),
               c(-0.024,-0.450,-1.104,-0.956), 
               c(-0.258,-0.083,-0.126,0.106),
               c(1.151,0.707,0.597,-0.495), 
               c(0.077,0.371,0.685,0.270), 
               c(-0.870,-0.496,-0.444,0.392,-0.934,-0.633), 
               c(-0.568,-1.189,-0.891,-0.856), 
               c(-0.727,-0.580,-0.473,-0.807), 
               c(-0.533,0.458,-0.383,0.313), 
               c(1.001,0.102,0.484,0.474,0.140), 
               c(0.855,0.509,0.205,0.296), 
               c(0.618,0.978,0.742,0.175), 
               c(-0.545,0.234,-0.434,-0.293), 
               c(-0.240,-0.150,0.355,-0.130))
\end{lstlisting}
The strata correspond to schools, and the unit of analysis is the teacher or class. The treatment equals $1$ for small classes (13--17 students per teacher) and $0$ for regular classes (22--25 students per teacher). The outcome is the standardized average mathematics score.

Reanalyze the Project STAR data below using the FRT.  Use $\hat{\tau}_\textsc{S}$, $W_\textsc{S}$ and $\tilde W$ in the FRT. Compare the $p$-values.

Remark: This book uses $Z$ for the treatment indicator but \citet{imbens2015causal} use $W.$

\paragraph{A multi-center trial}

\citet[][Table 1]{gould1998multi} reported the following data from a multi-center trial:

\begin{lstlisting}
> multicenter = read.csv("multicenter.csv")
> multicenter
   center n0 mean0  sd0 n1 mean1  sd1 n5 mean5  sd5
1       1  7  0.43 4.58  7 -5.43 5.53  8 -2.63 3.38
2       2 11  0.10 4.21 11 -2.59 3.95 12 -2.21 4.14
3       3  6  2.58 4.80  6 -3.94 4.25  7  1.29 7.39
4       4 10 -2.30 3.86 10 -1.23 5.17 10 -1.40 2.27
5       5 10  2.08 6.46 10 -6.70 7.45 10 -5.13 3.91
6       6  6  1.13 3.24  5  3.40 8.17  5 -1.59 3.19
7       7  5  1.20 7.85  6 -3.67 4.89  5 -1.40 2.61
8       8 12 -1.21 2.66 13  0.18 3.81 12 -4.08 6.32
9       9  8  1.13 5.28  8 -2.19 5.17  9 -1.96 5.84
10     10  9 -0.11 3.62 10 -2.00 5.35 10  0.60 3.53
11     11 15 -4.37 6.12 14 -2.68 5.34 15 -2.14 4.27
12     12  8 -1.06 5.27  9  0.44 4.39  9 -2.03 5.76
13     13 12 -0.08 3.32 12 -4.60 6.16 11 -6.22 5.33
14     14  9  0.00 5.20  9 -0.25 8.23  7 -3.29 5.12
15     15  6  1.83 5.85  7 -1.23 4.33  6 -1.00 2.61
16     16 14 -4.21 7.53 14 -2.10 5.78 12 -5.75 5.63
17     17 13  0.76 3.82 13  0.55 2.53 13 -0.63 5.41
18     18 15 -1.05 4.54 13  2.54 4.16 14 -2.80 2.89
19     19 15  2.07 4.88 15 -1.67 4.95 15 -3.43 4.71
20     20 11 -1.46 5.48 10 -1.99 5.63 10 -6.77 5.19
21     21  5  0.80 4.21  5 -3.35 4.73  5 -0.23 4.14
22     22 11 -2.92 5.42 10 -1.22 5.95 11 -4.45 6.65
23     23  9 -3.37 4.73  9 -1.38 4.17  7  0.57 2.70
24     24 12 -1.92 2.91 12 -0.66 3.55 12 -2.39 2.27
25     25  9 -3.89 4.76  9 -3.22 5.54  8 -1.23 4.91
26     26 15 -3.48 5.98 15 -2.13 3.25 14 -3.71 5.30
27     27 11 -1.91 6.49 12 -1.33 4.40 11 -1.52 4.68
28     28 10 -2.66 3.80 10 -1.29 3.18 10 -4.70 3.43
29     29 13 -0.77 4.73 13 -2.31 3.88 13 -0.47 4.95
\end{lstlisting}

This is a SRE with centers being the strata. 
The trial was conducted to study the efficacy and tolerability of finasteride, a drug for treating benign prostatic hyperplasia. 
Within each of the $29$ centers, patients were randomized into three arms: control, finasteride 1mg, and finasteride 5mg. The above dataset
provides summary statistics for the outcome, which is the change from baseline in total symptom score. The total symptom score is the sum of the responses to nine questions (score $0$ to $4$) about symptoms of various aspects of impaired urinary ability. The meanings of the columns are:

\begin{enumerate}
\item
\texttt{center}: ID of the centers;
\item
\texttt{n0, n1, n5}: sample sizes under the three arms;
\item
\texttt{mean0, mean1, mean5}: means of the outcomes under the three arms;
\item
\texttt{sd0, sd1, sd5}: standard deviations of the outcomes under the three arms. 
\end{enumerate}

The individual-level outcomes are not reported so we cannot implement the FRT. However, the Neymanian inference only requires the summary statistics. Report the point estimators and variance estimators for comparing ``finasteride 1mg'' and ``finasteride 5mg'' to ``control'', separately.

\paragraph{Data re-analyses}

Re-analyze the LaLonde data used in Chapter \ref{sec::example-neyman-formula}. Conduct both Fisherian and Neymanian inferences. 

The original experiment is a CRE. Now we pretend that the original experiment is an SRE.
First, re-analyze the data pretending that the experiment is stratified on the race (black, Hispanic, or other). Second, re-analyze the data pretending that the experiment is stratified on marital status. Third, re-analyze the data pretending that the experiment is stratified on the indicator of a high school diploma. 

Compare with the results obtained under a CRE.

\paragraph{Recommended reading}

\citet{miratrix:2013} provided a solid theory for post-stratification and compared it with stratification. A main theoretical result is that their difference is small asymptotically although they can differ in finite samples.

\chapter{Rerandomization and Regression Adjustment}
 \label{chapter:rerandomization-regression}

Stratification and post-stratification in Chapter  \ref{chapter::stratification-poststratification} are duals for discrete covariates in the design and analysis of randomized experiments.  How should we deal with multidimensional, possibly continuous, covariates? We can discretize continuous covariates, but this is not an ideal strategy with many covariates. Rerandomization and regression adjustment are duals for general covariates, which are the topics for this chapter.

The following table summarizes the topics of Chapters  \ref{chapter::stratification-poststratification} and \ref{chapter:rerandomization-regression}:

\begin{center}
\begin{tabular}{c|c|c}
\hline 
 & design & analysis \\ 
 \hline 
 discrete covariate & stratification & post-stratification \\
 \hline 
 general covariate & rerandomization & regression adjustment \\
 \hline 
 \end{tabular}
\end{center}

\section{Rerandomization}

\subsection{Experimental design}
Again we consider a finite population of $n$ experimental units, where $n_1$ of them receive the treatment and $n_0$ of them receive the control. Let
 $\bm{Z} = (Z_1,\ldots, Z_n)$ be the treatment vector for these units. Unit $i$ has covariate $X_i \in \mathbb{R}^K$ which can have continuous or binary components. Concatenate them as an $n\times K$ covariate matrix $\bm{X} = (X_1 , \ldots, X_n )  \tran$ and center them at mean zero $\bar{X}  = n^{-1} \sumn X_i = 0$ to simplify the presentation. 

The CRE balances the covariates in the treatment and control groups on average, for instance, the difference in means of the covariates
$$
\hat{\tau}_X  = n_1^{-1} \sumn Z_i X_i - n_0^{-1} \sumn (1-Z_i) X_i
$$
has mean zero under the CRE. However, it can result in undesired covariate balance across the treatment and control groups in the realized treatment allocation, that is, the realized value of $\hat{\tau}_X  $ is often not zero. Using the vector form of \citet{Neyman:1923} in Problem \ref{hw::vector-neyman1923} before, we can show that
$$
\cov(\hat{\tau}_X  ) 
= \frac{1}{n_1}  S_X^2  + \frac{1}{n_0}  S_X^2   
=  \frac{n}{n_1 n_0} S_X^2   ,
$$
where $ S_X^2   = (n-1)^{-1} \sumn  X_i X_i\tran $ is the finite-population covariance matrix of the covariates. The following Mahalanobis distance measures the difference between the treatment and control groups:
\begin{equation}
\label{eq::m-distance-def}
M = \hat{\tau}_X \tran \cov(\hat{\tau}_X  )^{-1} \hat{\tau}_X  = \hat{\tau}_X \tran  \left(  \frac{n}{n_1 n_0} S_X^2   \right)^{-1} \hat{\tau}_X  . 
\end{equation}
Technically, the above formula for $M$ in \eqref{eq::m-distance-def} is meaningful only if $S_X^2 $ is invertible, which means that the columns of the covariate matrix $\bm{X}$ are linearly independent. If a column can be represented by a linear combination of other columns, it is redundant and should be dropped before the experiment. 
A nice feature of $M$ is that it is invariant under non-degenerate linear transformations of $X$. Lemma \ref{lemma::invariance-mdistance} below summarizes the result with the proof relegated to Problem \ref{para::invariance-M}. 

\begin{lemma}\label{lemma::invariance-mdistance}
$M$ in \eqref{eq::m-distance-def} remains the same if we transform $X_i$ to $b_0  + BX_i$ for all units $i=1,\ldots, n$ where $b_0  \in \mathbb{R}^K$ and $B \in \mathbb{R}^{K\times K}$ is invertible. 
\end{lemma}

The finite population CLT \citep{li2017general} ensures that with large $n$, the Mahalanobis distance $M $ is approximately $ \chi^2_K$ under the CRE. Therefore, it is likely that $M$ has a large realized value under the CRE with asymptotic mean $K$ and variance $2K$. Rerandomization avoids covariate imbalance by discarding the treatment allocations with large values of $M$. Below I give a formal definition of the rerandomization using the Mahalanobis distance (ReM), which was proposed by \citet{cox:1982} and further studied by \citet{morgan2012rerandomization}.

\begin{definition}
[ReM]
Draw $\bm{Z}$ from CRE and accept it if and only if 
$$
M \leq a,
$$  
for some predetermined constant $a > 0$. 
\end{definition}

The problem of choosing $a$ is similar to the problem of choosing the number of strata in the SRE, which is non-trivial in practice. At one extreme, $a = \infty$, we just conduct the CRE. At the other extreme, $a=0$, there are very few feasible treatment allocations, and consequently, the experiment has little randomness, rendering randomization-based inference useless. As a compromise, we choose a small but not extremely small $a$, for example, $a=0.001$ or some upper quantile of a $\chi^2_K$ distribution.

ReM has many desirable properties. As mentioned above, it is invariant to linear transformations of the covariates. Moreover, it has nice geometric properties and elegant mathematical theory. This chapter will focus on ReM.

\subsection{Statistical inference}

An important question is how to analyze the data under ReM. \citet{bruhn2009pursuit} and \citet{morgan2012rerandomization} argued that we can always use the FRT as long as we simulate $\bm{Z}$ under the constraint $M\leq a$. This always yields finite-sample exact $p$-values under the sharp null hypothesis. See Problem \ref{hw::FRT-under-REM}.

It is a challenging problem to derive the finite sample properties of ReM without assuming the sharp null hypothesis. Instead, \citet{li2018asymptotic} derived the asymptotic distribution of the difference in means of the outcome $\hat{\tau}$ under ReM and the regularity conditions below.

\begin{condition}\label{condition::regularity-cre-rem}
As $n\rightarrow \infty$,
\begin{enumerate}
\item 
$n_1/n$ and $n_0/n$ have positive limits;
\item 
the finite population covariance of $\{  X_i, Y_i(1), Y_i(0), \tau_i \}$ has a finite limit;
\item
$\max_{1\leq i \leq n}  \{   Y_i(1) - \bar{Y}(1) \} ^2/n \rightarrow 0 $,
$\max_{1\leq i \leq n}  \{   Y_i(0) - \bar{Y}(0) \}^2/n \rightarrow 0 $,
and $\max_{1\leq i \leq n}   X_i \tran X_i/n \rightarrow 0 $,
\end{enumerate}
\end{condition}

Below is the main theorem for ReM, which relies on additional notation.    
Let 
$$
L_{K,a} \sim D_1\mid \bm{D}\tran \bm{D} \leq a
$$ 
where $\bm{D} = (D_1,\ldots, D_K)$ follows a $K$-dimensional standard Normal distribution; let $ \varepsilon$ follow a univariate standard Normal distribution; $L_{K,a}  \ind \varepsilon .$

\begin{theorem}\label{thm::remasymptotics}
Under ReM with $M\leq a$ and Condition \ref{condition::regularity-cre-rem}, we have\footnote{The notation ``$A \asim B$'' means that $A$ and $B$ have the same asymptotic distributions.} 
$$
\hat{\tau} - \tau \asim \sqrt{ \var(\hat \tau) }  \left\{   \sqrt{R^2} L_{K,a}  + \sqrt{1-R^2} \varepsilon \right\},
$$
where 
$$
\var( \hat{\tau} ) = \frac{S^2(1)}{n_1} + \frac{S^2(0)}{n_0} - \frac{S^2(\tau)}{n}
$$
is \citet{Neyman:1923}'s variance formula proved in Chapter \ref{chapter::neyman-cr}, and
$$
R^2 = \textup{corr}^2(\hat{\tau}, \hat{\tau}_X   )
$$
is the squared multiple correlation coefficient (see Section \ref{sec::measure-pearson-r2} for the definition)  between $\hat{\tau}$ and $\hat{\tau}_X $ under the CRE.
\end{theorem}

\begin{figure}
\centering
\includegraphics[width = 0.9\textwidth]{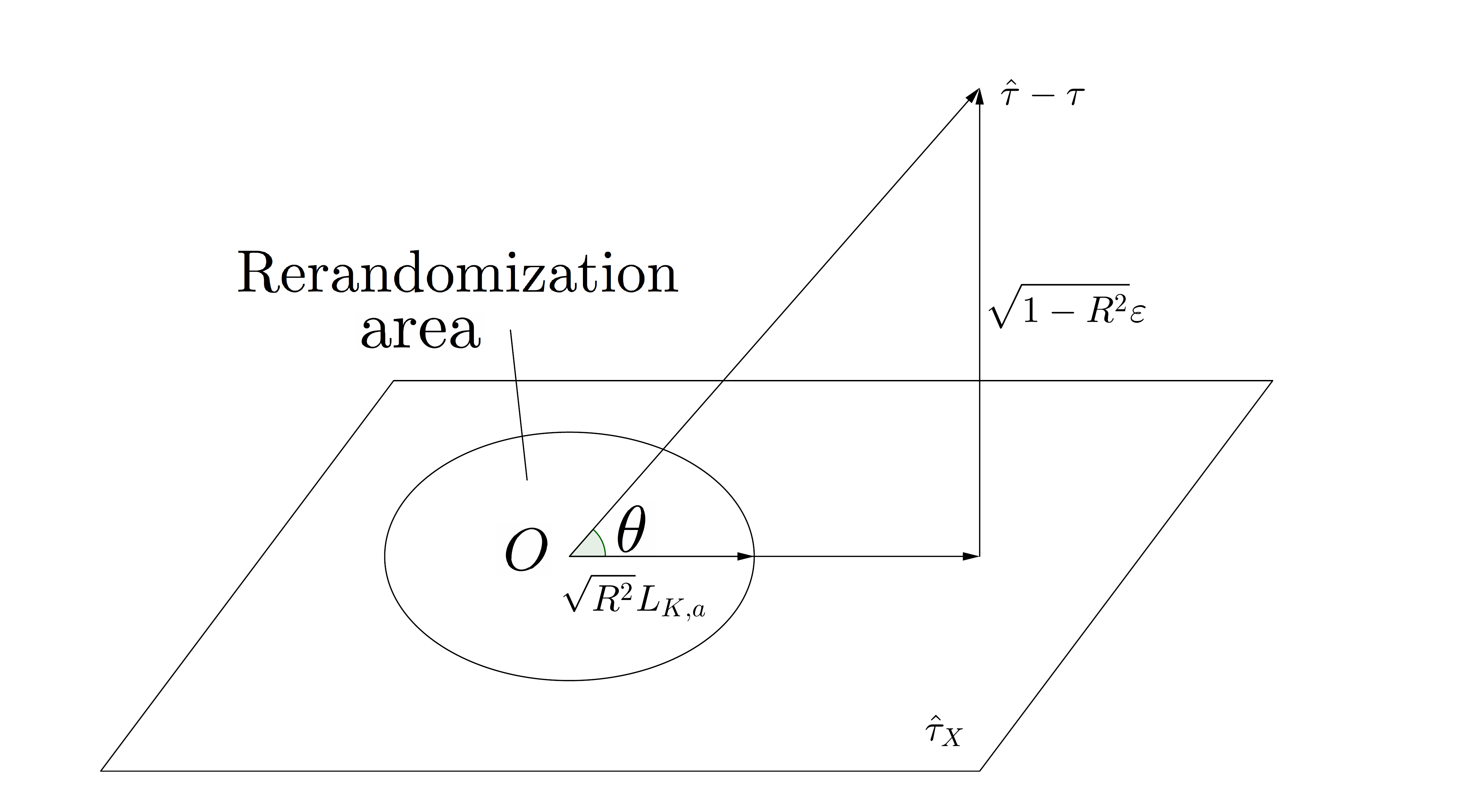}
\caption{Geometry of the asymptotic distribution of $\hat\tau$  under  ReM}\label{fig::geometry-rem}
\end{figure}

Although the proof of \citet{li2018asymptotic} is technical, the asymptotic distribution in Theorem \ref{thm::remasymptotics} has a clear geometric interpretation, as shown in Figure \ref{fig::geometry-rem}. It shows that $\hat\tau$ decomposes into a component that is a linear combination of $\hat\tau_X$ and a component that is orthogonal to $\hat\tau_X$. Geometrically, $\cos^2 \theta = R^2$, where $\theta$ is the angle between $\hat\tau$ and $\hat\tau_X$.
ReM affects the first component but does not change the second component. The truncated Normal distribution $L_{K,a}$ is due to the restriction of ReM on the first component.

When $a  = \infty$, the asymptotic distribution  simplifies to the one under the CRE:
$$
\hat{\tau} - \tau \asim \sqrt{ \var(\hat \tau) } \varepsilon .
$$
When the threshold $a$ is close to zero, the asymptotic distribution  simplifies to
$$
\hat{\tau} - \tau \asim \sqrt{ \var(\hat \tau) (1-R^2)} \varepsilon ;
$$
see \citet{wang2022rerandomization} for a rigorous proof.
So with a small threshold $a$, the efficiency gain due to ReM depends on $R^2$, which has the following equivalent form. 
\begin{proposition}
\label{prop::rsquared-forms}
Under the CRE, we have 
$$
R^2 = \textup{corr}^2(\hat{\tau}, \hat{\tau}_X   )
= \frac{  n_1^{-1}  S^2(1\mid X) + n_0^{-1} S^2(0\mid X) - n^{-1} S^2(\tau \mid X) }
{ n_1^{-1}  S^2( 1 ) + n_0^{-1} S^2( 0 ) - n^{-1} S^2( \tau  )},
$$
where $\{S^2( 1 ) , S^2( 0 ), S^2( \tau  )\}$ are the finite population variances of $\{ Y_i(1), Y_i(0), \tau_i \}_{i=1}^n$, and $\{S^2( 1 \mid x ) , S^2( 0 \mid x ), S^2( \tau  \mid x )\}$ are the corresponding finite population variances of their linear projections on $(1, X_i)$; see Section \ref{sec::sampleOLS} for the definition of linear projections. 
Under the constant causal effect assumption with $\tau_i = \tau$, the $R^2 $ reduces to $S^2(0\mid X) / S^2( 0 ) $, the finite population squared multiple correlation between $Y_i(0)$ and $X_i$.
\end{proposition}

I leave the proof of Proposition \ref{prop::rsquared-forms} to Problem \ref{para::r2-cre}. 

When $ 0 < a <\infty$, the asymptotic distribution of $\hat{\tau}$ has a more complicated form and is more concentrated at $\tau$ and thus the difference in means $\hat{\tau}$ is more precise under ReM than under the CRE.

If we ignore the design of ReM and still use the confidence interval based on \cite{Neyman:1923}'s variance formula and the Normal approximation, it is overly conservative and overcovers $\tau$ even if the individual causal effects are constant. \citet{li2018asymptotic} proposed to construct confidence intervals based on Theorem \ref{thm::remasymptotics}. We omit the discussion here but will come back to the inference issue in Section \ref{sec::unification-rem-regadj}.

\section{Regression adjustment}

What if we do not conduct rerandomization in the design stage but want to adjust for covariate imbalance in the analysis stage of the CRE? We will discuss several regression adjustment strategies. 

\subsection{Covariate-adjusted FRT}\label{sec::covariate-adjusted-frt-intro}

The covariates $\bm{X}$ are all fixed, and furthermore, under $\HF$, the observed outcomes are all fixed. Therefore, we can simulate the distribution of any test statistic $T = T(\bm{Z}, \bm{Y}, \bm{X})$ and calculate the $p$-value. The basic idea of the FRT remains the same in the presence of additional covariates.

There are two general strategies to construct the test statistic. Problem \ref{hw::frt-covariates} before hints at both of them. I summarize them below, using the terminology from \citet{zhao2020caFRT}.

\begin{definition}
[pseudo-outcome strategy for covariate-adjusted FRT]
\label{def::pseudo-y-frt}
We can construct the test statistic based on residuals from fitted statistical models. We can regress $Y_i$ on $X_i$ to obtain residual $\hat \varepsilon_i$, and then treat $\hat \varepsilon_i$ as the pseudo outcome to construct test statistics.
\end{definition}

\begin{definition}
[model-output strategy for covariate-adjusted FRT]
\label{def::model-out-frt}
We can use a regression coefficient as a test statistic. We can regress  $Y_i$ on $(Z_i, X_i)$ to obtain the coefficient of $Z_i$ as the test statistic.
\end{definition}

In the pseudo-outcome strategy in Definition \ref{def::pseudo-y-frt}, the regression of $Y_i$ on $X_i$ should not include the treatment $Z_i$ because we want to ensure that the pseudo outcome satisfies $\HF$ if the original outcome satisfies $\HF$. 
In the model-output strategy in Definition \ref{def::model-out-frt}, the regression of $Y_i$ on $(Z_i, X_i)$ includes the treatment $Z_i$ because we want to use the coefficient of $Z_i$ to measure the deviation from $\HF$.  
Computationally, in strategy one, we only need to run regression once, whereas in strategy two, we need to run regression many times. 

In the above,  ``regression'' is a generic term, which can be linear regression, logistic regression, or even machine learning algorithms. The FRT with any test statistics from these two strategies will be finite-sample exact under $\HF$ although they differ under alternative hypotheses. 
The rest of this section will review some test statistics based on OLS.

\subsection{Analysis of covariance and extensions}\label{sec::ancova}

Now we turn to direct estimation of the average causal effect $\tau$ that adjusts for the observed covariates. 

Historically, \citet{fisher1925statistical} proposed to use the analysis of covariance (ANCOVA) to improve estimation efficiency. This remains a standard strategy in many fields. He suggested running the OLS of $Y_i$ on $(1, Z_i, X_i)$ and obtaining the coefficient of $Z_i$ as an estimator for $\tau$. Let $\hat{\tau}_\textsc{F}$ denote Fisher's ANCOVA estimator.

A former Berkeley Statistics Professor, David A. Freedman, reanalyzed Fisher's ANCOVA under \citet{Neyman:1923}'s potential outcomes framework. \citet{freedman2008regression_a, freedman2008regression_b} found the following negative results:
\begin{enumerate}
[(1)]
\item
$\hat{\tau}_\textsc{F}$ is biased, but the simple difference in means $\hat{\tau}$ is unbiased.
\item 
The asymptotic variance of $\hat{\tau}_\textsc{F}$ may be even larger than that of $\hat{\tau}$ when $n_1\neq n_0$.
\item 
The standard error from the OLS is inconsistent for the true standard error of $\hat{\tau}_\textsc{F}$ under the CRE.   
\end{enumerate}

A former Berkeley Ph.D. student, Winston Lin, wrote a thesis in response to Freedman's critiques. \citet{lin2013} found the following positive results:
\begin{enumerate}
[(1)]
\item
The bias of $\hat{\tau}_\textsc{F}$ is small in large samples, and it goes to zero as the sample size approaches infinity.
\item
We can improve the asymptotic efficiency of both $\hat{\tau}$ and $\hat{\tau}_\textsc{F}$ by using the coefficient of $Z_i$ in the OLS of $Y_i$ on $(1, Z_i, X_i, Z_i\times X_i)$. Let $\hat{\tau}_\textsc{L}$ denote \citet{lin2013}'s estimator. Moreover, the EHW standard error is a conservative estimator for the true standard error of $\hat{\tau}_\textsc{L}$ under the CRE. 
\item
The EHW standard error\footnote{Without covariates, the HC2 correction yields identical variance estimator as \citet{Neyman:1923}'s classic one; see Problem \ref{hw::neyman-ols-nox}. For coherence, we can also use the HC2 correction for \citet{lin2013}'s estimator with covariate adjustment. When the number of covariates is small compared with the sample size and the covariates do not contain outliers, the variants of the EHW standard error perform similarly to the original one. When the number of covariates is large compared with the sample size or the covariates contain outliers, the variants can outperform the original one. In those cases, \citet{lei2018regression} recommend using the HC3 variant of the EHW standard error.} 
for $\hat{\tau}_\textsc{F}$ in the OLS fit of $Y_i$ on $(1, Z_i, X_i)$ is  a conservative estimator for the true standard error of $\hat{\tau}_\textsc{F}$ under the CRE. 
\end{enumerate}

\subsubsection{Some heuristics for \citet{lin2013}'s results}

\citet{Neyman:1923}'s result demonstrates that the variance of the difference-in-means estimator $\hat{\tau}$ depends on the variances of the potential outcomes. Intuitively, we can reduce the variance of the estimator by reducing the variances of the potential outcomes. A family of linearly adjusted estimators is
\begin{eqnarray} 
\hat{\tau}(\beta_1, \beta_0) &=& n_1^{-1} \sumn Z_i(Y_i - \beta_1\tran X_i) - n_0^{-1} \sumn (1-Z_i) (Y_i - \beta_0\tran X_i)  \label{eq::linear-adj-1}\\
&=& \left\{  \hat{\bar{Y}}(1) - \beta_1\tran \hat{\bar{X} }(1) \right\}  -  \left\{   \hat{\bar{Y}}(0) - \beta_0\tran \hat{\bar{X} }(0) \right\},
\label{eq::linear-adj-2}
\end{eqnarray}
where $\{ \hat{\bar{Y}}(1), \hat{\bar{Y}}(0) \}$ are the sample means of the outcomes, and $\{ \hat{\bar{X} }(1), \hat{\bar{X} }(0)\}$ are the sample means of the covariates. This covariate-adjusted estimator $\hat{\tau}(\beta_1, \beta_0)$ tries to reduce the variance of $\hat\tau$ by residualizing the potential outcomes. It reduces to $\hat\tau$ with $\beta_1 = \beta_0 = 0$. 
It has mean $\tau$ for any fixed values of $\beta_1$ and $\beta_0$ because $\bar{X}  = 0.$ We are interested in finding the $(\beta_1, \beta_0)$ that minimize the variance of $\hat{\tau}(\beta_1, \beta_0)$. This estimator is essentially the difference in means of the adjusted potential outcomes $\{ Y_i(1) - \beta_1\tran X_i  , Y_i(0) - \beta_0\tran  X_i \}_{i=1}^n $. Applying \citet{Neyman:1923}'s result, this estimator has variance
$$
\var\{ \hat{\tau}(\beta_1, \beta_0) \} = \frac{ {S}^2(1; \beta_1) }{ n_1 } + \frac{ {S}^2(0; \beta_0) }{ n_0 }
- \frac{  S^2(\tau; \beta_1, \beta_0)  }{ n },
$$
where $ {S}^2(z; \beta_z)$ $(z=1,0)$ and $S^2(\tau; \beta_1, \beta_0)$ are the finite population variances of the adjusted potential outcomes and individual causal effects, respectively; moreover,  a conservative variance estimate is 
$$
\hat{V}( \beta_1, \beta_0 ) = \frac{ \hat{S}^2(1; \beta_1) }{ n_1 } + \frac{ \hat{S}^2(0; \beta_0) }{ n_0 },
$$
where 
\begin{eqnarray*}
\hat{S}^2(1; \beta_1) &=& (n_1-1)^{-1} \sumn Z_i  ( Y_i - \gamma_1 - \beta_1\tran  X_i )^2,\\ 
\hat{S}^2(0; \beta_0) &=& (n_0-1)^{-1} \sumn (1-Z_i)  ( Y_i - \gamma_0 - \beta_0\tran  X_i )^2
\end{eqnarray*}
are the sample variances of the adjusted potential outcomes with $\gamma_1$ and $\gamma_0$ being the sample means of  $Y_i - \beta_1\tran  X_i$ under treatment and $Y_i   - \beta_0\tran  X_i $ under control. 
To minimize $\hat{V}( \beta_1, \beta_0 )$, we need to solve two OLS problems\footnote{
We can also minimize the true variance of $\hat{\tau}(\beta_1, \beta_0) $. 
See Problem \ref{hw::compare-true-variance} for more details. 
}:
$$
\min_{\gamma_1 , \beta_1 }
\sumn Z_i  ( Y_i - \gamma_1 - \beta_1\tran  X_i )^2,\quad
\min_{\gamma_0 , \beta_0 }
\sumn (1-Z_i)  ( Y_i - \gamma_0 - \beta_0\tran  X_i  )^2.
$$
We run OLS of $Y_i$ on $X_i$ for the treatment and control groups separately and obtain $( \hat{\gamma}_1, \hat{\beta}_1 )$ and $( \hat{\gamma}_0, \hat{\beta}_0 )$. The final estimator is
\begin{eqnarray*}
 \hat{\tau}( \hat{\beta}_1, \hat{\beta}_0) &=& n_1^{-1} \sumn Z_i(Y_i - \hat{\beta}_1\tran  X_i) 
- n_0^{-1} \sumn (1-Z_i)  (Y_i - \hat{\beta}_0\tran  X_i) \\
&=&\left\{  \hat{\bar{Y}}(1) - \hat \beta_1\tran \hat{\bar{X} }(1) \right\}  
-  \left\{   \hat{\bar{Y}}(0) - \hat \beta_0\tran \hat{\bar{X} }(0) \right\} . 
\end{eqnarray*}
From the properties of the OLS fits (see \eqref{eq::OLS-mean-data}), we know
$$
\hat{\bar{Y}}(1) =  \hat{\gamma}_1 + \hat{\beta}_1\tran  \hat{\bar{X} }(1),\quad
\hat{\bar{Y}}(0) =  \hat{\gamma}_0 + \hat{\beta}_0\tran  \hat{\bar{X} }(0).
$$
Therefore, we can rewrite the estimator as
\begin{equation}
\label{eq::lin-intercepts}
 \hat{\tau}( \hat{\beta}_1, \hat{\beta}_0)  =  \hat{\gamma}_1 - \hat{\gamma}_0  
\end{equation}
The equivalent form in \eqref{eq::lin-intercepts} suggests that we can obtain $ \hat{\tau}( \hat{\beta}_1, \hat{\beta}_0) $ from a single OLS fit below.

\begin{proposition}\label{prop::lin-ols}
The estimator $ \hat{\tau}( \hat{\beta}_1, \hat{\beta}_0) $ in \eqref{eq::lin-intercepts} equals the coefficient of $Z_i$ in the OLS fit of $Y_i$ on $(1, Z_i, X_i, Z_i\times X_i)$, which is \citet{lin2013}'s estimator $\hat{\tau}_\textsc{L}$ introduced before. 
\end{proposition}

I leave the proof of Proposition \ref{prop::lin-ols} to Problem \ref{para::lin-ols-covadj}, which is a pure linear algebra fact.

Based on the discussion above, a conservative variance estimator for $\hat{\tau}_\textsc{L}$ is 
\begin{eqnarray*}
\hat{V}(  \hat{\beta}_1, \hat{\beta}_0 ) 
&=& \frac{1}{n_1(n_1 - 1)} \sumn Z_i  ( Y_i - \hat{\gamma}_1 - \hat{\beta}_1\tran  X_i  ) ^2 \\
&& \qquad + \frac{1}{n_0(n_0 - 1)} \sumn (1-Z_i)  ( Y_i - \hat{\gamma}_0 - \hat{\beta}_0\tran  X_i )^2.
\end{eqnarray*}
Based on quite technical calculations, \citet{lin2013} further showed that the EHW standard error from the OLS in Proposition \ref{prop::lin-ols} is almost identical to $\hat{V}(  \hat{\beta}_1, \hat{\beta}_0 ) $ which is a conservative estimator of the true standard error of $\hat{\tau}_\textsc{L}$ under the CRE. Intuitively, this is because we do not assume that the linear model is correctly specified, and the EHW standard error is robust to model misspecification.

There is a subtle issue with the discussion above. The variance formula $\var\{ \hat{\tau}(\beta_1, \beta_0) \} $ works for fixed $(\beta_1, \beta_0)$, but the estimator $ \hat{\tau}( \hat{\beta}_1, \hat{\beta}_0) $ uses two estimated coefficients $( \hat{\beta}_1, \hat{\beta}_0) $. The additional uncertainty in the estimated coefficients may cause finite-sample bias in the final estimator. \citet{lin2013} showed that this issue goes away asymptotically because $ \hat{\tau}( \hat{\beta}_1, \hat{\beta}_0) $ behaves similarly to $ \hat{\tau}( \tilde{\beta}_1, \tilde{\beta}_0) $, where $\tilde{\beta}_1$ and $\tilde{\beta}_0$ are the limit of $\hat{\beta}_1$ and $ \hat{\beta}_0$, respectively. Heuristically, the difference between  $ \hat{\tau}( \hat{\beta}_1, \hat{\beta}_0) $ and $ \hat{\tau}( \tilde{\beta}_1, \tilde{\beta}_0) $ depends on 
$$
(\hat{\beta}_z - \tilde{\beta}_z)\tran \hat{\bar{X} }(z) ,\quad (z=0,1) 
$$
which is a product of two terms of small order. As a warning,  the asymptotic theory requires a large sample size and some regularity conditions on the potential outcomes and covariates. In finite samples, regression adjustment can be even harmful due to the additional uncertainty in the estimated coefficients $\hat{\beta}_1$ and $\hat{\beta}_0$.

\subsubsection{Understanding \citet{lin2013}'s estimator via predicting the potential outcomes}
\label{sec::understand-lin-predict}
 
\begin{table}
\centering
\caption{Predicting the potential outcomes}\label{tb::predict-po-ols}
\begin{tabular}{cccccc}
\hline 
$X$ & $Z$ & $Y(1)$ & $Y(0)$ &  $\hat{Y}(1)$ & $\hat{Y}(0)$ \\
\hline
$X_1 $ & 1 & $Y_1(1)$ & ? & $\hat{\mu}_1(X_1 )$ & $\hat{\mu}_0(X_1)$ \\
\vdots \\
$X_{n_1}$ & 1 & $Y_{n_1}(1)$ & ? & $\hat{\mu}_1(X_{n_1})$ & $\hat{\mu}_0(X_{n_1})$ \\
$X_{n_1 + 1}$ & 0 & ? & $Y_{n_1+1}(0)$ & $\hat{\mu}_1(X_{n_1+1})$ & $\hat{\mu}_0(X_{n_1+1})$ \\
\vdots \\
$X_{n}$ & 0 & ? & $Y_{n}(0)$ & $\hat{\mu}_1(X_{n})$ & $\hat{\mu}_0(X_{n})$ \\
\hline 
\end{tabular}
\end{table}

We can view \citet{lin2013}'s estimator as a {\it predictive estimator} based on OLS fits of the potential outcomes on the covariates. We build a prediction model for $Y(1)$ based on $X$ using the data from the treatment group: 
\begin{eqnarray}
\label{eq::pred-y1-linear}
\hat{\mu}_1(X) = \hat{\gamma}_1 + \hat{\beta}_1\tran X.
\end{eqnarray}
Similarly, we build a prediction model for $Y(0)$ based on $X$ using the data from the control group: 
\begin{eqnarray}
\label{eq::pred-y0-linear}
\hat{\mu}_0(X)= \hat{\gamma}_0 + \hat{\beta}_0\tran X . 
\end{eqnarray}
Table \ref{tb::predict-po-ols} illustrates the prediction of the potential outcomes. 
If we predict the missing potential outcomes, then we have the following predictive estimator:
\begin{eqnarray}\label{eq::predictive-estimator-generic}
\hat{\tau}_\text{pred} = n^{-1} \left\{   \sum_{Z_i=1}Y_i + \sum_{Z_i=0}\hat{\mu}_1(X_i) - \sum_{Z_i=1} \hat{\mu}_0(X_i) - \sum_{Z_i=0} Y_i     \right\} .
\end{eqnarray}
We can verify that with \eqref{eq::pred-y1-linear} and \eqref{eq::pred-y0-linear}, the predictive estimator equals \citet{lin2013}'s estimator:
\begin{eqnarray}
\label{eq::predictiveestimator}
\hat{\tau}_\text{pred} = \hat{\tau}_\textsc{L}.
\end{eqnarray}
If we predict all potential outcomes even if they are observed, we have the following {\it projective estimator}:
\begin{eqnarray}\label{eq::projective-estimator-generic}
\hat{\tau}_\text{proj} = n^{-1}  \sumn \{ \hat{\mu}_1(X_i) - \hat{\mu}_0(X_i)    \}.
\end{eqnarray}
We can verify that with \eqref{eq::pred-y1-linear} and \eqref{eq::pred-y0-linear}, the projective estimator equals \citet{lin2013}'s estimator:
\begin{eqnarray}
\label{eq::projectiveestimator}
\hat{\tau}_\text{proj} = \hat{\tau}_\textsc{L}.
\end{eqnarray}
I leave the proofs of \eqref{eq::predictiveestimator} and \eqref{eq::projectiveestimator} to Problem \ref{para::predictiveestimator}.

The terminology ``predictive'' and ``projective'' is from the survey sampling literature \citep{firth1998robust, ding2018causal}. 
The more general formulas \eqref{eq::predictive-estimator-generic} and \eqref{eq::projective-estimator-generic} are well-defined with other predictors of the potential outcomes. To make connections with \citet{lin2013}'s estimator, I focus on the linear predictors here. The predictors $\hat{\mu}_1(X)$ and $\hat{\mu}_0(X)$ can be quite general, including much more complicated machine learning algorithms. However, constructing a point estimator is just the first step in analyzing the CRE. A more important second step is to quantify the uncertainty associated with the estimator, which depends on the properties of the predictors of the potential outcomes. Nevertheless, without doing additional theoretical analysis, we can always use \eqref{eq::predictive-estimator-generic} and \eqref{eq::projective-estimator-generic}  as the test statistics in the FRT.

\subsubsection{Understanding \citet{lin2013}'s estimator via adjusting for covariate imbalance}
\label{sec::lin=adjustingforimbalance}

The linearly adjusted estimator has an equivalent form
\begin{eqnarray}\label{eq::linear-covariate-imbalance}
\hat{\tau}(\beta_1, \beta_0) = \hat{\tau} - \gamma\tran \hat{\tau}_X 
\end{eqnarray}
where $\gamma =\frac{n_0}{n} \beta_1 + \frac{n_1}{n} \beta_0$, so we can also write it as $\hat{\tau}(\gamma) = 
\hat{\tau}(\beta_1, \beta_0) .$ Similarly, \citet{lin2013}'s estimator has an equivalent form
\begin{eqnarray}\label{eq::lin-covariate-imbalance}
\hat{\tau}_\textsc{L} = \hat{\tau} - \hat \gamma\tran \hat{\tau}_X ,
\end{eqnarray}
where $\hat \gamma =\frac{n_0}{n} \hat \beta_1 + \frac{n_1}{n} \hat  \beta_0$. 
I leave the proofs of \eqref{eq::linear-covariate-imbalance} and \eqref{eq::lin-covariate-imbalance} to  Problem \ref{para::reg-covariate-imbalance}.  The forms \eqref{eq::linear-covariate-imbalance} and \eqref{eq::lin-covariate-imbalance} are the mathematical statements of ``adjusting for the covariate imbalance.'' They essentially subtract some linear combinations of the difference in means of the covariates. Since $\hat{\tau} $ and  $\hat{\tau}_X $ are correlated under the CRE, the covariate adjustment with an appropriately chosen $\gamma$ reduces the variance of $\hat{\tau} $. 
Another interesting feature of \eqref{eq::linear-covariate-imbalance} and \eqref{eq::lin-covariate-imbalance} is that the final estimators depend only on $\gamma$ or $\hat \gamma$, so the choice of the $\beta$-coefficients is not unique. \citet{li2019rerandomization} pointed out this simple fact.  Therefore, \citet{lin2013}'s estimator is just one of the optimal estimators. However, it can be easily implemented via the standard OLS with the EHW standard error. That's why this book focuses on it.

\subsection{Some additional remarks on regression adjustment}

\subsubsection{Duality between ReM and regression adjustment}
\citet{li2018asymptotic} pointed out that ReM and \citet{lin2013}'s regression adjustment are duals in using covariates in the design and analysis stages of the experiment. To be more specific, when $a$ is small,  the asymptotic distribution of $\hat{\tau}$ under ReM is almost identical to the asymptotic distribution of $\hat{\tau}_\textsc{L}$ under the CRE. So ReM uses covariates in the design stage and \citet{lin2013}'s regression adjustment uses covariates in the analysis stage, achieving nearly the same asymptotic efficiency gain when $a$ is small.

 \subsubsection{Equivalence of regression adjustment and post-stratification}
If we have discrete covariate $C_i$ with $K$ categories, we can create $K-1$ centered dummy variables 
$$
X_i = ( I(C_i=1) - \pi_{[1]}, \ldots,   I(C_i=K-1) - \pi_{[K-1]})
$$ 
where $\pi_{[k]}$ equals the proportion of units with $C_i = k$. 
In this case, \citet{lin2013}'s regression adjustment is equivalent to post-stratification, as summarized by the following proposition. 

\begin{proposition}
\label{prop::lin=ps-discrete}
$\hat{\tau}_\textsc{L}$ based in $X_i$ is numerically identical to the post-stratification estimator $\hat{\tau}_\textsc{PS}$ based on $C_i$ (see Chapter \ref{sec::poststra-cre}). 
\end{proposition}

I leave the proof of Proposition \ref{prop::lin=ps-discrete} as Problem \ref{para::lin=ps-discrete}.

\subsubsection{Difference-in-difference as a special case of covariate adjustment $\hat \tau(\beta_1, \beta_0)$}\label{sec::difference-in-difference-CRE}
An important covariate $X$ in many studies is the lagged outcome before the treatment. For instance, the covariate $X$ is the pre-test score if the outcome $Y$ is the post-test score in educational research; the covariate $X$ is the log wage before the job training program if the outcome $Y$ is the log wage after the job training program. With the lagged outcome $X$ as a covariate, a popular estimator is the {\it gain score} or {\it difference-in-difference} estimator with $\beta_1 = \beta_0=1$ in \eqref{eq::linear-adj-1} and \eqref{eq::linear-adj-2}: 
\begin{eqnarray*}
\hat{\tau}(1,1) &=& n_1^{-1} \sumn Z_i(Y_i -   X_i) - n_0^{-1} \sumn (1-Z_i) (Y_i -   X_i) \\
&=& \left\{   \hat{\bar{Y}}(1) - \hat{\bar{Y}}(0) \right\} -  \left\{   \hat{\bar{X} }(1) - \hat{\bar{X} }(0) \right\} .
\end{eqnarray*}
The first form of $\hat{\tau}(1,1) $ justifies the name {\it gain score} because it is essentially the difference in means of the gain score $g_i = Y_i - X_i$. 
The second form of $\hat{\tau}(1,1) $ justifies the name {\it difference-in-difference} because it is the difference between two differences in means. 
This estimator is different from \citet{lin2013}'s estimator: it fixes $\beta_1 = \beta_0=1$ in advance while \citet{lin2013}'s estimator involves two estimated $\beta$'s. The $\hat{\tau}(1,1) $ is unbiased
with a conservative variance estimator
\begin{eqnarray*}
\hat{V}( 1,1 ) 
&=& \frac{1}{n_1(n_1 - 1)} \sumn Z_i  \{ g_i - \hat{\bar{g}}(1)   \}^2 \\
& & \qquad + \frac{1}{n_0(n_0 - 1)} \sumn (1-Z_i)  \{ g_i - \hat{\bar{g}}(0)   \}^2,
\end{eqnarray*}
where $\hat{\bar{g}}(1)$ and $ \hat{\bar{g}}(0)$ are the sample means of the gain score $g_i = Y_i - X_i$ under the treatment and control, respectively. 
When the lagged outcome is a strong predictor of the outcome, the gain score $g_i = Y_i - X_i$ often has a much smaller variance than the outcome itself. In this case, $\hat{\tau}(1,1) $ often greatly reduces the variance of the simple difference in means of the outcome. 
See Problem \ref{hw::did-CRE} for more details.

In theory,  \citet{lin2013}'s estimator is always more efficient than $\hat{\tau}(1,1)$ with large samples. However, \citet{lin2013}'s estimator is biased in finite samples, whereas $\hat{\tau}(1,1)$ is always unbiased.

\subsection{Extension to the SRE}\label{sec::covariate-SRE}

It is possible that we have an experiment stratified on a discrete variable $C$ and we also observe additional covariates $X$. If all strata are large, then we can obtain 
\citet{lin2013}'s estimators within strata $\hat{\tau}_{\textsc{L},[k]}$ and obtain the final estimator as
$$
\hat{\tau}_{\textsc{L},\textsc{S}} = \sum_{k=1}^K \pi_{[k]} \hat{\tau}_{\textsc{L},[k]} . 
$$
A conservative variance estimator is 
$$
\hat{V}_{\textsc{L},\textsc{S}} =  \sum_{k=1}^K \pi_{[k]}^2 \hat{V}_{\textsc{ehw},[k]},
$$
where $\hat{V}_{\textsc{ehw},[k]}$ is the EHW variance estimator from the OLS fit of the outcome on the intercept, the treatment indicator, the covariates, and their interactions within stratum $k.$ Importantly, we need to center covariates by their stratum-specific means.

\section{Unification, combination, and comparison}
\label{sec::unification-rem-regadj}

\citet{li2019rerandomization} unified the literature and showed that we can combine rerandomization and regression adjustment. That is, if we rerandomize in the design stage, we can use \citet{lin2013}'s estimator with the EHW standard error in the analysis stage. The combination of rerandomization and regression adjustment improves covariate balance in the design stage and estimation efficiency in the analysis stage.

Table \ref{tb::rerandomization-regression} summarizes the literature from \citet{Neyman:1923} to \citet{li2019rerandomization}. Arrow 1 illustrates the efficiency gain of covariate adjustment in the CRE: asymptotically,  $\hat\tau_\textsc{L}$ has a smaller variance than $\hat\tau$. 
Arrow 2 illustrates the efficiency gain of ReM: asymptotically, $\hat{\tau}$ has narrower quantile ranges under the ReM than under the CRE.
Arrows 3 and 4 illustrate the benefits of the combination of ReM and the CRE.

\begin{table}
\centering
\caption{Design and analysis of experiments}\label{tb::design-analysis-2x2}
\label{tb::rerandomization-regression}
\begin{tabular}{c|cccc}
\hline 
  & & & analysis & \\
  \hline 
  & CRE & $\hat\tau$ \citep{Neyman:1923} &  $\stackrel{1}{\longrightarrow}$  &  $\hat\tau_\textsc{L}$ \citep{lin2013} \\
  design&  &  $2\Big\downarrow$ && $  \Big\downarrow 4$ \\
  & ReM & $\hat\tau$  \citep{li2018asymptotic} &  $\stackrel{3}{\longrightarrow}$  &$\hat\tau_\textsc{L}$ \citep{li2019rerandomization}  \\
  \hline 
\end{tabular}
\end{table}

\section{Simulation}
\label{sec::rem-reg-simulation}

\citet{angrist2009incentives}  conducted an experiment to evaluate different strategies to improve academic performance among college freshmen in a Canadian university. Here I use a subset of the original data, focusing on the control group and the treatment group offered academic support services and financial incentives for good grades. The outcome is the GPA at the end of the first year. I impute the missing outcomes with the observed average which is somewhat arbitrary (see Problem \ref{hw::missing-y-experiments}). Two covariates are the gender and baseline GPA.

\begin{lstlisting}
angrist   = read.dta("star.dta")
angrist2  = subset(angrist, control == 1||sfsp == 1)
## imputing missing outcomes
y = angrist2$GPA_year1
meany    = mean(y, na.rm = TRUE)
y = ifelse(is.na(y), meany, y)
z = angrist2$sfsp 
x = angrist2[, c("female", "gpa0")]
\end{lstlisting}

The following code gives the results based on the unadjusted and adjusted estimators. 
\begin{lstlisting}
> ## unadjusted estimator
> fit_unadj = lm(y ~ z)
> ace_unadj = coef(fit_unadj)[2]
> se_unadj  = sqrt(hccm(fit_unadj, type = "hc2")[2, 2])
> 
> ## regression adjustment
> x         = scale(x)
> fit_adj   = lm(y ~ z*x)
> ace_adj   = coef(fit_adj)[2]
> se_adj    = sqrt(hccm(fit_adj, type = "hc2")[2, 2])
> 
> res = c(ace_unadj, ace_adj, se_unadj, se_adj)
> dim(res) = c(2, 2)
> t.stat   = res[, 1]/res[, 2]
> p.value  = 2*pnorm(abs(t.stat), lower.tail = FALSE) 
> res      = cbind(res, t.stat, p.value)
> rownames(res) = c("Neyman", "Lin")
> colnames(res) = c("estimate", "s.e.", "t-stat", "p-value")
> round(res, 3)
       estimate  s.e. t-stat p-value
Neyman    0.054 0.076  0.719   0.472
Lin       0.075 0.072  1.036   0.300
\end{lstlisting}
The adjusted estimator has a smaller standard error although it gives the same insignificant result as the unadjusted estimator.

I also use this dataset to conduct simulation studies to evaluate the four design and analysis strategies summarized in Table \ref{tb::design-analysis-2x2}. I fit quadratic functions of the outcome on the covariates and use them to impute all the missing potential outcomes, separately for the treated and control groups. To show the improvement of ReM and regression adjustment, I also rescale the error terms by $0.1$ and $0.25$ to increase the signal-to-noise ratio. With the imputed Science Table, I generate 2000 treatments, obtain the observed data, and calculate the estimators. In the simulation, the ``true'' outcome model is nonlinear, but we still use linear adjustment for estimation. By doing this, we can evaluate the properties of the estimators even when the linear model is misspecified.

\begin{figure}
\centering
\includegraphics[width = \textwidth]{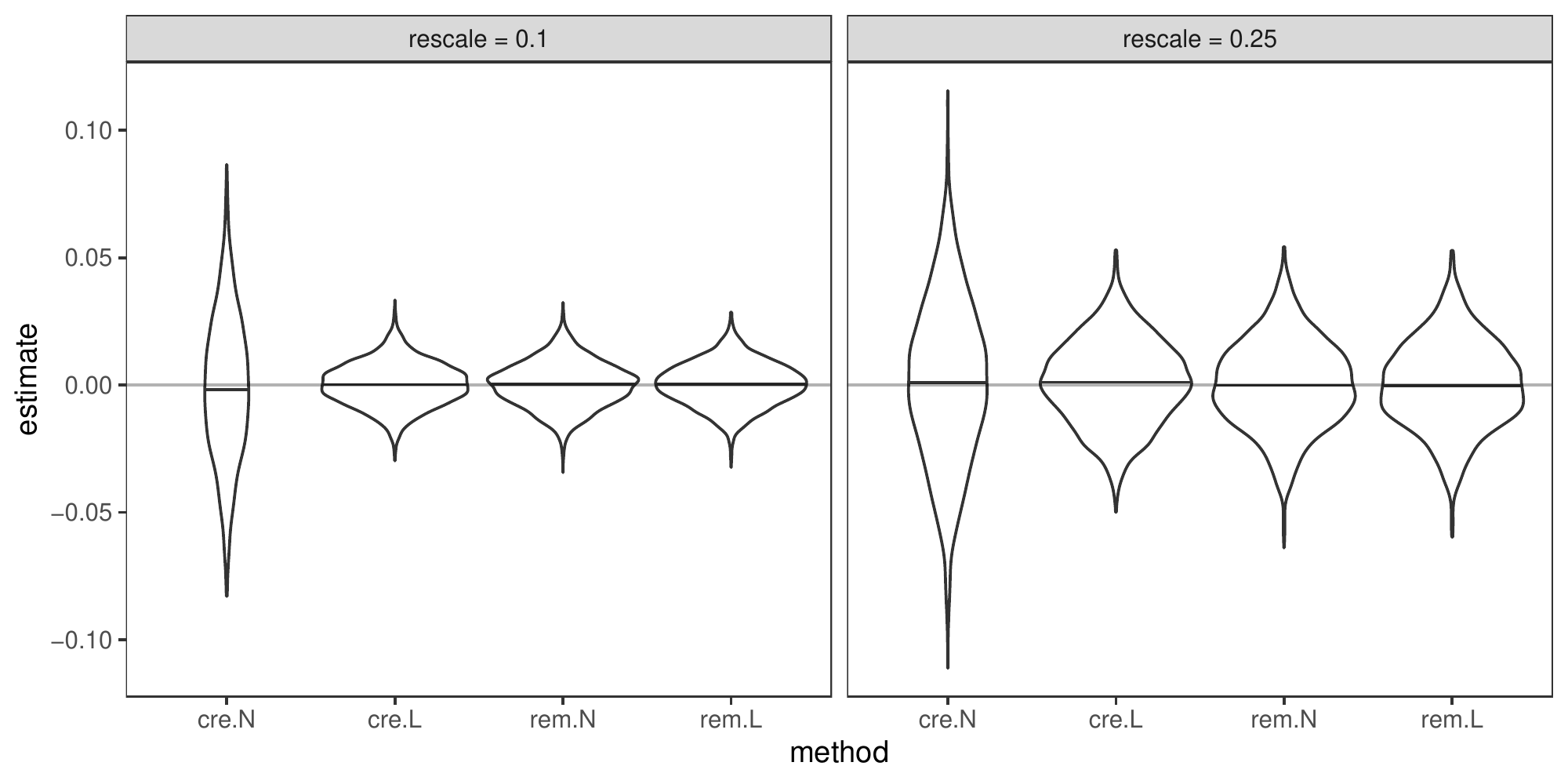}
\caption{Simulation with $2000$ Monte Carlo replicates and $a=0.05$ for ReM} \label{fg::rerandomizationandancova}
\end{figure}

Figure \ref{fg::rerandomizationandancova} shows the violin plots\footnote{A violin plot is similar to a box plot, with the addition of a rotated kernel density plot on each side.} of the four combinations, subtracting the true $\tau$ from the estimates. As predicted by the theory, all estimators are nearly unbiased, and both ReM and regression adjustment improve efficiency. They are more effective when the noise level is smaller. 

\section{Final remarks}

ReM uses the Mahalanobis distance as the balance criterion. We can consider general rerandomization with the balance criterion defined as a function of $\bm{Z}$ and $\bm{X}$.  For example, we can use the following criterion based on marginal tests for all coordinates of $X_i = (x_{i1}, \ldots, x_{iK})\tran$. We accept $\bm{Z}$ if and only if 
\begin{equation}
\label{eq::marginal-testing-rerandomization}
\left | \frac{  \hat{\tau}_{xk} }{   \sqrt{  \frac{n}{n_1 n_0} S_{xk}^2       }  }   \right | \leq  a \quad ( k=1,\ldots, K )
\end{equation}
for some predetermined constant $a > 0$, where $S_{xk}^2$ is the finite-population variance of covariate $x_{ik}$.  For example, $a$ is some upper quantile of a standard Normal distribution. See \citet{zhao2021no} for the theory for the rerandomization based on criterion \eqref{eq::marginal-testing-rerandomization} as well as other criteria.

With a continuous outcome, Fisher's ANCOVA has been the standard approach for many years.  \citet{lin2013}'s improvement has better theoretical properties even when the linear model is misspecified. With a binary outcome, it is common to use the coefficient of the treatment in the logistic regression of the observed outcome on the intercept, the treatment indicator, and covariates to estimate the causal effects. However, \citet{freedman2008randomization} showed that this logistic regression does not have nice properties under the potential outcomes framework. Even if the logistic model is correct, the coefficient estimates the conditional odds ratio (see Chapter \ref{sec::logistic-regression}) which may not be the parameter of interest; when the logistic model is incorrect, it is even harder to interpret the coefficient. From the discussion above, if the parameter of interest is the average causal effect, we can still use \citet{lin2013}'s estimator to analyze the binary outcome data in the CRE. \cite{guo2020generalized} extend  \citet{lin2013}'s theory to allow for using generalized linear models to construct estimators for the average causal effect under the potential outcomes framework.

Other extensions of \citet{lin2013}'s theory focus on high dimensional covariates. \citet{bloniarz2015lasso} focus on the regime with many covariates than the sample size, and suggest replacing the OLS fits with the least absolute shrinkage and selection operator (LASSO) fits \citep{tibshirani1996regression} of the outcome on the intercept, the treatment, covariates and their interactions. \citet{lei2018regression} focus on the regime with a diverging number of covariates without assuming sparsity, and under certain regularity conditions, they show that \citet{lin2013}'s estimator is still consistent and asymptotically Normal. \citet{wager2016high} propose to use machine learning methods to analyze high dimensional experimental data.

\section{Homework Problems}

\paragraph{FRT under ReM}\label{hw::FRT-under-REM}

Describe the FRT under ReM.

\paragraph{Invariance of the Mahalanobis Distance}\label{para::invariance-M}

Prove Lemma \ref{lemma::invariance-mdistance}.

\paragraph{Bias of the difference-in-means estimator under rerandomization}\label{hw::unbiasednessof-tauhat-rerandom}

Assume that we draw $\bm{Z} = (Z_1, \ldots, Z_n)$ from a CRE and accept it if and only if $\phi(\bm{Z}, \bm{X}) = 1$, where $\phi$ is a predetermined balance criterion. Show that if $n_1 = n_0$ and 
\begin{eqnarray}
\phi(\bm{Z}, \bm{X}) = \phi(\bm{1}_n-\bm{Z}, \bm{X}), \label{eq::symmetry}
\end{eqnarray}
then $\hat{\tau}$ is unbiased for $\tau$. 
Verify that rerandomization using the Mahalanobis distance satisfies \eqref{eq::symmetry} if $n_1 = n_0$. 
Give a counterexample that $\hat{\tau}$ is biased for $\tau$ when these two conditions do not hold.

Remark: $\phi(\bm{Z}, \bm{X}) $ can be a general balance criterion in this problem. ReM is a special case with $\phi(\bm{Z}, \bm{X})  =  I(M\leq a) $.

\paragraph{Equivalent form of $R^2$ in the CRE}\label{para::r2-cre}

Prove Proposition \ref{prop::rsquared-forms}.

\paragraph{More on linear projections of the potential outcomes onto covariates}\label{hw::projection-po-onto-x}

Show that
$$
S^2(1\mid X) = S_{Y(1)X} (S_X^2)^{-1} S_{XY(1)}
$$
where $S_{XY(1)}$ is the finite population covariance between $Y_i(1)$'s and $X_i$'s, $S_{Y(1)X} = S_{XY(1)}\tran$, and $S_X^2$ is the finite population covariance matrix of $X_i$'s. 
Find the analogous formulas for $S^2(0\mid X) $ and $S^2(\tau\mid X) .$

\paragraph{Comparing the true variances within the family of linearly adjusted estimator}\label{hw::compare-true-variance}

Show that the variance $\hat{\tau}(\beta_1, \beta_0)$ decomposes into
$$
\var\{ \hat{\tau}(\beta_1, \beta_0) \} 
=  
\var\{ \hat{\tau}( \tilde{\beta}_1, \tilde{\beta}_0) \}  
+
\var\{ \hat{\tau}(\beta_1, \beta_0) - \hat{\tau}(\tilde{\beta}_1, \tilde{\beta}_0) \}
$$
where $\tilde{\beta}_1$ and $\tilde{\beta}_0$ are the coefficients of $X_i$ in the OLS projection of $Y_i(1)$'s and $Y_i(0)$'s onto $(1, X_i)$'s, respectively.

Remark: \citet[][Example 9]{li2017general} derived the decomposition. It shows that based on the true variance, $\hat{\tau}( \tilde{\beta}_1, \tilde{\beta}_0)$ is an optimal choice within the family of linearly adjusted estimator because $\var\{ \hat{\tau}(\beta_1, \beta_0) - \hat{\tau}(\tilde{\beta}_1, \tilde{\beta}_0) \}$ is always non-negative.

\paragraph{Lin's estimator for covariate adjustment}\label{para::lin-ols-covadj}

Prove Proposition \ref{prop::lin-ols}. 

\paragraph{Predictive and projective estimators}\label{para::predictiveestimator}
Prove \eqref{eq::predictiveestimator} and \eqref{eq::projectiveestimator}.

 \paragraph{Equivalent form of the covariate-adjusted estimator}\label{para::reg-covariate-imbalance}

Prove \eqref{eq::linear-covariate-imbalance} and \eqref{eq::lin-covariate-imbalance}.

\paragraph{ANCOVA also adjusts for covariate imbalance}
\label{para::ancova-imbalance}

This problem gives a result for ANCOVA that is similar to \eqref{eq::lin-covariate-imbalance}. 

Show that
$$
\hat{\tau}_\textsc{F} =\hat\tau  - \hat\gamma_\textsc{F}\tran \hat\tau_X,
$$
where $ \hat\gamma_\textsc{F}$ is the coefficient of $X_i$ in the OLS fit of $Y_i$ on $(1,Z_i, X_i)$.

 \paragraph{Regression adjustment / post-stratification of CRE}\label{para::lin=ps-discrete}
 
 Prove Proposition \ref{prop::lin=ps-discrete}.

Remark:  Sometimes $\hat{\tau}_{\textsc{ps}}$ or $ \hat{\tau}_\textsc{L}$ may not be well-defined. In those cases, we treat $\hat{\tau}_{\textsc{ps}}$ and $ \hat{\tau}_\textsc{L}$ as equal. You can ignore this complexity in the proof.

\paragraph{More on the difference-in-difference estimator in the CRE}\label{hw::did-CRE}

This problem gives more details for the difference-in-difference estimator in the CRE in Section \ref{sec::difference-in-difference-CRE}. 

Show that $\hat{\tau}(1,1) $ is unbiased for $\tau$, calculate its variance, and show that $\hat{V}( 1,1 ) $ is a conservative estimator for the true variance of $\hat{\tau}(1,1) $. When does $E\{\hat{V}( 1,1 ) \} = \var\{  \hat{\tau}(1,1)  \}$ hold?

Compare the variances of $\hat{\tau}(0,0) $ and $\hat{\tau}(1,1) $ to show that 
$$
\var\{ \hat{\tau}(0,0) \}  \geq \var\{  \hat{\tau}(1,1)  \}
$$
if and only if 
$$
2 \frac{n_0}{n} \beta_1 + 2\frac{n_1}{n} \beta_0 \geq 1,
$$
where 
$$
\beta_1 = \frac{\sum_{i=1}^n (X_i-\bar{X})\{  Y_i(1)-\bar{Y}(1) \}  }{  \sum_{i=1}^n (X_i-\bar{X})^2 },\quad
\beta_0 = \frac{\sum_{i=1}^n (X_i-\bar{X})\{  Y_i(0)-\bar{Y}(0) \}  }{  \sum_{i=1}^n (X_i-\bar{X})^2 }
$$
are the coefficients of $X_i$ in the OLS fits of $Y_i(1)$ and $Y_i(0)$ on $(1,X_i)$, respectively.

Remark: \citet[][page 28]{gerber2012field} discussed a special case  of this problem with $n_1=n_0.$

\paragraph{Data re-analyses of the Penn Bonus Experiment}

Re-analyze the Penn Bonus Experiment data. The analysis in Chapter \ref{chapter::stratification-poststratification} uses the treatment indicator, the outcome, and the block indicator. Now we want to use all other covariates. 

Conduct regression adjustments within the strata of the experiment, and then combine these adjusted estimators to estimate the average causal effect. Report the point estimator, estimated standard error, and 95\% confidence interval. Compare them with those without regression adjustment.

\paragraph{Missing outcomes in randomized experiments}\label{hw::missing-y-experiments}

The data analysis in Section \ref{sec::rem-reg-simulation} uses a naive imputation method to deal with missing outcomes. It is somewhat arbitrary. 

Impute the missing outcomes under the treatment and control, respectively,  based on the observed means. Do the results change?

Impute the missing outcomes under the treatment and control, respectively, based on linear regressions of the observed outcomes on the covariates. Do the results change? 

Do you have other ways to deal with missing outcomes? Justify them and implement them to analyze the dataset.

\paragraph{Recommended reading}

The title of this chapter is the same as that of \citet{li2019rerandomization}, which studied the roles of rerandomization and regression adjustment in the design and analysis stages of randomized experiments, respectively.

\chapter{Matched-Pairs Experiment}
 \label{chapter::mpe}
 
The matched-pairs experiment (MPE) is the most extreme version of the SRE with only one treated unit and one control unit within each stratum. In this case, the strata are also called pairs. Although this type of experiment is a special case of the SRE discussed in Chapter \ref{chapter::stratification-poststratification}, it has its own estimation and inference strategy. Moreover, it has many new features and it is closely related to the ``matching'' strategy in observational studies which will be covered in Chapter \ref{chapter::matching-obs} later. So we discuss the MPE here,  in its own chapter.

\section{Design of the experiment and potential outcomes}
 
Consider an experiment with $2n$   units. If we have predictive covariates to the outcomes, we can pair units based on the similarity of covariates. With a scalar covariate, we can order units based on this covariate and then form pairs based on the adjacent units. With many covariates, we can define pairwise distances between units and then form pairs based on these distances. In this case, pair matching can be done using a greedy algorithm or an optimal nonbipartite matching algorithm. The greedy algorithm pairs the two units with the smallest distance, drops them from the pool of units, pairs the two remaining units with the smallest distance, etc. The optimal nonbipartite matching algorithm divides the $2n$ units into $n$ pairs of two units to minimize the sum of the within-pair distances. See \citet{greevy2004optimal} for more details of the computational aspect of the MPE. In this chapter, we assume that the pairs are formed based on the covariates, and discuss the subsequent design and analysis issues.

 Let $(i,j)$ index the unit $j$ in pair $i$, where $i=1,\ldots, n$ and $j=1,2$. Unit $(i, j)$  has potential outcomes $Y_{ij}(1)$ and $Y_{ij}(0)$ under the treatment and control, respectively. Within each pair, we randomly assign one unit to receive the treatment and the other to receive the control. Let
$$
Z_i = \begin{cases}
   1,& \text{if the first unit receives the treatment}, \\
   0, & \text{if the second unit receives the treatment}. 
\end{cases}  
$$ 
We can formally define MPE based on the treatment assignment mechanism.

\begin{definition}
[MPE]
We have 
\begin{eqnarray}
 (Z_i)_{i=1}^n \iidsim \text{Bernoulli}(1/2).
 \label{eq::mp-treatment}
\end{eqnarray} 
\end{definition}

The observed outcomes within pair $i$ are 
$$
Y_{i1} = Z_i Y_{i1}(1) + (1-Z_i) Y_{i1}(0)
= \begin{cases}
   Y_{i1}(1),& \text{if } Z_i = 1; \\
   Y_{i1}(0), & \text{if } Z_i = 0; 
\end{cases}
$$
and
$$
Y_{i2} = Z_i Y_{i2}(0) + (1-Z_i) Y_{i2}(1)
= \begin{cases}
   Y_{i2}(0),& \text{if } Z_i = 1; \\
   Y_{i2}(1), & \text{if } Z_i = 0 . 
\end{cases} 
$$
So the observed data are $( Z_i, Y_{i1} , Y_{i2} )_{i=1}^n$.

\section{FRT}
 
 Similar to the discussion before, we can always use the FRT to test the sharp null hypothesis: 
$$
\HF: Y_{ij}(1) = Y_{ij}(0)  \text{ for all }  i=1,\ldots  n  \text{ and } j=1,2. 
$$
When conducting the FRT, we need to simulate the distribution of $(Z_1, \ldots, Z_n)$ from \eqref{eq::mp-treatment}. 
I will discuss some canonical choices of test statistics based on the within-pair differences between the treated and control outcomes:
\begin{eqnarray*}
\hat{\tau}_i &=& \text{outcome under treatment} - \text{outcome under control} \text{ (within pair } i) \\
&=&  (2Z_i - 1) (Y_{i1}  - Y_{i2})\\
&=& S_i (Y_{i1}  - Y_{i2})  , 
\end{eqnarray*}
where the $S_i = 2Z_i - 1$ are IID random signs with mean $0$ and variance $1$, for $i=1,\ldots, n$. 
 Since the pairs with zero $\hat{\tau}_i $'s do not contribute to the randomization distribution, we drop those pairs in the discussion of the FRT. 
 
\begin{example} 
[paired $t$ statistic]\label{eg::paired-t-statistic}
The average of the within-pair differences is
$$
\hat{\tau} = n^{-1}  \sumn \hat{\tau}_i .
$$
Under $\HF$, 
$$
E(\hat{\tau} ) = 0
$$
and
$$
\var(  \hat{\tau}   ) 
= n^{-2} \sumn \var(  \hat{\tau}_i ) 
= n^{-2} \sumn \var(S_i) (Y_{i1}  - Y_{i2})^2 =  n^{-2} \sumn  \hat{\tau}_i^2 . 
$$
Based on the CLT for the sum of independent random variables, we have the  Normal approximation:
$$
\frac{  \hat{\tau}  }{ \sqrt{  n^{-2} \sumn  \hat{\tau}_i^2 } } \rightarrow \N01 
$$
in distribution.  We can use this Normal approximation to construct an asymptotic test. Many standard textbooks suggest using the following paired $t$ statistic in the MPE:
$$
t_\text{pair} = \frac{  \hat{\tau}  }{ \sqrt{  \{ n(n-1) \}^{-1} \sumn  (\hat{\tau}_i - \hat{\tau} ) ^2 } },
$$
which is almost identical to $\hat{\tau}$ with large $n$ and small $\hat{\tau} $ under $\HF$. 
\end{example} 

In classic statistics, the motivation for using $t_\text{pair}$ is under a different framework. When $\hat{\tau}_i \iidsim \textsc{N}(0, \sigma^2)$, we can show that $t_\text{pair} \sim t(n-1)$, i.e., the exact distribution of $t_\text{pair} $ is $t$ with degrees of freedom $n-1$, which is close to $\N01$ with a large $n$. The \ri{R} function \ri{t.test} with \ri{paired=TRUE} can implement this test. With a large $n$, these procedures give similar results. The discussion in Example  \ref{eg::paired-t-statistic} gives another justification of the classic paired $t$ test without assuming the Normality of the data.

  \begin{example}
 [Wilcoxon sign-rank statistic]\label{eg::mpe-sign-rank}
Based on the ranks $(R_1, \ldots, R_n)$ of $( | \hat{\tau}_1| , \ldots, | \hat{\tau}_n| )$, we can define a test statistic
$$
W = \sumn  I( \hat{\tau}_i > 0 ) R_i. 
$$
Under $\HF$, the $| \hat{\tau}_i|$'s are fixed so the $R_i$'s are also fixed, which implies that 
$$
E(W) = \frac{1}{2} \sumn R_i = \frac{1}{2} \sumn i = \frac{n(n+1)}{4}
$$
and
$$
\var(W) = \frac{1}{4} \sumn R_i^2 = \frac{1}{4} \sumn i^2 = \frac{n(n+1)(2n+1)}{24}. 
$$
The CLT for the sum of independent random variables ensures the following Normal approximation: 
$$
\frac{W - n(n+1)/4}{ \sqrt{n(n+1)(2n+1) / 24    } } \rightarrow \N01
$$
in distribution. 
We can use this Normal approximation to construct an asymptotic test.  The \ri{R} function \ri{wilcox.test} with \ri{paired=TRUE} can implement both the exact and asymptotic tests. 
 \end{example}

   \begin{example}
 [Kolmogorov--Smirnov-type statistic]\label{eg::mpe-ksstatistics}
Under $\HF$, the absolute values $( | \hat{\tau}_1| , \ldots, | \hat{\tau}_n| )$ are fixed but their signs are random. So $   (  \hat{\tau}_1 , \ldots,  \hat{\tau}_n  )$ and $ -   (  \hat{\tau}_1 , \ldots,  \hat{\tau}_n  )$ should have the same distribution. 
Let 
$$
\hat{F}(t)  = n^{-1} \sumn I( \hat{\tau}_i  \leq t  )
$$
be the empirical distribution of $  (  \hat{\tau}_1 , \ldots,  \hat{\tau}_n  )$, and
$$
1 - \hat{F}(-t-)  =    n^{-1} \sumn I( -  \hat{\tau}_i  \leq t  )
$$ 
be the empirical distribution of $ -   (  \hat{\tau}_1 , \ldots,  \hat{\tau}_n  )$, where $\hat{F}(-t-)$ is the left limit of the function $\hat{F}(\cdot)$ at $-t$. 
A Kolmogorov--Smirnov-type statistic is then 
$$
D = \max_t |  \hat{F}(t)  +  \hat{F}(-t-)  - 1  |.
$$
\citet{butler1969test} proposed this test statistic and derived its exact and asymptotic distributions. Unfortunately, this is not implemented in standard software packages. Nevertheless, we can simulate its exact distribution and compute the $p$-value based on the FRT. \footnote{
\citet{butler1969test}'s proposed this test statistic under a slightly different framework. Given IID draws of $    (  \hat{\tau}_1 , \ldots,  \hat{\tau}_n  )$ from a  distribution $F(y)$, if they are symmetrically distributed around $0$, then
$$
F(t) = \pr( \hat{\tau}_i \leq t ) =  \pr( -  \hat{\tau}_i \leq t )  = 1 - \pr( \hat{\tau}_i <- t) = 1 - F(-t-).
$$
Therefore, $\hat{F}(t)  +  \hat{F}(-t-)  - 1 $ measures the deviation from the null hypothesis of symmetry, which motivates the definition of $D$ in Example \ref{eg::mpe-ksstatistics}. A naive definition of the Kolmogorov--Smirnov-type statistic is to compare the empirical distributions of the outcomes under treatment and control as in Example \ref{eg::ks-CRE}. Using that definition, we effectively break the pairs. Although it can still be used in the FRT for the MPE, it does not capture the matched-pairs structure of the experiment.}
 \end{example}

  \begin{example}
 [sign statistic]\label{eg::mpe-sign}
 The sign statistic uses only the signs of the within-pair differences
 $$
 \Delta = \sumn I( \hat{\tau}_i > 0 ).
 $$
 Under $\HF$, 
 $$
 I( \hat{\tau}_i > 0 ) \iidsim \textup{Bernoulli}(1/2)
 $$
 and therefore 
 $$
  \Delta \sim \textup{Binomial}(n, 1/2).
 $$
 Based on this we have an exact Binomial test, which is implemented in the \ri{R} function 
\ri{binom.test} with \ri{p=1/2}. Using the CLT, we can also conduct a test based on the following Normal approximation of the Binomial distribution: 
$$
\frac{  \Delta - n/2 }{  \sqrt{n/4} } \rightarrow \N01 
$$
in distribution. 
 \end{example}

 \begin{example}
 [McNemar's statistic for a binary outcome]\label{eg::mcnenar-test-binary}
 If the outcome is binary, we can summarize the observed data from the MPE in a more compact way. Given a pair, the treated outcome can be either $1$ or $0$ and the control outcome can be either $1$ or $0$, yielding a two-by-two table as in Table \ref{tb::binarymatchedpairs}. 
 
\begin{table}
\centering
\caption{Counts of four types of pairs}\label{tb::binarymatchedpairs}
\begin{tabular}{ccc}
\hline 
 & control outcome 1 & control outcome 0\\
 \hline 
treated outcome 1 & $m_{11}$ & $m_{10}$ \\
treated outcome 0 & $m_{01}$ & $m_{00}$ \\
\hline 
\end{tabular}
\end{table} 
 
Under $\HF$, the numbers of concordant pairs $m_{11}$ and $m_{00}$ are fixed, and $m_{10} + m_{01}$ is also fixed. So the only random component is $m_{10}$ which has distribution 
$$
m_{10} \sim \textup{Binomial}( m_{10} + m_{01}, 1/2  ).
$$ 
This implies an exact test based on the Binomial distribution. The \ri{R} function \ri{mcnemar.test} gives an asymptotic test based on the Normal approximation of the Binomial distribution:
$$
\frac{   m_{10}  - (m_{10} + m_{01})/2  }{  \sqrt{  (m_{10} + m_{01}) / 4   } } 
= \frac{ m_{10}  -  m_{01}   }{  \sqrt{ m_{10} + m_{01}    }  }
\rightarrow \N01  
$$
in distribution. Both the exact FRT and the asymptotic test do not depend on $m_{11}$ or $m_{00}$. Only the numbers of discordant pairs matter in these tests. 
 \end{example}

\section{Neymanian inference}
 
The average causal effect within pair $i$ is 
 $$
 \tau_i = \frac{1}{2}  \left\{    Y_{i1}(1) + Y_{i2}(1) -     Y_{i1}(0)  - Y_{i2}(0)     \right\},
 $$
 and the average causal effect for all units is
$$
\tau = n^{-1} \sumn  \tau_i  = (2n)^{-1} \sumn \sum_{j=1}^2 \{  Y_{ij}(1) - Y_{ij}(0)   \} .
$$
It is intuitive that  $\hat{\tau}_i $ is unbiased for $ \tau_i $, so $\hat{\tau}  $ is unbiased for $\tau$. We can also calculate the variance of $\hat{\tau}  $. I relegate the exact formula to Problem \ref{para::variance-mp-design}  because the MPE is just a special case of the SRE.

However, we cannot follow the strategy under the SRE to estimate the variance of $\hat{\tau}  $ under the MPE. The within-pair sample variances of the outcomes are not well defined because, within each pair, we have only one treated and one control unit. The data do not allow us to estimate the variance of $\hat{\tau}_i $ within pair $i$.

Is it possible to estimate the variance of $\hat{\tau}$ in the MPE? Let us forget about the MPE and change the perspective to the classic IID sampling. If the $\hat{\tau}_i$'s are IID with mean $\mu$ and $\sigma^2$, then the variance of $\hat{\tau} = n^{-1} \sumn \hat{\tau}_i$ is $\sigma^2 / n$. An unbiased estimator for $\sigma^2$ is the sample variance $(n-1)^{-1} \sumn ( \hat{\tau}_i -  \hat{\tau} )^2$, so an unbiased estimator for $\var( \hat{\tau}  )$ is 
\begin{equation}\label{eq::mpe-variance-estimation}
\hat{V} = \{ n(n-1)  \}^{-1} \sumn ( \hat{\tau}_i -  \hat{\tau} )^2.
\end{equation}
The discussion also extends to the independent but not IID setting; see Problem \ref{hw::inid} in Chapter \ref{chapter::basic-prob-append}.
The above discussion seems a digression from the MPE which has completely different statistical assumptions. But at least it motivates a variance estimator $\hat{V} $, which uses the between-pair variance of $\hat{\tau}_i $ to estimate the variance of $\hat{\tau}  $.  Of course, it is derived under different assumptions. Does it also work for the MPE? Theorem \ref{thm::varianceest-mp} below is a positive result.
 
\begin{theorem}\label{thm::varianceest-mp}
Under the MPE, $\hat{V}$ defined in \eqref{eq::mpe-variance-estimation} is a conservative estimator for the true variance of $\hat{\tau}  $:
$$
E(  \hat{V}) - \var(   \hat{\tau}  ) = \{ n(n-1)  \}^{-1} \sumn  (\tau_i - \tau)^2 \geq 0  .
$$
If the $\tau_i$'s are constant across pairs, then $E(  \hat{V} )  = \var(\hat{\tau}) .$
\end{theorem}

Theorem \ref{thm::varianceest-mp} states that under the MPE, $\hat{V} $ is a conservative variance estimator in general and becomes unbiased if the average causal effects are constant across pairs. It is somewhat surprising because $\hat{V} $ depends on the between-pair variance of the $ \hat{\tau}_i $'s whereas $ \var(   \hat{\tau}  ) $ depends on the within-pair variance of each of $ \hat{\tau}_i $. The proof below might provide some insights into this surprising result.

\begin{myproof}{Theorem}{\ref{thm::varianceest-mp}}
Recall the basic algebraic fact that $\sumn (a_i - \bar a)^2 = \sumn a_i^2 - n \bar{a}^2$ which parallels the fact that $\var(W) = E(W^2) - (EW)^2$ for a random variable $W$. Use it in the following steps 2 and 5, we have 
\begin{eqnarray*}
n(n-1) E( \hat{V} ) &=& E\left\{   \sumn (\hat{\tau}_i - \hat{\tau})^2 \right\}  \\
&=&  E\left (    \sumn \hat{\tau}_i^2 - n \hat{\tau}^2  \right)   \\
&=& \sumn \{  \var(\hat{\tau}_i) + \tau_i^2 \} - n \{  \var(\hat{\tau})  + \tau^2 \}\\
&=& \sumn  \var(\hat{\tau}_i)  - n \var(\hat{\tau}) + \sumn  \tau_i^2 - n  \tau^2 \\
&=& n^2 \var(\hat{\tau})  - n \var(\hat{\tau})  + \sumn (\tau_i - \tau)^2.
\end{eqnarray*}
Therefore,
$$
E( \hat{V} )  = \var(\hat{\tau})  + \{ n(n-1)  \}^{-1}   \sumn (\tau_i - \tau)^2 \geq  \var(\hat{\tau})  .
$$
\end{myproof}

Similar to the discussions for other experiments, the Neymanian approach relies on the large-sample approximation:
$$
\frac{  \hat\tau  -  \tau }{  \sqrt{\var( \hat\tau )} } \rightarrow \N01
$$
in distribution if $n\rightarrow \infty$ and some regularity conditions hold. Due to the over-estimation of the variance, the Wald-type confidence interval
$$
 \hat\tau   \pm     z_{1-\alpha/2}  \sqrt{ \hat{V} }
$$
covers $\tau$ with probability at least $1-\alpha$.

Both the point estimator $\hat\tau$ and the variance estimator $\hat{V}$ can be conveniently obtained by OLS, as shown in the proposition below. 

\begin{proposition}\label{prop::mpe-variance}
$\hat\tau$ and
$\hat{V}$ are identical to the coefficient and variance estimator of the intercept from the OLS fit of the vector $(\hat{\tau}_1, \ldots, \hat{\tau}_n)\tran$ on the intercept only.
\end{proposition}

  I leave the proof of Proposition \ref{prop::mpe-variance} as Problem \ref{para::seols-matchedpairs}.

\section{Covariate adjustment}

Although we have matched on covariates in the design stage, it is possible that the matching is not perfect $(X_{i1}\neq X_{i2})$ and sometimes we have additional covariates beyond those used in the pair-matching stage. In those cases, we can adjust for the covariates to further improve estimation efficiency. Assume that each unit $(i, j)$ has covariates $X_{ij} $. Similar to the discussion in the CRE, there are two general strategies of covariate adjustment in the MPE.

\subsection{FRT}

I start with the covariate-adjusted FRT in the MPE. In parallel with Definition \ref{def::pseudo-y-frt},  we can construct test statistics based on the residuals from a model fitting of the outcome on the covariates, since those residuals are fixed numbers under the sharp null hypothesis. A canonical choice is to fit OLS of all observed $Y_{ij}$'s on $X_{ij}$'s to obtain the residuals $\hat \varepsilon_{ij}$'s. We can then construct test statistics pretending that the $\hat \varepsilon_{ij}$'s are the observed outcomes. \citet{rosenbaum2002covariance} advocated this strategy in particular to the MPE.

In parallel with Definition \ref{def::model-out-frt}, we can directly use some coefficients from model fitting as the test statistics. The discussion in the next subsection will suggest a choice of the test statistic for this strategy.

\subsection{Regression adjustment}

I now focus on estimating $\tau$. We can compute the within-pair differences in covariates $\hat{\tau}_{X,i} $ and their average $\hat{\tau}_X $ in the same way as the outcome. We can show that 
$$
 E(\hat{\tau}_{X,i}   ) = 0,\quad 
 E(\hat{\tau}_X ) = 0,
 $$
 and 
$$
\cov(\hat{\tau}_X   ) =n^{-2} \sumn \hat{\tau}_{X,i}   \hat{\tau}_{X,i}  \tran .
$$
 In a realized MPE, $\cov(\hat{\tau}_X   ) $ is not zero unless all the $\hat{\tau}_{X,i}  $'s are zero. With an unlucky draw of $(Z_1,\ldots, Z_n)$, it is possible that $\hat{\tau}_X $ differs substantially from zero. Similar to the discussion under the CRE in Chapter \ref{sec::lin=adjustingforimbalance}, adjusting for the imbalance of the covariate means is likely to improve estimation efficiency. 

Similar to the discussion in Section \ref{sec::lin=adjustingforimbalance}, we can consider a class of estimators indexed by $\gamma$:
$$
\hat{\tau}(\gamma) = \hat{\tau} - \gamma\tran \hat{\tau}_X 
$$
which has mean $0$ for any fixed $\gamma$. We want to choose $\gamma$ to minimize the variance of $\hat{\tau}(\gamma)$. Its variance is a quadratic function of $\gamma$:
\begin{eqnarray*}
\var\{ \hat{\tau}(\gamma) \} &=&   \var(\hat{\tau} - \gamma\tran \hat{\tau}_X ) \\
&=&  \var(\hat{\tau}  ) + \gamma\tran \cov(\hat{\tau}_X   ) \gamma - 2\gamma\tran \cov( \hat{\tau}_X ,  \hat{\tau}  ),
\end{eqnarray*}
which is minimized at 
$$
\tilde{\gamma} = \cov(\hat{\tau}_X   )^{-1}  \cov( \hat{\tau}_X ,  \hat{\tau}  ).
$$
We have obtained the formula for $\cov(\hat{\tau}_X   ) $ in the above, which can also be written as 
$$
\cov(\hat{\tau}_X   ) = n^{-2} \sumn |  \hat{\tau}_{X,i}  |  |\hat{\tau}_{X,i}  | \tran,
$$
where $| \cdot | $ denotes the component-wise absolute value of a vector. So $\cov(\hat{\tau}_X   ) $ is fixed and known from the observed data. However,  $\cov( \hat{\tau}_X ,  \hat{\tau}  )$ depends on unknown potential outcomes. Fortunately, we can obtain an unbiased estimator for it, as shown in Theorem \ref{thm::covariance-mp} below.

\begin{theorem}\label{thm::covariance-mp}
An unbiased estimator for $\cov( \hat{\tau}_X ,  \hat{\tau}  )$ is
$$
\hat{\theta} = \{n(n-1)  \}^{-1} \sumn ( \hat{\tau}_{X,i}  -  \hat{\tau}_{X} ) (   \hat{\tau}_i -  \hat{\tau}   ).
$$
\end{theorem}

The proof of Theorem \ref{thm::covariance-mp} is similar to that of Theorem \ref{thm::varianceest-mp}. I leave it to Problem \ref{para::covariance-est-proof}.

Therefore, we can estimate the optimal coefficient $\tilde{\gamma}$ by
\begin{eqnarray*}
\hat{\gamma} &=&  \left( n^{-2} \sumn \hat{\tau}_{X,i}   \hat{\tau}_{X,i}  \tran  \right)^{-1}
\left\{ \{n(n-1)  \}^{-1} \sumn (\hat{\tau}_{X,i}  -  \hat{\tau}_X )(   \hat{\tau}_i -  \hat{\tau}   ) \right\} \\
&\approx &  \left(  \sumn(  \hat{\tau}_{X,i}   - \hat{\tau}_X ) ( \hat{\tau}_{X,i}  -  \hat{\tau}_X  )\tran  \right)^{-1}
  \sumn (\hat{\tau}_{X,i}  -  \hat{\tau}_X ) (   \hat{\tau}_i -  \hat{\tau}   )  ,
\end{eqnarray*}
which is approximately the coefficient of the $\hat{\tau}_{X,i}$ in the OLS fit of the $\hat{\tau}_i $'s on the $\hat{\tau}_{X,i} $'s with an intercept. The final estimator is
$$
\hat{\tau}_{\text{adj}} = 
\hat{\tau}( \hat{\gamma} ) = \hat{\tau} - \hat{\gamma} \tran \hat{\tau}_X  ,
$$
which, by the property of OLS, is approximately the intercept in the OLS fit of the $\hat{\tau}_i $'s on the $\hat{\tau}_{X,i} $'s with an intercept.

A conservative variance estimator for $\hat{\tau}_{\text{adj}} $ is then
\begin{eqnarray*}
\hat{V} _{\text{adj}} 
&=& 
\hat{V} + \hat \gamma\tran \cov(\hat{\tau}_X   ) \hat \gamma - 2 \hat \gamma\tran \hat \theta  \\
&=& \hat{V}  - \hat \theta \tran \cov(\hat{\tau}_X   ) ^{-1}    \hat \theta .
\end{eqnarray*}

A subtle technical issue is whether $\hat\tau(\hat\gamma)$ has the same optimality as $\hat\tau(\tilde\gamma)$. We have encountered a similar issue in the discussion of \citet{lin2013}'s estimator. 
With large samples, we can show $\hat\tau(\hat\gamma) - \hat\tau(\tilde\gamma) =  - (\hat\gamma - \tilde\gamma)\tran \hat{\tau}_X $ is of higher order since it is the product of two ``small'' terms   $\hat\gamma - \tilde\gamma$ and $\hat{\tau}_X$.

Moreover,  \citet{fogarty2018regression} discussed the asymptotically equivalent regression formulation of the above covariate-adjusted procedure and gave a rigorous proof for the associated CLT. I summarize the regression formulation below without giving the regularity conditions.

\begin{proposition}\label{prop::colin2018regression}
Under the MPE, the covariate-adjusted estimator $\hat{\tau}_{\textup{adj}} $ and the associated variance estimator $\hat{V} _{\textup{adj}}$ can be conveniently approximated by the intercept and the associated variance estimator from the OLS fit of the vector of the $\hat{\tau}_i $'s on the $1$'s and the matrix of the $\hat{\tau}_{X,i} $'s. 
\end{proposition}

I leave the proof of Proposition \ref{prop::colin2018regression} as Problem \ref{para::seols-matchedpairs}. Interestingly, neither Proposition \ref{prop::mpe-variance} nor \ref{prop::colin2018regression} requires the EHW correction of the variance estimator. See \citet{fogarty2018regression} for more technical details.  Because we reduce the data from the MPE to the within-pair differences, it is unnecessary to center the covariates, which is different from the implementation of \citet{lin2013}'s estimator under the CRE.

\section{Examples}

\subsection{Darwin's data comparing cross-fertilizing and self-fertilizing on the height of corns}
\label{sec::frt-darwin-mpe}

This is a classical example from \citet{Fisher:1935}. It contains 15 pairs of corns with either cross-fertilizing or self-fertilizing, with the height being the outcome. The \ri{R} package \ri{HistData} provides the original data, where \ri{cross} and \ri{self} are the heights under cross-fertilizing and self-fertilizing, respectively, and \ri{diff} denotes their difference. 

\begin{lstlisting}
> library("HistData")
> ZeaMays
   pair pot  cross   self   diff
1     1   1 23.500 17.375  6.125
2     2   1 12.000 20.375 -8.375
3     3   1 21.000 20.000  1.000
4     4   2 22.000 20.000  2.000
5     5   2 19.125 18.375  0.750
6     6   2 21.500 18.625  2.875
7     7   3 22.125 18.625  3.500
8     8   3 20.375 15.250  5.125
9     9   3 18.250 16.500  1.750
10   10   3 21.625 18.000  3.625
11   11   3 23.250 16.250  7.000
12   12   4 21.000 18.000  3.000
13   13   4 22.125 12.750  9.375
14   14   4 23.000 15.500  7.500
15   15   4 12.000 18.000 -6.000
\end{lstlisting}

In total, the MPE has $2^{15} = 32768$ possible treatment assignment which is a tractable number in \ri{R}. The following function can enumerate all possible treatment assignments for the MPE:

\begin{lstlisting}
MP_enumerate = function(i, n.pairs) 
{
 if(i > 2^n.pairs)  print("i is too large.")
 a = 2^((n.pairs-1):0)
 b = 2*a
 2*sapply(i-1, 
          function(x) 
            as.integer((x %% b)>=a)) - 1
}
\end{lstlisting}

So we enumerate all the treatment assignments and calculate the corresponding $\hat{\tau}$'s and the one-sided exact $p$-value. 

\begin{lstlisting}
> difference = ZeaMays$diff 
> n.pairs    = length(difference)
> abs.diff   = abs(difference)
> t.obs      = mean(difference)
> t.ran      = sapply(1:2^15, 
+                     function(x){ 
+                       sum(MP_enumerate(x, 15)*abs.diff) 
+                       })/n.pairs
> pvalue     = mean(t.ran>=t.obs)
> pvalue
[1] 0.02633667
\end{lstlisting}

\begin{figure}[th]
\centering
\includegraphics[width = \textwidth]{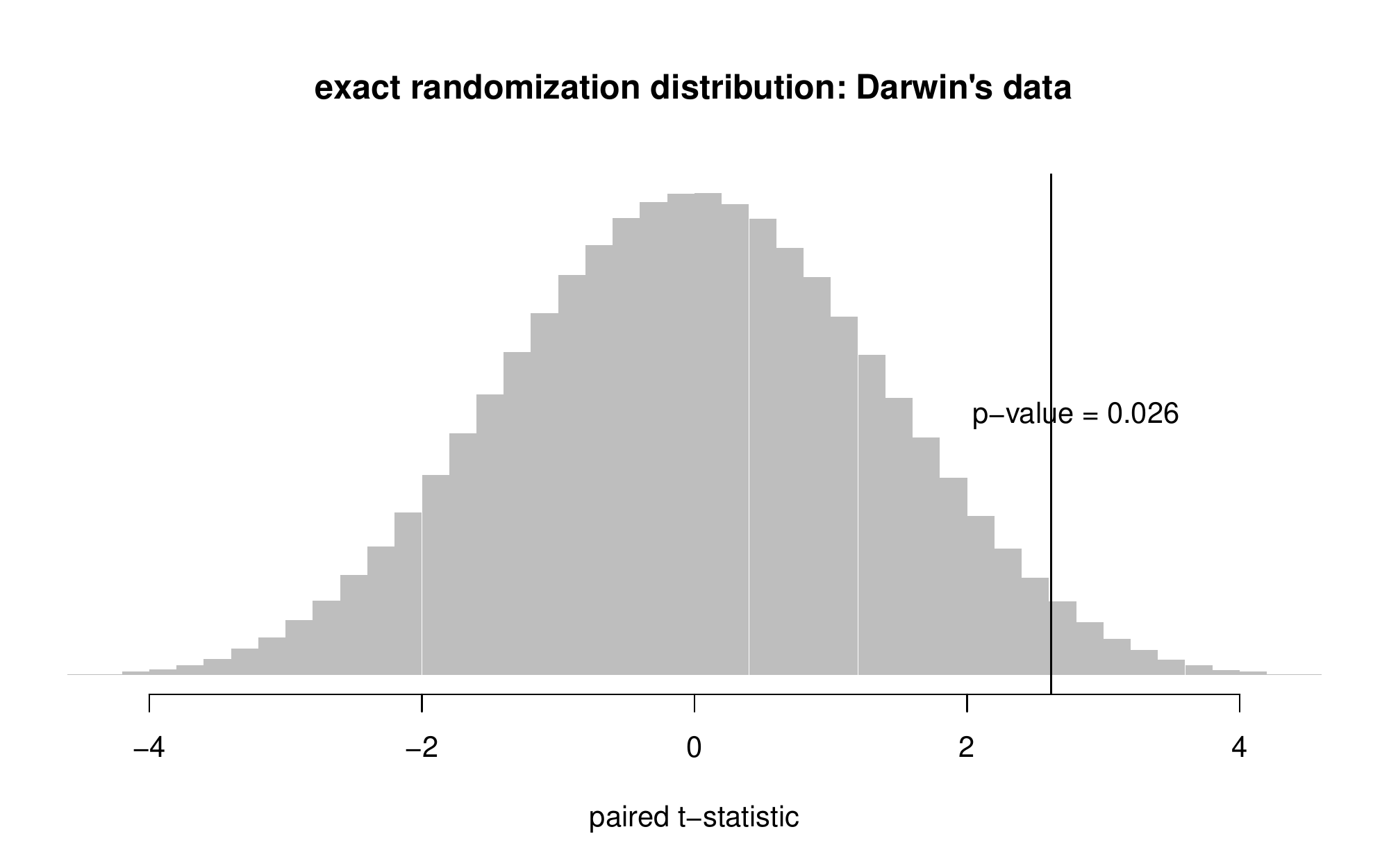}
\caption{Exact randomization distribution of $\hat{\tau}$ using Darwin's data}\label{fig::frt-darwin-mpe}
\end{figure}

Figure \ref{fig::frt-darwin-mpe} shows the  exact randomization distribution of $\hat{\tau}$ under $\HF$.

\subsection{Children's television workshop experiment data}
\label{sec::neyman-television-mpe}

I also re-analyze a subset of the data from the Children's Television Workshop experiment from \citet{ball1973reading} which was also analyzed by \citet[][Chapter 10]{imbens2015causal}. It contains 8 pairs of classes, with one of the two classes randomly assigned to be shown {\it The Electric Company} show during the standard reading-class period. It contains the pre-test score (covariate) and the post-test score (outcome). 
The following table summarizes the within-pair covariates and outcomes, as well as their differences:
\begin{lstlisting}
> dataxy = c(12.9, 12.0, 54.6, 60.6,
+            15.1, 12.3, 56.5, 55.5,
+            16.8, 17.2, 75.2, 84.8,
+            15.8, 18.9, 75.6, 101.9,
+            13.9, 15.3, 55.3, 70.6,
+            14.5, 16.6, 59.3, 78.4,
+            17.0, 16.0, 87.0, 84.2,
+            15.8, 20.1, 73.7, 108.6)
> dataxy = matrix(dataxy, 8, 4,  byrow = TRUE)           
> diffx = dataxy[, 2] - dataxy[, 1]
> diffy = dataxy[, 4] - dataxy[, 3]
> dataxy = cbind(dataxy, diffx, diffy)
> rownames(dataxy) = 1:8
> colnames(dataxy) = c("x.control", "x.treatment", 
+                      "y.control", "y.treatment",
+                      "diffx", "diffy")
> dataxy = as.data.frame(dataxy)
> dataxy
  x.control x.treatment y.control y.treatment diffx diffy
1      12.9        12.0      54.6        60.6  -0.9   6.0
2      15.1        12.3      56.5        55.5  -2.8  -1.0
3      16.8        17.2      75.2        84.8   0.4   9.6
4      15.8        18.9      75.6       101.9   3.1  26.3
5      13.9        15.3      55.3        70.6   1.4  15.3
6      14.5        16.6      59.3        78.4   2.1  19.1
7      17.0        16.0      87.0        84.2  -1.0  -2.8
8      15.8        20.1      73.7       108.6   4.3  34.9
\end{lstlisting}

The following \ri{R} code calculates $\hat\tau$ and $\hat V$. 
\begin{lstlisting}
> n      = dim(dataxy)[1]
> tauhat = mean(dataxy[, "diffy"])
> vhat   = var(dataxy[, "diffy"])/n
> tauhat
[1] 13.425
> sqrt(vhat)
[1] 4.636337
\end{lstlisting}
By Propositions \ref{prop::mpe-variance} and \ref{prop::colin2018regression},  we can  use the OLS to obtain the point estimators  and standard errors. Adjusting for covariates, we have
\begin{lstlisting}
> unadj = summary(lm(diffy ~ 1, data = dataxy))$coef
> round(unadj, 3) 
            Estimate Std. Error t value Pr(>|t|)
(Intercept)   13.425      4.636   2.896    0.023
\end{lstlisting}
Adjusting for covariates, we have
\begin{lstlisting} 
> adj = summary(lm(diffy ~ diffx, data = dataxy))$coef
> round(adj, 3)
            Estimate Std. Error t value Pr(>|t|)
(Intercept)    8.994      1.410   6.381    0.001
diffx          5.371      0.599   8.964    0.000
\end{lstlisting}

The above results assume large $n$, and $p$-values are justified if we believe the large-$n$ approximation. However, $n=8$ is not large. In total, we have $2^8 = 256$ possible treatment assignments, so the smallest possible $p$-value is $1/256 = 0.0039$, which is much larger than the $p$-value based on the Normal approximation of the covariate-adjusted estimator. In this example, it will be more reasonable to use the FRT with the studentized statistic, which is the \ri{t value} from the \ri{lm} function, to calculate exact $p$-values\footnote{See Chapter \ref{chapter::unification-fisher-neyman} for a justification.}. The following \ri{R} code calculates the $p$-values based on two $t$-statistics. 
\begin{lstlisting} 
> t.ran = sapply(1:2^8, function(x){ 
+   z.mpe = MP_enumerate(x, 8)
+   diffy.mpe = diffy*z.mpe
+   diffx.mpe = diffx*z.mpe
+   
+   c(summary(lm(diffy.mpe ~ 1))$coef[1, 3],
+     summary(lm(diffy.mpe ~ diffx.mpe))$coef[1, 3]) 
+ })
> p.unadj = mean(abs(t.ran[1, ]) >= abs(unadj[1, 3]))
> p.unadj
[1] 0.03125
> p.adj = mean(abs(t.ran[2, ]) >= abs(adj[1, 3]))
> p.adj
[1] 0.0078125
\end{lstlisting}

\begin{figure}[th]
\centering
\includegraphics[width = \textwidth]{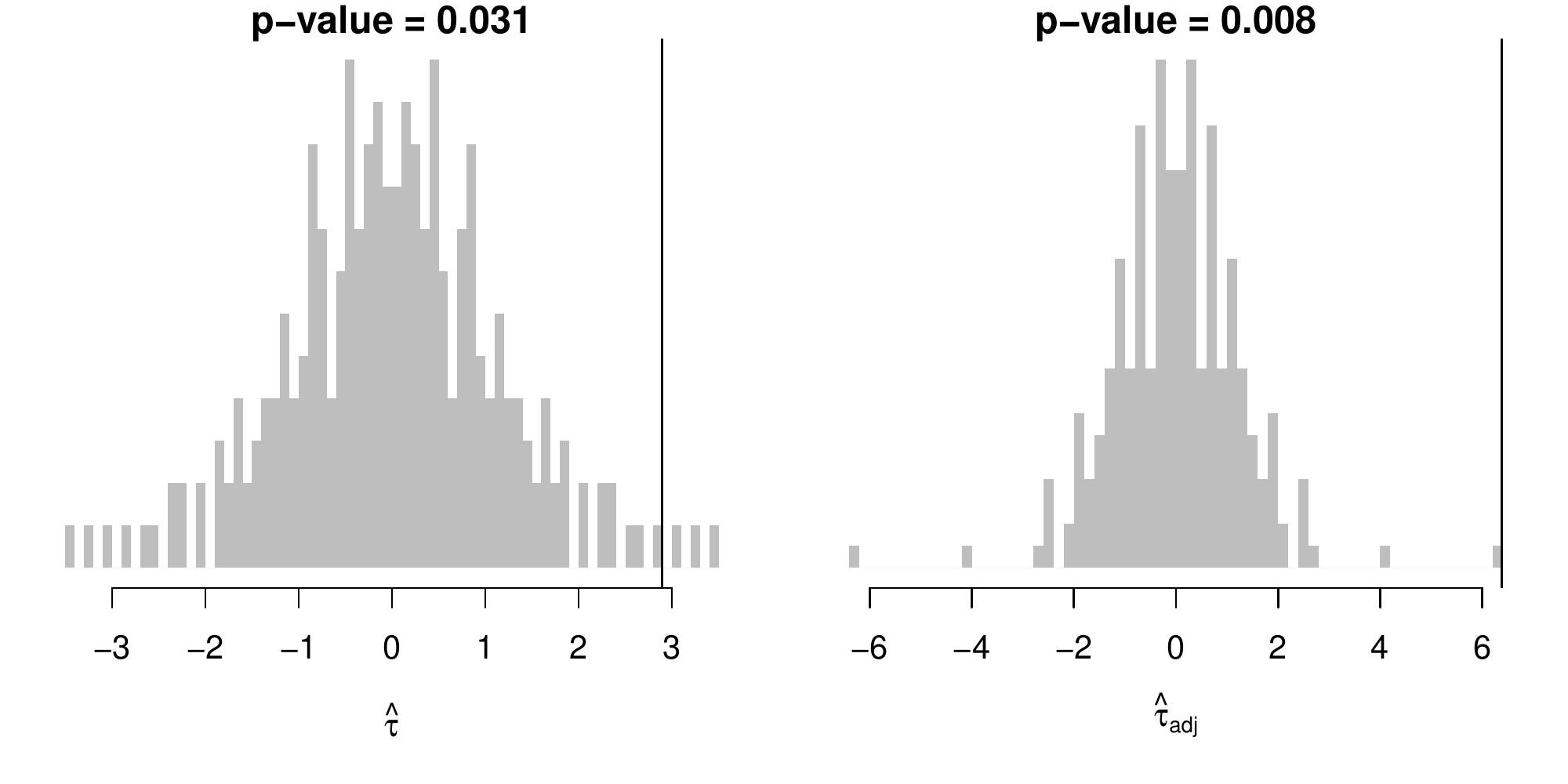}
\caption{Exact randomization distributions of the studentized statistics in Section \ref{sec::neyman-television-mpe} }\label{fig::frt-television-mpe}
\end{figure}

Figure \ref{fig::frt-television-mpe} shows the exact distributions of the two studentized statistics, as well as the two-sided $p$-values. The figure highlights the fact that the randomization distributions of the test statistics are discrete, taking at most 256 possible values. The Normal approximations are unlikely to be accurate, especially at the tails. We should report the $p$-values based on the FRT.

\section{Comparing the MPE and CRE}
\label{sec::discussion}

\citet{imai2008variance} compared the MPE and CRE. Heuristically, the conclusion is that the MPE gives more precise estimators if the matching is well done and the covariates are predictive of the outcome. However, without the outcome data in the design stage, it is hard to decide whether this is true or not. In the FRT, if covariates are predictive of the outcome, the MPE usually gives more powerful tests compared with the CRE. \citet{greevy2004optimal} illustrated this using simulation based on the Wilcoxon sign-rank statistic. However, this can be a subtle issue with finite samples. Consider an experiment with $2n$ units, with $n$ units receiving the treatment and $n$ units receiving the control. If we test the sharp null hypothesis at level $0.05$, then in the MPE, we need at least $2\times 5=10$ units since the smallest $p$-value is $1/2^5 = 1/32 < 0.05$ but $1/2^4 = 1/16 > 0.05$, but in the CRE, we need at least $2\times 4 = 8$ units since the smallest $p$-value is $1/\binom{8}{4} = 1/70 < 0.05$ but $1/\binom{6}{3} = 1/20 = 0.05$. So with $8$ units, it is impossible to reject the sharp null hypothesis in the MPE but it is possible in the CRE. Even if the covariates are perfect predictors of the outcome, the MPE is not superior to the CRE based on the FRT.

\section{Extension to the general matched experiment}\label{sec::general-matched-experiment}

It is straightforward to extend the MPE to the general matched experiment with varying numbers of control units. Assume that we have $n$ matched sets indexed by $i=1, \ldots, n$. For the matched set $i$, we have $1+M_i$ units. The $M_i$'s can vary. The total number of experimental units is $N = n + \sum_{i=1}^n M_i$. 
Let $ij$ index the unit $j$ within matched set $i$ ($i=1,\ldots, n$ and $j=1,\ldots, M_i+1$). Unit $ij$ has potential outcomes $Y_{ij}(1)$ and $Y_{ij}(0)$ under the treatment and control, respectively.

Within the matched set $i$ $(i=1, \ldots, n)$, the experimenter randomly selects exactly one unit to receive the treatment with the rest $M_i$ units receiving the control.
This general matched experiment is also a special case of the SRE with $n$ strata of size $1+M_i$ $(i=1, \ldots, n)$. 
Let $Z_{ij}$ be the treatment indicator for unit $ij$, which reveals one of the potential outcomes as
$$
Y_{ij} = Z_{ij} Y_{ij}(1) + (1-Z_{ij}) Y_{ij}(0) . 
$$ 

The average causal effect within the matched set $i$ equals
$$
\tau_i = (M_i + 1)^{-1} \sum_{j=1}^{1+M_i} \{ Y_{ij}(1)-Y_{ij}(0) \}. 
$$
Since the general matched experiment is a SRE, an unbiased estimator of $\tau_i $ is 
$$
\hat \tau_i = \sum_{j=1}^{M_i + 1} Z_{ij} Y_{ij} - M_i^{-1} \sum_{j=1}^{M_i+1} (1-Z_{ij}) Y_{ij}
$$
which is the difference in means of the outcomes within matched set $i$. 

Below we discuss the statistical inference with the general matched experiment.

\subsection{FRT}

As usual, we can always use the FRT to test the sharp null hypothesis
$$
\HF: Y_{ij}(1) = Y_{ij}(0)  \text{ for all }  i=1,\ldots, n; j=1,\ldots, M_i+1. 
$$
Because the general matched experiment is a special case of the SRE with many small strata, we can use the test statistics defined in Examples \ref{eg::hodgeslehmann}, \ref{eg::ks-stratified}, \ref{eg::mpe-sign-rank}, \ref{eg::mpe-ksstatistics}, \ref{eg::mpe-sign}, as well as the estimators and the corresponding $t$-statistics from the following two subsections.

\subsection{Estimating the average of the within-strata effects}

We first focus on estimating the average of the within-strata effects: 
$$
\tau = n^{-1} \sumn \tau_i .
$$
It has an unbiased estimator 
$$
\hat\tau = n^{-1} \sumn \hat \tau_i . 
$$
Interestingly, we can show that Theorem \ref{thm::varianceest-mp} holds for the general matched experiment, and so are other results for the MPE. In particular, we can use the OLS fit of the $\hat \tau_i $'s on the intercept to obtain the point and variance estimators for $\tau$. With covariates, we can use the OLS fit of the $\hat \tau_i $'s on the intercept and the $\hat\tau_{X,i}$'s, where
$$
\hat \tau_{X,i} = \sum_{j=1}^{M_i + 1} Z_{ij} X_{ij} - M_i^{-1} \sum_{j=1}^{M_i+1} (1-Z_{ij}) X_{ij}
$$
is the corresponding difference in means of the covariates within matched set $i$.

\subsection{A more general causal estimand}

Importantly, the $\tau$ above is the average of the $\tau_i$'s, which does not equal the average causal effect for the $N$ units in the experiment when the $M_i$'s vary. The average causal effect equals
$$
\tau' = N^{-1} \sumn \sum_{j=1}^{1+M_i} \{ Y_{ij}(1)-Y_{ij}(0) \} = \sumn \frac{1+M_i}{N} \tau_i.
$$ 
To unify the discussion, I consider the weighted causal effect
$$
\tau_w =  \sumn w_i \tau_i
$$
with $\sumn w_i = 1$, 
which includes $\tau$ as a special case with $w_i = n^{-1}$ and $\tau'$ as a special case with $w_i = (1+M_i)/N $ for $i=1,\ldots, n$. It is straightforward to obtain an unbiased estimator 
$$
\hat\tau_w =  \sumn w_i \hat\tau_i,
$$
and calculate its variance 
$$
\var(\hat\tau_w ) = \sumn w_i^2 \var (\hat\tau_i ). 
$$
However, estimating the variance of $\hat\tau_w $ is quite tricky because the $\hat\tau_i$'s are independent random variables without any replicates. This is a famous problem in theoretical statistics studied by \citet{hartley1969variance} and \citet{rao1970estimation}. \citet{fogarty2018mitigating} also discussed this problem without recognizing these previous works. I will give the final form of the variance estimator without detailing the motivation:
$$
\hat{V}_w = \sumn c_i (\hat\tau_i -\hat\tau_w )  ^2
$$
where 
$$
c_i = \frac{    \frac{w_i^2}{1-2w_i}   }{   1 + \sumn  \frac{w_i^2}{1-2w_i} }.
$$
As a sanity check, $c_i $ reduces to $ \{  n(n-1) \}^{-1}$ in the MPE with $M_i=1$ and $w_i = n^{-1}$. For simplicity, we focus on the case with $w_i < 1/2$ for all $i$'s, that is, there is no matched set containing more than half of the total weights. The following theorem extends Theorem \ref{thm::varianceest-mp}.  

\begin{theorem}
\label{thm::hadamdard-matched-experiment}
Under the general matched experiment with varying $M_i$'s, if $w_i < 1/2$ for all $i$'s, then  
 $$
 E(\hat{V}_w ) - \var( \hat\tau_w  ) = \sumn c_i (\tau_i - \tau_w )^2 \geq \var( \hat\tau_w  )  \geq 0
 $$
with equality holding if the $\tau_i $'s are constant. 
\end{theorem}

Although the theoretical motivation for $\hat{V}_w $ is quite complicated, it is not too difficult to verify Theorem \ref{thm::hadamdard-matched-experiment} directly. I relegate the proof to Problem \ref{para::variance-general-matched}.

\section{Homework Problems}

\paragraph{The true variance of $\hat{\tau}$ in the MPE}
\label{para::variance-mp-design}

Express $\var(\hat{\tau})$ in terms of the finite-population variances of the potential outcomes. 

\paragraph{A covariance estimator}\label{para::covariance-est-proof}
Prove Theorem \ref{thm::covariance-mp}.

\paragraph{Variance estimators via OLS}\label{para::seols-matchedpairs}
Prove Propositions \ref{prop::mpe-variance} and \ref{prop::colin2018regression}.

\paragraph{Point and variance estimator with binary outcome}\label{para::binary-matchedpairs-neyman}

This problem extends Example \ref{eg::mcnenar-test-binary} to Neymanian inference. 

Express $\hat\tau$ and $\hat V$ in terms of the counts in Table \ref{tb::binarymatchedpairs}.

\paragraph{Minimum sample size for the FRT}
\label{prob::minimum-n-0.001}

Extend the discussion in Section \ref{sec::discussion}.  Consider an experiment with $2n$ units, with $n$ units receiving the treatment and $n$ units receiving the control, and test the sharp null hypothesis at level $0.001$. What is the minimum value of $n$ for an MPE so that the smallest $p$-value does not exceed than $0.001$, and what is the corresponding minimum value of $n$ for a CRE?

\paragraph{Re-analyzing Darwin's data}

Chapter \ref{sec::frt-darwin-mpe} analyzed Darwin's data using the FRT based on the test statistic $\hat{\tau}$.

Re-analyze this dataset using the FRT with the Wilcoxon sign-rank statistic.

Re-analyze this dataset based on the Neymanian inference: report the unbiased point estimator, conservative variance estimator, and 95\% confidence interval.

\paragraph{Re-analyzing children's television workshop experiment data}

Chapter \ref{sec::neyman-television-mpe} analyzed the data from based on Neymanian inference.

Re-analyze this dataset using the FRT with different test statistics. 

Re-analyze this dataset using the FRT with covariate adjustment.

\paragraph{Re-analyzing \citet{angrist2009effects}'s data}\label{para::AL-mpe-neyman}

The original analysis of \citet{angrist2009effects}' was quite complicated. For this problem, please focus only on Table A1 of the original paper and view the schools as the experimental units. \citet{angrist2009effects} essentially conducted an MPE on the schools. Dropping pair 6 and all the pairs with noncompliance results in 14 complete pairs, with data shown below and also in \ri{AL2009.csv}: 
\begin{lstlisting}
   pair z  pr99  pr00  pr01  pr02
1     1 0 0.046 0.000 0.091 0.185
2     1 1 0.036 0.051 0.000 0.047
3     2 0 0.054 0.094 0.184 0.034
4     2 1 0.050 0.108 0.110 0.095
5     3 0 0.114 0.000 0.056 0.075
6     3 1 0.098 0.054 0.030 0.068
7     4 0 0.148 0.162 0.082 0.075
8     4 1 0.134 0.390 0.339 0.458
9     5 0 0.152 0.105 0.083 0.129
10    5 1 0.145 0.077 0.579 0.167
11    6 0 0.188 0.214 0.375 0.545
12    6 1 0.179 0.165 0.483 0.444
13    7 0 0.193 0.771 0.328 0.583
14    7 1 0.189 0.186 0.168 0.368
15    8 0 0.197 0.350 0.000 0.383
16    8 1 0.200 0.071 0.667 0.429
17    9 0 0.213 0.176 0.164 0.172
18    9 1 0.209 0.165 0.092 0.151
19   10 0 0.211 0.667 0.250 0.617
20   10 1 0.219 0.250 0.500 0.350
21   11 0 0.219 0.153 0.185 0.219
22   11 1 0.224 0.363 0.372 0.342
23   12 0 0.255 0.226 0.213 0.327
24   12 1 0.257 0.098 0.107 0.095
25   13 0 0.261 0.071 0.000    NA
26   13 1 0.263 0.441 0.448 0.435
27   14 0 0.286 0.161 0.126 0.181
28   14 1 0.285 0.389 0.353 0.309
\end{lstlisting}
The outcomes are the Bagrut passing rates in the years 2001 and 2002, with the Bagrut passing rates in 1999 and 2000 as pretreatment covariates. Re-analyze the data based on the Neymanian inference with and without covariates. In particular, how do you deal with the missing outcome in pair 25? 

\paragraph{Variance estimation in the general matched experiment} 
 \label{para::variance-general-matched}

This problem contains more details for Section  \ref{sec::general-matched-experiment}.

First, prove Theorem  \ref{thm::varianceest-mp} for the general matched experiment.
Second, prove Theorem \ref{thm::hadamdard-matched-experiment}. 
 
Remark: For the second part, we can first verify that $\hat\tau_i -\hat\tau_w$ has mean $\tau_i - \tau_w$ and variance
 $$
 \var(\hat\tau_i -\hat\tau_w) = \var(\hat\tau_w ) + (1-2w_i) \var(\hat\tau_i). 
 $$

\paragraph{Recommended readings}

\citet{greevy2004optimal} provided an algorithm to form matched pairs based on covariates. 
\citet{imai2008variance} discussed the estimation of the average causal effect without covariates,
and 
\citet{fogarty2018regression} discussed covariate adjustment in MPEs.

 \chapter{Unification of the Fisherian and Neymanian Inferences in Randomized Experiments}\label{chapter::unification-fisher-neyman}
 \chaptermark{Unification of Fisherian and Neymanian Inferences}

Chapters \ref{ch::frt-cre}--\ref{chapter::mpe} cover both the Fisherian and Neymanian inferences for different types of experiments. The Fisherian perspective focuses on the finite-sample exact $p$-values for testing the strong null hypothesis of no causal effects for any units whatsoever, whereas the Neymanian perspective focuses on unbiased estimation with a conservative large-sample confidence interval for the average causal effect. Both of them are justified by the physical randomization which is ensured by the design of the experiments. Because of this, they are both called {\it randomization-based inference} or {\it design-based inference}. Because they concern a finite population of units in the experiments, they are also called {\it finite-population inference}.  They are related but also have distinct features. 

In 1935, Neyman presented his seminal paper on randomization-based inference to the Royal Statistical Society. His paper \citep{neyman1935statistical} was attacked by Fisher in the discussion session. \cite{sabbaghi2014comments} reviewed this famous Neyman--Fisher controversy and presented some new results for this old problem. Instead of going to the philosophical issues, this chapter tries to provide a unified discussion.

\section{Testing strong and weak null hypotheses in the CRE} 
 \label{sec::strong-weak-cre-frt}

Let us revisit the treatment-control CRE.  The Fisherian perspective focuses on testing the strong null hypothesis
$$
\HF: Y_i(1) = Y_i(0) \text{ for all units } i=1,\ldots, n.
$$
The FRT delivers a finite-sample exact $p_\frt$ defined in \eqref{eq::exactpvalue}.

By duality of the confidence interval and hypothesis testing, the Neymanian perspective gives a test for the  weak null hypothesis
$$
\HN : \tau = 0  \Longleftrightarrow \HN: \bar{Y}(1) = \bar{Y}(0)
$$ 
based on 
$$
t= 
\frac{\hat{\tau} }{  \sqrt{  \hat{V}    } }  .
$$
By the CLT of $\hat\tau$ and the conservativeness of the variance estimator, we have 
$$
t  =   \sqrt{  \frac{ \var(\hat{\tau}) }{ \hat{V} } } \times  \frac{\hat{\tau} }{  \sqrt{  \var(\hat{\tau})   } }    \rightarrow  C \times   \N01 
$$
in distribution, where $C$ is smaller than or equal to $1$ but depends on the unknown potential outcomes. Using $\N01$ quantiles for the studentized statistic $t$, we have a conservative large-sample test for $\HN$. 

Furthermore, \citet{ding2017randomization} show that the FRT with the studentized statistic $t$ has the dual guarantees:
\begin{enumerate}
\item 
 $p_\frt$ is finite-sample exact under $\HF$;
\item
 $p_\frt$ is asymptotically conservative under $\HN$.
\end{enumerate}
Importantly, this is a feature of the studentized statistic $t$. \citet{ding2017randomization} showed that the FRT with other test statistics may not have the dual guarantees. In particular, the FRT with $\hat{\tau}$ may be asymptotically anti-conservative under $\HN$. I give some heuristics below to illustrate the importance of studentization in the FRT.

Under $\HN$, we have
$$
\hat{\tau} \asim \textsc{N}\left(0,    \frac{S^2(1)}{n_1} + \frac{S^2(0)}{n_0} -  \frac{S^2(\tau)}{n}    \right).
$$
The FRT pretends that the Science Table is $(Y_i, Y_i)_{i=1}^n$ induced by $\HF$, so the randomization distribution of $\hat{\tau} $ is
$$
(\hat{\tau} )^\pi \asim \textsc{N}\left(0,    \frac{s^2}{n_1} + \frac{s^2}{n_0}    \right),
$$
where $(\cdot)^\pi$ denotes the randomization distribution\footnote{$\pi$ is a standard notation for a random permutation.}
 and 
$
s^2 
$
is the sample variance of the observed outcomes. Based on \eqref{eq::frt-algebra-difference-in-means} in Chapter \ref{ch::frt-cre}, we can approximate the asymptotic variance of $(\hat{\tau} )^\pi$ under $\HF$ as
\begin{eqnarray*}
\frac{s^2}{n_1} + \frac{s^2}{n_0} 
& =& \frac{n}{n_1 n_0}  \left\{ 
 \frac{n_1 -1}{ n-1 } \hat S^2(1) + \frac{n_0 - 1}{n-1} \hat S^2(0) + \frac{n_1n_0}{n(n-1)} \hat\tau^2
 \right\} \\
&\approx & \frac{\hat S^2(1) }{n_0} +  \frac{\hat S^2(0) }{n_1} \\ 
&\approx & \frac{  S^2(1) }{n_0} +  \frac{  S^2(0) }{n_1},
\end{eqnarray*}
 which does not match the asymptotic variance of $\hat{\tau} $. Ideally, we should compute the $p$-value under $\HN$ based on the true distribution of $\hat{\tau} $, which, however, depends on the unknown potential outcomes. In contrast, we use the FRT to compute the $p_\frt$ based on the permutation distribution $(\hat{\tau} )^\pi $, which does not match the true distribution of $\hat{\tau} $ under $\HN$ even with large samples. Therefore, the FRT with $\hat\tau$ may not control the type one error rate under $\HN$ even with large samples.

Fortunately, the undesired property of the FRT with $\hat\tau$ goes away if we replace the test statistic $\hat{\tau} $ with the studentized version $t$. Under $\HN$, we have
$$
t \asim \textsc{N}(0,  C^2)
$$ 
where $C^2 \leq 1$ with equality holding if $Y_i(1) - Y_i(0) = \tau$ for all units $i=1, \ldots, n$. The FRT pretends that $Y_i(1) = Y_i(0) = Y_i$ for all $i$'s and generates the permutation distribution
$$
t^\pi \asim \textsc{N}(0,  1)
$$
where the variance equals $1$ because the Science Table used by the FRT has zero individual causal effects.  Under $\HN$, because the true distribution of $t$ is more dispersed than the corresponding permutation distribution, the $p_\frt$ based on $t$ is asymptotically conservative. 


\section{Covariate-adjusted FRTs in the CRE}
 \label{sec::xfrt-cre}
 
Extending the discussion in Section \ref{sec::strong-weak-cre-frt} to the case with covariates, \citet{zhao2020caFRT} recommend using the FRT with the studentized \citet{lin2013}'s estimator:
$$
t_\textsc{L}   =  \frac{  \hat{\tau}_\textsc{L} }{  \sqrt{\hat{V}_\textsc{L}}  },
$$ 
which is the robust $t$-statistic for the coefficient of $Z_i$ in the OLS fit of $Y_i$ on $( 1, Z_i, X_i, Z_i X_i) $. They show that the FRT with $t_\textsc{L} $ has multiple guarantees:
 \begin{enumerate}
\item
 $p_\frt$ is finite-sample exact under $\HF$;
\item
 $p_\frt$ is asymptotically conservative under $\HN$;
\item
 $p_\frt$ is asymptotically more powerful than the FRT with $t$ when $\HN$ does not hold and the covariates are predictive to the outcomes;
\item
the above properties hold even when the linear outcome model is misspecified. 
\end{enumerate}
Similarly, this is a feature of the studentized statistic $t_\textsc{L} $. \citet{zhao2020caFRT} show that other covariate-adjusted FRTs reviewed in Section \ref{sec::covariate-adjusted-frt-intro} may  be either anti-conservative under $\HN$ or less powerful than the FRT with $t_\textsc{L} $ when $\HN$ does not hold.

\section{A simulation study}
\label{sec::simulation-xfrt-typeoneerror}

Now I use simulation to evaluate the finite-sample properties of the $p_\frt$'s under the weak null hypothesis. I will use the following twelve test statistics.
\begin{enumerate}
\item
The first three test statistics are from the OLS implementation of \citet{Neyman:1923}, including the coefficient of the treatment, the $t$-statistic based on the classic standard error, and the $t$-statistic based on the EHW standard error.

\item
The next three test statistics are based on the pseudo-outcome strategy in Definition \ref{def::pseudo-y-frt}, advocated by \citet{rosenbaum2002covariance}. We first residualize the outcomes by regressing $Y_i$ on $(1,X_i)$, and then obtain the three test statistics similar to the first three.

\item
The next three test statistics are based on the OLS implementation of \citet{fisher1925statistical}, as discussed in Chapter \ref{sec::ancova}.

\item
The final three test statistics are based on the OLS implementation of \citet{lin2013}, as discussed in Chapter \ref{sec::ancova}.
\end{enumerate}

Consider a finite population of $n=100$ units subjected to a CRE of size $(n_1, n_0) =(20,80)$.
For each $i  $, we draw a univariate covariate $ X_i $ from Unif$(-1,1)$ and generate potential outcomes as $Y_i(1) \sim \textsc{N}(  X_i ^3 , 1)$ and $Y_i(0) \sim \textsc{N}( - X_i ^3 , 0.5^2)$.
Center the $Y_i(1)$'s and $Y_i(0)$'s  to ensure $\tau= 0$.  
Fix $\{Y_i(1), Y_i(0),  X_i \}_{ i = 1}^{n} $ in simulation. We draw a random permutation of $n_1$ 1's and $n_0$ 0's to obtain the observed outcomes and conduct FRTs. 
The procedure is repeated 500 times, with the $p$-values approximated by 500 independent permutations of the treatment vector in each replication.

\begin{sidewaysfigure} 
\centering
\includegraphics[width = \textwidth]{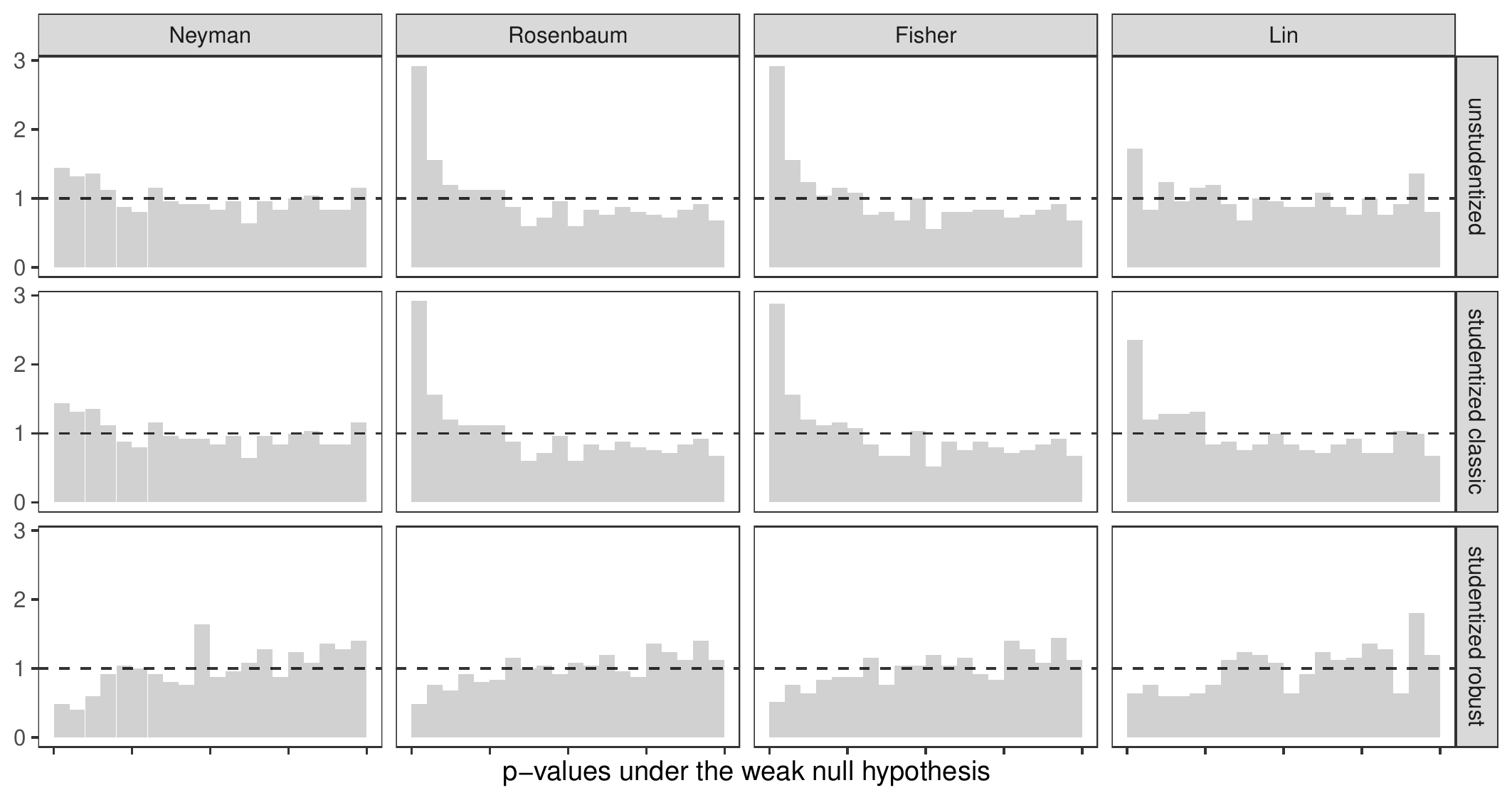}
\caption{Histograms of $p_\frt$ under the weak null hypothesis}\label{fig::xfrt-type1error}
\end{sidewaysfigure}

Figure \ref{fig::xfrt-type1error} shows the $p$-values under the weak null hypothesis.
The four robust $t$-statistics, as shown in the last row, are the only ones that preserve the correct type one error rates. 
In fact, they are conservative, which is coherent with the theory. 
All the other eight statistics yield type one error rates greater than the nominal levels and are thus not proper for testing the weak null hypothesis.

\section{General recommendations}

The recommendations for the SRE parallel those for the CRE if both the strong and weak null hypotheses are of interest. 
Recall the estimators in Chapters \ref{chapter::stratification-poststratification} and \ref{sec::covariate-SRE}. 
Without additional covariates, \citet{zhao2020caFRT} recommend  using the FRT with 
$$
t_\textsc{S} =  \frac{  \hat{\tau}_\textsc{S} }{  \sqrt{\hat{V}_\textsc{S}} } ;
$$ 
with additional  covariates, they recommend using the FRT with 
$$
t_\textsc{L,S} =  \frac{  \hat{\tau}_\textsc{L,S} }{  \sqrt{\hat{V}_\textsc{L,S}} } . 
$$

The analysis of ReM is trickier. \citet{zhao2020caFRT} show that the FRT with $t$ does not have the dual guarantees in Section \ref{sec::strong-weak-cre-frt}, but the FRT with $t_\textsc{L}$ still has the guarantees in Section \ref{sec::xfrt-cre}. This highlights the importance of both covariate adjustment and studentization in ReM.

Similar results hold for the MPE. Without covariates, we recommend using the FRT with the $t$-statistic for the intercept in the OLS fit of $\hat\tau_i$ on $1$; with covariates, we recommend using the FRT with the $t$-statistic for the intercept in the OLS fit of $\hat\tau_i$ on $1$ and $\hat\tau_{X,i}$. Figure \ref{fig::frt-television-mpe} in Chapter \ref{chapter::mpe} are based on these recommended FRTs.

Overall, the FRTs with studentized statistics are safer choices. When the large-sample Normal approximations to the studentized statistics are accurate, the  $p_\frt$'s are almost identical to the $p$-values based on Normal approximations. When the large-sample approximations are inaccurate, the FRTs at least guarantee valid $p$-values under the strong null hypotheses. This is the recommendation of this book.

\section{A case study}
\label{sec::case-study-chong}

Recall the SRE in \citet{chong2016iron} analyzed in Chapter \ref{sec::ps-chong}. 
I compare the ``soccer'' arm versus the ``control'' arm and the ``physician'' arm versus the ``control'' arm. We also compare the FRTs with and without using the covariate indicating the baseline anemia status. We use their dataset to illustrate the FRTs in the CRE and SRE.  The ten subgroup analyses within the same class levels use the FRTs with $t$ and $t_\textsc{L}$ for the CRE and the two overall analyses averaging over all class levels use the FRTs with $t_\textsc{S}$ and $t_\textsc{L,S}$ for the SRE. 

Table \ref{tb::pv-values-chong} shows the point estimators, standard errors, the $p$-value based on the Normal approximation of the robust $t$-statistics, and the $p$-value based on the FRTs. In most strata, covariate adjustment decreases the standard error since the baseline anemia status is predictive of the outcome. Table \ref{tb::pv-values-chong} also exhibits two exceptions: within class 2, covariate adjustment increases the standard error when comparing ``soccer'' and ``control''; in class 4, covariate adjustment increases the standard error when comparing ``physician'' and ``control''. This is due to the small group sizes within these strata, causing the asymptotic approximation inaccurate. Nevertheless, in these two scenarios, the differences in the standard error are in the third digit.  The $p$-values from the Normal approximation and the FRT are close with the latter being slightly larger in most cases. Based on the theory, the $p$-values based on the FRT should be trusted since it has an additional guarantee of being finite-sample exact under the sharp null hypothesis. This becomes important in this example since the group sizes are quite small within strata. 

\begin{table}
\centering
\caption{Re-analyzing \citet{chong2016iron}'s data. ``\textsc{N}'' corresponds to the unadjusted estimators and tests due to \citet{Neyman:1923}, and ``\textsc{L}'' corresponds to the covariate-adjusted estimators and tests due to \citet{lin2013}.}\label{tb::pv-values-chong}
\begin{subtable}[h]{0.45\textwidth}
\caption{soccer versus control}
    \begin{tabular}{rrrrr}
    \hline 
   &    est & s.e. &$p_\text{Normal}$ & $p_\frt$ \\
       class 1& &&&\\
\textsc{N}  &  0.051 &0.502 &     0.919  & 0.924 \\
\textsc{L}   &    0.050 & 0.489 &     0.919 &  0.929 \\
       class 2& &&&\\
\textsc{N}  &  -0.158 &0.451  &    0.726  & 0.722\\
\textsc{L}   &  -0.176 & 0.452   &   0.698  & 0.700 \\
       class 3& &&&\\
\textsc{N}  &  0.005 &0.403  &    0.990 &  0.989 \\
\textsc{L}   &   -0.096& 0.385 &     0.803&   0.806 \\
       class 4& &&&\\
\textsc{N}  &  -0.492& 0.447  &    0.271 &  0.288 \\
\textsc{L}   &    -0.511& 0.447  &    0.253 &  0.283 \\
       class 5& &&&\\
\textsc{N}  &  0.390& 0.369 &     0.291 &  0.314 \\
\textsc{L}   &   0.443 &0.318  &    0.164  & 0.186 \\
       all& &&&\\
\textsc{N}  &  -0.051 & 0.204  &    0.802 &  0.800 \\
\textsc{L}   &    -0.074& 0.200   &   0.712  & 0.712\\
\hline 
    \end{tabular}
\end{subtable}

\begin{subtable}[h]{0.45\textwidth} 
\caption{physician versus control}
    \begin{tabular}{rrrrr}
    \hline 
   &    est & s.e. &$p_\text{Normal}$ & $p_\frt$ \\
       class 1& &&&\\
\textsc{N}  &  0.567 &0.426  &    0.183 &  0.192 \\
\textsc{L}   &   0.588& 0.418  &   0.160  & 0.174 \\
       class 2& &&&\\
\textsc{N}  &   0.193& 0.438 &     0.659 &  0.666 \\
\textsc{L}   &  0.265& 0.409   &   0.517  & 0.523 \\
       class 3& &&&\\
\textsc{N}  &  1.305& 0.494  &    0.008 &  0.012\\
\textsc{L}   &  1.501& 0.462   &   0.001  & 0.003 \\
       class 4& &&&\\
\textsc{N}  &  -0.273& 0.413  &    0.508 &  0.515 \\
\textsc{L}   &  -0.313& 0.417   &   0.454  & 0.462\\
       class 5& &&&\\
\textsc{N}  & -0.050 &0.379  &    0.895 &  0.912 \\
\textsc{L}   &  -0.067& 0.279  &    0.811 &  0.816 \\
       all& &&&\\
\textsc{N}  &  0.406 & 0.202   &   0.045 &  0.047\\
\textsc{L}   &    0.463 & 0.190   &   0.015  & 0.017 \\
\hline 
    \end{tabular}
\end{subtable}
\end{table}

\begin{sidewaysfigure}
\centering
\includegraphics[width = \textwidth]{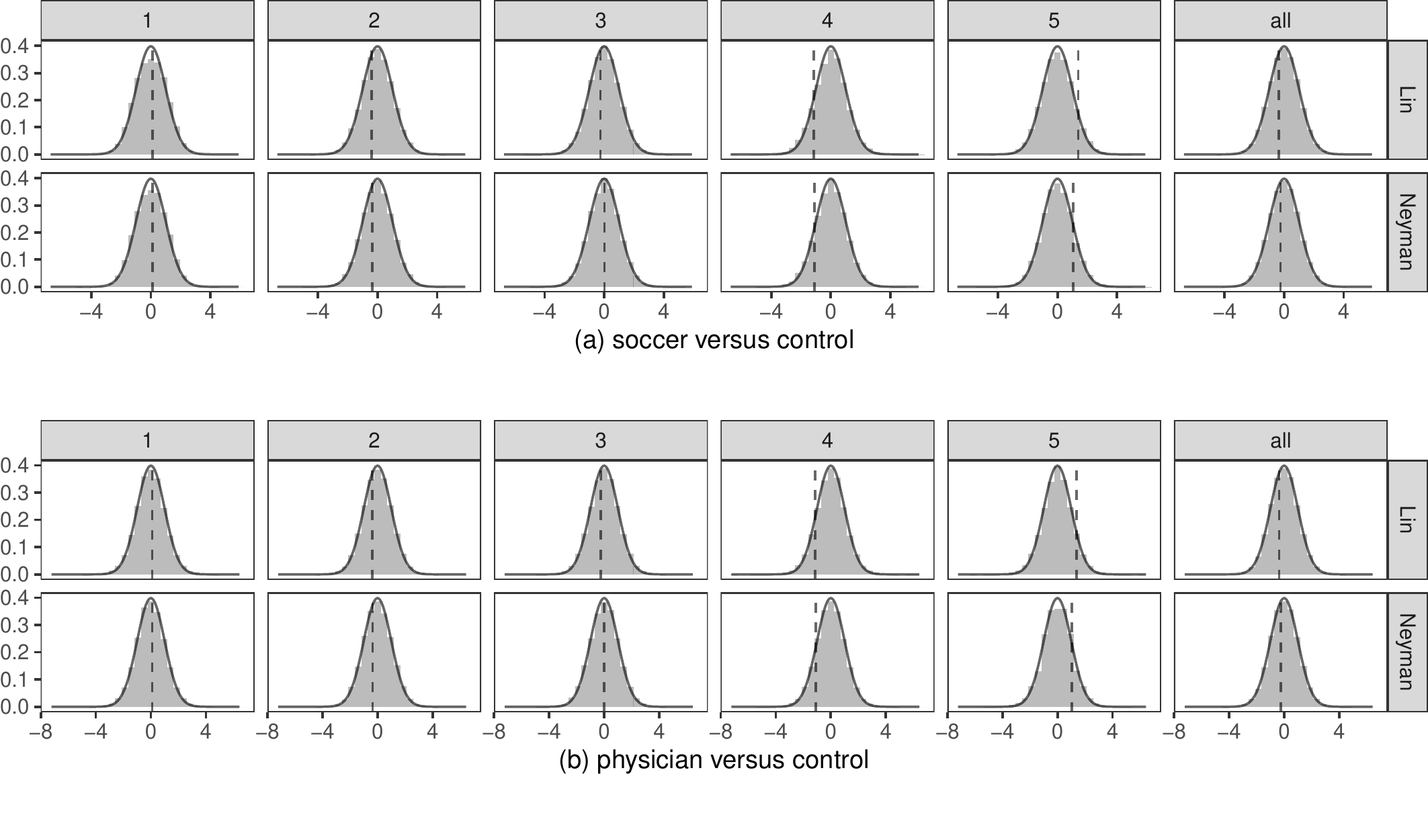}
\caption{Re-analyzing \citet{chong2016iron}'s data: randomization distributions with $5\times 10^4$ Monte Carlo draws and the $\N01$ approximations}
\label{fig::randomization-distributions-chong}
\end{sidewaysfigure}

Figure \ref{fig::randomization-distributions-chong} compares the histograms of the randomization distributions of the robust $t$-statistics with the asymptotic approximations. In the subgroup analysis, we can observe discrepancies between the randomization distributions and $\textsc{N}(0,1)$; averaged over all class levels, the discrepancy becomes unnoticeable. Overall, in this application, the $p$-values based on the Normal approximation do not differ substantially from those based on the FRTs. Two approaches yield coherent conclusions: the video with a physician telling the benefits of iron supplements improved academic performance and the effect was most significant among students in class 3; in contrast, the video with a famous soccer player telling the benefits of the iron supplements did not have any significant effect.

 \section{Homework Problems}

\paragraph{Re-analyzing \citet{angrist2009effects}'s data}\label{para::AL-mpe-fisher}

This is the Fisherian counterpart of Problem \ref{para::AL-mpe-neyman}.
Report the $p_\frt$'s from the FRTs with studentized statistics.

\paragraph{Replication of \citet{zhao2020caFRT}'s Figure 1}\label{para::zhaoding-xfrt}

\citet{zhao2020caFRT} use simulation to evaluate the finite-sample properties of the $p_\frt$'s from the FRTs with various test statistics. Based on their Figure 1, they recommend using the FRT with $t_\textsc{L,S} $ to analyze the SRE. Replicate their Figure 1.

\paragraph{Recommended readings}

Using the studentized statistics in permutation tests is not a new idea; see \citet{Janssen97} and \citet{Romano13}. Under the design-based framework, \citet{ding2017randomization} and \citet{wu2018randomization} made the recommendation for multiarmed experiments, and  \citet{zhao2020caFRT} extended the recommendation to covariate adjustment.

    \chapter{Bridging Finite and Super Population Causal Inference}\label{chapter::bridging}
 \chaptermark{Bridging finite and super population causal inference}

We have focused on the finite population perspective in randomized experiments. It treats all the potential outcomes as fixed numbers. Even if the potential outcomes are random variables, we can condition on them under the finite population perspective. The advantages of this perspective are
\begin{enumerate}
\item
it focuses on the design of the experiments;
\item
it requires minimal assumptions on the data-generating process of the outcomes.
\end{enumerate}
However, it is often criticized for having only {\it  internal validity} but not necessarily {\it external validity}, with informal definitions below. 

\begin{definition}
[internal validity]
The statistical analysis is valid for the study samples at hand. 
\end{definition}

\begin{definition}
[external validity]
The statistical analysis is valid for a broader population beyond the study samples. 
\end{definition}

Obviously, all experimenters want not only the internal validity but also the external validity of their experiments.  Since all statistical properties are conditional on the potential outcomes for the units we have, the results are only about the observed units under the finite population perspective. Then a natural question arises: do the finite population results generalize to a bigger population?

This is a fair critique of the finite population framework conditional on the potential outcomes. However, this can be a philosophical question. What we observed is a finite population, so any experimental design and analysis directly give us information about this finite population. Randomization only ensures internal validity given the potential outcomes of these units. The external validity of the results depends on the sampling process of the units. If the finite population is a representative sample of a larger population we are interested in, then of course the experimental results also have external validity. Otherwise, the results based on randomization inference may not generalize. To rigorously discuss this issue, we need to have a framework with two levels of randomness, one due to the sampling process of the units and the other due to the random assignment of the treatment. See \citet{miratrix2018worth} and \citet{abadie2020sampling} for formal discussions of this idea.\footnote{\citet{pearl2014external} discussed the {\it transportability} problem from the perspective of causal diagrams.}

For some statisticians, this is just a technical problem. We can change the statistical framework, assuming that the units are sampled from a superpopulation. Then all the statements are about the population of interest. This is a convenient framework, although it does not really solve the problem mentioned above. Below, I will introduce this framework for two purposes:
\begin{enumerate}
\item
it gives a different perspective on randomized experiments;
\item
it serves as a bridge between Parts \ref{part::rcts} and  \ref{part::observational-studies} of this book.
\end{enumerate}
The latter purpose is more important since the superpopulation framework allows us to derive more fruitful results for observational studies in which the treatment is not randomly assigned.

\section{CRE}\label{sec::cre-superpop}

Assume 
$$
\{  Z_i, Y_i(1), Y_i(0),  X_i  \}_{i=1}^n  \iidsim \{  Z, Y(1), Y(0), X \}
$$ 
from a superpopulation. So we can drop the subscript $i$ for quantities of this population.   With a little abuse of notation, we define the population average causal effect as
$$
\tau  = E\{  Y(1)  - Y(0)\}  = E\{  Y(1)  \} -  E\{ Y(0)\} .
$$
Under the superpopulation framework, we can formulate the CRE as below.

\begin{definition}
[CRE under the superpopulation framework]
\label{def::cre-superpopulation}
We have
$$
Z \ind \{  Y(1), Y(0), X \} .
$$
\end{definition}

Under Definition \ref{def::cre-superpopulation}, the average causal effect can be written as
\begin{eqnarray}
\tau &=& E\{  Y(1) \mid Z=1\}  - E\{  Y(0) \mid Z=0 \}   \nonumber  \\
&=& E( Y \mid Z=1)  - E(  Y \mid Z=0 ),
\label{eq::cre-identifiable}
\end{eqnarray}
which equals the difference in expectations of the outcomes. The first line in \eqref{eq::cre-identifiable} is the key step that leverages the value of randomization.  Since  $\tau  $ can be expressed as a function of the distributions of the observables, it is {\it nonparametrically identifiable}\footnote{In causal inference, we say that a parameter is nonparametrically identifiable if it can be determined by the distribution of the observed variables without imposing further parametric assumptions. See Definition \ref{def::nonparametric-identification} later for a more formal discussion.}. The formula \eqref{eq::cre-identifiable} immediately suggests a moment estimator $\hat{\tau}$, which is the difference in means of the outcomes defined before. Conditioning on $\bm{Z}$, this is then a standard two-sample problem comparing the means of two independent samples. We have
$$
E(  \hat{\tau} \mid \bm{Z}) = \tau
$$
and
$$
\var (  \hat{\tau} \mid \bm{Z}) =   \frac{ \var\{ Y(1) \}}{n_1} + \frac{ \var\{ Y(0) \}}{n_0}.
$$
Under IID sampling, the sample variances are unbiased for the population variances, so \citet{Neyman:1923}'s variance estimator is unbiased for $\var (  \hat{\tau} \mid \bm{Z}) $. The conservativeness problem goes away under this superpopulation framework.

We can also discuss covariate adjustment. Based on the OLS decompositions (see Chapter \ref{appendix::basic-linear-regression})
\begin{eqnarray}
Y(1) &=& \gamma_1 + \beta_1\tran X + \varepsilon(1),\label{eq::ols-decomposition-y1}\\
Y(0) &=& \gamma_0 + \beta_0\tran X + \varepsilon(0),\label{eq::ols-decomposition-y0}
\end{eqnarray}
we have
\begin{eqnarray*}
\tau  &=& E\{  Y(1)  - Y(0)\} \\
&=& \gamma_1  -  \gamma_0 + (\beta_1 - \beta_0)\tran E(X),
\end{eqnarray*}
since the residuals $  \varepsilon(1)$ and $\varepsilon(0)$ have mean zero due to the inclusion of the intercepts. 
We can use the OLS with the treated and control data to estimate the coefficients in \eqref{eq::ols-decomposition-y1} and \eqref{eq::ols-decomposition-y0}, respectively. The sample versions of the coefficients are $\hat{\gamma}_1, \hat{\beta}_1, \hat{\gamma}_0, \hat{\beta}_0$, so a covariate-adjusted estimator for  $\tau $ is
$$
\hat{\tau}_\text{adj} =  \hat \gamma_1  - \hat  \gamma_0 + (\hat \beta_1 - \hat \beta_0)\tran \bar{X}.
$$
If we center covariates with $ \bar{X} = 0$, the above estimator reduces to \citet{lin2013}'s estimator
$$
\hat{\tau}_\textsc{L} = \hat \gamma_1  - \hat  \gamma_0,
$$
which equals the coefficient of $Z$ in the pooled regression with treatment-covariates interactions; see Proposition \ref{prop::lin-ols}.

Unfortunately, the EHW variance estimator does not work for $\hat{\tau}_\textsc{L}$ because of the additional uncertainty in the sample mean of covariates $\bar{X} $ under the super population framework. \citet{berk2013covariance}, \citet{negi2020revisiting} and \citet{zhao2020caFRT} proposed a correction of the EHW variance estimator by adding an extra term
\begin{equation}\label{eq::lin-ehw-correction}
(\hat \beta_1 - \hat \beta_0)\tran S_X^2  (\hat \beta_1 - \hat \beta_0)  / n
\end{equation}
where $\hat \beta_1 - \hat \beta_0$ equals the coefficient of the interaction $Z_iX_i$ in obtaining \citet{lin2013}'s estimator and $S_X^2$ is the finite-population covariance matrix of the covariates. 
A conceptually simpler yet computationally intensive approach is to use the bootstrap to estimate the variance; see Chapter \ref{sec::bootstrap} and Problem \ref{hw::variance-bootstrap-superpopulation}.

\section{Simulation under the CRE: the super population perspective}
\label{sec::simulation-CRE-superp}

The following \ri{linestimator} function can compute \citet{lin2013}'s estimator, the EHW standard error, and the corrected standard error based on \eqref{eq::lin-ehw-correction} for the super population. 
\begin{lstlisting}
library("car")
linestimator = function(Z, Y, X){
  ## standardize X
  X      = scale(X)
  n      = dim(X)[1]
  p      = dim(X)[2]
  
  ## fully interacted OLS
  linreg = lm(Y ~ Z*X)
  est    = coef(linreg)[2]
  vehw   = hccm(linreg)[2, 2]
  
  ## super population correction
  inter  = coef(linreg)[(p+3):(2*p+2)]
  vsuper = vehw + sum(inter*(cov(X)%*%inter))/n
  
  c(est, sqrt(vehw), sqrt(vsuper))
}
\end{lstlisting}

I then use simulation to compare the EHW standard error and the corrected standard error. 
I choose sample size $n=500$ and use 2000 Monte Carlo repetitions. The potential outcomes are nonlinear in covariates and the error terms are not Normal. The true average causal effect equals $0.$
\begin{lstlisting}
res = replicate(2000, {
  n  = 500
  X  = matrix(rnorm(n*2), n, 2)
  Y1 = X[, 1] + X[, 1]^2 + runif(n, -0.5, 0.5)
  Y0 = X[, 2] + X[, 2]^2 + runif(n, -1, 1)
  Z  = rbinom(n, 1, 0.6)
  Y  = Z*Y1 + (1-Z)*Y0
  linestimator(Z, Y, X)
})
\end{lstlisting}

The following results confirm the theory. First, \citet{lin2013}'s estimator is nearly unbiased. 
Second, the average EHW standard error is smaller than the empirical standard derivation, resulting in undercoverage of the 95\% confidence intervals. 
Third, the average corrected standard error for super population is nearly identical to the empirical standard derivation, resulting in more accurate coverage of the 95\% confidence intervals.  
\begin{lstlisting}
> ## bias
> mean(res[1, ])
[1] -0.0001247585
> ## empirical standard deviation
> sd(res[1, ])
[1] 0.1507773
> ## estimated EHW standard error
> mean(res[2, ])
[1] 0.1388657
> ## coverage based on EHW standard error
> mean((res[1, ]-1.96*res[2, ])*(res[1, ]+1.96*res[2, ])<=0)
[1] 0.927
> ## estimated super population standard error
> mean(res[3, ])
[1] 0.1531519
> ## coverage based on population standard error
> mean((res[1, ]-1.96*res[3, ])*(res[1, ]+1.96*res[3, ])<=0)
[1] 0.9525
\end{lstlisting}

\section{Extension to the SRE}

We can extend the discussion in Section \ref{sec::cre-superpop} to the SRE since it is equivalent to independent CREs within strata. 
The notation below will be slightly different from that in Chapter \ref{chapter::stratification-poststratification}.

Assume that 
$$
\{ Z_i,  Y_i(1), Y_i(0) ,  X_i  \}  \iidsim \{ Z,  Y(1), Y(0), X  \} .
$$ 
With a discrete covariate $ X_i \in \{ 1,\ldots, K\}$, we can formulate the SRE as below.

\begin{definition}
[SRE under the super population framework]\label{def::sre-superpopulation}
We have 
$$
Z\ind\{  Y(1), Y(0) \} \mid   X  .
$$
\end{definition}

Under Definition \ref{def::sre-superpopulation},  the conditional average causal effect can be rewritten as
\begin{eqnarray*}
\tau_{[k]}  &=&  E\{   Y(1) -  Y(0)  \mid X= k \} \\
&=& E(Y\mid Z=1, X=k) - E(Y\mid Z=0, X=k),
\end{eqnarray*}
so the average causal effect can be rewritten as
\begin{eqnarray*}
\tau &=&   E\{   Y(1) -  Y(0)   \}  \\
&=& \sum_{k=1}^K \pr(X = k) E\{   Y(1) -  Y(0)  \mid X = k \} \\
&=& \sum_{k=1}^K \pr(X  = k) \tau_{[k]} .
\end{eqnarray*}
%
The discussion in Section \ref{sec::cre-superpop} holds within all strata, so we can derive the superpopulation analog for the SRE. When there are more than two treatment and control units within each stratum, we can use $\hat{V}_\textsc{S}$ as the variance estimator for $\var( \hat{\tau}_\textsc{S} )$.

%
%

We will see the exact form of Definition \ref{def::sre-superpopulation} in Part \ref{part::observational-studies} later.

 \section{Homework Problems}
 
\paragraph{OLS decomposition of the observed outcome under the CRE}\label{hw::ols-decompose-cre}

Based on \eqref{eq::ols-decomposition-y1} and \eqref{eq::ols-decomposition-y0}, show that the OLS decomposition of the observed outcome on the treatment, covariates, and their interaction is
$$
Y= \alpha_0 + \alpha_Z  Z +
\alpha_X \tran X + \alpha_{ZX} \tran X Z + \varepsilon 
$$
where 
$$
\alpha_0 = \gamma_0 ,\quad
\alpha_Z = \gamma_1 - \gamma_0,\quad
\alpha_X  =  \beta_0, \quad
\alpha_{ZX} = \beta_1 - \beta_0
$$
and
$$
\varepsilon  = Z \varepsilon (1) + (1-Z) \varepsilon (0). 
$$
That is,
$$
(\alpha_0, \alpha_Z, \alpha_X , \alpha_{ZX} )
=\arg\min_{a_0, a_Z, a_X, a_{ZX}   }  E  (  Y - a_0 - a_Z Z - a_X\tran X -a_{ZX} \tran X Z)^2.
$$

\paragraph{Variance estimation of \citet{lin2013}'s estimator under the super population framework}\label{hw::variance-bootstrap-superpopulation}
 
Under the super population framework, simulate $\{  X_i, Z_i, Y_i(1), Y_i(0) \}_{i=1}^n$ with $\beta_1 \neq \beta_0$ in \eqref{eq::ols-decomposition-y1} and \eqref{eq::ols-decomposition-y0}. Calculate \citet{lin2013}'s estimator in each simulated dataset. 
Compare the true variance and the following estimated variances:
\begin{enumerate}
\item
the EHW robust variance estimator;
\item
the EHW robust variance estimator with the correction term defined in \eqref{eq::lin-ehw-correction};
\item
the bootstrap variance estimator, with covariates centered at the beginning but not re-centered for each bootstrap sample;
\item
the bootstrap variance estimator, with covariates re-centered for each bootstrap sample.
\end{enumerate}


\paragraph{Recommended reading}

\citet{ding2017bridging} provide a unified discussion of the finite population and superpopulation inferences for the average causal effect.

\part{Observational studies}\label{part::observational-studies}
  \chapter{Observational Studies, Selection Bias, and Nonparametric Identification of Causal Effects}\label{chapter::observational-studies}
 \chaptermark{Observational Studies}

\citet{cochran1965planning} summarized  two common characteristics of observational studies:
\begin{enumerate}
\item
the objective is to elucidate cause-and-effect relationships;
\item 
it is not feasible to use controlled experimentation. 
\end{enumerate}

The first characteristic is identical to that of randomized experiments discussed in Part \ref{part::rcts} of this book, but the second differs fundamentally from randomized experiments.

\citet{dorn1953philosophy} suggested that the planner of an observational study should always ask the following question:
\begin{quote}
How would the study be conducted if it were possible to do
it by controlled experimentation?
\end{quote}
It is always helpful to follow \citet{dorn1953philosophy}'s suggestion because the potential outcomes framework has an intrinsic link to an experiment, either a real experiment or a thought experiment. Part \ref{part::observational-studies} of this book will discuss causal inference with observational studies. It will clarify the fundamental differences between observational studies and randomized experiments. Nevertheless, many ideas of causal inference with observational studies are deeply connected to those with randomized experiments.

\section{Motivating Examples}
 
 \begin{example}
 [job training program]\label{eg::job-training-obs}
 \citet{lalonde1986evaluating} was interested in the causal effect of a job training program on earnings. He compared the results based on a randomized experiment to the results based on observational studies. We have used the experimental data before, which is the \ri{lalonde} dataset in the \ri{Matching} package. We have also used an observational counterpart \ri{cps1re74.csv} in Chapter \ref{section:correlatioandregressionintro} and Problem \ref{hw::lalonde-obs-reg}.  \citet{lalonde1986evaluating} found that many traditional statistical or econometric methods for observational studies gave quite different estimates compared with the estimates based on the experimental data. \cite{dehejia1999causal} re-analyzed the data using methods motivated by causal inference, and found that those methods can recover the experimental gold standard well. Since then, this has become a canonical example in causal inference with observational studies. 
 \end{example}

  \begin{example}
 [smoking and homocysteine]\label{eg::nhanes2005-2006}
 \citet{bazzano2003relationship} compared the homocysteine\footnote{Homocysteine is a type of amino acid. A high level of homocysteine in the blood is regarded as a marker of cardiovascular disease.} levels in daily smokers and non-smokers based on the data from the National Health and Nutrition Examination Survey (NHANES) 2005--2006. \citet{rosenbaum2018sensitivity} documented the data as \ri{homocyst} in the   package \ri{senstrat}. The dataset has the following important covariates:
 
\begin{tabular}{ll}
\hline 
\texttt{female} &
1=female, 0=male
\\
\texttt{age3} &
three age categories: 20--39, 40--50, $\geq $60
\\
\texttt{ed3} &
three education categories: $<$ high school, high school, some college
\\
\texttt{bmi3}&
three BMI categories: $<$30, $[30,35)$, $\geq $ 35
\\
\texttt{pov2}&
TRUE=income at least twice the poverty level, FALSE otherwise\\
\hline 
\end{tabular} 

 \end{example}

   \begin{example} 
 [school meal program and body mass index]\label{eg::chan-bmi-ATE}
 \citet{chan2016globally} used a subsample of the data from NHANES 2007--2008 to study whether participation in school meal programs led to an increase in BMI for school children. They documented the data as \ri{nhanes_bmi} in the package \ri{ATE}. The dataset has the following important covariates:
 
\begin{tabular}{ll}
\hline 
\texttt{age}&
Age  
\\
\texttt{ChildSex}&
Sex (1: male, 0: female)
\\
\texttt{black}&
Race (1: black, 0: otherwise) 
\\
\texttt{mexam}&
Race (1: Hispanic: 0 otherwise)
\\
\texttt{pir200\_plus}&
Family above 200\% of the federal poverty level  \\
\texttt{WIC}&
Participation in the special supplemental nutrition program \\
\texttt{Food\_Stamp}&
Participation in food stamp program   \\
\texttt{fsdchbi}&
Childhood food security  \\
\texttt{AnyIns}&
Any insurance \\
\texttt{RefSex}&
Sex of the adult respondent (1: male, 0: female) \\
\texttt{RefAge}&
Age of the adult respondent\\
\hline 
\end{tabular} 

 \end{example}
 
A common feature of Examples \ref{eg::job-training-obs}--\ref{eg::chan-bmi-ATE} is that we are interested in estimating the causal effect of a nonrandomized treatment on an outcome. They are all from observational studies.

\section{Causal effects and selection bias under the potential outcomes framework} 
 \label{sec::selection-bias-randomization}

For unit $i$ $(i=1,\ldots, n)$, we have  pretreatment covariates $X_i$, a binary treatment indicator $Z_i$, and an observed outcome $Y_i$ with two potential outcomes  $Y_i(1)$ and $Y_i(0)$ under the treatment and control, respectively. For simplicity, we assume   
$$
\{X_i, Z_i, Y_i(1),  Y_i(0)  \}_{i=1}^n \iidsim \{  X, Z, Y(1),  Y(0) \} .
$$ 
So we can drop the subscript $i$ for quantities of this population.  
The causal effects of interest are the average causal effect
$$
\tau = E\{  Y(1) - Y(0)  \},
$$
the average causal effect on the treated units
$$
\tau_\textsc{T} =  E\{  Y(1) - Y(0)  \mid Z = 1 \},
$$
and the average causal effect on the control units:
$$
\tau_\textsc{C} =  E\{  Y(1) - Y(0)  \mid Z = 0 \}. 
$$ 

By the linearity of the expectation, we have
\begin{eqnarray*}
\tau_\textsc{T} &=&  E\{  Y(1)    \mid Z = 1 \} - E\{   Y(0)  \mid Z = 1 \}  \\
&=&  E(  Y    \mid Z = 1 ) - E\{   Y(0)  \mid Z = 1 \}
\end{eqnarray*}
 and
\begin{eqnarray*}
 \tau_\textsc{C} &=&  E\{  Y(1)    \mid Z = 0 \} - E\{   Y(0)  \mid Z = 0 \} \\
 &=&  E\{  Y(1)    \mid Z = 0 \} - E(  Y   \mid Z = 0 ).
\end{eqnarray*}
 In the above two formulas of $\tau_\textsc{T} $ and $\tau_\textsc{C} $, the quantities $E(Y    \mid Z = 1 ) $ and $E(   Y   \mid Z = 0 ) $ are directly estimable from the data, but the quantities $E\{   Y(0)  \mid Z = 1 \}$ and $E\{  Y(1)    \mid Z = 0 \}$ are not. The latter two are {\it counterfactuals} because they are the means of the potential outcomes corresponding to the treatment level that is the opposite of the actual received treatment.

The simple difference in means, also known as the {\it prima facie}\footnote{It is a Latin phrase, which means ``based on the first impression'' or ``accepted as correct until proved otherwise.'' 
\citet{holland1986statistics} used this phrase. 
} causal effect,
\begin{eqnarray*}
\tau_\textsc{PF} &=& E(  Y    \mid Z = 1 )  - E(   Y   \mid Z = 0 )   \\
&=& E\{  Y(1)    \mid Z = 1 \}  - E\{   Y(0)  \mid Z = 0 \} 
\end{eqnarray*}
 is generally biased for the causal effects defined above. For example,
 $$
 \tau_\textsc{PF} - \tau_\textsc{T} = E\{   Y(0)  \mid Z = 1 \} - E\{   Y(0)  \mid Z = 0 \} 
 $$
 and
 $$
  \tau_\textsc{PF} - \tau_\textsc{C} =  E\{   Y(1)  \mid Z = 1 \} - E\{   Y(1)  \mid Z = 0 \} 
 $$
 are not zero in general, and they quantify the {\it selection bias}. They measure the differences in the means of the potential outcomes across the treatment and control groups.

 Why is randomization so important? \citet{rubin1978bayesian} first used potential outcomes to quantify the benefit of randomization. We have used the fact in Chapter \ref{chapter::bridging} that 
\begin{equation}\label{eq::rubinrandomization}
 Z  \ind \{ Y (1) , Y (0) \}
\end{equation}
in the CRE,  which implies that the selection bias terms are both zero:
 $$
  \tau_\textsc{PF} - \tau_\textsc{T}  = 
 E\{   Y(0)  \mid Z = 1 \} - E\{   Y(0)  \mid Z = 0 \}  = 0
 $$
 and
 $$
   \tau_\textsc{PF} - \tau_\textsc{C} =
 E\{   Y(1)  \mid Z = 1 \} - E\{   Y(1)  \mid Z = 0 \}  = 0. 
 $$
 So under the CRE in \eqref{eq::rubinrandomization},  we have 
\begin{equation}
\label{eq::no-selection-bias-all-equal}
 \tau = \tau_\textsc{T} = \tau_\textsc{C} = \tau_\textsc{PF}.
\end{equation}
From the above discussion, the fundamental benefit of randomization is to balance the distributions of the potential outcomes across the treatment and control groups. This is a much stronger guarantee than balancing the distributions of the observed covariates.

Without randomization, the selection bias terms can be arbitrarily large, especially for unbounded outcomes. This highlights the fundamental difficulty of causal inference with observational studies.

\section{Sufficient conditions for nonparametric identification of causal effects}

\subsection{Identification}\label{sec::identification-ignorability}

 Causal inference with observational studies is challenging. It relies on strong assumptions. A strategy is to use the information from the pretreatment covariates and assume that conditioning on the observed covariates $X$, the selection bias terms are zero, that is,
\begin{eqnarray}
\label{eq::meanind0} E\{   Y(0)  \mid Z = 1, X \} &=& E\{   Y(0)  \mid Z = 0, X \}  ,\\
\label{eq::meanind1} E\{   Y(1)  \mid Z = 1 , X\} &=& E\{   Y(1)  \mid Z = 0, X \}   . 
\end{eqnarray}
The assumptions in \eqref{eq::meanind0} and \eqref{eq::meanind1} state that the differences in the means of the potential outcomes across the treatment and control groups are entirely due to the difference in the observed covariates. So given the same value of the covariates, the potential outcomes have the same means across the treatment and control groups. Mathematically, \eqref{eq::meanind0} and \eqref{eq::meanind1} ensure that the conditional versions of the effects in \eqref{eq::no-selection-bias-all-equal} are identical: 
 $$
 \tau(X) = \tau_\textsc{T}(X) =  \tau_\textsc{C}(X) = \tau_\textsc{PF}(X),
 $$
 where
\begin{eqnarray*}
 \tau(X)  &=& E\{ Y(1) - Y(0) \mid X  \}, \\
 \tau_\textsc{T}(X) &=&  E\{ Y(1) - Y(0) \mid Z=1,  X  \},\\
  \tau_\textsc{C}(X) &=& E\{ Y(1) - Y(0) \mid Z=0,  X  \},\\
   \tau_\textsc{PF}(X) &=& E(Y\mid Z=1, X) - E(Y\mid Z=0, X).
\end{eqnarray*}
In particular, $ \tau(X) $ is often called the {\it conditional average causal effect} (CATE).

A key result in this chapter is that the average causal effect $\tau$ is {\it nonparametrically identifiable} under \eqref{eq::meanind0} and \eqref{eq::meanind1}. The notion of nonparametrically identifiability does not appear frequently in classic statistics, but it is key to causal inference with observational studies. I first give an abstract definition below. 

\begin{definition}[identification]
\label{def::nonparametric-identification}
A parameter $\theta$ is identifiable if it can be written as a function of the distribution of the observed data under certain model assumptions. 
A parameter $\theta$ is nonparametrically identifiable if it can be written as a function of the distribution of the observed data without any parametric model assumptions. 
\end{definition}

Definition \ref{def::nonparametric-identification} is too abstract at the moment. I will use more concrete examples in later chapters to illustrate its meaning. It is often neglected in classic statistics problems. For instance, the mean $\theta = E(Y)$ is nonparametrically identifiable if we have IID draws of $Y_i$'s; the Pearson correlation coefficient $\theta = \rho_{YX}$ is nonparametrically identifiable if we have IID draws of the pairs $(X_i, Y_i)$'s. In those examples, the parameters are nonparametrically identifiable automatically. However, Definition \ref{def::nonparametric-identification} is fundamental in causal inference with observational studies. In particular, the parameter of interest $\tau = E\{ Y(1)  - Y(0)\}$ depends on some unobserved random variables, so it is unclear whether it is nonparametrically identifiable based on observed data. Under the assumptions in \eqref{eq::meanind0} and \eqref{eq::meanind1}, it is nonparametrically identifiable, with details below.

 Because $   \tau_\textsc{PF}(X)$ depends only on the observables, it is nonparametrically identified by definition. Moreover, \eqref{eq::meanind0} and \eqref{eq::meanind1} ensure that the three causal effects are the same as $   \tau_\textsc{PF}(X)$, so $\tau(X) $, $\tau_\textsc{T}(X) $ and $  \tau_\textsc{C}(X)$ are all nonparametrically identified. Consequently, the unconditional versions are also nonparametrically identified under \eqref{eq::meanind0} and \eqref{eq::meanind1} due to the law of total expectation:
 $$
 \tau = E\{ \tau(X)\},\quad
 \tau_\textsc{T} = E\{  \tau_\textsc{T}(X) \mid Z=1  \}, \quad
   \tau_\textsc{C} = E\{   \tau_\textsc{C}(X) \mid Z=0 \}.
 $$
 From now on, I will focus on $\tau$ unless stated otherwise (Chapter \ref{chapter::ATT-and-other} will be an exception). The following theorem summarizes the identification formulas of $\tau$. 
 
\begin{theorem}
\label{thm::identification-formula}
Under \eqref{eq::meanind0} and \eqref{eq::meanind1}, the average causal effect $\tau$ is identified by 
 \begin{eqnarray} 
 \tau &=& E\{ \tau(X)\} \\
 &=& E\{  E(Y\mid Z=1, X) - E(Y\mid Z=0, X) \} \label{eq::nonparametric-identification-tau} \\
 &=&  \int \{  E(Y\mid Z=1, X=x) - E(Y\mid Z=0, X=x)\} f(x) \diff x.
 \end{eqnarray}
\end{theorem}

 The formula \eqref{eq::nonparametric-identification-tau} was formally established by \citet{rosenbaum1983central}, which is also called the g-formula by Robins \citep[see][]{hernan2020causal}.

 With a discrete covariate, we can write the identification formula in Theorem \ref{thm::identification-formula} as 
\begin{eqnarray}
 \tau  &=&  \sum_x  E(Y\mid Z=1, X=x)  \pr(X=x)  \nonumber \\
 &&  - \sum_x E(Y\mid Z=0, X=x) \pr(X=x), \label{eq::nonparametric-identification-tau-discrete}
\end{eqnarray}
 and also the simple difference in means as
  \begin{eqnarray}  
 \tau_\textsc{PF} &=& \sum_x   E(Y\mid Z=1, X=x) \pr(X=x\mid Z=1) \nonumber  \\
 &&- \sum_x   E(Y\mid Z=0, X=x) \pr(X=x\mid Z=0)  \label{eq::difference-in-means-discrete-x}
  \end{eqnarray} 
 by the law of total probability. Comparing \eqref{eq::nonparametric-identification-tau-discrete} and \eqref{eq::difference-in-means-discrete-x}, we can see that although both formulas compare the conditional expectations $ E(Y\mid Z=1, X=x) $ and $ E(Y\mid Z=0, X=x)$, they average over different distributions of the covariates. The causal parameter $\tau$ averages the conditional expectations over the common distribution of the covariates, but the difference in means $ \tau_\textsc{PF} $ averages the conditional expectations over two different distributions of the covariates in the treated and control groups.

 Usually, we impose a stronger assumption.
 
 \begin{assumption}[ignorability]
 \label{assume:weak-ignorability} 
 We have 
 \begin{eqnarray}
 Y(z) \ind Z \mid X \quad (z=0,1) .
 \label{eq::ignorability}
\end{eqnarray}
 \end{assumption}

Assumption \ref{assume:weak-ignorability} has many names:
\begin{enumerate}
\item
{\it ignorability}  due to \citet{rubin1978bayesian}\footnote{The reason for using this name is unclear based on the discussion in this book. In fact, it is from the Bayesian perspective of causal inference. It is beyond the scope of this book.};

\item
{\it unconfoundedness} which is popular among epidemiologists;

\item
{\it selection on observables} which is popular among social scientists;

\item
{\it conditional independence} which is merely a description of the notation $\ind$ in the assumption.
\end{enumerate}

Sometimes, we impose an even stronger assumption.

\begin{assumption}[strong ignorability]
\label{assume:strong-ignorability}
We have 
\begin{eqnarray}
 \{  Y(1), Y(0)  \} \ind Z \mid X .
  \label{eq::strong-ignorability}
\end{eqnarray}
\end{assumption}

Assumption \ref{assume:strong-ignorability}  is called {\it strong ignorability} \citep{rubin1978bayesian, rosenbaum1983central}. If the parameter of interest is $\tau$, then the stronger Assumptions \ref{assume:weak-ignorability} and \ref{assume:strong-ignorability} are just imposed for notational simplicity thanks to the conditional independence notation $\ind$. They are not necessary for identifying $\tau$. However,  they can not be relaxed if the parameter of interest is the causal effects on other scales (for example, distribution, quantile, or some transformation of the outcome). 
The  {\it strong ignorability} assumption requires that the potential outcomes vector be independent of the treatment given the covariates, but the {\it ignorability} assumption only requires each potential outcome be independent of the treatment given covariates. The former is stronger than the latter. However, their difference is rather technical and of pure theoretical interests; see Problem \ref{hw::ignorability-difference}. In most reasonable statistical models, they both hold; see Section \ref{sec::plausibility-ignorability} below. We will not distinguish them in this book and will simply use {\it ignorability} to refer to both.

 \subsection{Plausibility of the ignorability assumption} \label{sec::plausibility-ignorability}

A fundamental problem of causal inference with observational studies is the plausibility of the ignorability assumption. The discussion in Chapter \ref{sec::identification-ignorability} may seem too mathematical in the sense that the ignorability assumption serves as a sufficient condition to ensure the nonparametric identification of the average causal effect. What is its scientific meaning? Intuitively, it rules out all unmeasured covariates that affect the treatment and outcome simultaneously. Those ``common causes'' of the treatment and outcome are called confounders. That is why the ignorability assumption is also called the unconfoundedness assumption. More transparently, we can interpret the ignorability assumption based on the outcome data-generating process. If
\begin{eqnarray*}
Y(1) &=& g_1(X, V_1),\\
Y(0) &=& g_0(X, V_0), \\
Z &=& 1\{ g(X,V)  \geq 0 \}
\end{eqnarray*} 
with $(V_1, V_0)\ind V$, then both Assumptions \ref{assume:weak-ignorability} and \ref{assume:strong-ignorability} hold. In the above data-generating process, the ``common causes'' $X$ of the treatment and the outcome are all observed, and the remaining random components are independent. If the data-generating process changes to
\begin{eqnarray*}
Y(1) &=& g_1(X, U, V_1),\\
Y(0) &=& g_0(X, U, V_0), \\
Z &=& 1\{ g(X, U, V)  \geq 0 \}
\end{eqnarray*} 
with $(V_1, V_0)\ind V$, then Assumptions \ref{assume:weak-ignorability} and \ref{assume:strong-ignorability} do not hold in general. The unmeasured ``common cause'' $U$ induces dependence between the treatment and the potential outcomes even conditional on the observed covariates $X$. If we do not have access to $U$ and analyze the data based only on $(Z,X,Y)$, the final estimator will be biased for the causal parameter in general. This type of bias is called the {\it omitted variable bias} in econometrics; see Problem \ref{hw::cochran+ovb} in a later chapter.

The ignorability assumption can be reasonable if we observe a rich set of covariates $X$ that affect the treatment and the outcome simultaneously. I start with this assumption to discuss identification and estimation strategies in Part \ref{part::observational-studies} of this book. However, it is fundamentally untestable. We may justify it based on the scientific background knowledge, but we are often not sure whether it holds or not. Parts \ref{part::challenges-os} and \ref{part::instrumentalvariables} of this book will discuss other strategies when the ignorability assumption is not plausible.

\section{Two simple estimation strategies and their limitations}

\subsection{Stratification or standardization based on discrete covariates}

If the covariate $X_i \in \{1, \ldots, K \}$ is discrete, then ignorability \eqref{eq::ignorability} reads as 
 $$
 Y(z) \ind Z \mid X = k \quad (z=0,1; k=1,\ldots, K),
 $$
which essentially assumes that the observational study is an SRE under the superpopulation framework in Chapter \ref{chapter::bridging}. Therefore, we can use the estimator
$$
\hat{\tau} = \sum_{k=1}^K \pi_{[k]} \left\{   \hat{\bar{Y}}_{[k]}(1) - \hat{\bar{Y}}_{[k]}(0)   \right\},
$$
which is identical to the stratified or post-stratified estimator discussed in Chapter \ref{chapter::stratification-poststratification}.

This method is still widely used in practice. Example \ref{eg::nhanes2005-2006} contains discrete covariates, and I relegate the analysis to
Problem \ref{hw::stratification-example-rosenbaum}. However, there are several obvious difficulties in implementing this method. First, it works well for the case with small $K$. For large $K$, it is very likely that many strata have $n_{[k]1}  = 0$ or $n_{[k]0} = 0$, leading to the poorly defined $\hat{\tau}_{[k]}$'s for those strata. This is related to the issue of overlap which will be discussed in Chapter \ref{chapter::overlap} later. Second, it is not obvious how to apply this stratification method to multidimensional continuous or mixed covariates $X$. A standard method is to create strata based on the initial covariates and then apply the stratification method. This may result in arbitrariness in the analysis.

\subsection{Outcome regression}\label{sec::outcome-regression}

The most commonly used method based on outcome regression is to run the OLS with an additive model of the observed outcome on the treatment indicator and covariates, which assumes
$$
E(Y\mid Z, X) = \beta_0 + \beta_z Z  + \beta_x \tran X .
$$
If the above linear model is correct, then we have
\begin{eqnarray*}
\tau(X) &=&  E(Y\mid Z=1, X) - E(Y\mid Z=0, X) \\
&=& (\beta_0 + \beta_z   + \beta_x\tran  X ) - ( \beta_0    + \beta_x\tran  X )  \\
&=& \beta_z,
\end{eqnarray*}
which implies that the causal effect is homogeneous with respect to the covariates. This, coupled with ignorability, implies that
$$
\tau = E\{ \tau(X)  \} = \beta_z.
$$
Therefore, if ignorability holds and the outcome model is linear, then the average causal effect equals the coefficient of $Z$. This is one of the most important applications of the linear model. However, the causal interpretation of the coefficient of $Z$ is valid only under two strong assumptions: ignorability and the linear model.

We have discussed in Chapter \ref{chapter:rerandomization-regression}, the above procedure is suboptimal even in CREs because it ignores the treatment effect heterogeneity induced by the covariates. 
If we assume
$$
E(Y\mid Z, X) = \beta_0 + \beta_z Z  + \beta_x\tran  X  + \beta_{zx}\tran  X Z,
$$
we have
\begin{eqnarray*}
\tau(X) &=& 
E(Y\mid Z=1, X) - E(Y\mid Z=0, X)  \\
&=& (\beta_0 + \beta_z   + \beta_x\tran  X +\beta_{zx}\tran  X ) - ( \beta_0    + \beta_x\tran  X ) \\
&=& \beta_z+  \beta_{zx}\tran  X,
\end{eqnarray*}
which, coupled with ignorability, implies that
$$
\tau = E\{ \tau(X)  \} =E(\beta_z+  \beta_{zx}\tran  X) = \beta_z + \beta_{zx}\tran E(X). 
$$
The estimator for $\tau$ is then $ \hat{\beta}_z + \hat{\beta}_{zx}\tran  \bar{X}$, where $\hat{\beta}_z$ is the regression coefficient of $Z$ and $\bar{X}$ is the sample mean of $X$. If we center the covariates to ensure $\bar{X}=0$, then the estimator is simply the regression coefficient of $Z$. To simplify the procedure, we usually center the covariates at the beginning; also recall \citet{lin2013}'s estimator introduced in Chapter \ref{chapter:rerandomization-regression}.  \citet{rosenbaum1983central} and \citet{hirano2001estimation} discussed this estimator.

In general, we can use other more complex models to estimate the causal effects. For example, if we build two predictors $\hat{\mu}_1(X)$ and $\hat{\mu}_0(X)$ based on the treated and control data, respectively, then we have an estimator for the conditional average causal effect 
$$
\hat\tau(X) = \hat{\mu}_1(X) - \hat{\mu}_0(X)
$$
and an estimator for the average causal effect: 
$$
\hat{\tau}^{\text{reg}} = n^{-1} \sumn \{ \hat{\mu}_1(X_i) - \hat{\mu}_0(X_i) \}  . 
$$
The estimator $\hat{\tau} $ above has the same form as the projective estimator discussed in Chapter \ref{chapter:rerandomization-regression}. It is sometimes called the {\it outcome regression} estimator. The OLS-based estimators above are special cases. We can use the nonparametric bootstrap (see Chapter \ref{sec::bootstrap}) to estimate the standard error of the outcome regression estimator.

I give another example for binary outcomes below. 

 \begin{example}
 [outcome regression estimator for binary outcomes]\label{eg::reg-binary}
 With a binary outcome, we may model $Y$ using a logistic model
$$
E(Y\mid Z, X) = \pr(Y =1\mid Z, X) = \frac{   e^{\beta_0 + \beta_z Z + \beta_x \tran X}  }
{ 1+ e^{\beta_0 + \beta_z Z + \beta_x \tran X} } ,
$$
then based on the estimators of the coefficients $ \hat \beta_0, \hat \beta_z, \hat{\beta}_x $, we have the following estimator for the average causal effect: 
$$
\hat{\tau} = n^{-1} \sumn \left\{  
\frac{   e^{\hat\beta_0 +\hat \beta_z + \hat\beta_x \tran X_i}  }
{ 1+ e^{\hat\beta_0 +\hat \beta_z  + \hat\beta_x \tran X_i} }  
- \frac{   e^{\hat\beta_0 + \hat\beta_x \tran X_i}  }
{ 1+ e^{\hat\beta_0   + \hat\beta_x \tran X_i} } 
\right \}  . 
$$
This estimator is not simply the coefficient of the treatment in the logistic  model.\footnote{If the logistic outcome model is correct, then $\hat\beta_z$ estimates the conditional odds ratio of the treatment on the outcome given covariates, which does not equal $\tau$. 
\citet{freedman2008randomization} gave a warning of using the logistic regression coefficient to estimate $\tau$ in CREs.
See Chapter \ref{appendix::basic-linear-regression} for more details of the logistic regression.} It is a nonlinear function of all the coefficients as well as the empirical distribution of the covariates. In econometrics, this estimator is called the  ``average partial effect'' or ``average marginal effect'' of the treatment in the logistic model. Many econometric software packages can report this estimator associated with the standard error.  Similarly, we can also derive the corresponding estimator based on a fully interacted logistic model; see Problem \ref{hw::outcome-two-logit}. 
 \end{example}

The above predictors for the conditional means of the outcome can also be other machine learning tools. In particular, 
\citet{hill2011bayesian} championed the use of tree methods for estimating $\tau$, and \citet{wager2018estimation} proposed to use them also for estimating $\tau(X) $. \citet{wager2018estimation}  also combined the tree methods with the idea of the {\it propensity score} in the next chapter. Since then, the intersection of machine learning and causal inference has been an active research area \citep[e.g.,][]{hahn2020bayesian, kunzel2019metalearners}.

The biggest problem with the above approach based on outcome regressions is its sensitivity to the specification of the outcome model. Problem \ref{hw::lalonde-obs-reg} gave such an example. Depending on the incentive of empirical research and publications, people might report their favorable causal effects estimates after searching over a wide set of candidate models, without confessing this model searching process. This is a major source of $p$-hacking in causal inference. Problem \ref{hw::lalonde-obs-reg} hinted at this issue. \citet{leamer1978specification} criticized this approach in empirical research.

\section{Homework Problems}

\paragraph{A simple identity}\label{hw::simple-identity-taus}

Show that $\tau = \pr(Z=1) \tau_\textsc{T} + \pr(Z=0)\tau_\textsc{C}$.

\paragraph{Nonparametric identification of other causal effects}\label{hw::nonparametric-identification-dce-qce}

Under ignorability, show that
\begin{enumerate}
\item
 the distributional causal effect
 $$
\textsc{DCE}_y = \pr\{  Y(1) > y  \} - \pr\{ Y(0) > y \} 
$$
is nonparametrically identifiable for all $y$;
\item
the quantile causal effect 
$$
\textsc{QCE}_q = \text{quantile}_q\{ Y(1) \} - \text{quantile}_q\{ Y(0) \},
$$
is nonparametrically identifiable for all $q$, where $\text{quantile}_q\{ \cdot \} $ is the $q$th quantile of a random variable. 
\end{enumerate}

Remark:
In probability theory, $\pr\{  Y(z) \leq y \}$ is the cumulative distribution function and $\pr\{  Y(z) > y \}$ is the survival function of the potential outcome $Y(z)$.
The distributional causal effect compares the survival functions of the potential outcomes under treatment and control. 
 
The  quantile causal effect  compares the marginal quantiles of the potential outcomes, which is different from the quantile of the individual causal effect
$$
\tau_q =  \text{quantile}_q\{ Y(1)  - Y(0)\} .
$$
In fact,  $\tau_q$ is not identifiable in the sense that the marginal distributions $\pr\{  Y(1) \leq  y  \} $ and $ \pr\{ Y(0) \leq  y \} $ can not uniquely determine $\tau_q$.

\paragraph{Outcome imputation estimator in the fully interacted logistic model}\label{hw::outcome-two-logit}

This problem extends Example \ref{eg::reg-binary}. 

Assume that a binary outcome follows a logistic model
$$
E(Y\mid Z, X) = \pr(Y =1\mid Z, X) = \frac{   e^{\beta_0 + \beta_z Z + \beta_x \tran X + \beta_{xz}  \tran X Z }  }
{ 1+ e^{\beta_0 + \beta_z Z + \beta_x \tran X + \beta_{xz}  \tran X Z} } .
$$
What is the corresponding outcome regression estimator for the average causal effect?

\paragraph{Data analysis: stratification and regression}\label{hw::stratification-example-rosenbaum}

Use the dataset \ri{homocyst} in the package \ri{senstrat}. 
The outcome is \ri{homocysteine}, the homocysteine level, and the treatment is \ri{z}, where $z=1$ for a daily smoker and $z=0$ for a never smoker. Covariates are \ri{female, age3, ed3, bmi3, pov2} with detailed explanations in the package, and \ri{st} is a stratum indicator, defined by all the combinations of the discrete covariates.

\begin{enumerate}
[(1)]
\item
How many strata have only treated or control units? What is the proportion of the units in these strata? Drop these strata and perform a stratified analysis of the observational study. Report the point estimator, variance estimator, and 95\% confidence interval for the average causal effect.

\item
Run the OLS of the outcome on the treatment indicator and covariates without interactions. Report the coefficient of the treatment and the robust standard error.

Drop the strata with only treated or control units. Re-run the OLS and report the result.

\item
Apply \citet{lin2013}'s estimator of the average causal effect. Report the coefficient of the treatment and the robust standard error.

If you do not drop the strata with only treated or control units, what will happen?

\item
Compare the results in the above three analyses. Which one is more credible?

\end{enumerate}

\paragraph{Ignorability versus strong ignorability}\label{hw::ignorability-difference}

Give an example such that the ignorability holds but the strong ignorability does not hold.

Remark: This is related to a classic probability problem of finding three random variables $A,B,C$ such that 
$$
A\ind C \text{ and } B\ind C \text{ but } (A,B) \nind C.
$$

\paragraph{Recommended reading}

\citet{cochran1965planning} is a classic reference on observational studies. It contains many useful insights but does not use the formal potential outcomes framework.  
Dylan Small founded a new journal called {\it Observational Studies} in 2015 \citep{small2015introduction} and reprinted an old article ``Observational Studies'' by W. G. Cochran as  \citet{cochran2015}.  Many leading researchers also made insightful comments on  \citet{cochran2015}.

\chapter{The Central Role of the Propensity Score in Observational Studies for Causal Effects}\label{chapter::pscore-key}
\chaptermark{Propensity Score}

\citet{rosenbaum1983central}   proposed the key concept {\it propensity score} and discussed its role in causal inference with observational studies. It is one of the most cited papers in statistics, and \citet{titterington2013biometrika} listed it as the second most cited paper published in {\it Biometrika} during the past 100 years. Its number of citations has grown very fast in recent years.

Under the IID sampling assumption, we have four random variables associated with each unit: $\{  X, Z, Y(1), Y(0) \} $. Following the basic probability rules, we can factorize the joint distribution as 
\begin{eqnarray*}
&&\pr \{  X, Z, Y(1), Y(0) \}  \\
 &=& \pr(X)  
\times   \pr\{ Y(1), Y(0) \mid X \}   
 \times  \pr\{  Z\mid X, Y(1), Y(0)  \} ,
\end{eqnarray*}
where
\begin{enumerate}
\item
 $\pr(X)$ is the covariate distribution, 
 \item
 $\pr\{ Y(1), Y(0) \mid X \}$ is the outcome distribution conditional on the covariates $X$,
 \item
 $\pr\{  Z\mid X, Y(1), Y(0)  \}$ is the treatment distribution conditional on the covariates $X$, also known as the treatment assignment mechanism. 
\end{enumerate}
Usually, we do not want to model the covariates because they are background information happening before the treatment and outcome. If we want to go beyond the outcome model, then we must focus on the treatment assignment mechanism, which leads to the definition of the propensity score.

\begin{definition}
[propensity score]\label{def::pscore}
Define
$$
e(X, Y(1), Y(0)   ) = \pr\{  Z=1\mid X, Y(1), Y(0)  \}
$$
as the {\it propensity score}. Under strong ignorability, we have
$$
 e(X, Y(1), Y(0)   )  = \pr\{  Z=1\mid X, Y(1), Y(0)  \} = \pr(Z=1 \mid X),
$$
so the propensity score reduces to 
$$
e(X) = \pr(Z=1 \mid X),
$$
the conditional probability of receiving the treatment given the observed covariates.  
\end{definition}

\citet{rosenbaum1983central} used  $e(X) = \pr(Z=1 \mid X)$ as the definition of the propensity score because they focused on observational studies under strong ignorability. It is sometimes helpful to view $e(X, Y(1), Y(0)   ) = \pr\{  Z=1\mid X, Y(1), Y(0)  \}$ as the general definition of the propensity score even when strong ignorability fails.  See Problem \ref{hw::principal-unobserved-cov} for more details. 

Following \citet{rosenbaum1983central}, this chapter will demonstrate that $e(X)$ is a key quantity in causal inference with observational studies under ignorability.

\section{The propensity score as a dimension reduction tool}

\subsection{Theory}

\begin{theorem}
\label{thm::pscore-dimreduction}
If $Z\ind \{ Y(1), Y(0)  \} \mid X$, then
$
Z\ind \{ Y(1), Y(0)  \} \mid e(X).
$
\end{theorem}

Theorem \ref{thm::pscore-dimreduction} states that if strong ignorability holds conditional on covariates $X$, then it also holds conditional on the scalar propensity score $e(X)$. 
The ignorability requires conditioning on many background characteristics $X$ of the units, but Theorem \ref{thm::pscore-dimreduction} implies that conditioning on the propensity score $e(X)$ removes all confounding induced by covariates $X.$ The original covariates $X$ can be general and have many dimensions, but the propensity score $e(X)$ is a one-dimensional scalar variable bounded between $0$ and $1$. Therefore, the propensity score reduces the dimension of the original covariates but still maintains the ignorability. As a statistical terminology, we can view the propensity score as a {\it dimensional reduction} tool. We will first prove Theorem \ref{thm::pscore-dimreduction}  below and then give an application of the dimension reduction property of the propensity score.

\begin{myproof}{Theorem}{\ref{thm::pscore-dimreduction}}
By the definition of conditional independence, we need to show that 
\begin{eqnarray}
\pr\{ Z=1\mid Y(1), Y(0), e(X)  \} = \pr\{ Z=1\mid e(X)  \} .
\label{eq::pscore-dimreduction}
\end{eqnarray}
The left-hand side of \eqref{eq::pscore-dimreduction} equals
\begin{eqnarray*}
&&\pr\{ Z=1\mid Y(1), Y(0), e(X)  \}  \\
&=& E\{ Z \mid Y(1), Y(0), e(X)  \}  \\
&=& E\Big[   E\{ Z \mid Y(1), Y(0), e(X) , X  \}  \mid   Y(1), Y(0), e(X)    \Big]  \\
&&\quad  \quad  \quad  \quad   \quad \quad \text{(tower property; see Section \ref{appendix::tower-property})} \\
&=& E\Big[   E\{ Z \mid Y(1), Y(0),   X  \}  \mid   Y(1), Y(0), e(X)    \Big]  \\
&=& E\Big\{    E(Z \mid  X  )  \mid   Y(1), Y(0), e(X)    \Big\}  \quad \text{(strong ignorability)} \\
&=& E\Big\{   e(X) \mid   Y(1), Y(0), e(X)    \Big\} \\
&=& e(X).
\end{eqnarray*}
The right-hand side of \eqref{eq::pscore-dimreduction} equals 
\begin{eqnarray*}
&& \pr\{ Z=1\mid e(X)  \} \\
  &=&  E\{ Z \mid e(X)  \} \\
 &=& E\Big[   E\{ Z \mid e(X), X  \}  \mid e(X)  \Big ] \quad \text{(tower property)} \\
 &=& E\Big\{    E( Z \mid  X )  \mid e(X)  \Big \}   \\
 &=& E\Big\{   e(X)  \mid e(X)  \Big \} \\
 &=& e(X). 
\end{eqnarray*}
So the left-hand side of \eqref{eq::pscore-dimreduction} equals the right-hand side of \eqref{eq::pscore-dimreduction}. 
\end{myproof}

\subsection{Propensity score stratification}
\label{sec::pscore-stratification}

Theorem \ref{thm::pscore-dimreduction} motivates a simple method for estimating causal effects: propensity score stratification. 
Starting from the simple case, we assume that the propensity score is known and only takes $K$ possible values $\{ e_1,\ldots, e_K \}$ with $K$ being much smaller than the sample size $n$. Theorem \ref{thm::pscore-dimreduction} reduces to
$$
Z\ind \{ Y(1), Y(0)  \} \mid e(X) = e_k  \quad (k=1,\ldots, K).
$$
Therefore, we have an SRE, that is, we have $K$ independent CREs within strata of the propensity score. We can analyze the observational data in the same way as the SRE stratified on $e(X)$.

In general, the propensity score is not known and is not discrete. We often fit a statistical model for $\pr(Z = 1\mid X)$ (for example, a logistic model of the binary $Z$ given $X$) to obtain the estimated propensity score $\hat{e}(X)$. This estimated propensity score can take as many values as the sample size, but we can discretize it to approximate the simple case above. For example, we can discretize the estimated propensity score by its $K$ quantiles to obtain $\hat{e}'(X)$:
$
\hat{e}'(X_i) = e_k,
$
the $k/K$-th quantile of $\hat{e}(X)$,
if $ \hat{e}(X_i)$ is between the $(k-1)/K$-th and $k/K$-th quantiles of the $\hat{e}(X_i)$'s. Then  
$$
Z\ind \{ Y(1), Y(0)  \} \mid \hat{e}'(X) = e_k  \quad (k=1,\ldots, K).
$$
holds approximately. So we can analyze the observational data in the same way as the SRE stratified on $\hat{e}'(X)$. 
The ignorability holds only approximately given $\hat{e}'(X)$. We can further use regression adjustment based on covariates to remove bias and improve efficiency. To be more specific, we can obtain \citet{lin2013}'s  estimator within each stratum and construct the final estimator by a weighted average (see Chapter \ref{sec::covariate-SRE}).

With an unknown propensity score, we need to fit a statistical model to obtain the estimated propensity score $\hat{e}(X)$. This makes the final estimator dependent on the model specification. However, the propensity score stratification estimator only requires the correct ordering of the estimated propensity scores rather than their exact values, which makes it relatively robust compared with other methods. This robustness property of propensity score stratification appeared in many numerical examples but its rigorous quantification is still missing in the literature.

An important practical question is how to choose $K$? If $K$ is too small, then ignorability does not hold conditional on $\hat{e}'(X)$ even approximately. If $K$ is too large, then we do not have enough units within each stratum of the estimated propensity score and many strata have only treated or control units. Therefore, we face a trade-off. Following \citet{cochran1968effectiveness}'s heuristics, \citet{rosenbaum1983central} and \citet{rosenbaum1984reducing} suggested $K=5$ which removes a large amount of bias in many settings. However, with extremely large datasets, propensity score stratification leads to biased estimators with a fixed $K$ \citep{lunceford2004stratification}. It is thus reasonable to increase $K$ as long as each stratum still has enough treated and control units. \citet{wang2020robust} suggested a greedy choice of $K$, which is the maximum number of strata such that the stratified estimator is well-defined. However, the rigorous theory for this procedure is not fully established.

Another important practical question is how to compute the standard errors of the estimators based on propensity score stratification. Some researchers have conditioned on the discretized propensity scores $\hat{e}'(X)$ and reported standard errors based on the SRE. This effectively ignores the uncertainty in the estimated propensity scores. This often leads to an overestimation of the true variance since a surprising result in the literature is that using the estimated propensity scores decreases the asymptotic variance for estimating the average causal effect \citep{su2023estimated}. 
Other researchers bootstrapped the whole procedure to account for full uncertainty. However, the theory for the bootstrap is still unclear due to the discreteness of this estimator.

 \subsection{Application}\label{sec::application-pscore-stratification}

To illustrate the propensity score stratification method, I revisited Example \ref{eg::chan-bmi-ATE}. 
\begin{lstlisting}
> nhanes_bmi = read.csv("nhanes_bmi.csv")[, -1]
> z = nhanes_bmi$School_meal
> y = nhanes_bmi$BMI
> x = as.matrix(nhanes_bmi[, -c(1, 2)])
> x = scale(x)
\end{lstlisting}
Example \ref{eg::chan-bmi-ATE} introduced the covariates in the dataset. The treatment \ri{School_meal} is the indicator for participation in the school meal plan, and the outcome \ri{BMI} is the BMI.

\begin{figure}
\includegraphics[width = \textwidth]{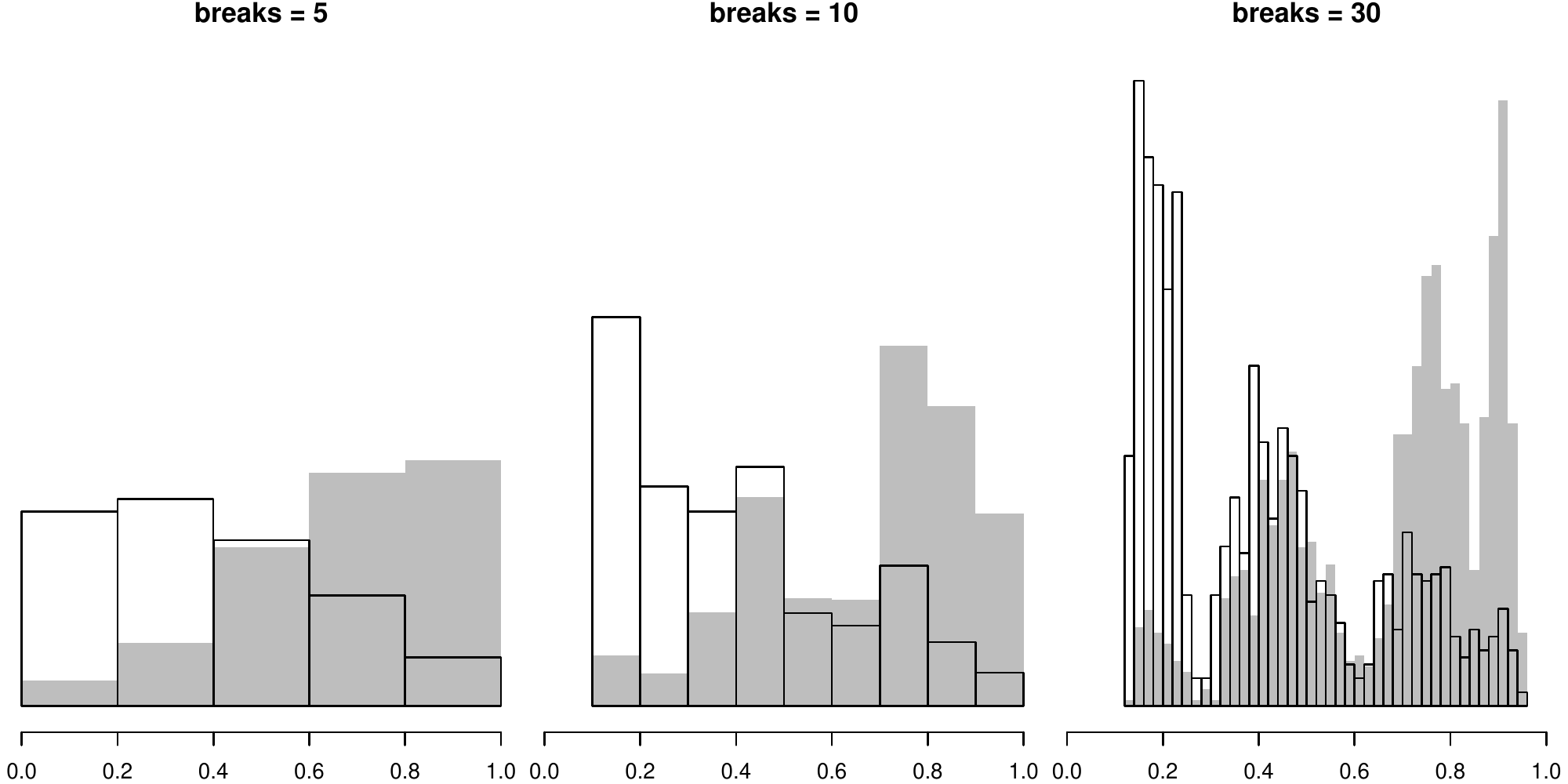}
\caption{Histograms of the estimated propensity scores based on the \ri{nhanes_bmi} data: white for the control group and grey for the treatment group}\label{fig::pscore_bmi}
\end{figure}

Figure \ref{fig::pscore_bmi} shows the histograms of the estimated propensity scores with different numbers of bins $(K=5,10,30)$. 
Based on propensity score stratification, we can calculate the point estimators and the standard errors for difference choice of $K \in \{5, 10, 20, 50, 80 \}$ as follows (with the function \ri{Neyman_SRE} defined in Chapter \ref{chapter::stratification-poststratification} for analyzing the SRE):
\begin{lstlisting}
> pscore = glm(z ~ x, family = binomial)$fitted.values
> n.strata = c(5, 10, 20, 50, 80)
> strat.res = sapply(n.strata, FUN = function(nn){
+   q.pscore = quantile(pscore, (1:(nn-1))/nn)
+   ps.strata = cut(pscore, breaks = c(0,q.pscore,1), 
+                   labels = 1:nn)
+   Neyman_SRE(z, y, ps.strata)})
> 
> rownames(strat.res) = c("est", "se")
> colnames(strat.res) = n.strata
> round(strat.res, 3)
         5     10     20     50     80
est -0.116 -0.178 -0.200 -0.265 -0.204
se   0.283  0.282  0.279  0.272     NA
\end{lstlisting}
Increasing $K$ from $5$ to $50$ reduces the standard error. However, we cannot go as extreme as $K=80$ because the standard error is not well-defined in some strata with only one treated or one control unit. The above estimators show a negative but insignificant effect of the meal program on  BMI.

We can also compare the above estimator with the three simple regression estimators: the one without adjusting for any covariates, and Fisher's and Lin's estimators. 
\begin{lstlisting}
> DiM = lm(y ~ z)
> Fisher = lm(y ~ z + x)
> Lin = lm(y ~ z + x + z*x)
> res.regression = c(coef(DiM)[2], hccm(DiM)[2, 2]^0.5,
+                    coef(Fisher)[2], hccm(Fisher)[2, 2]^0.5,
+                    coef(Lin)[2], hccm(Lin)[2, 2]^0.5)
> res.regression = matrix(res.regression,
+                         nrow = 2, ncol = 3)
> rownames(res.regression) = c("est", "se")
> colnames(res.regression) = c("naive", "fisher", "lin")
> round(res.regression, 3)   
    naive fisher    lin
est 0.534  0.061 -0.017
se  0.225  0.227  0.226
\end{lstlisting} 
The naive difference in means differs greatly from the other methods due to the large imbalance in covariates. Fisher's estimator still gives a positive point estimate although it is not significant. Lin's estimator and the propensity score stratification estimators give qualitatively the same results. The propensity score stratification estimators are stable across different choices of $K$.

\section{Propensity score weighting}

\subsection{Theory}

\begin{theorem}
\label{thm::pscore-weighting}
If $Z\ind \{ Y(1), Y(0)  \} \mid X$ and $0<e(X)<1$, then
$$
E\{  Y(1) \} =  E\left\{  \frac{ZY}{e(X)}  \right\},\quad
E\{  Y(0) \} =  E\left\{  \frac{(1-Z)Y}{1-e(X)}  \right\} ,
$$
and
$$
\tau = E\{  Y(1) -  Y(0) \} =  E\left\{  \frac{ZY}{e(X)} - \frac{(1-Z)Y}{1-e(X)}   \right\} .
$$
\end{theorem}

Before proving Theorem \ref{thm::pscore-weighting}, it is important to note the additional assumption $0<e(X)<1$. It is called the {\it overlap} or {\it positivity} condition. The formulas in Theorem \ref{thm::pscore-weighting} become infinity if $e(X)=0$ or $1$ for some values of $X$. It is not a requirement only for the identification formulas based on propensity score weighting. Although it was not stated explicitly in Theorem \ref{thm::identification-formula}, the conditional expectations $E(Y\mid Z=1, X)$ and $ E(Y\mid Z=0, X)$ in the identification formula of $\tau$ in \eqref{eq::nonparametric-identification-tau} is well defined only if $0< e(X) < 1$. The overlap condition can be viewed as a technical condition to ensure that the formulas in Theorems \ref{thm::identification-formula} and \ref{thm::pscore-weighting} are well defined. It can also cause some philosophical issues for causal inference with observational studies. When unit $i$ has $e(X_i)=1$, we always observe its potential outcome under the treatment, $Y_i(1)$, but can never observe its potential outcome under the control, $Y_i(0)$. In this case, the potential outcome $Y_i(0)$ may not even be well defined, making the definition of the causal effect ambiguous for unit $i$. \citet{king2006dangers} called $Y_i(0)$ an {\it extreme counterfactual} when $e(X_i)=1$, and discussed their dangers in causal inference. A similar problem arises if unit $i$ has $e(X_i)=0$.

In sum,  $Z\ind \{ Y(1), Y(0)  \} \mid X$ requires adequate covariates to ensure the conditional independence of the treatment and potential outcomes, and $0<e(X)<1$ requires residual randomness in the treatment conditional on the covariates. In fact, \citet{rosenbaum1983central}'s definition of strong ignorability includes both of these conditions. In modern literature, they are often stated separately.

\begin{myproof}{Theorem}{\ref{thm::pscore-weighting}}
I only prove the result for $E\{ Y(1) \}$ because the proof of the result for $E\{ Y(0) \}$ is similar. We have
\begin{eqnarray*}
&&E\left\{  \frac{ZY}{e(X)}  \right\} \\
&=& E\left\{  \frac{ZY(1)}{e(X)}  \right\} \\
&=& E\left[ E\left\{  \frac{ZY(1)}{e(X)}  \mid X \right\}  \right]  \quad \text{(tower property)} \\
&=& E\left[  \frac{1}{e(X)}   E\left\{   ZY(1)  \mid X \right\}  \right]   \\
&=& E\left[  \frac{1}{e(X)}   E( Z  \mid X ) E\left\{   Y(1)  \mid X \right\}  \right]  \quad \text{(strong ignorability)} \\\\
&=& E\left[  \frac{1}{e(X)}   e(X) E\left\{   Y(1)  \mid X \right\}  \right] \\
&=& E\left[ E\left\{   Y(1)  \mid X \right\}  \right] \\
&=& E\{   Y(1) \}. 
\end{eqnarray*}
\end{myproof}

\subsection{Inverse propensity score weighting estimators}\label{sec::ipw-estimators}

Theorem \ref{thm::pscore-weighting} motivates the following moment estimator for the average causal effect:
$$
\hat{\tau}^\textup{ht} = \frac{1}{n} \sumn \frac{Z_i Y_i}{ \hat{e}(X_i) }
- \frac{1}{n} \sumn \frac{ (1-Z_i) Y_i}{ 1 - \hat{e}(X_i) } ,
$$
where $\hat{e}(X_i)$ is the estimated propensity score. This is the inverse propensity score weighting (IPW) estimator, which is also called the Horvitz--Thompson (HT) estimator. \citet{horvitz1952generalization} proposed it in survey sampling and \citet{rosenbaum1987model} used in causal inference with observational studies.

However, the estimator $\hat{\tau}^\textup{ht}$ has many problems. In particular, it is not invariant to the location transformation of the outcome. 
Proposition \ref{prop::no-invariance-HT} states this problem precisely, with the proof relegated to Problem \ref{hw::hajek-weights}. 

\begin{proposition}
[lack of invariance for the HT estimator]
\label{prop::no-invariance-HT}
If we change $Y_i$ to $Y_i+c$ with a constant $c$, then the HT estimator $\hat{\tau}^\textup{ht}$ becomes $\hat{\tau}^\textup{ht} + c (\hat{ 1 }_\textsc{T} - \hat{ 1 }_\textsc{C})$, where 
$$
\hat{ 1 }_\textsc{T}=   \frac{1}{n} \sumn \frac{Z_i }{ \hat{e}(X_i) } ,\quad
\hat{ 1 }_\textsc{C} =\frac{1}{n} \sumn \frac{ (1-Z_i) }{ 1 - \hat{e}(X_i) }   
$$
can be viewed as two different estimates of the constant $1$. 
\end{proposition}

In Proposition \ref{prop::no-invariance-HT}, I use the funny notation $\hat{ 1 }_\textsc{T} $ and $ \hat{ 1 }_\textsc{C}$ because with the true propensity score these two terms both have expectation 1; see Problem \ref{hw::hajek-weights} for more details. 
In general, $\hat{ 1 }_\textsc{T} - \hat{ 1 }_\textsc{C}$ is not zero in finite samples. 
Since adding a constant to every outcome should not change the average causal effect, the HT estimator is not reasonable because of its dependence on $c.$ A simple fix to the problem is to normalize the weights by $\hat{ 1 }_\textsc{T}  $ and $\hat{ 1 }_\textsc{C} $ respectively, resulting in the following estimator 
$$
\hat{\tau}^\textup{hajek} = \frac{  \sumn \frac{Z_i Y_i}{ \hat{e}(X_i) }  }{  \sumn \frac{Z_i }{ \hat{e}(X_i) } }
- \frac{  \sumn \frac{ (1-Z_i) Y_i}{ 1 - \hat{e}(X_i) }  }{  \sumn \frac{ 1-Z_i}{ 1 - \hat{e}(X_i) }  }.
$$
This is the Hajek estimator due to \citet{hajek1971comment} in the context of survey sampling with varying probabilities. We can verify that the Hajek estimator is invariant to the location transformation. That is, if we replace $Y_i$ by $Y_i + c$, then $\hat{\tau}^\textup{hajek}$ remains the same; see Problem \ref{hw::hajek-weights}. Moreover, many numerical studies have found that $\hat{\tau}^\textup{hajek}$ is much more stable than $\hat{\tau}^\textup{ht}$ in finite samples.

\subsection{A problem of IPW and a fundamental problem of causal inference}
\label{sec::ipw-problem}

Many asymptotic analyses require a {\it strong overlap} condition
$$
0<  \alpha_\textsc{L} \leq e(X) \leq  \alpha_\textsc{U} < 1,
$$
that is, the true propensity score is bounded away from $0$ and $1$. However, \citet{d2017overlap} pointed out that this is a rather strong assumption, especially with many covariates. Chapter \ref{chapter::overlap} will discuss this problem in detail.

Even if the strong overlap condition holds for the true propensity score, the estimated propensity scores can be close to $0$ or $1$. 
When this happens, the weighting estimators blow up to infinity, which results in extremely unstable behavior in finite samples. We can either truncate the estimated propensity score by changing it to
$$
 \max\Big[  \alpha_\textsc{L},   \min\{   \hat{e}(X_i), \alpha_\textsc{U} \}   \Big] ,
$$
or trim the observations by dropping units with $\hat{e}(X_i)$ outside the interval $[\alpha_\textsc{L},   \alpha_\textsc{U}  ]$. \citet{crump2009dealing} suggested $  \alpha_\textsc{L} = 0.1$ and $\alpha_\textsc{U} = 0.9 $, and \citet{kurth2005results} suggested $  \alpha_\textsc{L} = 0.05$ and $\alpha_\textsc{U} = 0.95 $. \citet{yang2018asymptotic} established some asymptotic theory for trimming. Overall, although trimming often stabilizes the IPW estimators, it also injects additional arbitrariness into the procedure.

 \subsection{Application}\label{sec::application-pscore-weighting}

The following functions can compute the IPW estimators and their bootstrap standard errors. 
\begin{lstlisting}
ipw.est = function(z, y, x, truncps = c(0, 1))
{
  ## fitted propensity score
  pscore   = glm(z ~ x, family = binomial)$fitted.values
  pscore   = pmax(truncps[1], pmin(truncps[2], pscore))
  
  ace.ipw0 = mean(z*y/pscore - (1 - z)*y/(1 - pscore))
  ace.ipw  = mean(z*y/pscore)/mean(z/pscore) - 
    mean((1 - z)*y/(1 - pscore))/mean((1 - z)/(1 - pscore))
  
  return(c(ace.ipw0, ace.ipw))     
}


ipw.boot = function(z, y, x, n.boot = 500, truncps = c(0, 1))
{
  point.est  = ipw.est(z, y, x, truncps)
  
  ## nonparametric bootstrap
  n.sample   = length(z)
  x          = as.matrix(x)
  boot.est   = replicate(n.boot, {
    id.boot = sample(1:n.sample, n.sample, replace = TRUE)
    ipw.est(z[id.boot], y[id.boot], x[id.boot, ], truncps)
  })
  boot.se    = apply(boot.est, 1, sd)
  
  res = cbind(point.est, boot.se)
  colnames(res) = c("est", "se")
  rownames(res) = c("HT", "Hajek")
  
  return(res)
}
\end{lstlisting}

Revisiting Example \ref{eg::chan-bmi-ATE}, we can obtain the IPW estimators based on different truncations of the estimated propensity scores. The following results are the two weighting estimators with the bootstrap standard errors, with truncations at $(0,1)$, $(0.01, 0.99)$, $(0.05, 0.95)$, and $(0.1, 0.9)$:
\begin{lstlisting}
> trunc.list = list(trunc0 = c(0,1), 
+                   trunc.01 = c(0.01, 0.99), 
+                   trunc.05 = c(0.05, 0.95), 
+                   trunc.1 = c(0.1, 0.9))
> trunc.est = lapply(trunc.list,
+                    function(t){
+                      est = ipw.boot(z, y, x, truncps = t)
+                      round(est, 3)
+                    })
> trunc.est
$trunc0
         est    se
HT    -1.516 0.496
Hajek -0.156 0.258

$trunc.01
         est    se
HT    -1.516 0.501
Hajek -0.156 0.254

$trunc.05
         est    se
HT    -1.499 0.501
Hajek -0.152 0.255

$trunc.1
         est    se
HT    -0.713 0.425
Hajek -0.054 0.246
\end{lstlisting}
The HT estimator gives results far away from all other estimators we discussed so far. The point estimates seem too large and they are negatively significant unless we truncate the estimated propensity scores at $(0.1, 0.9)$. This is an example showing the instability of the HT estimator.

\section{The balancing property of the propensity score}

\subsection{Theory}

\begin{theorem}
\label{thm::pscoreASbscore}
The propensity score satisfies 
$$
Z\ind X \mid e(X) .
$$ 
Moreover, for any function $h(\cdot)$, we have 
\begin{eqnarray}
E\left\{   \frac{Zh(X)}{e(X)}   \right\} = E\left\{   \frac{(1-Z)h(X)}{1-e(X)}   \right\}
\label{eq::balancehX}
\end{eqnarray}
provided the existence of the moments on both sides of \eqref{eq::balancehX}. 
\end{theorem}

Theorem \ref{thm::pscoreASbscore} does not require the ignorability assumption. It is about the treatment $Z$ and covariates $X$ only. 
The first part of  Theorem \ref{thm::pscoreASbscore} states that conditional on the propensity score, the treatment indicator, and the covariates are independent. Therefore, within the same level of the propensity score, the covariate distributions are balanced across the treatment and control groups. 
The second part of Theorem \ref{thm::pscoreASbscore} states that an equivalent form of covariate balance based on the weighting form. 
I give a proof of Theorem \ref{thm::pscoreASbscore} below.

\begin{myproof}{Theorem}{\ref{thm::pscoreASbscore}}
First, we show $Z\ind X \mid e(X)$, that is,
\begin{eqnarray}
\pr\{Z=1\mid X, e(X)\} = \pr\{Z=1\mid  e(X)\}.
\label{eq::balance1}
\end{eqnarray} 
Following similar steps as the proof of Theorem \ref{thm::pscore-dimreduction}, we can show that the left-hand side of \eqref{eq::balance1} equals
$$
\pr\{Z=1\mid X, e(X)\} = \pr( Z=1\mid X ) = e(X),
$$
and the right-hand side of \eqref{eq::balance1} equals 
\begin{eqnarray*}
\pr\{Z=1\mid  e(X)\} &=& E\{  Z\mid  e(X) \}  \\
&=& E\Big[ E\{  Z\mid X,  e(X) \} \mid e(X) \Big] \\
&=& E\Big[ E\{  Z\mid X  \} \mid e(X) \Big] \\
&=& E\Big[ e(X) \mid e(X) \Big] \\
&=& e(X). 
\end{eqnarray*}
Therefore, \eqref{eq::balance1} holds. 

Second, we show \eqref{eq::balancehX}. We can use similar steps as the proof of Theorem \ref{thm::pscore-weighting}. But given Theorem \ref{thm::pscore-weighting}, we have a simpler proof. If we view $h(X)$ as an outcome, then its two potential outcomes are identical and ignorability holds: $Z\ind \{  h(X), h(X) \}  \mid X$. The difference between the left-hand and right-hand sides of \eqref{eq::balancehX} is the average causal effect of $Z$ on $h(X)$, which is zero. 
\end{myproof}

\subsection{Covariate balance check}
\label{sec::balance-check-bmi}

The proof of Theorem \ref{thm::pscoreASbscore} is simple. But Theorem \ref{thm::pscoreASbscore} has useful implications for the statistical analysis. Before getting access to the outcome data, we can check whether the propensity score model is specified well enough to ensure the covariate balance in the data. \citet{rubin2007design} viewed this as the design stage of the observational study, and \citet{rubin2008objective} argued that this can result in more objective causal inference because the design stage does not involve the values of the outcomes.\footnote{While this is a useful recommendation in practice, it is not entirely clear how to quantify the objectiveness.}

In the propensity score stratification, we have the discretized estimated propensity score $\hat{e}'(X)$ and approximately
$$
Z\ind X\mid \hat{e}'(X) = e_k\quad (k=1,\ldots, K).
$$
Therefore, we can check whether the covariate distributions are the same across the treatment and control groups within each stratum of the discretized estimated propensity score.

In propensity score weighting, we can view $h(X)$ as a pseudo outcome and estimate the average causal effect on $h(X)$. Because the true average causal effect on $h(X)$ is $0$, the estimate should not be significantly different from $0$.  A canonical choice of $h(X)$ is $X$.

\begin{figure}
\centering
\includegraphics[width = \textwidth]{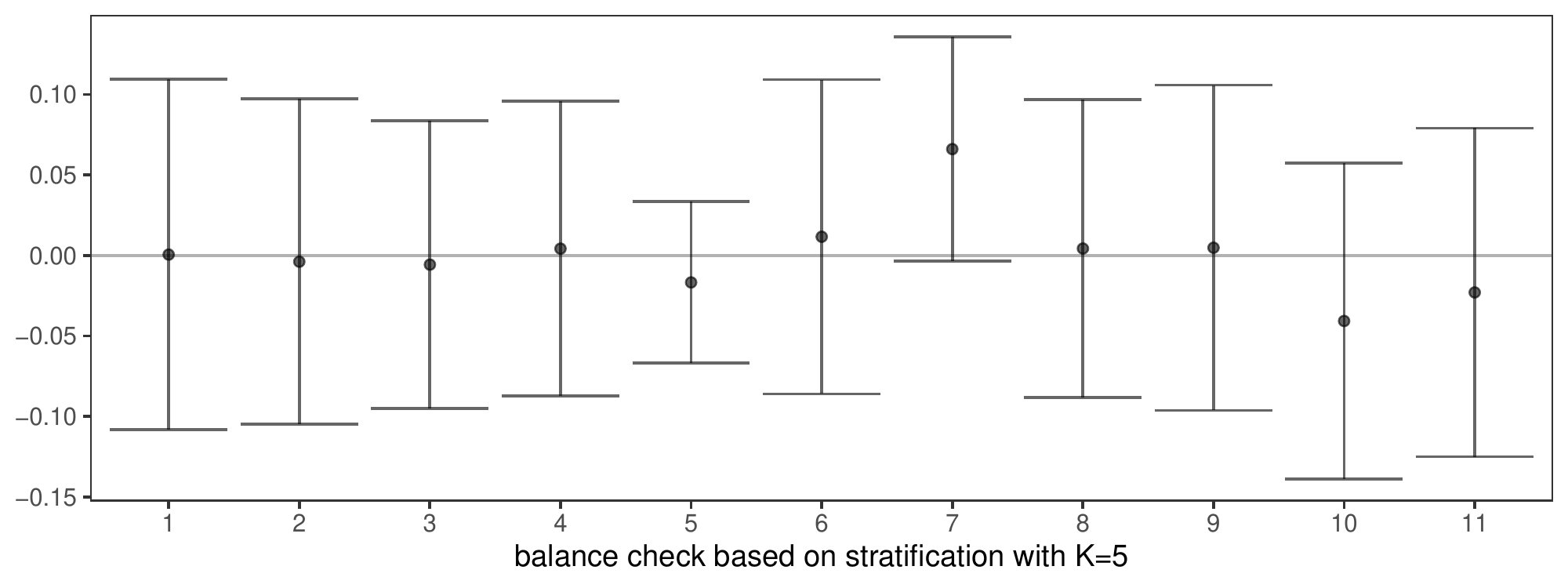}
\includegraphics[width = \textwidth]{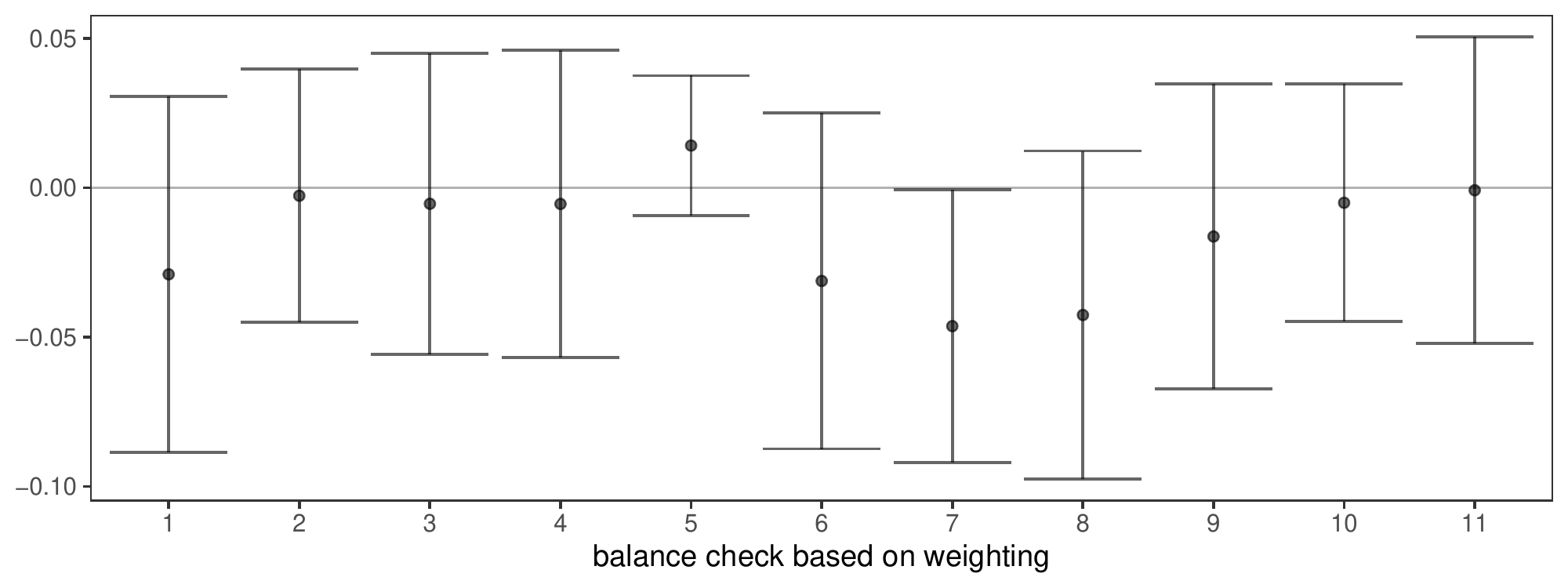}
\caption{Balance check: point estimates and 95\% confidence intervals of the average causal effect on covariates}\label{fig::balance-check}
\end{figure}

Let us revisit Example \ref{eg::chan-bmi-ATE} again. Based on propensity score stratification with $K=5$, all the covariates are well-balanced across the treatment and control groups. Similar results hold for the Hajek estimator. The only exception is \ri{Food_Stamp}, the 7th covariate in Figure \ref{fig::balance-check}. 
Figure \ref{fig::balance-check} shows the balance-checking results. 

 \section{Homework Problems}

 \paragraph{Another version of Theorem \ref{thm::pscore-dimreduction}} \label{hw::principal-unobserved-cov}
 Prove that 
\begin{equation}
\label{eq::principal-covariate-ind}
 Z\ind \{  Y(1), Y(0) , X\}  \mid e(X, Y(1), Y(0)   ) .
\end{equation}  
 
 Remark:
 This result holds without assuming strong ignorability. It implies that
 $$
  Z\ind \{  Y(1), Y(0) \}  \mid  \{ X,   e(X, Y(1), Y(0)  \} .
 $$
\citet{rosenbaum2020modern} and \citet{rosenbaum2023propensity} pointed out the result in \eqref{eq::principal-covariate-ind} and called $e(X, Y(1), Y(0)   ) $ the  {\it  principal unobserved covariate}.

\paragraph{Another version of Theorem \ref{thm::pscore-dimreduction}} 

Theorem \ref{thm::pscore-dimreduction} states a result under strong ignorability. An analogous result also holds under ignorability. 
That is, if ignorability holds conditional on covariates $X$, then it also holds conditional on the scalar propensity score $e(X)$. 

\begin{theorem}
\label{thm::pscore-dimension-reduction-weak-ig}
If $Z\ind Y(z)  \mid X$ for $z=0,1$, then
$
Z\ind Y(z)\mid e(X)
$
for $z=0,1$. 
\end{theorem}

Prove Theorem \ref{thm::pscore-dimension-reduction-weak-ig}.

 \paragraph{More results on the IPW estimators}
\label{hw::hajek-weights} 
 
This is related to the discussion of the HT estimator in Section \ref{sec::ipw-estimators}. First, prove Proposition \ref{prop::no-invariance-HT}.  
Second, prove 
 \begin{eqnarray*} 
E\left\{ \frac{1}{n} \sumn \frac{Z_i }{e(X_i) } \right\} =1,\quad
E\left\{  \frac{1}{n} \sumn \frac{ (1-Z_i) }{ 1 - e(X_i) }  \right\} = 1.
\end{eqnarray*}
Third, prove that if we add a constant $c$ to every observed outcome $Y_i$, the Hajek estimator
$
\hat{\tau}^\textup{hajek} 
$
remains the same.

\paragraph{Re-analysis of \citet{Rosenbaum::1983JRSSB}}

Table \ref{table::rr1983table1} is from \citet{Rosenbaum::1983JRSSB}, which concerned the causal effect of the coronary artery bypass surgery compared with the medical therapy on the functional improvement 6 months after cardiac catheterization.  They first estimated the propensity score based on 74 observed covariates and then formed 5 strata based on the discretized estimated propensity score. Because the treatment is binary and the outcome is also binary, they represented the data in a table. 
Based on Table \ref{table::rr1983table1} , estimate the average causal effect, and report the 95\% confidence interval of the average causal effect.

\begin{table}
\caption{Table 1 of \citet{Rosenbaum::1983JRSSB}}\label{table::rr1983table1}
\begin{tabular}{clcc}
\hline 
stratum by $\hat{e}(X)$ & treatment &  number of patients & proportion improved\\
\hline 1 & Surgical & 26 & 0.54 \\
& Medical & 277 & 0.35 \\
2 & Surgical & 68 & 0.70 \\
& Medical & 235 & 0.40 \\
3 & Surgical & 98 & 0.70 \\
& Medical & 205 & 0.35 \\
4 & Surgical & 164 & 0.71 \\
& Medical & 139 & 0.30 \\
5 & Surgical & 234 & 0.70 \\
& Medical & 69 & 0.39 \\
\hline
\end{tabular}
\end{table}

Remark:   If you are interested, you can read the whole paper of \citet{Rosenbaum::1983JRSSB} after reading Part \ref{part::challenges-os} of the book. It is a canonical paper on sensitivity analysis in causal inference.

\paragraph{Balancing score and propensity score: more theoretical results}
\label{hw::balancingscore-rr1983}

\citet{rosenbaum1983central} also introduced the notion of {\it balancing score}.

\begin{definition}
[balancing score]\label{def::balancing-score}
  $b(X)$ is a balancing score if 
  $$
  Z\ind X\mid b(X).
  $$
\end{definition}

In Definition \ref{def::balancing-score},  $b(X)$ can be a scalar or a vector. An obvious balancing score is $b(X) = X$, but it is not a useful one without any simplification of the original covariates. By Theorem \ref{thm::pscoreASbscore}, the propensity score is a special balancing score. More interestingly, \citet{rosenbaum1983central} showed that the propensity score is the coarsest balancing score, as in Theorem \ref{thm::balancingscore} below which includes Theorem \ref{thm::pscoreASbscore} as a special case.

\begin{theorem}
\label{thm::balancingscore}
$b(X)$ is a balancing score if and only if $b(X)$ is finer than $e(X)$ in the sense that $e(X) = f(b(X))$ for some function $f(\cdot)$. 
\end{theorem}

Theorem \ref{thm::balancingscore} is relevant in subgroup analysis. In particular, we may be interested in not only the average causal effect $\tau$ but also the subgroup effects. For instance, we may want to estimate the average causal effects among boys and girls, respectively. Without loss of generality, assume the first component of $X$ is the indicator for girls, and we are interested in estimating 
$$
\tau(x_1) = E\{ Y(1) - Y(0) \mid X_1 = x_1  \},\quad (x_1=1,0).
$$
Theorem \ref{thm::balancingscore} implies that under ignorability, we also have 
\begin{equation}
\label{eq::subgroup-ignorability}
Z\ind \{ Y(1) , Y(0) \} \mid e(X), X_1
\end{equation}
because $b(X) = \{e(X), X_1\}$ is finer than $e(X)$ and thus a balancing score. The conditional independence in \eqref{eq::subgroup-ignorability} ensures ignorability holds given the propensity score, within each level of $X_1$. Therefore, we can perform the same analysis based on the propensity score, within each level of $X_1$, yielding estimates for two subgroup effects.

With the above motivation in mind, now prove Theorem \ref{thm::balancingscore}.

\paragraph{Some basics of subgroup effects}

This problem is related to Problem \ref{hw::balancingscore-rr1983}, but you can work on it independently. 

Consider a standard observational study with covariates $X = (X_1, X_2)$, where $X_1$ denotes a binary subgroup indicator (e.g., statistics major or not statistics major) and $X_2$ contains the rest of the covariates. The parameter of interest is the subgroup causal effect 
$$
\tau(x_1) = E\{ Y(1) - Y(0) \mid X_1 = x_1  \},\quad (x_1=1,0).
$$
Show that
$$
\tau(x_1) =  E\left\{   \frac{  1(X_1 = x_1) ZY}{ e(X) }   - \frac{  1(X_1 = x_1) (1-Z)Y}{ 1 -  e(X) }    \right\} \Big /  \pr(X_1 = x_1)
$$
and give the corresponding HT and Hajek estimators for $\tau(x_1) $.

 \paragraph{Recommended reading}

The title of this chapter is the same as the title of the classic paper by \citet{rosenbaum1983central}. Most results in this chapter are directly drawn from their original paper.

\citet{rubin2007design} and \citet{rubin2008objective} highlighted the importance of the design stage of observational studies for more objective causal inference

\chapter{The Doubly Robust or the Augmented Inverse Propensity Score Weighting Estimator for the Average Causal Effect}\label{chapter::doubly-robust}
 \chaptermark{Doubly Robust Estimator}

Under ignorability $Z\ind \{ Y(1), Y(0) \} \mid X$ and overlap $0< e(X) < 1$, Chapter \ref{chapter::pscore-key} has shown two identification formulas of the average causal effect $\tau = E\{ Y(1) - Y(0)  \}$. First, the outcome regression formula  is 
\begin{eqnarray}
\tau  =  E\{ \mu_1(X) \} - E\{ \mu_0(X) \} \label{eq::iden-outreg}
\end{eqnarray}
where 
\begin{eqnarray*}
\mu_1(X)  &=& E\{  Y(1) \mid X \} = E(Y\mid Z=1, X),\\ 
\mu_0(X) &=& E\{  Y(0) \mid X \} = E(Y\mid Z=0, X)
\end{eqnarray*}
are the two conditional mean functions of the outcome given covariates under the treatment and control, respectively. Second, the IPW formula is 
\begin{eqnarray}
\tau = E\left\{  \frac{ZY}{e(X)}   \right\} - E\left\{  \frac{(1-Z)Y}{1-e(X)}   \right\} \label{eq::iden-ipw}
\end{eqnarray}
where
$$
e(X) = \pr(Z=1\mid X)
$$
is the propensity score introduced in Chapter \ref{chapter::pscore-key}.

The outcome regression estimator requires fitting a model for the outcome given the treatment and covariates. It is consistent if the outcome model is correctly specified. The IPW estimator requires fitting a model for the treatment given the covariates. It is consistent if the propensity score model is correctly specified.

Mathematically, we have many combinations of \eqref{eq::iden-outreg} and \eqref{eq::iden-ipw} that lead to different identification formulas of the average causal effect. Below I will discuss a particular combination that has appealing theoretical properties. This combination motivates an estimator that is consistent if either the propensity score or the outcome model is correctly specified. It is called the {\it doubly robust} estimator, championed by James Robins \citep{scharfstein1999adjusting, bang2005doubly}.

\section{The doubly robust estimator}

\subsection{Population version}

We posit a model for the conditional means of the outcome $\mu_1(X, \beta_1)$ and $\mu_0(X, \beta_0)$, indexed by the parameters $\beta_1$ and $\beta_0$. For example, if the conditional means are linear or logistic under the working model, then the parameters are just the regression coefficients. If the outcome  model is  correctly specified, then $\mu_1(X, \beta_1) = \mu_1(X)$ and $   \mu_0(X, \beta_0) = \mu_0(X)$. 
We posit a working model for the propensity score $e(X, \alpha)$, indexed by the parameter $\alpha$. For example, if the working model is logistic, then $\alpha$ is the regression coefficient. If the propensity score model is correctly specified, then $e(X,\alpha) = e(X)$. In practice, both models may be misspecified. Sometimes, we call them {\it working models}  due to the possibility of misspecification. 

Define 
\begin{eqnarray}
\tilde{\mu}_1^\textup{dr} &=& E\left[   \frac{Z\{ Y-\mu_1(X,\beta_1) \}}{  e(X,\alpha) }   + \mu_1(X, \beta_1)  \right] , \label{eq::dr-aug-out1}   \\
\tilde{\mu}_0^\textup{dr} &=& E\left[   \frac{(1-Z)\{ Y-\mu_0(X,\beta_0) \}}{  1-e(X,\alpha) }   + \mu_0(X, \beta_0)  \right]  , \label{eq::dr-aug-out0} 
\end{eqnarray}
which can also be written as 
\begin{eqnarray}
\tilde{\mu}_1^\textup{dr} &=& E\left[   \frac{Z Y}{  e(X,\alpha) }   - \frac{  Z -  e(X,\alpha)  }{ e(X,\alpha) } \mu_1(X, \beta_1)  \right]  ,\label{eq::dr-aug-ipw1} \\
\tilde{\mu}_0^\textup{dr} &=& E\left[   \frac{(1-Z) Y}{  1-e(X,\alpha) }    - \frac{ e(X,\alpha)-Z  }{1- e(X,\alpha)} \mu_0(X, \beta_0)  \right]    .\label{eq::dr-aug-ipw0} 
\end{eqnarray}
 
The formulas in \eqref{eq::dr-aug-out1} and \eqref{eq::dr-aug-out0} augment the outcome regression estimator by inverse propensity score weighting terms of the residuals. The formulas in \eqref{eq::dr-aug-ipw1} and \eqref{eq::dr-aug-ipw0} augment the IPW estimator by the imputed outcomes. For this reason, the doubly robust estimator is also called the {\it augmented inverse propensity score weighting} (AIPW) estimator.

The augmentation strengthens the theoretical properties in the following sense.

\begin{theorem}
\label{thm::doublyrobust}
Assume ignorability $Z \ind  \{  Y(1), Y(0) \} \mid X$ and overlap $0 < e(X) < 1$. 
\begin{enumerate}
\item
If either $e(X,\alpha) = e(X)$ or $\mu_1(X, \beta_1) = \mu_1(X)$, then $\tilde{\mu}_1^\textup{dr} = E\{  Y(1) \}$.
\item 
If either $e(X,\alpha) = e(X)$ or $\mu_0(X, \beta_0) = \mu_0(X)$, then $\tilde{\mu}_0^\textup{dr} = E\{  Y(0) \}$.
\item 
If either $e(X,\alpha) = e(X)$ or $\{\mu_1(X, \beta_1) = \mu_1(X),   \mu_0(X, \beta_0) = \mu_0(X)\}$, then $\tilde{\mu}_1^\textup{dr} - \tilde{\mu}_0^\textup{dr} = \tau$. 
\end{enumerate}
\end{theorem}

By Theorem \ref{thm::doublyrobust}, $\tilde{\mu}_1^\textup{dr} - \tilde{\mu}_0^\textup{dr} $ equals $\tau$ if either the propensity score model or the outcome model is correctly specified. That's why it is called the doubly robust estimator.

\begin{myproof}{Theorem}{\ref{thm::doublyrobust}}
I only prove the result for $\mu_1 = E\{  Y(1) \}$. The proof for the result for $\mu_0 = E\{  Y(0) \}$ is similar.

We have the decomposition
\begin{eqnarray*}
&&\tilde{\mu}_1^\textup{dr} - E\{  Y(1) \}     \\
&=& E\left[   \frac{Z\{ Y(1)-\mu_1(X,\beta_1) \}}{  e(X,\alpha) }   - \{ Y(1) -  \mu_1(X, \beta_1) \} \right]  
\qquad (\text{by definition})\\
&=&    E\left[   \frac{Z - e(X,\alpha) }{  e(X,\alpha) }   \{ Y(1) -  \mu_1(X, \beta_1) \} \right]  
\qquad (\text{combining terms})\\
&=&  E\left(  E\left[   \frac{Z - e(X,\alpha) }{  e(X,\alpha) }   \{ Y(1) -  \mu_1(X, \beta_1) \} \mid X \right] \right) 
\qquad (\text{tower property})\\
&=& E\left[  E\left\{   \frac{Z - e(X,\alpha) }{  e(X,\alpha) } \mid X \right\} \times E\left\{  Y(1) -  \mu_1(X, \beta_1)  \mid X \right\} \right] 
\qquad (\text{ignorability})\\
&=& E\left[   \frac{e(X) - e(X,\alpha) }{  e(X,\alpha) }  \times  \left\{  \mu_1(X) -  \mu_1(X, \beta_1)   \right\} \right]  . 
\end{eqnarray*}
Therefore, $\tilde{\mu}_1^\textup{dr} - E\{  Y(1) \}  = 0$
 if either $e(X,\alpha) = e(X)$ or $\mu_1(X, \beta_1) = \mu_1(X)$. 
\end{myproof}

\subsection{Sample version}

Based on the population versions of $\tilde{\mu}_1^\textup{dr} $ and $\tilde{\mu}_0^\textup{dr} $, we can obtain their sample analogs to construct a doubly robust estimator for $\tau$.

\begin{definition}
[doubly robust estimator for the average causal effect]\label{def::dr:ace}
Based on the data $(X_i, Z_i, Y_i)_{i=1}^n$, we can obtain a doubly robust estimator for $\tau$ by the following steps:\footnote{I used $\hat e(X_i)$ for $e(X_i, \hat{\alpha})$ and $ \hat \mu_z(X_i) $ for $ \mu_z(X_i, \hat{\beta}_1) $ before when I did not want to emphasize the parameters in the working models.} 
\begin{enumerate}
[(1)]
\item
obtain the fitted values of the propensity scores: $e(X_i, \hat{\alpha})$; 
\item
obtain the fitted values of the outcome means: $ \mu_1(X_i, \hat{\beta}_1) $ and $ \mu_0(X_i, \hat{\beta}_0) $; 
\item
construct the doubly robust estimator: $\hat{\tau}^\textup{dr} = \hat{\mu}_1^\textup{dr}  - \hat{\mu}_0^\textup{dr}$, where
$$
 \hat{\mu}_1^\textup{dr}  = \frac{1}{n} \sumn \left[ \frac{Z_i\{ Y_i-\mu_1(X_i,\hat{\beta}_1) \}}{  e(X_i, \hat{\alpha}) }   + \mu_1(X_i, \hat{\beta}_1) \right]
$$
and
$$
 \hat{\mu}_0^\textup{dr}  = \frac{1}{n} \sumn \left[  \frac{(1-Z_i)\{ Y_i-\mu_0(X_i, \hat{\beta}_0) \}}{  1-e(X_i,\hat{\alpha}) }   + \mu_0(X_i, \hat{\beta}_0)       \right] . 
$$
\end{enumerate}
\end{definition}

By Definition \ref{def::dr:ace}, we can also write the doubly robust estimator as
$$
\hat{\tau}^\textup{dr} =\hat{\tau}^\textup{reg}  + 
\frac{1}{n} \sumn  \frac{Z_i\{ Y_i-\mu_1(X_i,\hat{\beta}_1) \}}{  e(X_i, \hat{\alpha}) } 
-  \frac{1}{n} \sumn   \frac{(1-Z_i)\{ Y_i-\mu_0(X_i, \hat{\beta}_0) \}}{  1-e(X_i,\hat{\alpha}) }  . 
$$ 
Analogous to \eqref{eq::dr-aug-ipw1} and \eqref{eq::dr-aug-ipw0}, we can also rewrite it as 
$$
\hat{\tau}^\textup{dr} = \hat{\tau}^\textup{ipw}  
  - \frac{1}{n} \sumn    \frac{  Z_i -  e(X_i, \hat\alpha)  }{ e(X_i, \hat\alpha) } \mu_1(X_i, \hat\beta_1)
  + \frac{1}{n} \sumn    \frac{ e(X_i, \hat\alpha)-Z_i  }{1- e(X_i,\hat \alpha)} \mu_0(X_i, \hat\beta_0) .
$$

\citet{funk2011doubly} suggested to  approximate the variance of $\hat{\tau}^\textup{dr}$ via the nonparametric bootstrap by resampling from $(Z_i, X_i, Y_i)_{i=1}^n$.

\section{More intuition and theory for the doubly robust estimator}

Although the beginning of this chapter claims that the basic identification formulas based on outcome regression and IPW immediately yield infinitely many other identification formulas, the particular forms of the doubly robust estimators in \eqref{eq::dr-aug-out1} and \eqref{eq::dr-aug-out0} are not obvious to come up with. The original motivation for \eqref{eq::dr-aug-out1} and \eqref{eq::dr-aug-out0} was quite theoretical,  and relied on something called the {\it semiparametric efficiency theory} in advanced mathematical statistics \citep{bickel1993efficient}. It is beyond the level of this book. Below I will give two intuitive perspectives to construct \eqref{eq::dr-aug-out1} and \eqref{eq::dr-aug-out0}. Both Sections \ref{sec::improving-ipw} and \ref{sec::improving-outreg} below focus on the estimation of $E\{ Y(1) \}$ since the estimation of $E\{ Y(0) \}$ is similar by symmetry.

\subsection{Reducing the variance of the IPW estimator}\label{sec::improving-ipw}

The IPW estimator for $\mu_1$ based on
$$
\mu_1 = E\left\{   \frac{ZY}{e(X)}   \right\}
$$
completely ignores the outcome model of $Y$. It has the advantage of being consistent without assuming any outcome model. However, if the covariates are predictive of the outcome, the residual based on a working outcome model usually has a smaller variance than the outcome even when this working outcome model is wrong. With a possibly misspecified outcome model $\mu_1(X, \beta_1)$, a trivial decomposition holds:
$$
\mu_1  = E\{ Y(1) \} = E\{ Y(1)  -\mu_1(X, \beta_1) \} + E\{\mu_1(X, \beta_1)\}.
$$
If we apply the IPW formula to the first term in the above formula viewing $Y(1)  -\mu_1(X, \beta_1)$ as a ``pseudo potential outcome'' under the treatment, we can rewrite the above formula as
\begin{eqnarray}
\mu_1  &=& E\left\{ \frac{Z \{Y  -\mu_1(X, \beta_1) \}}{e(X)} \right\} + E\{\mu_1(X, \beta_1)\}  \label{eq::explain-ipw-efficiency-basic}  \\
&=& E\left\{ \frac{Z \{Y  -\mu_1(X, \beta_1) \}}{e(X)}  +  \mu_1(X, \beta_1) \right\} ,\label{eq::explain-ipw-efficiency}
\end{eqnarray}
which holds if the propensity score model is correct without assuming that the outcome model is correct. Using a working model to improve efficiency is an old idea from survey sampling. \citet{little2004robust} and \citet{lumley2011connections} pointed out its connection with the doubly robust estimator.

\subsection{Reducing the bias of the outcome regression estimator}\label{sec::improving-outreg}

The discussion in Section \ref{sec::improving-ipw} starts with the IPW estimator and improves its efficiency based on a working outcome model. Alternatively, we can also start with an outcome regression estimator based on
$$
\tilde{\mu}_1 = E\{\mu_1(X, \beta_1)\} 
$$
which may not be the same as $\mu_1$ since the outcome model may be wrong. The bias of this estimator is $ E\{\mu_1(X, \beta_1)  -  Y(1)\}   $, which can be estimated by an IPW estimator
$$
B = 
E\left\{ \frac{Z \{\mu_1(X, \beta_1) - Y\}}{e(X)} \right\} 
$$
if the propensity score model is correct. So a de-biased estimator is
$
\tilde{\mu}_1 - B  ,
$
which is identical to \eqref{eq::explain-ipw-efficiency}.

\section{Examples}
\label{sec::dr-examples}

\subsection{Summary of some canonical estimators for $\tau$}

The following \ri{R} code implements the outcome regression, HT, Hajek, and doubly robust estimators for $\tau$. These estimators can be conveniently implemented based on the fitted values of the \ri{glm} function. The default choice for the propensity score model is the logistic model, and the default choice for the outcome model is the linear model with \ri{out.family = gaussian}\footnote{The \ri{glm} function is more general than the \ri{lm} function. With \ri{out.family = gaussian}, \ri{glm} is identical to \ri{lm}.}. For binary outcomes, we can also specify \ri{out.family = binomial} to fit the logistic model.

\begin{lstlisting}
OS_est = function(z, y, x, out.family = gaussian, 
                  truncps = c(0, 1))
{
     ## fitted propensity score
     pscore   = glm(z ~ x, family = binomial)$fitted.values
     pscore   = pmax(truncps[1], pmin(truncps[2], pscore))
     
     ## fitted potential outcomes
     outcome1 = glm(y ~ x, weights = z, 
                    family = out.family)$fitted.values
     outcome0 = glm(y ~ x, weights = (1 - z), 
                    family = out.family)$fitted.values
     
     ## outcome regression estimator
     ace.reg  = mean(outcome1 - outcome0) 
     ## IPW estimators
     y.treat     = mean(z*y/pscore)
     y.control   = mean((1 - z)*y/(1 - pscore))
     one.treat   = mean(z/pscore)
     one.control = mean((1 - z)/(1 - pscore))
     ace.ipw0    = y.treat - y.control
     ace.ipw     = y.treat/one.treat - y.control/one.control
     ## doubly robust estimator
     res1      = y - outcome1
     res0      = y - outcome0
     r.treat   = mean(z*res1/pscore)
     r.control = mean((1 - z)*res0/(1 - pscore))
     ace.dr    = ace.reg + r.treat - r.control

     return(c(ace.reg, ace.ipw0, ace.ipw, ace.dr))     
}
\end{lstlisting}

It is tedious to calculate the analytic formulas for the variances of the above estimators. The bootstrap provides convenient approximations to the variances based on resampling from $\{ Z_i, X_i, Y_i \}_{i=1}^n$. Building upon the function \ri{OS_est} above, the following function returns point estimators as well as the bootstrap standard errors.

\begin{lstlisting}
OS_ATE = function(z, y, x, n.boot = 2*10^2,
                     out.family = gaussian, truncps = c(0, 1))
{
     point.est  = OS_est(z, y, x, out.family, truncps)
     
     ## nonparametric bootstrap
     n          = length(z)
     x          = as.matrix(x)
     boot.est   = replicate(n.boot, {
       id.boot = sample(1:n, n, replace = TRUE)
       OS_est(z[id.boot], y[id.boot], x[id.boot, ], 
              out.family, truncps)
     })

     boot.se    = apply(boot.est, 1, sd)
     
     res        = rbind(point.est, boot.se)
     rownames(res) = c("est", "se")
     colnames(res) = c("reg", "HT", "Hajek", "DR")
     
     return(res)
}
\end{lstlisting}

\subsection{Simulation}
\label{sec::dr-simulation}

I will use simulation to evaluate the finite-sample properties of the estimators under four scenarios:
\begin{enumerate}
\item\label{case::11}
both the propensity score and outcome models are correct;
\item\label{case::01}
the propensity score model is wrong but the outcome model is correct;
\item\label{case::10}
the propensity score model is correct but the outcome model is wrong;
\item\label{case::00}
both the propensity score and outcome models are wrong.
\end{enumerate}
I will report the average bias, the true standard error, and the average estimated standard error of the estimators over simulation.

In case \ref{case::11}, the data generating process is
\begin{lstlisting}
  x       = matrix(rnorm(n*2), n, 2)
  x1      = cbind(1, x)
  beta.z  = c(0, 1, 1)
  pscore  = 1/(1 + exp(- as.vector(x1%*%beta.z)))
  z       = rbinom(n, 1, pscore)
  beta.y1 = c(1, 2, 1)
  beta.y0 = c(1, 2, 1)
  y1      = rnorm(n, x1%*%beta.y1)
  y0      = rnorm(n, x1%*%beta.y0)
  y       = z*y1 + (1 - z)*y0
\end{lstlisting}
In case \ref{case::01}, I modify the propensity score model to be nonlinear:
\begin{lstlisting}
  x1      = cbind(1, x, exp(x))
  beta.z  = c(-1, 0, 0, 1, -1)
  pscore  = 1/(1 + exp(- as.vector(x1%*%beta.z)))
\end{lstlisting}
In case \ref{case::10}, I modify the outcome model to be nonlinear: 
\begin{lstlisting}
  beta.y1 = c(1, 0, 0, 0.2, -0.1)
  beta.y0 = c(1, 0, 0, -0.2, 0.1)
  y1      = rnorm(n, x1%*%beta.y1)
  y0      = rnorm(n, x1%*%beta.y0)
\end{lstlisting}
In case \ref{case::00}, I modify both the propensity score and the outcome model.

We set the sample size to be $n=500$ and generate $500$ independent data sets according to the data-generating processes above. 
In case \ref{case::11}, 
\begin{lstlisting}
          reg   HT Hajek   DR
ave.bias 0.00 0.02  0.03 0.01
true.se  0.11 0.28  0.26 0.13
est.se   0.10 0.25  0.23 0.12
\end{lstlisting}
All estimators are nearly unbiased. The two weighting estimators have larger variances. 
In case \ref{case::01},
\begin{lstlisting}
          reg    HT Hajek    DR
ave.bias 0.00 -0.76 -0.75 -0.01
true.se  0.12  0.59  0.47  0.18
est.se   0.13  0.50  0.38  0.18
\end{lstlisting}
The two weighting estimators are severely biased due to the misspecification of the propensity score model. The outcome regression and doubly robust estimators are nearly unbiased.

In case \ref{case::10},
\begin{lstlisting}
           reg   HT Hajek   DR
ave.bias -0.05 0.00 -0.01 0.00
true.se   0.11 0.15  0.14 0.14
est.se    0.11 0.14  0.13 0.14
\end{lstlisting}
The outcome regression estimator has a larger bias than the other three estimators due to the misspecification of the outcome model. The weighting and doubly robust estimators are nearly unbiased.

In case \ref{case::00},
\begin{lstlisting}
           reg   HT Hajek   DR
ave.bias -0.08 0.11 -0.07 0.16
true.se   0.13 0.32  0.20 0.41
est.se    0.13 0.25  0.16 0.26
\end{lstlisting}
All estimators are biased because both the propensity score and outcome models are wrong. The HT and doubly robust estimator has the largest bias. When both models are wrong, the doubly robust estimator appears to be doubly fragile.

In all the cases above, the bootstrap standard errors are close to the true ones when the estimators are nearly unbiased for the true average causal effect.

\subsection{Applications}
\label{sec::ATE-application}

Revisiting Example \ref{eg::chan-bmi-ATE}, we obtain the following estimators and bootstrap standard errors:
\begin{lstlisting}
       reg     HT  Hajek     DR
est -0.017 -1.516 -0.156 -0.019
se   0.230  0.492  0.246  0.233
\end{lstlisting}
The two weighting estimators are much larger than the other two estimators. Truncating the estimated propensity score at $[0.1, 0.9]$, we obtain the following estimators and bootstrap standard errors:
\begin{lstlisting}
       reg     HT  Hajek     DR
est -0.017 -0.713 -0.054 -0.043
se   0.223  0.422  0.235  0.231
\end{lstlisting}
The Hajek estimator becomes much closer to the outcome regression and doubly robust estimators, while the Horvitz--Thompson estimator is still an outlier.

\section{Some further discussion}

Recall the proof of Theorem \ref{thm::doublyrobust}, the key for the double robustness property is the product structure in 
$$
\tilde{\mu}_1^\textup{dr} - E\{  Y(1) \} 
= E\left[   \frac{e(X) - e(X,\alpha) }{  e(X,\alpha) }  \times  \left\{  \mu_1(X) -  \mu_1(X, \beta_1)   \right\} \right] ,
$$
which ensures that the estimation error is zero if either $e(X) = e(X,\alpha) $ or $ \mu_1(X) =  \mu_1(X, \beta_1) $. This delicate structure renders the doubly robust estimator possibly doubly fragile when both the propensity score and the outcome models are misspecified. The product of two errors multiply to yield potentially much larger errors. 
The simulation in Chapter \ref{sec::dr-simulation} confirms this point.

\citet{kang2007demystifying} criticized the doubly robust estimator based on simulation studies. They found that the finite-sample performance of the doubly robust estimator can be even wilder than the simple outcome regression and IPW estimators. 
Despite the critique from \citet{kang2007demystifying},  the doubly robust estimator has been a standard strategy in causal inference since the seminal work of \citet{scharfstein1999adjusting}. Recently, it was resurrected in the theoretical statistics and econometrics literature with a fancier name ``double machine learning'' \citep{chernozhukov2018double}. The basic idea is to replace the working models for the propensity score and outcome with machine learning tools which can be viewed as more flexible models than the traditional parametric models.

\section{Homework problems}

 \paragraph{A sanity check}\label{hw::discreteX-obs}
Consider the case in which the covariate is discrete $X \in \{1, \ldots, K\}$ and the parameter of interest is $\tau$. Without imposing any model assumptions, the estimated propensity score $\hat{e}(X)$ equals $\hat{e}_{[k]} = \hat{\pr}(Z=1\mid X=k) $, the proportion of units receiving the treatment, and the estimated outcome means are the sample means of the outcomes $\hat{\bar{Y}}_{[k]}(1)  = \hat{E}(Y\mid Z=1, X=k) $ and $\hat{\bar{Y}}_{[k]}(0)  = \hat{E}(Y\mid Z=0, X=k) $ under treatment, within stratum $X=k\ (k=1, \ldots, K)$. Show that the stratified estimator, outcome regression estimator, HT estimator, Hajek estimator, and doubly robust estimator are all identical numerically.

\paragraph{An alternative form of the doubly robust estimator for $\tau$}\label{hw::alternative-dr2-robins}

Motivated by \eqref{eq::explain-ipw-efficiency-basic}, we have an alternative form of the doubly robust estimator for $\mu_1 = E\{ Y(1) \} $: 
$$
\tilde{\mu}_1^\text{dr2} = 
\frac{ E\left[   \frac{Z\{ Y-\mu_1(X,\beta_1) \}}{  e(X,\alpha) } \right] }{E\left[   \frac{Z }{  e(X,\alpha) } \right]} 
 + E\{ \mu_1(X, \beta_1)  \} .
$$
Show that $\tilde{\mu}_1^\text{dr2} = \mu_1$ if either $e(X,\alpha)  = e(X)$ or $\mu_1(X, \beta_1) =  \mu_1(X)$. Give the analogous formula for estimating $\mu_0$. Give the sample analog of the doubly robust estimator for $\tau$ based on these formulas. 

Remark: 
This form of doubly robust estimator appeared in \citet{robins2007comment}.

\paragraph{An upper bound of the bias of the doubly robust estimator}\label{hw::upperbound-dr}

Consider the population version of the doubly robust estimator $\tilde{\mu}_1^\textup{dr} $ for $E\{  Y(1) \} $. 
Show that
$$
 |   \tilde{\mu}_1^\textup{dr} - E\{  Y(1) \}  |  \leq
\sqrt{
E\left[   \frac{ \{  e(X) - e(X,\alpha)\}^2  }{  e(X,\alpha)^2 }  \right] \times 
E\left[  \left\{  \mu_1(X) -  \mu_1(X, \beta_1)   \right\}^2 \right] .
}
$$
Find the analogous upper bound for the bias of $\tilde{\mu}_0^\textup{dr} $ for $E\{  Y(0) \} $.

Remark: You may find Section \ref{sec::cauchy-schwarz-inequality} useful for the proof.

\paragraph{Data analysis of Example \ref{eg::job-training-obs}}

Analyze the dataset \ri{cps1re74.csv} using the methods discussed so far.

\paragraph{Analyzing a dataset from the Karolinska Institute}\label{hw::Karolinska-dr}
 
 \citet{rubin2008objective} used the dataset \texttt{karolinska.txt} to illustrate the ideas of causal inference in observational studies. 
 The dataset has 158 cardia cancer patients diagnosed between 1988 and 1995 in Central and Northern Sweden, 79 diagnosed at large volume hospitals, defined as treating more than ten patients with cardia cancer during that period, and 79 diagnosed at the remaining small volume hospitals. The treatment \texttt{z}  is the indicator of whether a patient was diagnosed at a large volume hospital. The outcome \texttt{y} is whether the patient survived longer than 1 year after the diagnosis. The covariates \texttt{x} contain information about age, whether a patient was from a rural area, and whether a patient was male. 
 
\begin{lstlisting}
karolinska = read.table("karolinska.txt", header = TRUE)
z = karolinska$hvdiag
y = 1 - (karolinska$year.survival == 1)
x = as.matrix(karolinska[, c(3, 4, 5)])
\end{lstlisting}

Analyze the dataset using the methods discussed so far.

 \paragraph{Recommended reading}
 
 \citet{lunceford2004stratification} gave a review and comparison of many methods discussed in Chapters \ref{chapter::pscore-key} and \ref{chapter::doubly-robust}.

\chapter{The Average Causal Effect on the Treated Units and Other Estimands}
 \label{chapter::ATT-and-other}

Chapters \ref{chapter::observational-studies}--\ref{chapter::doubly-robust} focused on the identification and estimation of the average causal effect $\tau = E\{  Y(1) - Y(0) \}$ under the ignorability and overlap assumptions. Conceptually, it is straightforward to extend the discussion to the average causal effects on the treated and control units:
\begin{eqnarray*}
\tau_\textsc{T}  &=&  E\{  Y(1) - Y(0) \mid Z=1 \} ,  \\
\tau_\textsc{C}  &=&  E\{  Y(1) - Y(0) \mid Z=0 \} . 
\end{eqnarray*}
If $\tau_\textsc{T}$ and $\tau_\textsc{C}$ differ from $\tau$, then the average causal effects are heterogeneous across the treatment and control groups. Whether we should estimate $\tau_\textsc{T}$, $\tau_\textsc{C}$ or $\tau$ depends on the practical question of interest.

Because of the symmetry, this chapter focuses on $\tau_\textsc{T} $. Chapter \ref{sec::other-estmand-overlap} also discusses extensions to other estimands.

\section{Nonparametric identification of $\tau_\textsc{T} $ }

The average causal effect on the treated units equals 
$$
\tau_\textsc{T} = E(Y\mid Z=1) -  E\{    Y(0) \mid Z=1 \},
$$
where the first term $E(Y\mid Z=1) $ is directly identifiable from the data and the second term $E\{    Y(0) \mid Z=1 \}$ is counterfactual. The key assumption to identify the second term is the following ignorability and overlap assumptions.

\begin{assumption}
\label{assume::ignorability-y0}
$Z\ind Y(0) \mid X$ and $e(X) < 1$. 
\end{assumption}

Because the key is to identify $E\{    Y(0) \mid Z=1 \}$, we only need the ``one-sided'' ignorability and overlap assumptions. Under Assumption \ref{assume::ignorability-y0}, we have the following identification result for $\tau_\textsc{T} $.

\begin{theorem}
\label{thm::identification-ATT}
Under Assumption \ref{assume::ignorability-y0}, we have 
\begin{eqnarray*}
 E\{    Y(0) \mid Z=1 \}  &=& E\left\{  E(    Y \mid Z=0 , X) \mid  Z=1  \right\} \\
 &=&  \int  E(    Y \mid Z=0 , X=x) f(  x \mid Z=1) \diff x .
\end{eqnarray*}
\end{theorem}

By Theorem \ref{thm::identification-ATT} the counterfactual mean $ E\{    Y(0) \mid Z=1 \}  $ equals the conditional mean of the observed outcomes under the control, averaged over the distribution of the covariates under the treatment.
It implies that $\tau_\textsc{T} $ is nonparametrically  identified by
\begin{equation}
\tau_\textsc{T}  = E(Y\mid Z=1) -E\left\{  E(    Y \mid Z=0 , X) \mid  Z=1  \right\}  \label{eq::att-identification} 
\end{equation}

\begin{myproof}{Theorem}{\ref{thm::identification-ATT}} 
We have 
\begin{eqnarray*}
 E\{    Y(0) \mid Z=1 \} &=&  E\big[   E\{    Y(0) \mid Z=1 , X\} \mid  Z=1  \big] \\
 &=& E\big[   E\{    Y(0) \mid Z=0 , X\} \mid  Z=1  \big] \\
 &=& E\left\{  E(    Y \mid Z=0 , X) \mid  Z=1  \right\} \\
 &=& \int  E(    Y \mid Z=0 , X=x) f(  x \mid Z=1)  \diff x.
\end{eqnarray*}
\end{myproof}

With a discrete $X$, the identification formula in Theorem \ref{thm::identification-ATT} reduces to 
$$
 E\{    Y(0) \mid Z=1 \}  = \sum_{k=1}^K E(    Y \mid Z=0 , X=k) \pr(X=k \mid  Z=1  ),
$$
%
motivating the following stratified estimator for $\tau_\textsc{T} $:
$$
\hat{\tau}_\textsc{T} =   \hat{\bar{Y}}(1) - \sum_{k=1}^K \hat{\pi}_{[k]\mid 1} \hat{\bar{Y}}_{[k]}(0),
$$
where $\hat{\pi}_{[k]\mid 1} = n_{[k]1}/n_1$ is the proportion of category $k$ of $X$ among the treated units.

For continuous $X$, we need to fit an outcome model for $E(Y\mid Z=0, X)$ using the control units. If the fitted values for the control potential outcomes are $\hat{\mu}_0(X_i)$, then the outcome regression estimator is
$$
\hat{\tau}_\textsc{T}^\text{reg} =  \hat{\bar{Y}}(1) - n_1^{-1} \sumn Z_i \hat{\mu}_0(X_i) = n_1^{-1} \sumn Z_i  \{  Y_i - \hat{\mu}_0(X_i)  \}. 
$$

\begin{example}\label{eg::att-regression1}
If we specify a linear model for all units
$$
E(Y\mid Z, X) = \beta_0  + \beta_z  Z + \beta_x\tran X,
$$
then 
\begin{eqnarray*}
\tau_\textsc{T} &=& E\{   E(Y\mid Z=1, X)  -  E(    Y \mid Z=0 , X)  \mid Z=1\} \\
&=&  \beta_z.
\end{eqnarray*}
If we run OLS to obtain $( \hat{\beta}_0, \hat{\beta}_z,  \hat{\beta}_x )$, then we can use $\hat{\beta}_z$ to estimate $\tau_\textsc{T} $. 
%
%
Section \ref{sec::outcome-regression} shows that $\hat{\beta}_z$ is an estimator for $\tau$, and this example further shows that $\hat{\beta}_z$ is an estimator for $\tau_\textsc{T}$. This is not surprising because the linear model assumes constant causal effects across units. 
\end{example}

\begin{example}\label{eg::att-regression2}
The identification formula depends only on $E(Y\mid Z=0, X)$, so we need only to specify a model for the control units. When this model is linear, 
$$
E(Y\mid Z=0, X) =\beta_{0\mid 0}   + \beta_{x\mid 0} \tran X,
$$
we have 
\begin{eqnarray*}
\tau_\textsc{T} &=& E(Y\mid Z=1) - E(   \beta_{0\mid 0}   + \beta_{x\mid 0} \tran X  \mid Z=1 ) \\
&=& E(Y\mid Z=1) -   \beta_{0\mid 0}    -  \beta_{x\mid 0}   \tran E(X\mid Z=1). 
\end{eqnarray*}
If we run OLS with only the control units to obtain $( \hat{\beta}_{0\mid 0} ,  \hat{\beta}_{x\mid 0})$, then the estimator is
$$
\hat{\tau}_\textsc{T} = \hat{\bar{Y}}(1) -\hat{\beta}_{0\mid 0}  - \hat{\beta}_{x\mid 0}  \tran \hat{\bar{X}}(1).
$$
Using the property of the OLS (see \ref{eq::OLS-mean-data}), we have 
$$
\hat{\bar{Y}}(0) = \hat{\beta}_{0\mid 0} + \hat{\beta}_{x\mid 0}  \tran \hat{\bar{X}}(0).
$$ 
Therefore, the above estimator reduces to
$$
\hat{\tau}_\textsc{T} = \left\{  \hat{\bar{Y}}(1) - \hat{\bar{Y}}(0) \right\} - \hat{\beta}_{x\mid 0}  \tran \left\{  \hat{\bar{X}}(1) -  \hat{\bar{X}}(0) \right\}, 
$$
which is similar to \eqref{eq::ATT--regression1} with a different coefficient for the difference in means of the covariates. 

As an algebraic fact, we can show that this estimator equals the coefficient of $Z$ in the OLS fit of the outcome on the treatment, covariates, and their interactions, with the covariates centered as $ X_i  - \hat{\bar{X}}(1)$. See Problem \ref{para::regression-att-algebra} for more details.  
\end{example}

\section{Inverse propensity score weighting and doubly robust estimation of $\tau_\textsc{T} $}

\begin{theorem}
\label{thm::ipw-att}
Under Assumption \ref{assume::ignorability-y0}, we have
\begin{eqnarray}
E\{ Y(0) \mid Z=1 \} = E\left\{         \frac{e(X)}{e}  \frac{1-Z}{1-e(X)} Y \right\}
\label{eq::ipw-att}
\end{eqnarray}
and
\begin{eqnarray}\label{eq::att-ipw-formula}
\tau_\textsc{T} = E(Y\mid Z=1) -  E\left\{         \frac{e(X)}{e}  \frac{1-Z}{1-e(X)} Y \right\} ,
\end{eqnarray}
where $e = \pr(Z=1)$ is the marginal probability of the treatment. 
\end{theorem}

\begin{myproof}{Theorem}{\ref{thm::ipw-att}} 
The left-hand side of \eqref{eq::ipw-att} equals
\begin{eqnarray*}
E\{ Y(0) \mid Z=1 \} &=& E\{ ZY(0)   \} /e \\
&=& E\big[  E( Z \mid X) E\{ Y(0) \mid X  \} \big] /e \\
&=&E\big[ e(X)  E\{ Y(0) \mid X  \} \big] /e . 
\end{eqnarray*}
The right-hand side of \eqref{eq::ipw-att} equals 
\begin{eqnarray*}
E\left\{         \frac{e(X)}{e}  \frac{1-Z}{1-e(X)} Y \right\} 
&=& E\left[  E\left\{         \frac{e(X)}{e}  \frac{1-Z}{1-e(X)} Y(0) \mid X \right\}  \right]\\
&=& E\left[   \frac{e(X)}{e \{1-e(X)\} }   E\left\{       (1-Z)Y(0) \mid X \right\}  \right]\\
&=& E\left[   \frac{e(X)}{e \{1-e(X)\} }   E(      1-Z\mid X )   E\left\{       Y(0) \mid X \right\}  \right]\\
&=& E\big[ e(X)  E\{ Y(0) \mid X  \} \big] /e .  
\end{eqnarray*}
So \eqref{eq::ipw-att}  holds. 
\end{myproof}

We have two  IPW  estimators  
$$
\hat{\tau}_\textsc{T}^\text{ht} =  \hat{\bar{Y}}(1) - n_1^{-1} \sumn \hat{o}(X_i) (1-Z_i) Y_i
$$
and
$$
\hat{\tau}_\textsc{T}^\text{hajek} = \hat{\bar{Y}}(1) 
- \frac{  \sumn \hat{o}(X_i) (1-Z_i) Y_i }{ \sumn \hat{o}(X_i) (1-Z_i) } ,
$$
where $\hat{o}(X_i) = \hat{e}(X_i) / \{ 1 - \hat{e}(X_i)  \} $ is the fitted odds of the treatment given covariates.

We also have a doubly robust estimator for $E\{  Y(0)\mid Z=1\}$ which combines the propensity score and the outcome models. Define
\begin{eqnarray}
\label{eq::dr-att-mu0}
\tilde{\mu}_{0\textsc{T}}^\textup{dr} = E\left[ o(X,\alpha)  (1-Z) \{ Y - \mu_0(X, \beta_0) \}     + Z  \mu_0(X, \beta_0) \right] /e, 
\end{eqnarray}
where $o(X,\alpha)  = e(X,\alpha)/ \{ 1 - e(X,\alpha)\}.$

\begin{theorem}
\label{thm::dr-att}
Under Assumption \ref{assume::ignorability-y0}, if either $ e(X,\alpha) = e(X)$ or $\mu_0(X, \beta_0)=\mu_0(X) $, then $\mu_{0\textsc{T}}^\text{dr}  = E\{ Y(0) \mid Z=1 \}.$
\end{theorem}

\begin{myproof}{Theorem}{\ref{thm::dr-att}}
We have the decomposition
\begin{eqnarray*}
&&  e\left[  \tilde{\mu}_{0\textsc{T}}^\text{dr}  -  E\{ Y(0) \mid Z=1 \} \right] \\
&=& E\left[ o(X,\alpha) ( 1-Z )  \{ Y(0) - \mu_0(X, \beta_0) \} + Z  \mu_0(X, \beta_0) \right] -  E\{ Z Y(0)   \} \\
&=& E\left[ o(X,\alpha) ( 1-Z ) \{ Y(0) - \mu_0(X, \beta_0) \} - Z  \{ Y (0) - \mu_0(X, \beta_0) \} \right] \\
&=& E\left[ \left\{ o(X,\alpha)  ( 1-Z  ) -Z\right\} \{ Y (0) - \mu_0(X, \beta_0) \}   \right] \\
&=& E\left[    \frac{ e(X,\alpha) - Z  }{1-  e(X,\alpha)}   \{ Y (0) - \mu_0(X, \beta_0) \}   \right] \\
&=& E\left[  E\left\{   \frac{ e(X,\alpha) - Z  }{1-  e(X,\alpha)}  \mid X \right\}  \times E  \{ Y (0) - \mu_0(X, \beta_0)  \mid X\}   \right] \\
&=& E\left[   \frac{ e(X,\alpha) - e(X)  }{1-  e(X,\alpha)} \times    \{ \mu_0(X) - \mu_0(X, \beta_0)  \}   \right] .
\end{eqnarray*}
Therefore, $\tilde{\mu}_{0\textsc{T}}^\text{dr}  -  E\{ Y(0) \mid Z=1 \}  = 0$
 if either $ e(X,\alpha) = e(X)$ or $\mu_0(X, \beta_0)=\mu_0(X) $. 
\end{myproof}

Based on the population versions of $\tilde{\mu}_{0\textsc{T}}^\text{dr}$ in \eqref{eq::dr-att-mu0}, we can obtain its sample version to construct a doubly robust estimator for $\tau_\textsc{T}$. 

\begin{definition}
[doubly robust estimator for $\tau_\textsc{T}$]\label{def::dr-att}
Based on the data $(X_i, Z_i, Y_i)_{i=1}^n$, we can obtain a doubly robust estimator for $\tau_\textsc{T}$ by the following steps: 
\begin{enumerate}
[(1)]
\item
obtain the fitted values of the propensity scores $e(X_i, \hat{\alpha})$ and then obtain the fitted values of the odds $o(X_i, \hat{\alpha})  =  e(X_i, \hat{\alpha}) / (1- e(X_i, \hat{\alpha}))$; 
\item
obtain the fitted values of the outcome mean under control $ \mu_0(X_i, \hat{\beta}_0) $; 
\item
construct the doubly robust estimator: $\hat{\tau}_\textsc{T}^\textup{dr} = \hat{\bar{Y}}(1) - \hat{\mu}_{0\textsc{T}}^\textup{dr}$, where
$$
 \hat{\mu}_{0\textsc{T}}^\textup{dr}  
 = \frac{1}{n_1} \sumn \left[ o(X_i,\hat{\alpha}) (1-Z_i)\{ Y_i-\mu_0(X_i, \hat{\beta}_0) \}      +Z_i \mu_0(X_i, \hat{\beta}_0)       \right].
$$
\end{enumerate}
\end{definition}

By Definition \ref{def::dr-att}, we can rewrite $\hat{\tau}_\textsc{T}^\textup{dr}$ as
$$
e(X_i, \hat{\alpha})
= \hat{\tau}_\textsc{T}^\textup{reg}
 - \frac{1}{n_1} \sumn  o(X_i,\hat{\alpha})  (1-Z_i)\{ Y_i-\mu_0(X_i, \hat{\beta}_0) \}    
$$
or
$$
e(X_i, \hat{\alpha})
= \hat{\tau}_\textsc{T}^\textup{ht} 
- \frac{1}{n_1} \sumn  \{ o(X_i,\hat{\alpha}) (1-Z_i) + Z_i  \} \mu_0(X_i, \hat{\beta}_0)  .
$$
Similar to the discussion of $\hat{\tau}^\textup{dr}  $, we can  estimate the  variance of $\hat{\tau}_\textsc{T}^\textup{dr}  $  via the bootstrap by resampling from $(Z_i, X_i, Y_i)_{i=1}^n$. 
\citet{hahn1998role}, \citet{mercatanti2014debit}, \citet{shinozaki2015brief} and \citet{yang2018asymptotic} are references on the estimation of $\tau_\textsc{T}$.

\section{An example}
\label{sec::dr-att-example}

The following \ri{R} function implements two outcome regression estimators, two IPW estimators, and the doubly robust estimator for $\tau_\textsc{T}$. To avoid extreme estimated propensity scores, we can also truncate them from the above. 

\begin{lstlisting}
ATT.est = function(z, y, x, out.family = gaussian, Utruncps = 1)
{
  ## sample size
  nn  = length(z)
  nn1 = sum(z)
  
  ## fitted propensity score
  pscore   = glm(z ~ x, family = binomial)$fitted.values
  pscore   = pmin(Utruncps, pscore)
  odds     = pscore/(1 - pscore)
  
  ## fitted potential outcomes
  outcome0 = glm(y ~ x, weights = (1 - z), 
                 family = out.family)$fitted.values
  
  ## outcome regression estimator
  ace.reg0 = lm(y ~ z + x)$coef[2]
  ace.reg  = mean(y[z==1]) - mean(outcome0[z==1]) 
  ## propensity score weighting estimator
  ace.ipw0 = mean(y[z==1]) - 
                mean(odds*(1 - z)*y)*nn/nn1
  ace.ipw  = mean(y[z==1]) - 
                mean(odds*(1 - z)*y)/mean(odds*(1 - z))
  ## doubly robust estimator
  res0     = y - outcome0
  ace.dr   = ace.reg - mean(odds*(1 - z)*res0)*nn/nn1
  
  return(c(ace.reg0, ace.reg, ace.ipw0, ace.ipw, ace.dr))     
}
\end{lstlisting}

The following \ri{R} function further implements the bootstrap variance estimators. 

\begin{lstlisting}
OS_ATT = function(z, y, x, n.boot = 10^2,
                  out.family = gaussian, Utruncps = 1)
{
  point.est  = ATT.est(z, y, x, out.family, Utruncps)
  
  ## nonparametric bootstrap
  n   = length(z)
  x          = as.matrix(x)
  boot.est   = replicate(n.boot, {
    id.boot = sample(1:n, n, replace = TRUE)
    ATT.est(z[id.boot], y[id.boot], x[id.boot, ], 
            out.family, Utruncps)
  })
  
  boot.se    = apply(boot.est, 1, sd)
  
  res        = rbind(point.est, boot.se)
  rownames(res) = c("est", "se")
  colnames(res) = c("reg0", "reg", "HT", "Hajek", "DR")
  
  return(res)
}
\end{lstlisting}

Now we re-analyze the data in Example \ref{eg::chan-bmi-ATE} to estimate $\tau_\textsc{T}$. 
We obtain
\begin{lstlisting}
     reg0    reg     HT  Hajek     DR
est 0.061 -0.351 -1.992 -0.351 -0.187
se  0.227  0.258  0.705  0.328  0.287
\end{lstlisting}
without truncating the estimated propensity scores, and
\begin{lstlisting}
     reg0    reg     HT  Hajek     DR
est 0.061 -0.351 -0.597 -0.192 -0.230
se  0.223  0.255  0.579  0.302  0.276
\end{lstlisting}
by truncating the estimated propensity scores from the above at $0.9$. The HT estimator is sensitive to the truncation as expected. The regression estimator in Example \ref{eg::att-regression1} is quite different from other estimators. It imposes an unnecessary assumption that the regression functions in the treatment and control group share the same coefficient of $X$. The regression estimator in Example \ref{eg::att-regression2} is much closer to the Hajek and doubly robust estimators. The estimates above are slightly different from those in Section \ref{sec::ATE-application}, suggesting the existence of treatment effect heterogeneity across $\tau_\textsc{T}$ and $\tau$.

\section{Other estimands}\label{sec::other-estmand-overlap}

\citet{li2018balancing} gave a unified discussion of the causal estimands in observational studies. Starting from the conditional average causal effect $\tau(X)$, they proposed a general class of estimands
$$
\tau^h =  \frac{  E\{ h(X) \tau(X)  \} }{ E\{ h(X)   \} }
$$
indexed by a weighting function $h(X)$ with $E\{ h(X)   \}  \neq 0$. The normalization in the denominator is to ensure that a constant causal effect $\tau(X) = \tau$ averages to the same $\tau$.

Under ignorability, 
$$
\tau^h =  \frac{  E[  h(X)   \{ \mu_1(X) - \mu_0(X) \}   ] }{ E\{ h(X)   \} }
$$
which 
motivates the outcome regression estimator
$$
\hat \tau^h =  \frac{   \sumn h(X_i)  \{ \hat{\mu}_1(X_i)   -  \hat{\mu}_0(X_i) \}  }{  \sumn h(X_i) } .
$$
Moreover, we can show that $\tau^h$ has the following weighting form:

\begin{theorem}\label{thm::weighted-estimand-li}
Under the ignorability and overlap assumption, we have
$$
\tau^h = E\left\{ \frac{ZYh(X)}{ e(X)}  - \frac{(1-Z)Yh(X)}{1-e(X)}  \right\} / E\{ h(X)   \} .
$$
\end{theorem}

The proof of Theorem \ref{thm::weighted-estimand-li} is similar to those of Theorems \ref{thm::pscore-weighting} and \ref{thm::ipw-att} which is relegated to Problem \ref{problem::general-estimand-li}. Based on Theorem \ref{thm::weighted-estimand-li}, we can construct the corresponding IPW estimator for $\tau^h$.

By Theorem \ref{thm::weighted-estimand-li}, each unit is associated with the weight due to the definition of the estimand as well as the weight due to the inverse of the propensity score. Finally, the treated units are weighted by $h(X) / e(X)$ and the control units are weighted by $h(X)/ \{  1-e(X) \}.$ \citet[][Table 1]{li2018balancing} summarized several estimands, and I present a part of it below: 

\begin{tabular}{cccc}
\hline 
 population & $h(X)$ & estimand & weights \\ \hline 
combined & $1$ & $\tau$ & $1/e(X)$ and $1/\{  1-e(X) \} $  \\
treated & $e(X)$ & $\tau_\textsc{T}$ & 1 and $ e(X)/\{  1-e(X) \}$ \\
control & $1-e(X)$ & $\tau_\textsc{C}$ & $\{  1-e(X) \} /e(X) $ and 1\\
overlap & $e(X)\{  1-e(X) \} $ &$\tau_\textsc{O}$  &  $ 1-e(X)$ and $e(X) $ \\
\hline 
\end{tabular}

The overlap population and the corresponding estimand
$$
\tau_\textsc{O} = \frac{   E\left[ e(X)\{  1-e(X) \}   \tau(X) \right] }{   E\left[ e(X)\{  1-e(X) \}    \right] }
$$
is new to us. This estimand has the largest weight for units with $e(X) = 1/2$ and down weights the units with extreme propensity scores. 
A nice feature of this estimand is that its IPW estimator is stable without the possibly extremely small values of $e(X)$ and $1-e(X) $ in the denominator. If $ e(X) \ind  \tau(X) $ including the special case of $\tau(X) = \tau$, the parameter $\tau_\textsc{O} $ reduces to $\tau$. In general, however, the estimand $\tau_\textsc{O} $ may cause controversy because it changes the initial population and depends on the propensity score which may be misspecified in practice. \citet{li2018balancing} and \citet{li2019addressing} gave some justifications and numerical evidence.
This estimand will appear again in Chapter \ref{chapter::pscore-unification}.

We can also construct the doubly robust estimator for $\tau^h$. I relegate the details to Problem \ref{problem::general-estimand-dr}.

 \section{Homework Problems}

\paragraph{Comparing $\tau_\textsc{T},  \tau_\textsc{C}, $ and $\tau$}\label{hw::difference-tau-T-tau} 
 
Assume $Z\ind \{Y(1), Y(0)\} \mid X$.  Recall $e(X) = \pr(Z=1\mid X)$ is the propensity score,  $e = \pr(Z=1)$ is the marginal probability of the treatment, and $\tau(X) = E\{  Y(1)-Y(0) \mid X \}$ is the CATE. 
Show that 
$$
\tau_\textsc{T} - \tau = \frac{  \cov\{ e(X) ,  \tau(X)  \} }{e},\quad
\tau_\textsc{C} - \tau = - \frac{  \cov\{ e(X) ,  \tau(X)  \} }{1-e}.
$$

Remark: The results above also imply that $ \tau_\textsc{T} -  \tau_\textsc{C} =  \cov\{ e(X) ,  \tau(X)  \} / \{  e(1-e) \}$.  So the differences among $ \tau_\textsc{T} ,   \tau_\textsc{C} , \tau$ depend on the covariance between the propensity score and CATE.   
When $e(X) $ and $  \tau(X) $ are uncorrelated, their differences are 0.

\paragraph{An algebraic fact about a regression estimator for $\tau_\textsc{T}$} 
 \label{para::regression-att-algebra}
 
 This problem provides more details for Example \ref{eg::att-regression2}. 
 
Show that if we center the covariates by $X_i - \hat{\bar{X}}(1)$ for all units, then $\hat{\tau}_\textsc{T} $ equals the coefficient of $Z$ in the OLS fit of the outcome on the intercept, the treatment, covariates, and their interactions.

 \paragraph{Simulation for the average causal effect on the treated units}

Chapter \ref{chapter::doubly-robust} ran some simulation studies for $\tau$. Run similar simulation studies for $\tau_\textsc{T}$ with either correct or incorrect propensity score or outcome models. 

You can choose different model parameters, a larger number of simulation settings, and a larger number of bootstrap replicates. Report your findings, including at least the bias, variance, and variance estimator via the bootstrap. You can also report other properties of the estimators, for example, the asymptotic Normality and the coverage rates of the confidence intervals.

\paragraph{An alternative form of the doubly robust estimator for $\tau_\textsc{T}$}\label{hw::alternative-dr2-robins-att}

Motivated by \eqref{eq::dr-att-mu0}, we have an alternative form of doubly robust estimator for $ E\{ Y(0) \mid Z=1 \}  $: 
$$
\tilde{\mu}_{0\textsc{T}}^\text{dr2} = 
\frac{  E\left[ o(X,\alpha)   (1-Z) \{ Y - \mu_0(X, \beta_0) \}  \right] }{  E\left[ o(X,\alpha)   (1-Z) \right] } 
  + E\{  Z  \mu_0(X, \beta_0) \} /e.
$$

Show that under Assumption \ref{assume::ignorability-y0}, $\tilde{\mu}_{0\textsc{T}}^\text{dr2}= E\{ Y(0) \mid Z=1 \}$ if either $e(X,\alpha)  = e(X)$ or $\mu_0(X, \beta_0) =  \mu_0(X)$.  Give the sample analog of the doubly robust estimator for $\tau_\textsc{T}$.

 \paragraph{Average causal effect on the control units}\label{hw::formulas-atc}

Prove the identification formulas for $\tau_\textsc{C}$, analogous to \eqref{eq::att-identification} and \eqref{eq::att-ipw-formula}. Propose the doubly robust estimator for $\tau_\textsc{C}$.

\paragraph{Estimating individual effect and conditional average causal effect}\label{hw::individual-effect-prediction}

Assume that $\{ Z_i, X_i, Y_i(1), Y_i(0) \}_{i=1}^n \iidsim \{ Z, X, Y(1), Y(0) \}$. The individual effect is $\tau_i = Y_i(1) - Y_i(0)$ and the conditional average causal effect is $\tau(X_i) = E\{ Y_i(1) - Y_i(0) \mid X_i  \}$. Since we will discuss the individual effect, we do not drop the subscript $i$ since $\tau$ means the average causal effect, not the population version of $Y(1) - Y(0)$. 

\begin{enumerate}
\item
Under randomization with $Z_i\ind \{ Y_i(1), Y_i(0)  \}$ and $e = \pr(Z_i = 1)$, show that
$$
\delta_i = \frac{Z_i Y_i}{e} - \frac{(1-Z_i)Y_i}{1-e}
$$
is an unbiased predictor of the individual effect in the sense that
$$
E( \delta_i  -  \tau_i ) = 0\quad (i=1,\ldots, n).
$$
Further show that $E( \delta_i ) = \tau $ for all $i=1,\ldots, n$. 

\item
Under ignorability with $Z_i\ind \{ Y_i(1), Y_i(0)  \}\mid X_i$ and $e(X_i) = \pr(Z_i=1\mid X_i)$, show that 
$$
\delta_i = \frac{Z_i Y_i}{ e(X_i) } - \frac{(1-Z_i)Y_i}{1-e(X_i)}
$$
is an unbiased predictor of the individual effect and the conditional average causal effect in the sense that
$$
E( \delta_i  -  \tau_i ) = 0, \quad
E\{  \delta_i  -  \tau(X_i) \} = 0, \quad (i=1,\ldots, n).
$$
Further show that $E( \delta_i ) = \tau $ for all $i=1,\ldots, n$. 
\end{enumerate}

\paragraph{General estimand and $(\tau_\textsc{T}, \tau_\textsc{C})$}\label{problem::general-estimand-att-atc}

Assume unconfoundedness.
Show that $\tau^h = \tau_\textsc{T}$ if $h(X) = e(X)$, and $\tau^h = \tau_\textsc{C}$ if $h(X) = 1-  e(X)$.

\paragraph{More on $\tau_\textsc{O}$}\label{problem::ATO-treated-control}

Show that
$$
\tau_\textsc{O} 
=  \frac{   E  [   \{1-e(X)\} \tau(X) \mid Z=1   ]      }{  E  \{ 1-e(X) \mid Z=1   \}   }
=  \frac{   E  \{ e(X) \tau(X) \mid Z=0   \}      }{  E  \{ e(X) \mid Z=0   \}   }.
$$

\paragraph{IPW for the general estimand}\label{problem::general-estimand-li}

Prove Theorem \ref{thm::weighted-estimand-li}.

\paragraph{Doubly robust estimation for general estimand}\label{problem::general-estimand-dr}

For a given $h(X)$, we have the following formulas for constructing the doubly robust estimator for $\tau^h$: 
\begin{eqnarray*}
\tilde{\mu}_1^{h,\textup{dr}} &=& E\left[   \frac{Zh(X)  \{ Y-\mu_1(X,\beta_1)   \}}{  e(X,\alpha) }   + h(X) \mu_1(X, \beta_1)  \right] ,  \\
\tilde{\mu}_0^{h, \textup{dr}} &=& E\left[   \frac{(1-Z) h(X)  \{ Y-\mu_0(X,\beta_0) \}}{  1-e(X,\alpha) }   + h(X) \mu_0(X, \beta_0)  \right]  .  
\end{eqnarray*}
Show that under ignorability and overlap,
\begin{enumerate}
\item
if either $e(X,\alpha) = e(X)$ or $\mu_1(X, \beta_1) = \mu_1(X)$, then $\tilde{\mu}_1^{h,\textup{dr}}= E\{  h(X)  Y(1) \}$;
\item
if either $e(X,\alpha) = e(X)$ or $\mu_0(X, \beta_0) = \mu_0(X)$, then $\tilde{\mu}_0^{h, \textup{dr}}= E\{ h(X)  Y(0) \}$;
\item 
if either $e(X,\alpha) = e(X)$ or $\{\mu_1(X, \beta_1) = \mu_1(X),   \mu_0(X, \beta_0) = \mu_0(X)\}$, then 
$$
\frac{\tilde{\mu}_1^{h,\textup{dr}}  -\tilde{\mu}_0^{h, \textup{dr}}  }{  E\{ h(X)  \}  }
= \tau^h .
$$
\end{enumerate}

Remark: 
\citet{tao2019doubly} proved the above results. However, they hold only for a  given $h(X)$. The most interesting cases of $\tau_\textsc{T}, \tau_\textsc{C}$ and $\tau_\textsc{O}$ all have weight depending on the propensity score $e(X)$, which must be estimated in the first place. The above formulas do not apply to constructing the doubly robust estimators for $\tau_\textsc{T}$ and $\tau_\textsc{C}$; there does not exist a doubly robust estimator for $\tau_\textsc{O}$.

\paragraph{Analyzing a dataset from the Karolinska Institute}\label{hw::Karolinska-att}

Revisit Problem \ref{hw::Karolinska-dr}. Estimate $\tau_\textsc{T}$ based on the methods introduced in this Chapter.

 \paragraph{Recommended reading}
 
 \citet{shinozaki2015brief} focused on $\tau_\textsc{T}$, and \citet{li2018balancing} discussed general $\tau^h$.

\chapter{Using the Propensity Score in Regressions for Causal Effects}\label{chapter::pscore-unification}

Since \citet{rosenbaum1983central}'s seminal paper, many creative uses of the propensity score have appeared in the literature \citep[e.g.,][]{bang2005doubly, robins2007comment, van2011targeted, vansteelandt2014regression}. This chapter discusses two simple methods to use the propensity score: 
\begin{enumerate}
\item
including the propensity score as a covariate in regressions;
\item
running regressions weighted by the inverse of the propensity score. 
\end{enumerate}
I choose to  focus on these two methods because of the following reasons: 
\begin{enumerate}
\item
they are easy to implement, and involve only standard statistical software packages for regressions;
\item
their properties are comparable to many more complex methods;
\item
they can be easily extended to allow for flexible statistical models including machine learning algorithms.
\end{enumerate}

\section{Regressions with the propensity score as a covariate}
\label{sec::regress-on-pscore}

By Theorem \ref{thm::pscore-dimreduction}, if ignorability holds conditioning on $X$, then it also holds conditioning on $e(X)$: 
$$
Z\ind \{ Y(1) , Y(0)\} \mid e(X). 
$$ 
Analogous to \eqref{eq::nonparametric-identification-tau}, $\tau$ is also nonparametrically identified by
$$
\tau = E\Big[   E\{ Y\mid Z=1, e(X) \} - E\{ Y\mid Z=0, e(X) \} \Big] ,
$$
which motivates methods based on regressions of $Y$ on $Z$ and $e(X)$.

The simplest regression specification is the OLS fit of $Y$ on $\{1, Z, e(X)\}$, with the coefficient of $Z$ as an estimator, denoted by $\tau_{e}$. For simplicity, I will discuss the population OLS:
$$
\arg\min_{a,b,c}  E\{ Y - a - bZ - c  e(X)  \}^2
$$
with $\tau_{e}$ defined as the coefficient of $Z$.
It is consistent for $\tau$ if we have a correct propensity score model and the outcome model is indeed linear in $Z$ and $e(X)$. The more interesting result is that $\tau_{e}$ estimates $\tau_\textsc{O}$ introduced in Chapter \ref{sec::other-estmand-overlap} if we have a correct propensity score model even if the outcome model is completely misspecified. 

\begin{theorem}\label{thm::ols-pscore-tau-o}
If $Z\ind \{  Y(1), Y(0) \} \mid X$, then the coefficient of $Z$ in the population OLS fit of $Y$ on $\{1, Z, e(X)\}$ equals
$$
\tau_e=  \tau_\textsc{O} = \frac{  E\{ h_\textsc{O}(X) \tau(X) \} }{ E\{ h_\textsc{O}(X) \} },
$$
recalling that $ h_\textsc{O}(X)  = e(X) \{ 1- e(X)  \} $ and $\tau(X) = E\{ Y(1) - Y(0) \mid X  \}$. 
\end{theorem}

An unusual feature of Theorem \ref{thm::ols-pscore-tau-o} is that the overlap condition is not needed anymore. Even if some units have propensity score $e(X)$ equaling $0$ or $1$, their associated weight $ e(X) \{ 1- e(X) \}$ is zero so they do not contribute anything to the final parameter $ \tau_\textsc{O} $.

\begin{myproof}{Theorem}{\ref{thm::ols-pscore-tau-o}}
I will use the FWL theorem reviewed in Section \ref{appendix::fwl-theorem} to prove Theorem \ref{thm::ols-pscore-tau-o}. By the FWL theorem, we can obtain $\tau_e$ in two steps:
\begin{enumerate}
\item
we obtain the residual $\tilde{Z}$ from the OLS fit of $Z$ on $\{1, e(X)\}$;
\item
we obtain $\tau_e$ from the OLS fit of $Y$ on $\tilde{Z}$. 
\end{enumerate}


We can use the result on covariance in Chapter \ref{appendix::tower-property} to simplify the coefficient of $e(X)$ in the OLS fit of $Z$ on $\{1, e(X)\}$ is
\begin{eqnarray*}
\frac{ \cov\{Z, e(X)  \}  }{  \var\{ e(X) \} }
&=&  \frac{  E [ \cov\{Z, e(X)  \mid X\}  ] + \cov\{  E(Z\mid X), e(X) \}   }{  \var\{ e(X) \}  } \\
&=& \frac{0 +\var\{ e(X) \}  }{ \var\{ e(X) \} } = 1,
\end{eqnarray*}
so the intercept is $E(Z) - E\{e(X)\} = 0$ and the residual is $\tilde{Z} = Z - e(X)$. This makes sense since $Z - e(X)$ is uncorrelated with any function of $X$. 

Therefore, we can obtain $\tau_e$ from the univariate OLS fit of $Y$ on a centered variable $Z - e(X)$:
$$
\tau_e = 
\frac{ \cov\{  Z-e(X) , Y \}  }{  \var\{ Z - e(X) \} } .
$$
The denominator simplifies to 
\begin{eqnarray*}
\var\{ Z - e(X) \} &=& E\{ Z - e(X)\}^2 \\
&=& E\{ Z + e(X)^2 - 2Ze(X)\} \\
&=&E\{ e(X) + e(X)^2 - 2e(X)^2 \} \\
&=&E\{ h_\textsc{O}(X) \} .
\end{eqnarray*}
The numerator simplifies to
\begin{eqnarray*}
&&\cov\{ Z-e(X)  , Y\} \\
&=&  E[ \{ Z-e(X)  \} Y ] \\
&=& E[ \{ Z-e(X)  \} ZY(1) ] + E[ \{ Z-e(X)  \} (1-Z)Y(0) ] \\
&& \quad\quad\quad\quad (\text{by } Y = ZY(1)  + (1-Z)Y(0)) \\
&=& E[ \{ Z- Ze(X)  \} Y(1)]  - E[  e(X)  (1-Z)Y(0) ]  \\
&=&  E[  Z\{ 1- e(X)  \} Y(1) ] - E [ e(X)  (1-Z)Y(0) ] \\
&=& E[  e(X)\{ 1- e(X)  \} \mu_1(X)]  - E[  e(X)  \{ 1- e(X)  \}  \mu_0(X) ]\\
&&   \quad\quad\quad\quad (\text{tower property and ignorability}) \\
&=& E\{ h_\textsc{O}(X) \tau(X) \} .
\end{eqnarray*}
The conclusion follows. 
 \end{myproof}

 From the proof of Theorem \ref{thm::ols-pscore-tau-o}, we can simply run the OLS of $Y$ on the centered treatment $ Z-e(X)$. \citet{lee2018simple} proposed this procedure. Moreover, we can also include $X$ in the OLS fit which may improve efficiency in finite samples. However, this does not change the estimand, which is still $\tau_\textsc{O}$. I summarize these two results in the corollary below.
 
 \begin{corollary}\label{coro::regress-on-pscore}
If $Z\ind \{  Y(1), Y(0) \} \mid X$, then 
 \begin{enumerate}
 \item[(1)]
 the coefficient of $Z-e(X)$ in the population OLS fit of $Y$ on $Z-e(X)$ or $\{1, Z-e(X) \}$ equals $\tau_\textsc{O}$;
 \item[(2)]
 the coefficient of $Z$ in the population OLS fit of $Y$ on $\{1, Z, e(X), X\}$ equals $\tau_\textsc{O}$. 
 \end{enumerate}
 \end{corollary}

\begin{myproof}{Corollary}{\ref{coro::regress-on-pscore}}
(1) The first result is an intermediate step in the proof of Theorem \ref{thm::ols-pscore-tau-o}. The second result holds because regressing  $Y$ on $Z-e(X)$ or $\{1, Z-e(X) \}$ does not change the coefficient of $Z-e(X) $ since it has mean zero.

(2) We can use the FWL theorem again. We can first obtain the residual from the population OLS of $Z$ on $\{1, e(X), X\}$, which is $Z-e(X) $ because 
$$
Z-e(X) = Z-0 - 1\cdot e(X) - 0\tran X $$ 
and $Z-e(X)$ is uncorrelated with any functions of $X$. 
Then the coefficient of $Z$ in the population OLS fit of $Y$ on $\{1, Z, e(X), X\}$ equals the coefficient of $Z$ in the population OLS fit of $Y$  on $Z-e(X)$. 
  \end{myproof}

  Theorem \ref{thm::ols-pscore-tau-o} motivates a two-step estimator for $\tau_\textsc{O} $:
\begin{enumerate}
\item
fit a propensity score model to obtain $\hat e(X_i) $;
\item
run OLS of $Y_i$ on $(1,X_i, \hat e(X_i) )$ to obtain the coefficient of $Z_i$. 
\end{enumerate}  
Corollary \ref{coro::regress-on-pscore}(1) motivates a two-step estimator for $\tau_\textsc{O} $: 
\begin{enumerate}
\item
fit a propensity score model to obtain $\hat e(X_i) $;
\item
run OLS of $Y_i$ on $Z_i - \hat e(X_i)$ to obtain the coefficient of $Z_i$.   
\end{enumerate}
Corollary \ref{coro::regress-on-pscore}(1) motivates another two-step estimator for $\tau_\textsc{O} $: 
\begin{enumerate}
\item
fit a propensity score model to obtain $\hat e(X_i) $;
\item
run OLS of $Y_i$ on $(1, Z_i , \hat e(X_i),  X_i ) $ to obtain the coefficient of $Z_i$.   
\end{enumerate}
Although OLS is convenient for obtaining point estimators, the corresponding standard errors are incorrect due to the uncertainty in the first step estimation of the propensity score. We can use the bootstrap to approximate the standard errors.

\citet{robins1992estimating} discussed many OLS estimators based on the propensity score. The above results seem special cases of their general theory although they did not point out the connection with the estimand under the overlap weight, which was resurrected by \citet{li2018balancing}. \citet{lee2018simple} proposed to regress $Y$ on $Z-e(X)$ from a different perspective without making connections to the existing results in \citet{robins1992estimating} and \citet{li2018balancing}.

\citet{rosenbaum1983central} proposed to estimate the average causal effect based on the OLS fit of $Y$ on $\{1, Z, e(X), Ze(X)\}$. When this outcome model is correct, their estimator is consistent for the average causal effect. However, when the model is incorrect, the corresponding estimator has a much more complicated interpretation.  \citet{little2004robust} suggested constructing estimators based on the OLS of $Y$ on $Z$ and a flexible function\footnote{For example, we can include polynomial terms of $e(X)$ in the regression. \citet{little2004robust} suggested using splines.} of $e(X)$ and showed it enjoys certain double robustness property.  Due to the complexity of implementation, I omit the discussion.

\section{Regressions weighted by the inverse of the propensity score}

\subsection{Average causal effect}

We first re-examine the Hajek estimator of $\tau$: 
$$
\hat{\tau}^\textup{hajek} = \frac{  \sumn \frac{Z_i Y_i}{ \hat{e}(X_i) }  }{  \sumn \frac{Z_i }{ \hat{e}(X_i) } }
- \frac{  \sumn \frac{ (1-Z_i) Y_i}{ 1 - \hat{e}(X_i) }  }{  \sumn \frac{ 1-Z_i}{ 1 - \hat{e}(X_i) }  },
$$
which equals the difference between the weighted means of the outcomes in the treatment and control groups. Numerically, it is identical to the coefficient of $Z_i$ in the following weighted least squares (WLS) of $Y_i$ on $(1,Z_i)$.

\begin{proposition}
\label{prop::hajek-wls}
$\hat{\tau}^\textup{hajek}$ equals $\hat{\beta}$ from the following WLS:
$$
(\hat{\alpha}, \hat{\beta}) = \arg\min_{\alpha, \beta} \sum_{i=1}^n w_i(Y_i - \alpha - \beta Z_i)^2
$$
 with weights 
\begin{eqnarray}\label{eq::weight-hajek}
w_i =  \frac{Z_i}{   \hat{e}(X_i)  } + \frac{1-Z_i}{ 1-  \hat{e}(X_i) } 
= 
\begin{cases}
\frac{1}{\hat{e}(X_i) } & \text{ if } Z_i=1;\\
\frac{1}{ 1- \hat{e}(X_i) } & \text{ if } Z_i=0.
\end{cases} 
\end{eqnarray}
\end{proposition}

\citet{imbens2004nonparametric} pointed out the result in Proposition \ref{prop::hajek-wls}. I leave it as a Problem \ref{hw::ate-hajek-wls}.  
By Proposition \ref{prop::hajek-wls}, it is convenient to obtain $\hat{\tau}^\textup{hajek}$ based on WLS. However, due to the uncertainty in the estimated propensity score, the standard error reported by WLS is incorrect for the true standard error of $\hat{\tau}^\textup{hajek}$. The bootstrap provides a convenient approximation to the true standard error.

Why does the WLS give a consistent estimator for $\tau$? Recall that in the CRE with a constant propensity score, we can simply use the coefficient of $Z_i$ in the OLS fit of $Y_i$ on $(1,Z_i)$ to estimate $\tau$. In observational studies, units have different probabilities of receiving the treatment and control, respectively. If we weight the treated units by $1/e(X_i)$ and the control units by $1/\{ 1-e(X_i)\}$, then both treated and control groups can represent the whole population. Thus, by weighting, we effectively have a pseudo-randomized experiment. Consequently, the difference between the weighted means is consistent for $\tau$. The numerical equivalence of $\hat{\tau}^\textup{hajek}$ and WLS is not only a fun numerical fact itself but also useful for motivating more complex estimators with covariate adjustment. I give one extension below.

Recall that in the CRE, we can use the coefficient of $Z_i$ in the OLS fit of $Y_i$ on $(1,Z_i, X_i, Z_iX_i)$ to estimate $\tau$, where the covariates are centered with $\bar{X} = 0$. This is \citet{lin2013}'s estimator which uses covariates to improve efficiency. A natural extension to observational studies is to estimate $\tau$ using the coefficient of $Z_i$ in the WLS fit of $Y_i$ on $(1,Z_i, X_i, Z_iX_i)$ with weights defined in \eqref{eq::weight-hajek}. \citet{hirano2001estimation} used this estimator in an application.
The fully interacted linear model is equivalent to two separate linear models for the treated and control groups. 
If the linear models
$$
E(Y\mid Z=1, X) = \beta_{10} +  \beta_{1x}\tran X,\quad
E(Y\mid Z=0, X) = \beta_{00} +  \beta_{0x}\tran X,
$$
are correctly specified, then both OLS and WLS give consistent estimators for the coefficients and the estimators of the coefficient of $Z$ are consistent for $\tau$. More interestingly, the estimator of the coefficient of $Z$ based on WLS is also consistent for $\tau$ if the propensity score model is correct and the outcome model is incorrect. That is, the estimator based on WLS is doubly robust. \citet{robins2007comment} discussed this property and attributed this result to M. Joffe's unpublished paper. I will give more details below.

Let $e(X_i, \hat{\alpha})$ be the fitted propensity score and $( \mu_1(X_i, \hat{\beta}_1),  \mu_0(X_i, \hat{\beta}_0)) $ be the fitted values of the outcome means based on the WLS. The outcome regression estimator is 
$$
\hat{\tau}^\text{reg}_\text{wls} = \frac{1}{n} \sumn \mu_1(X_i, \hat{\beta}_1) - \frac{1}{n} \sumn \mu_0(X_i, \hat{\beta}_0)
$$
and the doubly robust estimator for $\tau$ is 
$$
\hat{\tau}^\text{dr}_\text{wls}  = 
\hat{\tau}^\text{reg}_\text{wls}  + \frac{1}{n} \sumn  \frac{Z_i \{ Y_i- \mu_1(X_i, \hat{\beta}_1)\}  }{  e(X_i, \hat{\alpha}) } 
- \frac{1}{n} \sumn  \frac{(1-Z_i) \{ Y_i- \mu_0(X_i, \hat{\beta}_0)\}  }{  1- e(X_i, \hat{\alpha}) }   .
$$
An interesting result is that this doubly robust estimator equals the outcome regression estimator, which reduces to the coefficient of $Z_i$ in the WLS fit of $Y_i$ on $(1,Z_i, X_i, Z_iX_i)$ if we use weights \eqref{eq::weight-hajek}.

\begin{theorem}
\label{thm::numerical-equivalence-wls-dr}
If $\bar{X}=0$ and $( \mu_1(X_i, \hat{\beta}_1),  \mu_0(X_i, \hat{\beta}_0)) = (  \hat \beta_{10} +   \hat \beta_{1x}\tran X_i ,   \hat \beta_{00} +   \hat\beta_{0x}\tran X_i )$ based on the WLS fit of $Y_i$ on $(1,Z_i, X_i, Z_iX_i)$ with  weights \eqref{eq::weight-hajek}, then
$$
\hat{\tau}^\textup{dr}_\textup{wls}  = \hat{\tau}^\textup{reg}_\textup{wls} 
= \hat \beta_{10} - \hat \beta_{00} ,
$$
which is the coefficient of $Z_i$ in the WLS fit. 
\end{theorem}

\begin{myproof}{Theorem}{\ref{thm::numerical-equivalence-wls-dr}}
The WLS fit of $Y_i$ on $(1,Z_i, X_i, Z_iX_i)$ is equivalent to two WLS fits based on the treated and control data. Both WLS fits include intercepts, so the weighted residuals have mean 0 (see \eqref{eq::OLS-mean-data}): 
$$
\sumn  \frac{Z_i ( Y_i- \hat \beta_{10} -  \hat \beta_{1x}\tran X_i ) }{  \hat{e}(X_i) }   =0 
$$
and
$$
\sumn  \frac{(1-Z_i) (  Y_i-\hat \beta_{00}  -   \hat\beta_{0x}\tran X_i   ) }{  1-\hat{e}(X_i) }  = 0 .
$$
So the difference between $\hat{\tau}^\text{dr} $ and $\hat{\tau}^\text{reg}$ is exactly zero. Both reduces to
\begin{eqnarray*}
 \frac{1}{n} \sumn (\hat \beta_{10} +   \hat \beta_{1x}\tran X_i)   - \frac{1}{n} \sumn (\hat \beta_{00} +   \hat\beta_{0x}\tran X_i)  
&=& \hat \beta_{10} - \hat \beta_{00}  + (\hat \beta_{1x} - \hat\beta_{0x})\tran \bar{X} \\
&=&\hat \beta_{10} - \hat \beta_{00}
\end{eqnarray*}
with centered covariates. So they both equal the coefficient of $Z_i$ in the WLS fit of $Y_i$ on $(1,Z_i, X_i, Z_iX_i)$. 
\end{myproof}

\citet{freedman2008weighting} discouraged the use of the WLS estimator above based on some simulation studies. They showed that when the outcome model is correct, the WLS estimator is worse than the OLS estimator since the WLS estimator has large variability in their simulation setting with homoskedastic outcomes. This may not be true in general. When the errors have variance proportional to the inverse of the propensity scores, the WLS estimator will be more efficient than the OLS estimator. 

\citet{freedman2008weighting}  also showed that the estimated standard error based on the WLS fit is not consistent for the true standard error because it ignores the uncertainty in the estimated propensity score. This can be easily fixed by using the bootstrap to approximate the variance of the WLS estimator.

Nevertheless, \citet{freedman2008weighting}  found that ``weighting may help under some circumstances'' because when the outcome model is incorrect, the WLS estimator is still consistent if the propensity score model is correct.

I end this section with Table \ref{table::regression-causal-cre-obs} summarizing the regression estimators for causal effects in both randomized experiments and observational studies.

\begin{table}
\caption{Regression estimators in  CREs and unconfounded observational studies. The weights $w_i$'s are defined in \eqref{eq::weight-hajek}.
Assume covariates are centered at $\bar{X} = 0$.}\label{table::regression-causal-cre-obs}
\begin{tabular}{|c|cc|}
\hline 
 & CRE & unconfounded observational studies \\
 \hline 
without $X$ &  $Y_i\sim (1, Z_i)$ & $Y_i\sim (1, Z_i)$ with weights $w_i$   \\
with $X$ & $Y_i\sim (1, Z_i ,X_i, Z_i X_i ) $ & $Y_i\sim (1, Z_i ,X_i, Z_i X_i )$ with weights $w_i$  \\
\hline 
\end{tabular}
\end{table}

\subsection{Average causal effect on the treated units}

The results for $\tau_\textsc{T}$ parallel those for $\tau$. First, the Hajek estimator for $\tau_\textsc{T}$
$$
\hat{\tau}_\textsc{T}^\text{hajek} = \hat{\bar{Y}}(1) 
- \frac{  \sumn \hat{o}(X_i) (1-Z_i) Y_i }{ \sumn \hat{o}(X_i) (1-Z_i) } ,
$$
with $\hat{o}(X_i) = \hat{e}(X_i) / \{ 1- \hat{e}(X_i) \}$, 
equals the coefficient of $Z_i$ in the following WLS fit  $Y_i$ on $(1,Z_i)$.

\begin{proposition}
\label{prop::hajek-att}
$\hat\tau_\textsc{T}^{\text{hajek}}$ is numerically identical to $\hat{\beta}$ in the following WLS:
$$
(\hat{\alpha}, \hat{\beta}) = \arg\min_{\alpha, \beta} \sum_{i=1}^n  w_{\textsc{T}i}  (Y_i - \alpha - \beta Z_i)^2
$$
 with weights 
\begin{eqnarray}\label{eq::weight-hajek-att}
w_{\textsc{T}i} = Z_i +(1-Z_i )  \hat{o}(X_i)
= 
\begin{cases}
1 & \text{ if } Z_i=1;\\
 \hat{o}(X_i)   & \text{ if } Z_i=0.
\end{cases} 
\end{eqnarray}
\end{proposition}

Similar to Proposition \ref{prop::hajek-wls}, Proposition \ref{prop::hajek-att} is a pure linear algebra result. I relegate its proof to Problem \ref{hw::ate-hajek-wls}.

Second, if we center covariates at $\hat{\bar{X}}(1) = 0$, then we can estimate $\tau_\textsc{T}$ using the coefficient of $Z_i$ in the WLS fit of $Y_i$ on $(1,Z_i, X_i, Z_iX_i)$ with weights defined in \eqref{eq::weight-hajek-att}. Similarly, this estimator equals the regression estimator 
$$
\hat{\tau}_{\textsc{T},\text{wls}}^\text{reg} = \hat{\bar{Y}}(1) - \frac{1}{n_1} \sumn Z_i \mu_0(X_i, \hat{\beta}_0),
$$
which also equals the doubly robust estimator
$$
\hat{\tau}_{\textsc{T},\text{wls}}^\text{dr} =\hat{\tau}_{\textsc{T},\text{wls}}^\text{reg} 
- \frac{1}{n_1} \sumn  \hat{o}(X_i)  (1-Z_i) \{  Y_i- \mu_0(X_i, \hat{\beta}_0) \} .
$$

\begin{theorem}
\label{thm::numerical-equivalence-wls-dr-att}
If $\hat{\bar{X}}(1) = 0$ and $ \mu_0(X_i, \hat{\beta}_0) =  \hat \beta_{00} +   \hat\beta_{0x}\tran X_i$ based on the WLS fit of $Y_i$ on $(1,Z_i, X_i, Z_iX_i)$ with  weights \eqref{eq::weight-hajek-att}, then
$$
\hat{\tau}_{\textsc{T},\textup{wls}}^\textup{dr}  = \hat{\tau}_{\textsc{T},\textup{wls}}^\textup{reg} 
= \hat \beta_{10} - \hat \beta_{00} ,
$$
which is the coefficient of $Z_i$ in the WLS fit. 
\end{theorem}

\begin{myproof}{Theorem}{\ref{thm::numerical-equivalence-wls-dr-att}}
Based on the WLS fits in the treatment and control groups, we have
\begin{eqnarray}
\label{eq::wls-treated-att}
\sumn Z_i ( Y_i -  \hat \beta_{10} -  \hat \beta_{1x}\tran X_i ) &=& 0 ,\\
\label{eq::wls-control-att}
 \sumn \hat{o}(X_i)  (1-Z_i) ( Y_i - \hat \beta_{00} - \hat\beta_{0x}\tran X_i ) &=& 0.
\end{eqnarray}
The second result \eqref{eq::wls-control-att}  ensures that $\hat{\tau}_{\textsc{T},\text{wls}}^\text{dr}  = \hat{\tau}_{\textsc{T},\text{wls}}^\text{reg} $. Both reduces to 
$$
 \hat{\bar{Y}}(1) 
-  \frac{1}{n_1} \sumn Z_i   (\hat \beta_{00} +   \hat\beta_{0x}\tran X_i)
= \frac{1}{n_1} \sumn  Z_i(Y_i - \hat \beta_{00} - \hat\beta_{0x}\tran X_i) .
$$
With covariates centered at $\hat{\bar{X}}(1) = 0$, the first result \eqref{eq::wls-treated-att} implies that $\hat{\bar{Y}}(1)  =  \hat \beta_{10} $ which further simplifies the estimator to $ \hat \beta_{10} - \hat \beta_{00}$. 
\end{myproof}

\section{Homework problems}

\paragraph{Hajek estimators as WLS estimators}\label{hw::ate-hajek-wls}

Prove Propositions \ref{prop::hajek-wls} and \ref{prop::hajek-att}. 

Remark: These are special cases of Problem \ref{hw::wls-uni-binary} on the univariate WLS.

\paragraph{Predictive estimator and doubly robust estimator}\label{hw::dr-projective-estimator}

Another outcome regression estimator is the predictive estimator
$$
\hat{\tau}^\text{pred} = 
\hat{\mu}_1^\text{pred} - \hat{\mu}_0^\text{pred} 
$$
where
$$
\hat{\mu}_1^\text{pred} = 
\frac{1}{n} \sumn \left\{  Z_i Y_i + (1-Z_i)  \mu_1(X_i, \hat{\beta}_1)   \right\}
$$
and
$$
\hat{\mu}_0^\text{pred}  = 
\frac{1}{n} \sumn \left\{  Z_i \mu_0(X_i, \hat{\beta}_1) + (1-Z_i) Y_i  \right\} .
$$
It differs from the outcome regression estimator discussed before in that it only predicts the counterfactual outcomes but not the observed outcomes. 

Show that the doubly robust estimator  equals $\hat{\tau}^\text{pred} $ if $(\mu_1(X_i, \hat{\beta}_1), \mu_0(X_i, \hat{\beta}_1) ) =  
( \hat \beta_{10} +   \hat \beta_{1x}\tran X_i, \hat \beta_{00} +   \hat\beta_{0x}\tran X_i)$ are from the WLS fits of $Y_i$ on $(1, X_i)$ based on the treated and control data, respectively, with weights
\begin{eqnarray}
\label{eq::odds-weight-predictive}
w_i =  Z_i/\hat{o}(X_i) + (1-Z_i) \hat{o}(X_i)
= \begin{cases}
\frac{1}{\hat{o}(X_i)} = \frac{ 1-  \hat{e}(X_i) }{ \hat{e}(X_i) } & \text{ if } Z_i=1;\\
 \hat{o}(X_i) = \frac{  \hat{e}(X_i)  }{ 1-\hat{e}(X_i)} & \text{ if } Z_i=0.
\end{cases} 
\end{eqnarray}

Remark: 
\citet{cao2009improving} and  \citet{vermeulen2015bias} motivated the weights in \eqref{eq::odds-weight-predictive} from other more theoretical perspectives.

\paragraph{Weighted logistic regression  with a binary outcome}
\label{hw::wlogit-dr}
With a binary outcome, we can replace linear outcome models with logistic outcome models. Show that with weights in the logistic regressions, the doubly robust estimators equal the outcome regression estimator. The result holds for both $\tau$ and $\tau_\textsc{T}$.

\paragraph{Causal inference with a misspecified linear regression}

Define the population OLS of $Y$ on $ (1, Z, X) $ as
$$
(\beta_0, \beta_1, \beta_2) = \arg\min_{b_0, b_1, b_2}   E(Y - b_0 - b_1 Z - b_2\tran X)^2. 
$$
Recall that $e(X) = \pr(Z=1\mid X)$ is the propensity score, and define $\tilde{e}(X) = \gamma_0+\gamma_1\tran X$ as the OLS projection of $Z$ on $X$ with
$$
(\gamma_0, \gamma_1) =  \arg\min_{c_0, c_1}   E(Z - c_0 -  c_1\tran X)^2. 
$$

\begin{enumerate}
\item
Show that
$$
\beta_1 = \frac{    E [  \tilde{w}(X) \{ \mu_1(X) - \mu_0(X) \} ]  }{   E \{ \tilde{w}(X) \} }
+ \frac{  E [  \{  e(X) - \tilde{e}(X)  \}  \mu_0(X)  ] }{  E \{ \tilde{w}(X) \}  }
$$
where $\tilde{w}(X) = e(X)\{ 1 - \tilde{e}(X)  \}$. 

\item
When $X$ contains the dummy variables for a discrete covariate, show that
$$
\beta_1 = \frac{    E [  w(X) \{ \mu_1(X) - \mu_0(X) \} ]  }{   E \{ w(X) \} }
$$
where $w(X) = e(X)\{ 1 - e(X)  \}$ is the overlap weight introduced in Chapter \ref{sec::other-estmand-overlap}. 

\end{enumerate}

Remark: \citet{vansteelandt2021assumption} gave the formula in part 1 without a detailed proof. The result in part 2 was derived many times in the literature \citep[e.g.,][]{angrist1998estimating, ding2021frisch}.

\paragraph{Analyzing a dataset from the Karolinska Institute}\label{hw::Karolinska-pscore}

Revisit Problem \ref{hw::Karolinska-dr}. Estimate $\tau_\textsc{O}$ and $\tau$ based on the methods introduced in this Chapter.

 \paragraph{Recommended reading}
\citet{kang2007demystifying} gave a critical review of the doubly robust estimator, using simulation to compare it with many other estimators. \citet{robins2007comment} gave a very insightful comment on \citet{kang2007demystifying}.

\chapter{Matching in  Observational Studies}\label{chapter::matching-obs}
 
Matching has a long history in empirical research. W. Cochran and D. Rubin popularized it in statistical causal inference.   \citet{cochran1973controlling} is an early review paper. \citet{rubin2006matched} collects Rubin's contributions to this topic. 
This chapter also discusses modern contributions by \citet{abadie2006large, abadie2008failure, abadie2011bias} based on the asymptotic analysis of matching estimators.

\section{A simple starting point: many more control units}
\label{sec::ideal-matching-mpe}

\begin{figure}[ht]
\centering 
\includegraphics[width = \textwidth]{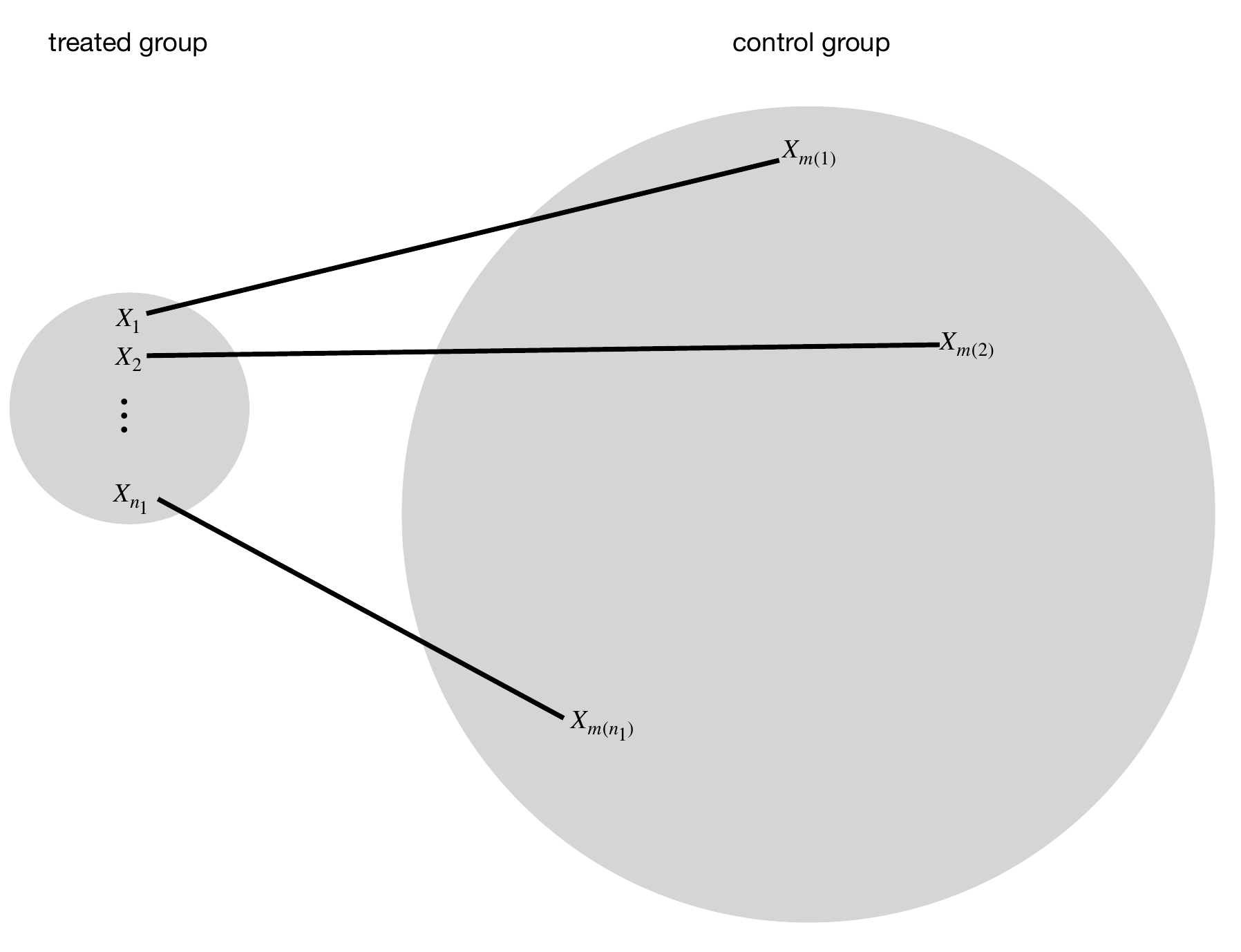}
\caption{Illustration of matching in observational studies}\label{fig::matching-causal}
\end{figure}

Figure \ref{fig::matching-causal} illustrates the basic idea of matching in treatment-control observational studies. 
Consider a simple case with the number of control units $n_0$ being much larger than the number of treated units $ n_1$. For unit $i=1,\ldots, n_1$ in the treated group, we find a unit $m(i)$ in the control group such that $X_i = X_{m(i)}$. In the ideal case, we have exact matches. Therefore, the units within a matched pair have the same propensity score $e(X_i) = e(X_{m(i)})$. Consequently, conditional on the event that one unit receives the treatment and the other receives the control, the probability of unit $i$ receiving the treatment and unit $m(i)$ receiving the control is
$$
\pr( Z_i = 1, Z_{m(i)} = 0 \mid  Z_i +Z_{m(i)} = 1, X_i, X_{m(i)} )  = 1/2
$$
by a symmetry argument.\footnote{The rigorous argument is
\begin{eqnarray*}
&&\pr( Z_i = 1, Z_{m(i)} = 0 \mid  Z_i +Z_{m(i)} = 1, X_i, X_{m(i)} ) \\
&=& \frac{ \pr( Z_i = 1, Z_{m(i)} = 0 \mid   X_i, X_{m(i)} ) }{ \pr( Z_i = 1, Z_{m(i)} = 0 \mid   X_i, X_{m(i)} )   + \pr( Z_i = 0, Z_{m(i)} = 1 \mid   X_i, X_{m(i)} ) } \\
&=& \frac{  e(X_i)  \{ 1 - e(X_{m(i)}) \}  }{ e(X_i)  \{ 1 - e(X_{m(i)}) \}   + \{1-e(X_i)\} e(X_{m(i)}) } \\
&=& \frac{1}{2}.
\end{eqnarray*}
}
That is, the treatment assignment is identical to the MPE conditioning on the covariates and the event that each pair has a treated unit and a control unit. So we can analyze the exactly matched observational study as if it is an MPE, using either the FRT or the Neymanian approach in Chapter \ref{chapter::mpe}. This gives us inference on the causal effect on the treated units.

We can also find multiple control units for each treated unit. In general, we can find $M_i$ matched control units for the treated unit $i$. When the $M_i$'s vary, it is called the {\it variable-ratio matching} \citep{ming2000substantial, ming2001note, pimentel2015variable}. With perfect matching, the treatment assignment mechanism is identical to the general matched experiment discussed in Section \ref{sec::general-matched-experiment}. We can use the analytic results in that section to analyze the matched observational study.

\citet{rosenbaum2002design} advocated the above analysis strategy. In most observational studies, however, $X_i = X_{m(i)}$ does not hold for all units. The above reasoning for FRT does not hold.  Recently, \citet{guo2022statistical} reported negative results on the consequence of inexact matching in FRT. This is a warning of this strategy.

\section{A more complicated but realistic scenario}

Even if the control group is large, we often do not have exact matches. What we can achieve is that $X_i \approx X_{m(i)}  $ or $X_i -X_{m(i)} $ is small under some distance metric. So we have only approximate matches. For example, we define
$$
m(i) = \arg\min_{k:Z_k = 0} d(X_i ,X_k), 
$$
where $d(X_i ,X_k)$ measures the distance between $X_i$ and $X_k$. Some canonical choices of the distance are the Euclidean distance
$$
d(X_i ,X_k) =  (X_i - X_k)\tran ( X_i - X_k),
$$
and the Mahalanobis distance
$$
d(X_i ,X_k) =(X_i - X_k)\tran \Omega^{-1} (X_i - X_k)
$$
with $\Omega$ being the sample covariance matrix of the $X_i$'s from the whole population or only the control group.

I review some subtle issues about matching below. See \citet{stuart2010matching} for a review paper. 
\begin{enumerate}
\item
(one-to-one or one-to-$M$ matching)
The above discussion focused on one-to-one matching. We can also extend the discussion to one-to-$M$ matching.

\item
I focus on matching with replacement but some practitioners prefer matching without replacement. If the pool of control units is large, these two methods will not differ too much for the final result. Matching with replacement is computationally more convenient, but matching without replacement involves computationally intensive discrete optimization. Matching with replacement usually gives matches of higher quality but it introduces dependence by using the same units multiple times. In contrast, the advantage of matching without replacement is the independence of matched units and the simplicity in the subsequent data analysis.

\item
Because of the residual covariate imbalance within matched pairs, it is crucial to use covariate adjustment when analyzing the data. In this case, covariate adjustment is not only for efficiency gain but also for bias correction.

\item
If $X$ is ``high dimensional'', it is likely that $d(X_i ,X_k)$ is too large for some unit $i$ in the treated group and for all choices of the units in the control group. If this happens, we may have to drop some units that are hard to find matches. By doing this, we effectively change the study population of interest.

\item
It is hard to avoid the above problem. For example, if $X_i \sim \text{N}(0, I_p), X_k \sim\text{N}(0, I_p) ,$ and $X_i \ind X_k$, then 
$$
(X_i - X_k)\tran (X_i - X_k) \sim    2 \chi^2_p
$$ 
which has mean $2p$ and variance $8p$ (see Chapter \ref{subsection::chi2-t-distribution}). 
Theory shows that with large $p$, imperfect matching causes a large bias in causal effect estimation. 
This suggests that if $p$ is large,  we must do some dimension reduction before matching. 
\citet{rosenbaum1983central} proposed to use matching based on the propensity score. With the estimated propensity score, we find pairs of units $\{ i, m(i)\}$ with small values of $ |  \hat{e}(X_i) -  \hat{e}(X_{m(i)})  |$ or $| \logit\{ \hat{e}(X_i)\} - \logit\{ \hat{e}(X_{m(i)}) \} | $\footnote{Define $\logit(w) = \log \frac{w}{1-w}$. The logit function is a map from $[0,1]$ to $(-\infty, \infty)$.}, i.e., we have a one-dimensional matching problem. 
\end{enumerate}

\section{Matching estimator for the average causal effect}

In a sequence of papers, Abadie and Imbens (AI) rigorously characterized the asymptotic properties of the matching estimator and proposed the corresponding large-sample confidence intervals for the average causal effect. 
They chose the standard setup for observational studies with  $\{ X_i , Z_i ,Y_i(1), Y_i(0)  \}_{i=1}^n \iidsim \{ X, Z, Y(1), Y(0) \}$.  

\subsection{Point estimation and bias correction}

AI focused on $1$-$M$ matching with replacement. For a treated unit $i$, we can simply impute the potential outcome under the treatment as $\hat{Y}_i(1) = Y_i$, and impute the potential outcome under the control as 
$$
\hat{Y}_i(0) = M^{-1} \sum_{k\in J_i} Y_k,
$$
where $J_i$ is the set of matched units from the control group for unit $i$. For example, we can compute $d(X_i, X_k)$ for all $k$ in the control group, and then define $J_i$ as the indices of $k$ with the $M$ smallest values of $d(X_i, X_k)$.

For a control unit $i$, we simply impute the potential outcome under the control as $\hat{Y}_i(0) = Y_i$, and impute the potential outcome under the treatment as
$$
\hat{Y}_i(1) = M^{-1} \sum_{k\in J_i} Y_k,
$$
where $J_i$ is the set of matched units from the treatment group for unit $i$.

The matching estimator is 
$$
\hat{\tau}^\textup{m} = n^{-1} \sumn \{\hat{Y}_i(1) -\hat{Y}_i(0) \}.
$$

AI showed that $\hat{\tau}^\textup{m}$ has non-negligible bias especially when $X$ is multidimensional and the number of control units is comparable to the number of treated units. Through some technical derivations,  they proposed the following estimator for the bias: 
$$
\hat{B} = n^{-1} \sumn  \hat{B}_i
$$
where
$$
\hat{B}_i = (2Z_i - 1) M^{-1} \sum_{k\in J_i} \{  \hat{\mu}_{1-Z_i}(X_i) -  \hat{\mu}_{1-Z_i}(X_k)     \}
$$
with $\{  \hat{\mu}_1(X_i), \hat{\mu}_0(X_i) \}$ being the predicted outcomes by, for example, OLS fits. 
For a treated unit with $Z_i  =1$, the estimated bias is
$$
\hat{B}_i = M^{-1} \sum_{k\in J_i} \{  \hat{\mu}_{0}(X_i) -  \hat{\mu}_{0}(X_k)     \}
$$
which corrects the discrepancy in predicted control potential outcomes due to the mismatch in covariates; for a control unit with $Z_i = 0$, the estimated bias is
 $$
\hat{B}_i =  - M^{-1} \sum_{k\in J_i} \{  \hat{\mu}_{1}(X_i) -  \hat{\mu}_{1}(X_k)     \}
$$
which corrects the discrepancy in predicted treated potential outcomes due to the mismatch in covariates.

The final bias-corrected matching estimator is
$$
\hat{\tau}^\textup{mbc} = \hat{\tau}^\textup{m}  - \hat{B},
$$
 which has the following linear expansion.
 
 \begin{proposition}
 \label{prop::matching-expansion-ate} We have 
 \begin{eqnarray}\label{eq::linearATE}
\hat{\tau}^\textup{mbc}  
= n^{-1} \sumn \hat{\psi}_i
\end{eqnarray} 
where 
 $$
  \hat{\psi}_i = \hat{\mu}_1(X_i) -  \hat{\mu}_0(X_i) + (2Z_i - 1) (1+K_i/M) \{   Y_i - \hat{\mu}_{Z_i}(X_i)  \}
 $$
 with $K_i$ being the times that unit $i$ is used as a match. 
 \end{proposition}

The linear expansion in Proposition  \ref{prop::matching-expansion-ate} follows from simple but tedious algebra. I leave its proof as Problem \ref{para::linear-bcm-ate}. The linear expansion motivates a simple variance estimator 
$$
\hat{V}^\textup{mbc}  = \frac{1}{n^2} \sumn (  \hat{\psi}_i - \hat{\tau}^\textup{mbc}  )^2,
$$
by viewing $\hat{\tau}^\textup{mbc}  $ as sample averages of the $  \hat{\psi}_i $'s.  \citet{abadie2008failure}  first showed that the simple bootstrap by resampling the original data does not work for estimating the variance of the matching estimators, but their proposed variance estimation procedure is not easy to implement.  \citet{otsu2017bootstrap} proposed to bootstrap the $  \hat{\psi}_i $'s in the linear expansion. However,  \citet{otsu2017bootstrap}'s bootstrap essentially yields the variance estimator $\hat{V}^\textup{mbc}  $, which is simple to calculate.

\subsection{Connection with the doubly robust estimators}
 \label{sec::connection-dr-att}
 
The bias-corrected matching estimators and the doubly robust estimators are closely related. They both equal the outcome regression estimator
with some modifications based on the residuals 
$$ 
 \hat{R}_i = 
 \begin{cases}
 Y_i - \hat{\mu}_1(X_i)  & \text{ if } Z_i=1;\\
 Y_i - \hat{\mu}_0(X_i)    & \text{ if } Z_i=0.
\end{cases} 
 $$
  For the average causal effect $\tau$, recall the outcome regression estimator  
$$
\hat{\tau}^{\textup{reg}} =  n^{-1} \sumn \{ \hat{\mu}_1(X_i) -  \hat{\mu}_0(X_i)\} 
$$
and the doubly robust estimator 
$$
\hat{\tau}^\textup{dr}  =  \hat{\tau}^{\textup{reg}} 
+ n^{-1} \sumn \left\{  \frac{Z_i \hat{R}_i }{ \hat{e}(X_i)} -  \frac{ (1-Z_i) \hat{R}_i}{ 1-  \hat{e}(X_i) } \right\} .
$$
Furthermore, we can verify that $\hat{\tau}^\textup{mbc} $ has a form similar to $\hat{\tau}^\textup{dr}  $.
\begin{proposition}
\label{prop::mbc-dr}
The bias-corrected matching estimator for $\tau$ equals 
$$
\hat{\tau}^\textup{mbc}   =\hat{\tau}^{\textup{reg}} 
+ n^{-1} \sumn \left\{ \left(1+{ K_i\over M} \right)  Z_i \hat{R}_i  - \left( 1+{ K_i\over M} \right) (1-Z_i)  \hat{R}_i \right\} .
$$
\end{proposition}

I leave the proof of Proposition \ref{prop::mbc-dr}  as Problem \ref{para::dr-mbc-form}.  From Proposition \ref{prop::mbc-dr}, we can view matching as a nonparametric method to estimate the propensity score, and the resulting bias-corrected matching estimator as a doubly robust estimator. For instance, $1+K_i /  M$ should be close to $1/\hat{e}(X_i)$. When a treated unit has a small $e(X_i)$, the resulting weight based on the estimated propensity score $1/\hat{e}(X_i)$ will be large, and at the same time, it will be matched to many control units, resulting in large $K_i$ and thus large $1+K_i /  M$. However, this connection also raised an obvious question regarding matching. With a fixed $M$, the estimator $1+K_i /  M$ for $1/e(X_i)$ will be very noisy. Allowing $M$ to grow with the sampling size is likely to improve the matching-based nonparametric estimator for the propensity score and thus improve the asymptotic properties of the matching and bias-corrected matching estimators. \citet{lin2023estimation} provided a formal theory that once we allow $M$ to grow at a proper rate, the bias-corrected matching estimator $\hat{\tau}^\textup{mbc}   $ can achieve similar properties as the doubly robust estimator.

 \section{Matching estimator for the average causal effect on the treated}

For the average causal effect on the treated 
$$
\tau_\textsc{T} = E(Y\mid Z=1) - E\{  Y(0)  \mid Z=1\},
$$ 
we only need to impute the missing potential outcomes under control for all the treated units, resulting in the following estimator
$$
\hat{\tau}_\textsc{T}^\textup{m} = n_1^{-1} \sumn Z_i \{  Y_i - \hat{Y}_i(0) \}.
$$
Again it is biased with multidimensional $X$.
\citet{otsu2017bootstrap} propose to estimate its bias by 
$$  
\hat{B}_\textsc{T} = n_1^{-1} \sumn Z_i  \hat{B}_{\textsc{T},i} 
$$
where
$$
\hat{B}_{\textsc{T},i} = M^{-1} \sum_{k \in J_i} \{  \hat{\mu}_0(X_i) -  \hat{\mu}_0(X_k)     \}
$$
corrects the bias due to the mismatch of covariates for a treated unit with $Z_i = 1.$

The final bias-corrected estimator is
$$
\hat{\tau}_\textsc{T}^\textup{mbc} = \hat{\tau}_\textsc{T}^\textup{m} - \hat{B}_\textsc{T},
$$ 
which has the following linear expansion.

\begin{proposition}\label{prop::linear-expansion-att}
We have 
\begin{eqnarray}\label{eq::linearATT}
\hat{\tau}_\textsc{T}^\textup{mbc}  =   n_1^{-1} \sumn \hat{\psi}_{\textsc{T}, i},
\end{eqnarray}
where 
$$
\hat{\psi}_{\textsc{T}, i} =   Z_i \{ Y_i - \hat{\mu}_0(X_i) \}
- (1-Z_i) K_i/M \{ Y_i - \hat{\mu}_0(X_i) \} . 
$$
\end{proposition}

I leave the proof to Problem \ref{para::linear-bcm-ate}. 
Motivated by \citet{otsu2017bootstrap}, we can view $\hat{\tau}_\textsc{T}^\textup{mbc}  $ as $n/n_1$ multiplied by the sample average of the $\hat{\psi}_{\textsc{T}, i}$'s, so an intuitive variance estimator is
$$
\hat{V}_\textsc{T}^\textup{mbc} =  \left( \frac{n}{n_1} \right)^2 \frac{1}{n^2} \sumn (\hat{\psi}_{\textsc{T}, i} - \hat{\tau}_\textsc{T}^\textup{mbc} n_1/n)^2 = \frac{1}{n_1^2} \sumn (\hat{\psi}_{\textsc{T}, i} - \hat{\tau}_\textsc{T}^\textup{mbc} n_1/n )^2.
$$

 Similar to the discussion in Section \ref{sec::connection-dr-att}, we can compare the doubly robust and bias-corrected matching estimators with the outcome regression estimator. 
 For the average causal effect on the treated units $\tau_\textsc{T}$, recall the outcome regression estimator 
$$
\hat{\tau}^{\textup{reg}}_\textsc{T} = n_1^{-1} \sumn Z_i \{ Y_i - \hat{\mu}_0(X_i) \} ,
$$
and the doubly robust estimator
$$
\hat{\tau}_\textsc{T}^\textup{dr} =  \hat{\tau}^{\textup{reg}}_\textsc{T} 
- n_1^{-1} \sumn \frac{  \hat{e}(X_i) }{ 1- \hat{e}(X_i) }  (1-Z_i)  \hat{R}_i .
$$
Furthermore, we can verify that $\hat{\tau}_\textsc{T}^\textup{mbc} $ has a form similar to $\hat{\tau}_\textsc{T}^\textup{dr} $. 
\begin{proposition}
\label{prop::mbc-dr-att}
The bias correction matching estimator for  $\tau_\textsc{T}$ equals
$$
\hat{\tau}_\textsc{T}^\textup{mbc}   = \hat{\tau}^{\textup{reg}}_\textsc{T} 
 - n_1^{-1} \sumn { K_i\over M}  (1-Z_i)  \hat{R}_i.
$$
\end{proposition}

I leave the proof of Proposition \ref{prop::mbc-dr-att}  as Problem \ref{para::dr-mbc-form-att}.  Proposition \ref{prop::mbc-dr-att} suggests that matching essentially uses $K_i/M$ to estimate the odds of the treatment given covariates.

\section{A case study}
\label{sec::matching-application}

\subsection{Experimental data}
Now I revisit the LaLonde data using \citet{sekhon2011multivariate}'s \ri{Matching} package. We have used this package several times for the dataset \ri{lalonde}, and now we will use its key function \ri{Match}.  The experimental part gives us the following results:
\begin{lstlisting}
> library("car")
> library("Matching")
> 
> ## Chapter 15.5.1
> ## experimental data
> data("lalonde")
> y = lalonde$re78
> z = lalonde$treat
> x = as.matrix(lalonde[, c("age", "educ", "black",
+                           "hisp", "married", "nodegr",
+                           "re74", "re75")])
> 
> ## analysis the randomized experiment
> neymanols = lm(y ~ z)
> fisherols = lm(y ~ z + x)
> xc = scale(x)
> linols = lm(y ~ z*xc)
> resols = c(neymanols$coef[2],
+            fisherols$coef[2],
+            linols$coef[2],
+            sqrt(hccm(neymanols, type = "hc2")[2, 2]),
+            sqrt(hccm(fisherols, type = "hc2")[2, 2]),
+            sqrt(hccm(linols, type = "hc2")[2, 2]))
> resols = matrix(resols, 3, 2)
> rownames(resols) = c("neyman", "fisher", "lin")
> colnames(resols) = c("est", "se")
> resols 
            est       se
neyman 1794.343 670.9967
fisher 1676.343 677.0493
lin    1621.584 694.7217
\end{lstlisting} 
All regression estimators show positive significant results on the job training program. We can analyze the data as if it is an observational study based on 1-1 matching, yielding the following results:
\begin{lstlisting}
> matchest.adj = Match(Y = y, Tr = z, X = x, BiasAdjust = TRUE)
> summary(matchest.adj)

Estimate...  2119.7 
AI SE......  876.42 
T-stat.....  2.4185 
p.val......  0.015583 

Original number of observations..............  445 
Original number of treated obs...............  185 
Matched number of observations...............  185 
Matched number of observations  (unweighted).  268
\end{lstlisting} 
Both the point estimator and standard error increase, but qualitatively, the conclusion remains the same.

\subsection{Observational data}
Then I revisit the observational counterpart of the data:
\begin{lstlisting}
> dat <- read.table("cps1re74.csv", header = TRUE)
> dat$u74 <- as.numeric(dat$re74==0)
> dat$u75 <- as.numeric(dat$re75==0)
> y = dat$re78
> z = dat$treat
> x = as.matrix(dat[, c("age", "educ", "black",
+                           "hispan", "married", "nodegree",
+                           "re74", "re75", "u74", "u75")])
 \end{lstlisting} 
If we use simple OLS estimators, the results are far from the experimental benchmark and are sensitive to the specification of the regression:
\begin{lstlisting}
> neymanols = lm(y ~ z)
> fisherols = lm(y ~ z + x)
> xc = scale(x)
> linols = lm(y ~ z*xc)
> resols = c(neymanols$coef[2],
+            fisherols$coef[2],
+            linols$coef[2],
+            sqrt(hccm(neymanols, type = "hc2")[2, 2]),
+            sqrt(hccm(fisherols, type = "hc2")[2, 2]),
+            sqrt(hccm(linols, type = "hc2")[2, 2]))
> resols = matrix(resols, 3, 2)
> rownames(resols) = c("neyman", "fisher", "lin")
> colnames(resols) = c("est", "se")
> resols 
             est        se
neyman -8506.495  583.4426
fisher  1067.546  628.4389
lin    -4265.801 3211.7718
  \end{lstlisting} 
  However, if we use 1-1 matching, the results almost recover those based on the experimental data:
  \begin{lstlisting}
> matchest = Match(Y = y, Tr = z, X = x, BiasAdjust = TRUE)
> summary(matchest)

Estimate...  1747.8 
AI SE......  916.59 
T-stat.....  1.9068 
p.val......  0.056543 

Original number of observations..............  16177 
Original number of treated obs...............  185 
Matched number of observations...............  185 
Matched number of observations  (unweighted).  248
   \end{lstlisting} 
   
Ignoring the ties in the matched data, we can also use the matched-pairs analysis, which again yields results similar to those based    on the experimental data:
  \begin{lstlisting}
> diff = y[matchest$index.treated] - 
+           y[matchest$index.control]
> round(summary(lm(diff ~ 1))$coef[1, ], 2)
  Estimate Std. Error    t value   Pr(>|t|) 
   1581.44     558.55       2.83       0.01 
> 
> diff.x = x[matchest$index.treated, ] - 
+               x[matchest$index.control, ]
> round(summary(lm(diff ~ diff.x))$coef[1, ], 2)
  Estimate Std. Error    t value   Pr(>|t|) 
   1842.06     578.37       3.18       0.00 
   \end{lstlisting}

   \subsection{Covariate balance checks}
   
Moreover, we can use simple OLS to check covariate balance. Before matching, the covariates are highly imbalanced, signified by many stars associated with the coefficients. 
  \begin{lstlisting}
> lm.before = lm(z ~ x)
> summary(lm.before) 

Residuals:
     Min       1Q   Median       3Q      Max 
-0.18508 -0.01057  0.00303  0.01018  1.01355 

Coefficients:
              Estimate Std. Error t value Pr(>|t|)    
(Intercept)  1.404e-03  6.326e-03   0.222   0.8243    
xage        -4.043e-04  8.512e-05  -4.750 2.05e-06 ***
xeduc        3.220e-04  4.073e-04   0.790   0.4293    
xblack       1.070e-01  2.902e-03  36.871  < 2e-16 ***
xhispan      6.377e-03  3.103e-03   2.055   0.0399 *  
xmarried    -1.525e-02  2.023e-03  -7.537 5.06e-14 ***
xnodegree    1.345e-02  2.523e-03   5.331 9.89e-08 ***
xre74        7.601e-07  1.806e-07   4.208 2.59e-05 ***
xre75       -1.231e-07  1.829e-07  -0.673   0.5011    
xu74         4.224e-02  3.271e-03  12.914  < 2e-16 ***
xu75         2.424e-02  3.399e-03   7.133 1.02e-12 *** 
      \end{lstlisting}

      However, after matching, the covariates are well-balanced, signified by the absence of stars for all coefficients.

   \begin{lstlisting}
> lm.after = lm(z ~ x, 
+               subset = c(matchest$index.treated, 
+                          matchest$index.control))
> summary(lm.after) 

Residuals:
     Min       1Q   Median       3Q      Max 
-0.66864 -0.49161 -0.03679  0.50378  0.65122 

Coefficients:
              Estimate Std. Error t value Pr(>|t|)  
(Intercept)  6.003e-01  2.427e-01   2.474   0.0137 *
xage         3.199e-03  3.427e-03   0.933   0.3511  
xeduc       -1.501e-02  1.634e-02  -0.918   0.3590  
xblack       6.141e-05  7.408e-02   0.001   0.9993  
xhispan      1.391e-02  1.208e-01   0.115   0.9084  
xmarried    -1.328e-02  6.729e-02  -0.197   0.8437  
xnodegree   -3.023e-02  7.144e-02  -0.423   0.6723  
xre74        6.754e-06  9.864e-06   0.685   0.4939  
xre75       -9.848e-06  1.279e-05  -0.770   0.4417  
xu74         2.179e-02  1.027e-01   0.212   0.8321  
xu75        -2.642e-02  8.327e-02  -0.317   0.7512
         \end{lstlisting}

\section{Discussion}

With many covariates, matching based on the original covariates may suffer from the curse of dimensionality.      \citet{rosenbaum1983central} suggested to use matching based on the estimated propensity score.     \citet{abadie2016matching} provided a form theory for this strategy.

\section{Homework Problems}

\paragraph{Linear expansions of the bias-corrected estimators}\label{para::linear-bcm-ate}

 Prove Propositions \ref{prop::matching-expansion-ate} and \ref{prop::linear-expansion-att}.

\paragraph{Doubly robust form of the bias-corrected matching estimator for $\tau$}\label{para::dr-mbc-form}

Prove Proposition \ref{prop::mbc-dr}.

\paragraph{Doubly robust form of the bias-corrected matching estimator for $\tau_\textsc{t}$}\label{para::dr-mbc-form-att}

Prove Proposition \ref{prop::mbc-dr-att}.

\paragraph{Revisit Example \ref{eg::chan-bmi-ATE}}

Analyze the dataset in Example \ref{eg::chan-bmi-ATE} using the matching estimator. Compare the results with previous results.
You should check the covariate balance before and after matching. You can also choose a different number of matches for the matching estimator.  Moreover, you can even apply various estimators to the matched data. 
Are your results sensitive to your choices?

\paragraph{Revisit Chapter \ref{sec::matching-application}}

Chapter \ref{sec::matching-application} analyzed the LaLonde observational study using matching. Matching performs well because it gives an estimator that is close to the experimental gold standard. Reanalyze the data using the outcome regression, propensity score stratification, two IPW, and the doubly robust estimators. Compare the results to the matching estimator and to the estimator from the experimental gold standard.

Note that you have many choices. For example, the number of strata for stratification and the threshold to trim to data based on the estimated propensity scores. You may consider fitting different propensity score and outcome models, e.g., including some quadratic terms of the basic covariates. You can even apply these estimators to the matched data. 

This is a classic dataset and hundreds of papers have used it. You can read some references \citep{dehejia1999causal, hainmueller2012entropy} and you can also be creative in your data analysis.

\paragraph{Data re-analyses}

\citet{ho2007matching} is an influential paper in political science, based on which the authors have developed an \ri{R} package \ri{MatchIt} \citep{ho2011matchit}. \citet{ho2007matching} analyzed two datasets, both of which are available from the Harvard Dataverse.

Re-analyze these two datasets using the methods discussed so far. You can also try other methods as long as you can justify them.

 \paragraph{Recommended reading}
 
The literature on matching estimators is massive, and three excellent review papers are \citet{sekhon2009opiates}, \citet{stuart2010matching}, and \citet{imbens2015matching}.

\part{Difficulties and challenges of observational studies}\label{part::challenges-os}

\chapter{Difficulties of Unconfoundedness in Observational Studies for Causal Effects}
 \label{chapter::difficulties-observational-studies}

Part \ref{part::observational-studies} of this book discusses causal inference with observational studies under two assumptions: unconfoundedness and overlap. Both are strong assumptions and are likely to be violated in practice. This chapter will discuss the difficulties of the unconfoundedness assumption.  Chapters \ref{chapter::evalue}--\ref{sec::rosenbaum-sensitivity-analysis} will discuss various strategies for sensitivity analysis in observational studies with unmeasured confounding. Chapter \ref{chapter::overlap}  will discuss the difficulties of the overlap assumption.

\section{Some basics of the causal diagram}

\citet{pearl1995causal} introduced the causal diagram as a powerful tool for causal inference in empirical research. 
\citet{pearl2000causality} is a textbook on the causal diagram. Here I introduce the causal diagram as an intuitive tool for illustrating the causal relationships among variables. 

For example, if we have the causal diagram
$$
\xymatrix{
 & X \ar[ld]\ar[rd] \\
Z\ar[rr] &&Y 
}
$$
and focus on the causal effect of $Z$ on $Y$,
we can read it as the following data-generating process: 
$$
\left\{ 
\begin{aligned}
      X & \sim F_X(x), \\
      Z  &=   f_Z(X, \varepsilon_Z) , \\
      Y(z) &=  f_Y(X, z, \varepsilon_Y(z)) ,
\end{aligned}
\right. 
$$
where $\varepsilon_Z \ind \varepsilon_Y(z)$ for both $z=0,1$. 
In the above,  covariates $X$ are generated from a distribution  $F_X(x)$, the treatment assignment is a function of $X$ with a random error term $\varepsilon_Z$, and the potential outcome $ Y(z) $ is a function of $X$, $z$ and a random error term $\varepsilon_Y(z)$. We can easily read from the equations that $Z\ind Y(z) \mid X$, i.e., the unconfoundedness assumption holds.

If we have a causal diagram 
$$
\xymatrix{
 X \ar[d]\ar[rrd] & & U\ar[d]\ar[lld] \\
Z\ar[rr] &&Y 
}
$$
we can read it as  the following data-generating process: 
$$
\left\{ 
\begin{aligned}
       X & \sim F_X(x), \\
       U& \sim F_U(u), \\ 
      Z  &=   f_Z(X, U,  \varepsilon_Z) , \\
      Y(z) &=  f_Y(X, U, z, \varepsilon_Y(z)) ,
\end{aligned}
\right. 
$$
where $\varepsilon_Z \ind \varepsilon_Y(z)$ for both $z=0,1$. We can easily read from the equations that $Z\ind Y(z) \mid (X,U)$ but $Z\nind Y(z) \mid X$, i.e., the unconfoundedness assumption holds conditional on $(X,U)$ but does not hold conditional on $X$ only. In this case, $U$ is an unmeasured confounder. In this diagram, $U$ is called an unmeasured confounder.

\section{Assessing the unconfoundedness assumption}

The unconfoundedness assumption 
$$
Z\ind Y(1) \mid X, \quad 
Z\ind Y(0) \mid X
$$
implies that 
\begin{eqnarray*}
\pr\{ Y(1)\mid  Z=1, X \}  &=& \pr\{ Y(1)\mid  Z=0, X \} , \\
\pr\{ Y(0)\mid  Z=1, X \} &=& \pr\{ Y(0)\mid  Z=0, X \} . 
\end{eqnarray*}
So the unconfoundedness assumption basically requires that the counterfactual distribution $\pr\{ Y(1)\mid  Z=0, X \}$ equals the observed distribution 
$\pr\{ Y(1)\mid  Z=1, X \}$, and the counterfactual distribution $\pr\{ Y(0)\mid  Z=1, X \} $ equals the observed distribution $\pr\{ Y(0)\mid  Z=0, X \}$. Because the counterfactual distributions are not directly identifiable from the data, the unconfoundedness assumption is fundamentally untestable without additional assumptions. I will discuss two strategies to assess the unconfoundedness assumption. Here, ``assess'' is a weaker notion than ``test''. The former is referred to as supplementary analysis that supports or undermines the initial analysis, but the latter is referred to as formal statistical testing.

\subsection{Using negative outcomes}

Assume that $Y^\textup{n}$ is an outcome similar to $Y$ and ideally, shares the same confounding structure as $Y$. If we believe $Z\ind Y(z)\mid X$, then we also tend to believe $Z\ind Y^\textup{n} (z)\mid X$.
Moreover, we know, a priori, the effect of $Z$ on $Y^\textup{n}$:
$$
\tau(Z\rightarrow Y^\textup{n} )
= 
E\{  Y^\textup{n}(1) - Y^\textup{n}(0) \} .
$$
An important example is that $\tau(Z\rightarrow Y^\textup{n} ) = 0$. A causal diagram satisfying these requirements is below:

$$
\xymatrix{
 & X \ar[ld]\ar[rd] \ar[rr] & & Y^\textup{n}  \\
Z\ar[rr] &&Y 
}
$$

\begin{example}
\citet{Cornfield::1959} studied the causal role of cigarette smoking on lung cancer based on observational studies. They controlled for many important background variables but it is still possible to have some unmeasured confounders biasing the observed effects. To strengthen the evidence for causation, they also reported the effect of cigarette smoking on car accidents, which was close to zero, the anticipated effect based on biology. So even if they could not rule out unmeasured confounding in the analysis, this supplementary analysis based on a negative outcome makes the evidence of the causal effect of cigarette smoking on lung cancer stronger. 
\end{example}

\begin{example}
\citet{imbens2015causal} suggested using the lagged outcome as a negative outcome. In most cases, it is reasonable to believe that the lagged outcome and the outcome have a similar confounding structure. Since the lagged outcome happens before the treatment, the average causal effect on it must be $0$. However, their suggestion should be used with caution since in most studies we simply treat lagged outcomes as an observed confounder. 

In some sense, the covariate balance check in Chapter \ref{chapter::pscore-key} is a special case of using negative controls. Similar to the problem of using lagged outcomes as negative controls, those covariates are usually a part of the unconfoundedness assumption. Therefore, the failure of the covariate balance check does not really falsify the unconfoundedness assumption but rather the model specification of the propensity score. 
\end{example}

\begin{example}
Observational studies of elderly persons have shown that vaccination against influenza remarkably reduces one's risk of pneumonia/influenza hospitalization and all-cause mortality in the following season, after adjustment for measured covariates. \citet{jackson2006evidence} were skeptical about the large magnitude and thus conducted supplementary analysis on negative outcomes. Vaccination often begins in the fall, but influenza transmission is often minimal until the winter. Based on biology, the effect of vaccination should be most prominent during influenza season. But \citet{jackson2006evidence} found a greater effect before the influenza season, suggesting that the observed effect is due to unmeasured confounding. 
\end{example}

\citet{jackson2006evidence} seems the most convincing one since the influenza-related outcomes before and during the influenza season should have similar confounding patterns. \citet{Cornfield::1959}'s additional evidence seems weaker since car accidents and lung cancer have very different causal mechanisms with respect to cigarette smoking. In fact, \citet{Fisher::1957}'s critique was that the relationship between cigarette smoking on lung cancer may be due to an unobserved genetic factor (see Chapter \ref{chapter::evalue}). Such a genetic factor might affect cigarette smoking and lung cancer simultaneously, but it seems unlikely that it also affects car accidents.

\citet{lipsitch2010negative} is a recent article on negative outcomes. 
\citet{rosenbaum1989role} discussed the role of known effects in causal inference.

\subsection{Using negative exposures}

Negative exposures are duals of negative outcomes. 
Assume $Z^\textup{n}$ is a treatment variable similar to $Z$ and shares the same confounding structure as $Z$. If we believe $Z \ind Y  (z) \mid X$, then we tend to believe
$
Z^\textup{n}\ind Y  (z) \mid X.
$
Moreover, we know, a priori, the effect of $Z^\textup{n}$ on $Y $ 
$$
\tau(Z^\textup{n}\rightarrow Y  )
= 
E\{  Y(1^\textup{n}) - Y (0^\textup{n}) \}.
$$
An important example is that $\tau(Z^\textup{n}\rightarrow Y  ) = 0.$ A causal diagram satisfying these requirements is below:

$$
\xymatrix{
Z^\textup{n} & & X \ar[ll] \ar[ld]\ar[rd] \\
&Z  \ar[rr] && Y 
}
$$

\begin{example}
\citet{sanderson2017negative} give many examples of negative exposures in determining the effect of intrauterine exposure on later outcomes by comparing the association of maternal exposure during pregnancy with the outcome of interest, with the association of paternal exposure with the same outcome. They review studies on the effect of maternal and paternal smoking on offspring outcomes, and studies on the effect of maternal and paternal BMI on later offspring BMI and autism spectrum disorder.  In these examples, we expect the association of maternal exposure with the outcome to be larger than that of paternal exposure with the outcome. 
\end{example}

\subsection{Summary}

The unconfoundedness assumption is fundamentally untestable without additional assumptions. Although negative outcomes and negative controls in observational studies cannot prove or disprove unconfoundedness, using them in supplementary analyses can strengthen the evidence for causation. However, it is often non-trivial to conduct this type of supplementary analysis because it involves more data and more importantly, a deeper understanding of the causal problems to find convincing negative outcomes and negative controls.

\section{Problems of over-adjustment}
\label{sec::pearl-bias}

We have discussed many methods for estimating causal effects under the unconfoundedness assumption:
$$
Z\ind \{ Y(1), Y(0) \} \mid X .
$$
This is an assumption conditioning on $X$. It is crucial to select the right set of $X$ that ensures the conditional independence. 
\citet{rosenbaum2002design} wrote that ``there is no reason to avoid adjustment for a variable describing subjects before treatment.'' Similarly, \citet{rubin2007design} wrote that ``typically, the more conditional an assumption, the more acceptable it is.'' Both argued that we should control for all observed pretreatment covariates. \citet{vanderweele2011new} called it the {\it pretreatment criterion}. 
Pearl disagreed with this recommendation and gave two counterexamples below.

\subsection{M-bias}
\label{section::m-bias}

M-bias appears in the following causal diagram with an M-structure:
$$
\xymatrix{
U_1 \ar[dd]\ar[dr] &&U_2 \ar[dd]\ar[dl] \\
&X& \\
Z   && Y 
}
$$

We can read from the diagram the data-generating process:
$$
\left\{ 
\begin{aligned}
&U_1 \ind U_2 ,\\
&X = f_X(U_1, U_2, \varepsilon_X),\\
&Z=   f_Z(U_1,  \varepsilon_Z) , \\
&Y=      Y(z) =  f_Y(U_2, \varepsilon_Y) ,
\end{aligned}
\right. 
$$
where $(\varepsilon_X, \varepsilon_Z, \varepsilon_Y) $ are independent random error terms. In the above causal diagram, $X$ is observed, but $U_1$ and $U_2$ are unobserved. 
If we change the value of $Z$, the value of $Y$ will not change at all. So the true causal effect of $Z$ on $Y$ must be $0$. From the data-generating equations, we can read that $Z\ind Y$, so the association between $Z$ and $Y$ is $0$, and, in particular, 
$$
\tau_\textsc{PF} = E(Y\mid Z=1) - E(Y\mid Z=0) = 0.
$$
This means that without adjusting for the covariate $X$, the simple estimator is unbiased for the true parameter. 

However, if we condition on $X$, then $U_1\nind  U_2 \mid X$, and consequently, $Z\nind Y\mid X$ and
$$
\int \{  E(Y\mid Z=1, X=x) - E(Y\mid Z=0, X=x) \} f(x) \diff x \neq 0
$$
in general. 
To gain intuition, we consider the case with Normal linear models\footnote{It is not ideal for our discussion of binary $Z$, but it simplifies the derivations. \citet{ding2015adjust} gave a detailed discussion with more natural models for binary $Z$. }:
$$
\left\{ 
\begin{aligned}
&X = a U_1 + b U_2 +  \varepsilon_X,\\
&Z=   cU_1 +  \varepsilon_Z , \\
&Y=      Y(z) =  dU_2 +  \varepsilon_Y ,
\end{aligned}
\right. 
$$
where $(U_1, U_2, \varepsilon_X, \varepsilon_Z, \varepsilon_Y)  \iidsim \N01$. 
We have 
$$
\cov(Z,Y) = \cov(  cU_1 +  \varepsilon_Z,  dU_2 +  \varepsilon_Y) = 0,
$$
but by the result in Problem \ref{hw::partial-correlation}, the partial correlation coefficient between $Z$ and $Y$ given $X$ is\footnote{The notation $\propto$ reads as ``proportional to'' and allows us to drop some unimportant constants.}
\begin{eqnarray*}
\rho_{ZY\mid X} &=& \frac{   \rho_{ZY} - \rho_{ZX} \rho_{YX}  }{  \sqrt{  1-\rho_{ZX}^2 }  \sqrt{1- \rho_{YX}^2  }  }   
\\
&\propto &  - \rho_{ZX} \rho_{YX}  \\
&\propto &  - \cov(Z, X) \cov(Y, X)  \\
&=&  - abcd,
\end{eqnarray*}
the product of the coefficients on the path from $Z$ to $Y$. 
So the unadjusted estimator is unbiased but the adjusted estimator has a bias proportional to $abcd$.

The following simple example illustrates M-bias.

\begin{lstlisting}
> ## M bias with large sample size
> n  = 10^6
> U1 = rnorm(n)
> U2 = rnorm(n)
> X  = U1 + U2 + rnorm(n)
> Y  = U2 + rnorm(n)
> ## with a continuous treatment Z
> Z  = U1 + rnorm(n)
> round(summary(lm(Y ~ Z))$coef[2, 1], 3)
[1] 0
> round(summary(lm(Y ~ Z + X))$coef[2, 1], 3)
[1] -0.2
> 
> ## with a binary treatment Z
> Z  = (Z >= 0)
> round(summary(lm(Y ~ Z))$coef[2, 1], 3)
[1] 0.002
> round(summary(lm(Y ~ Z + X))$coef[2, 1], 3)
[1] -0.42
\end{lstlisting}

\subsection{Z-bias}
\label{sec::z-bias}

Consider the following causal diagram:
$$
\xymatrix{
&&&U \ar[dl]_b \ar[dr]^c \\
X \ar[rr]^a &&Z\ar[rr]^\tau   && Y 
}
$$
with the data generating process\footnote{Again, we generate continuous $Z$ from a linear model to simplify the derivations.  \citet{ding2017instrumental} extended the theory to more general causal models, especially for binary $Z$. 
}
$$
\left\{ 
\begin{aligned}
&Z=   aX + bU +   \varepsilon_Z , \\
&Y(z)=   \tau z + cU +  \varepsilon_Y ,
\end{aligned}
\right. 
$$
where $(U, X, \varepsilon_Z, \varepsilon_Y) $ are IID  $ \N01$. In this data-generating process, we have
$X\ind U,$ $X\nind Z$, and $X$ affects $Y$ only through $Z$.

The unadjusted estimator is
\begin{eqnarray*}
\tau_\textup{unadj} &=& \frac{ \cov(Z, Y)  }{  \var(Z) }  \\
&=& \frac{ \cov(Z, \tau Z + cU)  }{ \var(Z) }  \\
&=& \tau + \frac{c \cov(aX + bU, U)  }{  \var(Z) }  \\
&=& \tau + \frac{c b }{  a^2 + b^2 + 1} ,
\end{eqnarray*}
which has bias $bc/(a^2 + b^2 + 1)$. The adjusted estimator from the OLS of $Y$ on $(Z,X)$ satisfies
$$
\left\{ 
\begin{aligned}
E\{  Z(Y- \tau_\textup{adj} Z  - \alpha X  )  \} &= 0,\\
E\{  X(Y - \tau_\textup{adj} Z - \alpha X)  \} &= 0.
\end{aligned}
\right. 
$$
Solve the above linear system of $(\tau_\textup{adj}, \alpha)$ to obtain the formula of $  \tau_\textup{adj}$:
\begin{equation}
\label{eq::zbias-result}
  \tau_\textup{adj}  = \tau +  \frac{    bc }{   b^2+1 },
\end{equation}
which has bias $bc/(b^2+1)$. I relegate the details for solving the linear system of $(\tau_\textup{adj}, \alpha)$ as Problem \ref{hw::z-bias-formula}. 

So the unadjusted estimator has a smaller bias than the adjusted estimator. More interestingly, the stronger the association between $X$ and $Z$ is (measured by $a$), the larger the bias of the adjusted estimator is.

The mathematical derivation is not extremely hard. But this type of bias seems rather mysterious. Here is the intuition. The treatment is a function of $X$, $U$, and other random errors. If we condition on $X$, it is merely a function of  $U$ and other random errors. Therefore, conditioning on $X$ makes $Z$ less random, and more critically, makes the unmeasured confounder $U$ play a more important role in $Z$. Consequently, the confounding bias due to $U$ is amplified by conditioning on $X$. This idealized example illustrates the danger of over-adjusting for some covariates.

 \citet{heckman2004using} observed this phenomenon in simulation studies. 
 \citet[][technical report in 2006]{wooldridge2009should} verified it in linear models. 
 \citet{pearl2010class, pearl2011invited} explained it using causal diagrams. 
 \citet{ding2017instrumental} provided a more general theory as well as some intuition for this phenomenon. 
 This type of bias is called Z-bias because, in Pearl's original papers, he used the symbol $Z$ for our variable $X$. Throughout the book, however, $Z$ is used for the treatment variable. In Part \ref{part::instrumentalvariables} of this book,  we will call $Z$ an instrumental variable if it satisfies the causal diagram presented in this subsection. This justifies the {\it instrumental variable bias} as another name for this type of bias.

The following simple example illustrates Z-bias.
\begin{lstlisting}
> ## Z bias with large sample size
> n  = 10^6
> X = rnorm(n)
> U = rnorm(n)
> Z = X + U + rnorm(n)
> Y = U + rnorm(n)
> 
> round(summary(lm(Y ~ Z))$coef[2, 1], 3)
[1] 0.333
> round(summary(lm(Y ~ Z + X))$coef[2, 1], 3)
[1] 0.501
> 
> ## stronger association between X and Z
> Z = 2*X + U + rnorm(n)
> round(summary(lm(Y ~ Z))$coef[2, 1], 3)
[1] 0.167
> round(summary(lm(Y ~ Z + X))$coef[2, 1], 3)
[1] 0.501
> 
> ## even stronger association between X and Z
> Z = 10*X + U + rnorm(n)
> round(summary(lm(Y ~ Z))$coef[2, 1], 3)
[1] 0.01
> round(summary(lm(Y ~ Z + X))$coef[2, 1], 3)
[1] 0.5
\end{lstlisting}

\subsection{What covariates should we adjust for in observational studies?}

We never know the true underlying data-generating process which can be quite complicated. However, the following causal diagram helps to clarify many ideas. It already rules out the possibility of $M$-bias discussed in Section \ref{section::m-bias}. 
$$
\xymatrix{
 &&X_R&& \\
X_Z \ar[ddr] && X \ar[ddl] \ar[ddr]&& X_Y \ar[ddl] \\
 &\\
 &Z \ar[rr] \ar[ddr] &&Y \ar[ddl]& \\
 & \\
 &&X_I&&
}
$$
The covariates above have different features: 
\begin{enumerate}
\item
$X$ affects both the treatment and the outcome. Conditioning on $X$ ensures the unconfoundedness assumption, so we should control for $X$.
\item
$X_R$ is pure random noise not affecting either the treatment or the outcome. Including it in the analysis does not bias the estimate but introduces unnecessary variability in finite samples.
\item
$X_Z$ is an instrumental variable that affects the outcome only through the treatment. In the diagram above, including it in the analysis does not bias the estimate but increases the variability of the estimate. However, with unmeasured confounding, including it in the analysis amplifies the bias as shown in Section \ref{section::m-bias}. 
\item
$X_Y$ affects the outcome only but not the treatment. Without conditioning on it, the unconfoundedness assumption still holds. Since they are predictive of the outcome, including them in analysis often improves precision.
\item
$X_I$ is affected by the treatment and outcome. It is a post-treatment variable, not a pretreatment covariate. We should not include it if the goal is to infer the effect of the treatment on the outcome. We will discuss issues with post-treatment variables in causal inference in Part \ref{part::post-treatmentvariable} of this book. 
\end{enumerate}

If we believe the above causal diagram, we should adjust for at least $X$ to remove bias and more ideally, further adjust for $X_Y$ to reduce variance.

\section{Homework Problems}

\paragraph{More details for the formula of Z-bias}\label{hw::z-bias-formula}
Verify \eqref{eq::zbias-result}.

\paragraph{Cochran's formula or the omitted variable bias formula}
\label{hw::cochran+ovb}

Sir David Cox calls the following result {\it Cochran's formula} \citep{cochran1938omission, cox2007generalization} and econometricians call it the {\it omitted variable bias formula} \citep{angrist2008mostly}. 
A special case appeared in \citet{fisher1925statistical}.
It is also a counterpart of the Frisch--Waugh--Lovell theorem in Chapter \ref{appendix::fwl-theorem}.  

The formula has two versions. All vectors below are column vectors. 

\begin{enumerate}
[(1)]
\item
(Population version) Assume $(y_i, x_{1i}, x_{2i})_{i=1}^n$ are IID, where $y_i$ is a scalar, $x_{i1}$ has dimension $K$, and $x_{i2}$ has dimension $L$. 

We have the following OLS decompositions of random variables
\begin{eqnarray}
y_i &=& \beta_1 \tran x_{i1} + \beta_2  \tran x_{2i} + \varepsilon_i , \label{eq::long}\\
y_i&=&\gamma  \tran x_{i1} + e_i, \label{eq::short}\\
x_{i2} &=& \delta  \tran x_{i1} + v_i . \label{eq::inter}
\end{eqnarray}
Equation \eqref{eq::long} is called the long regression, and Equation \eqref{eq::short} is called the short regression. In Equation \eqref{eq::inter}, $\delta$ is a matrix because it is a regression of a vector on a vector. You can view \eqref{eq::inter} as regression of each component of $x_{i2}$ on $x_{i1}$. 

Show that $\gamma = \beta_1 + \delta \beta_2.$

\item
(Sample version) We have an $n\times 1$ vector $Y$, an $n\times K$ matrix $X_1$, and an $n\times L$ matrix $X_2$. We do not assume any randomness. All results below are purely linear algebra. 

We can obtain the following OLS fits: 
\begin{eqnarray*}
Y &=& X_1 \hat{\beta}_1 + X_2 \hat{\beta}_2+ \hat{\varepsilon},\\
Y &=& X_1 \hat{\gamma} + \hat{e} ,\\
X_2 &=& X_1 \hat{\delta} + \hat{v},
\end{eqnarray*}
where $\hat{\varepsilon},  \hat{e}, \hat{v}$ are the residuals. 
Again, the last OLS fit means the OLS fit of each column of $X_2$ on $X_1$, and therefore the residual $\hat{v}$ is an $n\times L$ matrix.

Show that $\hat{\gamma} = \hat{\beta}_1 +  \hat{\delta} \hat{\beta}_2$.

\end{enumerate}

Remark: The product terms  $\delta \beta_2$ and $\hat{\delta} \hat{\beta}_2$ are often referred to as the omitted-variable bias at the population level and sample level, respectively.

 \paragraph{Recommended reading} 
 
\citet{imbens2020potential} reviews and compares the roles of potential outcomes and causal diagrams for causal inference.

\chapter{E-Value: Evidence for Causation in Observational Studies with Unmeasured Confounding}
\chaptermark{E-Value}\label{chapter::evalue}

All the methods discussed in Part \ref{part::observational-studies}  rely crucially on the ignorability assumption. They require controlling for all confounding between the treatment and outcome. However, we cannot use the data to validate the ignorability assumption. Observational studies are often criticized due to the possibility of unmeasured confounding. The famous Yule--Simpson Paradox reviewed in Chapter \ref{chapter::correlationassociation} demonstrates that an unmeasured binary confounder can completely overturn an observed association between the treatment and outcome. However, to overturn a larger observed association, this unmeasured confounder must have a stronger association with the treatment and the outcome. In other words, not all observational studies are created equal. Some provide stronger evidence for causation than others.

The following three chapters will discuss various sensitivity analysis techniques that can quantify the evidence of causation based on observational studies in the presence of unmeasured confounding. This chapter starts with the E-value, introduced by \citet{vanderweele2017sensitivity} based on the theory in \citet{ding2016sensitivity}. It is more useful for observational studies using logistic regressions to estimate the conditional risk ratio of a treatment on a binary outcome. Chapter \ref{chapter::sensitivity-ACE} discusses sensitivity analysis for the average causal effect based on outcome regression,  IPW, and doubly robust estimation.  Chapter \ref{sec::rosenbaum-sensitivity-analysis} discusses Rosenbaum's framework for sensitivity analysis for matched observational studies.

\section{Cornfield-type sensitivity analysis}

Although we do not assume ignorability given $X$:
$$
Z\nind \{  Y(1), Y(0) \} \mid X ,
$$
we still assume latent ignorability given $X$ and an unmeasured confounder $U$: 
$$
Z\ind \{  Y(1), Y(0) \} \mid (X, U) . 
$$
The technique in this chapter works the best for a binary outcome $Y$ although it can be extended to other non-negative outcomes \citep{ding2016sensitivity}. Focus on binary $Y$ now. The true conditional causal effect on the risk ratio scale is defined as 
$$
\RR_{ZY\mid x}^\true = \frac{  \pr\{ Y(1) = 1\mid X=x  \} }{ \pr\{ Y(0) = 1\mid X=x  \} },
$$
and the observed conditional risk ratio equals 
$$
\RR_{ZY\mid x}^\obs = \frac{  \pr( Y  = 1\mid Z=1, X=x  ) }{ \pr( Y  = 1\mid  Z=0, X=x  ) }.
$$
In general, with an unmeasured confounder $U$,  they are different:
$$
\RR_{ZY\mid x}^\true  \neq \RR_{ZY\mid x}^\obs 
$$ 
because
$$
\RR_{ZY\mid x}^\true = \frac{  \int  \pr( Y = 1\mid Z=1, X=x, U=u  )  f(u \mid X=x) \diff u }
{ \int \pr( Y = 1\mid Z=0,  X=x , U=u  )  f(u \mid X=x) \diff u }
$$
and
$$
\RR_{ZY\mid x}^\obs = \frac{ \int   \pr( Y  = 1\mid Z=1, X=x , U=u ) f(u \mid Z=1, X=x) \diff u   }
{ \int  \pr( Y  = 1\mid  Z=0, X=x, U=u  ) f(u \mid Z=0, X=x) \diff u  } 
$$
are averaged over different distributions of $U$.

Historically,  \citet{doll1950smoking} found that the risk ratio of cigarette smoking on lung cancer was $9$ even after adjusting for many observed covariates $X$\footnote{Their original analysis was based on a case-control study and estimated the odds ratio of cigarette smoking on lung cancer. But the risk ratio is close to the odds ratio since lung cancer is a rare outcome; see Proposition \ref{prop::2x2-independence}.}. \citet{Fisher::1957} criticized their result to be noncausal because it is possible that a hidden gene simultaneously causes cigarette smoking and lung cancer although the true causal effect of cigarette smoking on lung cancer is absent. This is the {\it common cause} hypothesis, also discussed by \citet{reichenbach1991direction}. \citet{Cornfield::1959} took a more constructive perspective and asked: how strong this unmeasured confounder must be to explain away the observed association between cigarette smoking and lung cancer? Below we will use  \citet{ding2016sensitivity}'s general formulation of the problem.

Consider the following causal diagram:
$$
\xymatrix{
 &U \ar[dl] \ar[dr] \\
 Z   && Y 
}
$$
which conditions on $X$. So $Z\ind Y\mid (X, U)$. Conditioning on $X$ and $U$, we observe no association between $Z$ and $Y$; but conditioning on only $X$, we observe an association between $Z$ and $Y$.  Although we can allow $U$ to be general as \citet{ding2016sensitivity}, we assume that $U$ is binary to simplify the presentation.

Define two sensitivity parameters:
$$
\RR_{ZU\mid x} = \frac{ \pr(U=1\mid Z=1,X=x) }{ \pr(U=1\mid Z=0,X=x)  } 
$$
measures the treatment-confounder association, 
and
$$
\RR_{UY\mid x} = \frac{  \pr(Y=1\mid U=1, X=x) }{  \pr(Y=1\mid U=0, X=x) },
$$
measures the confounder-outcome association, conditional on covariates $X=x$.  We can show the main result below.

\begin{theorem}\label{thm::bounding-factor}
Under $Z\ind Y\mid (X, U)$, assume 
\begin{equation}\label{eq::largerthan1-evalue}
\RR_{ZY\mid x}^\obs > 1, \quad
 \RR_{ZU\mid x} > 1,\quad 
 \RR_{UY\mid x} > 1.
\end{equation}
We have 
$$
\RR_{ZY\mid x}^\obs \leq \frac{ \RR_{ZU\mid x}  \RR_{UY\mid x} }{  \RR_{ZU\mid x} + \RR_{UY\mid x} - 1} .
$$
\end{theorem}

In Theorem \ref{thm::bounding-factor}, we assume \eqref{eq::largerthan1-evalue}  without loss of generality. If $\RR_{ZY\mid x}^\obs < 1$, we can relabel the treatment and control levels to ensure $\RR_{ZY\mid x}^\obs > 1$. If $  \RR_{ZU\mid x}  < 1$, we can redefine the unmeasured confounder $U$ as $1-U$ to ensure $   \RR_{ZU\mid x}  > 1$. If $\RR_{ZY\mid x}^\obs > 1$ and $ \RR_{ZU\mid x} > 1$, then $ \RR_{UY\mid x}  > 1$ holds automatically. I relegate this subtle technical detail to Problem \ref{hw::technical-largerthan1}.

Theorem \ref{thm::bounding-factor} shows the upper bound of the observed risk ratio of the treatment on the outcome if the conditional independence $Z\ind Y\mid (X, U)$ holds. Under this conditional independence assumption, the association between the treatment and the outcome is purely due to the association between the treatment and the confounder $\RR_{ZU\mid x} $, and the association between the confounder and the outcome, $\RR_{UY\mid x}$. The upper bound equals $\RR_{ZU\mid x} \times  \RR_{UY\mid x} /  (\RR_{ZU\mid x} + \RR_{UY\mid x} - 1)$. 
A similar inequality appeared in \citet{lee2011bounding}. 
It is also related to Cochran's formula or the omitted-variable bias formula for linear models, which was reviewed in Problem \ref{hw::cochran+ovb}.

Reversely, to generate a certain value of the observed risk ratio $\RR_{ZY\mid x}^\obs$, the two confounding measures $\RR_{ZU\mid x} $ and $\RR_{UY\mid x}$ cannot be arbitrary. Their function $\RR_{ZU\mid x}  \times  \RR_{UY\mid x} /  (\RR_{ZU\mid x} + \RR_{UY\mid x} - 1)$ must be at least at large as $\RR_{ZY\mid x}^\obs$. 

I will give the proof of Theorem \ref{thm::bounding-factor}  below.

\begin{myproof}{Theorem}{\ref{thm::bounding-factor}}
Define 
$$ 
f_{1,x} = \pr(U=1\mid Z=1,X=x),\quad
f_{0,x}  = \pr(U=1\mid Z=0,X=x) .
$$ 
We can decompose $\RR_{ZY\mid x}^\obs$ as
\begin{eqnarray*}
&&\RR_{ZY\mid x}^\obs  \\
&=& \frac{  \pr( Y  = 1\mid Z=1, X=x  ) }{ \pr( Y  = 1\mid  Z=0,  X=x  ) } \\
&=& \frac{ \left[ \splitfrac{ \pr(U=1\mid Z=1,  X=x) \pr( Y  = 1\mid Z=1, U=1,  X=x  ) }{+ \pr(U=0\mid Z=1,  X=x) \pr( Y  = 1\mid Z=1, U=0,  X=x  )} \right]}
{ \left[ \splitfrac{\pr(U=1\mid Z=0,  X=x) \pr( Y  = 1\mid  Z=0, U=1,  X=x  )}{  +  \pr(U=0\mid Z=0,  X=x) \pr( Y  = 1\mid  Z=0, U=0,  X=x  ) } \right]  } \\
&=& \frac{ \left[ \splitfrac{ \pr(U=1\mid Z=1,  X=x) \pr( Y  = 1\mid U=1,  X=x  ) }{+ \pr(U=0\mid Z=1,  X=x) \pr( Y  = 1\mid U=0,  X=x  )}\right] }
{  \left[ \splitfrac{ \pr(U=1\mid Z=0,  X=x) \pr( Y  = 1\mid   U=1,  X=x  )  }{+  \pr(U=0\mid Z=0,  X=x) \pr( Y  = 1\mid  U=0,  X=x  ) }\right]  } \\
&=& \frac{  f_{1,x} \RR_{UY\mid x} + 1- f_{1,x}  }{ f_{0,x} \RR_{UY\mid x} + 1- f_{0,x}  } \\
&=& \frac{  (\RR_{UY\mid x} - 1) f_{1,x} + 1  }{      \frac{  \RR_{UY\mid x} - 1}{  \RR_{ZU\mid x} }    f_{1,x} + 1  } .
\end{eqnarray*}
We can verify that $\RR_{ZY\mid x}^\obs$ is increasing in $f_{1,x} $ using the result in Problem \ref{hw::technical-for-evalue-increasing}. So letting $f_{1,x} = 1$, we have
$$
\RR_{ZY\mid x}^\obs \leq \frac{  (\RR_{UY\mid x} - 1) + 1  }{      \frac{  \RR_{UY\mid x} - 1}{  \RR_{ZU\mid x} }    + 1  }
= \frac{ \RR_{ZU\mid x}  \RR_{UY\mid x} }{  \RR_{ZU\mid x} + \RR_{UY\mid x} - 1} .
$$
\end{myproof}

In the proof of Theorem \ref{thm::bounding-factor}, we have obtain an identity
\begin{equation}
\label{eq::identity-evalue}
\RR_{ZY\mid x}^\obs = \frac{  (\RR_{UY\mid x} - 1) f_{1,x} + 1  }{      \frac{  \RR_{UY\mid x} - 1}{  \RR_{ZU\mid x} }    f_{1,x} + 1  } .
\end{equation}
But this identity involves three parameters 
$$
\{  f_{1,x}, \RR_{ZU\mid x},  \RR_{UY\mid x}  \} ;
$$
see Problem \ref{hw::Schlesselman1978formula} for a related formula. In contrast, the upper bound in Theorem \ref{thm::bounding-factor} involves only two parameters 
$$
\{  \RR_{ZU\mid x},  \RR_{UY\mid x}  \}
 $$
 which measure the strength of the confounder.
 Mathematically, the identity \ref{eq::identity-evalue} is stronger than the inequality in Theorem \ref{thm::bounding-factor}. However, \ref{eq::identity-evalue}  involves more sensitivity parameters compared with Theorem \ref{thm::bounding-factor}. Therefore, we face a trade-off of accuracy and convenience in sensitivity analysis.

\section{E-value}

Lemma \ref{lemma::betaw1w2} below is useful for deriving interesting corollaries of Theorem \ref{thm::bounding-factor}. I relegate its proof to Problem \ref{hw::proof-lemma-evalue}. 

\begin{lemma}
\label{lemma::betaw1w2}
Define $\beta(w_1, w_2) = w_1 w_2/(w_1 + w_2 - 1)$ for $w_1 > 1$ and $w_2 > 1$.  
\begin{enumerate}
[(1)]
\item
$\beta(w_1, w_2)$ is symmetric in $w_1$ and $w_2$;
\item
$\beta(w_1, w_2)$ is increasing in both $w_1$ and $w_2$;
\item\label{beta2}
$\beta(w_1, w_2) \leq w_1$ and $\beta(w_1, w_2) \leq w_2$;
\item\label{beta3}
$\beta(w_1, w_2) \leq w^2/(2w-1)$, where $w = \max(w_1, w_2)$.
\end{enumerate}
\end{lemma}

Using Theorem \ref{thm::bounding-factor} and Lemma \ref{lemma::betaw1w2}\eqref{beta2}, we have
$$
 \RR_{ZU\mid x} \geq \RR_{ZY\mid x}^\obs, \quad  \RR_{UY\mid x}  \geq \RR_{ZY\mid x}^\obs , 
$$
or, equivalently, 
$$
\min( \RR_{ZU\mid x} ,  \RR_{UY\mid x} ) \geq \RR_{ZY\mid x}^\obs.
$$
Therefore, to explain away the observed relative risk, both confounding measures $\RR_{ZU\mid x} $ and $  \RR_{UY\mid x}$ must be at least as large as $\RR_{ZY\mid x}^\obs$.
\citet{Cornfield::1959} first derived the inequality $\RR_{ZU\mid x} \geq \RR_{ZY\mid x}^\obs$, also called the {\it Cornfield inequality} \citep{gastwirth1998cornfield}. \citet{Schlesselman::1978} derived the inequality $\RR_{UY\mid x}  \geq \RR_{ZY\mid x}^\obs$. 
These are related to to the {\it data processing inequality} in information theory\footnote{In information theory, the {\it mutual information}  
$$
I(A,B) = \iint p(a,b) \log_2 \frac{  p(a,b) }{p(a)p(b)} \text{d} a \text{d} b
$$ 
measures the dependence between two random variables $A$ and $B$, where $p(\cdot)$ denotes the joint or marginal density of $(A,B)$. The {\it data processing inequality}  is a famous result: if $Z\ind Y \mid U$, then $I(Z,Y) \leq I(Z,U)$ and $I(Z,Y) \leq I(U,Y)$.
Lihua Lei and Bin Yu both pointed out to me the connection between Cornfield's inequality and the data processing inequality. 
}.

If we define $w = \max( \RR_{ZU\mid x} ,  \RR_{UY\mid x} )$, then we can use Theorem \ref{thm::bounding-factor} and Lemma \ref{lemma::betaw1w2}\eqref{beta3} to obtain
\begin{eqnarray*}
&&w^2/(2w-1) \geq \beta( \RR_{ZU\mid x} , \RR_{UY\mid x}) \geq \RR_{ZY\mid x}^\obs \\ 
&\Longrightarrow & w^2 - 2 \RR_{ZY\mid x}^\obs w + \RR_{ZY\mid x}^\obs \geq 0,
\end{eqnarray*}
which is a quadratic inequality. One root $  \RR_{ZY\mid x}^\obs - \sqrt{ \RR_{ZY\mid x}^\obs (\RR_{ZY\mid x}^\obs - 1)} $ is always smaller than or equal to $1$, so we have
$$
w = \max( \RR_{ZU\mid x} ,  \RR_{UY\mid x} ) \geq \RR_{ZY\mid x}^\obs + \sqrt{ \RR_{ZY\mid x}^\obs (\RR_{ZY\mid x}^\obs - 1)} . 
$$
Therefore, to explain away the observed relative risk, the maximum of the confounding measures $\RR_{ZU\mid x} $ and $  \RR_{UY\mid x}$ must be at least as large as $\RR_{ZY\mid x}^\obs + \sqrt{ \RR_{ZY\mid x}^\obs (\RR_{ZY\mid x}^\obs - 1)} $.  
Based on this result, \citet{vanderweele2017sensitivity} introduced the following notion of  E-value for measuring the {\it evidence} of causation with observational studies.

\begin{definition}
[E-Value]
With the observed conditional risk ratio $\RR_{ZY\mid x}^\obs$, define the E-Value as 
$$
\RR_{ZY\mid x}^\obs + \sqrt{ \RR_{ZY\mid x}^\obs (\RR_{ZY\mid x}^\obs - 1)} . 
$$ 
\end{definition}

The E-value is defined for the parameter  $\RR_{ZY\mid x}^\obs$. In practice, $\RR_{ZY\mid x}^\obs$ is estimated with sampling error. We can calculate the E-value based on the estimated $\RR_{ZY\mid x}^\obs$, as well as the corresponding E-values for the lower and upper confidence limits of $\RR_{ZY\mid x}^\obs$.

Fisher's $p$-value measures the evidence for causal effects in randomized experiments. We have discussed the $p$-value based on the FRT in Part \ref{part::rcts} of this book. However, in observational studies with large sample sizes, $p$-values can be a poor measure of evidence for causal effects. Even if the true causal effects are $0$, a tiny amount of unmeasured confounding can bias the estimate, which can result in extremely small $p$-values given the small sampling uncertainty. The sampling uncertainty is usually secondary in observational studies with large sample sizes, but the uncertainty due to unmeasured confounding is often the first-order problem that does not diminish with increased sample sizes.  \citet{vanderweele2017sensitivity} argued that the E-value is a better measure of the evidence for causal effects in observational studies.

\section{A classic example}
\label{sec::evalue-example}

I revisit a classic example below.

\begin{figure}[t]
\centering
\includegraphics[width = 0.8 \textwidth]{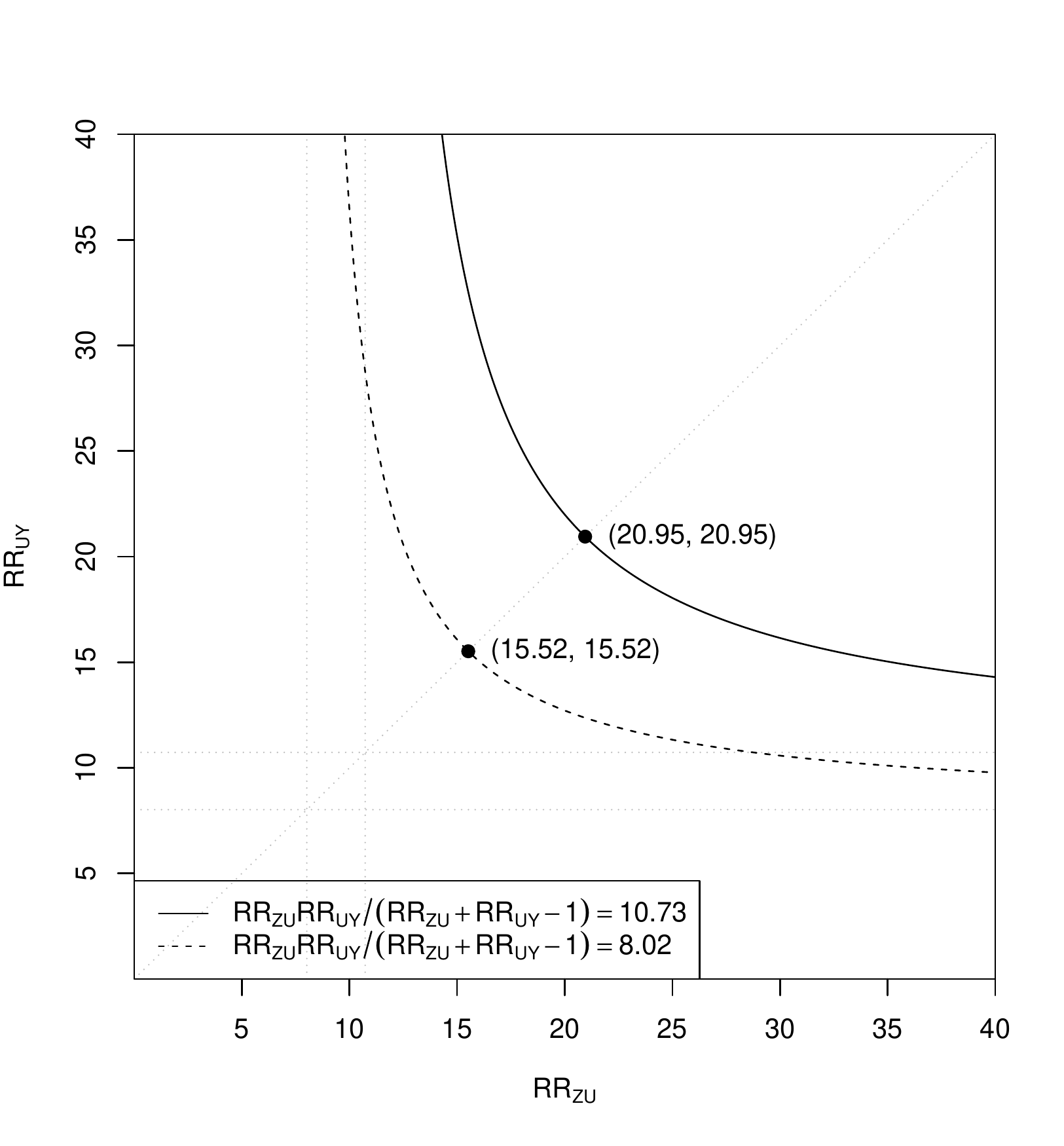}
\caption{Magnitude of confounding to explain away the observed risk ratio based on the data from \citet{Hammond::1968}}\label{fg::hammond-e-value}
\end{figure}

\begin{example}\label{eg::hammond-evalue}
\citet{Hammond::1968} used the U.S. population to study the cigarette smoking and lung cancer relationship. Ignoring covariates, their data can be represented by a two-by-two table below:

\begin{center}
\begin{tabular}{ccc}
\hline 
 & Lung cancer & No lung cancer \\
 \hline 
 Smoker & 397 & 78557 \\
 Non-smoker & 51 & 108778  \\ 
 \hline 
\end{tabular}
\end{center}

Based on the data, they obtained an estimate of the risk ratio $10.73$ with a 95\% confidence interval $[8.02, 14.36]$ (see Section \ref{section::asymptotic-inference-2x2} for the formulas). To explain away the point estimate, the E-value is 
$$
10.73 + \sqrt{10.73\times (10.73-1)} = 20.95;
$$
to explain away the lower confidence limit, the E-value is 
$$
8.02 +  \sqrt{8.02\times (8.02 - 1)} = 15.52.
$$

Figure \ref{fg::hammond-e-value} shows the joint values of the two confounding measures to explain away the point estimate and lower confidence limit of the risk ratio. In particular, to explain away the point estimate, they must lie in the area above the solid curve; to explain away the lower confidence limit, they must lie in the area above the dashed curve.\footnote{Based on some simple algebra, the solid curve and dashed curve are both hyperbolas.} 
\end{example}

The simple \ri{R} code below  computes the numbers in Example \ref{eg::hammond-evalue}:
 \begin{lstlisting}
> ## e-value based on RR
> evalue = function(rr)
+ {
+   rr + sqrt(rr*(rr - 1))
+ }
> 
> p1  = 397/(397+78557)
> p0  = 51/(51+108778)
> rr  = p1/p0
> logrr = log(p1/p0)
> se    = sqrt(1/397+1/51-1/(397+78557)-1/(51+108778))
> upper = exp(logrr+1.96*se)
> lower = exp(logrr-1.96*se)
> 
> ## point estimate
> rr
[1] 10.72978
> ## lower CI
> lower
[1] 8.017414
> ## e-value based on rr
> evalue(rr)
[1] 20.94733
> ## e-value based on lower CI
> evalue(lower)
[1] 15.51818
 \end{lstlisting}

\section{Extensions}

\subsection{E-value and Bradford Hill's criteria for causation}

The E-value provides evidence for causation. But evidence is not a proof. With a larger E-value, we need a stronger unmeasured confounder to explain away the observed risk ratio; the evidence for causation is stronger. With a smaller E-value, we need a weaker unmeasured confounder to explain away the observed risk ratio; the evidence for causation is weaker. Coupled with the discussion in Section \ref{sec::evalue-monotone}, a larger observed risk ratio has stronger evidence for causation. This is closely related to Sir Bradford Hill's first criterion for causation: {\it strength} of the association \citep{hill1965environment}. Theorem \ref{thm::bounding-factor} provides a mathematical quantification of his heuristic argument.

In a famous paper,  \citet{hill1965environment} proposed a set of nine criteria to provide evidence for causation between a presumed cause and outcome.

\begin{definition}\label{def::hill-causation}
Bradford Hill gave nine criteria for causation:
\begin{enumerate}
\item
strength;
\item
consistency;
\item
specificity;
\item
temporality;
\item
biological gradient;
\item
plausibility;
\item
coherence;
\item
experiment;
\item
analogy.
\end{enumerate}
\end{definition}

The E-value is a way to justify his first criterion. 
That is, stronger association often provides stronger evidence for causation because to explain away stronger association, we need stronger confounding measures. 
We have discussed randomized experiments in Part \ref{part::rcts}, which corroborates his eighth criterion. Due to the space limit, I omit the detailed discussion of his other criteria and encourage the readers to read  \citet{hill1965environment}. Recently, this paper has been re-printed as \citet{hill2020environment} with insightful comments from many leading researchers in causal inference.

\subsection{E-value after logistic regression}

With a binary outcome, it is common for epidemiologists to use a logistic regression of the outcome $Y_i$ on the treatment indicator $Z_i$ and covariates $X_i$:
$$
\pr(Y_i = 1\mid Z_i, X_i) = \frac{  e^{\beta_0 + \beta_1 Z_i + \beta_2\tran X_i}  }{1  + e^{\beta_0 + \beta_1 Z_i + \beta_2\tran X_i}} .
$$
In the logistic model above, the coefficient of $Z_i$ is the log of the conditional odds ratio between the treatment and the outcome given the covariates:
$$
\beta_1 = \log \frac{   \pr(Y_i=1\mid Z_i =1, X_i=x)/  \pr(Y_i=0\mid Z_i =1, X_i=x)  }
{  \pr(Y_i=1\mid Z_i =0, X_i=x)/  \pr(Y_i=0\mid Z_i =0, X_i=x)  }  .
$$
Importantly, the logistic model assumes a common odds ratio across all values of the covariates. Moreover, when the outcome is rare in that $ \pr(Y_i=1\mid Z_i =1, X_i=x)$ and $\pr(Y_i=1\mid Z_i =0, X_i=x)$ are close to $0$, the conditional odds ratio approximates the conditional risk ratio (see Proposition \ref{prop::2x2-independence}(3)):
$$
\beta_1  \approx \log  \frac{   \pr(Y_i=1\mid Z_i =1, X_i=x)  }
{  \pr(Y_i=1\mid Z_i =0, X_i=x)  } = \log   \RR_{ZY\mid x}^\obs.
$$
Therefore, based on the estimated logistic regression coefficient and the corresponding confidence limits, we can calculate the E-value immediately. This is the leading application of the E-value.

\begin{example}
\label{eg::NCHS2003}
The \ri{NCHS2003.txt} contains the National Center for Health Statistics birth certificate data, with the following binary indicator variables useful for us:

\begin{tabular}{ll}
\hline 
\texttt{PTbirth} &
 pre-term birth \\
\texttt{preeclampsia} &
 pre-eclampsia \\
\texttt{ageabove35}& 
 an older mother with age $\geq 35$ (the treatment) \\
\texttt{somecollege}&
 college education \\
\texttt{mar}& 
 marital status \\
\texttt{smoking}& 
 smoking status \\
\texttt{drinking}& 
 drinking status \\
\texttt{hispanic}& 
 mother's ethnicity  \\
\texttt{black}& 
 mother's ethnicity  \\
\texttt{nativeamerican}& 
 mother's ethnicity \\
\texttt{asian}& 
 mother's ethnicity\\
 \hline 
\end{tabular}

This version of the data is from \citet{valeri2014estimation}. This example focuses on the outcome \ri{PTbirth} and Problem \ref{hw::eg::NCHS2003} focuses on the outcome \ri{ pre-eclampsia}, a multisystem hypertensive disorder of pregnancy. The following \ri{R} code computes the E-values after fitting a logistic regression. Based on the E-values, we conclude that to explain away the point estimate, the maximum confounding measure must be larger than 1.94, and to explain away the lower confidence limit, the maximum confounding measure must be larger than 1.91. Although these confounding measures are not as strong as those in Section \ref{sec::evalue-example}, they appear to be fairly large in epidemiologic studies. 
\end{example}

The simple \ri{R} code below computes the numbers in Example \ref{eg::NCHS2003}: 
 \begin{lstlisting}
> NCHS2003 = read.table("NCHS2003.txt", header = TRUE, sep = "\t")
> ## outcome: PTbirth
> y_logit = glm(PTbirth ~ ageabove35 + 
+                 mar + smoking + drinking + somecollege + 
+                 hispanic + black + nativeamerican + asian,
+               data = NCHS2003, 
+               family = binomial)
> log_or   = summary(y_logit)$coef[2, 1:2]
> est      = exp(log_or[1])
> lower.ci = exp(log_or[1] - 1.96*log_or[2])
> est
Estimate 
1.305982 
> evalue(est)
Estimate 
1.938127 
> lower.ci
Estimate 
1.294619 
> evalue(lower.ci)
Estimate 
1.912211 
\end{lstlisting}

\subsection{Non-zero true causal effect}
 
Theorem \ref{thm::bounding-factor} assumes no true causal effect of the treatment on the outcome. \citet{ding2016sensitivity} proved a general theorem allowing for a non-zero true causal effect.

\begin{theorem}
\label{thm::general-evalue}
Modify the definition of  $\RR_{UY\mid x}   $ as
$$
 \RR_{UY\mid x}  =  \max_{z=0,1} \frac{  \pr(Y=1\mid Z=z, U=1, X=x) }{  \pr(Y=1\mid Z=z, U=0, X=x) }.
$$
Assume \eqref{eq::largerthan1-evalue}. 
We have 
$$
\RR_{ZY\mid x}^\true \geq 
\RR_{ZY\mid x}^\obs \Big /  \frac{ \RR_{ZU\mid x}  \RR_{UY\mid x} }{  \RR_{ZU\mid x} + \RR_{UY\mid x} - 1} .
$$
\end{theorem}

Theorem \ref{thm::bounding-factor} is a special case of  Theorem \ref{thm::general-evalue} with $\RR_{ZY\mid x}^\true   = 1.$
See the original paper of \citet{ding2016sensitivity} for the proof of Theorem \ref{thm::general-evalue}.  Without assuming any additional assumptions, Theorem \ref{thm::general-evalue} states a lower bound of the true risk ratio $\RR_{ZY\mid x}^\true$ given the observed risk ratio $\RR_{ZY\mid x}^\obs$ and the two sensitivity parameters $\RR_{ZU\mid x} $ and $ \RR_{UY\mid x}$.

When the treatment is apparently preventive to the outcome, the observed risk ratio is smaller than 1. In this case, Theorems \ref{thm::bounding-factor} and \ref{thm::general-evalue} are not directly useful, and we must re-label the treatment levels and calculate the E-value based on $1/ \RR_{ZY\mid x}^\obs$.

\section{Critiques and responses}

Since the original paper was published,  the E-value has become a standard number reported in many epidemiologic studies. Nevertheless, it also attracted critiques \citep{ioannidis2019limitations}. I will review some limitations of E-values below.

\subsection{E-value is just a monotone transformation of the risk ratio}
\label{sec::evalue-monotone}

\begin{figure}
\includegraphics[width=\textwidth]{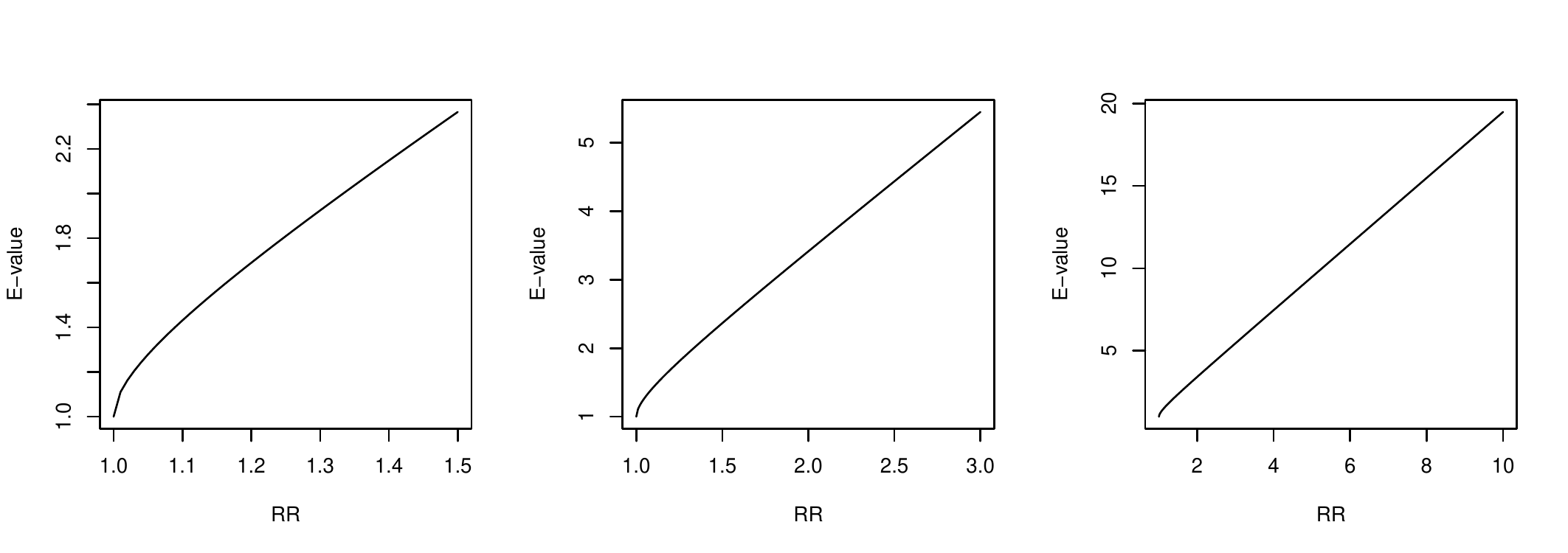}
\caption{E-value as a monotone transformation of the risk ratio: three figures have different ranges of the risk ratio.}\label{fig::evalue-rr}
\end{figure}

From Figure \ref{fig::evalue-rr}, we can see that if the risk ratio is large, then the E-value $\RR_{ZY\mid x}^\obs + \sqrt{ \RR_{ZY\mid x}^\obs (\RR_{ZY\mid x}^\obs - 1)}$ is nearly $2 \RR_{ZY\mid x}^\obs$ which is linear in the risk ratio. For a small risk ratio, the E-value is more nonlinear in $ \RR_{ZY\mid x}^\obs$. Critics often say that the E-value is merely a monotone transformation of the point estimator or the confidence limits of the risk ratio. So it does not provide any additional information.

This is partially true. Indeed, the E-value is entirely based on the point estimator or the confidence limits of the risk ratio. However, it has a meaningful interpretation based on Theorem \ref{thm::bounding-factor}: to explain away the observed risk ratio, the maximum of the confounding measures must be at least as large as the E-value.

%
%

\subsection{Calibration of the E-value}

The E-value equals the maximum value of the association between the confounder and the treatment and that between the confounder and the outcome to completely explain away an observed association. An obvious problem is that this confounder is fundamentally latent. So it is not trivial to decide whether a certain E-value is large or small. 
Another related problem is that the E-value depends on how many observed covariates $X$ we have controlled for since it quantifies the strength of the residual confounding given $X$. Therefore, E-values across studies are not directly comparable. The E-value provides evidence for causation but this evidence should be assessed carefully based on background knowledge of the problem of interest.

The following leave-one-covariate-out approach is an intuitive approach to calibrating the E-value. With $X = (X_1, \ldots, X_p)$, we can pretend that the component $X_j$ were not observed and compute the $Z$-$X_j$ and $X_j$-$Y$ risk ratios given other observed covariates $(j=1, \ldots, p)$. These risk ratios provide the range for the confounding measures due to $U$ if we believe that the unmeasured $U$ is not as strong as some of the observed covariates. However, I am not aware of any formal justification for this approach.

\subsection{It works the best for  a binary outcome and the risk ratio}

Theorem \ref{thm::bounding-factor} works well for a binary outcome and the risk ratio.  \citet{ding2016sensitivity} also proposed sensitivity analysis methods for other causal parameters, including the risk difference for binary outcomes, the mean ratio for non-negative outcomes, and the hazard ratio for survival outcomes. However, they are not as elegant as the E-value for binary outcomes based on the risk ratio. The next chapter will propose a simple sensitivity analysis method for the average causal effect that can include the outcome regression, IPW, and doubly robust estimators in Part \ref{part::observational-studies} as special cases.

\section{Homework Problems}

\paragraph{A technical lemma for the proof of Theorem \ref{thm::bounding-factor}}
\label{hw::technical-for-evalue-increasing}

Show that 
$$
f(x) = \frac{ax+1}{bx+1}
$$
is increasing in $x$ if $a>b$ and decreasing in $x$ is $a<b$.

\paragraph{Technical assumption for Theorem \ref{thm::bounding-factor}}
\label{hw::technical-largerthan1}

Revisit the proof of Theorem \ref{thm::bounding-factor}. Assume $Z\ind Y \mid (X, U)$. Show that if $ \RR_{ZU\mid x} > 1$ but $ \RR_{UY\mid x} < 1$, then $\RR_{ZY\mid x}^\obs < 1$.

Remark: 
This result is intuitive. Condition on $X.$ It says that if the $Z$-$U$ relationship is positive and the $U$-$Y$ relationship is negative, then the $Z$-$Y$ relationship is negative given the conditional independence of $Z$ and $Y$ given $U$. 
Based on this result, if we assume $\RR_{ZY\mid x}^\obs > 1$ and $ \RR_{ZU\mid x} > 1$, then $ \RR_{UY\mid x} > 1$ must be true. 
Therefore, the third condition in \eqref{eq::largerthan1-evalue} is in fact redundant.

\paragraph{Lemma \ref{lemma::betaw1w2}}\label{hw::proof-lemma-evalue}

Prove Lemma \ref{lemma::betaw1w2}.

\paragraph{\citet{Schlesselman::1978}'s formula}\label{hw::Schlesselman1978formula}

For simplicity, we condition on $X$ implicitly in this problem. 
Consider a binary treatment $Z$, outcome $Y$, and unmeasured confounder $U$.
Assume a common risk ratio of the treatment on the outcome within both $U=0$ and $U=1$:
$$
 \RR_{ZY|U=0} = \RR_{ZY|U=1}, 
$$
and also a common risk ratio of the confounder on the outcome within both $Z=0$ and $Z=1$:
$$  
\RR_{UY|Z=0} = \RR_{UY|Z=1} , \text{ denoted by }\gamma.
$$

Show that  
$$
\frac{\RR_{ZY}^{\obs}}{  \RR_{ZY}^{\true} } = \frac{1 + (  \gamma  -1)\pr(U=1\mid Z=1)  }
{ 1 + (  \gamma -1)\pr(U=1\mid Z=0) } .
$$

Remark: First verify that if  $\RR_{ZY|U=0} = \RR_{ZY|U=1}$ then 
$$
\RR_{ZY}^{\true}  = \RR_{ZY|U=0} = \RR_{ZY|U=1} . 
$$
This identity shows the {\it collapsibility} of the risk ratio. 
In epidemiology, the risk ratio is a {\it collapsible} measure of association.

\citet{Schlesselman::1978}'s formula does not assume conditional independence $Z\ind Y\mid U$, but assumes homogeneity of the $Z$-$Y$ and $U$-$Y$ risk ratios. It is a classic formula for sensitivity analysis. It is an identity that is simple to implement with pre-specified
$$
\{   \gamma  ,  \pr(U=1\mid Z=1), \pr(U=1\mid Z=0)  \} .
$$
However, it involves more sensitivity parameters than Theorem \ref{thm::bounding-factor}. Even though Theorem \ref{thm::bounding-factor} only gives an inequality, it is not a loose inequality compared to \citet{Schlesselman::1978}'s formula under stronger assumptions. With Theorem \ref{thm::bounding-factor},  \citet{Schlesselman::1978}'s formula is only of historical interest.

\paragraph{E-value after logistic regression: data analysis}\label{hw::eg::NCHS2003}

This problem uses the same dataset as Example \ref{eg::NCHS2003}. 

Report the E-value for the outcome \ri{preeclampsia}.

\paragraph{Cornfield-type inequalities for the risk difference}\label{hw::cornfield-rd}

Consider binary $Z, Y, U$, and condition on $X$ implicitly. Assume latent ignorability given $U$.  Show that under $Z\ind Y\mid U$, we have
\begin{eqnarray}
\RD_{ZY}^\obs = \RD_{ZU} \times \RD_{UY}
\label{eq::cornfield-rd}
\end{eqnarray}
where $\RD_{ZY}^\obs$ is the observed risk difference of $Z$ on $Y$, and $\RD_{ZU}$ and $ \RD_{UY}$ are the treatment-confounder and confounder-outcome risk differences, respectively (recall the definition of the risk difference in Chapter \ref{sec::twobytwotables-withresume}).

Remark: Without loss of generality, assume that $\RD_{ZY}^\obs, \RD_{ZU} ,\RD_{UY}$ are all positive. Then \eqref{eq::cornfield-rd} implies that 
$$
\min(\RD_{ZU}  , \RD_{UY}) \geq \RD_{ZY}^\obs
$$
and
$$
\max(\RD_{ZU}  , \RD_{UY}) \geq  \sqrt{ \RD_{ZY}^\obs} .
$$
These are the Cornfield inequalities for the risk difference with a binary confounder.
They show that for an unmeasured confounder to explain away an observed risk difference $\RD_{ZY}^\obs$, the treatment-confounder and confounder-outcome risk differences must both be larger than $\RD_{ZY}^\obs$, and the maximum of them must be larger than the square root of $\RD_{ZY}^\obs$.

\citet{Cornfield::1959} obtained, but did not appreciate the significance of \eqref{eq::cornfield-rd}. \citet{gastwirth1998cornfield} and \citet{poole2010origin} discussed the first Cornfield condition for the risk difference, and \citet{ding2014generalized} discussed the second. 

\citet{ding2014generalized}  also derived more general results without assuming a binary $U$. Unfortunately, the results for a general $U$ are weaker than those above for a binary $U$, that is, the inequalities become looser with more levels of $U$. This motivated \citet{ding2016sensitivity} to focus on the Cornfield inequalities for the risk ratio, which do not deteriorate with more levels of $U$.

\paragraph{Recommended reading}

\citet{ding2016sensitivity}  extended and unified the Cornfield-type sensitivity analysis, which is the theoretical basis for the notion of E-value.

\chapter{Sensitivity Analysis for the Average Causal Effect with Unmeasured Confounding}
\chaptermark{Sensitivity Analysis for Average Causal Effect}\label{chapter::sensitivity-ACE}

Cornfield-type sensitivity analysis works best for binary outcomes on the risk ratio scale, conditional on the observed covariates. Although  \citet{ding2016sensitivity} also proposed Cornfield-type sensitivity analysis methods for the average causal effect, they are not general enough and are not convenient to apply. Below I give a more direct approach to sensitivity analysis based on the conditional expectations of the potential outcomes. The advantage of this approach is that it can deal with commonly used estimators for the average causal effect under the sensitivity analysis framework. 
The idea appeared in the early work of \citet{robins1999association} and \citet{scharfstein1999adjusting}. This chapter is based on \citet{luding2023flexible}'s recent formulation.

 The approach is closely related to the idea of deriving worse-case bounds on the average potential outcomes. I will first review the simpler idea of bounds, and then extend the approach to sensitivity analysis.

\section{Introduction}

Recall the canonical setup of an observational study with  $\{  Z_i, X_i,  Y_i(1), Y_i(0) \}_{i=1}^n \iidsim  \{ Z, X, Y(1), Y(0) \}$ and focus on the average causal effect
$$
\tau = E\{ Y(1) - Y(0) \}.
$$
It decomposes to
\begin{eqnarray*}
\tau &=& \left[  E(Y\mid Z=1) \pr(Z=1) + E\{ Y(1)\mid Z=0 \} \pr(Z=0) \right]  \\
&&- \left[   E\{ Y(0)\mid Z=1\} \pr(Z=1) + E(Y\mid Z=0)  \pr(Z=0)  \right] . 
\end{eqnarray*}
So the fundamental difficulty is to estimate the counterfactual means 
$$
E\{ Y(1)\mid Z=0 \}  ,\quad \quad E\{ Y(0)\mid Z=1\}.
$$
There are in general two extreme strategies to estimate them.

We have discussed the first strategy in Part \ref{part::observational-studies}, which relies on ignorability. 
Assuming
\begin{eqnarray*}
E\{ Y(1)\mid Z=1, X \} &=& E\{ Y(1)\mid Z=0, X \} ,\\ 
E\{ Y(0)\mid Z=1, X \} &=& E\{ Y(0)\mid Z=0, X \}, 
\end{eqnarray*}
we can identify the counterfactual means by the observables:
$$
E\{ Y(1)\mid Z=0 \}  
= E\left\{  E( Y\mid Z=1, X )  \mid Z=0\right\}
$$
and, similarly, 
$$
E\{ Y(0)\mid Z=1\} = E\left\{  E( Y\mid Z=0, X )  \mid Z=1\right\} . 
$$

The second strategy in the next section assumes nothing except that the outcomes are bounded between $\underline{y}$ and $\overline{y}$. 
This is natural for binary outcomes with $\underline{y} = 0$ and $\overline{y} = 1$. 
With this assumption, the two counterfactual means are also bounded between $\underline{y}$ and $\overline{y}$, which implies the worst-case bounds on $\tau$. 
Table \ref{table::manski-bounds} illustrates the basic idea and Chapter \ref{sec::manski-bounds} below reviews this strategy in more detail.

\begin{table}
\centering
\caption{Science Table with bounded outcome $ [\underline{y}, \overline{y}]$, where $\underline{y}$ and $\overline{y}$ are two constants}\label{table::manski-bounds}
\begin{tabular}{ccccccc}
\hline 
$Z$ & $Y(1)$ & $Y(0)$ &  Lower $Y(1)$ & Upper $Y(1)$ & Lower $Y(0)$ & Upper $Y(0)$ \\
\hline 
$1$ & $Y_1(1)$ & ? & $Y_1(1)$  & $Y_1(1)$ & $\underline{y}$ & $\overline{y}$ \\
$\vdots$&$\vdots$&$\vdots$&$\vdots$&$\vdots$&$\vdots$&$\vdots$   \\
$1$ & $Y_{n_1}(1)$ & ? &  $Y_{n_1}(1)$  & $Y_{n_1}(1)$ & $\underline{y}$ & $\overline{y}$ \\
\hline 
$0$ & ? & $Y_{n_1+1}(0)$ & $\underline{y}$ & $\overline{y}$& $Y_{n_1+1}(0)$ & $Y_{n_1+1}(0)$ \\ 
$\vdots$&$\vdots$&$\vdots$&$\vdots$&$\vdots$&$\vdots$&$\vdots$ \\ 
$0$ & ? & $Y_{n}(0)$ & $\underline{y}$ & $\overline{y}$& $Y_{n}(0)$ & $Y_{n}(0)$ \\
\hline 
\end{tabular}
\end{table}

\section{Manski-type worse-case bounds on the average causal effect without assumptions}\label{subsection::manski-bounds}
\label{sec::manski-bounds}

Assume that the outcome is bounded between $\underline{y}$ and $\overline{y}$.
From the decomposition
$$
E\{  Y(1) \} = E\{  Y(1) \mid Z=1\} \pr(Z=1)  +  E\{  Y(1) \mid Z=0\} \pr(Z=0)  ,
$$
we can derive that $E\{  Y(1) \}$ has lower bound
$$
E\{  Y \mid Z=1\} \pr(Z=1)  +   \underline{y} \pr(Z=0) 
$$
and upper bound
$$
E\{  Y \mid Z=1\} \pr(Z=1)  +  \overline{y} \pr(Z=0).
$$
Similarly, from the decomposition
$$
E\{  Y(0) \} = E\{  Y(0) \mid Z=1\} \pr(Z=1)  +  E\{  Y(0) \mid Z=0\} \pr(Z=0)  ,
$$
we can derive that $E\{  Y(0) \}$ has lower bound
$$
\underline{y}\pr(Z=1)  +  E\{  Y \mid Z=0\} \pr(Z=0) 
$$
and upper bound
$$
 \overline{y} \pr(Z=1)  +  E\{  Y \mid Z=0\} \pr(Z=0). 
$$
Combining these bounds, we can derive that the average causal effect $\tau = E\{  Y(1) \} - E\{  Y(0) \}$ has the lower bound
$$
E\{  Y \mid Z=1\} \pr(Z=1)  +   \underline{y} \pr(Z=0) - \overline{y} \pr(Z=1)  -  E\{  Y \mid Z=0\} \pr(Z=0) 
$$
and the upper bound
$$
 E\{  Y \mid Z=1\} \pr(Z=1)  +  \overline{y} \pr(Z=0) - \underline{y}\pr(Z=1)  -   E\{  Y \mid Z=0\} \pr(Z=0). 
$$ 
The length of the bounds is $\overline{y}-\underline{y}$. The bounds are not informative but are better than the a priori bounds $[\underline{y}-\overline{y}, \overline{y}-\underline{y}]$ with length $2(\overline{y}-\underline{y})$. Without further assumptions, the observed data distribution does not uniquely determine $\tau$. In this case, we say that $\tau$ is {\it partially identified}, with the formal definition below.

\begin{definition}
[partial identification]\label{def::partial-identification}
A parameter $\theta$ is partially identified if the observed data distribution is compatible with multiple values of $\theta$. 
\end{definition}

Compare Definitions \ref{def::nonparametric-identification} and \ref{def::partial-identification}. If the parameter $\theta$ is uniquely determined by the observed data distribution, then it is identifiable; otherwise, it is only partially identifiable.  Therefore, $\tau$ is identifiable with the ignorability assumption, but only partially identifiable without the ignorability assumption.

\citet{cochran1953} used the idea of worse-case bounds in surveys with missing data but abandoned the idea because it often gives very conservative results. Similarly, the above worst-case bounds on $\tau$ are often uninteresting from a practical perspective because they often cover $0$.  Moreover, this strategy does not apply to the settings with unbounded outcomes.

Manski applied the idea to causal inference \citep{manski1990nonparametric} and many other econometric models \citep{manski2003partial}. 
This idea of bounding causal parameters with minimal assumptions is powerful when coupled with other qualitative assumptions. \citet{manski2003partial} surveyed many strategies. 
For instance, we may believe that the treatment does not harm any units, so the monotonicity assumption holds:
$
Y(1)\geq Y(0) . 
$
Then the lower bound on $\tau$ is zero but the upper bound is unchanged. 
Another type of assumption is 
$Z = I\{  Y(1) \geq Y(0) \}$, that is, the treatment selection is based on the difference between the latent potential outcomes. This assumption can also improve the bounds on $\tau$. 
A more detailed discussion of this approach is beyond the scope of this book.

\section{Sensitivity analysis for the average causal effect}

The first strategy is optimistic and assumes that the potential outcomes do not differ across treatment and control groups, conditional on the observed covariates. The second strategy is pessimistic and does not infer the counterfactual means based on the observed data at all. The following strategy is in-between.

\subsection{Identification formulas}

Define
\begin{eqnarray*}
\frac{ E\{ Y(1)\mid Z=1, X \} }{ E\{ Y(1)\mid Z=0, X \} } = \varepsilon_1(X), \\
\frac{ E\{ Y(0)\mid Z=1, X \} }{ E\{ Y(0)\mid Z=0, X \} } = \varepsilon_0(X) ,
\end{eqnarray*}
which are the sensitivity parameters. For simplicity, we can further assume that they are constant independent of $X$. In practice, we need to fix them or vary them in a pre-specified range. Recall that $\mu_1(X) = E(Y\mid Z=1, X)$ and $\mu_0(X) = E(Y\mid Z=0, X)$ are the conditional mean functions of the observed outcomes under treatment and control, respectively. We can identify the two counterfactual means and the average causal effect as follows. 

\begin{theorem}
\label{thm::outcome-reg-sensitivity}
With known $\varepsilon_1(X)$ and $ \varepsilon_0(X) $, we have 
\begin{eqnarray*}
E\{ Y(1)\mid Z=0 \}   &=& E\left\{\mu_1(X)   / \varepsilon_1(X)  \mid Z=0\right\},\\ 
E\{ Y(0)\mid Z=1 \}   &=& E\left\{\mu_0(X)    \varepsilon_0(X)  \mid Z=1\right\}
\end{eqnarray*} 
and therefore
\begin{eqnarray}
\tau & = & E\{ ZY + (1-Z) \mu_1(X)   / \varepsilon_1(X)\}  \nonumber  \\
 && - E\{ Z \mu_0(X)    \varepsilon_0(X) + (1-Z)Y\}  \label{eq::predictive}\\
& = & E\{ Z \mu_1(X) + (1-Z) \mu_1(X)   / \varepsilon_1(X)\}  \nonumber \\
&&- E\{ Z \mu_0(X)    \varepsilon_0(X) + (1-Z) \mu_0(X)\}. \label{eq::projective}
\end{eqnarray}
\end{theorem}

I leave the proof of Theorem \ref{thm::outcome-reg-sensitivity} to Problem \ref{hw::proof-theorem-1-sa}. 
With the fitted outcome model, \eqref{eq::predictive} and \eqref{eq::projective} motivate the following predictive and projective estimators for $\tau$:
\begin{eqnarray*}
\hat\tau^{\textup{pred}} &=&\left\{  n^{-1}\sumn Z_i Y_i + n^{-1} \sumn (1-Z_i) \hat{\mu}_1(X_i)   / \varepsilon_1(X_i)  \right\} \\
&&- \left\{ n^{-1}  \sumn Z_i \hat{\mu}_0(X_i)    \varepsilon_0(X_i) + n^{-1} \sumn (1-Z_i) Y_i \right\},
\end{eqnarray*}
and
\begin{eqnarray*}
\hat{\tau}^{\textup{proj}}&=&\left\{ n^{-1}\sumn Z_{i}\hat{\mu}_{1}(X_{i})+n^{-1}\sumn(1-Z_{i})\hat{\mu}_{1}(X_{i})/\varepsilon_{1}(X_{i})\right\} \\
&&-\left\{ n^{-1}\sumn Z_{i}\hat{\mu}_{0}(X_{i})\varepsilon_{0}(X_{i})+n^{-1}\sumn(1-Z_{i})\hat{\mu}_{0}(X_{i})\right\}.
\end{eqnarray*} 
The terminology ``predictive'' and ``projective'' is from the survey sampling literature \citep{firth1998robust, ding2018causal}; see also Chapter \ref{sec::understand-lin-predict}. The estimators $\hat\tau^{\textup{pred}} $ and $\hat{\tau}^{\textup{proj}}$ differ slightly: the former uses the observed outcomes when available, whereas the latter replaces the observed outcomes with the fitted values.

More interesting, we can also identify $\tau$ by an inverse probability weighting formula.

\begin{theorem}\label{thm::ipw-sensitivity}
With known $\varepsilon_1(X)$ and $ \varepsilon_0(X) $, we have 
\begin{eqnarray*} 
E\{ Y(1) \} = 
E\left\{ w_1(X) \frac{Z}{e(X)}  Y  \right\}  ,\quad
 E\{Y(0)\} = E\left\{ w_0(X) \frac{1-Z}{1-e(X)} Y \right\},
\end{eqnarray*}
where
$$
w_1(X) = e(X) + \{1-e(X)\}/ \varepsilon_1(X) ,\quad
w_0(X) = e(X)  \varepsilon_0(X) + 1- e(X). 
$$
\end{theorem}

I leave the proof of Theorem \ref{thm::ipw-sensitivity} to Problem \ref{hw::proof-theorem-2-sa}. 
Theorem \ref{thm::ipw-sensitivity} modifies the classic  IPW  formulas with two extra factors $w_1(X) $ and $w_0(X) $, which depend on both the propensity score and the sensitivity parameters. With the fitted propensity scores, Theorem \ref{thm::ipw-sensitivity} motivates the following estimators for $\tau$: 
\begin{eqnarray*} 
\hat\tau^{\textup{ht}} &=&  n^{-1} \sumn \frac{ \{ \hat{e}(X_i)\varepsilon_1(X_i) + 1-\hat{e}(X_i) \} Z_iY_i     }{  \varepsilon_1(X_i)  \hat e(X_i)  } \\
&&- n^{-1} \sumn  \frac{ \{  \hat e(X_i)  \varepsilon_0(X_i) + 1- \hat e(X_i)  \} (1-Z_i) Y_i }{ 1- \hat e(X_i)} 
\end{eqnarray*}
and
\begin{eqnarray*} 
\hat\tau^{\textup{haj}} &=&   \sumn \frac{ \{ \hat{e}(X_i)\varepsilon_1(X_i) + 1-\hat{e}(X_i) \} Z_iY_i     }{  \varepsilon_1(X_i)  \hat  e(X_i)  }  \Big / \sumn \frac{Z_i}{ \hat{e}(X_i)  }   \\
&&- n^{-1} \sumn  \frac{ \{  \hat  e(X_i)  \varepsilon_0(X_i) + 1- \hat e(X_i)  \} (1-Z_i) Y_i }{ 1- \hat e(X_i)} \Big / \sumn \frac{1-Z_i}{ 1- \hat{e}(X_i)  } .
\end{eqnarray*}

More interestingly, with fitted propensity score and outcome models, the following estimator for $\tau$ is doubly robust:
\begin{eqnarray*}
\hat{\tau}^{\textup{ht}} = \hat\tau^{\textup{ht}} 
- n^{-1} \sumn \{Z_i- \hat e(X_i)\}  \left\{   \frac{  \hat  \mu_1(X_i) }{  \hat e(X_i) \varepsilon_1(X_i)  }    + \frac{  \hat  \mu_0(X_i)\varepsilon_0(X_i)  }{ 1-  \hat  e(X_i) } \right\}.
\end{eqnarray*}
That is, with known $\varepsilon_1(X_i) $ and $\varepsilon_0(X_i) $, the estimator  $\hat{\tau}^{\textup{dr}} $ is consistent for $\tau$ if either the propensity score model or the outcome model is correctly specified. We can use the bootstrap to approximate the variance of the above estimators. See \citet{luding2023flexible} for technical details.

When $\varepsilon_1(X_i)  =  \varepsilon_0(X_i) = 1$, the above estimators reduce to the predictive estimator, IPW estimator, and the doubly robust estimators introduced in Part \ref{part::observational-studies}.

\subsection{Example}
\label{sec::eg-sensitivity-tau}

\subsubsection{R functions for sensitivity analysis}

The following \ri{R} function can compute the point estimates for sensitivity analysis. 

\begin{lstlisting}
OS_est_sa = function(z, y, x, out.family = gaussian, 
                     truncps = c(0, 1), e1 = 1, e0 = 1)
{
     ## fitted propensity score
     pscore   = glm(z ~ x, family = binomial)$fitted.values
     pscore   = pmax(truncps[1], pmin(truncps[2], pscore))
     
     ## fitted potential outcomes
     outcome1 = glm(y ~ x, weights = z, 
                    family = out.family)$fitted.values
     outcome0 = glm(y ~ x, weights = (1 - z), 
                    family = out.family)$fitted.values
     
     ## outcome regression estimator
     ace.reg  = mean(z*y) + mean((1-z)*outcome1/e1) - 
                   mean(z*outcome0*e0) - mean((1-z)*y) 
     ## IPW estimators
     w1 = pscore + (1-pscore)/e1
     w0 = pscore*e0 + (1-pscore)
     ace.ipw0 = mean(z*y*w1/pscore) - 
                   mean((1 - z)*y*w0/(1 - pscore))
     ace.ipw  = mean(z*y*w1/pscore)/mean(z/pscore) - 
                   mean((1 - z)*y*w0/(1 - pscore))/mean((1 - z)/(1 - pscore))
     ## doubly robust estimator
     aug = outcome1/pscore/e1 + outcome0*e0/(1-pscore)
     ace.dr   = ace.ipw0 + mean((z-pscore)*aug)

     return(c(ace.reg, ace.ipw0, ace.ipw, ace.dr))     
}
\end{lstlisting}

I relegate the calculation of the standard errors to Problem \ref{hw::sensitivity-se}. 

\subsubsection{Revisit Example \ref{eg::chan-bmi-ATE}}

With 
$$
\varepsilon_1(X) = \varepsilon_0(X) \in \{1/2, 1/1.7, 1/1.5, 1/1.3, 1, 1.3, 1.5, 1.7, 2\},
$$ 
we obtain an array of doubly robust estimates of $\tau$ based on the following \ri{R} code:
\begin{lstlisting}
> nhanes_bmi = read.csv("nhanes_bmi.csv")[, -1]
> z = nhanes_bmi$School_meal
> y = nhanes_bmi$BMI
> x = as.matrix(nhanes_bmi[, -c(1, 2)])
> x = scale(x)
> 
> 
> E1 = c(1/2, 1/1.7, 1/1.5, 1/1.3, 1, 1.3, 1.5, 1.7, 2)
> E0 = c(1/2, 1/1.7, 1/1.5, 1/1.3, 1, 1.3, 1.5, 1.7, 2)
> EST = outer(E1, E0)
> ll1 = length(E1)
> ll0 = length(E0)
> for(i in 1:ll1)
+   for(j in 1:ll0)
+     EST[i, j] = OS_est_sa(z, y, x, e1 = E1[i], e0 = E0[j])[4]
\end{lstlisting}

\begin{table}[ht]
\centering
\caption{Sensitivity analysis for the average causal effect}\label{table::sensitivity-ACE}
\begin{tabular}{rrrrrrrrrr}
  \hline
 & 1/2 & 1/1.7 & 1/1.5 & 1/1.3 & 1 & 1.3 & 1.5 & 1.7 & 2 \\ 
  \hline
1/2 & 11.62 & 10.44 & 9.40 & 8.03 & 4.96 & 0.97 & -1.69 & -4.35 & -8.34 \\ 
  1/1.7 & 9.22 & 8.05 & 7.00 & 5.64 & 2.57 & -1.42 & -4.08 & -6.75 & -10.74 \\ 
  1/1.5 & 7.63 & 6.45 & 5.41 & 4.05 & 0.97 & -3.02 & -5.68 & -8.34 & -12.33 \\ 
  1/1.3 & 6.03 & 4.86 & 3.81 & 2.45 & -0.62 & -4.61 & -7.27 & -9.94 & -13.93 \\ 
  1 & 3.64 & 2.47 & 1.42 & 0.06 & -3.01 & -7.01 & -9.67 & -12.33 & -16.32 \\ 
  1.3 & 1.80 & 0.63 & -0.42 & -1.78 & -4.85 & -8.85 & -11.51 & -14.17 & -18.16 \\ 
  1.5 & 0.98 & -0.19 & -1.24 & -2.60 & -5.67 & -9.66 & -12.33 & -14.99 & -18.98 \\ 
  1.7 & 0.36 & -0.82 & -1.86 & -3.23 & -6.30 & -10.29 & -12.95 & -15.61 & -19.60 \\ 
  2 & -0.35 & -1.52 & -2.57 & -3.93 & -7.00 & -10.99 & -13.65 & -16.32 & -20.31 \\ 
   \hline
\end{tabular}
\end{table}

Table \ref{table::sensitivity-ACE} presents the point estimates.  
The signs of the estimates are not sensitive to sensitivity parameters larger than 1, but they are quite sensitive to sensitivity parameters smaller than 1. When the participants of the meal plan tend to have higher BMI (that is, $\varepsilon_1(X) >1$ and $ \varepsilon_0(X)  > 1$), the average causal effect of the meal plan on BMI is negative. However, this conclusion can be quite sensitive if the participants of the meal plan tend to have lower BMI.

\section{Homework Problems}

\paragraph{Proof of Theorem \ref{thm::outcome-reg-sensitivity}}\label{hw::proof-theorem-1-sa}

Prove Theorem \ref{thm::outcome-reg-sensitivity}. 

\paragraph{Proof of Theorem \ref{thm::ipw-sensitivity}}\label{hw::proof-theorem-2-sa}

Prove Theorem \ref{thm::ipw-sensitivity}.

\paragraph{Standard errors in sensitivity analysis}\label{hw::sensitivity-se}

Chapter \ref{sec::eg-sensitivity-tau} only presents the point estimates. Report the corresponding bootstrap standard errors.

\paragraph{Sensitivity analysis for the average causal effect on the treated units $\tau_\textsc{T}$}\label{hw::att-sentivity-analysis}

This problem extends Chapter \ref{chapter::ATT-and-other} to allow for unmeasured confounding for estimating
$$
\tau_\textsc{T} = E\{  Y(1) - Y(0)  \mid Z=1  \} 
= E(Y\mid Z=1) - E\{  Y(0)  \mid Z=1  \} .
$$
We can easily estimate $E(Y\mid Z=1)$ by the sample moment, $\hat{\mu}_{\textsc{t}1} =   \sumn Z_i Y_i / \sumn Z_i$. The only counterfactual term is $E\{  Y(0)  \mid Z=1  \}$. Therefore, we only need the sensitivity parameter $ \varepsilon_0(X) $. We have the following two identification formulas with a known  $\varepsilon_0(X) .$

\begin{theorem}
\label{thm::att-reg-ipw}
With known $\varepsilon_0(X) $, we have 
 \begin{eqnarray*}
E\{  Y(0)  \mid Z=1  \}  &=& E\left\{  Z \mu_0(X)  \varepsilon_0(X)    \right\} /e \\
&=& E\left\{  e(X)\varepsilon_0(X) \frac{ 1-Z}{1-e(X)}  Y      \right\} /e,
\end{eqnarray*}
where $e = \pr(Z=1)$
\end{theorem}

Prove Theorem \ref{thm::att-reg-ipw}. 

Remark:  Theorem \ref{thm::att-reg-ipw} motivates using $\hat{\tau}_{\textsc{t}}^* = \hat{\mu}_{\textsc{t}1}  - \hat{\mu}_{\textsc{t}0}^*$ to estimate $\tau_{\textsc{t}}$, where  
 \begin{eqnarray*}
\hat{\mu}_{\textsc{t}0}^\text{reg} &=& n_1^{-1} \sumn Z_i  \varepsilon_0(X_i) \hat  \mu_0(X_i) ,   \\
\hat{\mu}_{\textsc{t}0}^\text{ht} &=& n_1^{-1} \sumn   \varepsilon_0(X_i) \hat  o(X_i) (1-Z_i) Y_i ,   \\
\hat{\mu}_{\textsc{t}0}^\text{haj} &=&  \sumn   \varepsilon_0(X_i) \hat  o(X_i) (1-Z_i) Y_i  /  \sumn   \hat  o(X_i) (1-Z_i) , 
\end{eqnarray*}
with $  \hat  o(X_i) = \hat  e(X_i) / \{1- \hat  e(X_i)\}$ being the estimated conditional odds of the treatment given the observed covariates. 
Moreover, we can construct the doubly robust estimator 
$\hat{\tau}_{\textsc{t}}^\text{dr} = \hat{\mu}_{\textsc{t}1}  - \hat{\mu}_{\textsc{t}0}^\text{dr}$ for $\tau_{\textsc{t}}$, where
$$
\hat{\mu}_{\textsc{t}0}^\text{dr} = 
\hat{\mu}_{\textsc{t}0}^\text{ht}  - n_1^{-1} \sumn \varepsilon_0(X_i) \frac{\hat e(X_i) - Z}{1-\hat e(X_i)}  \hat \mu_0(X_i)  . 
$$
\citet{luding2023flexible} provide more details.

\paragraph{R code}

Implement the estimators in Problem \ref{hw::att-sentivity-analysis}. 
Analyze the data used in Chapter \ref{sec::eg-sensitivity-tau}.

\paragraph{Recommended reading}

 \citet{Rosenbaum::1983JRSSB} and \citet{imbens2003sensitivity} are two classic papers on sensitivity analysis which, however, involve more complicated procedures.

\chapter{Rosenbaum-Style $p$-Values for Matched Observational Studies with Unmeasured Confounding}
\chaptermark{Sensitivity Analysis in Matching}\label{sec::rosenbaum-sensitivity-analysis}

\citet{rosenbaum1987sensitivity} introduced a sensitivity analysis technique for matched observational studies. Although it works for general matched studies \citep{rosenbaum2002design}, the theory is most elegant for one-to-one matching. 
Different from Chapters \ref{chapter::evalue} and \ref{chapter::sensitivity-ACE}, Rosenbaum-type sensitivity analysis works best for matched observational studies for testing the sharp null hypothesis of no individual treatment effect.

\section{A model for sensitivity analysis with matched data}

Consider a matched observational study, with $(i,j)$ indexing unit $j$ in pair $i$ $(i=1,\ldots, n; j=1,2)$. 
With exactly matched pairs, unit $(i, 1)$ and unit $(i, 2)$ have the same covariates $X_i$. 
Assume IID sampling, and extend Definition \ref{def::pscore} to define the propensity score as
$$
e_{ij} = \pr\{  Z_{ij} = 1\mid X_i, Y_{ij}(1), Y_{ij}(0) \} .
$$
Let $\mathbb{S}_i = \{Y_{i1}(1), Y_{i1}(0), Y_{i2}(1), Y_{i2}(0)\}$ denote the set of all potential outcomes within pair $i$. 
Conditioning on the event that $Z_{i1} + Z_{i2} = 1$, we have
\begin{eqnarray*}
\pi_{i1} &=& \pr\{ Z_{i1} = 1 \mid X_i, \mathbb{S}_i, Z_{i1} + Z_{i2} = 1 \} \\
&=& \frac{ \pr\{ Z_{i1} = 1, Z_{i2} = 0  \mid X_i, \mathbb{S}_i \} }
{ \pr\{ Z_{i1} + Z_{i2} = 1 \mid X_i, \mathbb{S}_i \} } \\
&=& \frac{ \pr\{ Z_{i1} = 1, Z_{i2} = 0  \mid X_i, \mathbb{S}_i \} }{ 
\pr\{ Z_{i1} = 1, Z_{i2} = 0 \mid X_i, \mathbb{S}_i \} + \pr\{ Z_{i1} = 0, Z_{i2} = 1 \mid X_i, \mathbb{S}_i \}
}\\
&=& \frac{  e_{i1} (1-e_{i2}) }{ e_{i1} (1-e_{i2}) + (1-e_{i1}) e_{i2}} 
\end{eqnarray*}
Define $o_{ij} = e_{ij} / (1 - e_{ij})$ as the odds of the treatment for unit $(i,j)$, and we have
$$
\pi_{i1}  = \frac{o_{i1}}{  o_{i1} + o_{i2} } . 
$$
Under ignorability, $e_{ij}$ is only a function of $X_i$, and therefore, $e_{i1} = e_{i2}$ and $\pi_{i1} = 1/2 $. Thus the treatment assignment mechanism conditional on covariates and potential outcomes is equivalent to that from an MPE with equal treatment and control probabilities. This is a strategy to analyze matched observational studies we discussed in Chapter \ref{sec::ideal-matching-mpe}.

In general, $e_{ij}$ is also a function of the unobserved potential outcomes, and it can range from $0$ to $1$. 
\citet{rosenbaum1987sensitivity}'s model for sensitivity analysis imposes bounds on the odds ratio $o_{i1}/o_{i2} $. 

\begin{assumption}[Rosenbaum's sensitivity analysis model]\label{assume::rosenbaum-sensitivity}
The odds ratios are bounded by 
$$
o_{i1}/o_{i2} \leq \Gamma,\quad
o_{i2}/o_{i1} \leq \Gamma, \quad (i=1, \ldots, n)
$$
for some pre-specified $\Gamma \geq  1$. Equivalently, 
$$ 
\frac{1}{1+\Gamma } \leq 
\pi_{i1} \leq \frac{\Gamma}{1+\Gamma} \quad (i=1, \ldots, n)
$$
for some pre-specified $\Gamma \geq  1$.
\end{assumption}

Under Assumption \ref{assume::rosenbaum-sensitivity}, we have a biased MPE with unequal and varying treatment and control probabilities across pairs. When $\Gamma = 1$,  we have $\pi_{i1} = 1/2 $ and thus a standard MPE. Therefore, $\Gamma > 1$ measures the deviation from the ideal MPE due to the omitted variables in matching.

 \section{Worst-case $p$-values under Rosenbaum's sensitivity analysis model}

Consider testing the sharp null hypothesis 
$$
\HF: Y_{ij}(1) = Y_{ij}(0) \text{ for } i = 1,\ldots, n \text{ and } j=1,2
$$
based on the within-pair differences
$
\hat{\tau}_i = (2Z_{i1} - 1) (Y_{i1} - Y_{i2}) \  (i=1,\ldots, n). 
$
Under $\HF$, $|\hat{\tau}_i|$ is fixed but $S_i = I(\hat{\tau}_i > 0)$ is random if $\hat{\tau}_i \neq 0$. Consider the following class of test statistics:
$$
T = \sumn S_i q_i,
$$
where $q_i\geq 0$ is a function of $ (|\hat{\tau}_1|, \ldots, |\hat{\tau}_n|   ).$ Special cases include the sign statistic, the pair $t$ statistic (up to some constant shift), and the Wilcoxon signed-rank statistic:
$$
T = \sumn S_i,\quad T = \sumn S_i |\hat{\tau}_i|,\quad T = \sumn S_i R_i,
$$
where $(R_1,\ldots, R_n)$ are the ranks of $ (|\hat{\tau}_1|, \ldots, |\hat{\tau}_n|   ).$ 

What is the null distribution of the test statistic $T$ under Assumption \ref{assume::rosenbaum-sensitivity} with a general $\Gamma$? It can be quite complicated because Assumption \ref{assume::rosenbaum-sensitivity}  does not fully specify the exact values of the $\pi_{i1}$'s. Fortunately, we do not need to know the exact distribution of $T$ but rather the worst-case distribution of $T$ that yields the largest $p$-value under Assumption \ref{assume::rosenbaum-sensitivity}. This worst-case distribution corresponds to
$$
S_i \iidsim  \text{Bernoulli}\left(  \frac{ \Gamma}{ 1+\Gamma} \right).
$$
The corresponding distribution of $T$ has mean 
$$
E_\Gamma( T ) = \frac{\Gamma}{1+\Gamma} \sumn q_i, 
$$
and variance 
$$
\var_\Gamma( T ) = \frac{\Gamma}{(1+\Gamma)^2} \sumn q_i^2,
$$
with a Normal approximation
$$
\frac{T -\frac{\Gamma}{1+\Gamma} \sumn q_i }{ \sqrt{ \frac{\Gamma}{(1+\Gamma)^2} \sumn q_i^2  }} \rightarrow  \N01 
$$
in distribution. 
In practice, we can report a sequence of $p$-values as a function of $\Gamma$.

\section{Examples}

\subsection{Revisiting the LaLonde data}\label{sec::rosenbaum-gamma-examples}

We conduct Rosenbaum-style sensitivity analysis in the matched LaLonde data. 
Using the \ri{Matching} package, we can construct the matched dataset. 
\begin{lstlisting}
library("Matching")
library("sensitivitymv")
library("sensitivitymw")
dat <- read.table("cps1re74.csv",header=T)
dat$u74 <- as.numeric(dat$re74==0)
dat$u75 <- as.numeric(dat$re75==0)
y = dat$re78
z = dat$treat
x = as.matrix(dat[, c("age", "educ", "black",
                      "hispan", "married", "nodegree",
                      "re74", "re75", "u74", "u75")])
matchest = Match(Y = y, Tr = z, X = x)
ytreated = y[matchest$index.treated]
ycontrol = y[matchest$index.control]
datamatched = cbind(ytreated, ycontrol)
\end{lstlisting}

\begin{figure}
\centering
\includegraphics[width =  \textwidth]{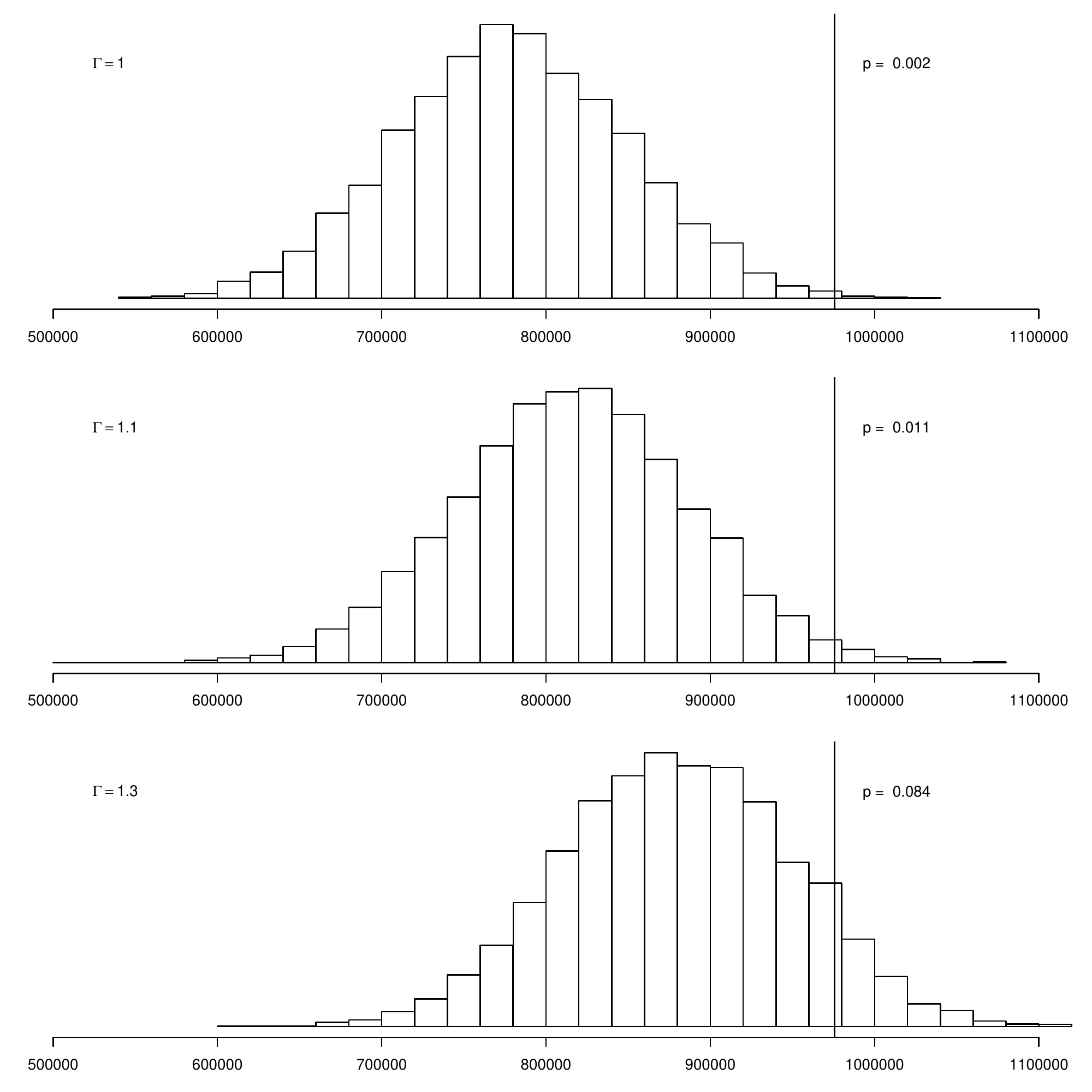}
\caption{The worst-case distributions of $T = \sumn S_i | \hat{\tau}_i |$ with $S_i$ IID Bernoulli$(\Gamma/(1+\Gamma))$, based on the matched LaLonde data.}\label{fg::lalonde_gamma}
\end{figure}

\begin{figure}
\centering
\includegraphics[width = \textwidth]{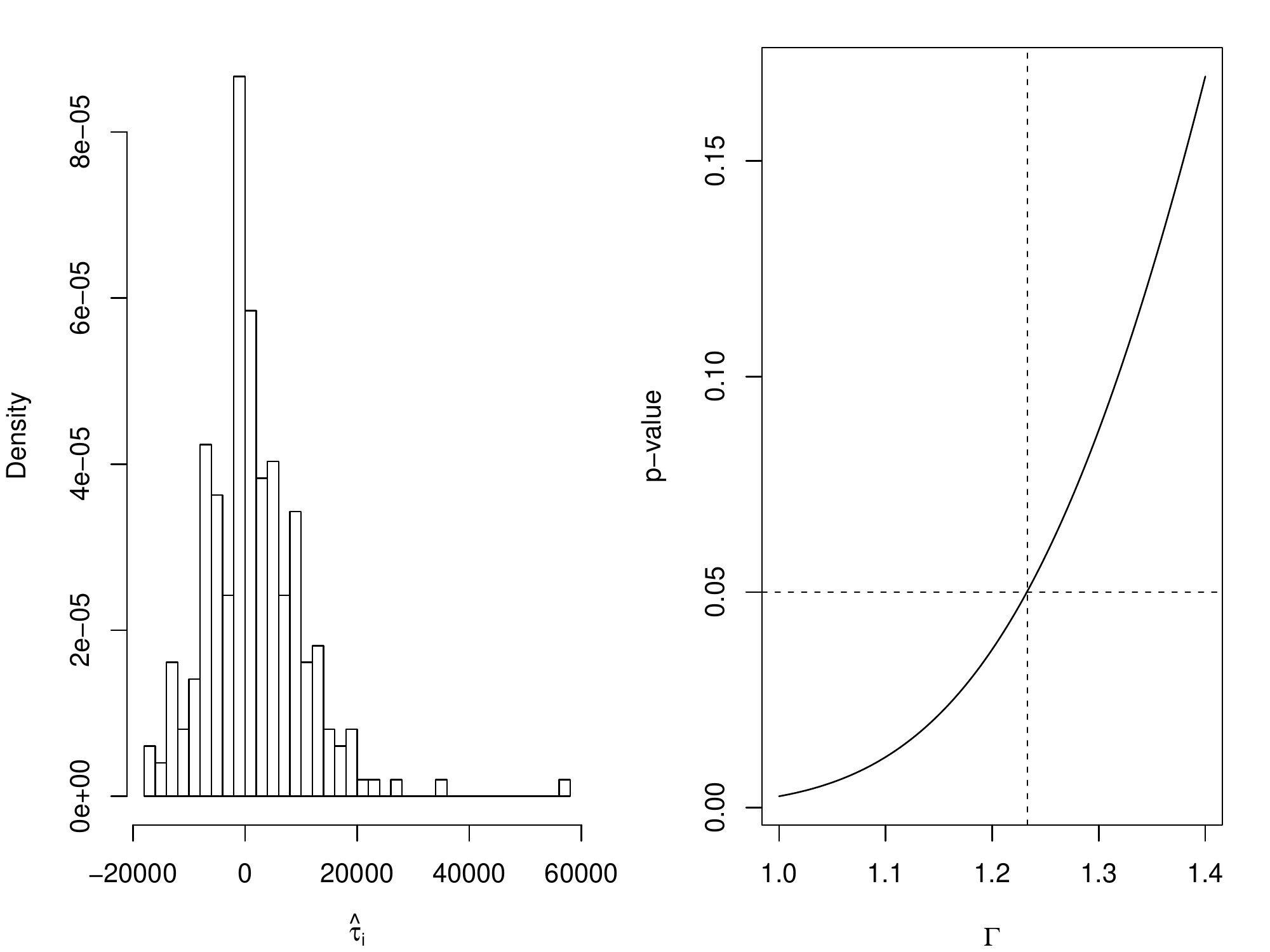}
\caption{$p$-value as a function of $\Gamma$, based on the matched LaLonde data.}\label{fg::p-gamma-lalonde}
\end{figure}

We consider using the test statistic $T = \sumn S_i | \hat{\tau}_i |$. Under the ideal MPE with $\Gamma = 1$, we can simulate the distribution of $T$ and obtain the $p$-value $0.002$, as shown in the first subfigure in Figure \ref{fg::lalonde_gamma}. With a slightly larger $\Gamma = 1.1$, the worst-case distribution of $T$ shifts to the right, and the $p$-value increases to $0.011$. If we further increase $\Gamma$ to $1.3$, then the worst-case distribution of $T$ shifts further and the $p$-value exceeds $0.05$. Figure \ref{fg::p-gamma-lalonde} shows the histogram of the $\hat{\tau}_i$'s and the $p$-value as a function of $\Gamma$; $\Gamma  = 1.233$ measures the maximum confounding that we can still reject the null hypothesis at level $0.05$. 

We can also use the \ri{senmw} function in the \ri{sensitivitymw} package to obtain a sequence of $p$-values against $\Gamma$.
\begin{lstlisting}
Gamma  = seq(1, 1.4, 0.001)
Pvalue = Gamma
for(i in 1:length(Gamma))
{
  Pvalue[i] = senmw(datamatched, gamma = Gamma[i], 
                    method = "t")$pval
}
\end{lstlisting}
Figure   \ref{fg::p-gamma-lalonde} shows the plot of the $p$-value against $\Gamma$.

\subsection{Two examples from Rosenbaum's packages}

\begin{figure}
\centering 
\begin{subfigure}[b]{\textwidth}
                \centering
                \includegraphics[width=\textwidth]{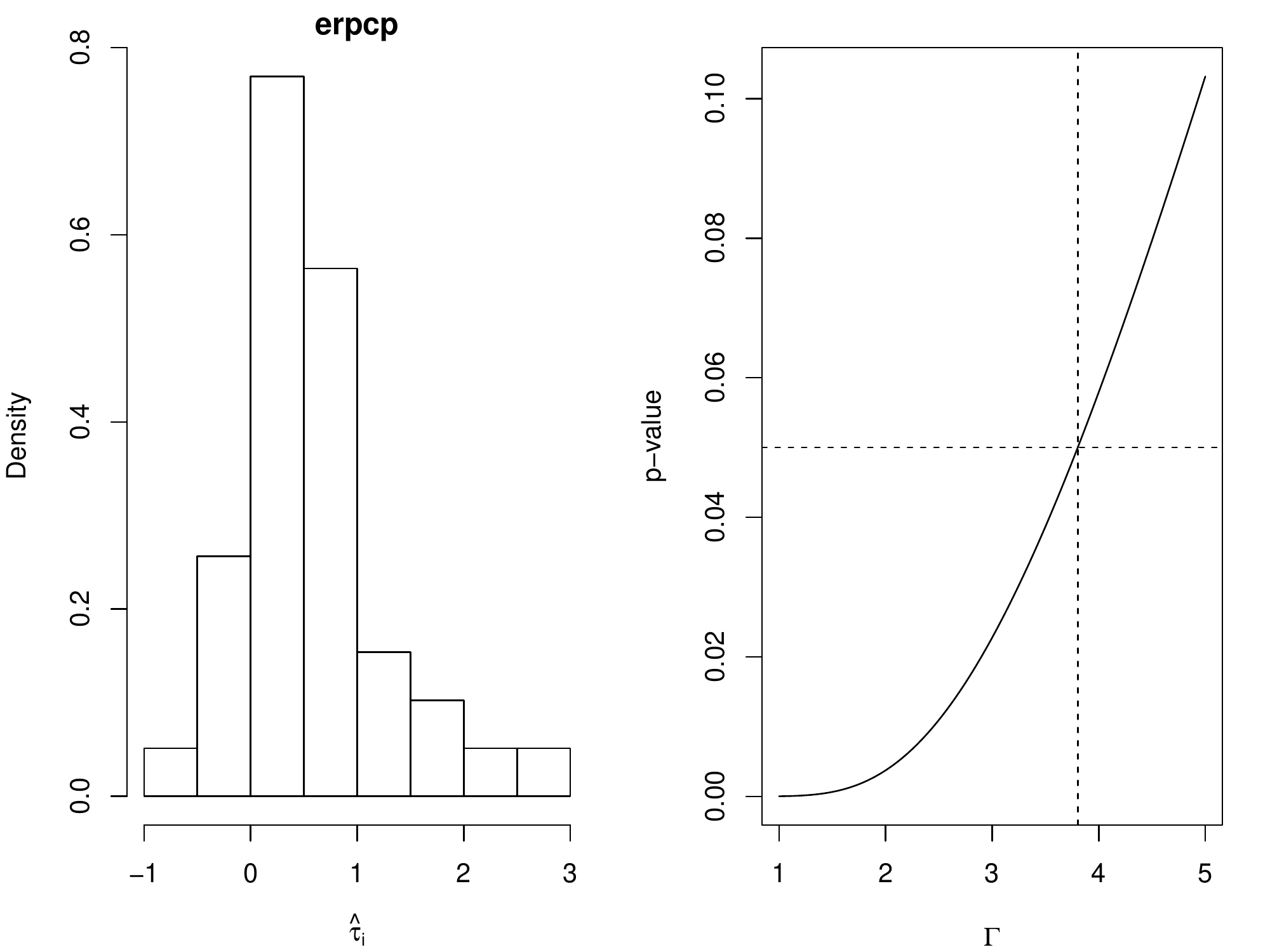}
                \caption{erpcp data}
                \label{fig::erpcp}
        \end{subfigure}%

\begin{subfigure}[b]{\textwidth}
                \centering
                \includegraphics[width=\textwidth]{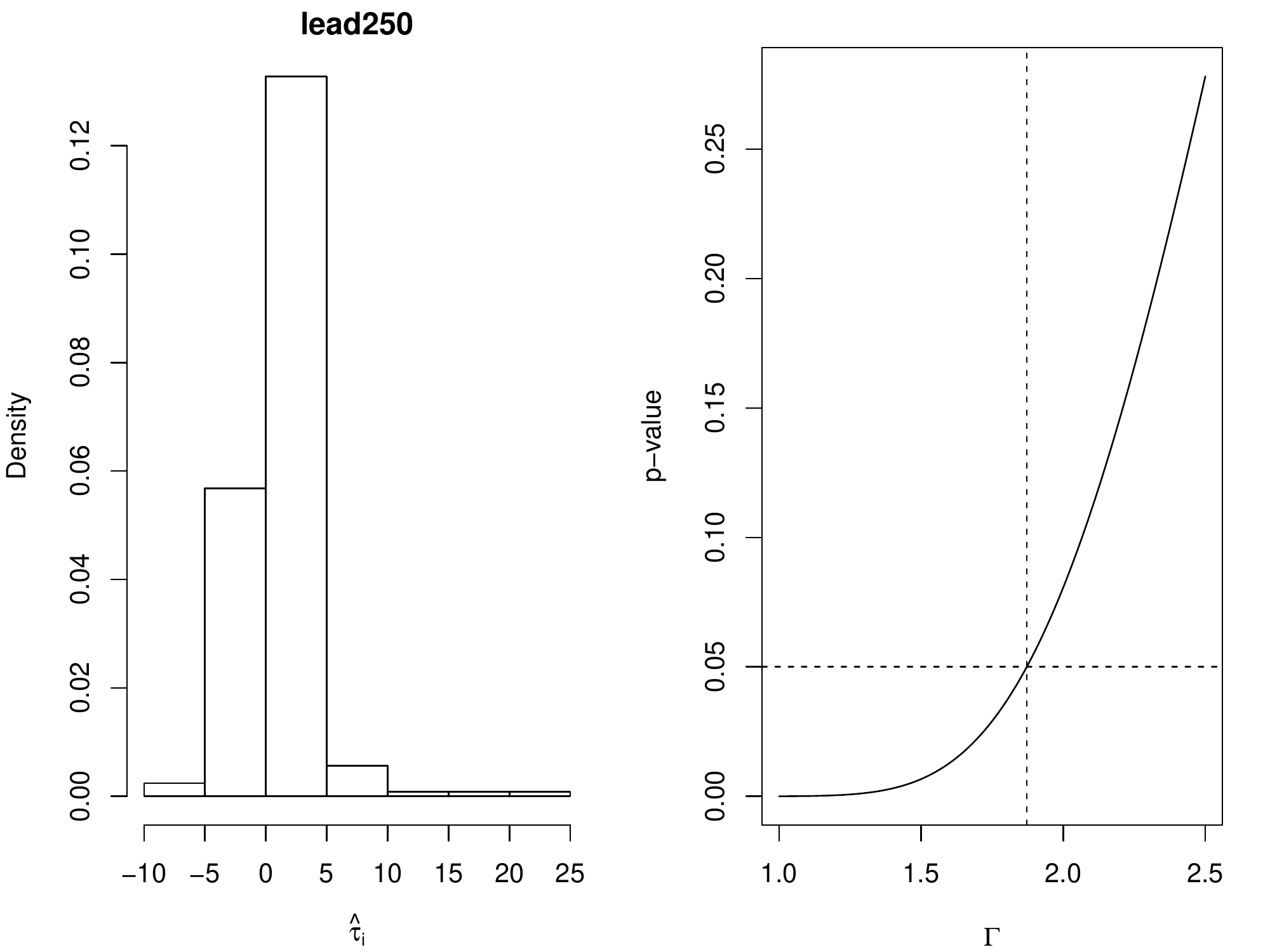}
                \caption{ead250 data}
                \label{fig::ead250}
        \end{subfigure}%
\caption{Two examples: histogram of the $\hat{\tau}_i$'s and $p$-value as a function of $\Gamma$}\label{fg::another2sensitivity}
\end{figure}

 The \texttt{erpcp} dataset is from the \ri{R} package \texttt{sensitivitymw}. It contains $n=39$ matched pairs of a welder and a control, based on observed covariates age and smoking. The outcome is the DNA elution rate.  Figure \ref{fg::another2sensitivity}(a) shows the histogram of the within-pair differences of the outcomes and the $p$-value against $\Gamma$ based on the pair $t$-statistic. The following \ri{R} code generates Figure \ref{fig::erpcp}.

\begin{lstlisting}
par(mfrow = c(1, 2), mai = c(0.8, 0.8, 0.3, 0.3))
data(erpcp)
hist(erpcp[, 1] - erpcp[, 2], main = "erpcp",
     xlab = expression(hat(tau)[i]),
     freq = FALSE)

Gamma  = seq(1, 5, 0.005)
Pvalue = Gamma
for(i in 1:length(Gamma))
{
  Pvalue[i] = senmw(erpcp, gamma = Gamma[i], method = "t")$pval
}
gammastar = Gamma[which(Pvalue >= 0.05)[1]]
gammastar

plot(Pvalue ~ Gamma, type = "l", 
     xlab = expression(Gamma), 
     ylab = "p-value")
abline(h = 0.05, lty = 2)
abline(v = gammastar, lty = 2)
\end{lstlisting}

 The \texttt{lead250} dataset is from the \ri{R} package \texttt{sensitivitymw}. It contains $n=250$ matched pairs of a daily smoker and a control of nonsmoker, based on observed covariates gender, age, race education level, and household income from NHANES. The outcome is the blood lead level in $\mu g/l$. Figure \ref{fg::another2sensitivity}(b) shows the histogram of the within-pair differences of the outcomes and the $p$-value against $\Gamma$ based on the pair $t$-statistic. The following \ri{R} code generates Figure \ref{fig::ead250}.
 
 \begin{lstlisting}
par(mfrow = c(1, 2), mai = c(0.8, 0.8, 0.3, 0.3))
data(lead250)
hist(lead250[, 1] - lead250[, 2],
     main = "lead250",
     xlab = expression(hat(tau)[i]),
     freq = FALSE)

Gamma  = seq(1, 2.5, 0.001)
Pvalue = Gamma
for(i in 1:length(Gamma))
{
  Pvalue[i] = senmw(lead250, gamma = Gamma[i], method = "t")$pval
}
gammastar = Gamma[which(Pvalue >= 0.05)[1]]
gammastar

plot(Pvalue ~ Gamma, type = "l", 
     xlab = expression(Gamma), ylab = "p-value")
abline(h = 0.05, lty = 2)
abline(v = gammastar, lty = 2)
\end{lstlisting}

\section{Homework Problems}

\paragraph{A model for Assumption \ref{assume::rosenbaum-sensitivity}}\label{hw::rosenbaum-equivalent}

Assumption \ref{assume::rosenbaum-sensitivity} has the following equivalent form.

\begin{assumption}
\label{assume::rosenbaum-equivalent}
The propensity score satisfies the following model
$$
\log \frac{  \pi_{ij} }{  1-\pi_{ij}  } = g(X_i) + \gamma U_{ij} ,\quad (i=1,\ldots, n; j=1,2)
$$
where $g(\cdot)$ is an unknown function and $U_{ij} \in [0, 1]$ is a a bounded unobserved covariate for unit $(i, j)$. 
\end{assumption}

Show that if Assumption \ref{assume::rosenbaum-equivalent} holds, then Assumption \ref{assume::rosenbaum-sensitivity} must hold with $\Gamma = e^\gamma$.

Remark: \citet[][page 108]{rosenbaum2002design} also shows that if Assumption \ref{assume::rosenbaum-sensitivity} holds, then Assumption \ref{assume::rosenbaum-equivalent} must hold for some $U_{ij}$'s. Assumption \ref{assume::rosenbaum-equivalent} may be easier to interpret because $\gamma$ measures the log of the conditional odds ratio of $U$ on the treatment; see Chapter \ref{sec::logistic-regression} for the interpretation of the coefficients in the logistic regression.

\paragraph{Application of Rosenbaum's approach}
Re-analyze Example \ref{eg::chan-bmi-ATE} using Rosenbaum's approach based on matching.

\paragraph{Recommended reading}

\citet{rosenbaum2015two} provides a tutorial for his two \ri{R} packages for sensitivity analysis with matched observational studies.

\chapter{Overlap in Observational Studies: Difficulties and Opportunities}
  \label{chapter::overlap}

\section{Implications of overlap}

In Part \ref{part::observational-studies} of this book, causal inference with observational studies relies on two critical assumptions: ignorability  
$$
Z\ind \{ Y(1), Y(0)  \} \mid X
$$
 and overlap  
 $$
 0 <  e(X) <  1 .
 $$ 
 \citet{d2017overlap} pointed out the tension between these two assumptions: typically, more covariates make the ignorability assumption more plausible (ignoring M-bias discussed in Chapter \ref{section::m-bias}), but more covariates make the overlap assumption less plausible because the treatment becomes more predictable given more covariates. 

If some units have $e(X) = 0$ or $e(X) = 1$, then we have philosophic difficulty in thinking about the counterfactual potential outcomes \citep{king2006dangers}. In particular, if a unit deterministically receives the treatment, then it may not be meaningful to conceive its potential outcome under the control; if a unit deterministically receives the control, then it may not be meaningful to conceive its potential outcome under the treatment.
Even if the true propensity score is not exactly $0$ or $1$, the estimated propensity score can be very close to $0$ or $1$ in finite samples, which makes the estimators based on IPW numerically unstable. 
I have discussed this issue in Chapter \ref{chapter::pscore-key}.

Many statistical analyses in fact require a strict version of overlap:

\begin{assumption}
[strict overlap]
\label{assume::strong-overlap}
$\eta \leq e(X) \leq 1-\eta$ for some $\eta \in (0,1/2).$
\end{assumption}

However, \citet[][Corollary 1]{d2017overlap} showed that Assumption \ref{assume::strong-overlap} has very strong implications when the number of covariates is large. For simplicity, I present only one of their results. Let $X_k\ (k=1,\ldots, p)$ be the $k$th component of the covariate $X = (X_1, \ldots, X_p)$, and 
$$
e= \pr(Z=1)
$$ 
be the proportion of the treated units.

\begin{theorem}\label{thm::overlap-implications}
Assumption \ref{assume::strong-overlap} implies that $\eta \leq e \leq 1-\eta$ and
\begin{eqnarray} 
&&p^{-1} \sum_{k=1}^p \big | E(X_k\mid Z=1) - E(X_k\mid Z=0)  \Big |   \nonumber  \\
& \leq &  
p^{-1/2} C^{1/2}  \left\{    e\lambda_1^{1/2} + (1-e)  \lambda_0^{1/2}  \right\} , \label{eq::mean-balance-overlap}
\end{eqnarray}
where 
$$
C  = \frac{  (e-\eta)(1-e-\eta)   }{  e^2(1-e)^2 \eta (1-\eta)      }
$$
is a positive constant depending only on $(e, \eta)$, and $\lambda_1$ and $\lambda_0$ are the maximum eigenvalues of the covariance matrices $ \cov(X\mid Z=1)$ and $\cov(X\mid Z=0) $, respectively. 
\end{theorem}

What is the order of the maximum eigenvalues in Theorem \ref{thm::overlap-implications}? \citet{d2017overlap} showed that it is usually smaller than $O(p)$ unless the components of $X$ are highly correlated. If the components of $X$ are highly correlated, then some components are redundant after including other components. If the components of $X$ are not highly correlated, then the right-hand side converges to zero. So the average difference in means of the covariates is close to zero, that is, the treatment and control groups are nearly balanced in means averaged over all dimensions of the covariates. Mathematically,  the left-hand side of  \eqref{eq::mean-balance-overlap} converging to zero rules out the possibility that all dimensions of $X$ have non-vanishing differences in means across the treatment and control groups. 
It is a strong requirement in observational studies with many covariates.

\subsection{Trimming in the presence of limited overlap}\label{sec::trimming-overlap}

When Assumption \ref{assume::strong-overlap}  does not hold, it is common to trim the units based on the estimated propensity scores \citep{crump2009dealing, yang2018asymptotic}. Trimming drops units within regions of little overlap, which changes the population and estimand. The restrictive implications of overlap in Assumption \ref{assume::strong-overlap} suggest that trimming must be employed more often and one may need to trim a large proportion of units to achieve desirable overlap in high dimensions.

\subsection{Outcome modeling in the presence of limited overlap}\label{sec::outcome-modeling-overlap}

The somewhat negative results in  \citet{d2017overlap} also highlight the limitation of focusing only on the propensity score in the presence of limited overlap. This in some sense challenges \citet{rubin2008objective}'s view that ``for objective causal inference, design trumps analysis.'' \citet{rubin2008objective} argued strongly for the role of the ``design'' stage in observational studies. The ``design'' stage focuses on the propensity score which may not satisfy the overlap condition with many covariates.

With high dimensional covariates, outcome modeling becomes more important. In particular, if the outcome means only depend on a function of the original covariates in that
$$
E\{  Y(z) \mid X \} = f_z( r(X)   ),\quad (z=0,1)
$$
then it suffices to control for $r(X)$, a lower-dimensional summary of the original covariates. See Problem \ref{hw::outcome-modeling-overlap} for more details. Due to the dimension reduction, the strict overlap condition on $r(X)$ can be much weaker than the strict overlap condition on $X$. This is conceptually straightforward, but the corresponding theory and methods are missing.

\section{Causal inference with no overlap: regression discontinuity}
\label{sec::rdd}

Let us start from the simple case with a univariate $X$. 
An extreme treatment assignment mechanism is a deterministic one:
 $$
 Z = I(X \geq x_0),
$$
where $x_0$ is a predetermined threshold. 
An interesting consequence of this assignment is that the ignorability  assumption holds automatically:
$$
Z\ind \{  Y(1), Y(0) \} \mid X
$$
because $Z$ is a deterministic function of $X$ and a constant is independent of any random variables. However, the overlap assumption is violated by definition:
$$
e(X) = \pr(Z=1\mid X) = 1(X \geq x_0)
=\begin{cases}
1 & \text{ if } X \geq x_0, \\
0& \text{ if }  X < x_0.
\end{cases}
$$
So our analytic strategies discussed in Part \ref{part::challenges-os} are no longer applicable here. We must change our perspective.

The discussion above seems contrived, with a deterministic treatment assignment. Interestingly, it has many applications in practice and is called {\it regression discontinuity}. Below, I first review some canonical examples and then give a mathematical formulation of this type of study.

\subsection{Examples and graphical diagnostics}
\label{sec::rdd-examples}

\begin{figure}
\centering
\includegraphics[width = \textwidth]{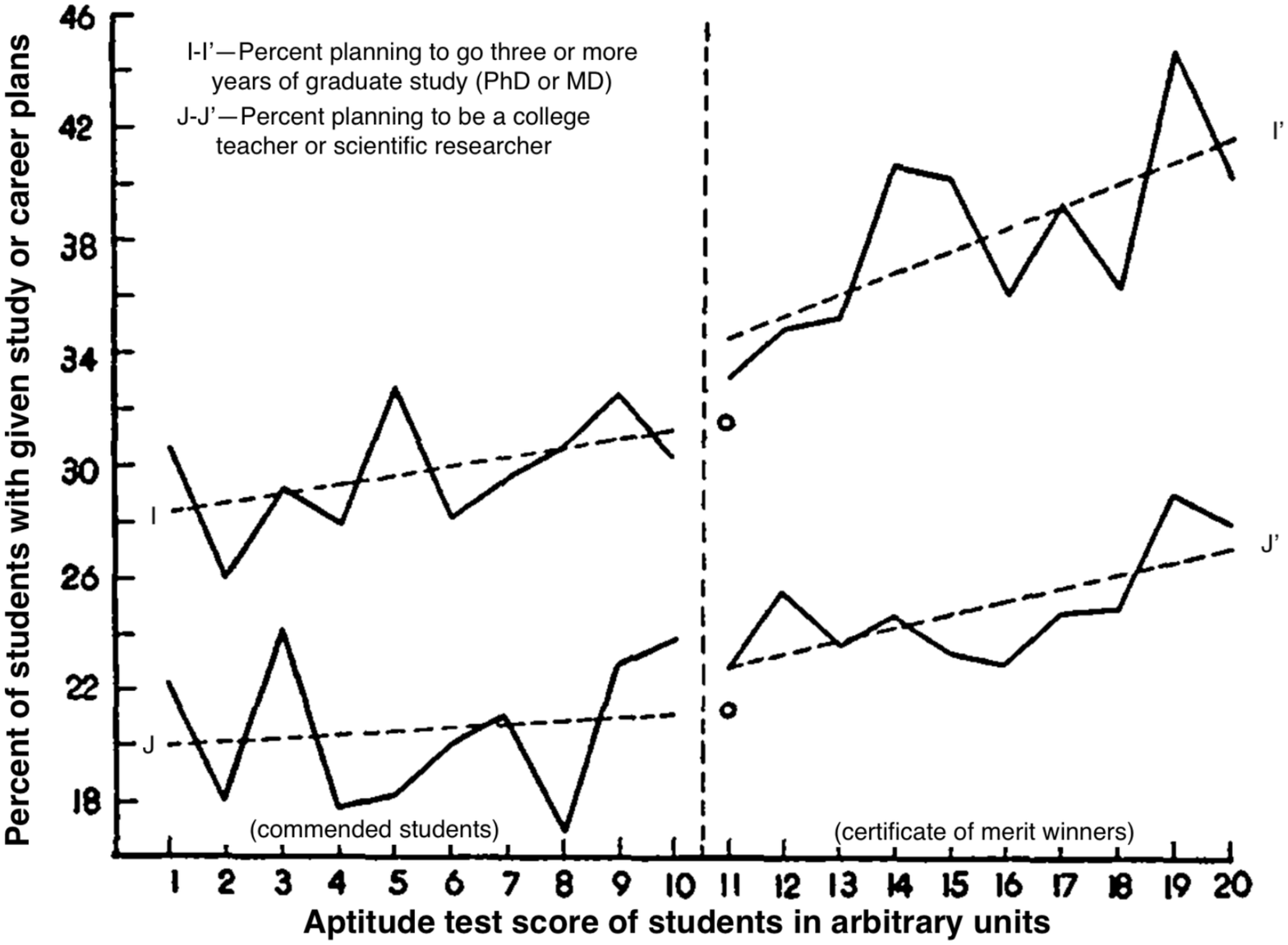}
\caption{A graph from \citet{thistlethwaite1960regression} with minor modifications of the unclear text in the original paper}\label{fig::thistlethwaite1960regression}
\end{figure}

\begin{example}
\citet{thistlethwaite1960regression} first proposed the idea of {\it regression-discontinuity analysis}. Their motivating example was to study the effect of students' winning the Certificate of Merit on later career plans, where the Certificate of Merit was determined by whether the Scholarship Qualifying Test score was above a certain threshold. Their initial analysis was mainly graphical. Figure \ref{fig::thistlethwaite1960regression} shows one of their graphs. 
\end{example}

\begin{example}
\citet{bor2014regression}  used regression discontinuity to study the effect of when to start HIV patients with antiretroviral on their mortality, where the treatment is determined by whether the patients' CD4 counts were below  200 cells/$\mu$L (note that CD4 cells are white blood cells that fight infection.)
\end{example}

\begin{figure}
\centering
\includegraphics[width = \textwidth]{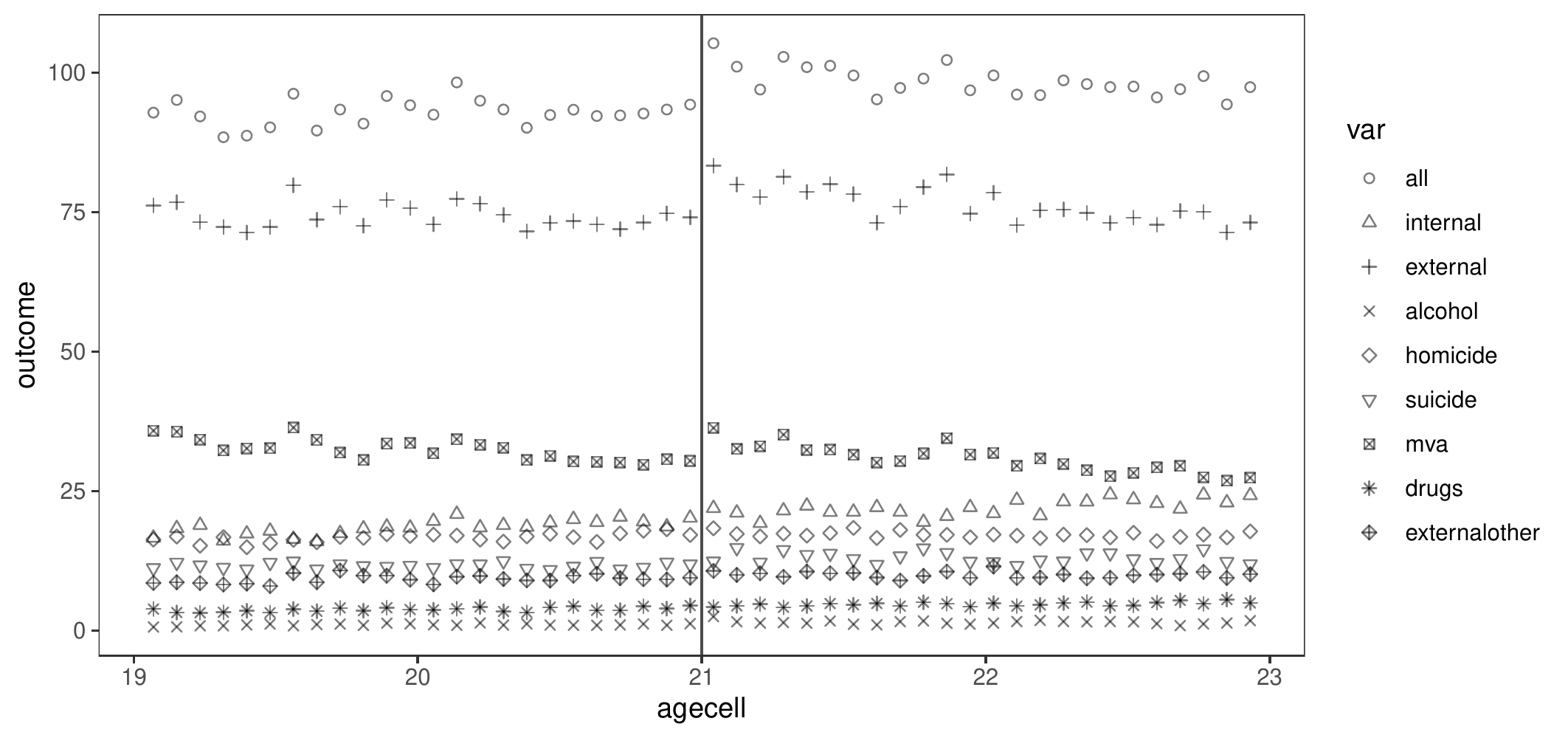}
\caption{Minimum legal drinking age example from  \citet{carpenter2009effect}}\label{fig::mlda-9outcomes}
\end{figure}

\begin{example}
\citet{carpenter2009effect} studied the effect of alcohol consumption on mortality, which leverages the minimum legal drinking age as a discontinuity for alcohol consumption.  They derived mortality data from the National Center for Health Statistics,  including the decedent's date of birth and date of death. They computed the age profile of deaths per 100,000 person-years with outcomes measured by the following nine variables:
\begin{tabular}{cc}
\hline 
\texttt{all} & all deaths, the sum of  \texttt{internal} and \texttt{external}\\
\texttt{internal} & deaths due to internal causes\\
\texttt{external} & deaths due to external causes, the sum of the rest\\
\texttt{homicide} & homicides \\
\texttt{suicide} & suicides\\
\texttt{mva} & motor vehicle accidents\\
\texttt{alcohol} & deaths with a mention of alcohol\\
\texttt{drugs} & deaths with a mention of drug use\\
\texttt{externalother} & deaths due to other external causes\\
\hline 
\end{tabular}

Figure \ref{fig::mlda-9outcomes} plots the number of deaths per 100,000 person-years for nine measures based on the data used by \citet{angrist2014mastering}. From the jumps at age 21, it seems obvious that there is an increase in mortality at age 21, primarily due to motor vehicle accidents. I leave the formal analysis as Problem \ref{hw::MLDA}. 
\end{example}

\subsection{A mathematical formulation of regression discontinuity}

The technical term for the variable $X$ that determines the treatment is the {\it running variable}. 
Intuitively, regression discontinuity can identify a {\it local average causal effect} at the cutoff point $x_0$: 
$$
\tau(x_0) = E\{ Y(1) - Y(0)  \mid X=x_0 \}. 
$$
In particular, for the potential outcome under treatment, we have
\begin{eqnarray}
E\{ Y(1) \mid X=x_0 \} &=& \lim_{\varepsilon \rightarrow 0+ } E\{ Y(1) \mid X=x_0+\varepsilon \} \label{eq::srd-continuity} \\
&=& \lim_{\varepsilon \rightarrow 0+} E\{ Y(1) \mid Z=1, X=x_0+\varepsilon \}  \label{eq::srd-z} \\
&=&  \lim_{\varepsilon \rightarrow 0+} E( Y \mid Z=1, X=x_0+\varepsilon ),
\end{eqnarray} 
where \eqref{eq::srd-continuity} holds if $ E\{ Y(1) \mid X=x\}$ is continuous from the right at $x_0$ and \eqref{eq::srd-z} follows by the definition of $Z$. Similarly, for the potential outcome under control, we have
$$
E\{ Y(0) \mid X=x_0 \} =  \lim_{\varepsilon \rightarrow 0+} E( Y \mid Z=0, X=x_0-\varepsilon )
$$
if $ E\{ Y(0) \mid X=x\}$ is continuous from the left at $x_0$. 
So the local average causal effect at $x_0$ can be identified by the difference of the two limits. I summarize the key identification result below.

\begin{theorem}
\label{thm::identification-rdd}
Assume that the treatment is determined by  $Z = I(X \geq x_0)$
where $x_0$ is a predetermined threshold. 
Assume that $ E\{ Y(1) \mid X=x\}$ is continuous from the right at $x_0$ and $ E\{ Y(0) \mid X=x\}$ is continuous from the left at $x_0$. Then the local average treatment effect at $X=x_0$ is identified by
$$
\tau(x_0) = \lim_{\varepsilon \rightarrow 0+} E( Y \mid Z=1, X=x_0+\varepsilon ) - \lim_{\varepsilon \rightarrow 0+} E( Y \mid Z=0, X=x_0-\varepsilon ).
$$
\end{theorem}

Since the right-hand side of the above equation only involves observables, the parameter $\tau(x_0)$ is nonparametrically identified. However, the form of the identification formula is totally different from what we derived before. In particular, the identification formula involves limits of two conditional expectation functions.

\subsection{Regressions near the boundary}\label{sec::rdd-numerical}

\begin{figure}
\centering
\includegraphics[width = \textwidth]{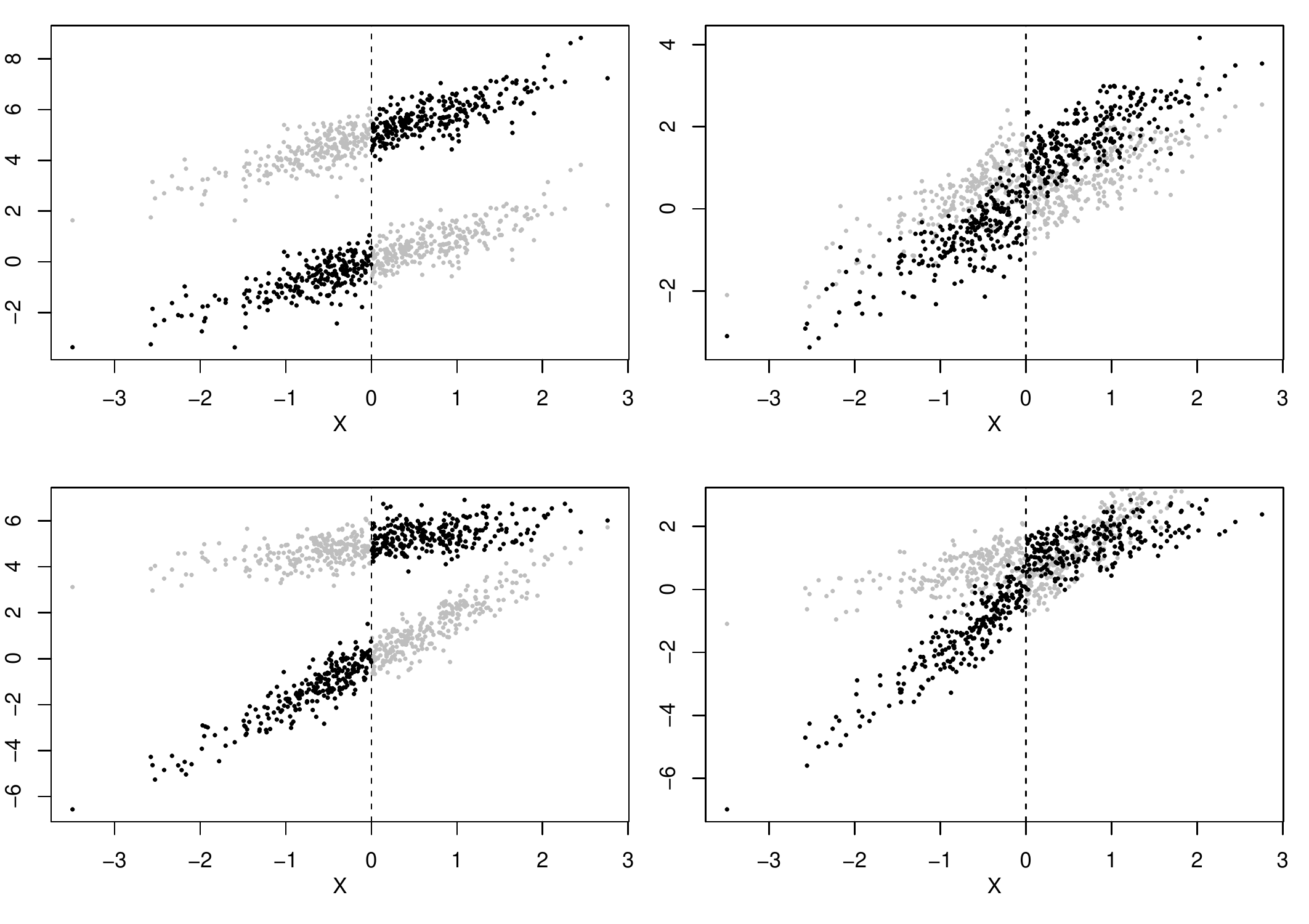}
\caption{Examples of regression discontinuity. The black points are the observed outcomes whereas the grey points are the unobserved counterfactual outcomes. 
In the first column, the data-generating processes result in visible jumps at the cutoff points; in the second column, the jumps are not so visible. In the first row, the data generating processes have constant $\tau(x)$; in the second row, $\tau(x)$ varies with $x$.  
}\label{fig::4examples-rdd}
\end{figure}

If we are lucky,  graphical diagnostics sometimes can clearly show the causal effect at the cutoff point. However, many outcomes are noisy so graphical diagnostics are not enough in finite samples. Figure \ref{fig::4examples-rdd} shows two simulated examples with obvious jumps at the cutoff point and two examples without obvious jumps, although the underlying data-generating processes all have discontinuities. 

Assume that $E( Y \mid Z=1, X=x ) = \gamma_1 + \beta_1 x$ and $E( Y \mid Z=0, X=x ) = \gamma_0 + \beta_0 x $ are linear in $x$. We can run OLS based on the treated and control data to obtain the fitted lines $\hat\gamma_1 + \hat\beta_1 x$ and $\hat\gamma_0 + \hat\beta_0 x$, respectively. We can then estimate the average causal effect at the point $X = x_0$ as
$$
\hat\tau(x_0) = (\hat\gamma_1 -\hat\gamma_0 ) + (\hat\beta_1 -\hat\beta_0) x_0 .
$$
Numerically, $\hat\tau(x_0) $ is identical to the coefficient of $Z_i$ in the OLS
\begin{eqnarray}
Y_i \sim \{ 1, Z_i, X_i - x_0, Z_i (X_i - x_0)  \} , \label{eq::ols-rdd-1}
\end{eqnarray}
and it is also identical to the coefficient of $Z_i$ in the OLS
\begin{eqnarray}
Y_i \sim \{ 1, Z_i,  R_i, L_i \},\label{eq::ols-rdd-2}
\end{eqnarray}
where 
$$
R_i = \max(X_i - x_0, 0) ,\qquad L_i= \min (X_i - x_0, 0)
$$ 
indicate the right and left parts of  $X_i - x_0$, respectively. I leave the algebraic details to Problem \ref{hw::rdd-ols-fits}.

However, this approach may be sensitive to the violation of the linear model assumption. Theory suggests running regression using only the local observations near the cutoff point\footnote{This is called {\it local linear regression} in nonparametric statistics, which belongs to a broader class of {\it local polynomial regression} \citep{fan2018local}. }. However, the rules for choosing the ``local points'' are quite involved. Fortunately, the \ri{rdrobust} function in the \ri{rdrobust} package in \ri{R} implements various choices of  ``local points.'' Since choosing the ``local points'' is the key to regression discontinuity, it seems more sensible to report estimates and confidence intervals based on various choices of the ``local points.'' This can be viewed as a sensitivity analysis for regression discontinuity.

\subsection{An example}
\label{section::lee-RDD-example}

\citet{lee2008randomized} gave a famous example of using regression discontinuity to study the incumbency advantage in the U.S. House. 
He wrote that ``incumbents are, by definition, those politicians who were successful in the previous election. If what makes them successful is somewhat persistent over time, they should be expected to be somewhat more successful when running for re-election.'' Therefore, this is a fundamentally challenging causal inference problem. The regression discontinuity is a clever study design to study this problem.

\begin{figure}
\centering
\includegraphics[width = \textwidth]{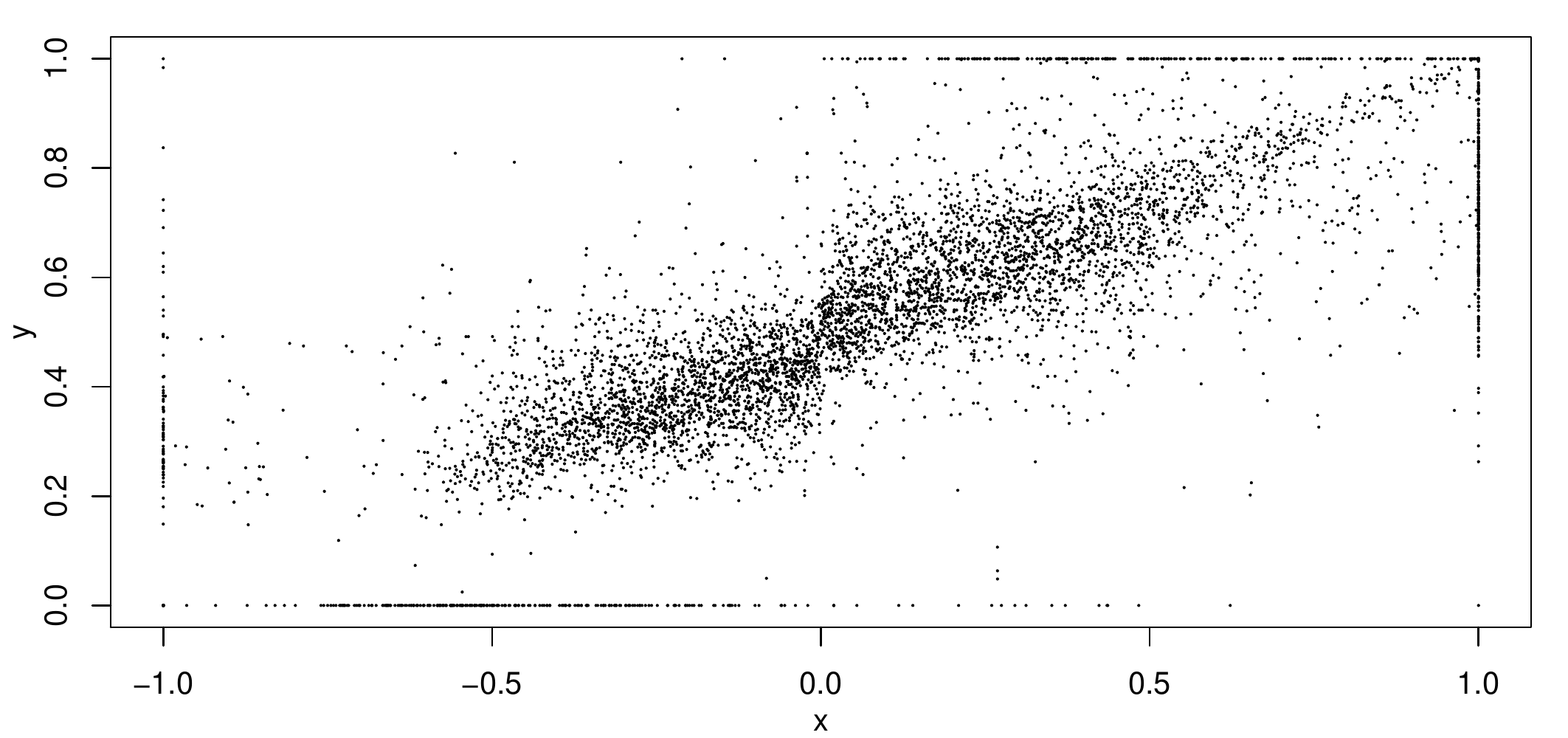}
\caption{Raw data of \citet{lee2008randomized} }\label{fig::lee2008}
\end{figure}

The running variable is the lagged vote in the previous election centered at $0$, and the outcome is the vote in the current election, with units being the congressional districts. The treatment is the binary indicator for being the current incumbent party in a district, determined by the lagged vote. The following \ri{R} code generates  Figure \ref{fig::lee2008} that shows the raw data. 

\begin{lstlisting}
house = read.csv("house.csv")[, -1]
plot(y ~ x, data = house, pch = 19, cex = 0.1)
abline(v = 0, col = "grey")
\end{lstlisting}

The \ri{rdrobust} function gives three sets of the point estimate and confidence intervals. They all suggest a positive incumbency advantage. 

\begin{lstlisting}
> library(rdrobust)
> RDDest = rdrobust(house$y, house$x)
Warning message:
In rdrobust(house$y, house$x) :
  Mass points detected in the running variable.
> cbind(RDDest$coef, RDDest$ci)
                    Coeff   CI Lower   CI Upper
Conventional   0.06372533 0.04224798 0.08520269
Bias-Corrected 0.05937028 0.03789292 0.08084763
Robust         0.05937028 0.03481238 0.08392818
\end{lstlisting}

\begin{figure}[h]
\centering
\includegraphics[width = \textwidth]{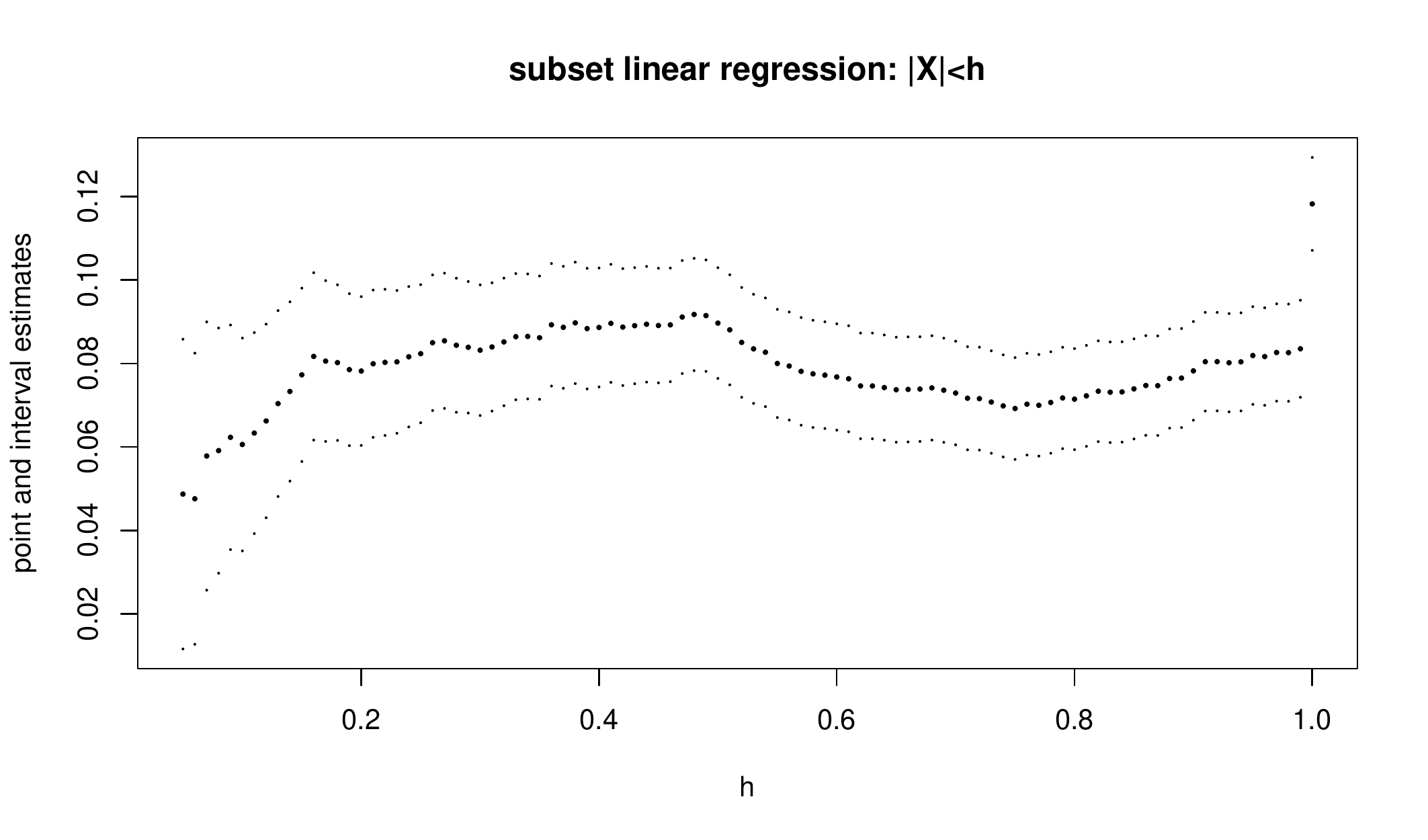}
\caption{Estimates and confidence intervals based on local linear regressions}\label{fig::rdd-lee}
\end{figure}

We can also obtain the point estimates and the confidence intervals based on OLS with different choices of the local points defined by $|X| < h$ based on the following \ri{R} code.
\begin{lstlisting}
house$z = (house$x >= 0)
hh = seq(0.05, 1, 0.01)
local.lm = sapply(hh, function(h){
  Greg = lm(y ~ z + x + z*x, data = house,
            subset = (abs(x)<=h))
  cbind(coef(Greg)[2], confint(Greg, 'zTRUE'))
})
plot(local.lm[1, ] ~ hh, type = "p",
     pch = 19, cex = 0.3,
     ylim = range(local.lm),
     xlab = "h",
     ylab = "point and interval estimates",
     main = "subset linear regression: |X|<h")
lines(local.lm[2, ] ~ hh, type = "p",
      pch = 19, cex = 0.1) 
lines(local.lm[3, ] ~ hh, type = "p",
      pch = 19, cex = 0.1)
\end{lstlisting}
Figure \ref{fig::rdd-lee} shows the point estimates and the confidence intervals as a function of $h$. While the point estimates and the confidence intervals are sensitive to the choice of $h$, the qualitative result remains the same as above. 
 
\subsection{Problems of regression discontinuity}\label{sec::problems-RD}

What can go wrong with the regression discontinuity analysis? The technical challenge is to specify the neighborhood near the cutoff point. We have discussed this issue above.

In addition, Theorem \ref{thm::identification-rdd} holds under a continuity condition. It may be violated in practice. For instance, if the mortality rate jumps at the age of 21, then we may not be able to attribute the jumps in Figure \ref{fig::mlda-9outcomes} to the change in drinking behavior due to the legal drinking age. However, it is hard to check the violation of the continuity condition empirically.

\citet{mccrary2008manipulation} proposed an indirect test for the validity of the regression discontinuity. He suggested checking the density of the running variable at the cutoff point. The discontinuity in the density of the running variable at the cutoff point may suggest that some units were able to manipulate their treatment status perfectly. The \ri{DCdensity} function in the \ri{R} package \ri{rdd} implements this test. I omit the details.

\section{Homework Problems}

\paragraph{A theorem for the role of outcome modeling in the presence of limited overlap}\label{hw::outcome-modeling-overlap}

This problem extends the discussion in Section \ref{sec::outcome-modeling-overlap}. I state a formal theorem below.

\begin{theorem}
\label{thm::outcome-overlap}
Assume 
$$
Z\ind Y(z) \mid X, \quad (z=0,1)
$$
and
$$
E\{  Y(z) \mid X \} = f_z( r(X)   ),\quad (z=0,1)
$$
for some function $r(X)$. 
The average causal effect $\tau = E\{ Y(1) - Y(0)  \}$ can be identified by
$$
\tau =E\{   E(Y\mid Z=1, r(X)) - E(Y\mid Z=0, r(X)) \} 
$$
or
$$
\tau = E\left\{   \frac{ZY}{  e(r(X)) }  \right\}  - E\left\{   \frac{(1-Z)Y}{ 1 -  e(r(X)) }  \right\}
$$
where $e(r(X)) = \pr\{ Z=1\mid r(X) \}$. 
\end{theorem}

Prove Theorem \ref{thm::outcome-overlap}.

Remark: When $r(X) = X$, Theorem \ref{thm::outcome-overlap} reduces to the standard IPW formula for the average causal effect. When $r(X)$ has a lower dimension than $X$, Theorem \ref{thm::outcome-overlap} has broader applicability if the overlap condition on $e(X)$ fails whereas the overlap condition on $e(r(X)) $ may still hold.

 \paragraph{Linear potential outcome models}\label{hw::rdd-ols-fits}

 This problem gives more details for the numerical equivalence in Section \ref{sec::rdd-numerical}. 
 
 Show that $\hat\tau(x_0) $ equals the coefficients of $Z_i$ in OLS fits \eqref{eq::ols-rdd-1} and \eqref{eq::ols-rdd-1}. 
 

Remark: It is helpful to start with the figures of $Z_i(X_i - x_0)$, $L_i$, and $R_i$ with $X_i$ on the x-axis. The conclusion holds by a reparametrization of the OLS regressions.

\paragraph{Simulation for regression discontinuity}

In Figure \ref{fig::4examples-rdd}, the potential outcomes are simulated from linear models. Change them to nonlinear models, and compare different point estimators and confidence intervals, including the biases and variances of the point estimators, and the coverage properties of confidence intervals.

\paragraph{Re-analysis of the data on the minimum legal drinking age}\label{hw::MLDA}

Figure \ref{fig::mlda-9outcomes} shows the jumps at the cutoff point. 
Analyze the data \ri{mlda.csv} of \citet{carpenter2009effect}.

\paragraph{Re-analysis of \citet{lee2008randomized}'s data}\label{hw::lee2008-robust}

Figure \ref{fig::rdd-lee} plots the confidence intervals based on the standard errors assuming homoskedasticity. Generate a figure with confidence intervals based on the EHW standard errors in OLS.

\paragraph{Recommended reading}

\citet{d2017overlap} discussed the implications of overlap with high dimensional covariates. 

\citet{thistlethwaite1960regression}'s original paper on regression discontinuity was re-printed as \citet{thistlewaiteregression2016obs} with many insightful comments.  Coincidentally, \citet{thistlethwaite1960regression} and \citet{Rubin:1974} were both published in the {\it Journal of Educational Psychology}.

\part{Instrumental variables}\label{part::instrumentalvariables}
   
\chapter{An Experimental Perspective of the Instrumental Variable}
 \label{chapter::iv-experiment}

The {\it instrumental variable method} has been a powerful tool in econometrics. It identifies causal effects in studies without unconfoundedness between the treatment and the outcome. It relies on an additional variable, called the {\it instrumental variable} (IV), that satisfies certain conditions. These conditions may not be easy to digest when you learn them for the first time. In some sense, the IV method is like magic. This chapter presents a not-so-magic perspective based on the encouragement design. This again echos \citet{dorn1953philosophy}'s suggestion that the planner of an observational study should always ask himself the following question:\footnote{This quote also appeared in Chapter \ref{chapter::observational-studies} before.}
\begin{quote}
How would the study be conducted if it were possible to do
it by controlled experimentation?
\end{quote} 
The experimental analog of the IV method is {\it noncompliance} in the {\it encouragement design} \citep{zelen1979new, powers1984effects, holland1986statistics}.

\section{Encouragement Design and Noncompliance}

Consider an experiment with units indexed by $i = 1,\ldots, n$. Let $Z_i$ be the treatment assigned, with $1$ for the treatment and $0$ for the control. Let $D_i$ be the treatment received, with $1$ for the treatment and $0$ for the control.  When $Z_i \neq D_i$ for some unit $i$, the noncompliance problem arises.  Noncompliance is a very common problem, especially in encouragement designs involving human beings as experimental units. In those cases, the experimenters cannot force the units to take the treatment but rather only encourage them to do so. 
Let $Y_i$ be the outcome of interest. 

Consider complete randomization of $Z$ and ignore covariates $X$ for now. We have the potential values for the treatment received $\{  D_i(1), D_i(0) \}$ and  the potential values for the outcome $\{ Y_i(1), Y_i(0)  \}$, all with respect to the treatment assignment levels $1$ and $0$. 
Their observed values are $D_i = Z_i D_i(1) + (1-Z_i) D_i(0)$ and $Y_i = Z_i Y_i(1) + (1-Z_i) Y_i(0)$, respectively. 
For notational simplicity, we assume   $\{  Z_i, D_i(1), D_i(0) ,Y_i(1), Y_i(0)  \}_{i=1}^n \iidsim \{  Z, D(1), D(0) ,Y(1), Y(0)  \}$ and sometimes drop the subscript $i$ when it should not cause confusions.

We start with the CRE. 
\begin{assumption}[randomization]
\label{assume::random}
$Z\ind \{D(1), D(0) ,Y(1), Y(0)   \}$. 
\end{assumption}

Randomization allows for the identification of the average causal effects on $D$ and $Y$:
$$
\tau_D = E\{  D(1) - D(0) \} = E(D\mid Z=1) - E(D \mid Z=0)
$$
and
$$
\tau_Y = E\{  Y(1) - Y(0) \} = E(Y\mid Z=1) - E(Y\mid Z=0).  
$$
We can use simple difference-in-means estimators $\hat{\tau}_D$ and $\hat{\tau}_Y$ to estimate $\tau_D$ and $\tau_Y$, respectively.

Reporting the estimate $\hat{\tau}_Y$ with the associated standard error is called the intention-to-treat (ITT) analysis. It estimates the effect of the treatment assignment on the outcome, and the CRE in Assumption \ref{assume::random} justifies this analysis. However, it may not answer the scientific question, that is, the causal effect of the treatment received on the outcome.

\section{Latent Compliance Status and Effects}

\subsection{Nonparametric identification}

Following \citet{imbens1994identification} and \citet{angrist1996identification}, we stratify the population based on the joint potential values of $U_i = \{  D_i(1), D_i(0) \}$. Because $D$ is binary, we have four possible combinations:
\begin{eqnarray*}
U_i=\left \{ \begin{array} {ll}
\at  , & \text{if} \hspace{2mm} D_i(1)=1 \hspace{2mm} \text{and} \hspace{2mm} D_i(0)=1; \\
\cp  , & \text{if} \hspace{2mm} D_i(1)=1 \hspace{2mm} \text{and} \hspace{2mm} D_i(0)=0; \\
\df  ,& \text{if} \hspace{2mm} D_i(1)=0 \hspace{2mm} \text{and} \hspace{2mm} D_i(0)=1; \\
\nt  , & \text{if} \hspace{2mm} D_i(1)=0 \hspace{2mm} \text{and} \hspace{2mm} D_i(0)=0,
\end{array}
\right.  
\end{eqnarray*}
where ``$\at$'' is for ``always taker,'' ``$\cp$'' is for ``complier,'' ``$\df$'' is for ``defier,'' and ``$\nt$'' is for ``never taker.'' 
Because we cannot observe $D_i(1)$ and $D_i(0)$ simultaneously, $U_i$ is a latent variable for the compliance behavior of unit $i$.

Based on $U$, we can use the law of total probability to decompose the average causal effect on $Y$ into four terms:
\begin{eqnarray} 
\tau_Y &=& E\{ Y(1) - Y(0) \mid U=\at  \} \pr(U = \at )  \nonumber  \\
&&+ E\{ Y(1) - Y(0) \mid U=\cp   \} \pr(U = \cp ) \nonumber  \\
&&+ E\{ Y(1) - Y(0) \mid U=\df  \} \pr(U = \df ) \nonumber  \\
&&+ E\{ Y(1) - Y(0) \mid U=\nt  \} \pr(U = \nt ) . \label{eq::decompose-ate-y-cace}
\end{eqnarray}  
Therefore, $\tau_Y $ is a weighted average of four latent subgroup effects. We will look into more details of the latent groups below.

Assumption \ref{assume::Mono} below restricts the third term in \eqref{eq::decompose-ate-y-cace} to be zero. 
\begin{assumption}[monotonicity]
\label{assume::Mono}
$\pr(U_i = \df) = 0$ or $D_i(1)\geq D_i(0)$ for all $i$, that is, there are no defiers. 
\end{assumption}

Assumption \ref{assume::Mono} holds automatically with {\it one-sided noncompliance} when the units assigned to the control arm have no access to the treatment, i.e., $D_i(0) = 0$ for all units. Under randomization, Assumption \ref{assume::Mono} has a testable implication that 
\begin{equation}
\label{eq::test-mono}
\pr(D=1\mid Z=1) \geq \pr(D=1\mid Z=0).
\end{equation}
But Assumption \ref{assume::Mono} is much stronger than the inequality \eqref{eq::test-mono}. The former restricts $D_i(1)$ and $D_i(0)$ at the individual level whereas the latter restricts them only on average. Nevertheless, when the testable implication \eqref{eq::test-mono} holds, we cannot use the observed data to refute Assumption \ref{assume::Mono}.

Assumption \ref{assume::ER} below restricts the first and last terms in \eqref{eq::decompose-ate-y-cace} to be zero based on the mechanism of the treatment assignment on the outcome through only the treatment received. 
\begin{assumption}[exclusion restriction]
\label{assume::ER}
$Y_i(1) = Y_i(0)$ for always takers with $U_i=\at$ and never takers with $U_i = \nt$. 
\end{assumption}

As equivalent statement of Assumption \ref{assume::ER} is that $D_i(1) =  D_i(0)$ must imply $Y_i(1) =  Y_i(0)$ for all $i$.  It requires that the treatment assignment affects the outcome only if it affects the treatment received. In double-blind RCTs\footnote{In general, it is better to blind the experiment to avoid various biases arising from placebo effects, patients' expectations, etc. In double-blind RCTs, both doctors and patients do not know the treatment; in single-blind RCTs, the patients do not know the treatment but the doctors know. Sometimes, it is impossible to conduct double or even single-blind trials; those trials are called open trials.}, it is biologically plausible because the outcome only depends on the actual treatment received. That is, if the treatment assignment does not change the treatment received, it does not change the outcome either.  It can be violated if the treatment assignment has {\it direct effects}\footnote{I use ``direct effects'' informally here. See more detailed discussion of the concept in Chapters \ref{chapter::mediation} and \ref{chapter::cde} later.} on the outcome, not through the treatment received. For example, some RCTs are not double-blinded, and the treatment assignment can have some unknown pathways to the outcome.

Under Assumptions  \ref{assume::Mono} and \ref{assume::ER}, the decomposition \eqref{eq::decompose-ate-y-cace} only has the second term :
\begin{eqnarray}
\label{eq::aceY}
\tau_Y  = E\{ Y(1) - Y(0) \mid U=\cp   \} \pr(U = \cp ).
\end{eqnarray}
Similarly, we can decompose the average causal effect on $D$ into four terms:
\begin{eqnarray*}
\tau_D &=& E\{ D(1) - D(0) \mid U=\at  \} \pr(U = \at ) \\
&&+ E\{ D(1) - D(0) \mid U=\cp   \} \pr(U = \cp ) \\
&&+ E\{ D(1) - D(0) \mid U=\df  \} \pr(U = \df ) \\
&&+ E\{ D(1) - D(0) \mid U=\nt  \} \pr(U = \nt )  \\
&=& 0\times \pr(U = \at ) 
+ 1 \times \pr(U = \cp ) 
+ (-1)\times   \pr(U = \df )
+ 0 \times  \pr(U = \nt ) ,
\end{eqnarray*}
which, under Assumption \ref{assume::Mono}, reduces to
\begin{eqnarray}
\label{eq::aceD}
\tau_D  = \pr(U = \cp ) .
\end{eqnarray}
By \eqref{eq::aceD}, the proportion of the compliers $\pi_\cp = \pr(U = \cp)$ equals the average causal effect of the treatment assigned on $D$, an identifiable quantity under the CRE. Although we do not know all the compliers based on the observed data,  we can identify their proportion in the whole population based on \eqref{eq::aceD}. 
Combining \eqref{eq::aceY} and \eqref{eq::aceD}, we have the following result.

\begin{theorem}
\label{thm::CACE-identification}
Under Assumptions \ref{assume::Mono}--\ref{assume::ER}, we have 
$$
E\{ Y(1) - Y(0) \mid U=\cp   \} = \frac{ \tau_Y  }{ \tau_D}
$$
if $\tau_D \neq 0$. 
\end{theorem}

Following \citet{imbens1994identification} and \citet{angrist1996identification}, we define a new causal effect below.

\begin{definition}[CACE or LATE]\label{def::CACE}
Define 
$$
\tau_\cp  =  E\{ Y(1) - Y(0) \mid U=\cp   \}  
$$
as the ``complier average causal effect (CACE)'' or the ``local average treatment effect (LATE)''. It has alternative forms:
\begin{eqnarray*}
\tau_\cp  &=&  E\{ Y(1) - Y(0) \mid D(1) = 1 , D(0) = 0  \} \\
&=&    E\{ Y(1) - Y(0) \mid D(1)  >  D(0)    \} . 
\end{eqnarray*}
\end{definition}

The CACE is the average causal effect of $Z$ on $Y$ among compliers with $ D(1) = 1 , D(0) = 0$, or, equivalently, units with $ D(1)  >  D(0)$ under the monotonicity.
Based on Definition \ref{def::CACE}, we can rewrite Theorem \ref{thm::CACE-identification} as
$$
\tau_\cp  = \frac{ \tau_Y  }{ \tau_D} ,
$$
that is, the CACE or LATE equals the ratio of the average causal effects on $Y$ over that on $D$. Under Assumption \ref{assume::random}, we further identify the CACE below.

\begin{corollary}
\label{corollary::identification-CACE}
Under Assumptions \ref{assume::random}--\ref{assume::ER}, we have 
$$
\tau_\cp = \frac{ E(Y\mid Z=1) - E(Y\mid Z=0) }{ E(D\mid Z=1) - E(D \mid Z=0) } .
$$
\end{corollary}

Therefore, under CRE, monotonicity, and exclusion restriction, we can nonparametrically identify the CACE as the ratio of the difference in means of the outcome over the difference in means of the treatment received.

\subsection{Estimation}
\label{sec::estimation-IV}

Based on Corollary \ref{corollary::identification-CACE},  we can estimate $\tau_\cp$ by a simple ratio 
$$
\hat{\tau}_\cp  = \frac{  \hat{\tau}_Y }{  \hat{\tau}_D  } ,
$$ 
which is called the Wald estimator \citep{wald1940fitting} or the IV estimator. In the above discussion, $Z$ acts as the IV for $D$.

We can obtain the variance estimator based on the following heuristics (see Example \ref{thm::delta-method}):
$$
\hat{\tau}_\cp - \tau_\cp  = ( \hat{\tau}_Y  -  \tau_\cp\hat{\tau}_D)/\hat{\tau}_D 
\approx  ( \hat{\tau}_Y  -  \tau_\cp\hat{\tau}_D)/ \tau_D 
= \hat{\tau}_A / \tau_D ,
$$ 
where $\hat{\tau}_A$ is the difference-in-means of the adjusted outcome $A_i = Y_i -  \tau_\cp D_i$. So the asymptotic variance of $\hat{\tau}_\cp$ is close to the  variance  of $\hat{\tau}_A$ divided by $\tau_D^2$. 
The variance estimation proceeds in the following steps: 
\begin{enumerate}
\item
obtain the adjusted outcomes $\hat{A}_i =  Y_i -  \hat{\tau}_\cp D_i\ (i=1, \ldots, n)$;

\item
obtain the Neyman-type variance estimate based on the adjusted outcomes:
$$
\hat{V}_{\hat{A}} = \frac{\hat{S}_{\hat{A}}^2(1)}{n_1} + \frac{\hat{S}_{\hat{A}}^2(0)}{n_0},
$$
where $\hat{S}_{\hat{A}}^2(1)$ and $\hat{S}_{\hat{A}}^2(0)$ are the sample variances of the $\hat{A}_i $'s under the treatment and control, respectively; 

\item
obtain the final variance estimator   $ \hat{V}_{\hat{A}} /  \hat{\tau}_D ^2 .$ 
\end{enumerate}

See Problem \ref{problem::var-iv-estimate} for the justification of the above variance estimator. Alternatively, we can also use the bootstrap to approximate the variance of $\hat{\tau}_\cp $.
The following functions can calculate the point estimator $\hat{\tau}_\cp $ and the standard error with the formula $\hat{V}_{\hat{A}} $ and the bootstrap. 

\begin{lstlisting}
## IV point estimator
IV_Wald = function(Z, D, Y)
{
       tau_D = mean(D[Z==1]) - mean(D[Z==0])
       tau_Y = mean(Y[Z==1]) - mean(Y[Z==0])
       CACE  = tau_Y/tau_D
       
       c(tau_D, tau_Y, CACE)
}

## IV se via the delta method
IV_Wald_delta = function(Z, D, Y)
{
       est         = IV_Wald(Z, D, Y)
       AdjustedY   = Y - D*est[3]
       VarAdj      = var(AdjustedY[Z==1])/sum(Z) + 
                          var(AdjustedY[Z==0])/sum(1 - Z)
       
       c(est[3], sqrt(VarAdj)/abs(est[1]))
}

##IV se via the bootstrap
IV_Wald_bootstrap = function(Z, D, Y, n.boot = 200)
{
       est      = IV_Wald(Z, D, Y)
       
       CACEboot  = replicate(n.boot, {
         id.boot = sample(1:length(Z), replace = TRUE)
         IV_Wald(Z[id.boot], D[id.boot], Y[id.boot])[3]
       })
       
       c(est[3], sd(CACEboot))
}
\end{lstlisting}

Under the null hypothesis that $\tau_\cp = 0$, we can approximate the variance by $\hat{V}_Y /  \hat{\tau}_D ^2$, where $\hat{V}_Y$ is the Neyman-type variance estimate for the difference in means of $Y$. This variance estimator is inconsistent if the true $\tau_\cp$ is not zero. Therefore, it works for testing but not for estimation. Nevertheless, it gives insights for the ITT estimator and the Wald estimator. The ITT estimator $\hat{\tau}_Y$ has estimated standard error  $\sqrt{ \hat{V}_Y}$. The Wald estimator $ \hat{\tau}_Y / \hat{\tau}_D$ essentially equals the ITT estimator multiplied by $1/\hat{\tau}_D > 1$, which is larger in magnitude but at the same time, its estimated standard error increases by the same factor. 
Based on the variance estimators $\hat{V}_Y $ and $\hat{V}_Y /  \hat{\tau}_D ^2$, the confidence intervals for $\tau_Y$ and $\tau_\cp$ are
$$
\hat{\tau}_Y \pm z_{1-\alpha/2} \sqrt{ \hat{V}_Y}
$$
and
$$
\hat{\tau}_Y  / \hat{\tau}_D \pm z_{1-\alpha/2} \sqrt{ \hat{V}_Y}  / \hat{\tau}_D
= 
\left( \hat{\tau}_Y \pm z_{1-\alpha/2} \sqrt{ \hat{V}_Y}  \right) / \hat{\tau}_D 
$$
respectively, where $z_{1-\alpha/2}$ is the $1-\alpha/2$ upper quantile of the standard Normal random variable. 
These confidence intervals give the same qualitative conclusions since they will both cover zero or not. In some sense, the IV analysis provides the same qualitative information as the ITT analysis of $Y$ although it involves more complicated procedures.

\section{Covariates}

\subsection{Covariate adjustment in the CRE}

We now consider completely randomized experiments with covariates and assume 
$$
Z\ind \{ D(1), D(0), Y(1), Y(0), X \}.
$$
With covariates $X$, we can obtain \citet{lin2013}'s estimators $\hat{\tau}_{D,\textsc{L}}$ and $\hat{\tau}_{Y,\textsc{L}}$ for both $D$ and $Y$, resulting in $\hat{\tau}_{\cp, \textsc{L}} = \hat{\tau}_{Y,\textsc{L}} / \hat{\tau}_{D,\textsc{L}}$. 
 We can approximate the asymptotic variance of $\hat{\tau}_{\cp, \textsc{L}} $ using the bootstrap. The following functions can calculate the point estimator $\hat{\tau}_{\cp, \textsc{L}}p $ and the standard error based on the bootstrap. 
 
\begin{lstlisting}
## covariate adjustment in IV analysis
IV_Lin = function(Z, D, Y, X)
{
  X     = scale(as.matrix(X))
  tau_D = lm(D ~ Z + X + Z*X)$coef[2]
  tau_Y = lm(Y ~ Z + X + Z*X)$coef[2]
  CACE  = tau_Y/tau_D
  
  c(tau_D, tau_Y, CACE)
}


##IV_adj se via the bootstrap
IV_Lin_bootstrap = function(Z, D, Y, X, n.boot = 200)
{
  X         = scale(as.matrix(X))
  est       = IV_Lin(Z, D, Y, X)
  CACEboot  = replicate(n.boot, {
    id.boot = sample(1:length(Z), replace = TRUE)
    IV_Lin(Z[id.boot], D[id.boot], Y[id.boot], X[id.boot, ])[3]
  })
  
  c(est[3], sd(CACEboot))
}
\end{lstlisting}

\subsection{Covariates in conditional randomization or unconfounded observational studies}
\label{section::unconfounded-IV}

If randomization holds conditionally, i.e.,
$$
Z\ind \{ D(1), D(0), Y(1), Y(0) \} \mid  X,
$$
then we must adjust for covariates to avoid bias. The analysis is also straightforward since we already have discussed many estimators in Part \ref{part::observational-studies} for estimating the effects of $ Z$ on $D$ and $Y$, respectively. We can just use them in the ratio formula $\hat{\tau}_\cp = \hat{\tau}_Y / \hat{\tau}_D$ and use the bootstrap to approximate the asymptotic variance. 
I do not implement the corresponding estimator and variance estimator but relegate it as Problem \ref{hw::iv-conditional-x}.

\section{Weak IV}

\subsection{Some simulation}
 
Even if $\tau_D >0$, there is a positive probability that $\hat{\tau}_D$ is zero, so the variance of $\hat{\tau}_\cp$ is infinity (see Problem \ref{problem::infinite-var-iv}). The variance from the Normal approximation discussed before is not the variance of  $\hat{\tau}_\cp$ but rather the variance of its asymptotic distribution. This is a subtle technical point.  When $ \tau_D$ is close to $0$, which is referred to as the weak IV case, the ratio estimator $\hat{\tau}_\cp = \hat{\tau}_Y / \hat{\tau}_D$ has poor finite-sample properties.  In this scenario, $\hat{\tau}_\cp$ has finite-sample bias and non-Normal asymptotic distribution, and the corresponding Wald-type confidence intervals have poor coverage properties\footnote{The theory often assumes that $\tau_D$ has the order $n^{-1/2}$. Under this regime, the proportion of compliers goes to 0 as $n$ goes to infinity. The IV method can only identify a subgroup average causal effect with the proportion shrinking to 0.  This is a contrived regime for theoretical analysis. It is hard to justify this assumption in practice. The following discussion does not assume it.}. In the simple case with a binary outcome $Y$, we know that $\tau_Y$ must be bounded between $-1$ and $1$, but there is no guarantee that $\hat{\tau}_\cp$ is bounded between $-1$ and $1$.

\begin{figure}
\centering 
\begin{subfigure}[b]{\textwidth}
                \centering
                \includegraphics[width=\textwidth]{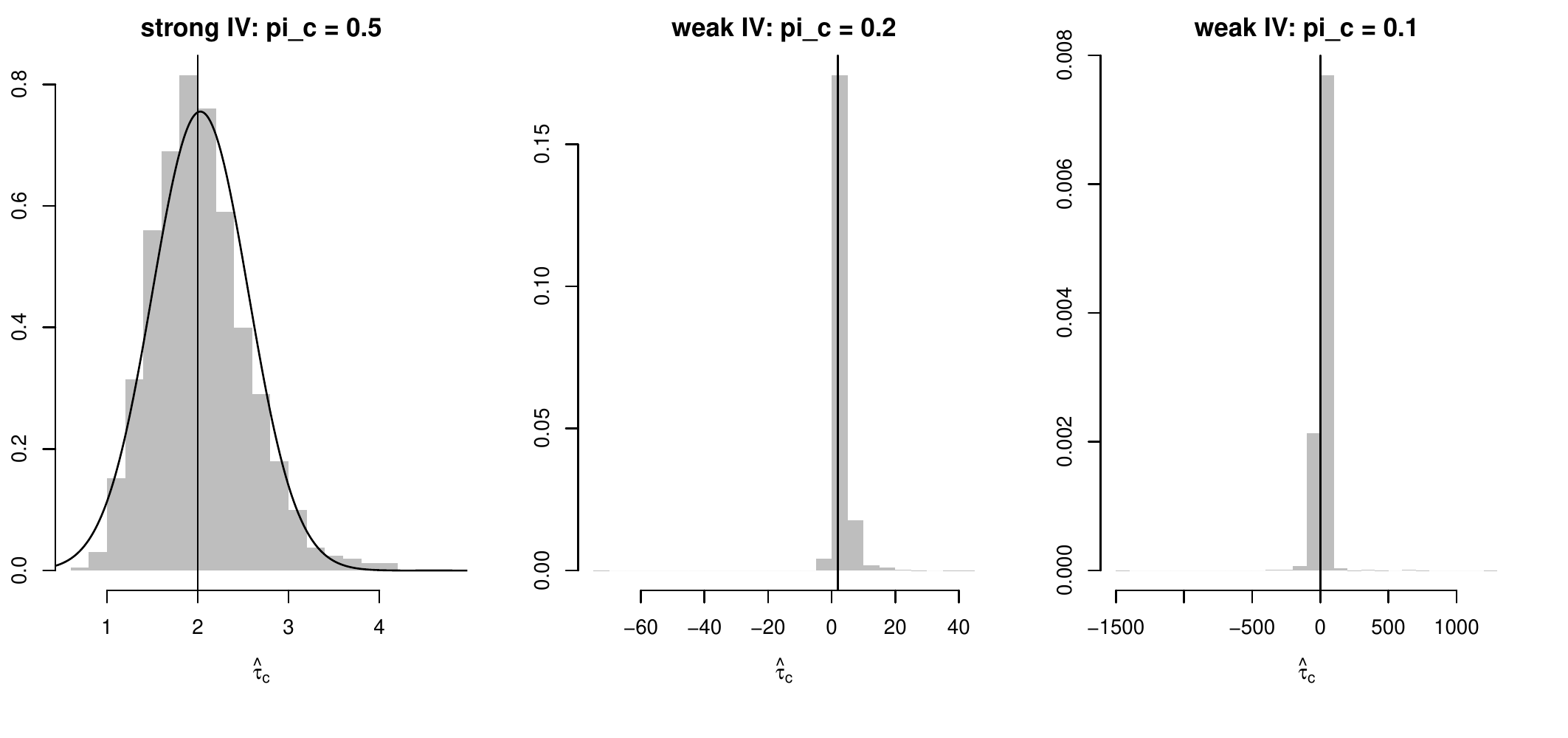}
                \caption{Without covariate adjustment}
                \label{fig::no-x-iv}
        \end{subfigure}%

\begin{subfigure}[b]{\textwidth}
                \centering
                \includegraphics[width=\textwidth]{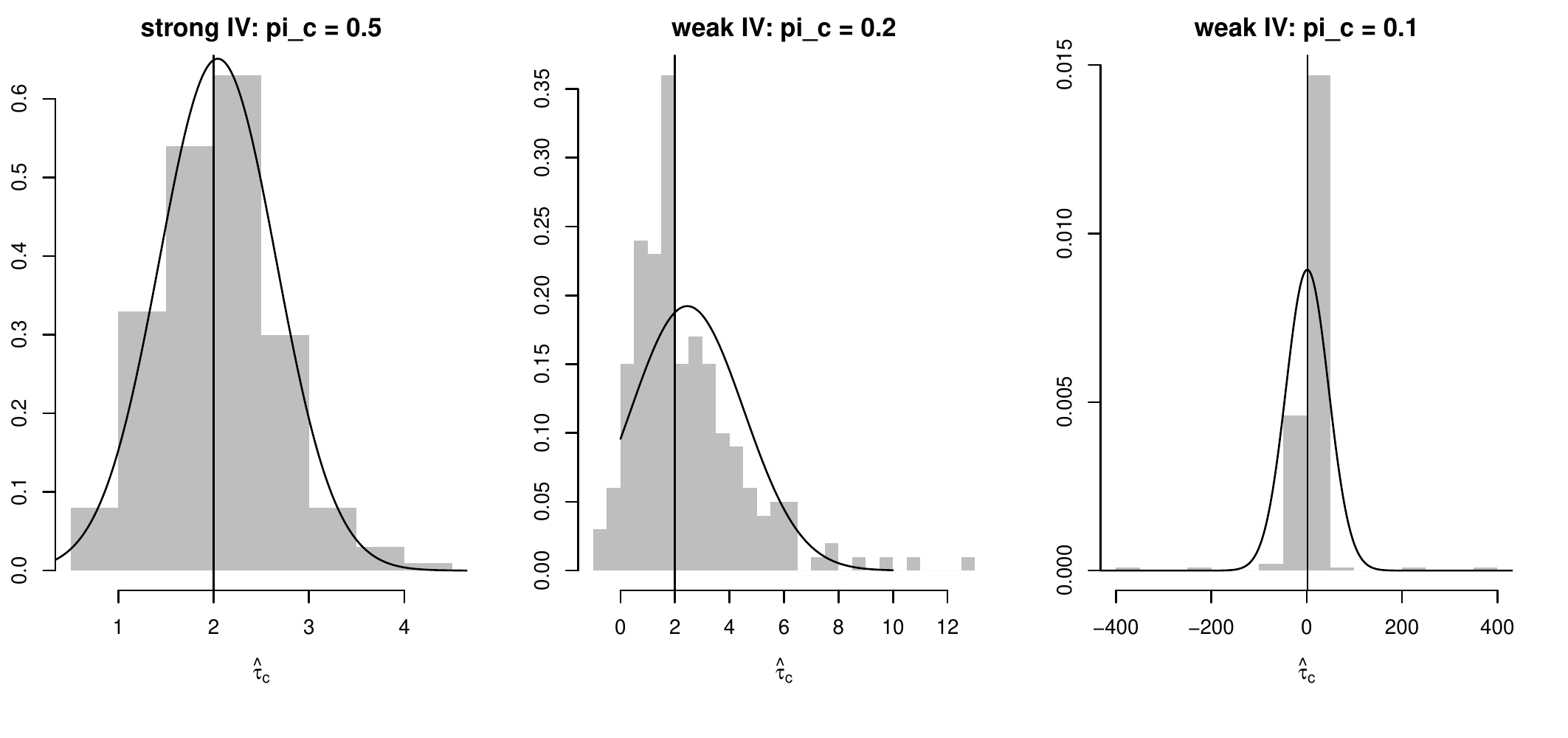}
                \caption{With covariate adjustment}
                \label{fig::x-iv}
        \end{subfigure}%
\caption{Simulation with strong and weak IVs}\label{fg::strong-weak-ivs}
\end{figure}

Figures \ref{fig::no-x-iv} and \ref{fig::x-iv} show the histograms of $\hat{\tau}_\cp$ and $\hat{\tau}_{\cp, \textsc{L}}$ over simulation with different $\pi_\cp$. I leave the detailed data-generating processes to the \ri{R} code. From the figures, the distributions of the estimators $\hat{\tau}_\cp$ and $\hat{\tau}_{\cp, \textsc{L}}$ are far from Normal when $\pi_\cp$ equals $0.2$ and $0.1$. Statistical inference based on asymptotic Normality is unlikely to be reliable. 

\subsection{A procedure robust to weak IV}

How do we deal with a weak IV?
From a testing perspective, there is an easy solution. Because $\tau_\cp = \tau_Y / \tau_D$, so the following two null hypotheses are equivalent:
$$
H_0: \tau_\cp  = 0 \Longleftrightarrow H_0': \tau_Y = 0
$$
if $\tau_D > 0$. 
Therefore, we simply test $H_0'$, i.e., the average causal effect of $Z$ on $Y$ is zero. This echoes our discussion in Section \ref{sec::estimation-IV} about the relationship between the ITT analysis and the IV analysis.

From an estimation perspective, we can focus on the confidence interval although the point estimator has poor finite-sample properties. Because $\tau_\cp = \tau_Y / \tau_D$, this is similar to the classical Fieller--Creasy problem in statistics. Below we discuss a strategy for constructing confidence intervals for $\tau_\cp$ motivated by \citet{fieller1954some}; see Chapter \ref{sec::fc-problem}. 
By the duality between confidence intervals and hypothesis tests (See Chapter \ref{section::duality-ci-testing}), we can construct a confidence set for $\tau_\cp$ by inverting a sequence of null hypotheses
$$
H_0(b): \tau_\cp = b . 
$$ 
Given the true value $\tau_\cp = b$, we have
$$
 \tau_Y  - b  \tau_D  = 0.
$$
The null hypothesis $H_0(b)$ is equivalent to the null hypothesis of zero average causal effect on the outcome $A_i( b) = Y_i - b  D_i $:
$$
H_0(b): \tau_{A(b)} = 0 .
$$

Let $ \hat{\tau}_A(b)$ be a generic point estimator for $\tau_{A(b)} $ with the associated variance estimator $\hat{V}_A( b )$. The point and variance estimators depend on the settings. 
In the CRE without covariates, $ \hat{\tau}_A(b)$ is the difference in means of the outcome $A_i( b)$ and $\hat{V}_A( b )$ is the Neyman-type variance estimator. 
In the CRE with covariates, $ \hat{\tau}_A(b)$ is \citet{lin2013}'s estimator for the outcome $A_i( b)$ and $\hat{V}_A( b )$ is the EHW variance estimator in the associated OLS fit of $Y_i - b D_i$ on $(Z_i, X_i, Z_i X_i)$ with the correction term\footnote{If we adopt the finite-population perspective, then we do not need this correction.} discussed in Chapter \ref{sec::cre-superpop}. 
In unconfounded observational studies,  we can obtain the estimator for the average causal effect on $A_i( b)$ and the associated variance estimator based on many existing strategies in Part \ref{part::observational-studies} of the book.

Based on $ \hat{\tau}_A(b)$ and $\hat{V}_A( b )$, we can construct a Wald-type test for $H_0(b)$. Inverting tests, we can construct the following confidence set for $\tau_\cp$:
$$
\left\{  b:    \frac{   \hat{\tau}_A^2(b)  }{  \hat{V}_A(b)  } \leq z_{1-\alpha/2}^2 \right\}
$$
where $z_{1-\alpha/2}$ is the $1-\alpha/2$ upper quantile of the standard Normal random variable. 
This is close to the Anderson--Rubin-type confidence interval in econometrics \citep{anderson1950asymptotic}.  Due to its connection to \citet{fieller1954some}, I will call it the Fieller--Anderson--Rubin (FAR) confidence interval. 
These weak-IV confidence intervals reduce to the asymptotic confidence intervals when the IV is strong. But they have additional guarantees when the IV is weak. I recommend using them in practice.

\begin{example}\label{eg::FAR-CI-noX}
To gain intuition about the FAR confidence interval, we look into the simple case of the CRE without covariates. The quadratic inequality in the confidence interval reduces to 
\begin{eqnarray*}
&& (\hat{\tau}_Y  - b   \hat{\tau}_D)^2 \\ 
& \leq & z_{1-\alpha/2}
\left[      n_1^{-1} \{  \hat{S}_Y^2(1)  + b^2\hat{S}_D^2(1) - 2 b \hat{S}_{YD}(1)  \}  \right. \\
&& \left. 
 + n_0^{-1} \{  \hat{S}_Y^2(0)  + b^2\hat{S}_D^2(0) - 2 b \hat{S}_{YD}(0)  \}    \right] ,
\end{eqnarray*}
where $\{ \hat{S}_Y^2(1) ,  \hat{S}_D^2(1) , \hat{S}_{YD}(1)  \}$ and $ \{ \hat{S}_Y^2(0), \hat{S}_D^2(0),\hat{S}_{YD}(0) \} $ are the sample variances and covariances of $Y$ and $D$ under the treatment and control, respectively. 
The confidence set can be a close interval, two disconnected intervals, an empty set, or the whole real line. I relegate the detailed discussion to Problem \ref{para::FAR-CI-cases}. 
\end{example}

\subsection{Implementation and simulation}\label{sec::FARci-simulation}

The following functions can compute a sequence of $p$-values as a function of $b$. The function \ri{FARci} does not use covariates, and the function \ri{FARciX} uses covariates which relies on the function \ri{linestimator} defined in Chapter \ref{sec::simulation-CRE-superp}. 

\begin{lstlisting}
FARci = function(Z, D, Y, Lower, Upper, grid)
{
  CIrange = seq(Lower, Upper, grid)
  Pvalue  = sapply(CIrange, function(t){
    Y_t       = Y - t*D
    Tauadj    = mean(Y_t[Z==1]) - mean(Y_t[Z==0])
    VarAdj    = var(Y_t[Z==1])/sum(Z) + 
      var(Y_t[Z==0])/sum(1 - Z)
    Tstat     = Tauadj/sqrt(VarAdj)
    (1 - pnorm(abs(Tstat)))*2
  })
  
  return(list(CIrange = CIrange, Pvalue  = Pvalue))
}

FARciX = function(Z, D, Y, X, Lower, Upper, grid)
{
  CIrange = seq(Lower, Upper, grid)
  X       = scale(X)
  Pvalue  = sapply(CIrange, function(t){
    Y_t       = Y - t*D
    linest    = linestimator(Z, Y_t, X)
    Tstat     = linest[1]/linest[3]
    (1 - pnorm(abs(Tstat)))*2
  })
  
  return(list(CIrange = CIrange, Pvalue  = Pvalue))
}
\end{lstlisting}

\begin{figure}
\centering 
\begin{subfigure}[b]{0.45\textwidth}
                \centering
                \includegraphics[width=\textwidth]{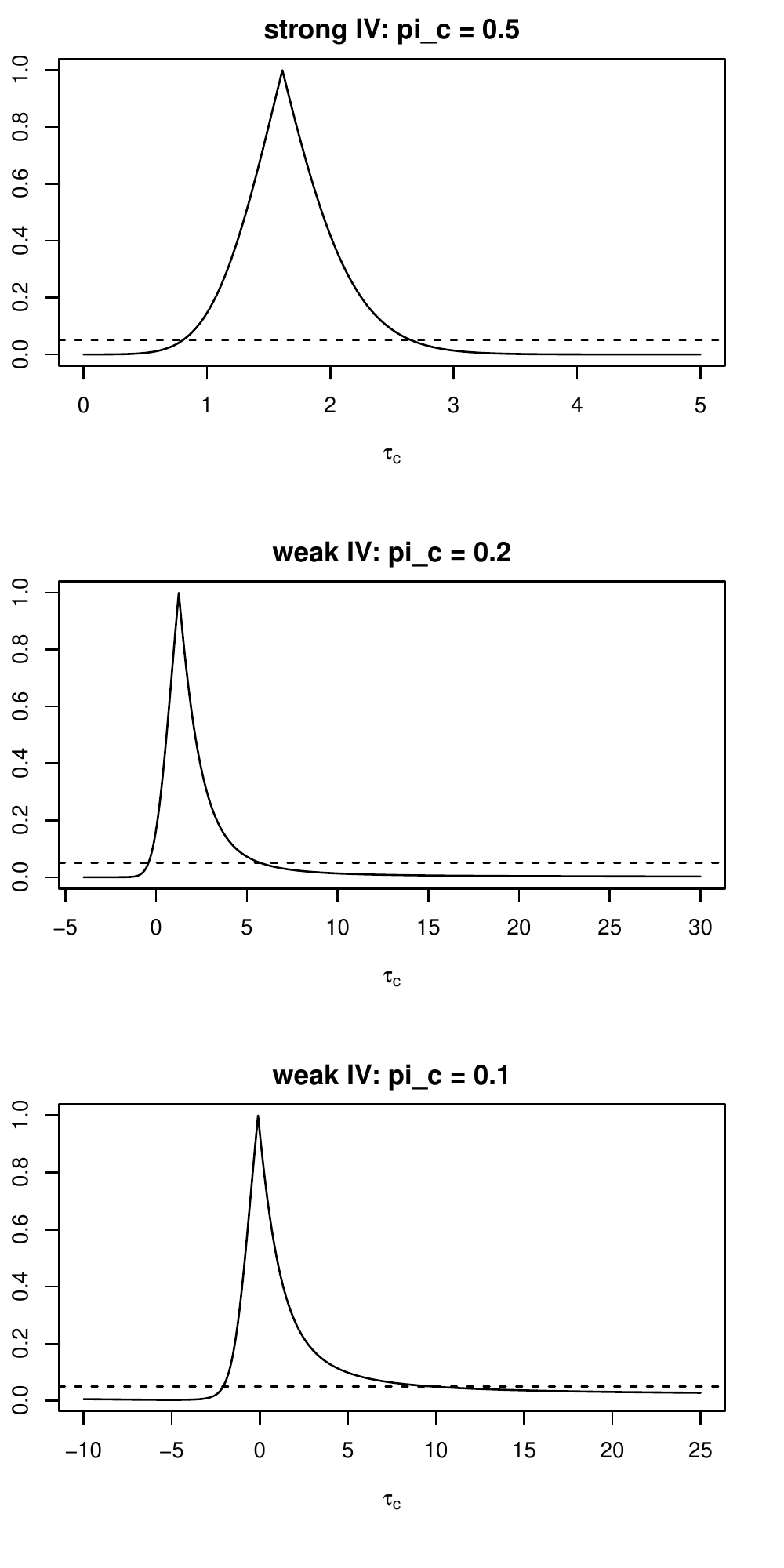}
                \caption{Realization 1}
                \label{fig::FARci-realization1}
        \end{subfigure}%
\begin{subfigure}[b]{0.45\textwidth}
                \centering
                \includegraphics[width=\textwidth]{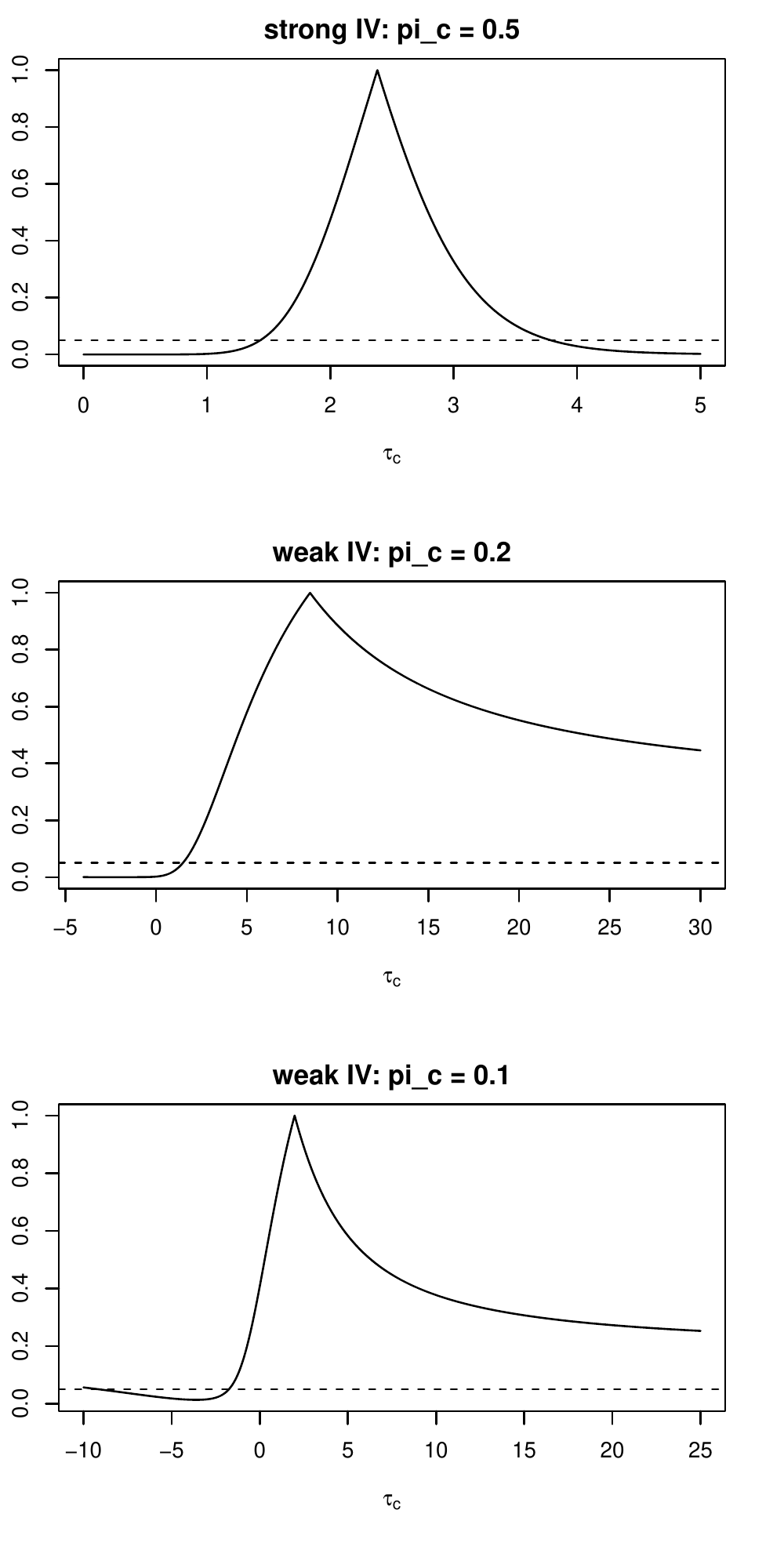}
                \caption{Realization 2}
                \label{fig::FARci-realization2}
        \end{subfigure}%
\caption{FAR confidence interval based on simulated data with different $\pi_\cp$: two realizations}\label{fg::FAR-ci-simulation}
\end{figure}

Figure \ref{fg::FAR-ci-simulation} displays the $p$-value as a function of $b$ in simulated data with different $\pi_\cp$. Figures \ref{fig::FARci-realization1} and \ref{fig::FARci-realization2} are based on two realizations of the same data-generating process, with details relegated to the \ri{R} code.  In Figure \ref{fig::FARci-realization1}, the FAR confidence sets are all closed intervals, whereas, in Figure \ref{fig::FARci-realization2}, the confidence sets are not closed intervals with $\pi_\cp = 0.2$ and $\pi_\cp = 0.1$. The shapes of the confidence sets remain stable across two realizations with $\pi_\cp = 0.5$.

\section{Application}
\label{sec::cace-application}

The \ri{mediation} package contains a dataset \ri{jobs} from the Job Search Intervention Study (JOBS II), which was a randomized field experiment that investigated the efficacy of a job training intervention on unemployed workers. The variable \ri{treat} is the indicator for whether a participant was randomly selected for the JOBS II training program, and the variable \ri{comply} is the indicator for whether a participant actually participated in the JOBS II program. An outcome of interest is \ri{job_seek} for measuring the level of job-search self-efficacy with values from 1 to 5. Covariates include \ri{sex}, \ri{age}, \ri{marital}, \ri{nonwhite}, \ri{educ}, and \ri{income}. 

 \begin{lstlisting}
> jobsdata = read.csv("jobsdata.csv")
> Z = jobsdata$treat
> D = jobsdata$comply
> Y = jobsdata$job_seek
> getX     = lm(treat ~ sex + age + marital 
+               + nonwhite + educ + income,
+               data = jobsdata)
> X = model.matrix(getX)[, -1]
\end{lstlisting}

We can estimate $\tau_\cp$ by $\hat{\tau}_\cp$ and obtain the standard error based on $\hat{V}_{\hat{A}}$ and the bootstrap. We can further conduct covariate adjustment to obtain $\hat{\tau}_{\cp,\textsc{L}}$ and obtain the standard error based on the bootstrap. The results are below. The point estimator and the standard error are stable across methods. 
 \begin{lstlisting}
> ## without covariates
> res = rbind(IV_Wald_delta(Z, D, Y),
+             IV_Wald_bootstrap(Z, D, Y, n.boot = 10^3))
> ## with covariates 
> res = rbind(res, 
+             IV_Lin_bootstrap(Z, D, Y, X, n.boot = 10^3)) 
> res = cbind(res, res[, 1] - 1.96*res[, 2],
+             res[, 1] + 1.96*res[, 2])
> row.names(res) = c("delta", "bootstrap", "with covariates")
> colnames(res)  = c("est", "se", "lower CI", "upper CI")
> round(res, 3)
                  est    se lower CI upper CI
delta           0.109 0.081   -0.050    0.268
bootstrap       0.109 0.083   -0.054    0.271
with covariates 0.118 0.082   -0.042    0.278
\end{lstlisting}

We can also construct the FAR confidence sets by inverting tests.  They are similar to the confidence intervals above. 
 \begin{lstlisting}
                   lower CI upper CI
without covariates   -0.050    0.267
with covariates      -0.047    0.282
\end{lstlisting}

\begin{figure}[ht]
\centering
\includegraphics[width = \textwidth]{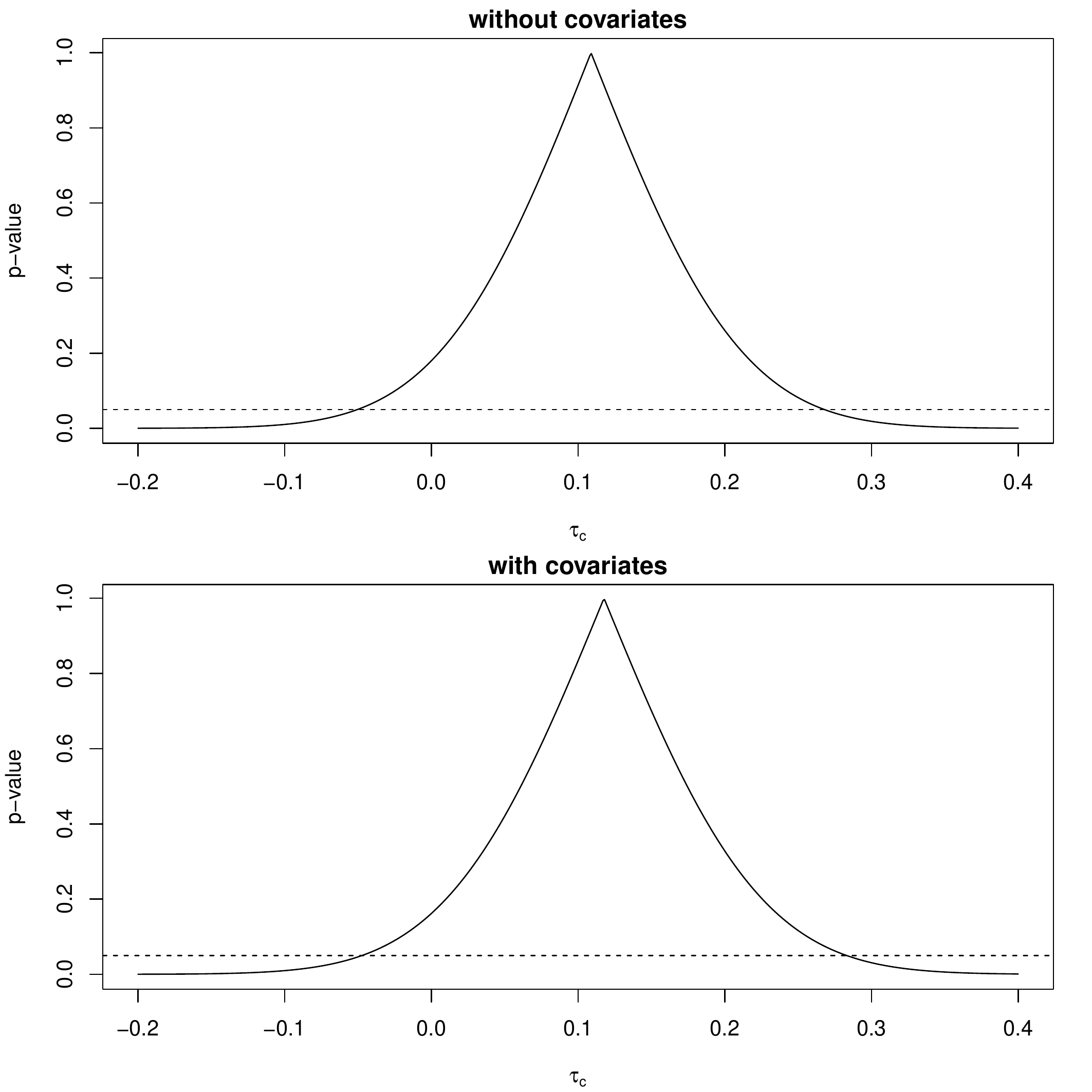}
\caption{Confidence interval of $\tau_\cp$ by inverting tests: upper panel without covariate adjustment, and lower panel with covariate adjustment}\label{fig::FAR-p-value}
\end{figure}

Figure \ref{fig::FAR-p-value} plots the $p$-values for a sequence of tests. 

\section{Interpreting the CACE}\label{eq::interpret-iv-double}

The notation for potential outcomes $\{ D(1), D(0), Y(1), Y(0)  \}$ is with respect to the hypothetical intervention of the treatment assigned $Z$. So $\tau_\cp$ is the average causal effect of the treatment assigned on the outcome for compliers. Fortunately, $D=Z$ for compliers, so we can also interpret $\tau_\cp$ as the average causal effect of the treatment received on the outcome for compliers. This partially answers the scientific question.

Some papers use different notation. For instance, \citet{angrist1996identification} use $Y_i(z,d)$ for the potential outcome of unit $i$ under a two-by-two factorial experiment\footnote{The name ``factorial experiment'' is from the experimental design literature \citep{dasgupta2014causal}. The experimenter randomizes multiple factors to each unit. The treatment in the factorial experiment has multiple levels.
} with the treatment assigned $z$ and treatment received $d$. \citet[][Section 3.1]{angrist2022empirical} comments on the intellectual history of this choice of notation. With the notation, the exclusion restriction assumption then has the following form.

\begin{assumption}[exclusion restriction]
\label{assume::ER-double-index}
$Y_i(z,d) = Y_i(d)$ for all $i$, that is, the potential outcome is only a function of $d$.
\end{assumption}

Based on the causal diagram below, Assumption \ref{assume::ER-double-index} rules out the direct arrow from $Z$ to $Y$. In such a case, $Z$ is an IV for $D$. 
\begin{center}
$
\xymatrix{
 && & U  \ar[dl] \ar[dr] &\\
Z \ar[rr] &&D \ar[rr]&&Y
}
$
\end{center}
Under Assumption \ref{assume::ER-double-index}, the augmented notation $Y_i(z,d)$ reduces to $Y_i(d)$, which justifies the name of ``exclusion restriction.'' Therefore, $Y_i(1,d) = Y_i(0,d)$ for $d=0,1$, which, coupled with Assumption \ref{assume::Mono}, implies that
\begin{eqnarray*}
Y_i(z=1) - Y_i(z=0) &=& Y_i(1, D_i(1)) - Y_i(0, D_i(0)) \\
&=& \left \{ \begin{array} {ll}
0 , & \text{if} \hspace{2mm}  U_i = \at, \\
0 , & \text{if}  \hspace{2mm}  U_i = \nt,\\
Y_i(d=1) - Y_i(d=0) ,& \text{if} \hspace{2mm} U_i = \cp.
\end{array}
\right.  
\end{eqnarray*}
In the above, I emphasize the potential outcomes are with respect to $z$, $d$, or both, to avoid confusion. 
The previous decomposition of $\tau_Y$ holds and we have the following result from \citet{imbens1994identification} and \citet{angrist1996identification}.

Recall the average causal effect on $D$, $\tau_D = E\{D(1)  - D(0) \},$ define the average causal effect on $Y$ as $\tau_Y = E\{ Y(D(1)) - Y(D(0))  \}$, and define the complier average causal effect as 
$$
\tau_\cp =  E\{  Y(d=1) - Y(d=0) \mid U = \cp \}.
$$

\begin{theorem}\label{thm::iv-double-index}
Under Assumptions \ref{assume::Mono}--\ref{assume::ER-double-index}, we have 
$$
Y(D(1)) - Y(D(0)) = \{  D(1)  - D(0) \} \times \{ Y(d=1) - Y(d=0) \}
$$
and  
$
\tau_\cp   =   \tau_Y /  \tau_D  .
$
\end{theorem}

The proof is almost identical to the proof of Theorem \ref{thm::CACE-identification} with modifications of the notation. I leave it as Problem \ref{para::iv-alternative-form-theorem}. From the notation $Y_i(d)$, it is more convenient to interpret $\tau_\cp$ as the average causal effect of the treatment received on the outcome for compliers.

\section{Homework problems}

\paragraph{Variance of the Wald estimator}\label{problem::infinite-var-iv}

Show that $\var(\hat{\tau}_\cp) = \infty$.

\paragraph{Asymptotic variance of the Wald estimator and its estimation}\label{problem::var-iv-estimate}

Consider the large-sample regime with $n\rightarrow \infty$. 
First show that $\sqrt{n}(\hat{\tau}_\cp - \tau_\cp) \rightarrow \textsc{N}(0, V)$ in distribution and find $V$. Then show that $\hat{V}_{\hat{A}} /V \rightarrow 1$ in probability.

\paragraph{Proof of the main theorem of  \citet{imbens1994identification} and \citet{angrist1996identification}} 
 \label{para::iv-alternative-form-theorem}
 
 Prove Theorem \ref{thm::iv-double-index}.

\paragraph{More on the FAR confidence set}\label{para::FAR-CI-cases}

The confidence set in Example \ref{eg::FAR-CI-noX} can be a close interval, two disconnected intervals, an empty set, or the whole real line. Find the precise condition for each case.

 \paragraph{More simulation for the FAR confidence set}\label{para::FAR-CI-simulation-X}

Figure \ref{fg::FAR-ci-simulation} shows the FAR confidence sets without using covariates. Conduct parallel simulation for 
 the FAR confidence sets adjusting for covariates in CRE.

\paragraph{Binary IV and ordinal treatment received}\label{para::iv-ordinal-treatment}

\citet{angrist1995two} discussed a more general setting with a binary IV $Z$, an ordinal treatment received $D \in \{0,1,\ldots, J\}$, and an outcome $Y$. The ordinal treatment received has potential outcomes $D(1)$ and $D(0)$ with respect to the binary IV, and the outcome has potential outcomes $Y(z,d)$ with respect to both the binary IV and the ordinal treatment received. Extend the discussion in Section \ref{eq::interpret-iv-double} and the corresponding IV assumptions as below.

\begin{assumption}\label{assumption::iv-ordinal-treatment}
We have
(1) randomization that $Z\ind \{   D(z),  Y(z,d)  : z=0,1; d = 0,1,\ldots, J \}$; 
(2) monotonicity that $D(1) \geq D(0)$; and
(3) exclusion restriction that $Y(z,d) = Y(d)$ for all $z=0,1$ and $ d = 0,1,\ldots, J $. 
\end{assumption}

They proved Theorem \ref{thm::iv-ordinal-treatment} below.

\begin{theorem}\label{thm::iv-ordinal-treatment}
Under Assumption \ref{assumption::iv-ordinal-treatment}, we have 
$$
\frac{  E(Y\mid Z=1) - E(Y\mid Z=0) }{ E(D\mid Z=1) - E(D\mid Z=0)  }
= \sum_{j=1}^J  w_j E\{  Y(j) - Y(j-1)  \mid D(1) \geq j > D(0) \}
$$
where
$$
w_j = \frac{   \pr\{  D(1) \geq j > D(0)   \}   }{   \sum_{j'=1}^J  \pr\{  D(1) \geq j' > D(0)   \}   }.
$$
\end{theorem}

Prove Theorem \ref{thm::iv-ordinal-treatment}.

Remark: 
When $J=1$, Theorem \ref{thm::iv-ordinal-treatment} reduces to Theorem \ref{thm::iv-double-index}. With a general $J$, it states that the standard IV formula identifies a weighted average of some latent subgroup effects. The weights are proportional to the probability of the latent groups defined by $  D(1) \geq j > D(0) $, and the latent subgroup effect $E\{  Y(j) - Y(j-1)  \mid D(1) \geq j > D(0) \}$ compares the adjacent levels of the treatment received. However, this weighted average may not be easy to interpret because the latent groups overlap.

The proof can be tedious. 
A trick is to write the treatment received and outcome under treatment assignment $z$ as 
$$
D(z) = \sum_{j=0}^J j 1\{ D(z) = j\} ,\quad 
Y(D(z)) = \sum_{j=0}^J  Y(j)1\{ D(z) = j\} 
$$
to obtain
$$
D(1) - D(0) = \sum_{j=0}^J j [ 1\{ D(1) = j\} -1\{ D(0) = j\} ]
$$
and
$$
 Y(D(1))  - Y(D(0))   = 
\sum_{j=0}^J Y(j)   [ 1\{ D(1) = j \}   - 1\{ D(0) = j \}  ].
$$
Then use the following {\it Abel's lemma}, also called {\it summation by parts}: 
$$
\sum_{j=0}^J  f_j (g_{j+1} - g_j) 
= 
f_J g_{J+1} - f_0 g_0  - \sum_{j=1}^J g_j  (f_j - f_{j-1})
$$
for appropriately specified sequences $(f_j)$ and $(g_j)$.

\paragraph{Data analysis: a flu shot encouragement design \citep{mcdonald1992effects}}\label{problem::flushot}

The dataset in \ri{fludata.txt} is from a randomized encouragement design of \citet{mcdonald1992effects}, which was also re-analyzed by \citet{hirano2000assessing}. It contains the following variables:

\begin{tabular}{ll}
\hline 
\texttt{assign} & binary encouragement to receive the flu shot\\
\texttt{receive} & binary indicator for receiving the flu shot \\
\texttt{outcome} & binary outcome for flu-related hospitalization\\
\texttt{age} & age of the patient\\
\texttt{sex} & sex of the patient\\
\texttt{race} & race of the patient\\
\texttt{copd} &  chronic obstructive pulmonary disease\\
\texttt{dm} & diabetes\\
\texttt{heartd} & heart  disease\\
\texttt{renal} & renal  disease\\
\texttt{liverd} & liver  disease\\
\hline 
\end{tabular}

Analyze the data with and without adjusting for the covariates.

\paragraph{IV estimation conditional on covariates}\label{hw::iv-conditional-x}

Implement the estimators and corresponding variance estimators mentioned in Chapter \ref{section::unconfounded-IV} in \ri{R}. 

Remark: The problem is useful for Problem \ref{hw::analyze-karolinska-iv}.

\paragraph{Data analysis: the Karolinska data}\label{hw::analyze-karolinska-iv}

Revisit Problem \ref{hw::Karolinska-dr}. 
\citet{rubin2008objective} used the Karolinska data as an example for the IV method. In \texttt{karolinska.txt}, whether a patient was diagnosed at a large volume hospital can be viewed as an IV for whether a patient was treated at a large volume hospital. This is plausible conditional on other observed covariates. See \citet{rubin2008objective}'s analysis for more details.

Re-analyze the data assuming that the IV is randomly assigned conditional on observed covariates.

 \paragraph{Data analysis: a job training program}

The file \texttt{jobtraining.rtf} contains the description of the data files \texttt{X.csv} and \texttt{Y.csv}. 

The dataset \texttt{X.csv} contains the pretreatment covariates. You can view the sampling weight variable \texttt{wgt} as a covariate too. Many previous analyses made this simplification although this is always a controversial issue in statistical analysis of survey data. 
Conduct analyses with and without covariates. 

The dataset \texttt{Y.csv} contains the sampling weight, treatment assigned, treatment received, and many post-treatment variables. Therefore, this dataset contains many outcomes depending on your questions of interest. The data also have many complications. First, some outcomes are missing. Second, unemployed individuals do not have wages. Third, the outcomes are repeatedly observed over time. When you analyze the data, please give details about your choice of the questions of interest and estimators. 

Remark: 
\citet{schochet2008does} analyzed the original data.
\citet{frumento2012evaluating} provided a more sophisticated analysis based on the framework in Chapter \ref{chapter::principal-stratification} later.

 \paragraph{Recommended reading}

\citet{angrist1996identification} bridged the econometric IV perspective and statistical causal inference based on potential outcomes and demonstrated its usefulness with an application. 

Some other early references on IV are \citet{permutt1989simultaneous},  \citet{sommer1991estimating}, \citet{baker1994paired}, and \citet{cuzick1997adjusting}.

\chapter{Disentangle Mixture Distributions and Instrumental Variable Inequalities}
 \label{chapter::iv-inequalities}

 The IV model in Chapter \ref{chapter::iv-experiment} imposes  Assumptions \ref{assume::random}--\ref{assume::ER}:
 \begin{enumerate}
 \item
 $Z\ind \{D(1), D(0) ,Y(1), Y(0)   \}$;
 \item
 $\pr(U = \df) = 0$;
 \item 
 $Y(1) = Y(0)$ for $U=\at$ or  $\nt$. 
 \end{enumerate}

Table \ref{table::obs-latent} summarizes the observed groups and the corresponding latent groups under the monotonicity assumption.

\begin{table}[h]
\centering
\caption{Observed groups and latent groups under Assumption \ref{assume::Mono} }\label{table::obs-latent}
\begin{tabular}{cccc}
\hline 
$Z$ & $D$ & $D(1)$ & latent groups \\
\hline 
$Z=1$ & $D=1$ & $D(1) = 1$ & $U = \cp$ or $\at$ \\
$Z=1$ & $D=0$ & $D(1) = 0$ & $U = \nt$ \\ 
$Z=0$ & $D=1$ & $D(0) = 1$ & $U = \at$  \\ 
$Z=0$ & $D=0$ & $D(0) = 0$ & $U = \cp$ or $\nt$ \\
\hline 
\end{tabular}
\end{table}

Interestingly, Assumptions \ref{assume::random}--\ref{assume::ER} together have some testable implications. \citet{balke1997bounds} called them the {\it instrumental variable inequalities}. This chapter will give an intuitive derivation of a special case of these inequalities. The proof is a direct consequence of identifying the means of the potential outcomes for all latent groups defined by $U$. 

\section{Disentangle Mixture Distributions}
\label{sec::disentangle-mixtures}

We summarize the main results in Theorem \ref{thm::cace-mixture-components} below. 
Define 
$$
\pi_u = \pr(U = u) \quad  \quad  (u=\at, \nt, \cp) 
$$ as the proportion of type $U=u$, and
$$
\mu_{zu} = E\{  Y(z) \mid U = u\}\quad  \quad (z=0,1;u=\at, \nt, \cp) .
$$ 
as the mean of the potential outcome $Y(z) $ for type $U=u$. 
Exclusion restriction implies that $\mu_{1\nt} = \mu_{0\nt} $ and $\mu_{1\at} = \mu_{0\at} $. Let $\mu_{\nt}$ and $\mu_{\at} $ denote them, respectively.

\begin{theorem}
\label{thm::cace-mixture-components}
Under Assumptions \ref{assume::random}--\ref{assume::ER}, we can identify the proportions of the latent types by
\begin{eqnarray*}
 \pi_\nt &=& \pr(D=0\mid Z=1)  ,\\
  \pi_\at &=&\pr(D=1\mid Z=0) , \\
  \pi_\cp &=& E(D\mid Z=1) - E(D\mid Z=0) ,
\end{eqnarray*}
and the type-specific means of the potential outcomes by
\begin{eqnarray*}
  \mu_{\nt} &=& E(Y\mid Z=1, D=0) ,\\
 \mu_{\at} &=& E(Y\mid Z=0, D=1),\\
 \mu_{1\cp} &=& \pi_\cp^{-1}\left\{   E(DY\mid Z=1 )  -  E(DY\mid Z=0)    \right\}, \\
  \mu_{0\cp}  &=& \pi_\cp^{-1} \big[   E\{ (1-D)Y\mid Z=0 \}  -  E\{(1-D)Y\mid Z=1 \}    \big]. 
\end{eqnarray*}
\end{theorem}

\begin{myproof}{Theorem}{\ref{thm::cace-mixture-components}}
Part I: We first identify the proportions of the latent compliance types.
We can identify the proportion of the never takers by
\begin{eqnarray*}
\pr(D=0\mid Z=1) 
&=& \pr(U=\nt \mid Z=1)  \\
&=& \pr(U=\nt) \\
&=&  \pi_\nt ,
\end{eqnarray*}
and the proportion of the always takers by
\begin{eqnarray*}
\pr(D=1\mid Z=0) 
&=& \pr(U=\at \mid Z=0)  \\
&=& \pr(U=\at) =  \pi_\at . 
\end{eqnarray*}
Therefore, the proportion of compliers is
\begin{eqnarray*}
\pi_\cp &=& \pr(U=\cp) \\
&=& 1-  \pi_\nt -  \pi_\at  \\
&=& 1 - \pr(D=0\mid Z=1)  - \pr(D=1\mid Z=0) \\
&=& E(D\mid Z=1) - E(D\mid Z=0) \\
&=& \tau_D,
\end{eqnarray*}
which is coherent with our previous discussion. Although we do not know individual latent compliance types for all units, we can identify the proportions of never-takers, always-takers, and compliers.

Part II: We then identify the means of the potential outcomes within latent compliance types. 
The observed group $(Z=1,D=0)$ only has never takers, so 
$$
E(Y\mid Z=1, D=0) = E\{  Y(1) \mid Z=1, U=\nt \} = E\{  Y(1) \mid   U=\nt \}  =
\mu_{\nt}  . 
$$
The observed group $(Z=0,D=1)$ only has always takers, so 
$$
E(Y\mid Z=0, D=1) = E\{  Y(0) \mid Z=0, U=\at \} = E\{  Y(0) \mid   U=\at \}  =
\mu_{\at} . 
$$
The observed group $(Z=1,D=1)$ has both compliers and always takers, so
\begin{eqnarray*}
E(Y\mid Z=1, D=1) &=& E\{ Y(1)\mid Z=1, D(1)=1\} \\
&=& E\{ Y(1)\mid D(1)=1\}  \\   
&=& \pr\{  D(0)=1 \mid D(1)=1 \} E\{ Y(1)\mid D(1)=1, D(0)=1\}  \\
&&+ \pr\{  D(0)=0 \mid D(1)=1 \}E\{ Y(1)\mid D(1)=1, D(0)=0\}   \\
&=& \frac{\pi_\cp}{\pi_\cp + \pi_\at} \mu_{1\cp} +  \frac{\pi_\at}{\pi_\cp + \pi_\at} \mu_{\at}.
\end{eqnarray*}
Solve the linear equation above to obtain
\begin{eqnarray*}
\mu_{1\cp} &=& \pi_\cp^{-1} \left\{  (\pi_\cp + \pi_\at)  E(Y\mid Z=1, D=1)  - \pi_\at    E(Y\mid Z=0, D=1)    \right\} \\
&=& \pi_\cp^{-1} \left\{  \pr(D=1\mid Z=1)  E(Y\mid Z=1, D=1) \right.\\
&&\left. \qquad  -\pr(D=1\mid Z=0) E(Y\mid Z=0, D=1)    \right\} \\
&=&  \pi_\cp^{-1}\left\{   E(DY\mid Z=1 )  -  E(DY\mid Z=0)    \right\}.
\end{eqnarray*}
The observed group $(Z=0,D=0)$ has both compliers and never takers, so 
\begin{eqnarray*}
E(Y\mid Z=0,D=0) &=& E\{ Y(0)\mid Z=0, D(0)=0\} \\
&=& E\{ Y(0)\mid D(0)=0\} \\ 
&=& \pr\{  D(1)=1 \mid D(0)=0 \} E\{ Y(0)\mid D(1)=1, D(0)=0\}  \\
&&+ \pr\{  D(1)=0 \mid D(0)= 0\}E\{ Y(0)\mid D(1)=0, D(0)=0\}   \\
&=& \frac{\pi_\cp}{\pi_\cp + \pi_\nt} \mu_{0\cp} +  \frac{\pi_\nt}{\pi_\cp + \pi_\nt} \mu_{\nt}.
\end{eqnarray*}
Solve the linear equation above to obtain 
\begin{eqnarray*}
\mu_{0\cp}  &=& \pi_\cp^{-1} \left\{  (\pi_\cp + \pi_\nt)  E(Y\mid Z=0, D=0)  - \pi_\nt    E(Y\mid Z=1, D=0)    \right\} \\
&=& \pi_\cp^{-1} \left\{ \pr(D=0\mid Z=0)  E(Y\mid Z=0, D=0)  \right.\\
&&\left. \qquad  -\pr(D=0\mid Z=1)   E(Y\mid Z=1, D=0)    \right\} \\
&=& \pi_\cp^{-1} \big[   E\{ (1-D)Y\mid Z=0 \}  -  E\{(1-D)Y\mid Z=1 \}    \big]. 
\end{eqnarray*}
\end{myproof}

Based on the formulas of $\mu_{1\cp} $ and $\mu_{0\cp} $ in Theorem \ref{thm::cace-mixture-components}, we can simplify $\tau_\cp = \mu_{1\cp}  - \mu_{0\cp}  $ as 
$$
\tau_\cp =  \{  E(Y\mid Z=1) - E(Y\mid Z=0) \}/\pi_\cp ,
$$
which is the same as the formula in Theorem \ref{thm::CACE-identification} before. 

Theorem \ref{thm::cace-mixture-components} focuses on identifying the means of the potential outcomes, $\mu_{zu} $. \citet{imbens1997estimating} derived more general identification formulas for the distribution of the potential outcomes; I leave the details to Problem \ref{hw::disentangle-mixture-distrubution}.

\section{Testable implications: Instrumental Variable Inequalities}

Is there any additional value of the detour to derive the formula of $\tau_\cp$ through Theorem \ref{thm::cace-mixture-components}? The answer is yes. For binary outcome, the following inequalities must be true: 
$$
0\leq \mu_{1\cp} \leq 1, \quad 
0\leq \mu_{0\cp} \leq 1, 
$$
which implies four inequalities
\begin{eqnarray*}
E(DY\mid Z=1 )  -  E(DY\mid Z=0) &\geq & 0 ,\\
E(DY\mid Z=1 )  -  E(DY\mid Z=0) &\leq  & E(D\mid Z=1) - E(D\mid Z=0)  ,\\ 
E\{ (1-D)Y\mid Z=0 \}  -  E\{(1-D)Y\mid Z=1 \}   &\geq & 0,\\
E\{ (1-D)Y\mid Z=0 \}  -  E\{(1-D)Y\mid Z=1 \}  &\leq & E(D\mid Z=1) - E(D\mid Z=0) . 
\end{eqnarray*}
Rearranging terms, we obtain the following unified inequalities.

\begin{theorem}[Instrumental Variable Inequalities]\label{thm::iv-inequality}
With a binary outcome $Y$, Assumptions \ref{assume::random}--\ref{assume::ER} imply 
\begin{eqnarray}\label{eq::iv-inequalities}
E(Q\mid Z=1) - E(Q\mid Z=0) \geq 0,  
\end{eqnarray}
where $Q = DY, D(1-Y), (D - 1)Y$ and $D+Y-DY$.
\end{theorem}

Under the IV assumptions \ref{assume::random}--\ref{assume::ER}, the difference in means for $Q = DY, D(1-Y), (D - 1)Y$ and $D+Y-DY$ must all be non-negative. Importantly, these implications only involve the distribution of the observed variables. Rejection of the IV inequalities leads to rejection of the  IV assumptions.

\citet{balke1997bounds} derived more general IV inequalities with and without assuming monotonicity. The proving strategy above is due to  \citet{jiang2020measurement} for a slightly more complex setting. Theorem \ref{thm::iv-inequality} states the testable implications only for a binary outcome. Problem \ref{hw::ivineq-binary} gives an equivalent form, and Problem \ref{hw::ivineq-general-Y} gives the result for a general outcome.

\section{Examples}\label{section::binary-y-ivineq}

For a binary outcome, we can estimate all the parameters by the method of moments below.

\begin{lstlisting}
IVbinary = function(n111, n110, n101, n100, 
                    n011, n010, n001, n000){
  
  n_tr = n111 + n110 + n101 + n100
  n_co = n011 + n010 + n001 + n000
  n    = n_tr + n_co
  
  ## proportions of the latent strata
  pi_n = (n101 + n100)/n_tr
  pi_a = (n011 + n010)/n_co
  pi_c = 1 - pi_n - pi_a
  
  ## four observed means of the outcomes (Z=z,D=d)
  mean_y_11 = n111/(n111 + n110)
  mean_y_10 = n101/(n101 + n100)
  mean_y_01 = n011/(n011 + n010)
  mean_y_00 = n001/(n001 + n000)
  
  ## means of the outcomes of two strata
  mu_n1 = mean_y_10
  mu_a0 = mean_y_01
  ## ER implies the following two means
  mu_n0 = mu_n1
  mu_a1 = mu_a0
  ## stratum (Z=1,D=1) is a mixture of c and a
  mu_c1 = ((pi_c + pi_a)*mean_y_11 - pi_a*mu_a1)/pi_c
  ## stratum (Z=0,D=0) is a mixture of c and n
  mu_c0 = ((pi_c + pi_n)*mean_y_00 - pi_n*mu_n0)/pi_c
  
  ## identifiable quantities from the observed data
  list(pi_c = pi_c, 
       pi_n = pi_n, 
       pi_a = pi_a, 
       mu_c1= mu_c1,
       mu_c0= mu_c0,
       mu_n1= mu_n1,
       mu_n0= mu_n0,
       mu_a1= mu_a1,
       mu_a0= mu_a0,
       tau_c= mu_c1 - mu_c0)
}
\end{lstlisting}

We then re-visit two canonical examples with binary data. 

\begin{example}\label{eg::1-iv-inequalities}
\citet{improve2014endovascular} assess the effectiveness of the emergency endovascular versus the open surgical repair strategies for patients with a clinical diagnosis of ruptured aortic aneurism. Patients are randomized to either the emergency endovascular or the open repair strategy. The primary outcome is the survival status after 30 days. Let $Z$ be the treatment assigned, with $Z=1$ for the endovascular strategy and $Z=0$ for the open repair. Let $D$ be the treatment received. 
Let $Y$ be the survival status, with $Y=1$ for dead, and $Y=0$ for alive. Table \ref{tab::data1} summarizes the observed data.
Using the \ri{IVbinary} function above, we can obtain the following estimates:
\begin{lstlisting}
> investigators_analysis = IVbinary(n111 = 107,
+                                   n110 = 42,
+                                   n101 = 68,
+                                   n100 = 42,
+                                   n011 = 24,
+                                   n010 = 8,
+                                   n001 = 131,
+                                   n000 = 79)
>                         
> investigators_analysis
$pi_c
[1] 0.4430582

$pi_n
[1] 0.4247104

$pi_a
[1] 0.1322314

$mu_c1
[1] 0.7086064

$mu_c0
[1] 0.6292042

$mu_n1
[1] 0.6181818

$mu_n0
[1] 0.6181818

$mu_a1
[1] 0.75

$mu_a0
[1] 0.75

$tau_c
[1] 0.07940223
\end{lstlisting}
There is no evidence of violating the IV assumptions. 
\end{example}

\begin{example}\label{eg::2-iv-inequalities}
In \citet{hirano2000assessing}, physicians are randomly selected to receive a letter encouraging them to inoculate patients at risk for flu. The treatment is the actual flu shot, and the outcome is an indicator of flu-related hospital visits. However, some patients do not comply with their assignments. Let $Z_i$ be the indicator of encouragement to receive the flu shot, with $Z=1$ if the physician receives the encouragement letter, and  $Z=0$ otherwise. Let $D$ be the treatment received.
Let $Y$ be the outcome, with $Y=0$ if for a flu-related hospitalization during the winter, and $Y=1$ otherwise. See Problem \ref{problem::flushot} for more details of the data. Table \ref{tab::data2} summarizes the observed data.
Using the \ri{IVbinary} function above, we can obtain the following estimates:
\begin{lstlisting}
> flu_analysis = IVbinary(n111 = 31,
+                         n110 = 422,
+                         n101 = 84,
+                         n100 = 935,
+                         n011 = 30,
+                         n010 = 233,
+                         n001 = 99,
+                         n000 = 1027)
> flu_analysis
$pi_c
[1] 0.1183997

$pi_n
[1] 0.6922554

$pi_a
[1] 0.1893449

$mu_c1
[1] -0.004548064

$mu_c0
[1] 0.1200094

$mu_n1
[1] 0.08243376

$mu_n0
[1] 0.08243376

$mu_a1
[1] 0.1140684

$mu_a0
[1] 0.1140684

$tau_c
[1] -0.1245575
\end{lstlisting}
Since $\hat\mu_{1\cp} < 0$, there is evidence of violating the IV assumptions. 
\end{example}

 \begin{table}[t]
\caption{Binary data and IV inequalities}
\begin{center}
\begin{subtable}{0.5\textwidth}
\caption{ \citet{improve2014endovascular}'s study}
\begin{tabular}{cccccc} 
\hline 
      &   \multicolumn{2}{c}{$Z=1$}  &  &  \multicolumn{2}{c}{$Z=0$}  \\  \cline{2-3} \cline{5-6} 
      & $D=1$  &  $D=0$           &       &  $D=1$  &  $D=0$\\
 $Y=1$ &   107    &      68         &      &    24       &    131        \\
 $Y=0$ &   42    &     42            &      &     8         &     79           \\  
 \hline
\end{tabular}
\label{tab::data1}
\end{subtable}%
\vspace{0.1in}
\begin{subtable}{0.5\textwidth}
\caption{ \citet{hirano2000assessing}'s study}
\begin{tabular}{cccccc}
\hline 
      &   \multicolumn{2}{c}{$Z=1$}  &  &  \multicolumn{2}{c}{$Z=0$}  \\  \cline{2-3} \cline{5-6} 
      & $D=1$  &  $D=0$           &       &  $D=1$  &  $D=0$\\
 $Y=1$ &   31    &     85         &      &     30       &    99       \\
 $Y=0$ &   424   &     944            &      &     237         &    1041           \\  
 \hline
\end{tabular}
\label{tab::data2}
\end{subtable}
\end{center}
\end{table}%

\section{Homework problems}

\paragraph{More detailed data analysis}\label{hw::more-data-binary}

Examples \ref{eg::1-iv-inequalities} and \ref{eg::2-iv-inequalities} ignored the uncertainty in the estimates. Calculate the confidence intervals for the true parameters.

\paragraph{Risk ratio for compliers}\label{hw::rr-compliers}

With a binary outcome, we can define the risk ratio for compliers as
$$
\textsc{rr}_\cp = \frac{  \pr\{  Y(1) = 1\mid U = \cp \}  }{ \pr\{  Y(0) = 1\mid U = \cp \}   } .
$$

Show that under Assumptions \ref{assume::random}--\ref{assume::ER}, we can identify it by
$$
\textsc{rr}_\cp =  \frac{  E(DY\mid Z=1)  - E(DY\mid Z=0)  }{  E\{  (D-1)Y\mid Z=1 \}   - E\{ (D-1)Y\mid Z=0 \} }.
$$

Remark: Using Theorem \ref{thm::cace-mixture-components}, we can identify any comparisons between $E\{  Y(1) \mid U = \cp \} $ and $E\{  Y(0) \mid U = \cp \} $.

\paragraph{Disentangle the mixtures: distributional results}\label{hw::disentangle-mixture-distrubution}

This problem extends Theorem \ref{thm::cace-mixture-components}. Define
$$
f_{zu}(y) = \pr\{  Y(z) = y  \mid U=u \},  \quad  (z=0,1;u=\at, \nt, \cp)
$$
as the density of $Y(z)$ for latent stratum $U=u$, and define 
$$
g_{zd}(y) = \pr(Y = y\mid Z=z, D=d)
$$
as the density of the outcome within the observed group $(Z=z, D=d)$. 
Exclusion restriction implies that $f_{1\nt}(y) = f_{0\nt} (y) $ and $f_{1\at}(y) = f_{0\at}(y) $. Let $f_{\nt}(y)$ and $ f_{\at} (y) $ denote them, respectively.

Prove  Theorem \ref{thm::cace-mixture-components-distribution} below.

\begin{theorem}
\label{thm::cace-mixture-components-distribution}
Under Assumptions \ref{assume::random}--\ref{assume::ER}, we can identify the type-specific densities of the potential outcomes by
\begin{eqnarray*}
  f_{\nt}(y) &=& g_{10}(y) ,\\
  f_{\at} (y) &=& g_{01}(y),\\
f_{1\cp}(y) &=& \pi_\cp^{-1} \{   \pr(D=1\mid Z=1 )g_{11}(y)  -  \pr(D=1\mid Z=0) g_{01}(y)   \} , \\
f_{0\cp} (y) &=& \pi_\cp^{-1} \{  \pr(D=0\mid Z=0)  g_{00}(y)  -  \pr(D=0\mid Z=1)  g_{10}(y)    \} . 
\end{eqnarray*}
\end{theorem}

\paragraph{Alternative form of Theorem \ref{thm::iv-inequality}}
\label{hw::ivineq-binary}

The inequalities in \eqref{eq::iv-inequalities} can be re-written as
\begin{eqnarray*}
\pr(D=1, Y=y\mid Z=1) &\geq & \pr(D=1, Y=y\mid Z=0), \\
\pr(D=0, Y=y\mid Z=0) &\geq & \pr(D=0, Y=y\mid Z=1)
\end{eqnarray*}
for both $y=0,1$.

\paragraph{IV inequalities for a general outcome}
\label{hw::ivineq-general-Y}

For a general outcome $Y$, show that Assumptions \ref{assume::random}--\ref{assume::ER} imply 
\begin{eqnarray*}
\pr(D=1, Y\geq y\mid Z=1) &\geq & \pr(D=1, Y\geq y\mid Z=0), \\
\pr(D=1, Y <  y\mid Z=1) &\geq & \pr(D=1, Y <  y\mid Z=0), \\
\pr(D=0, Y\geq y\mid Z=0) &\geq & \pr(D=0, Y\geq y\mid Z=1),\\
\pr(D=0, Y <  y\mid Z=0) &\geq & \pr(D=0, Y <  y\mid Z=1)
\end{eqnarray*}
for all $y$. 
 
Remark:  \citet{imbens1997estimating} and \citet{kitagawa2015test} discussed similar results.  We can test the first inequality based on an analog of the Kolmogorov--Smirnov statistic:
$$
\text{KS}_1 = \max_{y} \Big |
\frac{  \sumn Z_i D_i  1(Y_i \leq y)  }{ \sumn Z_i D_i  } - \frac{  \sumn (1-Z_i) D_i  1(Y_i \leq y)  }{ \sumn (1-Z_i) D_i  }
\Big |.
$$

\paragraph{Example for the IV inequalities}

Give an example in which all the IV inequalities hold and another example in which not all the IV inequalities hold. You need to specify the joint distribution of $(Z,D,Y)$ with binary variables.

 \paragraph{Violations of the key assumptions}\label{eq::violations-of-IV}

Theorem \ref{thm::CACE-identification} relies on randomization, monotonicity, and exclusion restriction. The latter two are not testable even in randomized experiments. When they are violated, the IV estimator no longer identifies the CACE. This problem gives two cases below, which are restatements of  Propositions 2 and 3 in \citet{angrist1996identification}. Recall $\pi_u = \pr(U=u)$ and $\tau_u = E\{  Y(1) - Y(0) \mid U=u \}$ for $u=\at, \nt, \cp, \df$.

\begin{theorem}
\label{theorem::wald-violation-assumptions}
(a) 
Under Assumptions  \ref{assume::random} and \ref{assume::Mono} without the exclusion restriction, we have
$$
\frac{ E(Y\mid Z=1) - E(Y\mid Z=0) }{ E(D\mid Z=1) - E(D \mid Z=0) } - \tau_\cp 
= \frac{  \pi_\at \tau_\at + \pi_\nt \tau_\nt  }{  \pi_\cp } .
$$

(b)
Under Assumptions  \ref{assume::random} and \ref{assume::ER} without the monotonicity, we have 
$$
\frac{ E(Y\mid Z=1) - E(Y\mid Z=0) }{ E(D\mid Z=1) - E(D \mid Z=0) } - \tau_\cp 
= \frac{  \pi_\df (\tau_\cp + \tau_\df )  }{   \pi_\cp - \pi_\df   }.
$$
\end{theorem}
 
Prove Theorem \ref{theorem::wald-violation-assumptions}.

 \paragraph{Problems of other analyses}\label{problem::as-treated-analysis}

In the process of deriving the IV inequalities in Section \ref{sec::disentangle-mixtures}, we disentangled the mixture distributions by identifying the proportions of the latent strata as well as the conditional means of their potential outcomes. These results help to understand the drawbacks of other seemingly reasonable analyses. I review three estimators below and suppose Assumptions \ref{assume::random}--\ref{assume::ER} hold.

\begin{enumerate}
\item 
The {\it as-treated analysis} compares the means of the outcomes among units receiving the treatment and control, yielding
$$
\tau_\textsc{at} = E(Y\mid D=1) - E(Y\mid D=0).
$$
Show that 
$$
\tau_\textsc{at} = \frac{  \pi_\at \mu_\at + \pr(Z=1)\pi_\cp \mu_{1\cp}   }{   \pr(D=1) } 
- \frac{  \pi_\nt \mu_\nt + \pr(Z=0) \pi_\cp \mu_{0\cp}   }{   \pr(D=0) } .
$$

\item 
The {\it per-protocol analysis} compares the units that comply with the treatment assigned in treatment and control groups, yielding
$$
\tau_\textsc{pp} = E(Y\mid Z=1, D=1) - E(Y\mid Z=0, D=0).
$$
Show that 
$$
\tau_\textsc{pp} =\frac{  \pi_\at \mu_\at + \pi_\cp \mu_{1\cp} }{  \pi_\at   + \pi_\cp }
- \frac{  \pi_\nt \mu_\nt + \pi_\cp \mu_{0\cp} }{  \pi_\nt   + \pi_\cp }.
$$

\item 
We may also want to compare the outcomes among units receiving the treatment and control, conditioning on their treatment assignment, yielding
\begin{eqnarray*}
\tau_{Z=1} &=& E(Y\mid Z=1, D=1) - E(Y\mid Z=1, D=0),\\ 
\tau_{Z=0} &=& E(Y\mid Z=0, D=1) - E(Y\mid Z=0, D=0).
\end{eqnarray*}
Show that they reduce to
$$
\tau_{Z=1}  = \frac{  \pi_\at \mu_\at + \pi_\cp \mu_{1\cp} }{  \pi_\at   + \pi_\cp } - \mu_\nt,\quad
\tau_{Z=0} = \mu_\at - \frac{  \pi_\nt \mu_\nt + \pi_\cp \mu_{0\cp} }{  \pi_\nt   + \pi_\cp }.
$$
\end{enumerate}

\paragraph{Bounds on the average causal effect on the whole population}\label{para::bounds-ACE}

Extend the discussion in Section \ref{sec::disentangle-mixtures} based on the notation in Section \ref{eq::interpret-iv-double}. With the potential outcome $Y(d)$, define the average causal effect of the treatment received on the outcome as
$$
\delta = E\{  Y(d=1) - Y(d=0) \},
$$
and modify the definition of $\mu_{zu} $ as 
$$
m_{du} = E\{  Y(d) \mid U = u\},\quad  (z=0,1;u=\at, \nt, \cp) 
$$  
due to the change of the notation. 
They satisfy
$$
\delta = \sum_{u = \at, \nt, \cp}  \pi_u  (m_{1u} - m_{0u}).
$$

Section \ref{sec::disentangle-mixtures} identifies $\pi_\at$, $\pi_\nt$, $\pi_\cp$, $m_{1\at} = \mu_{1 \at}$, $m_{0\nt} = \mu_{0\nt }$, $m_{1\cp} = \mu_{1\cp}$ and $m_{0\cp} = \mu_{0\cp}$. But the data do not contain any information about $m_{0 \at}$ and $m_{1\nt }$. Therefore, we cannot identify $\delta$. With a bounded outcome, we can bound $\delta$.  

\begin{theorem}
\label{thm::bound-on-ACE}
Under Assumptions \ref{assume::Mono}--\ref{assume::ER-double-index} with a bounded outcome in $[\underline{y}, \overline{y}]$, we have $\underline{\delta} \leq  \delta  \leq \overline{\delta}$, where
$$
\underline{\delta} =  \delta'  - \overline{y} \pr(D=1\mid Z=0) + \underline{y} \pr(D=0\mid Z=1)
$$
and
$$
\overline{\delta} = \delta'  -   \underline{y} \pr(D=1\mid Z=0) +  \overline{y}  \pr(D=0\mid Z=1) 
$$
with $\delta' = E(DY\mid Z=1) - E(Y-DY\mid Z=0) .$
\end{theorem}

Prove Theorem  \ref{thm::bound-on-ACE}.

Remark: In the special case with a binary outcome, the bounds simplify to 
 $$
\underline{\delta} =   E(DY\mid Z=1) - E( D+Y-DY\mid Z=0)  
$$
and
$$
\overline{\delta} =  E(DY+1-D\mid Z=1) - E(Y-DY\mid Z=0) .
 $$

 \paragraph{One-sided noncompliance and statistical inference}\label{para::15-onesidednoncompliance}
 
Consider a randomized encouragement design where the units assigned to the control have no access to the treatment. For unit $i$, let $Z_i$ be the binary treatment assigned, $D_i$ be the binary treatment received, and $Y_i$ be the outcome of interest. One-sided noncompliance happens when 
$$
Z_i=0 \Longrightarrow  D_i = 0 \quad (i=1,\ldots, n). 
$$
Suppose that Assumption \ref{assume::random} holds.

\begin{enumerate}
[(1)]
\item 
Does monotonicity Assumption \ref{assume::Mono} hold in this case? 
How many latent strata defined by $\{ D_i(1), D_i(0)  \}$ are there in this problem? How do we identify their proportions by the observed data distribution?  

\item 
State the assumption of exclusion restriction. Under exclusion restriction, show that $E\{  Y(z) \mid U = u \}$ can be identified by the observed data distributions.
Give the formulas for all possible values of $z$ and $u$. How do we identify the CACE in this case?

\item
If we observe pretreatment covariates $X_i$ for all units $i$, how do we use the covariate information to improve the estimation efficiency of the CACE?

 \item 
 Under Assumption \ref{assume::random}, the exclusion restriction Assumption \ref{assume::ER} has testable implications, which are the IV inequalities for one-sided noncompliance. State the IV inequalities.
 
\item
\citet{sommer1991estimating} provided the following dataset: 
 
 \begin{center}
\begin{tabular}{cccccc} 
\hline 
      &   \multicolumn{2}{c}{$Z=1$}  &  &  \multicolumn{2}{c}{$Z=0$}  \\  \cline{2-3} \cline{5-6} 
      & $D=1$  &  $D=0$           &       &  $D=1$  &  $D=0$\\
 $Y=1$ &   9663    &     2385         &      &     0       &    11514       \\
 $Y=0$ &   12   &     34            &      &     0         &    74           \\  
 \hline 
\end{tabular}
 \end{center}
 
The treatment assigned $Z$ is whether or not the child was assigned to the vitamin A supplement, the treatment received indicator $D$ is whether or not the child received the vitamin A supplement, and the binary outcome $Y$ is the survival indicator. The original RCT was conducted in Indonesia. 
Re-analyze the data. 
 \end{enumerate}

Remark: \citet{bloom1984accounting} first discussed one-sided noncompliance and proposed the estimator $\hat\tau_\cp = \hat\tau_Y / \hat\tau_D$, which is sometimes called the Bloom estimator.  \citet{bloom1984accounting}'s notation is different from this chapter.

  \paragraph{One-sided noncompliance with partial adherence}
\label{para::one-sided-partial-adherence}
 
\citet[][Table 3]{sanders2021incorporating} reported the following data from an RCT aiming to estimate the efficacy of smoking cessation interventions among individuals with psychotic disorders. 

\begin{center}
\begin{tabular}{cccc}
\hline  
group assigned& treatment received &group size& \# positive outcomes \\
\hline  
Control & None& 151 &25 \\
Treatment& None& 35 & 7 \\
Treatment& Partial& 42& 17 \\
Treatment& Full& 70& 40\\
\hline 
\end{tabular}
\end{center}

Three tiers of treatment received are defined as follows: ``full'' treatment corresponds to attending all $8$ treatment sessions, ``partial'' corresponds to attending $5$ to $7$ sessions, and ``none'' corresponds to $<5$ sessions. The outcome is
defined as the binary indicator of smoking reduction of $50\%$ or greater relative to baseline, measured at three months.

In this problem, the treatment assignment $Z$ is binary but the treatment received $D$ takes three values $0, 0.5, 1$ for ``none'', ``partial'', and ``full.''  The three-leveled $D$ causes complications, but it can only be 0 under the control assignment. How many latent strata $U = \{ D(1), D(0) \} $ do we have in this problem? Can we identify their proportions?

How do we extend the exclusion restriction to this problem? What can be the causal effects of interest? Can we identify them? 

Analyze the data based on the questions above.

 \paragraph{Recommended reading}

\citet{balke1997bounds} derived more general IV inequalities.

\chapter{An Econometric Perspective of the Instrumental Variable}
 \label{chapter::iv-econometric}

Chapters \ref{chapter::iv-experiment} and \ref{chapter::iv-inequalities} discuss the IV method from the experimental perspective. Figure \ref{fig::causal-diagram-IV} illustrates the intuition behind the discussion.

\begin{figure}[h]
\centering 
$
\xymatrix{
 && & U  \ar[dl] \ar[dr] &\\
Z \ar[rr] &&D \ar[rr]&&Y
}
$
\caption{Causal diagram for IV}\label{fig::causal-diagram-IV}
\end{figure}

In an encouragement design with noncompliance, $Z$ is randomized, so it is independent of the confounder $U$ between the treatment received $D$ and the outcome $Y$.  Importantly, the treatment assignment $Z$ does not have any direct effect on the outcome $Y$. It acts as an IV for the treatment received $D$ in the sense that it affects the outcome $Y$ only through the treatment received $D$. This IV is generated by the experimenter.

In many applications, randomization is infeasible. Then how can we draw causal inference in the presence of unmeasured confounding between the treatment $D$ and outcome $Y$? A clever idea from econometrics is to find {\it natural experiments} to mimic the setting of encouragement designs. To identify the causal effect of $D$ on $Y$ with unmeasured confounding, we can find another variable $Z$ that satisfies the assumptions of the diagram in Figure \ref{fig::causal-diagram-IV}.  The variable $Z$ must satisfy the following conditions:
\begin{enumerate}
\item
it should be close to being randomized so that it is independent of the unmeasured confounding $U$;
\item
it should change the distribution of $D$;
\item
it should affect the outcome $Y$ only indirectly through $D$ but not directly.
\end{enumerate}
If all these conditions hold, then $Z$ is a valid IV for estimating the effect of $D$ on $Y$.

This chapter will provide the traditional econometrics perspective on IV. It is based on linear regression. \citet{imbens1994identification} and \citet{angrist1996identification} made a fundamental contribution by clarifying the connection between this perspective and the experimental perspective in Chapters \ref{chapter::iv-experiment} and \ref{chapter::iv-inequalities}. I will start with examples and then give more algebraic details.

\section{Examples of studies with IVs}

Finding IV for causal inference is more an art than a science. The algebraic details in later sections are not the most complicated ones in statistics. However, it is fundamentally challenging to find IVs in empirical research. Below are some famous examples.

\begin{example} 
In an encouragement design, $Z$ is the randomly assigned treatment, $D$ is the final treatment received, and $Y$ is the outcome. The IV assumptions encoded by Figure \ref{fig::causal-diagram-IV} are plausible in double-blind RCTs as discussed in Chapter \ref{chapter::iv-experiment}. This is the ideal case for IV. 
\end{example}

 \begin{example}
 \citet{hearst1986delayed}   reported that men with low lottery numbers in the Vietnam Era draft lottery had higher mortality rates afterward.  They attributed this to the negative effect of military service.
 \citet{angrist1990lifetime} further reported that men with low lottery numbers in the Vietnam Era draft lottery had lower subsequent earnings. He attributed this to the negative effect of military service. 
 These explanations are plausible because the lottery numbers were randomly generated, men with low lottery numbers were more likely to have military service, and the lottery numbers were unlikely to affect the subsequent mortality or earnings. That is, Figure \ref{fig::causal-diagram-IV} is plausible. 
\citet{angrist1996identification} reanalyzed the data using the IV framework. Here, the lottery number is the IV, military service is the treatment, and mortality or earnings is the outcome. 
\end{example}

\begin{example}
\citet{angrist1991does} studied the return of schooling in years on earnings, using the quarter of birth as an IV. This IV is plausible because of the pseudo-randomization of the quarter of birth.  It affected the years of schooling because (1) most states in the U.S. required the students to enter school in the calendar year in which they turned six, and (2) compulsory schooling laws typically required students to remain in school before their sixteenth birthday. More importantly, it is plausible that the quarter of birth did not affect earnings directly. 
\end{example}

\begin{example}
\citet{angrist1998children} studied the effect of family size on mothers' employment and work, using the sibling sex composition as an IV. This IV is plausible because of the pseudo-randomization of the sibling sex composition. Moreover, parents in the U.S. with two children of the same sex are more likely to have a third child than those parents with two children of different sex. It is also plausible that the sibling sex composition does not affect the mother's employment and work directly. 
\end{example}

\begin{example}\label{example::Card1993}
\citet{card1993using} studied the effect of schooling on wages, using the geographic variation in college proximity as an IV. In particular, $Z$ contains dummy variables for whether a subject grew up near a two-year college or a four-year college. Although this study is classic, it might be a poor example for IV because parents' choices of where to live might not be random, and moreover, where a subject grew up might matter for the subsequent wage. 
\end{example}

\begin{example}
\citet{voight2012plasma} studied the causal effect of plasma high-density lipoprotein (HDL) cholesterol on the risk of heart attack based on {\it Mendelian randomization}. They used some single-nucleotide polymorphisms (SNPs) as genetic IV for HDL, which are random with respect to the unmeasured confounders between HDL and heart attack by Mendel's second law, and affect heart attack only through HDL. 
I will give more details of Mendelian randomization in Chapter \ref{chapter::iv-mr}. 
\end{example}

\section{Brief Review of the Ordinary Least Squares}

Before discussing the econometric view of IV, I will first review the OLS (see Chapter \ref{appendix::basic-linear-regression}). This is a standard topic in statistics. However, it has different mathematical formulations, and the choice of formulation matters for the interpretation. 

The first view is based on projection.  Given a random variable $Y$ and a random variable or vector $D$  with finite second moments, define the population OLS coefficient as
\begin{eqnarray*}
\beta  &=& \arg\min_b E(Y - D  \tran b  )^2 \\
&=& E(DD\tran)^{-1} E(DY),
\end{eqnarray*}
and then define the population residual as $\varepsilon = Y -  D  \tran \beta $. By definition, $Y$ decomposes into 
\begin{eqnarray}\label{eq::olsdecomposition}
Y = D  \tran \beta  + \varepsilon,
\end{eqnarray}
which must satisfy 
$$
E(  D \varepsilon) = 0.
$$ 
Based on  $(D_i,Y_i)_{i=1}^n \iidsim (D, Y)$, the OLS estimator of $\beta$ is the  moment estimator
$$
\hat{\beta} = \left(  \sumn D_i D_i\tran \right)^{-1}   \sumn D_i Y_i  . 
$$
Because
\begin{eqnarray*}
\hat{\beta}  &=& \left(  \sumn D_i D_i\tran \right)^{-1}  \sumn D_i ( D_i\tran \beta + \varepsilon_i   )  \\
&=& \beta + \left(  \sumn D_i D_i\tran \right)^{-1} \sumn D_i  \varepsilon_i   ,
\end{eqnarray*}
we can show that $\hat{\beta}$ is consistent for $\beta$ by the law of large numbers and the fact $E(\varepsilon D) = 0.$ 
The classic EHW robust variance estimator for $\cov(\hat\beta)$ is
$$
\hat{V}_\textsc{ehw} = \left(  \sumn D_i D_i\tran \right)^{-1}
\left(  \sumn \hat{\varepsilon}_i^2 D_i D_i\tran \right)
\left(  \sumn D_i D_i\tran \right)^{-1}
$$
where $\hat{\varepsilon}_i  = Y_i - D_i\tran \hat{\beta}$ is the residual.

The second view is to treat
\begin{eqnarray}\label{eq::ols-true-model}
Y = D  \tran \beta  + \varepsilon,
\end{eqnarray}
as a true model for the data-generating process. 
That is, given the random variables $(D, \varepsilon)$, we generate $Y$ based on the linear equation \eqref{eq::ols-true-model}.
Importantly, in the data-generating process, $\varepsilon$ and $D$ may be correlated with $E( D \varepsilon) \neq  0$.  Figure \ref{fig::endogenous-regressor} gives such an example. This is the fundamental difference compared with the first view where $E(\varepsilon D) = 0$ holds by the definition of the population OLS. 
Consequently, the OLS estimator can be inconsistent:
$$
\hat{\beta} \rightarrow \beta + E(DD\tran)^{-1} E(D\varepsilon) \neq \beta 
$$
in probability, as the sample size $n$ approaches infinity.

I end this section with definitions of {\it endogenous} and {\it exogenous} regressors based on \eqref{eq::ols-true-model}, although their definitions are not unique in econometrics.

\begin{definition}
\label{def::endogenous-exogenous}
When $E(\varepsilon D) \neq  0$, the regressor $D$ is called {\it endogenous}; when $E(\varepsilon D) =  0$, the regressor $D$ is called {\it exogenous}. 
\end{definition}

The terminology in Definition \ref{def::endogenous-exogenous} is standard in econometrics. 
When $E(\varepsilon D) \neq  0$, we also say that we have {\it endogeneity}; when $E(\varepsilon D) =  0$, we also say that we have {\it exogeneity}.

In the first view of OLS, the notions of endogeneity and exogeneity do not play any roles because $E(\varepsilon D) =  0$ by definition.  Statisticians holding the first view usually find the notions of endogeneity and exogeneity strange, and consequently, find the idea of IV unnatural. To understand the econometric view of IV, we must switch to the second view of OLS.

\begin{figure}[h]
\centering
\begin{subfigure}[]{0.4\textwidth}
                \centering
$
\xymatrix{
& U \ar[dl] \ar[dr]& \\
D \ar[dr] & & \varepsilon \ar[dl] \\
&Y&
}
$                \caption{$E(D\varepsilon) \neq 0$}
        \end{subfigure}%
        
        \begin{subfigure}[]{0.4\textwidth}
                \centering
$
\xymatrix{
&\varepsilon  \ar[dl] \ar[dr] &\\
D \ar[rr]&&Y
}
$
                \caption{marginalized over $\varepsilon$}
        \end{subfigure}%
        
\caption{ Different representations of the endogenous regressor $D$. In the upper panel, $U$ represents unmeasured common causes of $D$ and $\varepsilon$.}\label{fig::endogenous-regressor}
\end{figure}

\section{Linear Instrumental Variable Model}
\label{sec::linear-iv-model}

When $D$ is endogenous, the OLS estimator is inconsistent. We must use additional information to construct a consistent estimator for $\beta$. 
I will focus on the following linear IV model:

\begin{definition}
[linear IV model]\label{def::linear-iv-model}
We have 
$$
Y = D  \tran \beta  + \varepsilon ,
$$
with
\begin{eqnarray}\label{eq::iv-moment-condition}
E(\varepsilon Z) = 0 . 
\end{eqnarray}
\end{definition}

The linear IV model in Definition \ref{def::linear-iv-model} can be illustrated by the following causal graph: 
$$
\xymatrix{
&&&\varepsilon  \ar[dl] \ar[dr] &\\
Z\ar[rr]&&D \ar[rr]&&Y
}
$$

The above linear IV model allows that $E(\varepsilon D) \neq 0$ but requires an alternative moment condition \eqref{eq::iv-moment-condition}. With $E(\varepsilon) = 0$ by incorporating the intercept, the new condition states that $Z$ is uncorrelated with the error term $\varepsilon$. But any randomly generated noise is uncorrelated with $\varepsilon$, so an additional condition must hold to ensure that $Z$ is useful for estimating $\beta$. Intuitively, the additional condition requires that $Z$ is correlated to $D$, with more technical details stated below.

The mathematical requirement \eqref{eq::iv-moment-condition} seems simple. However, it is a key challenge in empirical research to find such a variable $Z$ that satisfies \eqref{eq::iv-moment-condition}. Since the condition \eqref{eq::iv-moment-condition} involves the unobservable $\varepsilon$, it is generally untestable.

\section{The Just-Identified Case}
\label{sec::iv-just-identify}

We first consider the case in which $Z$ and $D$ have the same dimension and $E(ZD\tran)$ has full rank. The condition $E(\varepsilon Z) = 0$ implies that
$$
E\{  Z (Y -  D \tran \beta) \} = 0 .
$$
Solve the linear equations to obtain
$$
E(  ZY)  = E(ZD\tran) \beta  \\
 \Longrightarrow 
\beta = E(ZD\tran) ^{-1} E(  ZY)  
$$
if $E(ZD\tran)$ is not degenerate. 
The OLS is a special case if $E(\varepsilon D) = 0$, i.e., $D$ acts as an IV for itself. The resulting moment estimator is
\begin{eqnarray}\label{eq::iv-estimator}
\hat{\beta}_\textsc{iv} = \left(  \sumn Z_i D_i\tran \right)^{-1}  \sumn Z_iY_i.
\end{eqnarray}

It can be insightful to work out the details for the case with a  scalar $D$ and $Z$. See Example \ref{eg::scalar-iv} below.

\begin{example}\label{eg::scalar-iv}
In the simple case with an intercept and scalar $D$ and $Z$, we have  the model
$$
\left\{ \begin{array}{l}
Y = \alpha +  \beta  D + \varepsilon ,\\
E(\varepsilon) = 0, \quad \cov(\varepsilon, Z) = 0,
\end{array}
\right.
$$
Under this model, we have 
$$
\cov( Z, Y ) = \beta \cov(Z, D) 
$$
which implies
$$
 \beta = \frac{ \cov( Z, Y )   }{ \cov(Z, D) }.
$$
Standardize the numerator and denominator by $\var(Z)$ to obtain 
$$
\beta =  \frac{ \cov( Z, Y ) / \var(Z)  }{ \cov(Z, D) / \var(Z) },
$$
which equals the ratio between the coefficients of $Z$ in the OLS fits of $Y$ and $D$ on $Z$. 
If $Z$ is binary, these coefficients are differences in means (see Problem \ref{problem::pols-binary}), and $\beta$ reduces to
$$
\beta = \frac{ E(Y\mid Z=1) - E(Y\mid Z=0)  }{ E(D\mid Z=1) - E(D\mid Z=0) } .
$$
This is identical to the identification formula in Theorem \ref{thm::CACE-identification}. That is, with a binary IV $Z$ and a binary treatment $D$, the IV estimator recovers the CACE under the potential outcomes framework. This is a key result in  \citet{imbens1994identification} and \citet{angrist1996identification}.  
\end{example}

\section{The Over-Identified Case}
\label{sec::sec::iv-over-identify}

The discussion in Section \ref{sec::iv-just-identify} focuses on the {\it just-identified} case.  When $Z$ has a lower dimension than $D$ and $E(ZD\tran)$ does not have full column rank, the equation $E(  ZY)  = E(ZD\tran) \beta$ has infinitely many solutions. This is the {\it under-identified} case in which the coefficient $\beta$ cannot be uniquely determined even with $Z$. It is a challenging case beyond the scope of this book. To ensure identifiability, we need at least as many IVs as the endogenous regressors. 

When $Z$ has a higher dimension than $D$ and $E(ZD\tran)$ has full column rank, we have many ways to determine $\beta$ from $E(  ZY)  = E(ZD\tran) \beta$. What is more,  the sample analog 
$$
n^{-1} \sumn Z_iY_i = n^{-1} \sumn Z_i D_i\tran \beta
$$
may not have any solution because the number of equations is larger than the number of unknown parameters.

A computational trick for the over-identified case is the {\it two-stage least squares}  (TSLS) estimator \citep{theil1953estimation, basmann1957generalized}. It is a clever computational trick, which has two steps.

\begin{definition}
[Two-stage least squares]\label{def::2sls}
Define the TSLS estimator of the coefficient of $D$ with $Z$ being the IV as follows. 
\begin{enumerate}
\item
Run OLS of $D$ on $Z$, and obtain the fitted value $\hat{D}_i\ (i=1,\ldots, n)$. If $D_i$ is a vector, then we need to run component-wise OLS to obtain $\hat{D}_i$. Put the fitted vectors in a matrix $\hat{D}$ with rows $\hat{D}_i\tran $;
\item
Run OLS of $Y$ on $\hat{D}$, and obtain the coefficient $\hat{\beta}_\textsc{tsls}$.  
\end{enumerate}
\end{definition}

To see why TSLS works, we need more algebra. Write it more explicitly as
\begin{eqnarray} 
\hat{\beta}_\textsc{tsls} &=& \left(  \sumn \hat{D}_i\hat{D}_i\tran \right) ^{-1} 
\sumn \hat{D}_i Y_i  \label{eq::2sls-definition} \\
&=& \left(  \sumn \hat{D}_i\hat{D}_i\tran \right) ^{-1} 
\sumn \hat{D}_i (D_i \tran \beta + \varepsilon_i)  \nonumber  \\
&=& \left(  \sumn \hat{D}_i\hat{D}_i\tran \right) ^{-1} 
\sumn \hat{D}_i D_i \tran \beta + \left(  \sumn \hat{D}_i\hat{D}_i\tran \right) ^{-1} \sumn \hat{D}_i  \varepsilon_i .   \nonumber 
\end{eqnarray}  
The first stage OLS fit ensures $D_i  = \hat{D}_i + \check D_i$ with orthogonal fitted values and residuals, that is,
\begin{equation} \label{eq::ols-orthogonal}
\sumn \hat{D}_i  \check D_i\tran = 0
\end{equation}
is a zero square matrix with the same dimension as $D_i$. The orthogonality \eqref{eq::ols-orthogonal} implies 
$$
\sumn \hat{D}_i D_i \tran = \sumn \hat{D}_i\hat{D}_i\tran,
$$ 
which further implies that 
\begin{eqnarray}
\label{eq::2sls-ols}
\hat{\beta}_\textsc{tsls} = \beta + \left(  \sumn \hat{D}_i\hat{D}_i\tran \right) ^{-1} \sumn \hat{D}_i  \varepsilon_i.
\end{eqnarray}
The first stage OLS fit also ensures 
\begin{equation}
\label{eq::ols-donz}
\hat{D}_i = \hat{\Gamma}\tran  Z_i 
\end{equation}
which implies that
\begin{eqnarray}
\label{eq::2sls-consistency}
\hat{\beta}_\textsc{tsls} = \beta + \left\{  \hat{\Gamma}\tran  \left(  n^{-1}  \sumn Z_i Z_i\tran  \right) \hat{\Gamma}  \right\}  ^{-1}  
\hat{\Gamma}\tran  \left( n^{-1} \sumn Z_i  \varepsilon_i \right) . 
\end{eqnarray}
Based on \eqref{eq::2sls-consistency}, we can see the consistency of the TSLS estimator by the law of large numbers and the fact that the term $n^{-1} \sumn Z_i  \varepsilon_i $ has probability limit $E(Z\varepsilon) = 0$. We can also use \eqref{eq::2sls-consistency} to show that when $Z$ and $D$ have the same dimension,  $\hat{\beta}_\textsc{tsls} $ is numerically identical to $\hat{\beta}_\textsc{iv} $ defined in Section \ref{sec::iv-just-identify}, which is left  as Problem \ref{prob::algebra-tsls}.

Based on \eqref{eq::2sls-ols}, we can obtain the standard error as follows. We first obtain the residual $\hat{\varepsilon}_i = Y_i - \hat{\beta}_\textsc{tsls}\tran D_i$, and then obtain the robust variance estimator as
$$
\hat{V}_\textsc{tsls} = \left( \sumn  \hat{D}_i \hat{D}_i\tran  \right)^{-1}
 \left( \sumn \hat{\varepsilon}_i ^2  \hat{D}_i \hat{D}_i\tran  \right)
\left( \sumn  \hat{D}_i \hat{D}_i\tran  \right)^{-1} . 
$$
Importantly, the $\hat{\varepsilon}_i$'s  are
not the residual from the second stage OLS $  Y_i - \hat{\beta}_\textsc{tsls}\tran \hat{D}_i $, so $\hat{V}_\textsc{tsls} $ differs from the robust variance estimator from the second stage OLS.

\section{A Special Case: A Single IV for a Single Endogenous Treatment}

This section focuses on a simple case with a single IV and a single endogenous treatment. It has wide applications. 
Consider the following {\it structural equations}:
\begin{eqnarray}
\label{eq::strucrual-iv}
\left\{
\begin{array}{l}
Y_i = \beta_0 + \beta_1 D_i + \beta_2\tran X_i + \varepsilon_{i},\\
D_i = \gamma_0 + \gamma_1 Z_i + \gamma_2 \tran X_i + \varepsilon_{2i},
\end{array}
\right.
\end{eqnarray}
where $D_i$ is a scalar endogenous regressor representing the treatment variable of interest (i.e., $E(\varepsilon_{i} D_i) \neq 0$), $Z_i$ is a scalar IV for $D_i$ (i.e., $E(\varepsilon_{i} Z_i) = 0$), and $X_i$ contains other exogenous regressors (i.e., $E(\varepsilon_{i} X_i) = 0$). This is a special case with $D$ replaced by $(1, D, X)$ and $Z$ replaced by $(1, Z, X)$.

\subsection{Two-stage least squares}

The TSLS estimator in Definition \ref{def::2sls} simplifies to the following form. 

\begin{definition}
[TSLS with a single endogenous regressor]\label{def::tsls-single}
Based on \eqref{eq::strucrual-iv}, the TSLS estimator has the following two steps.
\begin{enumerate}
\item
Run OLS of $D$ on $(1, Z,X)$, obtain the fitted values $\hat{D}_i\ (i=1,\ldots, n)$, and vectorize them as $\hat{D}$; 

\item 
Run OLS of $Y$ on $(1, \hat{D}, X)$, and obtain the coefficient $\hat{\beta}_\textsc{tsls}$, and in particular, $\hat{\beta}_{1,\textsc{tsls}}$, the coefficient of $\hat{D}$. 
\end{enumerate}
\end{definition}

\subsection{Indirect least squares}

The structural equation \eqref{eq::strucrual-iv} implies 
\begin{eqnarray*}
Y_i &=&\beta_0 + \beta_1 (\gamma_0 + \gamma_1 Z_i + \gamma_2 \tran X_i + \varepsilon_{2i}) + \beta_2\tran X_i + \varepsilon_{i} \\
&=& (\beta_0 + \beta_1\gamma_0 ) + \beta_1  \gamma_1 Z_i  + ( \beta_2 + \beta_1 \gamma_2 )\tran X_i  + ( \varepsilon_{i} + \beta_1 \varepsilon_{2i} ).
\end{eqnarray*}
Define $\Gamma_0 = \beta_0 + \beta_1\gamma_0, \Gamma_1  =  \beta_1  \gamma_1, \Gamma_2 = \beta_2 + \beta_1 \gamma_2$, and $\varepsilon_{1i} = \varepsilon_{i} + \beta_1 \varepsilon_{2i}$. We have the following equations
\begin{eqnarray}
\label{eq::reduced-iv}
\left\{
\begin{array}{l}
Y_i =\Gamma_0 + \Gamma_1 Z_i + \Gamma_2\tran X_i + \varepsilon_{1i},\\
D_i = \gamma_0 + \gamma_1 Z_i + \gamma_2 \tran X_i + \varepsilon_{2i},
\end{array}
\right.
\end{eqnarray}
which is called the {\it reduced form}, in contrast to the {\it structural form} in \eqref{eq::strucrual-iv}. The parameter of interest equals the ratio of two coefficients 
$$
\beta_1 = \Gamma_1 / \gamma_1.
$$ 
In the reduced form, the left-hand side are dependent variables $Y$ and $D$, and the right-hand side are the exogenous variable $Z$ and $X$ satisfying 
$$
E( Z \varepsilon_{1i} ) = E( Z \varepsilon_{2i} )  = 0, \quad 
E( X \varepsilon_{1i} ) = E( X \varepsilon_{2i} )  = 0.
$$ 
More importantly, OLS gives consistent estimators for the coefficients in the reduced form \eqref{eq::reduced-iv}.

The reduced form \eqref{eq::reduced-iv} suggests that the ratio of two OLS coefficients $\hat{\Gamma}_1$ and $\hat{\gamma}_1$ is a reasonable estimator for $\beta_1$. This is called the {\it indirect least squares} (ILS) estimator:
$$
\hat{\beta}_{1,\textsc{ils}} =  \hat{\Gamma}_1 / \hat{\gamma}_1 .
$$
Interestingly, it is numerically identical to the TSLS estimator under \eqref{eq::strucrual-iv}.

\begin{theorem}
\label{thm::ils=2sls}
With a single endogenous treatment and a single IV in \eqref{eq::strucrual-iv}, we have 
$$
\hat{\beta}_{1,\textsc{ils}}   = \hat{\beta}_{1,\textsc{tsls}} . 
$$
\end{theorem}

Theorem \ref{thm::ils=2sls} is an algebraic fact. 
\citet[][Section A.3]{imbens2014instrumental} pointed it out without giving a proof. 
I relegate its proof to Problem \ref{para::tsls-ils}. The ratio formula makes it clear that the TSLS estimator has poor finite sample properties with a weak instrument variable, i.e., $\gamma_1 $ is close to zero.

\subsection{Weak IV}

The following inferential procedure is simpler, more transparent, and more robust to weak IV. It is more computationally intensive though. 
The reduced form \eqref{eq::reduced-iv}  also implies that
\begin{eqnarray}
\label{eq::anderson-rubin-reg}
Y_i  - bD_i = (\Gamma_0 -b \gamma_0 ) +  (\Gamma_1  - b \gamma_1 ) Z_i
+ (\Gamma_2 - b \gamma_2  ) \tran X_i  + ( \varepsilon_{1i} -  b \varepsilon_{2i} )
\end{eqnarray}
for any $b$. 
At the true value $b = \beta_1$, the coefficient of $Z_i$ must be $0$. This simple fact suggests a confidence interval for $\beta_1$ by inverting tests for $H_0(b): \beta_1 = b$:
$$
\left\{ b:  | t_Z(b) | \leq z_{1-\alpha/2} \right\},
$$
where $t_Z(b) $ is the $t$-statistic for the coefficient of $Z$ based on the OLS fit of \eqref{eq::anderson-rubin-reg} with the EHW standard error, and $z_{1-\alpha/2}$ is the $1-\alpha/2$ upper quantile of the standard Normal random variable. 
This confidence interval is more robust than the Wald-type confidence interval based on the TSLS estimator. It is similar to the  FAR confidence set discussed in Chapter \ref{chapter::iv-experiment}. This procedure makes the TSLS estimator unnecessary. What is more, we only need to run the OLS fit of $Y$ based on the reduced form if the goal is to test $\beta_1 = 0$ under \eqref{eq::strucrual-iv}.

\section{Application}\label{sec::tsls-application}

Revisit Example \ref{example::Card1993}. 
\citet{card1993using} used the National Longitudinal Survey of Young Men to estimate the causal effect of education on earnings. The data set contains 3010 men with ages between 14 and 24 in the year 1966, and \citet{card1993using} leveraged the geographic variation in college proximity as an IV for education. Here, $Z$ is the indicator of growing up near a four-year college, $D$ measures the years of education, and the outcome $Y$ is the log wage in the year 1976, ranging from 4.6 to 7.8. Additional covariates are race, age, and squared age, a categorical variable indicating living with both parents, single mom, or both parents, and variables summarizing the living areas in the past.

\begin{lstlisting}
> library("car")
> ## Card Data
> card.data = read.csv("card1995.csv")
> Y = card.data[, "lwage"]
> D = card.data[, "educ"]
> Z = card.data[, "nearc4"]
> X = card.data[, c("exper", "expersq", "black", "south", 
+                   "smsa", "reg661", "reg662", "reg663", 
+                   "reg664", "reg665", "reg666", 
+                   "reg667", "reg668", "smsa66")]
> X = as.matrix(X)
\end{lstlisting}

Based on TSLS, we can obtain the following the point estimator and $95\%$ confidence interval. 
\begin{lstlisting}
> Dhat    = lm(D ~ Z + X)$fitted.values
> tslsreg = lm(Y ~ Dhat + X)
> tslsest = coef(tslsreg)[2]
> ## correct se by changing the residuals
> res.correct       = Y - cbind(1, D, X)%*%coef(tslsreg)
> tslsreg$residuals = as.vector(res.correct)
> tslsse = sqrt(hccm(tslsreg, type = "hc0")[2, 2])
> res = c(tslsest, tslsest - 1.96*tslsse, tslsest + 1.96*tslsse)
> names(res) = c("TSLS", "lower CI", "upper CI")
> round(res, 3)
    TSLS lower CI upper CI 
   0.132    0.026    0.237 
\end{lstlisting}

\begin{figure}
\centering 
\includegraphics[width = \textwidth]{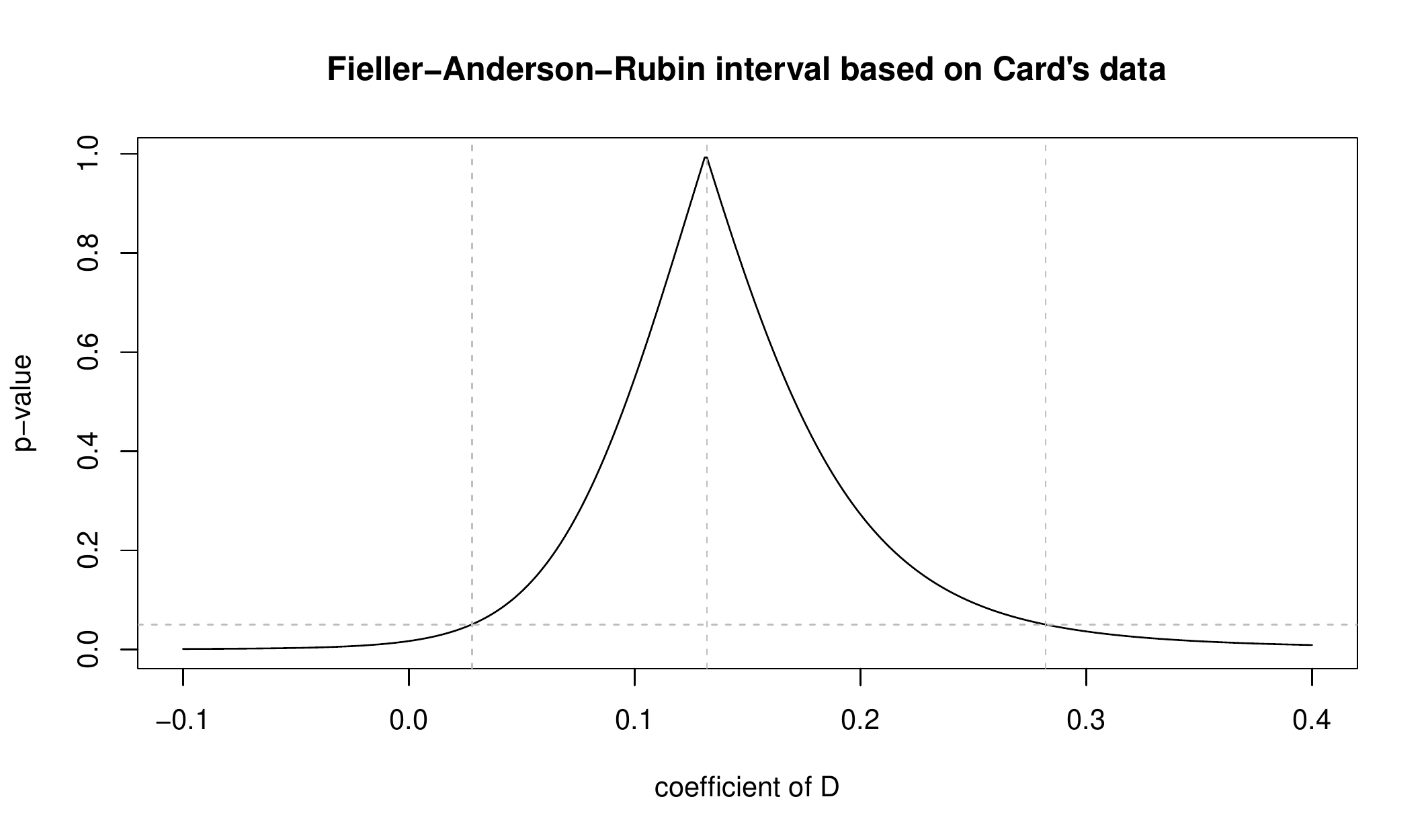}
\caption{Re-analysis of \citet{card1993using}'s data by inverting tests} \label{fig::card-far-ci} 
\end{figure}

Using the strategy of the FAR confidence set, we can compute $p$-value as a function of $b$. 
\begin{lstlisting}
> BetaAR   = seq(-0.1, 0.4, 0.001)
> PvalueAR = sapply(BetaAR, function(b){
+   Y_b   = Y - b*D
+   ARreg = lm(Y_b ~ Z + X)
+   coefZ = coef(ARreg)[2]
+   seZ   = sqrt(hccm(ARreg)[2, 2])
+   Tstat = coefZ/seZ
+   (1 - pnorm(abs(Tstat)))*2
+ })
\end{lstlisting}
Figure \ref{fig::card-far-ci} shows the $p$-values for a sequence of tests for the coefficient of $D$ based on following \ri{R} code:
\begin{lstlisting}
> plot(PvalueAR ~ BetaAR, type = "l",
+      xlab = "coefficient of D",
+      ylab = "p-value",
+      main = "Fieller-Anderson-Rubin interval based on Card's data")
> point.est = BetaAR[which.max(PvalueAR)]
> abline(h = 0.05, lty = 2, col = "grey")
> abline(v = point.est, lty = 2, col = "grey")
> ARCI = range(BetaAR[PvalueAR >= 0.05])
> abline(v = ARCI[1], lty = 2, col = "grey")
> abline(v = ARCI[2], lty = 2, col = "grey")
\end{lstlisting}
We report the point estimate as the value of $b$ with the largest $p$-value as well as the confidence interval as the region of $b$ with $p$-values larger than 0.05. 
\begin{lstlisting}
> FARres = c(point.est, ARCI)
> names(FARres) = c("FAR est", "lower CI", "upper CI")
> round(FARres, 3)
 FAR est lower CI upper CI 
   0.132    0.028    0.282
\end{lstlisting}
Comparing the TSLS and FAR methods, the lower confidence limits are very close but the upper confidence limits are slightly different due to the possibly heavy right tail of the distribution of the TSLS estimator. Overall, the TSLS and FAR methods give similar results in this example because the IV is not weak.

\section{Homework}

\paragraph{More algebra for TSLS in Section \ref{sec::sec::iv-over-identify}}\label{prob::algebra-tsls}

\begin{enumerate}
\item 
Show that the $\hat \Gamma$ in \eqref{eq::ols-donz} equals
$$
\hat{\Gamma} = \left(  \sumn Z_i Z_i\tran  \right)^{-1} \sumn Z_i D_i\tran .
$$

\item
Show $\hat{\beta}_\textsc{tsls} $ defined in \eqref{eq::2sls-definition} reduces to $ \hat{\beta}_\textsc{iv}$ defined in \eqref{eq::iv-estimator} if $Z$ and $D$ have the same dimension and
$$
n^{-1} \sumn Z_i Z_i\tran, \quad n^{-1} \sumn Z_i D_i\tran
$$
are both invertible. 
\end{enumerate}

\paragraph{Equivalence between TSLS and ILS}\label{para::tsls-ils}
Prove Theorem \ref{thm::ils=2sls}.

Remark: Use the FWL theorem in Chapter \ref{appendix::basic-linear-regression}.

\paragraph{Control function in the linear instrumental variable model}\label{para::control-function-iv}

Definition \ref{def::cf} below parallels Definition \ref{def::2sls} above.

\begin{definition}
[control function]\label{def::cf}
Define the control function estimator $\hat{\beta}_\textsc{cf}$ as follows.
\begin{enumerate}
\item
Run OLS of $D$ on $Z$, and obtain the residuals $\check D_i\ (i=1,\ldots, n)$. If $D_i$ is a vector, then we need to run component-wise OLS to obtain $\check D_i$. Put the residual vectors in a matrix $\check{D}$ with rows $\check{D}_i\tran $;
\item
Run OLS of $Y$ on $D$ and $\check{D}$, and obtain the coefficient of $D$, $\hat{\beta}_\textsc{cf}$.  
\end{enumerate}
\end{definition}
 
 Show that $\hat{\beta}_\textsc{cf} = \hat{\beta}_\textsc{tsls}$.

 Remark:
To prove the result, you can use the results in Problems \ref{hw::ols-orthogonal} and \ref{hw::ols-transform-regressors}.  
  In Definition \ref{def::cf}, $\check{D}$ from Step 1 is called the control function for Step 2. 
 \citet{hausman1978specification} pointed out this result. \citet{wooldridge2015control} provided a more general discussion of the control function methods in more complex models.

%
%
%
%
%

\paragraph{Data analysis: \citet{efron1991compliance}}

\citet{efron1991compliance} was one of the early studies dealing with noncompliance under the potential outcomes framework. The original randomized experiment, the Lipid Research Clinics Coronary Primary Prevention Trial (LRC-CPPT), was designed to evaluate the effect of the drug cholestyramine on cholesterol levels.
In the dataset \texttt{EF.csv}, the first column contains the binary indicators for treatment and control, the second column contains the proportions of the nominal cholestyramine dose actually taken, and the last three columns are cholesterol levels.
Note that the individuals did not know whether they were assigned to cholestyramine or to the placebo, but differences in adverse side effects could induce differences in compliance behavior by treatment status. All individuals were assigned the same nominal dose of the drug or placebo, for the same time period.
Column 3, $C_3$, was taken before communication about the benefits of a low-cholesterol diet, Column 4, $C_4,$ was taken after this suggestion, but before the random assignment to cholestyramine or placebo, and Column 5, $C_5$,  an average of post-randomization cholesterol readings, averaged over two-month readings for a period of time averaging 7.3 years for all the individuals in the study. \citet{efron1991compliance} used the change in cholesterol level as the final outcome of interest, defined as $C_5 - 0.25C_3 - 0.75C_4$. Their original paper contains more detailed descriptions.

The data structure here is more complicated than the noncompliance problem discussed in Chapters \ref{chapter::iv-experiment} and \ref{chapter::iv-inequalities}. 
\citet{jin2008principal} re-analyzed the data based on the idea in Chapter \ref{chapter::principal-stratification} later. 
You can analyze the data based on your understanding of the problem, but you need to justify your choice of the method. There is no gold-standard solution for this problem.

 \paragraph{Recommended reading}

\citet{imbens2014instrumental} gave an econometrician's perspective of IV.

\chapter{Application of the Instrumental Variable Method: Fuzzy Regression Discontinuity}
 \label{chapter::iv-frdd}

The regression discontinuity method introduced in Chapter \ref{chapter::overlap} and the IV method introduced in Chapters \ref{chapter::iv-experiment}--\ref{chapter::iv-econometric} are two important examples of {\it natural experiments}. The study designs are not as ideal as randomized experiments in Part \ref{part::rcts}, but they have features similar to randomized experiments. That's why they are called natural experiments.

Compounding the regression discontinuity method with the IV method yields the {\it fuzzy regression discontinuity} method, another important natural experiment. I will start with examples and then provide a mathematical formulation.

\section{Motivating examples}
\label{sec::frdd-motivation}

Chapter \ref{chapter::overlap} introduces the regression discontinuity method. The following three examples are slightly different because the treatments received are not deterministic functions of the running variables. Rather, the running variables discontinuously change the probabilities of the treatments received at the cutoff points. 

\begin{example}\label{eg::indianroad}
In 2000, the Government of India launched the Prime Minister's Village Road Program, and by 2015, this program had funded the construction of all-weather roads to nearly 200,000 villages.  Based on village-level data, 
\citet{asher2020rural} use the regression discontinuity method to estimate the effect of new feeder roads on various economic variables. The national program guidelines prioritized larger villages according to arbitrary thresholds based on the 2001 Population Census. The treatment variable equals 1 if the village received a new road before the year in which the outcomes were measured. The difference between the population size of a village and the threshold did not determine the treatment variable but affected its probability discontinuously at the cutoff point 0. 
\end{example}

\begin{example}\label{eg::italianuniversity}
\citet{li2015evaluating} used the data on the first-year students enrolled in 2004 to 2006 from two Italian universities to evaluate the causal effect of a university grant on the dropout rate. The students were eligible for this grant if their standardized family income was below  15,000 euros. For simplicity, we use the running variable defined as 15,000 minus the standardized family income. To receive this grant, the students must apply first. Therefore, the eligibility and the application status jointly determined the final treatment status. The running variable alone did not determine the treatment status although it changed the treatment probability at the cutoff point 0. 
\end{example}

\begin{example}
\citet{amarante2016cash} estimated the impact of in-utero exposure to a social assistance program on children's birth outcomes. They used a  regression discontinuity induced by the Uruguayan {\it Plan de Atención Nacional a la Emergencia Social}. It was a temporary social assistance program targeted to the poorest 10 percent of
households, implemented between April 2005 and December 2007. 
Households with a predicted low-income score below a predetermined threshold were assigned to the program. The predicted income score did not determine whether the mother received at least one program transfer during the pregnancy but it changed the probability of the final treatment received. The birth outcomes included birth weight, weeks of gestation, etc. 
\end{example}

The above examples are called fuzzy regression discontinuity in contrast to the (sharp) regression discontinuity in Chapter \ref{chapter::overlap}.  I will analyze the data in  Examples \ref{eg::indianroad} and \ref{eg::italianuniversity} in Chapter \ref{sec::application-frdd} below. 

\section{Mathematical formulation}
\label{sec::frdd-math}

\begin{figure}
\centering
\includegraphics[width = \textwidth]{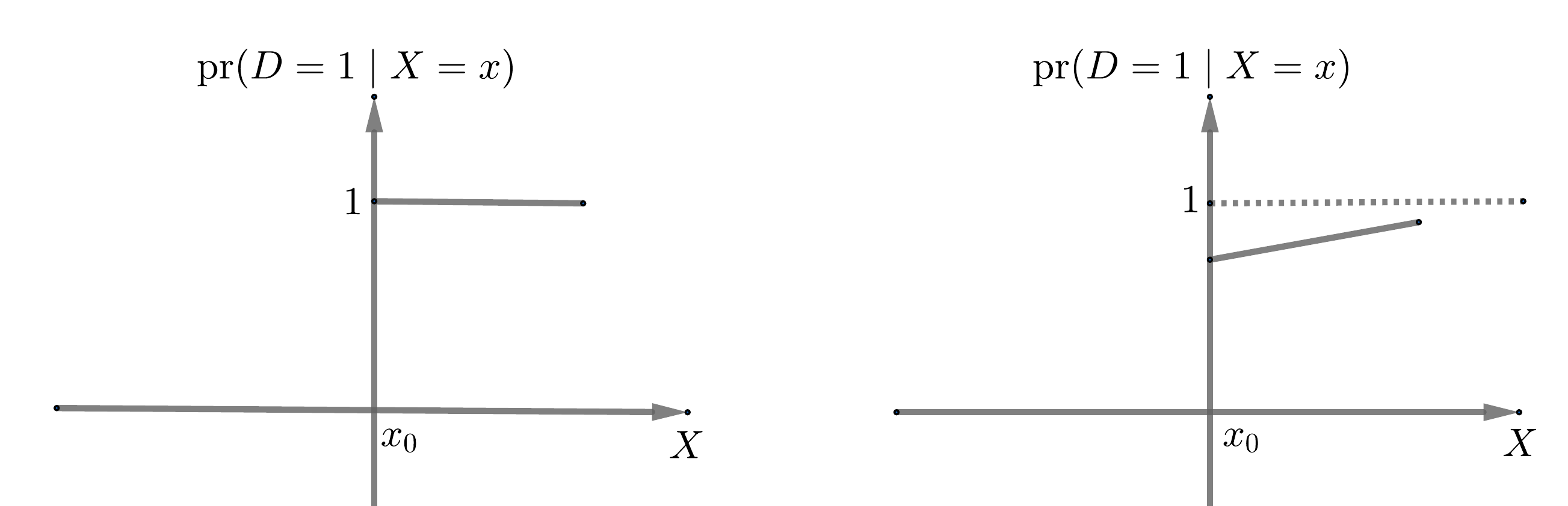}
\caption{The treatment assignments of sharp regression discontinuity (left) and fuzzy regression discontinuity (right)}\label{fig::rdd-compare}
\end{figure}

Let $X_i$ denote the running variable which determines 
$$
Z_i = I(X_i \geq x_0)
$$ 
with the cutoff point $x_0$. The treatment received $D_i$ may not equal $Z_i$, but $\pr(D_i=1\mid X_i = x)$ has a jump at $x_0$. Figure \ref{fig::rdd-compare} compares the treatment received probabilities of the sharp regression discontinuity and fuzzy regression discontinuity. It shows a special case of fuzzy regression discontinuity with $\pr(D=1\mid X < x_0) = 0$, which is coherent to Example \ref{eg::italianuniversity}.

Let $Y_i$  denote the outcome of interest. Viewing $Z_i$ as the treatment assigned, we can define potential outcomes $\{ D_i(1), D_i(0), Y_i(1), Y_i(0) \}$. The sharp regression discontinuity of $Z$ allows for the identification of
\begin{eqnarray*}
\tau_D(x_0)  &=& E\{ D(1) - D(0)  \mid X=x_0 \}\\
&=& \lim_{\varepsilon \rightarrow 0+} E( D \mid Z=1, X=x_0+\varepsilon ) - \lim_{\varepsilon \rightarrow 0+} E( D \mid Z=0, X=x_0-\varepsilon )
\end{eqnarray*}
and
\begin{eqnarray*}
\tau_Y(x_0)  &=& E\{ Y(1) - Y(0)  \mid X=x_0 \}\\
&=& \lim_{\varepsilon \rightarrow 0+} E( Y \mid Z=1, X=x_0+\varepsilon ) - \lim_{\varepsilon \rightarrow 0+} E( Y \mid Z=0, X=x_0-\varepsilon )
\end{eqnarray*}
based on Theorem \ref{thm::identification-rdd}. 
Using $Z$ as an IV for $D$ and imposing the IV assumptions at $X = x_0$, we can identify the local complier average causal effect by applying Theorem \ref{thm::CACE-identification}.

\begin{theorem}
\label{thm::identification-frdd}
Assume that the treatment is determined by  $Z = I(X \geq x_0)$
where $x_0$ is a predetermined threshold. 
Assume monotonicity
$$
D_i(1) \geq D_i(0)
$$ 
and exclusion restriction 
$$
D_i(1) = D_i(0) \Longrightarrow  Y_i(1) = Y_i(0) 
$$
in the infinitesimal neighborhood of $x_0$. The local complier average causal effect, defined as 
$$
 \tau_\cp(x_0)   
= E\{ Y(1) - Y(0)  \mid D(1)  >  D(0),  X=x_0 \},
$$
can be identified by
$$
 \tau_\cp(x_0)   
=  \frac{ E\{ Y(1) - Y(0)  \mid X=x_0 \}  }{E\{ D(1) - D(0)  \mid X=x_0 \}} .
$$
Further assume that $E\{ D(1) \mid X=x   \}$ and $E\{ Y(1) \mid X=x   \}$ are continuous from the right at $x=x_0$, and $E\{ D(0) \mid X=x   \}$ and $E\{ Y(0) \mid X=x   \}$ are continuous from the left at $x=x_0$. 
The local complier average causal effect  can be identified by 
$$
 \tau_\cp(x_0) 
 = \frac{ \lim_{\varepsilon \rightarrow 0+} E( Y \mid Z=1, X=x_0+\varepsilon ) - \lim_{\varepsilon \rightarrow 0+} E( Y \mid Z=0, X=x_0-\varepsilon ) }
{\lim_{\varepsilon \rightarrow 0+} E( D \mid Z=1, X=x_0+\varepsilon ) - \lim_{\varepsilon \rightarrow 0+} E( D \mid Z=0, X=x_0-\varepsilon )}
$$
if  $E(D\mid X=x)$ has a non-zero jump at $x=x_0$. 
\end{theorem}

 Theorem \ref{thm::identification-frdd} is a superposition of Theorems \ref{thm::identification-rdd} and \ref{thm::CACE-identification}. I leave its proof as Problem \ref{para::proof-theorem-frd}.

In both sharp and fuzzy regression discontinuity,  the key is to specify the neighborhood around the cutoff point. Practically, a smaller neighborhood leads to a smaller bias but a larger variance, while a larger neighborhood leads to a larger bias but a smaller variance. That is, we face a bias-variance trade-off. Some automatic procedures exist based on some statistical criteria, which rely on some strong conditions. It seems wiser to conduct sensitivity analysis over a range of the choice of the neighborhood. 

Assume that we have specified the neighborhood of $x_0$ determined by a bandwidth $h$. 
For data with $X_i \in [x_0-h, x_0 + h]$, we can estimate $\tau_D(x_0) $ by  
$$
\hat{\tau}_D(x_0) = \text{the coefficient of } Z_i
\text{ in the OLS fit of } D_i \text{ on } \{1, Z_i, R_i, L_i\},
$$
and estimate $\tau_Y(x_0) $ 
$$
\hat{\tau}_Y(x_0) = \text{the coefficient of } Z_i
\text{ in the OLS fit of } Y_i \text{ on } \{1, Z_i, R_i, L_i\},
$$
recalling the definitions $R_i = \max(X_i - x_0, 0)$ and $L_i= \min (X_i - x_0, 0)$. Then we can estimate the local complier average causal effect by  
$$
\hat\tau_\cp(x_0)  = \hat{\tau}_Y(x_0) / \hat{\tau}_D(x_0).
$$ 
This is an indirect least squares estimator. By Theorem \ref{thm::ils=2sls}, it is numerically identical to
$$
\text{the coefficient of } D_i
\text{ in the TSLS fit of } Y_i \text{ on } \{1, D_i, R_i, L_i\}  
$$
with $D_i$ instrumented by $Z_i$. In sum, after specifying $h$, the estimation of  $ \tau_\cp(x_0)$ reduces to a TSLS procedure with the local data around the cutoff point.

\section{Application}
 \label{sec::application-frdd}

\subsection{Re-analyzing \citet{asher2020rural}'s data}
\label{sec::indian-road-data}

\begin{figure}
\centering
\includegraphics[width = 0.9\textwidth]{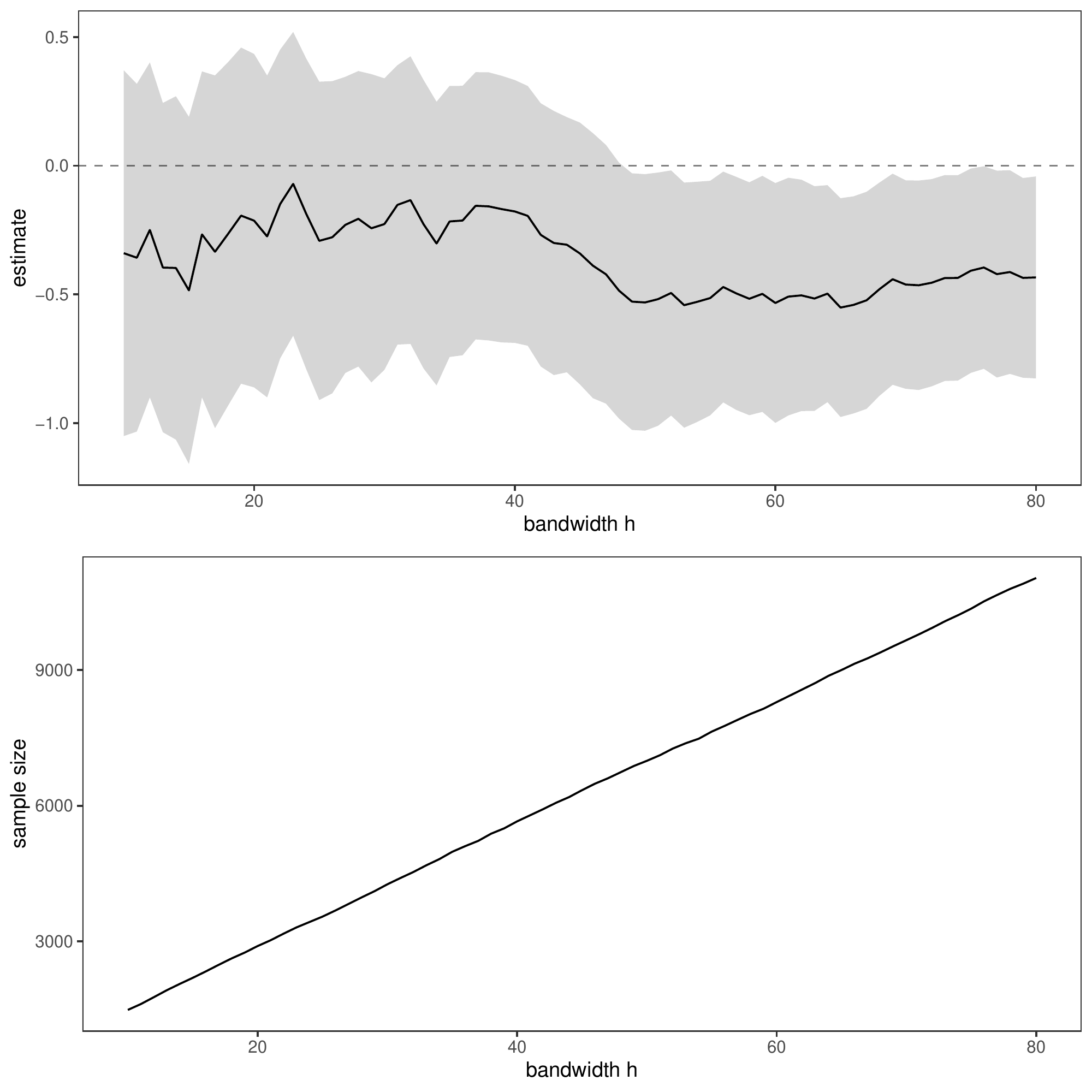}
\caption{Re-analyzing \citet{asher2020rural}'s data, with point estimates and standard errors from TSLS.}
\label{fig::indian_road}
\end{figure}

Revisit Example \ref{eg::indianroad}. We can compute the point estimates and standard errors for a sequence of $h$ based on the outcome \ri{occupation_index_andrsn}. 
\begin{lstlisting}
library("car")
road_dat = read.csv("indianroad.csv")
table(road_dat$t, road_dat$r2012)
road_dat$runv = road_dat$left + road_dat$right
## sensitivity analysis 
seq.h  = seq(10, 80, 1)
frd_sa = lapply(seq.h, function(h){
  road_sub = subset(road_dat, abs(runv)<=h)
  road_sub$r2012hat = lm(r2012 ~ t + left + right,
                         data = road_sub)$fitted.values
  tslsreg = lm(occupation_index_andrsn ~ r2012hat + left + right,
               data = road_sub)
  res = with(road_sub,
             {
               occupation_index_andrsn -
                 cbind(1, r2012, left, right)%*%coef(tslsreg)
             })
  tslsreg$residuals = as.vector(res)
  
  c(coef(tslsreg)[2],
    sqrt(hccm(tslsreg, type = "hc2")[2, 2]),
    length(res))
})
frd_sa = do.call(rbind, frd_sa)
\end{lstlisting}
Figure \ref{fig::indian_road} shows the results. The treatment effect is not significant unless $h$ is large. 

The package \ri{rdrobust} selects the bandwidth automatically. The results suggest that receiving a new road did not affect the outcome significantly. 

\begin{lstlisting}
> library("rdrobust")
> frd_road = with(road_dat,
+                 {
+                   rdrobust(y = occupation_index_andrsn,
+                            x = runv,
+                            c = 0,
+                            fuzzy = r2012)
+                 })
> res = cbind(frd_road$coef, frd_road$se)
> round(res, 3)
                Coeff Std. Err.
Conventional   -0.253     0.301
Bias-Corrected -0.283     0.301
Robust         -0.283     0.359
\end{lstlisting}

\subsection{Re-analyzing \citet{li2015evaluating}'s data}
 \label{sec::italian-university}

 Revisit Example  \ref{eg::italianuniversity}. 
Recall that the running variable is 15,000 minus the standardized income. In the analysis, I restrict the data to a subset with this running between $[-5,000,5,000]$, and then divide the running variable by $5,000$ so that the running variable is bounded between $[-1,1]$ at cutoff point zero. 
We can compute the point estimates and standard errors for a sequence of $h$. 
\begin{lstlisting}
library("car")
italy = read.csv("italy.csv")
italy$left  = pmin(italy$rv0, 0)
italy$right = pmax(italy$rv0, 0)
## sensitivity analysis 
seq.h  = seq(0.1, 1, 0.01)
frd_sa = lapply(seq.h, function(h){
  italy_sub = subset(italy, abs(rv0)<=h)
  italy_sub$Dhat = lm(D ~ Z + left + right,
                      data = italy_sub)$fitted.values
  tslsreg = lm(outcome ~ Dhat + left + right,
               data = italy_sub)
  res = with(italy_sub,
             {
               outcome -
                 cbind(1, D, left, right)%*%coef(tslsreg)
             })
  tslsreg$residuals = as.vector(res)
  
  c(coef(tslsreg)[2],
    sqrt(hccm(tslsreg, type = "hc2")[2, 2]),
    length(res))
})
frd_sa = do.call(rbind, frd_sa)
 \end{lstlisting}

\begin{figure}
\centering
\includegraphics[width = 0.9\textwidth]{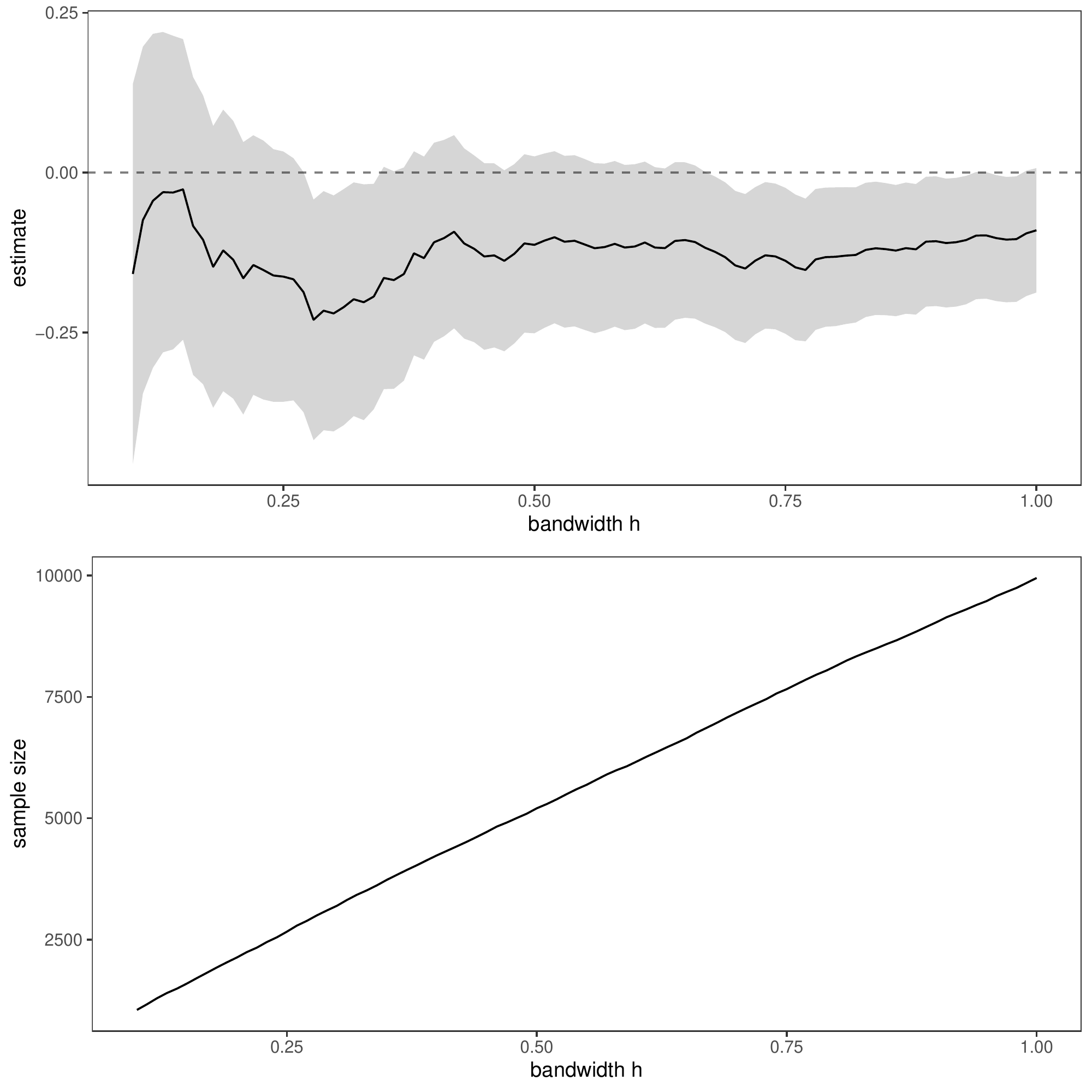}
\caption{Re-analyzing \citet{li2015evaluating}'s data, with point estimates and standard errors from TSLS.}
\label{fig::italy_university}
\end{figure}

  Figure \ref{fig::italy_university} shows the results and suggests that the university grant did not affect the dropout rate significantly. 

The results based on the package \ri{rdrobust}  reach the same conclusion. 
 
 \begin{lstlisting}
> library("rdrobust")
> frd_italy = with(italy,
+                  {
+                    rdrobust(y = outcome,
+                             x = rv0,
+                             c = 0,
+                             fuzzy = D)
+                  })
> res = cbind(frd_italy$coef, frd_italy$se)
> round(res, 3)
                Coeff Std. Err.
Conventional   -0.149     0.101
Bias-Corrected -0.155     0.101
Robust         -0.155     0.121
 \end{lstlisting}

 \section{Discussion}

Both Chapter \ref{chapter::overlap} and this chapter formulate regression discontinuity based on the continuity of the conditional expectations of the potential outcomes given the running variables. This perspective is mathematically simpler but it only identifies the local effects precisely at the cutoff point of the running variable. \citet{hahn2001identification} started this line of literature. 

An alternative, not so dominant perspective is based on {\it local randomization} \citep{cattaneo2015randomization, li2015evaluating}. If we view the running variable as a noisy measure of some underlying truth and the cutoff point is somewhat arbitrarily chosen, the units near the cutoff point do not differ systematically. This suggests that in a small neighborhood of the cutoff point, the units receive the treatment and the control in a random fashion just as in a randomized experiment. Similar to the issue of choosing $h$ in the first perspective, it is crucial to decide how local should the randomized experiment be under the regression discontinuity. It is not easy to quantify the intuition mathematically, and again conducting sensitivity analysis with a range of $h$ seems a reasonable approach in the second perspective as well.

See \citet{sekhon2017interpreting} for a more conceptual discussion of regression discontinuity.

\section{Homework Problems}

\paragraph{Proof of Theorem \ref{thm::identification-frdd}}\label{para::proof-theorem-frd}

Prove Theorem \ref{thm::identification-frdd}.

\paragraph{Data analysis}
 
Section \ref{sec::indian-road-data} reports the estimate of the effect on \ri{occupation_index_andrsn}. Four other outcome variables are \ri{transport_index_andrsn}, \ri{firms_index_andrsn}, \ri{consumption_index_andrsn}, and \ri{agriculture_index_andrsn}, with meanings defined in the original paper. Estimate the effects on these outcomes.

\paragraph{Reflection on the analysis of \citet{li2015evaluating}'s data}
\label{problem::italy-srd}

In \citet{li2015evaluating}, a key variable determining the treatment status is the binary application status $A$, which has potential outcomes $A(1)$ and $A(0)$ corresponding to the treatment $Z=1$ and control $Z=0$. By definition,
$$
D(1) = A(1), \quad D(0) = 0,
$$
so the compliers $\{ D(1), D(0) \} = (1,0)$ is equivalent to $A(1) = 1$. So 
$$
\tau_\cp(x_0) = E\{ Y(1) - Y(0) \mid A(1) = 1, X = x_0 \}. 
$$
Section \ref{sec::italian-university} used the whole data set to estimate $\tau_\cp(x_0)$. 

An alternative analysis is to use the units with $A=1$ only. Then the treatment status is determined by $X$. However, this analysis can be problematic because
\begin{eqnarray}
&& \lim_{\varepsilon \rightarrow 0+} E\{ Y \mid A  = 1, X = x_0 + \varepsilon \} -\lim_{\varepsilon \rightarrow 0+}  E\{ Y \mid A  = 1, X = x_0- \varepsilon  \}  \nonumber  \\
&=& E\{ Y(1) \mid A(1)  = 1,   X = x_0   \} -  E\{ Y(0) \mid A(0)  = 1,   X = x_0   \} .\label{hw::frdd-florence-A=1}
\end{eqnarray}
Prove \eqref{hw::frdd-florence-A=1} and explain why this analysis can be problematic.

Remark: 
The left-hand side of \eqref{hw::frdd-florence-A=1} is the identification formula of the local average treatment effect at $X=x_0$, conditional on $A=1$. The right-hand side of \eqref{hw::frdd-florence-A=1} is the difference in means of the potential outcomes for the subgroups of units with $(A(1)  = 1,   X = x_0 )$ and $(A(0)  = 1,   X = x_0)$, respectively. 
See Chapter \ref{chapter::principal-stratification} later for related discussions.

 \paragraph{Recommended reading}

\citet{imbens2008regression} gave practical guidance to regression discontinuity based on the potential outcomes framework. \citet{lee2010regression} reviewed regression discontinuity and its applications in economics.

\chapter{Application of the Instrumental Variable Method: Mendelian Randomization}
 \label{chapter::iv-mr}

\citet{katan1986apoupoprotein} was concerned with the observational studies suggesting that low serum cholesterol levels were associated with the risk of cancer. As we have discussed, however, observational studies suffer from unmeasured confounding. Consequently, it is difficult to interpret the observed association as causality. In the particular problem studied by \citet{katan1986apoupoprotein}, it is even possible that early stages of cancer reversely cause low serum cholesterol levels. Disentangling the causal effect of the serum cholesterol level on cancer seems a hard problem using standard epidemiologic studies. \citet{katan1986apoupoprotein} argued that Apolipoprotein E genes are associated with serum cholesterol levels but do not directly affect the cancer status. So if low serum cholesterol levels cause cancer, we should observe differences in cancer risks among people with and without the genotype that leads to different serum cholesterol levels. Using our language for causal inference, \citet{katan1986apoupoprotein}  proposed to use Apolipoprotein E genes as IVs.

\citet{katan1986apoupoprotein} did not conduct any data analysis but just proposed a conceptual design that could address not only {\it unmeasured confounding} but also {\it reverse causality}. Since then, more complicated and sophisticated studies have been conducted thanks to the modern genome-wide association studies. These studies used genetic information as IVs for exposures in epidemiologic studies to estimate the causal effects of exposures on outcomes. They were all motivated by {\it Mendel's second law}, the law of {\it random assortment}, which suggests that the inheritance of one trait is independent of the inheritance of other traits. Therefore, the method of using genetic information as IV is called {\it Mendelian Randomization} (MR).

\section{Background and motivation}

\begin{figure}
\centering
\includegraphics[width = 0.8 \textwidth]{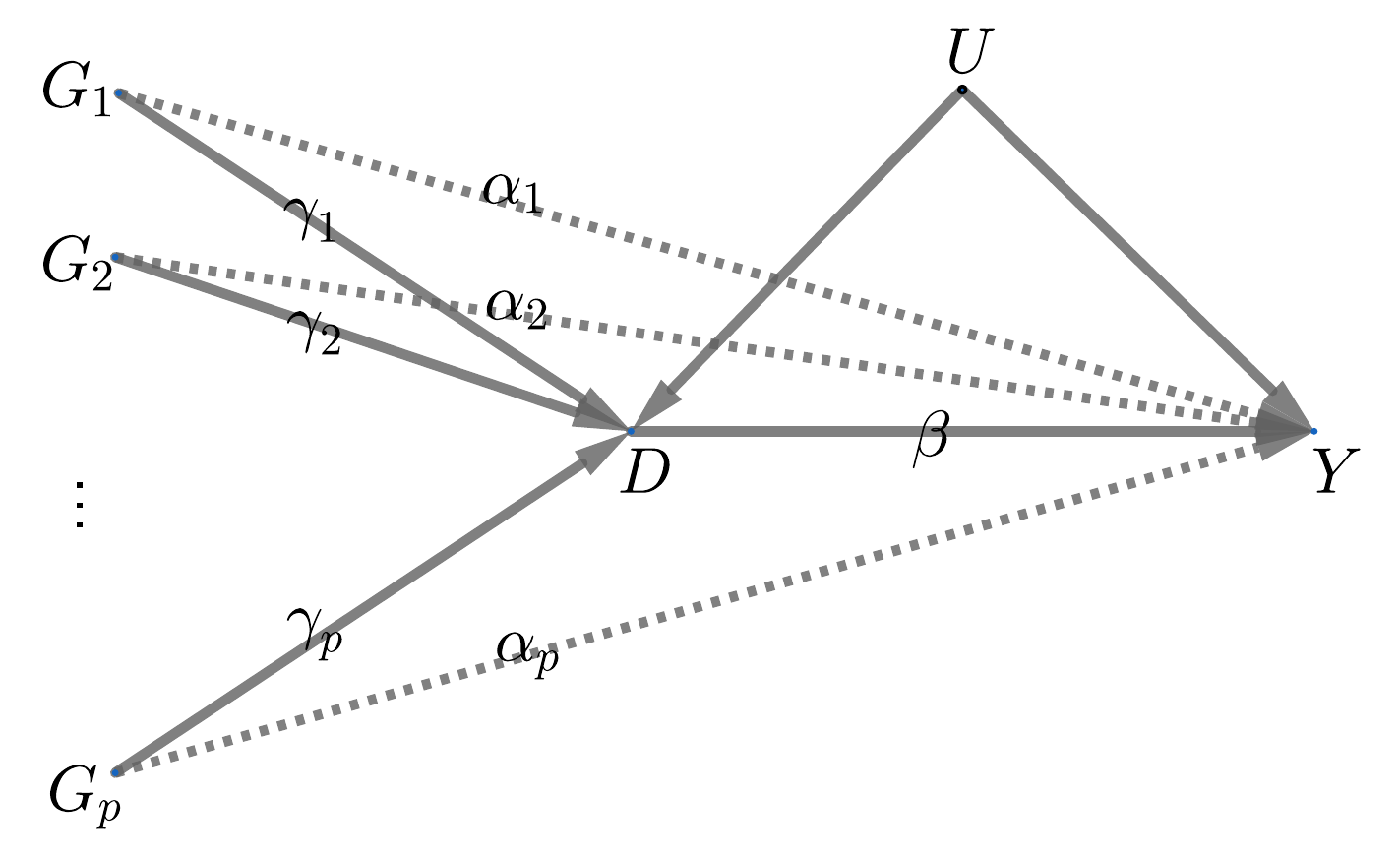}
\caption{Causal graph for MR}\label{fig::mr-dag}
\end{figure}

Graphically, Figure \ref{fig::mr-dag} shows the causal diagram on the treatment $D$, outcome $Y$, unmeasured confounder $U$, as well as the genetic IVs $G_1, \ldots, G_p$. In many MR studies, the genetic IVs are single nucleotide polymorphisms (SNPs). Because of pleiotropy, \footnote{Pleiotropy occurs when one gene influences two or more seemingly unrelated phenotypic traits. } it is possible that the genetic IVs have a direct effect on the outcome of interest, so Figure \ref{fig::mr-dag} also allows for the violation of the exclusion restriction assumption.  

The standard linear IV model assumes away the direct effect of the IVs on the outcome. Definition \ref{def::linear-iv-model-mr} below gives both the structural and reduced forms. 

\begin{definition}
[linear IV model]\label{def::linear-iv-model-mr}
The standard linear IV model
\begin{eqnarray}
Y  &=& \beta_0 +  \beta D + \beta_u U + \varepsilon_Y , \\
D  &=& \gamma_0 +   \gamma_1 G_1 + \cdots + \gamma_p G_p + \gamma_u U + \varepsilon_D ,
\end{eqnarray}
has reduced form
\begin{eqnarray}
Y  &=&\beta_0 +  \beta  \gamma_0 + \beta \gamma_1 G_1 + \cdots + \beta  \gamma_p G_p  + (\beta_u + \beta_0 \gamma_u )U + \varepsilon_Y , \\
D  &=& \gamma_0 +  \gamma_1 G_1 + \cdots + \gamma_p G_p + \gamma_u U + \varepsilon_D ,
\end{eqnarray}
\end{definition}

Definition \ref{def::linear-model-invalid-iv-mr} below allows for the violation of exclusion restriction. Then, $G_1, \ldots, G_p$ are not valid IVs.

\begin{definition}
[linear model with possibly invalid IVs]\label{def::linear-model-invalid-iv-mr}
The linear model 
\begin{eqnarray}
Y  &=& \beta_0 + \beta D + \alpha_1 G_1 + \cdots + \alpha_p G_p  + \beta_u U + \varepsilon_Y ,   \\
D  &=& \gamma_0 + \gamma_1 G_1 + \cdots + \gamma_p G_p + \gamma_u U + \varepsilon_D ,
\end{eqnarray}
 has reduced form
\begin{eqnarray}
Y  &=& ( \beta_0 +\beta  \gamma_0 ) +  (\alpha_1 + \beta \gamma_1)  G_1 + \cdots + (\alpha_p +  \beta  \gamma_p) G_p \nonumber  \\
&&+  ( \beta_u + \beta \gamma_u )  U + \varepsilon_Y ,   \\
D  &=& \gamma_0 +  \gamma_1 G_1 + \cdots + \gamma_p G_p + \gamma_u U + \varepsilon_D .
\end{eqnarray}
\end{definition}

Definitions \ref{def::linear-iv-model-mr} and \ref{def::linear-model-invalid-iv-mr} are slightly different from the linear IV  model in Chapter \ref{chapter::iv-econometric}. They include the confounder $U$ explicitly. But this slight difference does not change the discussion fundamentally.

In Definition \ref{def::linear-iv-model-mr} with exclusion restriction, we have 
$$
\Gamma_j = \beta \gamma_j,\quad (j=1,\ldots,p).
$$
In Definition \ref{def::linear-model-invalid-iv-mr} without exclusion restriction, we have 
$$
\Gamma_j = \alpha_j +   \beta \gamma_j,\quad (j=1,\ldots,p). 
$$

If we have individual data, we can apply the classic TSLS estimator to estimate $\beta$ under the linear IV model in Definition \ref{def::linear-iv-model-mr}. However, most MR studies do not have individual data but rather summary statistics from multiple genome-wide association studies. A canonical MR study with summary statistics has the following data structure.

\begin{assumption}
[MR study with summary statistics]\label{assume::MR-summary}
(a)
We have the regression coefficients of the treatment on the genetic IVs $\hat{\gamma}_1   , \ldots, \hat{\gamma}_p$ as well as the standard errors $\se_{D1}, \ldots, \se_{Dp}$. Assume 
\begin{eqnarray}
\label{eq::small-gammas}
\hat{\gamma}_1  \rightarrow \gamma_1 , \ldots, \hat{\gamma}_p \rightarrow \gamma_p
\end{eqnarray}
in probability, and ignore the uncertainty in the standard errors. 

(b)
We have the regression coefficients of the outcome on the genetic IVs $\hat{\Gamma}_1   , \ldots, \hat{\Gamma}_p$ as well as with standard errors $\se_{Y1}, \ldots, \se_{Yp}.$ Assume 
\begin{eqnarray}
\label{eq::big-gammas}
\hat{\Gamma}_1 \rightarrow \Gamma_1 , \ldots, \hat{\Gamma}_p \rightarrow \Gamma_p
\end{eqnarray}
in probability, and ignore the uncertainty in the standard errors.  

(c)
Assume $\hat{\gamma}_1   , \ldots, \hat{\gamma}_p, \hat{\Gamma}_1   , \ldots, \hat{\Gamma}_p$ are jointly Normal and independent.
\end{assumption}


The asymptotic Normality of the regression coefficients can be justified by the CLT. The standard errors are accurate estimates of the true standard errors in large samples. Therefore, the only subtle assumption is the joint independence of the regression coefficients. The independence of the $\hat{\gamma}_j$'s and the $\hat{\Gamma}_j$'s are reasonable because they are often calculated based on different samples. 

The independence among the $\hat{\gamma}_j$'s  can be reasonable if the $G_j$'s are independent and the true linear model for $D$ holds with homoskedastic error terms. See Chapter \ref{section::linear-model-var}. However, this assumption can be tricky if the error term of the linear model is heteroskedastic. Without the independence of the $G_j$'s, it is hard to justify the independence. If the regression coefficients are from nonlinear models, then it is even harder to justify the independence. The independence among the $\hat{\Gamma}_j$'s follows from a similar argument. 

This chapter focuses on the statistical inference of $\beta$ based on the above summary statistics in Assumption \ref{assume::MR-summary}.

\section{MR based on summary statistics}

\subsection{Fixed-effect estimator}\label{sec::mr-fixed-effect}

Under Definition \ref{def::linear-iv-model-mr},  $\alpha_j = 0$ which implies that $\beta = \Gamma_j / \gamma_j$ for all $j$. A simple approach is based on the so-called {\it meta-analysis} \citep{bowden2018improving}, that is, combining multiple estimates $\hat{\beta}_j =  \hat \Gamma_j / \hat\gamma_j$ for the common parameter $\beta$, also called the fixed effect. Using delta method (see Example \ref{thm::delta-method}), the ratio estimator $\hat{\beta}_j$ has approximate squared standard error 
$$
\se_j^2 = (\se_{Yj}^2 + \hat{\beta}_j^2\se_{Dj}^2  )    / \hat\gamma_j^2.
$$
Therefore, the best linear combination to estimate $\beta$ is  the Fisher weighting based on the inverses of the variances (see Problem \ref{hw::fisher-weighting}):
$$
\hat{\beta}_{\text{fisher}0} =  \frac{ \sum_{j=1}^p  \hat{\beta}_j  / \se_j^2  }{ \sum_{j=1}^p 1  / \se_j^2 }
$$
which has variance $(\sum_{j=1}^p 1  / \se_j^2)^{-1} $. Ignoring the uncertainty due to  $\hat\gamma_j$ quantified by $\se_{Dj}$, the estimator reduces to
$$
\hat{\beta}_{\text{fisher}1} =  \frac{ \sum_{j=1}^p  \hat{\beta}_j  \hat\gamma_j^2 /  \se_{Yj}^2 }{ \sum_{j=1}^p   \hat\gamma_j^2 /  \se_{Yj}^2 }
=\frac{ \sum_{j=1}^p  \hat \Gamma_j   \hat\gamma_j /  \se_{Yj}^2 }{ \sum_{j=1}^p   \hat\gamma_j^2 /  \se_{Yj}^2 },
$$
which has variance $( \sum_{j=1}^p   \hat\gamma_j^2 /  \se_{Yj}^2 )^{-1}$. Inference based on $\hat{\beta}_{\text{fisher}1} $ is suboptimal although it is more widely used in practice \citep{bowden2018improving}. 
Both $\hat{\beta}_{\text{fisher}0}$ and $\hat{\beta}_{\text{fisher}1}$ are called the {\it fixed-effect estimators}.

Focus on the suboptimal yet simpler estimator $\hat{\beta}_{\text{fisher}1}$. Under Definition \ref{def::linear-model-invalid-iv-mr},  we can show that 
$$
\hat{\beta}_{\text{fisher}1}  \rightarrow 
\frac{ \sum_{j=1}^p   \Gamma_j    \gamma_j /  \se_{Yj}^2 }{ \sum_{j=1}^p    \gamma_j^2 /  \se_{Yj}^2 }
= \beta + \frac{ \sum_{j=1}^p   \alpha_j    \gamma_j /  \se_{Yj}^2 }{ \sum_{j=1}^p    \gamma_j^2 /  \se_{Yj}^2 }
$$
in probability. 
If $\alpha_j = 0$ for all $j$, $\hat{\beta}_{\text{fisher}1} $ is consistent. Even if this does not hold, it is still possible that $\hat{\beta}_{\text{fisher}1}  $ is consistent as long as the inner product between $\alpha_j  $ and $  \gamma_j$ weighted by $ 1/  \se_{Yj}^2$ is zero. This holds if we have many genetic instruments and violation of the exclusion restriction, captured by $\alpha_j$, is an independent random draw from a distribution with mean zero.

\subsection{Egger regression}

Start with Definition \ref{def::linear-iv-model-mr}. 
With the true parameters, we have
$$
\Gamma_j = \beta \gamma_j \quad (j=1,\ldots, p);
$$
with the estimates, the above identity holds only approximately
$$
\hat \Gamma_j \approx  \beta \hat \gamma_j \quad (j=1,\ldots, p).
$$
This seems a classic least-squares problem of $\{\hat \Gamma_j \}_{j=1}^p$ on $\{\hat \gamma_j \}_{j=1}^p$. We can fit a WLS of $\hat \Gamma_j$ on $\hat \gamma_j$, with or without an intercept, possibly weighted by $w_j$, to estimate $\beta$. The following results hold thanks to the algebraic properties of the WLS reviewed in Chapter \ref{sec::appendix-wls}.

Without an intercept, the coefficient of $\hat \gamma_j$ is
$$
\hat{\beta}_{\text{egger}1}  = \frac{  \sum_{j=1}^p  \hat \gamma_j  \hat \Gamma_j w_j     }{ \sum_{j=1}^p  \hat \gamma_j ^2 w_j },
$$
which reduces to $\hat{\beta}_{\text{fisher}1} $ if $w_j =  1/\se_{Yj}^2$.  The WLS is called the {\it Egger regression}. 
It is more general than the fixed-effect estimators in Section \ref{sec::mr-fixed-effect}.  
With an intercept, the coefficient of $\hat \gamma_j$ is 
$$
\hat{\beta}_{\text{egger}0}  =  \frac{  \sum_{j=1}^p  (\hat \gamma_j  - \hat \gamma_w) (\hat \Gamma_j - \hat \Gamma_w) w_j     }
{ \sum_{j=1}^p  (\hat \gamma_j  - \hat \gamma_w) ^2 w_j }
$$
where $\hat \gamma_w = \sum_{j=1}^p \hat \gamma_j w_j /  \sum_{j=1}^p   w_j  $ and $\hat \Gamma_w = \sum_{j=1}^p \hat \Gamma_j w_j /  \sum_{j=1}^p   w_j  $  are the weighted averages of the $\hat \gamma_j$'s and $\hat \Gamma_j$'s, respectively.

Even without assuming that all $\gamma_j$'s are zero under Definition \ref{def::linear-model-invalid-iv-mr}, we have
$$
\hat{\beta}_{\text{egger}0} \rightarrow 
\frac{  \sum_{j=1}^p  (  \gamma_j  -   \gamma_w) ( \Gamma_j -   \Gamma_w) w_j     }
{ \sum_{j=1}^p  (  \gamma_j  -  \gamma_w) ^2 w_j }
=\beta + \frac{  \sum_{j=1}^p  (  \gamma_j  -   \gamma_w) ( \alpha_j -   \alpha_w) w_j     }
{ \sum_{j=1}^p  (  \gamma_j  -  \gamma_w) ^2 w_j }
$$
in probability, where $\gamma_w, \Gamma_w$ and $\alpha_w$ are the corresponding weighted averages of the true parameters. So $\hat{\beta}_{\text{egger}0} $ is consistent for $\beta$ as long as the WLS coefficient of $\alpha_j$ on $\gamma_j$ is zero. This is weaker than $\alpha_j=0$ for all $j$. This weaker assumption holds if $\gamma_j$ and $\alpha_j$ are realizations of independent random variables, which is called the Instrument Strength Independent of Direct Effect (InSIDE) assumption \citep{bowden2015mendelian}. More interestingly, the intercept from the Egger regression is
$$
\hat{\alpha}_{\text{egger}0}  = \hat \Gamma_w - \hat{\beta}_{\text{egger}0}  \hat \gamma_w,
$$
which, under the InSIDE assumption, converges to
$$
\Gamma_w -  \beta  \gamma_w = \alpha_w
$$
in probability. So the intercept estimates the weighted average of the direct effects.

\section{An example}
\label{sec::mr-examples}

First, the following function can implement Fisher weighting.
 \begin{lstlisting}
fisher.weight = function(est, se)
{
  n = sum(est/se^2)
  d = sum(1/se^2)
  res = c(n/d, sqrt(1/d))
  names(res) = c("est", "se")
  res
}
\end{lstlisting}

I use the \ri{bmi.sbp} data in the \ri{mr.raps} package \citep{zhao2020statistical} to illustrate the Fisher weighting based on different variance estimates. The results based on $\hat{\beta}_{\text{fisher}0} $ and $\hat{\beta}_{\text{fisher}1} $ are similar. 
 \begin{lstlisting}
> bmisbp = read.csv("mr_bmisbp.csv")
> bmisbp$iv      = with(bmisbp, beta.outcome/beta.exposure)
> bmisbp$se.iv   = with(bmisbp, se.outcome/beta.exposure)
> bmisbp$se.iv1  = with(bmisbp,
+                       sqrt(se.outcome^2 + iv^2*se.exposure^2)/beta.exposure)
> fisher.weight(bmisbp$iv, bmisbp$se.iv)
       est         se 
0.31727680 0.05388827 
> fisher.weight(bmisbp$iv, bmisbp$se.iv1)
       est         se 
0.31576007 0.05893783
\end{lstlisting}

The Egger regressions with or without the intercept also give very similar results. But the standard errors are quite different from those based on  Fisher weighting. 
\begin{lstlisting}
> mr.egger = lm(beta.outcome ~ 0 + beta.exposure,
+               data = bmisbp,
+               weights = 1/se.outcome^2)
> summary(mr.egger)$coef
               Estimate Std. Error  t value    Pr(>|t|)
beta.exposure 0.3172768  0.1105994 2.868704 0.004681659
> 
> mr.egger.w = lm(beta.outcome ~ beta.exposure,
+                 data = bmisbp,
+                 weights = 1/se.outcome^2)
> summary(mr.egger.w)$coef 
                  Estimate  Std. Error    t value    Pr(>|t|)
(Intercept)   0.0001133328 0.002079418 0.05450217 0.956603931
beta.exposure 0.3172989306 0.110948506 2.85987564 0.004811017
 \end{lstlisting}

\begin{figure}
\centering
\includegraphics[width = \textwidth]{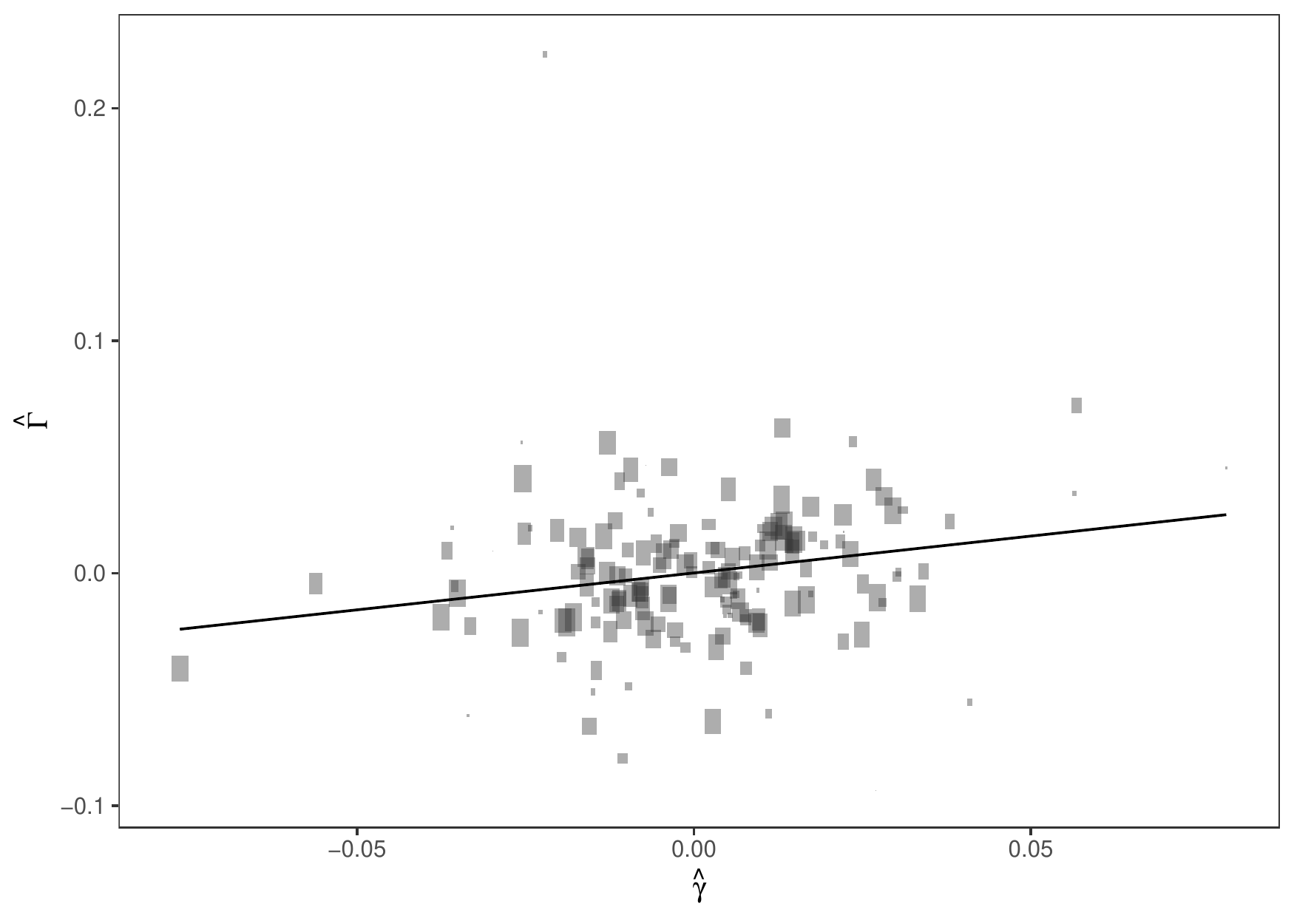}
\caption{Scatter plot proportional to the inverse of the variance, with the Egger regression line}\label{fig::bmi-sdp-egger}
\end{figure}

Figure \ref{fig::bmi-sdp-egger} shows the raw data as well as the fitted Egger regression line with the intercept. 
 
\section{Critiques of the analysis based on MR}

MR is an application of the idea of IV. It relies on strong assumptions. I provide three sets of critiques from the conceptual, biological, and technical perspectives.

Conceptually, the treatments in most studies based on MR are not well defined from the potential outcomes perspective. For instance, the treatments are often defined as the cholesterol level or the BMI. They are composite variables and can correspond to complex, non-unique definitions of the hypothetical experiments. The SUTVA often does not hold for these treatments. Recall the discussion in Chapter \ref{chapter::potential-outcomes}.

Biologically, the fundamental assumptions for the IV analysis may not hold. Mendel's second law ensures that the inheritances of different traits are independent. However, it does not ensure that the candidate IVs are independent of the hidden confounders between the treatment and the outcome. It is possible that these IVs have direct effects on the confounders. It is also possible that some unmeasured genes affect both the IVs and the confounders. Mendel's second law does not ensure the exclusion restriction assumption either. It is possible that the IVs have other causal pathways to the outcome, beyond the pathway through the treatment of interest.

Technically, the statistical assumptions for MR are quite strong.  Clearly, the linear IV model is a strong modeling assumption. The independence of the $\hat\gamma_j$'s and the $\hat\Gamma_j$'s is also quite strong. Other issues in the data-collecting process can further complicate the interpretation of the IV assumptions. For instance, the treatments and outcomes are often measured with errors, and the genome-wide associate studies are often based on the case-control design with the samples conditional on the outcomes (see Chapter \ref{sec::case-control}).

\citet{vanderweele2014methodological} is an excellent review paper that discusses the methodological challenges in MR.

\section{Homework Problems}

\paragraph{Data analysis}\label{para::data-bmi-bmi}
Analyze the \ri{bmi.bmi} data in the \ri{R} package \ri{mr.raps}. See the package and \citet[][Section 7.2]{zhao2020statistical} for more details.

 \paragraph{Recommended reading}

\citet{davey2003mendelian} reviewed the potentials and limitations of MR.

\part{Causal Mechanisms with Post-Treatment Variables}\label{part::post-treatmentvariable}

\chapter{Principal Stratification}
 \label{chapter::principal-stratification}

Parts \ref{part::rcts}--\ref{part::instrumentalvariables} of this book focus on the causal effects of a treatment on an outcome, possibly adjusting for some observed pretreatment covariates. Many applications also have some post-treatment variable $M$ which happens after the treatment and is related to the outcome. An important question is how to use the post-treatment variable $M$ appropriately. I will start with several motivating examples and then introduce \citet{frangakis2002principal}'s formulation of this problem based on potential outcomes.

\section{Motivating Examples}

\begin{example}[noncompliance]\label{eg::noncompliance}
In randomized experiments with noncompliance, we can use $M$ to represent the treatment received, which is affected by the treatment assignment $Z$ and affects the outcome $Y$. 
In this example, $M$ has the same meaning as $D$ in Chapter \ref{chapter::iv-experiment}.
\end{example}

\begin{example}[truncation by death]\label{eq::truncationbydeath}
In randomized experiments to patients with severe diseases, some patients may die before the measurement of the outcome $Y$, e.g., the quality of life. The post-treatment variable $M$ in this example is the binary indicator of the survival status. 
\end{example}

\begin{example}[unemployment]\label{eg::unemployment}
In job training programs, units are randomly assigned to treatment and control groups, and report their employment status $M$ and wage $Y$. Then the post-treatment variable is the binary indicator of the employment status $M$. 
\end{example}

\begin{example}
[surrogate endpoint]\label{eg::surrogate-endpoints}
In clinical trials, the outcomes of interest (e.g., 30 years of survival) require a long and costly follow-up. Practitioners instead collect data on some other variables early in the follow-up that are easy to measure. These variables are called the ``surrogate endpoints.'' A concrete example is from clinical trials on HIV patients, where the candidate surrogate endpoint $M$ is the CD4 cell count (recall that CD4 cells are white blood cells that fight infection). 
\end{example}

Examples \ref{eg::noncompliance}--\ref{eg::surrogate-endpoints} above have the similarity that a variable $M$ occurs after the treatment and is related to the outcome. It is possible that $M$ is on the causal pathway from $Z$ to $Y$. Figure \ref{fig::intermediate-diagram}(a) illustrates this mechanism. Example \ref{eg::noncompliance} corresponds to Figure \ref{fig::intermediate-diagram}(a). It is also possible that $M$ is not on the causal pathway from $Z$ to $Y$. Figure \ref{fig::intermediate-diagram}(b) illustrates this mechanism.  Examples \ref{eq::truncationbydeath} and \ref{eg::unemployment} correspond to Figure \ref{fig::intermediate-diagram}(b). Example \ref{eg::surrogate-endpoints} can correspond to Figure \ref{fig::intermediate-diagram}(a) or (b), depending on the choice of the surrogate endpoint.

\begin{figure}[h]
\centering
\begin{subfigure}[]{\textwidth}
                \centering
$$
\begin{xy}
\xymatrix{
&& & U \ar[dl]   \ar[dr] \\
Z \ar[rr] \ar@/_1.5pc/[rrrr] & & M\ar[rr] & & Y\\
}
\end{xy}
$$ 
               \caption{$M$ is on the causal pathway from $Z$ to $Y$, with $Z$ randomized and $U$ representing unmeasured confounding}
        \end{subfigure}%
        
        \begin{subfigure}[]{\textwidth}
                \centering
$$
\begin{xy}
\xymatrix{
&& & U \ar[dl]   \ar[dr] \\
Z \ar[rr] \ar@/_1.5pc/[rrrr] & & M  & & Y\\
}
\end{xy}
$$
                \caption{$M$ is not on the causal pathway from $Z$ to $Y$, with $Z$ randomized and $U$ representing unmeasured confounding}
        \end{subfigure}%
        
\caption{Causal diagrams with a post-treatment variable $M$}\label{fig::intermediate-diagram}
\end{figure}

In practice, the underlying causal diagrams can be much more complex than those in Figure \ref{fig::intermediate-diagram}. This chapter follows \citet{frangakis2002principal}'s formulation which does not assume the underlying causal diagrams.

\section{The Problem of Conditioning on the Post-Treatment Variable}

A naive method to deal with the post-treatment variable $M$ is to condition on its observed value as if it were a pretreatment covariate. However, $M$ is fundamentally different from $X$, because $M$ is affected by the treatment in general whereas $X$ is not. It is also a ``rule of thumb'' that data analyzers should not condition on any post-treatment variables in evaluating the average causal effect of the treatment on the outcome \citep{cochran1957analysis, rosenbaum1984consquences}. Based on potential outcomes, \citet{frangakis2002principal} gave the following insightful explanation. 

For simplicity, we focus on the CRE in this chapter.

\begin{assumption}
[CRE with an intermediate variable]
\label{assume::complete-randomization-withM}
We have
$$
Z \ind  \{ M(1), M(0), Y(1), Y(0), X \} . 
$$
\end{assumption}

Conditioning on $M= m$, we compare
\begin{equation}
\pr(Y\mid Z=1, M=m)\label{eq::proby|z1m}
\end{equation}
and
\begin{equation}
\pr(Y\mid Z=0, M=m).\label{eq::proby|z0m}
\end{equation}
This comparison seems intuitive because it measures the difference in the outcome distributions in the treated and control groups given the same value of the post-treatment variable. When $M$ is a pre-treatment covariate, this comparison yields a reasonable subgroup effect. However, when $M$ is a post-treatment variable, the interpretation of this comparison is problematic. Under Assumption \ref{assume::complete-randomization-withM}, we can re-write the probabilities in \eqref{eq::proby|z1m} and \eqref{eq::proby|z0m} as 
\begin{eqnarray*}
\pr(Y\mid Z=1, M=m)  &=& \pr\{ Y(1)\mid Z=1, M(1) = m  \} \\
&=& \pr\{ Y(1)\mid   M(1) = m  \}
\end{eqnarray*}
and
\begin{eqnarray*}
\pr(Y\mid Z=0, M=m) &=&  \pr\{ Y(0)\mid Z=0, M(0) = m  \} \\
&=& \pr\{ Y(0)\mid   M(0) = m  \} . 
\end{eqnarray*}
Therefore, under CRE, comparing \eqref{eq::proby|z1m} and \eqref{eq::proby|z0m} is equivalent to comparing the distributions of $Y(1)$ and $Y(0)$ for different subsets of units because the units with $M(1) = m$ are different from the units with $M(0) = m$ if the $Z$ affects $M$. Consequently, the comparison conditioning on $M=m$ does not have a causal interpretation in general unless $M(1) = M(0)$.\footnote{Based on the causal diagrams, we can reach the same conclusion.
In Figure \ref{fig::intermediate-diagram}, even though $Z\ind U$ by randomization of $Z$, conditioning on $M$ introduces the ``collider bias'' that causes $Z\nind U$. 
}

Revisit Example \ref{eg::noncompliance}. Comparing $\pr(Y\mid Z=1, M=1)$ and $\pr(Y\mid Z=0, M=1)$ is equivalent to comparing the treated potential outcomes for compliers and always-takers with the control potential outcomes for always-takers, under the monotonicity assumption that $M(1) \geq M(0)$. Part 3 of Problem \ref{problem::as-treated-analysis} has pointed out the drawbacks of this analysis.

Revisit Example \ref{eq::truncationbydeath}. If the treatment improves the survival status, the treatment can save more weak patients than the control. In this case, units with $M(1)=1$ are weaker than units with $M(0)=1$, so the naive comparison gives biased results that are in favor of the control. 

\section{Conditioning on the Potential Values of the Post-Treatment Variable}

\citet{frangakis2002principal} proposed to condition on the joint potential value of the post-treatment variable $U = \{ M(1), M(0) \}$ and compare
$$
\pr\{  Y(1) \mid M(1) = m_1, M(0) = m_0 \}
$$
and
$$
\pr\{  Y(0) \mid M(1) = m_1, M(0) = m_0 \}
$$
for some $(m_1, m_0)$. This is a comparison of the potential outcomes under treatment and control for the same subset of units with $M(1) = m_1$ and $ M(0) = m_0$. \citet{frangakis2002principal} called this strategy {\it principal stratification}, viewing $\{ M(1), M(0) \}$ as a pre-treatment covariate.  Based on this idea, we can define 
$$
\tau(m_1, m_0) = E\{  Y(1) - Y(0) \mid M(1) = m_1, M(0) = m_0\}
$$
as the principal stratification average causal effect for the subgroup with $M(1) = m_1, M(0) = m_0$. For a binary $M$, we have four subgroups
\begin{equation}
\label{eq::psace}
\left\{ \begin{array}{rcl}
\tau(1,1) &=& E\{  Y(1) - Y(0) \mid M(1) = 1, M(0) =1\} ,\\
\tau(1,0) &=&E\{  Y(1) - Y(0) \mid M(1) = 1, M(0) =0\},\\
\tau(0,1) &=&E\{  Y(1) - Y(0) \mid M(1) = 0, M(0) =1\},\\
\tau(0,0) &=&E\{  Y(1) - Y(0) \mid M(1) = 0, M(0) = 0 \}.
\end{array} 
\right.
\end{equation}
Since  $\{ M(1), M(0) \}$ is unaffected by the treatment, it is a covariate so $\tau(m_1, m_0) $ is a subgroup causal effect. For subgroups with $M(1) = M(0)$, the treatment does not change the intermediate variable, so  $\tau(1,1) $ and $\tau(0,0)$ measure the {\it dissociative effects}. 
For other subgroups with $m_1 \neq m_0$, the principal stratification average causal effects $\tau(m_1, m_0) $ measure the {\it associative effects}. The terminology is from \citet{frangakis2002principal}, which does not assume that $M$ is on the causal pathway from $Z$ to $Y$. When we have Figure \ref{fig::intermediate-diagram}(a), we can interpret the dissociative effects as direct effects of $Z$ on $Y$ that act independent of $M$, although we cannot simply interpret the associative effects as direct or indirect effects of $Z$ on $Y$.

\setcounter{example}{0}
\begin{example}[noncompliance]
With noncompliance, \eqref{eq::psace} consists of the average causal effects for always takers, compliers, defiers, and never takers \citep{imbens1994identification, angrist1996identification}. 
\end{example}

\begin{example}[truncation by death]
Because the outcome is well-defined only if the patient survives, three subgroup causal effects in \eqref{eq::psace}  are not meaningful, and the only well-defined subgroup effect is
\begin{eqnarray}
\label{eq::survivor}
\tau(1,1) = E\{  Y(1) - Y(0) \mid M(1) = 1, M(0) =1\} .
\end{eqnarray}
It is called the survivor average causal effect \citep{rubin2006causal}. It is the average causal effect of the treatment on the outcome for those units who survive regardless of the treatment status. 
\end{example}

\begin{example}[unemployment]
The unemployment problem is isomorphic to the truncation-by-death problem because the wage is well-defined only if the unit is employed in the first place. Therefore, the only well-defined subgroup effect is \eqref{eq::survivor}, the employed average causal effect. Previously, \citet{heckman1979sample} proposed a model, now called the Heckman Selection Model, to deal with unemployment in modeling the wage, viewing the wages of those unemployed as missing values\footnote{Heckman won the Nobel prize of economics in 2000 ``for his development of theory and methods for analyzing selective samples.''
His model contains two stages. First, the employment status is determined by a latent linear model
$$
M_i = 1(X_i\tran \beta + u_i \geq 0).
$$
Second, the latent log wage is determined by a linear model
$$
Y_i^* = W_i\tran \gamma + v_i
$$
and $Y_i^*$ is observed as $Y_i$ only if $M_i = 1$. In his two-stage model, the covariates $X_i$ and $W_i$ may differ, and the errors $(u_i , v_i)$ are correlated bivariate Normal. 
}. However, \citet{zhang2003estimation} and \citet{zhang2009likelihood} argued that $\tau(1,1)$ is a more meaningful quantity under the potential outcomes framework. 
\end{example}

\begin{example}[surrogate endpoint]
Intuitively, we want to assess the effect of the treatment on the outcome via the effect of the treatment on the surrogate endpoint. Therefore, a good surrogate endpoint should satisfy two conditions: first, if the treatment does not affect the surrogate, then it does not affect the outcome either; second, if the treatment affects the surrogate, then it affects the outcome too. The first condition is called the ``causal necessity'' by \citet{frangakis2002principal}, and the second condition is called the ``causal sufficiency'' by \citet{gilbert2008evaluating}. Based on \eqref{eq::psace} for a binary surrogate endpoint, causal necessity requires that $\tau(1,1)$ and $\tau(0,0)$ are zero, and causal sufficiency requires that $\tau(1,0)$ and $\tau(0,1)$ are not zero. 
\end{example}

\section{Statistical Inference and Its Difficulty}
\label{sec::difficulty-of-PS}

In Example \ref{eg::noncompliance}, if we have randomization, monotonicity, and exclusion restriction, then we can identify the complier average causal effect $\tau(1, 0)$. This is the key result derived in Chapter \ref{chapter::iv-experiment}.

However, in other examples, we cannot impose the exclusion restriction assumption. For instance, $\tau(1,1)$ is the main parameter of interest in Examples \ref{eq::truncationbydeath} and \ref{eg::unemployment}, and $\tau(1,1)$ and $\tau(0,0)$ are both of interest in Example \ref{eg::surrogate-endpoints}. Without the exclusion restriction assumption, it is very challenging to identify the principal stratification average causal effect. Sometimes, we cannot even impose the monotonicity assumption, and thus cannot identify the proportions of the latent strata in the first place.

\subsection{Special case: truncation by death with binary outcome}
I use the simple setting with a binary treatment, binary survival status, and binary outcome to illustrate the idea and especially the difficulty of statistical inference based on principal stratification.

In addition to Assumption \ref{assume::complete-randomization-withM}, we impose the monotonicity.

\begin{assumption}
[monotonicity]\label{assume::monotonicity-ps}
$M(1) \geq M(0)$.
\end{assumption}

Theorem \ref{thm::cace-mixture-components} demonstrates that under Assumptions \ref{assume::complete-randomization-withM} and \ref{assume::monotonicity-ps}, we can identify the proportions of the three latent strata by
\begin{eqnarray*} 
 \pi_{(1,1)} &=& \pr(M=1\mid Z=0), \\
 \pi_{(0,0)} &=& \pr(M=0\mid Z=1), \\
 \pi_{(1,0)} &=& \pr(M=1\mid Z=1) - \pr(M=1\mid Z=0). 
\end{eqnarray*}

Our goal is to identify the survivor average causal effect $\tau(1,1)$. First, we can easily identify $E\{ Y(0)\mid M(1) = 1, M(0)=1 \}$ because the observed group $(Z=0,M=1)$ consists of only survivors:
$$
E\{ Y(0)\mid M(1) = 1, M(0)=1 \} = E(Y\mid Z=0, M=1). 
$$
The key is then to identify $E\{ Y(1)\mid M(1) = 1, M(0)=1 \}$. The observed group $(Z=1,M=1)$ is a mixture of two strata $(1,1)$ and $(1,0)$, therefore we have 
\begin{eqnarray} 
E(Y\mid Z=1,M=1) 
&=& \frac{ \pi_{(1,1)}}{ \pi_{(1,1)} + \pi_{(1,0)}} E\{  Y(1)\mid   M(1)=1, M(0)=1 \} \nonumber  \\
&&+ \frac{ \pi_{(1,0)}}{ \pi_{(1,1)} + \pi_{(1,0)}} E\{  Y(1)\mid  M(1)=1, M(0)=0 \}. \nonumber  \\
&& \label{eq::bound-equation-basic}
\end{eqnarray} 
We have two unknown parameters but only one equation in \eqref{eq::bound-equation-basic}. So we cannot uniquely determine $E\{  Y(1)\mid   M(1)=1, M(0)=1 \}$ from the above equation \eqref{eq::bound-equation-basic}. Nevertheless, \eqref{eq::bound-equation-basic} contains some information about the quantity of interest. 
That is, $E\{  Y(1)\mid   M(1)=1, M(0)=1 \}$  is partially identified by Definition \ref{def::partial-identification}.

For a binary outcome $Y$, we know that $E\{  Y(1)\mid  M(1)=1, M(0)=0 \}$ is bounded between 0 and 1, and consequently, $E\{  Y(1)\mid   M(1)=1, M(0)=1 \}$ is bounded between the solutions to the following two equations:
\begin{eqnarray*}
E(Y\mid Z=1,M=1) &=& \frac{ \pi_{(1,1)}}{ \pi_{(1,1)} + \pi_{(1,0)}} E\{  Y(1)\mid   M(1)=1, M(0)=1 \}  \\
&&\qquad + \frac{ \pi_{(1,0)}}{ \pi_{(1,1)} + \pi_{(1,0)}}  
\end{eqnarray*}
and
$$
E(Y\mid Z=1,M=1) = \frac{ \pi_{(1,1)}}{ \pi_{(1,1)} + \pi_{(1,0)}} E\{  Y(1)\mid   M(1)=1, M(0)=1 \} .
$$
Therefore,  $E\{  Y(1)\mid   M(1)=1, M(0)=1 \}$ has lower bound
$$
\frac{  \{  \pi_{(1,1)} + \pi_{(1,0)} \} E(Y\mid Z=1,M=1) - \pi_{(1,0)}}{ \pi_{(1,1)}},
$$
and upper bound
$$
\frac{  \{  \pi_{(1,1)} + \pi_{(1,0)} \} E(Y\mid Z=1,M=1)  }{ \pi_{(1,1)}} . 
$$
We can then derive the bounds on $\tau(1,1)$, summarized in Theorem \ref{thm::bounds-on-SACE-binary-Y} below.

\begin{theorem}
\label{thm::bounds-on-SACE-binary-Y}
Under Assumptions \ref{assume::complete-randomization-withM} and \ref{assume::monotonicity-ps} with a binary $Y$, we have
\begin{eqnarray*}
&&\frac{  \{  \pi_{(1,1)} + \pi_{(1,0)} \} E(Y\mid Z=1,M=1) - \pi_{(1,0)}}{ \pi_{(1,1)}} - E(Y\mid Z=0, M=1) \\
&\leq & \tau(1,1) \\
& \leq &\frac{  \{  \pi_{(1,1)} + \pi_{(1,0)} \} E(Y\mid Z=1,M=1)  }{ \pi_{(1,1)}}  - E(Y\mid Z=0, M=1). 
\end{eqnarray*}
\end{theorem}

We can use \citet{imbens2004confidence}'s confidence interval for $\tau(1,1)$ which involves two steps: first, we obtain the estimated lower and upper bounds $[\hat{l}, \hat{u}]$ with estimated standard errors $(\text{se}_l, \text{se}_u)$; second, we construct the confidence interval as $[ \hat{l} - z_{1-\alpha}  \text{se}_l, \hat{u} + z_{1-\alpha}    \text{se}_u]$, where $z_{1-\alpha} $ is the $1-\alpha$ quantile of the standard Normal distribution. The validity of \citet{imbens2004confidence}'s confidence interval relies on some regularity conditions. In most truncation by death problems, those conditions hold because the lower and upper bounds are quite different, and they are bounded away from the extreme values $-1$ and $1$. I omit the technical details here.

To summarize, this is a challenging problem since we cannot identify the parameter based on the observed data even with an infinite sample size. We can derive large-sample bounds for $\tau(1,1)$ but the statistical inference based on the bounds is not standard. If we do not have monotonicity, the large-sample bounds have even more complex forms; see \citet{zhang2003estimation} and \citet[][Appendix A]{jiang2016principal}.

\subsection{An application}\label{section::ps-application}

I use the data in \citet{yang2016using} from the Acute Respiratory Distress Syndrome Network study which involves 861 patients with lung injury and acute respiratory distress syndrome. Patients were randomized to receive mechanical ventilation with either lower tidal volumes or traditional tidal volumes. The outcome is the binary indicator for whether patients could breathe without assistance by day 28. Table \ref{tb::truncationbydeath} summarizes the observed data.

\begin{table} 
\centering 
\caption{Data truncated by death with * indicating the outcomes for dead patients}\label{tb::truncationbydeath}
\begin{tabular}{cccc}
\hline 
\multicolumn{4}{c}{ Treatment $Z=1$} \\
\hline 
 & $Y=1$ & $Y=0$ &  total\\
$M=1$ &    54 & 268& 322 \\
$M=0$  & * & * & 109 \\
\hline 
\end{tabular}
~~~~~
\begin{tabular}{cccc}
\hline 
\multicolumn{4}{c}{ Control $Z=0$} \\
\hline 
 & $Y=1$ & $Y=0$ &total \\
$M=1$ &    59 & 218&  277 \\
$M=0$  & * & *& 152 \\
\hline 
\end{tabular}
\end{table}


We first obtain the point estimators of the latent strata:
\begin{eqnarray*} 
 \hat{\pi}_{(1,1)} &=& \frac{277}{277+152} = 0.646, \\
  \hat{\pi}_{(0,0)} &=&  \frac{109}{109+322} = 0.253,  \\ 
\hat{\pi}_{(1,0)} &=& 1 -0.646 -  0.253 =   0.101. 
\end{eqnarray*} 
The sample means of the outcome for surviving patients are
\begin{eqnarray*} 
\hat{E}(Y\mid Z=1,M=1) &=& \frac{54}{302} = 0.168, \\ 
\hat{E}(Y\mid Z=0, M=1) &=& \frac{59}{277} = 0.213.
\end{eqnarray*} 
The estimates for the bounds on $E\{  Y(1)\mid   M(1)=1, M(0)=1 \}$ are
$$
\left[ \frac{  ( 0.646 + 0.101 )\times  0.168-  0.101 }{  0.101 },  
\frac{ ( 0.646 + 0.101 )\times  0.168}{0.101} \right]
= [  0.037, 0.194],
$$
so the bounds on $\tau(1,1)$ are 
$$
[0.037- 0.213, 0.194- 0.213] = 
[  -0.176, -0.019 ] . 
$$
Incorporating the sampling uncertainty based on the bootstrap, the confidence interval based on \citet{imbens2004confidence}'s method is $[-0.267,  0.039]$, which covers 0.

\subsection{Extensions}

\citet{zhang2003estimation} started the literature of large-sample bounds. \citet{imai2008sharp} and \citet{lee2009training} were two follow-up papers. \citet{cheng2006bounds} derived the bounds with multiple treatment arms. \citet{yang2016using} used a secondary outcome and \citet{yang2018using} used detailed survival information to sharpen the bounds on the survivor average causal effect.

\section{Principal score method}

Without additional assumptions, we can only derive bounds on the causal effects within principal strata, but cannot identify them in general. We must impose additional assumptions to achieve nonparametric identification of the $\tau(m_1, m_0)$'s. There is no consensus on the choice of the assumptions. Those additional assumptions are not testable, and their plausibility depends on the application. 

A line of research based on the {\it principal score} under the {\it principal ignorability} assumption parallels causal inference with unconfounded observational studies. For simplicity, I focus on the case with strong monotonicity.

\subsection{Principal score method under strong monotonicity}

\begin{assumption}
[strong monotonicity]\label{assume::strong-monotonicity}
 $M(0) = 0$. 
\end{assumption}

Similar to the ignorability assumption, we now assume the {\it principal ignorability} assumption.

\begin{assumption}
[principal ignorability]\label{assume::principal-ignorability-SM}
We have 
$$
E\{  Y(0) \mid M(1) = 1, X \} = E\{  Y(0) \mid M(1) = 0, X \} .
$$ 
\end{assumption}

Assumption \ref{assume::principal-ignorability-SM} implies that
$
E\{  Y(0) \mid M(1)  , X \} = E\{  Y(0) \mid   X \} 
$
or, equivalently, 
\begin{eqnarray}
\label{eq::mean-independent}
E\{  Y(0)  M(1)  \mid  X \}  =  E\{  Y(0) \mid   X \}  E\{     M(1)  \mid  X \}  ,
\end{eqnarray}
that is, $Y(0)$ and $M(1)$ are mean independent (or uncorrelated) given $X$. 

Assumptions \ref{assume::complete-randomization-withM}, \ref{assume::strong-monotonicity} and \ref{assume::principal-ignorability-SM} ensure nonparametric identification of the causal effects within principal strata, as summarized by Theorem \ref{thm::identification-PCEs-pscore} below.

\begin{theorem}
\label{thm::identification-PCEs-pscore}
Under Assumptions \ref{assume::complete-randomization-withM}, \ref{assume::strong-monotonicity} and \ref{assume::principal-ignorability-SM}, 
\begin{enumerate}
[(a)]
\item 
the conditional and marginal probabilities of $M(1)$, 
$\pi(X) =  \pr\{ M(1)=1\mid X \}  $ and $\pi =  \pr\{ M(1)=1\} $, can be identified by
$$
\pi(X)   =  \pr(M=1\mid Z=1, X ) 
$$ 
and 
$$
\pi  =   \pr(M=1\mid Z=1 ) 
$$ 
respectively; 
\item
the principal  stratification average causal effects can be identified by 
$$
\tau(1,0) = E(Y\mid Z=1, M=1)  - E\{  \pi(X) Y  \mid Z=0 \}/\pi 
$$
and
$$
\tau(0,0) = E(Y\mid Z=1, M=0)  - E\{ (1-\pi(X)) Y  \mid Z=0 \}/(1-\pi) . 
$$
\end{enumerate}
\end{theorem}

The conditional probability $\pi(X) =  \pr\{ M(1)=1\mid X \} $ is called the {\it principal score}. Theorem \ref{thm::identification-PCEs-pscore} states that $\tau(1,0) $ and  $\tau(0,0)$ can be identified by the difference in means with appropriate weights depending on the principal score.

 \begin{myproof}{Theorem}{\ref{thm::identification-PCEs-pscore}}
 I will only prove the result for $\tau(1,0) $. 
We have
 $$
 E(Y\mid Z=1, M=1)  = E\{ Y(1) \mid Z=1, M(1) = 1 \} =  E\{ Y(1) \mid   M(1) = 1 \} .
 $$
Moreover, 
 \begin{eqnarray*}
&& E  \{\pi(X) Y\mid Z=0   \}  /\pi  \\
 &=& E  \{\pi(X) Y(0)  \mid Z=0 \}  /\pi  \\
 &=&  E  \{\pi(X) Y(0)   \}  /\pi \qquad   (\text{Assumption } \ref{assume::complete-randomization-withM}) \\
 &=&  E \left[    E\{\pi(X) Y(0) \mid X  \} \right]/\pi  \qquad (\text{tower property}) \\ 
 &=& E \left[  \pi(X)  E\{ Y(0) \mid X  \} \right]/\pi  \\ 
 &=& E \left[  E\{    M(1) \mid X\} E\{ Y(0) \mid X  \} \right]/\pi \\
 &=&  E \left[  E\{    M(1)  Y(0) \mid X  \} \right]/\pi  \qquad  (\text{by } \eqref{eq::mean-independent}) \\
 &=&  E\{    M(1) Y(0)  \}/\pi   \\
 &=&  E\{ Y(0) \mid M(1) = 1 \}  .
 \end{eqnarray*}
%
I relegate the proof of the result for $\tau(0,0) $ as Problem \ref{hw::proof-part2-identification-PCEs}. 
 \end{myproof}

Theorem \ref{thm::identification-PCEs-pscore} motivates the following simple estimators for  $\tau(1,0)$ and $\tau(0,0) $, respectively:
\begin{enumerate}
\item
fit a logistic regression of $M$ on $X$ using only data from the treated group to obtain $\hat \pi(X_i)$;
\item
estimate $\pi$ by $\hat \pi =  \sumn Z_i M_i / \sumn Z_i$; 
\item
obtain moment estimators:
$$
\hat \tau(1,0) =  \frac{  \sumn Z_iM_iY_i  }{ \sumn Z_iM_i }   -   \frac{   \sumn (1-Z_i) \hat \pi(X_i) Y_i  }{ \hat \pi   \sumn (1-Z_i) }   
$$
and
$$
\hat \tau(0,0) =  \frac{  \sumn Z_i(1-M_i)Y_i  }{ \sumn Z_i (1-M_i) }  - \frac{  \sumn (1-Z_i)(1-\hat \pi(X_i)\} Y_i  }{  (1- \hat \pi )\sumn (1-Z_i)  }
;
$$
\item
use the bootstrap to approximate the variances of $\hat \tau(1,0) $ and $\hat \tau(0,0) $. 
\end{enumerate}

 \subsection{An example}

The following function can compute the point estimates of $\hat \tau(1,0) $ and $\hat \tau(0,0) $. 

  \begin{lstlisting}
psw = function(Z, M, Y, X) {
  ## probabilities of 10 and 00
  pi.10 = mean(M[Z==1])
  pi.00 = 1 - pi.10
  
  ## conditional probabilities of 10 and 00
  ps.10 = glm(M ~ X, family = binomial, 
              weights = Z)$fitted.values
  ps.00 = 1 - ps.10
  
  ## PCEs 10 and 00 
  tau.10 = mean(Y[Z==1&M==1]) - mean(Y[Z==0]*ps.10[Z==0])/pi.10
  tau.00 = mean(Y[Z==1&M==0]) - mean(Y[Z==0]*ps.00[Z==0])/pi.00
  c(tau.10, tau.00)
}
   \end{lstlisting}

The following function can compute the point estimators as well as the bootstrap standard errors.
  \begin{lstlisting}
psw.boot = function(Z, M, Y, X, n.boot = 500){
  ## point estimates
  point.est = psw(Z, M, Y, X)
  ## bootstrap standard errors
  n = length(Z)  
  boot.est = replicate(n.boot, {
    id.boot = sample(1:n, n, replace = TRUE)
    psw(Z[id.boot], M[id.boot], Y[id.boot], X[id.boot, ])
  })
  boot.se    = apply(boot.est, 1, sd)
  ## results
  res        = rbind(point.est, boot.se)
  rownames(res) = c("est", "se")
  colnames(res) = c("tau10", "tau00")
  return(res)
}
    \end{lstlisting}

Revisit the data used in Chapter \ref{sec::cace-application}. Previously, we assumed exclusion restriction for the IV analysis. Now, we drop the exclusion restriction but assume principal ignorability. The results are below.

 \begin{lstlisting}
> jobsdata = read.csv("jobsdata.csv")
> getX     = lm(treat ~ sex + age + marital 
+               + nonwhite + educ + income,
+               data = jobsdata)
> X = model.matrix(getX)[, -1]
> Z = jobsdata$treat
> M = jobsdata$comply
> Y = jobsdata$job_seek
> table(Z, M)
   M
Z     0   1
  0 299   0
  1 228 372
> psw.boot(Z, M, Y, X)
        tau10       tau00
est 0.1694979 -0.09904909
se  0.1042405  0.15612983
\end{lstlisting}

The point estimator       $\hat \tau(1,0) $ does not differ much from the one based on the   IV analysis. The point estimator $\hat \tau(0,0) $ is close to zero with a large standard error. In this example, exclusion restriction seems a reasonable assumption.

\subsection{Extensions}

 \citet{follmann2000effect}, \citet{hill2002differential}, \citet{jo2009use}, \citet{jo2011use} and \citet{stuart2015assessing} started the literature of using the principal score to identify causal effects within principal strata. \citet{ding2017principal} provided a theoretical foundation for this strategy. They proved Theorem \ref{thm::identification-PCEs-pscore} as well as a more general version under monotonicity; see Problem \ref{para::prinpscore-mono}.

\section{Other methods}

To estimate principal stratification average causal effects without the exclusion restriction assumption,  \citet{zhang2009likelihood} proposed to use the Normal mixture models. However,   the inference based on the Normal mixture models can be quite fragile. A strategy is to use additional information to improve the inference under some restrictions \citep{ding2011identifiability, mealli2013using, mattei2013exploiting, jiang2016principal}.

Conceptually, the principal stratification framework works for general $M$. A multi-valued $M$ generates many latent principal strata, and a continuous $M$ generates infinitely many latent principal strata. In those cases, identifying the probability of the principal strata is non-trivial in the first place let alone identifying the principal stratification average causal effects.  \citet{jiang2020identification} reviewed some useful strategies.

\section{Homework problems}

\paragraph{Complete the proof of Theorem \ref{thm::identification-PCEs-pscore}}\label{hw::proof-part2-identification-PCEs}

Prove the result for $\tau(0,0) $ in Theorem \ref{thm::identification-PCEs-pscore}.

\paragraph{Principal score method under monotonicity}\label{para::prinpscore-mono}

This problem extends Theorem \ref{thm::identification-PCEs-pscore}, with Assumption \ref{assume::strong-monotonicity} replaced by Assumption \ref{assume::monotonicity-ps} and Assumption \ref{assume::principal-ignorability-SM} replaced by Assumption \ref{assume::principal-ignorability-M} below. 

\begin{assumption}
[principal ignorability]\label{assume::principal-ignorability-M}
We have 
$$
E\{  Y(1) \mid M(1) = 1, M(0) =0, X \} = E\{  Y(1) \mid M(1) = 1, M(0) =1, X \}
$$
and
$$
E\{  Y(0) \mid M(1) = 1, M(0) =0, X \} = E\{  Y(0) \mid M(1) = 0, M(0) =0, X \}. 
$$
\end{assumption}

\begin{theorem}
\label{thm::identification-PCEs-pscore-general-ding-lu}
Under Assumptions \ref{assume::complete-randomization-withM}, \ref{assume::monotonicity-ps} and \ref{assume::principal-ignorability-M}, 
\begin{enumerate}
[(a)]
\item
 the conditional and marginal principal scores can be identified by
\begin{eqnarray*}
\pi_{(0,0)}(X) &=& \pr(M=0\mid Z=1, X),\\
\pi_{(1,1)}(X) &=& \pr(M=1\mid Z=0, X),\\
\pi_{(1,0)}(X) &=& \pr(M=1\mid Z=1, X) - \pr(M=1\mid Z=0, X).
\end{eqnarray*}
and
\begin{eqnarray*}
\pi_{(0,0)} &=& \pr(M=0\mid Z=1),\\
\pi_{(1,1)} &=& \pr(M=1\mid Z=0),\\
\pi_{(1,0)} &=& \pr(M=1\mid Z=1) - \pr(M=1\mid Z=0)
\end{eqnarray*}
respectively;
\item
the principal  stratification average causal effects can be identified by 
\begin{eqnarray*}
\tau(1,0) &=& E\left\{w_{1, (1,0)}(X) Y \mid Z=1, M=1\right\}  \\
&& -E\left\{w_{0, (1,0)}(X) Y \mid Z=0, M=0\right\} , \\
\tau(0,0) &=& E(Y \mid Z=1, M=0)-E\left\{w_{0, (0,0)}( X ) Y \mid Z=0, M=0\right\}, \\
\tau(1,1) &=& E\left\{w_{1,  (1,1)}( X ) Y \mid Z=1, M=1\right\}-E(Y \mid Z=0, M=1)
\end{eqnarray*}
with 
\begin{eqnarray*}
w_{1, (1,0)}(X) &=&\frac{\pi_{(1,0)}(X)}{\pi_{(1,0)}(X)+\pi_{(1,1)}(X)} \Big / \frac{\pi_{(1,0)}}{\pi_{(1,0)}+\pi_{(1,1)}} , \\
w_{0, (1,0)}(X) &=& \frac{\pi_{(1,0)}(X)}{\pi_{(1,0)}(X)+\pi_{(0,0)}(X)} \Big / \frac{\pi_{(1,0)}}{\pi_{(1,0)}+\pi_{(0,0)}} ,  \\
w_{0, (0,0)}(X) &=&\frac{\pi_{(0,0)}(X)}{\pi_{(1,0)}(X)+\pi_{(0,0)}(X)} \Big / \frac{\pi_{(0,0)}}{\pi_{(1,0)}+\pi_{(0,0)}} , \\
w_{1, (1,1)}(X) &=&\frac{\pi_{(1,1)}(X)}{\pi_{(1,0)}(X)+\pi_{(1,1)}(X)} \Big / \frac{\pi_{(1,1)}}{\pi_{(1,0)}+\pi_{(1,1)}} .
\end{eqnarray*}
\end{enumerate}
\end{theorem}

Remark:
Based on Theorem \ref{thm::identification-PCEs-pscore-general-ding-lu}, we can construct weighting estimators. Theorem \ref{thm::identification-PCEs-pscore-general-ding-lu} is Proposition 2 in \citet{ding2017principal}, which also provided more details for the estimation.

\paragraph{Principal score method in observational studies}
\label{hw::principalscore-observational}

This problem extends Theorem \ref{thm::identification-PCEs-pscore}, with Assumption \ref{assume::complete-randomization-withM} replaced by the ignorability assumption below.

\begin{assumption}
\label{assume::ignorabilitywithM}
$
Z\ind \{ M(1), M(0), Y(1) , Y(0) \} \mid X.
$
\end{assumption}

Recall the definition of the propensity score $e(X) = \pr(Z=1\mid X)$. 
We have the following identification result.

\begin{theorem}
\label{thm::identification-PCEs-pscore-obs}
Under Assumptions \ref{assume::ignorabilitywithM}, \ref{assume::strong-monotonicity} and \ref{assume::principal-ignorability-SM}, 
\begin{enumerate}
[(a)]
\item 
the conditional and marginal probabilities of $M(1)$, 
$\pi(X) =  \pr\{ M(1)=1\mid X \}  $ and $\pi =  \pr\{ M(1)=1\} $, can be identified by
$$
\pi(X)   =  \pr(M=1\mid Z=1, X ) 
$$ 
and 
$$
\pi  =  E\{\pr(M=1\mid Z=1, X ) \}
$$ 
respectively; 
\item
the principal  stratification average causal effects can be identified by 
$$
\tau(1,0) = E\left\{   \frac{M}{\pi}   \frac{Z}{e(X)} Y \right\}
- E\left\{   \frac{ \pi(X)  }{\pi}   \frac{1-Z}{1-e(X)} Y    \right\} 
$$
and
$$
\tau(0,0) = E\left\{   \frac{1-M}{1-\pi}   \frac{Z}{e(X)} Y \right\}
- E\left\{   \frac{ 1- \pi(X)  }{1 - \pi}   \frac{1-Z}{1-e(X)} Y    \right\} .
$$
\end{enumerate}
\end{theorem}

Prove Theorem \ref{thm::identification-PCEs-pscore-obs}.

\paragraph{General principal score method}
\label{hw::general-principal-score-jiangetal}

Extend Theorems \ref{thm::identification-PCEs-pscore-general-ding-lu} and \ref{thm::identification-PCEs-pscore-obs} under Assumptions  \ref{assume::ignorabilitywithM}, \ref{assume::monotonicity-ps} and \ref{assume::principal-ignorability-M}.

Remark: See \citet{jiang2020multiply}.

\paragraph{Recommended reading}

\citet{frangakis2002principal}  proposed the principal stratification framework. 
\citet{zhang2003estimation} derived large-sample bounds on the survivor average causal effect. 
\citet{jiang2020identification} reviewed various strategies to identify the causal effects within principal strata.
\citet{jiang2020multiply} gave a unified discussion of this strategy for observational studies and proposed multiply robust estimators for causal effects within principal strata.

\chapter{ Mediation Analysis: Natural Direct and Indirect Effects}
 \label{chapter::mediation}

With an intermediate variable $M$ between the treatment $Z$ and outcome $Y$,  the causal effects within principal strata defined by $U = \{  M(1), M(0) \}$ can assess the treatment effect heterogeneity across latent groups $U$. When $M$ is indeed on the causal pathway from $Z$ to $Y$, causal effects within some principal strata, $\tau(1,1)$ and $ \tau(0,0)$, can give information about the direct effect of $Z$ on $Y$. However, these direct effects are only for two latent groups. The causal effects within the other two principal strata, $\tau(1,0)$ and $ \tau(0,1)$, contain both direct and indirect effects. Fundamentally, principal stratification does not provide any information about the indirect effect of $Z$ on $Y$ through $M$ because it does not even assume that $M$ can be intervened.

In the above discussion, I use the notions of ``direct effect'' and ``indirect effect'' in a casual way. When $M$ lies on the pathway from $Z$ to $Y$, researchers often want to assess the extent to which the effect of $Z$ on $Y$ is through $M$ and the extent to which the effect is through other pathways. This is called {\it mediation analysis}. It is the topic of this chapter.

\section{Motivating Examples}

In mediation analysis, we have a treatment $Z$, an outcome $Y$, a mediator $M$, and some background covariates $X$. Figure \ref{fg::mediationDAG} illustrates their relationship. Below we give some concrete examples.

\begin{figure}[ht]
\centering
$$
\begin{xy}
\xymatrix{
&& X \ar[dll] \ar[d] \ar[drr] \\
Z \ar[rr] \ar@/_1.5pc/[rrrr] & & M\ar[rr] & & Y}
\end{xy}
$$
\caption{Causal diagram for mediation analysis}\label{fg::mediationDAG}
\end{figure}

\begin{example}
\citet{Vanderweele::2012genetic} conducted mediation analysis to assess the extent to which the effect of variants on chromosome 15q25.1 on lung cancer is mediated through smoking and the extent to which it operates through other causal pathways\footnote{Recall Chapter \ref{chapter::evalue} for Fisher's hypothesis of a common genetic factor that affects the smoking behavior and lung cancer simultaneously.}. The exposure levels correspond to changes from 0 to 2 C alleles, smoking intensity is measured by the square root of cigarettes per day, and the outcome is the lung cancer indicator. \citet{Vanderweele::2012genetic}'s study contained many sociodemographic covariates.  
\end{example}

\begin{example}
\citet{rudolph2018causal} studied the causal mechanism from neighborhood poverty to adolescent substance use, mediated by the school and peer environment. They used data from the National Comorbidity Survey Replication Adolescent Supplement, a nationally representative survey of U.S. adolescents conducted during 2001--2004. The treatment is the binary indicator of neighborhood disadvantage, defined as living in the lowest tertile of neighborhood socioeconomic status based on data from the 2000 U.S. Census. 
Four binary mediators are measures of school and peer environments, and six binary outcomes are measures of substance use. 
Baseline covariates included the adolescent's sex, age, race, immigration generation, family income, etc.
\end{example}

\begin{example}\label{eg::mediation-jobs}
The \ri{mediation} package in \ri{R} contains a dataset called \ri{jobs}, which is from JOBS II, a randomized field experiment that investigates the efficacy of a job training intervention on unemployed workers. 
We used this dataset in Chapter \ref{sec::cace-application}. 
The program is designed to not only increase reemployment among the unemployed but also enhance the mental health of the job seekers. It is therefore of interest to assess the indirect effect of the intervention on mental health through job search efficacy and its direct effect acting through other pathways. We will revisit this example in Chapter \ref{section::BK-jobs} later. 
\end{example}

\section{Nested Potential Outcomes}

\subsection{Natural Direct and Indirect Effects}
Below we drop the index $i$ for unit $i$ and assume all random variables are IID draws from a superpopulation. For simplicity, we focus on a binary treatment $Z$.

We first consider the hypothetical intervention on $z$ and define potential mediators and outcomes corresponding to the intervention on $z$:
$$
\{ M(z) , Y(z): z=0,1\}.
$$
We then consider hypothetical intervention on both $z$ and $m$ and define potential outcomes corresponding to the interventions on $z$ and $m$:
$$
\{ Y(z, m): z=0, 1; m \in \mathcal{M}  \},
$$
where $ \mathcal{M} $ contains all possible values of $m$. 
\citet{Robins::1992} and \citet{Pearl::2001} further consider the nested potential outcomes corresponding to intervention on $z$ and $m = M(z')  $:
$$
\{ Y(z, M_{z'}) : z = 0,1; z'=0,1   \}
$$
where I write $M(z')$ as $ M_{z'}$ to avoid excessive parentheses. The notation $Y(z, M_{z'})$ is the hypothetical outcome if the treatment were set at level $z$ and the mediator were set at its potential level $M(z')$ under treatment $z'$. Importantly, $z$ and $z'$ can be different.
With a binary treatment, we have four nested potential outcomes in total:
$$
\{ Y( 1, M_1),    Y( 1, M_0),  Y( 0, M_1),  Y( 0, M_0) \}.
$$
The nested potential outcome  $Y(1, M_1)$ is the hypothetical outcome if the treatment were set at $z=1$ and the mediator were set at what would happen under $z=1$. Similarly, $Y(0, M_0)$ is the outcome if the treatment were set at $z=0$ and the mediator were set at what would happen under $z=0$. It would be surprising if $Y(1, M_1) \neq  Y(1)$ or $Y(0, M_0) \neq  Y(0)$.
Therefore, we make the following composition assumption throughout this chapter.

\begin{assumption}
[composition]\label{assume::composition}
$Y(z, M_z) =   Y(z)$ for $z=0,1$. 
\end{assumption}

The composition assumption cannot be proved. It is indeed an assumption. Without causing philosophical debates, we can even define $Y(1)$ as  $Y(1, M_1)$, and define $ Y(0)$ as $Y(0, M_0) $.  Then by definition, Assumption \ref{assume::composition} holds.

The nested potential outcome $Y(1,M_0)$ is the hypothetical outcome if the unit received treatment $1$ but its mediator were set at its natural value $M_0$ without the treatment. Similarly, $Y(0,M_1)$ is the hypothetical outcome if the unit received control $0$ but its mediator were set at its natural value $M_1$ under the treatment. They are two cross-world counterfactual terms and are useful for defining the direct and indirect effects.

\begin{definition}
[total, direct and indirect effects]
\label{def::mediation}
Define the total effect of the treatment on the outcome as
$$
\tau = E\{ Y(1)  - Y(0) \}.
$$
Define the natural direct effect as
$$
\NDE = E\{  Y(1,M_0) - Y(0, M_0)  \}. 
$$
Define the natural indirect effect as 
$$
\NIE = E\{  Y(1, M_1) - Y(1, M_0) \} .
$$
\end{definition}

The total effect is the standard average causal effect of $Z$ on $Y$. The natural direct effect measures the effect of the treatment on the outcome if the mediator were set at the natural value $M_0$ without the intervention. The natural indirect effect measures the effect of the treatment through changing the mediator if the treatment itself were set at $z=1$.  Under the composition assumption, the natural direct and indirect effects reduce to 
$$
\NDE = E\{  Y(1,M_0) - Y(0)  \} ,\quad
\NIE = E\{  Y(1) - Y(1, M_0) \},
$$
and therefore, we can decompose the total effect as the sum of the natural direct and indirect effects.

\begin{proposition}
\label{prop::mediation-decomposition}
By Definition \ref{def::mediation} and Assumption \ref{assume::composition}, we have 
$$
\tau = \NDE + \NIE .
$$
\end{proposition}

Mathematically, we can also define the natural indirect effect as $ E\{  Y(0, M_1) - Y(0, M_0) \}$ where the treatment is fixed at $0$. However, this definition does not lead to the decomposition in Proposition \ref{prop::mediation-decomposition}. 

\begin{figure}[ht]
\centering
\includegraphics[width =  \textwidth]{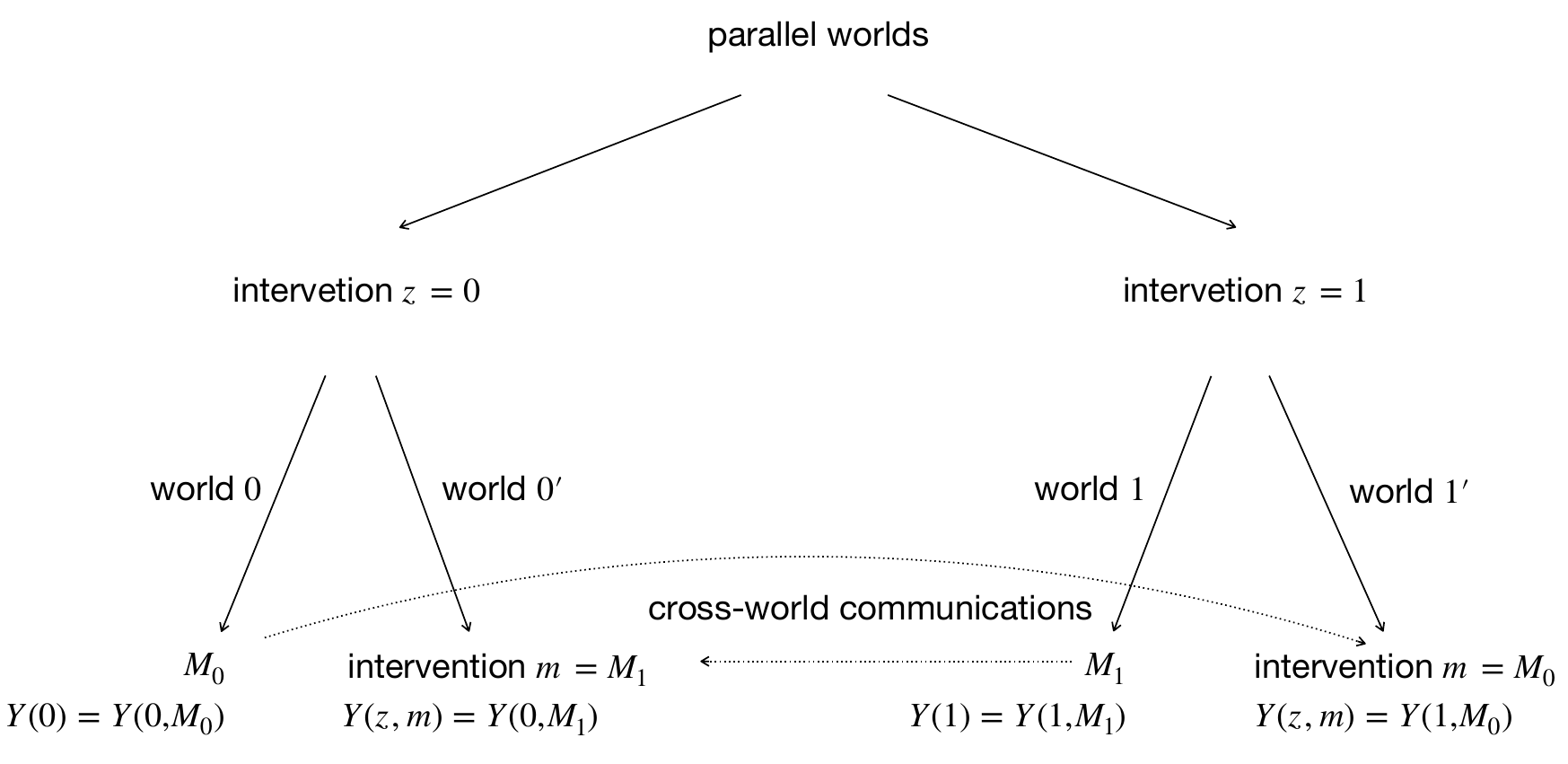}
\caption{Cross-world potential outcomes $Y(1,M_0)$ and $Y(0, M_1)$}\label{fg::crossworld}
\end{figure}

Unfortunately, the nest potential outcome $Y(1,M_0)$ is not an easy quantity to understand due to the cross-world nature of the interventions: the treatment is set at $z=1$ but the mediator is set at its natural value $M_0$ under treatment $z=0$. Clearly, these two interventions on the treatment cannot simultaneously happen in any realized experiment. To understand the cross-world potential outcome $Y(1,M_0)$, we need to imagine the existence of parallel worlds as shown in Figure \ref{fg::crossworld}. Let's focus on $Y(1,M_0)$. When the treatment is set at $z=1$, the mediator must take value $M_1$. If at the same time, we want to set the mediator at $m=M_0$, we must know the value of $M_0$ for the same unit from another experiment in the parallel world. This can be an unrealistic physical experiment because it requires that the same unit is intervened at two different levels of the treatment. Under some strong assumptions about the homogeneity of units, we may use another unit's mediator value under control as a proxy for $M_0$.

\subsection{Metaphysics or Science}
\label{sec::popper-science}

Causal inference is hard, and there is no agreement even on its mathematical notation. \citet{Robins::1992} and \citet{Pearl::2001} used the nested potential outcomes to define the natural direct and indirect effects. However, \citet{frangakis2002principal} called $Y(1,M_0)$ and $Y(0, M_1)$ {\it a priori counterfactuals} because we can not observe them in any physical experiments. In this sense, they do not exist a priori. According to \citet{popper1963conjectures}, a way to distinguish science and metaphysics is the falsifiability of the statements. That is, if a statement is not falsifiable based on any physical experiments or observations, then it is not a scientific but rather a metaphysical statement. Because we can not observe $Y(1,M_0)$ and $Y(0, M_1)$ in any experiments, we can not falsify any statements involving them except for the trivial ones (e.g., some outcomes are binary, or continuous, or bounded). Therefore, a strict Popperian statistician would view mediation analysis as metaphysics.

More strikingly, \citet{dawid2000causal} criticized the potential outcomes framework to be metaphysical, and he called Rubin's ``Science Table'' defined in Chapter \ref{sec::po-sciencetable} a ``metaphysical array.'' This is a critique on not only the a priori counterfactuals $Y(1,M_0)$ and $Y(0, M_1)$ but also the simpler potential outcomes $Y(1)$ and $Y(0)$.  \citet{dawid2000causal} argued that because we can never observe $Y(1)$ and $Y(0)$ jointly, then introducing the notation $\{ Y(1), Y(0)\}$ is a metaphysical activity. He is correct about the metaphysical nature of the joint distribution of $\pr\{ Y(1) , Y(0) \}$, but he is incorrect about the marginal distributions. Based on the observed data, we indeed can falsify some statements about the marginal distributions, although we cannot falsify any statements about the joint distribution.\footnote{
By the probability theory, given the marginal distributions of $\pr( Y(1) \leq y_1 ) $ and $\pr( Y(0) \leq y_0 ) $, we can bound the joint distribution of $\pr( Y(1) \leq y_1 , Y(0) \leq y_0 ) $ by  the Frechet--Hoeffding inequality:
\begin{eqnarray*}
&&\max\{  0,  \pr( Y(1) \leq y_1) +\pr( Y(0) \leq y_0 ) - 1  \}  \\
&\leq& 
\pr( Y(1) \leq y_1 , Y(0) \leq y_0 ) \\
&\leq& 
\min\{ \pr( Y(1) \leq y_1) , \pr( Y(0) \leq y_0 )  \}.
\end{eqnarray*}
 This is often a loose inequality.  Unfortunately, we do not have any information beyond this inequality without imposing additional assumptions. 
} Therefore, even according to \citet{popper1963conjectures}, Rubin's Science Table is not metaphysical because it has some nontrivial falsifiable implications although not all implications are falsifiable. This is the fundamental difference between $\{ Y(1), Y(0)\}$ and $\{ Y(1,M_0), Y(0, M_1) \} $.

\section{The Mediation Formula}

\citet{Pearl::2001}'s mediation formula relies on the following four assumptions. The first three essentially assume that the treatment and the mediator are both randomized conditional on observed covariates.

\begin{assumption}
\label{assume::mediation1}
There is no treatment-outcome confounding: 
$$
Z\ind Y(z,m)\mid X
$$ 
for all $z$ and $m$.
\end{assumption}

\begin{assumption}
\label{assume::mediation2}
There is no mediator-outcome confounding: 
$$
M \ind Y(z,m) \mid (X, Z)
$$ 
for all $z$ and $m$.
\end{assumption}

Assumptions \ref{assume::mediation1} and \ref{assume::mediation2} together are often called {\it sequential ignorability}. They
 are equivalent to the assumption that $(Z, M)$ are jointly randomized conditioning on $X$:
\begin{eqnarray}\label{eq::joint-randomization}
(Z, M) \ind Y(z,m)\mid X
\end{eqnarray} 
for all $z$ and $m$. I leave the proof of \eqref{eq::joint-randomization} as Problem \ref{para::sr-jr}.

\begin{assumption}
\label{assume::mediation3}
There is no treatment-mediator confounding: 
$$
Z \ind M(z) \mid X
$$ 
for all $z$. 
\end{assumption}

The last assumption is the cross-world independence. 

\begin{assumption}
\label{assume::mediation4}
There is no cross-world independence between the potential outcomes and potential mediators: 
$$
Y(z,m) \ind M(z') \mid X
$$ 
for all $z,z'$ and $m$. 
\end{assumption}

Assumptions \ref{assume::mediation1}--\ref{assume::mediation3} are very strong, but at least they hold under experiments with randomized treatment and mediator. Assumption \ref{assume::mediation4} is stronger because no physical experiment can ensure it. Because we can never observe $Y(z,m)$ and $ M(z') $ in any experiment if $z\neq z'$, Assumption \ref{assume::mediation4}  can never be validated so it is fundamentally metaphysical. 

I give an example below in which Assumptions \ref{assume::mediation1}--\ref{assume::mediation4} all hold. 

\begin{example}
\label{eg::mediation-assumptions}
Given $X$, we generate
\begin{eqnarray*}
Z &=& 1\{  g_Z(X, \varepsilon_Z)  \geq 0 \} ,\\
M(z) &=& 1\{   g_M(X, z, \varepsilon_M)  \geq 0 \} ,\\
Y(z,m) &=& g_Y(X, z, m, \varepsilon_Y ),
\end{eqnarray*}
for $z,m = 0,1$,
where 
$
\varepsilon_Z, \varepsilon_M,
\varepsilon_Y 
$
are all independent random errors. Consequently, we generate the observed values of $M$ and $Y$ from
\begin{eqnarray*}
M &=& M(Z) = 1\{   g_M(X, Z, \varepsilon_M)  \geq 0 \},\\
Y &=& Y(Z,M) = g_Y(X, Z, M, \varepsilon_Y ) . 
\end{eqnarray*}
We can verify that Assumptions \ref{assume::mediation1}--\ref{assume::mediation4}  hold under this data-generating process.
On the contrary, if we allow $  \varepsilon_M$ and $\varepsilon_Y$ to be $ \varepsilon_M(z)$ and $\varepsilon_Y(z,m)$, then Assumptions \ref{assume::mediation1}--\ref{assume::mediation4} can fail. 
See Problem \ref{para::mediation-assumptions-DGP} for more details.
\end{example}

\citet{Pearl::2001} proved the following key result for mediation analysis. 

\begin{theorem}
\label{thm::mediation}
Under Assumptions \ref{assume::mediation1}--\ref{assume::mediation4},  we have 
$$
E\{ Y(z, M_{z'})  \mid X=x \} =  \sum_m E(Y\mid Z=z, M=m, X=x) \pr(M =  m \mid Z=z', X=x)   
$$
and therefore,
\begin{eqnarray*}
&&E\{ Y(z, M_{z'})   \}  \\
&=&  \sum_x E\{ Y(z, M_{z'})  \mid X=x \} \pr(X =  x )  \\
&=&  \sum_x \sum_m E(Y\mid Z=z, M=m, X=x) \pr(M =  m \mid Z=z', X=x)   \pr(X =  x ).
\end{eqnarray*}
\end{theorem}

Theorem \ref{thm::mediation} assumes that both $M$ and $X$ are discrete. With general $M$ and $X$, the mediation formulas become
$$
E\{ Y(z, M_{z'}) \mid X=x  \} =   \int  E(Y\mid Z=z, M=m, X=x) f ( m \mid Z=z', X=x) \d m
$$
and
\begin{eqnarray*}
&&E\{ Y(z, M_{z'})   \}  \\
 &=& \int E\{ Y(z, M_{z'}) \mid X=x  \}   f  (x) \d x  \\
 &=& \int \int  E(Y\mid Z=z, M=m, X=x) f ( m \mid Z=z', X=x)     f  (x) \d m \d x   . 
\end{eqnarray*}
From Theorem \ref{thm::mediation}, the identification formulas for the means of the nested potential outcomes depend on the conditional mean of the outcome given the treatment, mediator, and covariates, as well as the conditional mean of the mediator given the treatment and covariates. We need to evaluate these two conditional means at different treatment levels if the nested potential outcome involves cross-world interventions. 

If we drop the cross-world independence assumption, we can modify the definition of the natural direct and indirect effects and the same formulas hold. See Problem \ref{problem::no-crossworld-modification} for more details.  

I give the proof below.

\begin{myproof}{Theorem}{\ref{thm::mediation}}
By the tower property, $E\{ Y(z, M_{z'})   \}  =  E[ E\{ Y(z, M_{z'})  \mid X\}  ]$, so we need only to prove the formula for $E\{ Y(z, M_{z'})  \mid X=x\}$. 
Starting with the law of total probability, we have
\begin{eqnarray*}
&& E\{ Y(z, M_{z'})  \mid X=x\}  \\
&=& 
\sum_m E\{ Y(z, M_{z'})  \mid  M_{z'} = m, X=x\} \pr(  M_{z'} = m\mid X=x) \\
&=&
\sum_m E\{ Y(z, m)  \mid  M_{z'} = m, X=x\} \pr(  M_{z'} = m\mid X=x) \\ 
&=& 
\sum_m \underbrace{ E\{ Y(z, m)  \mid    X=x\} }_{\text{Assumption }\ref{assume::mediation4} } 
\underbrace{ \pr(  M  = m\mid Z=z' , X=x) }_{\text{Assumption }\ref{assume::mediation3} } \\ 
&=&
\sum_m \underbrace{E( Y  \mid  Z=z, M=m,   X=x)}_{ \text{Assumptions }\ref{assume::mediation1}\text{ and }\ref{assume::mediation2} } 
\pr(  M  = m\mid Z=z' , X=x) .
\end{eqnarray*}
\end{myproof}

The above proof is trivial from a mathematical perspective. It illustrates the necessity of Assumptions \ref{assume::mediation1}--\ref{assume::mediation4}.

Conditional on $X=x,$ the mediation formulas for $Y(1, M_1)$ and $Y(0, M_0)$ simplify to
\begin{eqnarray*}
&& E\{  Y(1, M_1)\mid X=x \} \\ 
&=& \sum_m E( Y  \mid  Z=1, M=m,   X=x) \pr(M=m \mid Z=1, X=x)\\
&=& E( Y  \mid  Z=1 ,  X=x)
\end{eqnarray*}
and
\begin{eqnarray*}
&&E\{  Y(0, M_0)\mid X=x \}  \\
&=& \sum_m E( Y  \mid  Z=0, M=m,   X=x) \pr(M=m \mid Z=0, X=x)\\
&=& E( Y  \mid  Z=0 ,  X=x)
\end{eqnarray*}
based on the law of total probability; the mediation formula for $Y(1, M_0)$ simplifies to
\begin{eqnarray*}
&&E\{  Y(1, M_0)\mid X=x \}  \\
&=&  \sum_m E( Y  \mid  Z=1, M=m,   X=x) \pr(M=m \mid Z=0, X=x),
\end{eqnarray*}
where the conditional expectation of the outcome is given $Z=1$ but the conditional distribution of the mediator is given $Z=0$. 
This leads to the identification formulas of the natural direct and indirect effects.

\begin{corollary}
\label{corollary::mediation-nde-nie}
Under Assumptions \ref{assume::mediation1}--\ref{assume::mediation4}, 
the conditional natural direct and indirect effects are identified by
\begin{eqnarray*}
\NDE(x) &=& E\{ Y(1, M_0) - Y(0, M_0) \mid X=x \} \\
&=& \sum_m \{ E( Y  \mid  Z=1, M=m,   X=x) - E( Y  \mid  Z=0, M=m,   X=x) \} \\
&& \qquad  \times \pr(M=m \mid Z=0, X=x)
\end{eqnarray*}
and
\begin{eqnarray*}
\NIE(x) &=& E\{ Y(1, M_1) - Y(1, M_0) \mid X=x \} \\
&=& \sum_m   E( Y  \mid  Z=1, M=m,   X=x)\\
&&  \qquad  \times  \{ \pr(M=m \mid Z=1, X=x) -  \pr(M=m \mid Z=0, X=x) \};
\end{eqnarray*}
the unconditional ones can be identified by $\NDE = \sum_ x \NDE(x) \pr(X=x)$ and $\NIE = \sum_ x \NIE(x) \pr(X=x)$.
\end{corollary}

As a special case, with a binary $M$, the formula of the $\NIE$ reduces to a product form below.

\begin{corollary}
\label{corollary::mediation-binary-M}
Under Assumptions \ref{assume::mediation1}--\ref{assume::mediation4}, for a binary mediator $M$, we have 
$$
\NIE(x)  =      \tau_{Z\rightarrow M} (x) \tau_{M\rightarrow Y} (1,x)
$$
and
$
\NIE = E\{  \NIE(X)   \},
$
where
$$
\tau_{Z\rightarrow M}(x) =  \pr(M=1 \mid Z=1, X=x) -  \pr(M=1 \mid Z=0, X=x) . 
$$
and
$$
\tau_{M\rightarrow Y} (z,x) = E( Y  \mid  Z=z, M=1,   X=x) - E( Y  \mid  Z=z, M=0,   X=x)
$$
\end{corollary}

I leave the proof of Corollary \ref{corollary::mediation-binary-M} as Problem \ref{para::nie-binary-m}. 
Corollary \ref{corollary::mediation-binary-M} gives a simple formula in the case of a binary $M$. 
With randomized $Z$ conditional on $X$, we can view $\tau_{Z\rightarrow M}(x) $ as the conditional average causal effect of $Z$ on $M$. With randomized $M$ conditional on $(X,Z)$, we can view $\tau_{M\rightarrow Y} (z,x) $ as the conditional average causal effect of $M$ on $Y$. The conditional natural indirect effect equals their product. This is coherent with our intuition that the indirect effect acts from $Z$ to $M$ and then from $M$ to $Y$.

\section{The Mediation Formula Under Linear Models}

Theorem \ref{thm::mediation} gives the nonparametric identification formula for mediation analysis. It allows us to derive various formulas for mediation analysis under different models. I will introduce the famous Baron--Kenny method under linear models below.
\citet{vanderweele2015explanation} gives explicit formulas for the natural direct and indirect effects for many commonly used models. I  relegate the details of other models to Section \ref{chapter::mediation-problmes}. 

\subsection{The Baron--Kenny Method}

\begin{figure}[ht]
\centering
$$
\begin{xy}
\xymatrix{
&  & X \ar[ddll] \ar[dd]^{\beta_2} \ar[ddrr]^{\theta_4} \\
&&\\
Z \ar[rr]^{\beta_1} \ar@/_2.8pc/[rrrr]_{\theta_1} & & M\ar[rr]^{\theta_2} & & Y & \text{indirect effect: } \beta_1\theta_2 \\
&&&&& \text{  direct effect: } \theta_1 }
\end{xy}
$$
\caption{The Baron--Kenny method for mediation under linear models}\label{fg::mediationDAG-BK}
\end{figure}

The Baron--Kenny method assumes the following linear models for the mediator and outcome given the treatment and covariates.

\begin{assumption}
[linear models for the Baron--Kenny method]
\label{assume::mediation-bk-model}
Both the mediator and outcome follow linear models:
$$
\left\{\begin{array}{rcl}
E(M \mid Z, X) &=&  \beta_0 + \beta_1 Z + \beta_2\tran X, \\
E(Y\mid Z,M,X) &=& \theta_0 + \theta_1 Z + \theta_2 M + \theta_4\tran X.
\end{array}\right.
$$
\end{assumption}

Under these linear models,  the formulas for the natural direct and indirect effects simplify to functions of the coefficients. 

\begin{corollary}
[Baron--Kenny formulas for mediation]
\label{corollary::bk-formulas}
Under Assumptions \ref{assume::mediation1}--\ref{assume::mediation4} and \ref{assume::mediation-bk-model},  we have 
$$
\NDE = \theta_1
,\quad 
\NIE = \theta_2 \beta_1 .
$$
\end{corollary}

The formulas in Corollary \ref{corollary::bk-formulas} are intuitive based on Figure \ref{fg::mediationDAG-BK}. The direct effect equals the coefficient on the path $Z\rightarrow Y$, and the indirect effect equals the product of the coefficients on the path $Z\rightarrow M\rightarrow Y$. I give the proof below based on Theorem \ref{thm::mediation}.

\begin{myproof}{Corollary}{\ref{corollary::bk-formulas}}
The conditional NDE equals 
$$
\NDE(x)  
= \sum_m  \theta_1  \pr(M=m \mid Z=0, X=x) = \theta_1
$$
and the conditional NIE equals 
\begin{eqnarray*}
\NIE(x) &=&  \sum_m  (\theta_0 + \theta_1  + \theta_2 m + \theta_4\tran x )\\
&&  \qquad  \times  \{ \pr(M=m \mid Z=1, X=x) -  \pr(M=m \mid Z=0, X=x) \} \\
&=& \theta_2 \{ E(M=m \mid Z=1, X=x) -  E(M=m \mid Z=0, X=x) \} \\
&=&  \theta_2 \beta_1 ,
\end{eqnarray*}
which do not depend on $x$. Therefore, they are also the formulas for the unconditional natural direct and indirect effects.
\end{myproof}

If we obtain OLS estimators of these coefficients, we can estimate the direct and indirect effects by
$$
\hat{\NDE} = \hat{\theta}_1,\qquad
\hat{\NIE} = \hat{\theta}_2 \hat{\beta}_1,
$$
which is called the Baron--Kenny method \citep{Judd::1981, baron1986moderator} although it had several antecedents \citep[e.g.,][]{hyman1955survey, alwin1975decomposition, Judd::1981, sobel1982asymptotic}.

Standard software packages report the standard error of $\hat{\NDE}$ from OLS. \citet{sobel1982asymptotic, sobel1986some} used the delta method to obtain the standard error of $\hat{\NIE}$. 
Based on the formula in Example \ref{example::asymptotic-normal-product},  the asymptotic variance of $\hat{\theta}_2  \hat{\beta}_1 $ equals 
$
\var(\hat{\theta}_2)  \beta_1^2 + \theta_2^2 \var(\hat{\beta}_1) .
$
So the estimated variance is
$$
\hat{\var}(\hat{\theta}_2)  \hat{\beta}_1^2 + \hat{\theta}_2^2 \hat{\var}(\hat{\beta}_1) .
$$
Testing the null hypothesis of $\NIE$ based on $\hat{\theta}_2 \hat{\beta}_1$ and the estimated variance above is called {\it Sobel's test} in the literature of mediation analysis.

\subsection{An Example}\label{section::BK-jobs}

We can easily implement the Baron--Kenny method via the following code. 

 \begin{lstlisting}
library("car")
BKmediation = function(Z, M, Y, X)
{
  ## two regressions and coefficients
  mediator.reg   = lm(M ~ Z + X)
  mediator.Zcoef = mediator.reg$coef[2]
  mediator.Zse   = sqrt(hccm(mediator.reg)[2, 2])
  
  outcome.reg    = lm(Y ~ Z + M + X)
  outcome.Zcoef  = outcome.reg$coef[2]
  outcome.Zse    = sqrt(hccm(outcome.reg)[2, 2])
  outcome.Mcoef  = outcome.reg$coef[3]
  outcome.Mse    = sqrt(hccm(outcome.reg)[3, 3])
  
  ## Baron-Kenny point estimates
  NDE = outcome.Zcoef
  NIE = outcome.Mcoef*mediator.Zcoef
  
  ## Sobel's variance estimate based the delta method
  NDE.se = outcome.Zse
  NIE.se = sqrt(outcome.Mse^2*mediator.Zcoef^2 + 
                  outcome.Mcoef^2*mediator.Zse^2)
  
  res = matrix(c(NDE, NIE, 
                 NDE.se, NIE.se,
                 NDE/NDE.se, NIE/NIE.se), 
               2, 3)
  rownames(res) = c("NDE", "NIE")
  colnames(res) = c("est", "se", "t")
  
  res
}
 \end{lstlisting}

 Revisiting Example  \ref{eg::mediation-jobs}, we obtain the following estimates for the direct and indirect effects:
 
 \begin{lstlisting}
> jobsdata = read.csv("jobsdata.csv")
> Z = jobsdata$treat
> M = jobsdata$job_seek
> Y = jobsdata$depress2
> getX     = lm(treat ~ econ_hard + depress1 + 
+                 sex + age + occp + marital + 
+                 nonwhite + educ + income,
+               data = jobsdata)
> X = model.matrix(getX)[, -1]
> res = BKmediation(Z, M, Y, X)
> round(res, 3)
       est    se      t
NDE -0.037 0.042 -0.885
NIE -0.014 0.009 -1.528
\end{lstlisting}

Both the estimates for the direct and indirect effects are negative although they are insignificant.

\section{Sensitivity analysis}

Mediation analysis relies on strong and untestable assumptions. One crucial assumption is that there is no unmeasured confounding among the treatment, mediator, and outcome. Various sensitivity analysis methods appeared in the literature. In particular, \citet{ding2016sharp} proposed Cornfield-type sensitivity bounds, and \citet{zhang2022interpretable} proposed a sensitivity analysis method tailored to the Baron--Kenny method based on linear structural equation models. These are beyond the scope of this book.

\section{Homework problems}\label{chapter::mediation-problmes}

\paragraph{\citet{Robins::1992}'s clarification of the units based on nested potential outcomes}\label{hw::robins1992-latentgroups}

Assume $(Z, M, Y)$ are all binary. Based on the joint value of 
$$
M(1), M(0),  Y(1, M_1), Y(1, M_0), Y(0, M_1), Y(0, M_0) ,
$$
how many types of units do we have?

If we assume monotonicity of $Z$ on $M$, that is,
$$
M(1) \geq  M(0),
$$
then how many types of units do we have?

If we further assume monotonicity of $Z$ and $M$ on $Y$, that is,
$$
Y(1, m) \geq Y(0, m),  \quad Y(z, 1) \geq Y(z, 0)
$$
for all $z$ and $m$, then how many types of units do we have?

\paragraph{Sequential randomization and joint randomization}\label{para::sr-jr}

Show \eqref{eq::joint-randomization} is equivalent to Assumptions \ref{assume::mediation1} and \ref{assume::mediation2}.

\paragraph{Verifying the assumptions for mediation analysis}\label{para::mediation-assumptions-DGP}

Show that Assumptions \ref{assume::mediation1}--\ref{assume::mediation4} hold under the data generating process in Example \ref{eg::mediation-assumptions}. 
Also show that if we allow $  \varepsilon_M$ and $\varepsilon_Y$ to be $ \varepsilon_M(z)$ and $\varepsilon_Y(z,m)$, then Assumptions \ref{assume::mediation1}--\ref{assume::mediation4} can fail. 
 
 \paragraph{Another set of assumptions for the mediation formula}\label{para::alternative-si}

\citet{Imai::2010} invoked the following set of assumptions to derive the mediation formula. 

\begin{assumption}
\label{assume::mediation-imai}
We have 
$$
\{Y(z, m), M(z') \} \ind  Z\mid X
$$
and
$$
Y(z, m) \ind M(z' ) \mid (Z=z' , X)
$$
for all $z, z', m$.
\end{assumption}

\begin{theorem}
\label{thm::mediation-imai}
Under Assumption \ref{assume::mediation-imai}, the mediation formula in Theorem \ref{thm::mediation} holds. 
\end{theorem}

Prove Theorem \ref{thm::mediation-imai}.

\paragraph{Difference method and product method are identical}\label{hw::difference=product}

In the main text, we first obtain the OLS fits
$$
\left\{\begin{array}{rcl}
\hat M_i &=&  \hat\beta_0 + \hat\beta_1 Z_i + \hat\beta_2\tran X_i, \\
\hat Y_i &=&\hat \theta_0 + \hat\theta_1 Z_i +\hat \theta_2 M_i + \hat\theta_4\tran X_i.
\end{array}\right.
$$
The estimate for the NIE is the product $\hat \theta_2 \hat\beta_1 $, which is called the product method. 

Alternatively, we can first obtain OLS fits:
$$
\left\{\begin{array}{rcl}
\hat Y_i &=&  \hat\alpha_0 + \hat\alpha_1 Z_i + \hat\alpha_1 \tran X_i, \\
\hat Y_i &=&\hat \theta_0 + \hat\theta_1 Z_i +\hat \theta_2 M_i + \hat\theta_4\tran X_i.
\end{array}\right.
$$
Another estimate for the NIE is the difference $\hat\alpha_1 -  \hat\theta_1$, which is called the difference method. 

Show that $\hat\alpha_1 -  \hat\theta_1 = \hat \theta_2 \hat\beta_1 .$

Remark: Recall Cochran's formula in Problem \ref{hw::cochran+ovb}.

\paragraph{Natural indirect effect with a binary mediator}\label{para::nie-binary-m}

Prove Corollary \ref{corollary::mediation-binary-M}.

\paragraph{With treatment-outcome interaction on the outcome}\label{para::interaction}
\citet{vanderweele2015explanation} suggested using the following linear models:
$$
\left\{\begin{array}{rcl}
E(M \mid Z, X) &=&  \beta_0 + \beta_1 Z + \beta_2\tran X, \\
E(Y\mid Z,M,X) &=& \theta_0 + \theta_1 Z + \theta_2 M + \theta_3 ZM + \theta_4\tran X,
\end{array}\right.
$$
where the outcome model has the interaction term between the treatment and the mediator.

Under the above linear models,  show that 
$$
\NDE = \theta_1 + \theta_3\{ \beta_0 + \beta_2\tran E(X)\}, \qquad 
\NIE = (\theta_2 + \theta_3 ) \beta_1. 
$$
How do we estimate $\NDE$ and $\NIE$ with IID data?

Remark:
Consider the simple case with a binary $Z$ and binary $M$.
Under the linear models, the average causal effect of $Z$ of $M$ equals $\beta_1$, and the average causal effect of $M$ on $Y$ equals $\theta_2 + \theta_3 E(Z)$. Therefore, it is possible that both of these effects are positive, but the natural indirect effect is negative. For instance:
$$
\beta_1 = 1,\quad \theta_2 = 1, \quad  \theta_3= -1.5, \quad E(Z) = 0.5.
$$ 
This is somewhat paradoxical and can be called the {\it mediator paradox}. \citet{chen2007criteria} reported a related {\it surrogate endpoint paradox} or {\it intermediate variable paradox}.

\paragraph{Mediation analysis with continuous mediator and binary outcome}\label{para::continM-binaryY}

Consider the following Normal linear model for the mediator and logistic model for the binary outcome:
$$
\left\{\begin{array}{rcl}
M \mid Z, X &\sim &  \textsc{N}(\beta_0 + \beta_1 Z + \beta_2\tran X, \sigma_M^2), \\
\logit \{ \pr(Y = 1 \mid Z,M,X ) \} &=& \theta_0 + \theta_1 Z + \theta_2 M +   \theta_4\tran X,
\end{array}\right.
$$
where $\logit (w)  = \log\{ w/(1-w) \} $ with inverse $\expit(w) = (1 + e^{-w})^{-1}.$
Express $\NDE$ and $\NIE$ in terms of the model parameters and the distribution of $X$. How do we estimate $\NDE$ and $\NIE$ with IID data?

\paragraph{Mediation analysis with binary mediator and continuous outcome}\label{para::logit-binary-m}
Consider the following logistic model for the binary mediator and linear model for the outcome:
$$
\left\{\begin{array}{rcl}
\logit\{ \pr(M  = 1\mid Z, X) \}&=&  \beta_0 + \beta_1 Z + \beta_2\tran X, \\
E(Y\mid Z,M,X) &=& \theta_0 + \theta_1 Z + \theta_2 M +   \theta_4\tran X . 
\end{array}\right.
$$

Under these models, show that 
$$
\NDE = \theta_1  ,\qquad
\NIE = \theta_2 E \left\{  \expit(\beta_0 + \beta_1  + \beta_2\tran X)
-  \expit(\beta_0   + \beta_2\tran X)  \right\}. 
$$
How do we estimate $\NDE$ and $\NIE$ with IID data?

\paragraph{Mediation analysis with binary mediator and outcome}
\label{para::mediation-binaryMY}

Consider the following logistic models for the binary mediator and outcome:
$$
\left\{\begin{array}{rcl}
\logit\{ \pr(M  = 1\mid Z, X) \}&=&  \beta_0 + \beta_1 Z + \beta_2\tran X, \\
\logit \{ \pr(Y = 1 \mid Z,M,X ) \} &=& \theta_0 + \theta_1 Z + \theta_2 M +   \theta_4\tran X.
\end{array}\right.
$$
Express $\NDE$ and $\NIE$ in terms of the model parameters and the distribution of $X$. How do we estimate $\NDE$ and $\NIE$ with IID data?

\paragraph{Modify the definitions to drop the cross-world independence}\label{problem::no-crossworld-modification}

Define 
$$
Y(z, F_{M_{z'}\mid X}) = \int Y(z,m) f(M_{z'} = m\mid X) \d m
$$
as the potential outcome under treatment $z$ and a random draw from the distribution of $M_{z'}\mid X$. 
With a discrete $M$, the definition simplifies to
$$
Y(z, F_{M_{z'}\mid X}) =\sum_m Y(z,m) \pr(M_{z'} = m\mid X).
$$
The key difference between $Y(z,M_{z'})$ and $Y(z, F_{M_{z'}\mid X}) $ is that $M_{z'}$ is the potential mediator for the same unit whereas $F_{M_{z'}\mid X}$ is a random draw from the conditional distribution of the potential mediator in the whole population. 
Define the natural direct and indirect effects as
$$
\NDE = E\{Y(1, F_{M_{0}\mid X})  - Y(0, F_{M_{0}\mid X})   \},\quad
\NIE  =  E\{Y(1, F_{M_{1}\mid X})  - Y(1, F_{M_{0}\mid X})   \}.
$$

Show that under Assumptions \ref{assume::mediation1}--\ref{assume::mediation3}, the identification formulas for $\NDE$ and $\NIE$ remain the same as in the main text.

Remark: Modifying the definitions of the nested potential outcomes allows us to relax the strong cross-world independence assumption but weakens the interpretation of the natural direct and indirect effects. See \citet{vanderweele2015explanation} for more discussion and \citet{vanderweele2017mediation} for an application to a more complex setting with time-varying treatment and mediator.

\paragraph{Connections between principal stratification and mediation analysis}

\citet{vanderweele2008simple} and
\citet{forastiere2018principal} reviewed and compared principal stratification and mediation analysis.

\chapter{Controlled Direct Effect}
 \label{chapter::cde}

The formulation of mediation analysis in Chapter \ref{chapter::mediation} relies on the nested potential outcomes, and fundamentally, some nested potential outcomes are not observable in any physical experiments. If we stick to the Popperian philosophy of science reviewed in Chapter  \ref{sec::popper-science}, we should only define causal parameters in terms of quantities that are observable under some experiments. This chapter discusses an alternative view of causal inference with an intermediate variable. In this view, we only define the direct effect but can not define the indirect effect.

\section{Definition of the controlled direct effect}

We view $Z$ and $M$ as two treatment factors that can be manipulated, and define potential outcomes $Y(z,m)$ for $z=0,1$ and $m\in \mathcal{M}$. Based on these potential outcomes, we can define the {\it controlled direct effect} (CDE) below.

\begin{definition}
[CDE]
Define
$$
\CDE (m)  = E\{  Y(1,m)  - Y(0,m)\}.
$$
\end{definition}

By definition, $\CDE (m) $ is the average causal effect of the treatment if the intermediate variable is fixed at $m$. 
The parameter $\CDE(m)$ can capture the direct effect of the treatment holding the mediator at $m$. However, this formulation cannot capture the indirect effect. In particular, the parameter
$
E\{  Y(z,1) - Y(z,0)\}
$
only measures the effect of the mediator on the outcome holding the treatment at $z.$ This is not a meaningful definition of the indirect effect.

\section{Identification and estimation of the controlled direct effect}

To identify $\CDE(m)$, we need the following assumption, which requires that $Z$ and $M$ are jointly randomized given $X$.

\begin{assumption}\label{assume::si-cde}
Sequential ignorability requires
$$
Z\ind Y(z,m) \mid X, \quad M \ind  Y(z,m) \mid (Z,X)
$$
or, equivalently (see Problem \ref{para::sr-jr}),
$$
(Z,M) \ind  Y(z,m) \mid X.
$$
\end{assumption}

I will focus on the case with a binary $Z$ and $M$. 
Mathematically, we can just view this problem as an observational study with four treatment levels 
$$
(z,m) \in \{  (0,0), (0,1), (1,0), (1,1) \} .
$$
The following theorem extends the results for observational studies with a binary treatment, identifying
$$
\mu_{zm} = E\{  Y(z,m) \}
$$
based on outcome regression, IPW, and doubly robust estimation.

Define 
$$ 
\mu_{zm}(x) = E(Y\mid Z=z,M=m, X=x) 
$$
 as the outcome mean conditional on the treatment, mediator, and covariates. Define 
\begin{eqnarray*}
e_{zm}(x)  &=&  \pr(Z=z,M=m\mid X=x)  \\
&=& \pr(Z=z\mid X=x) \pr( M=m\mid Z=z, X=x)
\end{eqnarray*} 
as the probability of the joint value of $Z$ and $M$ conditional on the covariates. 

\begin{theorem}
\label{thm::cde-identification}
Under Assumption \ref{assume::si-cde}, we have
$$
\mu_{zm} = E\{  \mu_{zm}(X) \}
= E \left\{   \frac{I(Z=z, M=m) Y}{  e_{zm}(X)  }  \right\} .
$$
Moreover, based on the working models $e_{zm}(X, \alpha) $ and $\mu_{zm}(X, \beta) $ for $e_{zm}(X) $ and $\mu_{zm}(X) $, respectively, we have the doubly robust   formula 
$$
\mu_{zm}^{\textup{dr}}  = E\{  \mu_{zm}(X, \beta) \} + E \left[   \frac{I(Z=z, M=m) \{ Y - \mu_{zm}(X, \beta)  \}}{  e_{zm}(X, \alpha)  }  \right],
$$
which equals $\mu_{zm} $ if either $e_{zm}(X, \alpha) = e_{zm}(X)$ or $\mu_{zm}(X, \beta)  = \mu_{zm}(X)$. 
\end{theorem}

The proof of Theorem \ref{thm::cde-identification} is similar to those for the standard unconfounded observational studies. Problem \ref{para::obs-multi-valued-Z} gives a general result. 
Based on the outcome mean model, we can obtain $\hat{\mu}_{zm}(x) $ for $\mu_{zm}(x)$. Based on the treatment model, we can obtain $\hat{e}_z(x)$ for $\pr(Z=z\mid X = x)$; based on the intermediate variable model, we can obtain $\hat{e}_m(z,x)$ for $\pr(M=m\mid Z=z, X=x)$. We can then estimate $\mu_{zm} $ by outcome regression
$$
\hat{\mu}_{zm}^\text{reg} = n^{-1} \sumn  \hat{\mu}_{zm}(X_i),
$$
by IPW 
\begin{eqnarray*}
\hat{\mu}_{zm}^\text{ht} &=& n^{-1}\sumn  \frac{  I(Z_i=z, M_i=m) Y_i  }{  \hat{e}_z(X_i)  \hat{e}_m(z,X_i)  } , \\
\hat{\mu}_{zm}^\text{haj} &=&  \sumn  \frac{  I(Z_i=z, M_i=m) Y_i  }{  \hat{e}_z(X_i)  \hat{e}_m(z,X_i)  } \Big /  \sumn  \frac{  I(Z_i=z, M_i=m)    }{  \hat{e}_z(X_i)  \hat{e}_m(z,X_i)  }, 
\end{eqnarray*}
or by augmented IPW
$$
\hat{\mu}_{zm}^\text{dr} = \hat{\mu}_{zm}^\text{reg}  +  n^{-1} \sumn  \frac{  I(Z_i=z, M_i=m) \{ Y_i - \hat{\mu}_{zm}(X_i)\}   }{  \hat{e}_z(X_i)  \hat{e}_m(z,X_i)  } .
$$
We can then estimate $\CDE(m)$ by $\hat{\mu}_{1m}^* - \hat{\mu}_{0m}^*\ (*=\text{reg}, \text{ht}, \text{haj}, \text{dr} )$  and use the bootstrap to approximate the standard error.

If we are willing to assume a linear outcome model, the controlled direct effect simplifies to the coefficient of the treatment. Example \ref{eg::cde-linear} below gives the details. 
\begin{example}
\label{eg::cde-linear}
Under Assumption \ref{assume::si-cde} and a linear outcome model,
$$
E(Y\mid Z,M,X) = \theta_0 + \theta_1 Z + \theta_2 M +   \theta_4\tran X,
$$
we can show that
$
\CDE(m) 
$
equals the coefficient $ \theta_1$, which coincides with the natural direct effect in the Baron--Kenny method. I relegate the proof to Problem \ref{problem::cde-linear}. 
\end{example}

\section{Discussion}

The formulation of the controlled direct effect does not involve nested or a priori counterfactual potential outcomes, and its identification does not require the cross-world counterfactual independence assumption. The parameter $\CDE(m)$ can capture the direct effect of the treatment holding the mediator at $m$. However, this formulation cannot capture the indirect effect.  
I summarize the causal frameworks for intermediate variables in Table \ref{table::causal-intermediate-frameworks}. 
 
\begin{table}
\centering 
\caption{Causal frameworks for intermediate variables}\label{table::causal-intermediate-frameworks}
\begin{tabular}{cccc}
\hline 
chapter & framework & direct effect & indirect effect   \\
\hline 
\ref{chapter::principal-stratification} & principal stratification & $\tau(1,1),\ \tau(0,0)$ & ?  \\
\ref{chapter::mediation} & mediation analysis & $\NDE$ & $\NIE$  \\
\ref{chapter::cde}& controlled direct effect & $\CDE(m)$ & ?  \\
\hline 
\end{tabular}
\end{table} 
 
The mediation analysis framework can decompose the total effect into natural direct and indirect effects, but it requires nested potential outcomes and cross-world independence. 
The principal stratification and controlled direct effect frameworks cannot define indirect effects but they do not involve nested potential outcomes and cross-world independence.
Moreover, the principal stratification framework does not necessarily require that $M$ lies on the causal pathway from the treatment to the outcome. However, its identification and estimation involves disentangling mixture distributions, which is a nontrivial task in statistics.

  \section{Homework problems}

\paragraph{$\CDE$ and $\NDE$}\label{problem::cde-nde}

Show that under cross-world independence 
$
Y(z,m) \ind M(z') \mid X
$
for all $z,z'$ and $m$,  the conditional controlled direct effect $\CDE(m\mid x) = E\{  Y(1,m)  - Y(0,m)\mid X=x\}$ and the conditional natural direct effect $\NDE(x) = E\{   Y(1,M_0) - Y(0, M_0) \mid X=x  \}$ have the following relationship:
$$
\NDE(x) = \sum_m \CDE(m\mid x) \pr(M_0 = m\mid X=x)
$$
for a discrete $M$. Without the cross-world independence, does this relationship still hold in general?

  \paragraph{Observational studies with a multi-valued treatment}\label{para::obs-multi-valued-Z}
  
Theorem \ref{thm::cde-identification} is a special case of the following theorem for unconfounded observational studies with multiple treatment levels \citep{imai2004causal, cattaneo2010efficient}. Below, I state the general problem and theorem.

Consider an observational study with a multi-valued treatment $Z \in \{ 1, \ldots, K \}$, covariates $X$, and outcome $Y$. Unit $i$ has $K$ potential outcomes $Y_i(1), \ldots, Y_i(K)$ corresponding to the $K$ treatment levels. In general, we can define causal effect in terms of contrasts of the potential outcomes:
$$
\tau_C  = \sum_{k=1}^K C_k   E\{  Y(k) \}
$$
where $  \sum_{k=1}^K C_k  = 0$. The canonical choice of the pairwise comparison
$$
\tau_{k,k'} = E\{   Y(k) - Y(k') \}.
$$
Therefore, the key is to identify and estimate the means of the potential outcomes $\mu_k  =  E\{  Y(k) \}$ under the ignorability and overlap assumptions below based on IID data of $(Z_i,  X_i, Y_i)_{i=1}^n$.

\begin{assumption}
\label{assume::ignorability-multi-valued-Z}
$Z\ind \{ Y (1), \ldots, Y (K) \} \mid X$ and $\pr(Z = k \mid X) > 0$ for $k=1,\ldots, K.$
\end{assumption}

Define the generalized propensity score as
$$
e_k(X) = \pr(Z = k \mid X), 
$$
and define the conditional outcome mean as
$$
\mu_k(X) = E(   Y \mid Z=k,  X )
$$
for $k=1, \ldots, K.$ We have the following theorem.

\begin{theorem}
\label{thm::treatment-multi-valued-Z}
Under Assumption \ref{assume::ignorability-multi-valued-Z}, we have
$$
\mu_{k} = E\{  \mu_{k}(X) \}
= E \left\{   \frac{I(Z = k) Y}{  e_{k}(X)  }  \right\} .
$$
Moreover, based on the working models $e_{k}(X, \alpha) $ and $\mu_{k}(X, \beta) $ for $e_{k}(X) $ and $\mu_{k}(X) $, respectively, we have the doubly robust  formula 
$$
\mu_{k}^{\textup{dr}}  = E\{  \mu_{k}(X, \beta) \} + E \left[   \frac{I(Z=k) \{ Y - \mu_{k}(X, \beta)  \}}{  e_{k}(X, \alpha)  }  \right],
$$
which equals $\mu_{k} $ if either $e_{k}(X, \alpha) = e_{k}(X)$ or $\mu_{k}(X, \beta)  = \mu_{k}(X)$. 
\end{theorem}

Prove Theorem \ref{thm::treatment-multi-valued-Z}.

Remark: Theorem \ref{thm::cde-identification} is a special case of Theorem \ref{thm::treatment-multi-valued-Z} if we view the $(Z,M)$ in Theorem \ref{thm::cde-identification} as a treatment with four levels. The $  \CDE (m) $ is a special case of $\tau_C $.

  \paragraph{$\CDE$ in the linear outcome model}\label{problem::cde-linear}
  
  Show that under Assumption \ref{assume::si-cde}, if $E(Y\mid Z, M, X ) = \theta_0 + \theta_1 Z + \theta_2 M +   \theta_4\tran X$, then
  $$
  \CDE (m) = \theta_1
  $$
  for all $m$;
  if $E(Y\mid Z, M, X ) = \theta_0 + \theta_1 Z + \theta_2 M + \theta_3 ZM +    \theta_4\tran X$, then
  $$
\CDE(m) =   \theta_1 + \theta_3 m. 
  $$

    \paragraph{$\CDE$ in the logistic outcome model}\label{problem::cde-logit}
  
  Show that for a binary outcome, under Assumption \ref{assume::si-cde}, if 
  $$
  \logit \{ \pr(Y = 1\mid Z, M, X ) \} = \theta_0 + \theta_1 Z + \theta_2 M +   \theta_4\tran X ,
  $$ 
  then
  $$
  \CDE (m) = E\{ \expit( \theta_0 + \theta_1  + \theta_2 m +   \theta_4\tran X ) - \expit( \theta_0    + \theta_2 m +   \theta_4\tran X ) \} ; 
  $$
  if 
  $$
  \logit \{ \pr(Y = 1\mid Z, M, X ) \} = \theta_0 + \theta_1 Z + \theta_2 M + \theta_3 ZM +    \theta_4\tran X,
  $$ 
  then
  $$
\CDE(m) =  E\{ \expit( \theta_0 + \theta_1  + \theta_2 m  +\theta_3 m +   \theta_4\tran X ) - \expit( \theta_0    + \theta_2 m +   \theta_4\tran X ) \}.
  $$

  \paragraph{Recommended reading} 

\citet{nguyen2020clarifying} provided a friendly review of of the topics in Chapters \ref{chapter::mediation} and \ref{chapter::cde}.

\chapter{Time-Varying Treatment and Confounding}
 \label{chapter::timevarying}

Studies with time-varying treatments are common in biomedical and social sciences. James Robins championed the research in biostatistics. 
A classic example is that HIV patients may take azidothymidine, an antiretroviral medication, on and off over time \citep{robins2000marginal, hernan2000marginal}. Similar problems also exist in other fields.  In education, a classic example is that students may receive different types of instructions over time \citep{hong2008causal}. In political science, a classic example is that candidates continuously recalibrate their campaign strategy based on time-varying polls and opponent actions \citep{blackwell2013framework}.

Causal inference with time-varying treatments is not a simple extension of causal inference with a treatment at a single time point. The main challenge is time-varying confounding. Even if we assume all time-varying confounders are observed, we still face statistical challenges in adjusting for those confounders. On the one hand, we should stratify on these confounders to adjust for confounding; on the other hand, stratifying on post-treatment variables will cause bias. Due to these two conflicting goals, causal inference with time-varying treatments and confounding requires more sophisticated statistical methods. It is the main topic of this chapter.

To minimize the notational burden, I will use the setting with treatments at two time points to convey the most important ideas. Extensions to treatments at multiple time points can be conceptually straightforward although technical complexities will arise in finite samples. I will discuss the complexities and relegate general results to Problems \ref{problem::g-formula-multiple-time}--\ref{problem::snm-multiple-time}.

\section{Basic setup and sequential ignorability}

Start with treatments at two time points. The temporal order (not the causal diagram) of the variables with two time points is below:
$$
X_0 \rightarrow Z_1 \rightarrow X_1 \rightarrow Z_2 \rightarrow Y 
$$
where 
\begin{itemize}
\item
$X_0$ denotes the baseline pre-treatment covariates;
\item
$Z_1$ denotes the treatment at time point 1;
\item
$X_1$ denotes the time-varying covariates between the treatments at time points 1 and 2;
\item
$Z_2$ denotes the treatment at time point 2;
\item
$Y$ denotes the outcome. 
\end{itemize}

With binary treatments $(Z_1, Z_2 )$, each unit has four potential outcomes 
$$
Y(z_1, z_2)  \text{ for } z_1, z_2  = 0, 1.
$$ 
The observed outcome equals 
$$
Y = Y(Z_1,Z_2) = \sum_{z_1=0,1} \sum_{z_2=0,1}  1(Z_1=z_1) 1(Z_2=z_2) Y(z_1, z_2).
$$
I will focus on the canonical setting with sequential ignorability, that is, the treatments are sequentially randomized given the observed history.

\begin{assumption}
[sequential ignorability]
\label{assume::sr-time-varying} 
(1)
$Z_1$ is randomized given $X_0$:
$$
Z_1 \ind Y(z_1, z_2) \mid X_0 \text{ for } z_1, z_2  = 0, 1.
$$

(2)
$Z_2$ is randomized given $( Z_1, X_1, X_0)$:
$$
Z_2\ind Y(z_1, z_2) \mid (Z_1, X_1, X_0) \text{ for } z_1, z_2  = 0, 1.
$$
\end{assumption}

 Figure \ref{fig::si-simple} is a simple causal diagram corresponding to Assumption  \ref{assume::sr-time-varying}, which does not contain any unmeasured confounding. 
 
\begin{figure}
\bigskip 
$$
 \xymatrix{
 Z_1 \ar[r]  \ar@/^1.3pc/[rr]  \ar@/^2.0pc/[rrr] & X_1  \ar@/^1pc/[rr]\ar[r]  & Z_2  \ar[r] & Y
 }
 $$
 \caption{Assumption  \ref{assume::sr-time-varying} holds without unmeasured confounding $U$ between $X_1$ and $Y$. The causal diagram conditions on the pre-treatment covariates $X_0$.}\label{fig::si-simple}
\end{figure}

  Figure \ref{fig::si-complex} is a more complex causal diagram corresponding to Assumption  \ref{assume::sr-time-varying}. Sequential ignorability rules out only the confounding between the treatment $( Z_1, Z_2) $ and the outcome $Y$, but allows for unmeasured confounding between the time-varying covariate $X_1$ and the outcome $Y$.
  The possible existence of $U$ causes many subtle issues even under sequential ignorability. 

\begin{figure}
\bigskip 
$$
 \xymatrix{
 Z_1 \ar[r] \ar@/^1.3pc/[rr] \ar@/^2.0pc/[rrr] & X_1  \ar@/^1pc/[rr]\ar[r]  & Z_2  \ar[r] & Y \\
                                             &                                             &U\ar[lu]\ar[ru]
 }
 $$
 \caption{Assumption  \ref{assume::sr-time-varying} holds with unmeasured confounding between $X_1$ and $Y$. The causal diagram conditions on the pre-treatment covariates $X_0$.}\label{fig::si-complex}
\end{figure}

\section{g-formula and outcome modeling}
\label{section:si-outcome-regression}

Recall the outcome-based identification formula with a treatment at a single time point:
$$
E\{  Y(z) \}  = E\{  E(Y\mid Z=z, X) \} .
$$
With discrete $X$, it reduces to
$$
E\{  Y(z) \}  = \sum_x E(Y\mid Z=z, X=x) \pr(X=x) ;
$$
with continuous $X$, it reduces to
$$
E\{  Y(z) \}  =  \int E(Y\mid Z=z, X=x) f (x) \text{d} x.
$$
The following result extends it to the setting with treatments at two time points.

 \begin{theorem}
\label{thm::sr-outcome}
Under Assumption \ref{assume::sr-time-varying}, 
\begin{equation}\label{eq::identification-si-recursive}
E\{ Y(z_1, z_2) \} 
=  E\Big [   E\{    E(Y\mid  z_2, z_1, X_1, X_0)  \mid z_1, X_0 \}  \Big ].
\end{equation}
\end{theorem}

In Theorem \ref{thm::sr-outcome}, I simplify the notation `$Z_2=z_2$'' to ``$z_2$''. To void complex formulas in this chapter, I will use the lowercase letter to represent the event that the random variable takes the corresponding value.  
With discrete $X_0$ and $X_1$, the identification formula \eqref{eq::identification-si-recursive} reduces to
\begin{equation}\label{eq::identification-si-g-formula-1}
E\{ Y(z_1, z_2) \}  = \sum_{x_0} \sum_{x_1} E( Y \mid  z_2, z_1, x_1,  x_0) \pr(x_1\mid    z_1, x_0) \pr(x_0);
\end{equation}
with continuous $X_0$ and $X_1$, the identification formula \eqref{eq::identification-si-recursive} reduces to
\begin{equation}\label{eq::identification-si-g-formula-2}
E\{ Y(z_1, z_2) \}  =  \int \int E( Y \mid  z_2,  z_1, x_1,  x_0) f(x_1\mid    z_1, x_0) f(x_0) \text{d}x_1 \text{d}x_0.
\end{equation}
Compare \eqref{eq::identification-si-g-formula-1} with the formula based on the law of total probability to gain more insights:
\begin{eqnarray}
E(Y)  &=& \sum_{x_0} \sum_{z_1} \sum_{x_1} \sum_{z_2} E( Y \mid  z_2, z_1, x_1,  x_0)  \nonumber   \\
&& \qquad   \pr(z_2 \mid  z_1, x_1,  x_0)
\pr(x_1\mid    z_1, x_0)  \pr(z_1\mid x_0)  \pr(x_0). \label{eq::identification-law-total-prob}
\end{eqnarray}
Erasing the probabilities of $Z_2$ and $Z_1$ in \eqref{eq::identification-law-total-prob}, we can obtain the formula \eqref{eq::identification-si-g-formula-2}. This is intuitive because the potential outcome $Y(z_1, z_2)$ has the meaning of fixing $Z_1$ and $Z_2$ at $z_1$ and $z_2$, respectively.

Robins called \eqref{eq::identification-si-g-formula-1} and \eqref{eq::identification-si-g-formula-2} the g-formulas. 
Now I will prove Theorem  \ref{thm::sr-outcome}.

\begin{myproof}{Theorem}{\ref{thm::sr-outcome}}
By the tower property, 
$$
E\{ Y(z_1, z_2) \}  = E \Big[  E\{ Y(z_1, z_2) \mid X_0 \}  \Big ],
$$
so I will focus on $E\{ Y(z_1, z_2) \mid X_0 \}$. 
By Assumption \ref{assume::sr-time-varying}(1) and the tower property,
\begin{eqnarray*}
E\{ Y(z_1, z_2) \mid X_0 \} &=& E\{ Y(z_1, z_2) \mid  z_1, X_0 \} \\
&=& E\Big [ E\{ Y(z_1, z_2) \mid   z_1, X_1, X_0 \} \mid   z_1, X_0  \Big].
\end{eqnarray*}
By Assumption \ref{assume::sr-time-varying}(2),
\begin{eqnarray*}
E\{ Y(z_1, z_2) \mid X_0 \}  
&=& E\Big [ E\{ Y(z_1, z_2) \mid z_2,   z_1,X_1, X_0 \} \mid  z_1, X_0  \Big] \\
&=& E\Big [ E\{ Y \mid z_2, z_1, X_1, X_0 \} \mid   z_1, X_0  \Big] .
\end{eqnarray*}
The formula \eqref{eq::identification-si-recursive} follows. 
\end{myproof}

\subsection{Plug-in estimation based on outcome modeling}
\label{section::si-plug-in-outcome}

The g-formulas \eqref{eq::identification-si-g-formula-1} and \eqref{eq::identification-si-g-formula-2} suggest that to estimate the means of the potential outcomes, we need to model $E(Y\mid z_2,z_1, x_1, x_0)$, $\pr(x_1\mid z_1,x_0)$ and $\pr(x_0)$. With these fitted models, we can plug them into the g-formulas. 

With some special functional forms, this task can be simplified. Example \ref{example::linear-si} below gives the results under a linear model for the outcome. 

\begin{example}\label{example::linear-si}
Assume a linear outcome model
$$
E(Y\mid z_2,z_1,x_1, x_0) = \beta_0+ \beta_1 z_2 + \beta_2 z_1   + \beta_3  x_1  +\beta_4 x_0.
$$
We can verify that
\begin{eqnarray*}
E\{ Y(z_1, z_2) \}  &=& \sum_{x_0} \sum_{x_1} (\beta_0+ \beta_1 z_2 + \beta_2 z_1 + \beta_3 x_1  +\beta_4 x_0) \pr(x_1\mid   z_1, x_0) \pr(x_0)  \\
&=& \beta_0 + \beta_1 z_2 + \beta_2  z_1  +  \beta_3 \sum_{x_0} E(X_1\mid z_1, x_0) \pr(x_0)  +\beta_4 E(X_0) .
\end{eqnarray*}
Define 
\begin{equation}\label{eq::si-x1-identification}
E\{ X_1(z_1)\} = \sum_{x_0} E(X_1\mid z_1, x_0) \pr(x_0) 
\end{equation}
to simplify the formula as
$$
E\{ Y(z_1, z_2) \}  = \beta_0 + \beta_1 z_2 + \beta_2 z_1  + \beta_3 E\{ X_1(z_1)\}   +\beta_4 E(X_0) .
$$
In \eqref{eq::si-x1-identification}, I introduce the potential outcome of $X_1$ under the treatment $Z_1=z_1$ at time point 1. It is reasonable because the right-hand side of \eqref{eq::si-x1-identification} is the identification formula of $E\{ X_1(z_1)\} $ under  ignorability $X_1(z_1)\ind Z_1 \mid X_0$ for $z_1  = 0, 1$. We do not really need the potential outcome $X_1(z_1)$ and the ignorability, but it is a convenient notation and matches our previous discussion.

Define $\tau_{Z_1\rightarrow X_1} = E\{ X_1(1)  - X_1(0)\}$. We can verify that
\begin{eqnarray*}
E\{ Y(1,0)  - Y(0,0)\} &=& \beta_2 +  \beta_3 \tau_{Z_1\rightarrow X_1},\\
E\{ Y(0,1)  - Y(0,0)\} &=& \beta_1,\\
E\{ Y(1,1)  - Y(0,0)\} &=& \beta_1 +  \beta_2 + \beta_3 \tau_{Z_1\rightarrow X_1}.
\end{eqnarray*}
Therefore, we can estimate the effect of $(Z_1,Z_2)$ on $Y$ based on the above formulas by first estimating the regression coefficients $\beta$s and the average causal effect of $Z_1$ on $X_1$ using standard methods.

If we further assume a linear model for $X_1$
$$
E(X_1\mid Z_1, X_0) = \gamma_0 + \gamma_1 z_1 + \gamma_2 x_0,
$$
then $\tau_{Z_1\rightarrow X_1} = \gamma_1$ and  
\begin{eqnarray} 
E\{ Y(1,0)  - Y(0,0)\} &=& \beta_2 +  \beta_3 \gamma_1,  \label{eq::gformula-linear1}\\
E\{ Y(0,1)  - Y(0,0)\} &=& \beta_1, \label{eq::gformula-linear2}\\
E\{ Y(1,1)  - Y(0,0)\} &=& \beta_1 +  \beta_2 + \beta_3 \gamma_1. \label{eq::gformula-linear3} 
\end{eqnarray}
The formulas in \eqref{eq::gformula-linear1}--\eqref{eq::gformula-linear3} are intuitive based on Figure \ref{fig::si-simple-coefficients} with regression coefficients on the arrows. In \eqref{eq::gformula-linear1}, the effect of $Z_1$ equals the sum of the coefficients on the paths $Z_1\rightarrow Y$ and $Z_1\rightarrow X_1\rightarrow Y$; in \eqref{eq::gformula-linear2}, the effect of $Z_2$ equals the coefficient on the path $Z_2\rightarrow Y$; in \eqref{eq::gformula-linear3}, the total effect of $(Z_1,Z_2)$ equals the sum of the coefficients  on the paths $Z_1\rightarrow Y$, $Z_1\rightarrow X_1\rightarrow Y$ and $Z_2\rightarrow Y$. 
\end{example}

\begin{figure}[t]
\bigskip 
$$
 \xymatrix{
 Z_1 \ar[r]^{\gamma_1}   \ar@/^1.3pc/[rr]  \ar@/^2.0pc/[rrr]^{\beta_2} & X_1  \ar@/^1pc/[rr]^{\beta_3}  \ar[r] & Z_2  \ar[r]^{\beta_1} & Y
 }
 $$
 \caption{A linear causal diagram with coefficients on the arrows, conditional on the pre-treatment covariates $X_0$.}\label{fig::si-simple-coefficients}
\end{figure}

However, \citet{robinswasserman1997estimation} pointed out a surprising drawback of the plug-in estimation based on outcome modeling. They showed that with model misspecification in this strategy, data analyzers may falsely reject the null hypothesis of zero causal effect of $(Z_1,Z_2)$ on $Y$ even when the true effect is zero in the data-generating process. They called it the {\it g-null paradox}. Perhaps surprisingly, they show that the g-null paradox may even arise in the simple linear outcome model in Example \ref{example::linear-si}.  \citet{mcgrath2021revisiting} revisited this paradox. See Problem \ref{problem::g-null-paradox} for more details.

\subsection{Recursive estimation based on outcome modeling}
\label{section::recursive}

The plug-in estimation in Section \ref{section::si-plug-in-outcome} involves modeling the time-varying confounder $X_1$ and causes the unpleasant g-null paradox. It is not a desirable method. 

Recall the outcome regression estimator with a treatment at a single time based on $E\{  Y(z)  \} = E\{  E(Y\mid Z=z, X) \}$. We first fit a model of $Y$ on $X$ using the subset of the data with $Z=z$, and obtain the fitted values $\hat{Y}_i(z)$ for all units. We then obtain the estimator 
$$
\hat E\{  Y(z)  \}  = n^{-1} \sumn \hat{Y}_i(z).
$$

Similarly, the recursive expectation formula in \eqref{eq::identification-si-recursive} motivates a  simpler method for estimation. Start from the inner conditional expectation, denoted by 
$$
\tilde Y_2(z_1, z_2) = E(Y\mid Z_2 = z_2,Z_1 = z_1, X_1,  X_0).
$$ 
We can fit a model of $Y$ on $(X_1,X_0)$ using the subset of the data with $(Z_2 = z_2, Z_1 = z_1)$, and obtain the fitted values $\hat{Y}_{2i}(z_1, z_2) $ for all units. Move on to outer conditional expectation, denoted by 
$$
\tilde{Y}_1(z_1, z_2)  =  E\{  \tilde{Y}_2(z_1, z_2)  \mid Z_1=z_1, X_0 \}. 
$$
We can fit a model of $\hat{Y}_2(z_1, z_2) $ on $X_0$ using the subset of data with $Z_1=z_1$, and obtain the fitted values $\hat{Y}_{1i}(z_1, z_2) $ for all units. The final estimator for $E\{ Y(z_1, z_2) \} $ is then
$$
\hat E\{ Y(z_1, z_2) \}  = n^{-1} \sumn \hat{Y}_{1i}(z_1, z_2) .
$$

The above recursive estimation does not involve fitting a model for $X_1$ and avoids the g-null paradox. See Problem \ref{problem::recursive-null} for a special case. However, the estimator based on recursive regression is not easy to implement because it involves modeling variables that do not correspond to the natural structure of the causal diagram, e.g., $\tilde{Y}_2(z_1, z_2)$.

\section{Inverse propensity score weighting}\label{section::si-ipw}

Recall the IPW identification formula with a treatment at a single time point:
$$
E\{  Y(z) \} = E  \left\{   \frac{ 1(Z=z) Y  }{  \pr(Z=z\mid X) }  \right\}.
$$
The following result extends it to the setting with a treatment at two time points. Define 
$$
e(z_1, X_0) = \pr(Z_1=z_1\mid X_0) 
$$
and
$$
e(z_2, Z_1, X_1, X_0) =  \pr(Z_2 = z_2 \mid Z_1, X_1,  X_0)
$$
as the propensity scores at time points 1 and 2, respectively. 

\begin{theorem}
\label{thm::sr-ipw}
Under Assumption \ref{assume::sr-time-varying}, 
\begin{equation}
E\{ Y(z_1, z_2) \} 
= E\left\{    \frac{  1(Z_1=z_1) 1(Z_2=z_2) Y  }{   e(z_1, X_0)  e(z_2,Z_1, X_1,  X_0)     }   \right\} .
\label{eq::si-ipw}
\end{equation}
\end{theorem}

Theorem \ref{thm::sr-ipw} reveals the omitted overlap assumption:
$$
0< e(z_1, X_0)   <1,\quad
0< e(z_2, Z_1, X_1,  X_0)   < 1
$$
for all $z_1$ and $z_2$. If some propensity scores are 0 or 1, then the identification formula \eqref{eq::si-ipw} blows up to infinity.

\begin{myproof}{Theorem}{\ref{thm::sr-ipw}}
Conditioning on $(Z_1, X_1, X_0)$ and using Assumption \ref{assume::sr-time-varying}(2), we can simplify the right-hand side of \eqref{eq::si-ipw} as 
\begin{eqnarray} 
&&E\left\{    \frac{  1(Z_1=z_1) 1(Z_2=z_2) Y(z_1,z_2)  }{    \pr(Z_1=z_1\mid X_0)  \pr(Z_2 = z_2 \mid Z_1, X_1, X_0)    }   \right\} \nonumber  \\
&=& E\left\{    \frac{  1(Z_1=z_1) \pr(Z_2=z_2\mid Z_1, X_1, X_0) E(Y(z_1,z_2) \mid Z_1, X_1, X_0)  }{    \pr(Z_1=z_1\mid X_0)  \pr(Z_2 = z_2 \mid Z_1, X_1, X_0)    }   \right\}  \nonumber    \\
&=& E\left\{    \frac{  1(Z_1=z_1)    }{    \pr(Z_1=z_1\mid X_0)     }  E(Y(z_1,z_2) \mid Z_1, X_1, X_0)  \right\}   \nonumber   \\
&=& E\left\{    \frac{  1(Z_1=z_1)    }{    \pr(Z_1=z_1\mid X_0)     }  Y(z_1,z_2)  \right\}, \label{eq::ipw-condition1}
\end{eqnarray}
where \eqref{eq::ipw-condition1} follows from the tower property.

Conditioning on $X_0$ and using Assumption \ref{assume::sr-time-varying}(1), we can simplify the right-hand side of \eqref{eq::ipw-condition1} as 
\begin{eqnarray*} 
&& E\left\{    \frac{  \pr(Z_1=z_1\mid X_0)    }{    \pr(Z_1=z_1\mid X_0)     }  E(Y(z_1,z_2) \mid X_0)  \right\}\\
&=& E\left\{    E(Y(z_1,z_2) \mid X_0)  \right\}\\
&=&  E\{Y(z_1,z_2)\},
\end{eqnarray*}
where, again, the last line follows from the tower property. 
\end{myproof}

 The estimator based on IPW is much simpler which only involves modeling two binary treatment indicators. First, we can fit a model of $Z_1$ on $X_0$ to obtain the fitted values $\hat{e}_{1}(z_1, X_{0i})$ and fit a model of $Z_2$ on $(Z_1, X_1, X_0)$ to obtain the fitted values $\hat{e}_{2}(z_2, Z_{1i}, X_{1i}, X_{0i})$ for all units. Then, we obtain the following IPW estimator:
$$
\hat{E}^\textup{ht}\{ Y(z_1, z_2) \}  = n^{-1} \sumn 
  \frac{  1(Z_{1i}=z_1) 1(Z_{2i}=z_2) Y_i  }{   \hat{e}_{1}(z_1, X_{0i}) \hat{e}_{2} (z_2, Z_{1i}, X_{1i}, X_{0i})   }  .
$$
Similar to the discussion in Chapter \ref{chapter::pscore-key}, the HT estimator is not invariant to the location shift of the outcome and suffers from instability in finite samples. A modified Hajek-type estimator is $\hat{E}^\textup{haj}\{ Y(z_1, z_2) \}  = \hat{E}^\textup{ht}\{ Y(z_1, z_2) \}   / \hat 1^\textup{ht}(z_1,z_2)$, where 
$$
\hat 1^\textup{ht}(z_1,z_2)  = n^{-1}  \sumn 
  \frac{  1(Z_{1i}=z_1) 1(Z_{2i}=z_2)  }{   \hat{e}_{1} (z_1, X_{0i})\hat{e}_{2} (z_2, Z_{1i}, X_{1i}, X_{0i})   }  .
$$

\section{Multiple time points}

Extending the estimation strategies in Sections \ref{section:si-outcome-regression} and \ref{section::si-ipw} is not immediate with multiple time points. Even with a binary treatment and $K$ time points, the number of treatment combinations grows exponentially with $K$ (for example, $2^5 = 32$ and $2^{10} = 1024$). Consequently, the outcome regression and IPW estimators in Sections \ref{section:si-outcome-regression} and \ref{section::si-ipw} are not feasible in finite samples because they require enough data for every combination of the treatment levels.

\subsection{Marginal structural model}

A powerful approach is based on the marginal structural model (MSM) \citep{robins2000marginal, hernan2000marginal}. 
For simplicity of notation, I will only present the MSM with $K=2$ although its main use is in the general case. 

\begin{definition}
[MSM]
\label{def::si-msm}
The marginal mean of $Y(z_1, z_2)$ equals
$$
E\{  Y(z_1, z_2) \} = f(z_1,z_2; \beta). 
$$
\end{definition}

A leading example of Definition \ref{def::si-msm} is $E\{  Y(z_1, z_2) \} = \beta_0 + \beta_1 z_1 + \beta_2 z_2$. It is also straightforward to include the baseline covariates in the model. Definition \ref{def::si-msm-with-x0} below extends Definition \ref{def::si-msm}. 

 \begin{definition}
[MSM with baseline covariates]
\label{def::si-msm-with-x0}
The mean of $Y(z_1, z_2)$ conditional on $X_0$ equals
$$
E\{  Y(z_1, z_2) \mid X_0 \} = f(z_1,z_2, X_0; \beta). 
$$
\end{definition}

A leading example of Definition \ref{def::si-msm-with-x0} is 
\begin{equation}\label{eq::linear-msm}
E\{  Y(z_1, z_2)  \mid X_0\} = \beta_0 + \beta_1 z_1 + \beta_2 z_2 + \beta_3\tran X_0 .
\end{equation}
If we observe all the potential outcomes, we can solve $\beta$ from the following minimization problem:
\begin{equation}
\label{eq::beta-ols-msm}
\beta = \arg\min_b \sum_{z_2}\sum_{z_1}  E \{  Y(z_1, z_2) -  f(z_1,z_2, X_0; b)   \}^2 .
\end{equation}
For simplicity, I focus on the least squares formulation. We can also extend the discussion to general models; see Problem \ref{hw::logistic-msm} for an example of the logistic model.

Under sequential ignorability, we can solve $\beta$ from the following minimization problem that only involves observables.

\begin{theorem}
[IPW under MSM]
\label{thm::si-msm-ipw}
Under Assumption \ref{assume::sr-time-varying} and Definition \ref{def::si-msm-with-x0}, the $\beta$ in \eqref{eq::beta-ols-msm} equals 
$$
\beta = \arg\min_b \sum_{z_2}\sum_{z_1}  E\left[   \frac{  1(Z_1=z_1) 1(Z_2=z_2)   }{   e(z_1, X_0) e(z_2, Z_1, X_1, X_0)     }      
\{ Y -  f(z_1,z_2, X_0; b)  \}^2 \right] .
$$
\end{theorem}

The proof of Theorem \ref{thm::si-msm-ipw} is similar to that of Theorem \ref{thm::sr-ipw}. I relegate it to Problem \ref{problem::ipw-msm}.

 Theorem \ref{thm::si-msm-ipw} implies a simple estimation strategy based on weighted regressions. For instance, under \eqref{eq::linear-msm}, we can fit WLS of $Y_i$ on $(1,Z_{1i}, Z_{2i}, X_{0i})$ with weights $\hat{e}_{1}^{-1}(Z_{1i}, X_{0i})  \hat{e}_{2i}^{-1}(Z_{2i}, Z_{i1}, X_{1i}, X_{0i})$.

\subsection{Structural nested model}

A key problem of IPW is that it is not applicable if the overlap assumption is violated.  To address this challenge, Robins proposed the structural nested model. Again, to simplify the presentation, I only review the version with two time points.

\begin{definition}
[structural nested model]
\label{def::snm}
The conditional effect at time point 1 is
$$
E\{  Y(z_1,0) - Y(0,0) \mid  Z_1=z_1,  X_0 \} = g_1( z_1, X_0; \beta)   
$$
for all  $z_1$, 
and the conditional effect at time point 2 is 
$$
E\{  Y(z_1,z_2) - Y(z_1,0) \mid Z_2=z_2,Z_2=z_1,  X_1,  X_0 \} = g_2(z_2, z_1,  X_1,  X_0; \beta) 
$$
for all $ z_1, z_2.$
\end{definition}


In Definition \ref{def::snm},  two logical restrictions are 
$$
g_1( 0, X_0; \beta)  = 0
$$
and
$$
g_2(0, z_1,  X_1,  X_0; \beta) =0   \text{ for all } z_1.
$$ 
 Two leading choices of Definition \ref{def::snm} are below.

\begin{example}
\label{example::snm-leading1}
Assume
$$
    \begin{cases}
g_1( z_1, X_0; \beta)  =  \beta_1 z_1, \\
g_2(z_2, z_1,  X_1,  X_0; \beta) =  (\beta_2 + \beta_3 z_1  ) z_2 .
    \end{cases}    
  $$
\end{example}

\begin{example}
\label{example::snm-leading2}
Assume
$$
    \begin{cases}
g_1( z_1, X_0; \beta) = (\beta_1 + \beta_2\tran X_0) z_1, \\
g_2(z_2, z_1,  X_1,  X_0; \beta) = (\beta_3 + \beta_4 z_1  + \beta_5\tran X_1 ) z_2.
    \end{cases}    
    $$
\end{example}

Compare Definitions \ref{def::si-msm-with-x0} and \ref{def::snm}. The structural nested model allows for adjusting for the baseline covariates as well as the time-varying covariates whereas the marginal structural model only allows for adjusting for the baseline covariates. The estimation under Definition \ref{def::snm} is more involved. A strategy is to estimate the parameter based on estimating equations. 

I first introduce two important building blocks for discussing the estimation. 
Define 
$$
U_2(\beta) = Y- g_2(Z_2, Z_1, X_1, X_0;\beta) 
$$
and
$$
U_1(\beta) = Y- g_2(Z_2, Z_1, X_1, X_0;\beta)  - g_1( Z_1, X_0;\beta) .
$$
They are not directly computable from the data because they depend on the true value of the parameter $\beta$. At the true value, they have the following properties.

\begin{lemma}
\label{lemma::snm-1}
Under Assumption \ref{assume::sr-time-varying} and Definition \ref{def::snm}, we have
\begin{eqnarray*}
E\{  U_2(\beta)  \mid Z_2, Z_1, X_1, X_0 \}  &=&  E\{  U_2(\beta)  \mid  Z_1, X_1, X_0 \}  \\
&=&  E\{   Y(Z_1,0) \mid  Z_1, X_1, X_0 \}
\end{eqnarray*}
 and
 \begin{eqnarray*}
E\{  U_1(\beta)  \mid  Z_1, X_0 \}  &=&  E\{  U_1(\beta)  \mid  X_0 \}  \\
&=&  E\{   Y(0,0) \mid   X_0 \}.
\end{eqnarray*}
\end{lemma}

Lemma \ref{lemma::snm-1} involves a subtle  notation $ Y(Z_1,0) $ because $Z_1$ is random. It should be read as $  Y(Z_1,0) = Z_1 Y(1,0) +(1-Z_1) Y(0,0)$.
Based on the definitions and Lemma \ref{lemma::snm-1}, 
$U_1(\beta)$ acts as the control potential outcome before receiving any treatment and $U_2(\beta)  $ acts as the control potential outcome after receiving the treatment at time point 1.

\begin{myproof}{Lemma}{\ref{lemma::snm-1}}
{\bf Part 1}. 
We have
\begin{eqnarray*}
&&E\{  U_2(\beta)  \mid Z_2=1, Z_1, X_1, X_0 \}  \\
&=&    E\{  Y(Z_1,1)- g_2(1, Z_1, X_1, X_0;\beta)   \mid Z_2=1, Z_1, X_1, X_0 \}  \\
&=& E\{  Y(Z_1,0)   \mid Z_2=1, Z_1, X_1, X_0 \} 
\end{eqnarray*}
and
\begin{eqnarray*}
&&E\{  U_2(\beta)  \mid Z_2=0, Z_1, X_1, X_0 \} \\
 &=&    E\{  Y(Z_1,0)- g_2(0, Z_1, X_1, X_0;\beta)   \mid Z_2=0, Z_1, X_1, X_0 \}   \\
&=& E\{  Y(Z_1,0)   \mid Z_2=0, Z_1, X_1, X_0 \}
\end{eqnarray*}
so
\begin{eqnarray*}
E\{  U_2(\beta)  \mid Z_2, Z_1, X_1, X_0 \}  &=&  E\{  Y(Z_1,0)   \mid Z_2, Z_1, X_1, X_0 \} \\
&=& E\{  Y(Z_1,0)   \mid   Z_1, X_1, X_0 \}
\end{eqnarray*}
where the last identity follows from sequential ignorability. Since the last term does not depend on $Z_2$, we also have
$$
E\{  U_2(\beta)  \mid Z_2, Z_1, X_1, X_0 \}  = E\{  U_2(\beta)  \mid  Z_1, X_1, X_0 \} . 
$$

{\bf Part 2}. 
Using the above results, we have
 \begin{eqnarray*}
&&E\{  U_1(\beta)  \mid  Z_1, X_0 \}  \\
&=&  E\{  U_2(\beta)  - g_1( Z_1, X_0;\beta)  \mid  Z_1, X_0 \} \qquad (\text{Definition } \ref{def::snm}) \\
&=& E\left[    E\{  U_2(\beta)  - g_1( Z_1, X_0;\beta)  \mid X_1,  Z_1, X_0 \}   \mid  Z_1, X_0 \right]  \qquad (\text{tower property})  \\
&=&  E\left[    E\{ Y(Z_1,0) - g_1( Z_1, X_0;\beta)  \mid X_1,  Z_1, X_0 \}   \mid  Z_1, X_0 \right]   \qquad (\text{part 1}) \\
&=&  E\{ Y(Z_1,0) - g_1( Z_1, X_0;\beta)  \mid  Z_1, X_0 \} \qquad (\text{tower property})  \\
&=& E\{ Y(0,0)  \mid  Z_1, X_0 \} \qquad (\text{Definition } \ref{def::snm}) \\
&=&  E\{ Y(0,0)  \mid  X_0 \}  \qquad (\text{sequential ignorability}).
\end{eqnarray*}
Since the last term does not depend on $Z_1$, we also have
$$
E\{  U_1(\beta)  \mid  Z_1, X_0 \}   = E\{  U_1(\beta)  \mid   X_0 \}.   
$$
\end{myproof}

With Lemma \ref{lemma::snm-1}, we can prove Theorem \ref{thm::estimating-equation-snm} below. 

\begin{theorem}
\label{thm::estimating-equation-snm}
Under Assumption \ref{assume::sr-time-varying} and Definition \ref{def::snm},
$$
E\Big[  h_2(Z_1, X_1, X_0)  \{ Z_2 - e(1, Z_1, X_1, X_0)  \}  U_2(\beta)   \Big] = 0
$$
and
$$
E\Big[    h_1(X_0)  \{  Z_1 - e(1, X_0) \}  U_1(\beta) \Big] = 0 . 
$$
for any functions $h_1$ and $h_2$, provided that the moments exist. 
\end{theorem}

\begin{myproof}{Theorem}{\ref{thm::sr-ipw}}
Use the tower property by conditioning on $(Z_2, Z_1, X_1, X_0)$ and Lemma \ref{lemma::snm-1} to obtain
\begin{eqnarray*}
&& E\left[  h_2(Z_1, X_1, X_0)  \{  Z_2 - e(1, Z_1, X_1, X_0)  \}   E\{U_2(\beta) \mid  Z_2, Z_1, X_1, X_0\}   \right] \\
&=& E\left[  h_2(Z_1, X_1, X_0)  \{  Z_2 - e(1, Z_1, X_1, X_0)  \}   E\{U_2(\beta) \mid  Z_1, X_1, X_0\}   \right] .
\end{eqnarray*}
Use the  tower property by conditioning on $( Z_1, X_1, X_0)$ to show that the last identity equals 0 because $E\{  Z_2 - e(1, Z_1, X_1, X_0) \mid  Z_1, X_1, X_0\} = 0. $
 
Similarly,  use the tower property by conditioning on $(Z_1, X_0)$ and Lemma \ref{lemma::snm-1} to obtain
 \begin{eqnarray*}
&& E\left[    h_1(X_0)   \{  Z_1 - e(1, X_0) \}  E \{U_1(  \beta) \mid Z_1, X_0\}  \right]  \\
&=& E\left[    h_1(X_0)   \{  Z_1 - e(1, X_0) \} E \{U_1(  \beta) \mid X_0\}  \right] .
 \end{eqnarray*}
Use the  tower property by conditioning on $X_0$ to show that the last identity equals 0 because $E\{  Z_1 - e(1, X_0) \mid X_0 \} = 0.$
\end{myproof}

To use Theorem \ref{thm::estimating-equation-snm}, we must specify $h_1$ and $h_2$ to ensure that there are enough equations for solving $\beta$. Example \ref{example::g-estimation-snm} below revisits Example \ref{example::snm-leading1}.

\begin{example}
\label{example::g-estimation-snm}
Under Example \ref{example::snm-leading1}, we can choose $h_1 = 1$ and $h_2 = (1,Z_1)$ to obtain
 \begin{eqnarray*}
 E\left[    \{ Z_2 - e(1, Z_1, X_1, X_0)  \}  \{Y-  (\beta_2 + \beta_3 Z_1  ) Z_2\}   \right] &=& 0,
\\
 E\left[   Z_1 \{ Z_2 - e(1, Z_1, X_1, X_0)  \}  \{Y-  (\beta_2 + \beta_3 Z_1  ) Z_2\}   \right] &=& 0,
 \\
 E\left[    \{  Z_1 - e(1, X_0) \}  \{Y- (\beta_2 + \beta_3 Z_1  ) Z_2 - \beta_1  Z_1 \}  \right] &=& 0 . 
 \end{eqnarray*}
 We can then solve for the $\beta$'s from the above linear equations; see Problem \ref{problem::snm-estimation-example}. A natural question is whether alternative choices of $(h_1, h_2)$ can lead to more efficient estimators. The answer is yes. For example, we can choose many $(h_1, h_2)$ and use the generalized method of moment \citep{hansen1982large}.  The technical details are beyond this book.  
\end{example}

\citet{naimi2017introduction} and \citet{vansteelandt2014structural} provided tutorials on the structural nested models.

\section{Homework problems}

\paragraph{g-null paradox}\label{problem::g-null-paradox}

Consider the simple causal diagram in Figure \ref{fig::si-g-null-paradox} without pre-treatment covariates $X_0$ and without the arrows from $(Z_1, Z_2)$ to $Y$. So the effect of $(Z_1,Z_2)$ on $Y$ is zero.

\begin{figure}
\bigskip 
$$
 \xymatrix{
 Z_1 \ar[r]   & X_1   \ar[r]  & Z_2  & Y \\
                                             &                                             &U\ar[lu]\ar[ru]
 }
 $$
 \caption{With unmeasured confounding between $X_1$ and $Y$. The causal diagram ignores the pre-treatment covariates $X_0$.}\label{fig::si-g-null-paradox}
\end{figure}

Revisit Example \ref{example::linear-si}. Show that the expectation $E\{ Y(z_1, z_2) \} $ does not depend on $(z_1, z_2)$ if
$$
\beta_1  = \beta_2 = 0  \text{ and }   \beta_3 = 0
$$
or 
$$
\beta_1 = \beta_2 = 0 \text{ and } E\{ X_1(z_1)\}  \text{ does not depend on } z_1.
$$
holds.

Remark: 
However, $\beta_3 = 0$ in the first condition rules out the dependence of $Y$ on $X_1$, contradicting the existence of unmeasured confounder $U$ between $X_1$ and $Y$; the independence of $E\{ X_1(z_1)\} $ on $z_1$ rules out the dependence of $X_1$ on $Z_1$, contradicting with the existence of the arrow from $Z_1$ on $X_1$. That is, if there is an unmeasured confounder $U$ between $X_1$ and $Y$ and there is an arrow from $Z_1$ on $X_1$, then the formula of $E\{ Y(z_1, z_2) \} $ in Example \ref{example::linear-si} must depend on $(z_1, z_2)$, which leads to a contradiction with the absence of arrows from $(Z_1, Z_2)$ to $Y$.

\paragraph{Recursive estimation under the null model}\label{problem::recursive-null}

Consider the recursive estimation method in \ref{section::recursive} under the causal diagram in Problem \ref{problem::g-null-paradox}. 
Show that based on linear models, the estimator converges to 0.

\paragraph{IPW under MSM}\label{problem::ipw-msm}

Prove Theorem \ref{thm::si-msm-ipw}.

\paragraph{A nonlinear example of Definition \ref{def::si-msm-with-x0}}\label{hw::logistic-msm}

Another leading example of Definition \ref{def::si-msm-with-x0} is 
\begin{equation}\label{eq::logistic-msm}
\text{logit} \left[ \pr\{  Y(z_1, z_2)  =1  \mid X_0\}  \right] = \beta_0 + \beta_1 z_1 + \beta_2 z_2 + \beta_3\tran X_0 .
\end{equation}
If we observe all potential outcomes, we can solve $\beta$ by minimizing the expectation of the negative log-likelihood function (see Chapter \ref{sec::mle-logistic} for a simpler version):
\begin{equation}
\label{eq::beta-logistic}
\beta = \arg\min_b   \sum_{z_2}\sum_{z_1}  E\left\{   \log (1 + e^{\ell })
- Y(z_1, z_2)  \ell  \right\}
\end{equation}
where $\ell = \beta_0 + \beta_1 z_1 + \beta_2 z_2 + \beta_3\tran X_0.$
Under sequential ignorability, we can solve $\beta$ from the following minimization problem that only involves observables.

\begin{theorem}
[IPW under MSM]
\label{thm::si-msm-ipw-logistic}
Under Assumption \ref{assume::sr-time-varying} and Definition \ref{def::si-msm-with-x0}, the $\beta$ in \eqref{eq::beta-logistic} equals 
$$
\beta = \arg\min_b \sum_{z_2}\sum_{z_1}  E\left[   \frac{  1(Z_1=z_1) 1(Z_2=z_2)   }{   e(z_1, X_0) e(z_2, Z_1, X_1, X_0)     }      
\left\{   \log (1 + e^{\ell })
- Y(z_1, z_2)  \ell  \right\} \right] .
$$
\end{theorem}

Prove Theorem \ref{thm::si-msm-ipw-logistic}. 

Remark: 
Theorem \ref{thm::si-msm-ipw-logistic} implies a simple estimation strategy based on weighted regressions. For instance, under \eqref{eq::logistic-msm}, we can fit weighted logistic regression of $Y_i$ on $(1,Z_{1i}, Z_{2i}, X_{0i})$ with weights $\hat{e}_{1}^{-1}(Z_{1i}, X_{0i})  \hat{e}_{2i}^{-1}(Z_{2i}, Z_{i1}, X_{1i}, X_{0i})$.

\paragraph{Structural nested model with a single time point}\label{problem::snm-single-time}

Recall the standard setting of observational studies with IID data drawn from $\{ X,Z, Y(1), Y(0)\}$. Define the propensity score as $e(X) = \pr(Z=1\mid X)$. Assume
$$
Z\ind Y(0)\mid X
$$
and the following structural nested model.

\begin{definition}
[structural nested model with a single time point]
\label{definition::snm-single}
The conditional mean of the individual effect is
$$
E\{  Y(z) - Y(0)\mid Z=z, X \} = g(z,X;\beta).
$$
\end{definition}

In Definition \ref{definition::snm-single}, a logical restriction is $g(0,X;\beta) = 0$. Prove the following results.

\begin{enumerate}
\item
We have
$$
E\{  Y - g(Z,X;\beta) \mid X, Z\} = E\{  Y - g(Z,X;\beta) \mid X\}  = E\{  Y(0) \mid X\}. 
$$

\item
We have
\begin{eqnarray}\label{eq::Estimating-equation-snm-1}
E\Big [  h(X)  \{ Z-e(X) \}   \{  Y - g(Z,X;\beta) \}  \Big ] = 0
\end{eqnarray}
for any function $h$, provided that the moment exists.
\end{enumerate}

Remark:  Equation \eqref{eq::Estimating-equation-snm-1} is the basis for parameter estimation. 
Consider a special case of Definition \ref{definition::snm-single} with $g(z,X;\beta) = \beta z$. Choose $h(X) = 1$ to obtain
$$
E\{   (Z-e(X))   (Y-  \beta Z) \}  = 0.
$$
Solve for $\beta$ to obtain
$$
\beta = \frac{   E\{ (Z-e(X)) Y \}   }{   E\{ (Z-e(X))  Z\}   }.
$$
Therefore, $\beta$ equals the coefficient of $Z$ in the TSLS of $Y$ on $Z$ with $Z-e(X)$ being the IV for $Z$. With some basic calculations, we can also show that
$$
\beta = \frac{   \cov\{ Z-e(X), Y\}   }{ \cov\{ Z-e(X) \}  }.
$$
Therefore, $\beta$ equals the coefficient of $Z-e(X)$ in the OLS of $Y$ on $Z-e(X)$, which appeared in Chapter \ref{sec::regress-on-pscore} before. 

Consider another special case of Definition \ref{definition::snm-single} with $g(z,X;\beta) = (\beta_0 + \beta_1\tran X) z$. Choose $h(X) = (1,X)$ to obtain
$$
E\left\{   \begin{pmatrix}
Z-e(X)\\
(Z-e(X)) X
\end{pmatrix}  
(Y-  \beta_0 Z - \beta_1\tran XZ) \right\}  = 0.
$$
That is, $(\beta_0,\beta_1)$ equal the coefficients in the TSLS of $Y$ on $(Z, XZ)$ with $(Z-e(X),  (Z-e(X)) X )$ being the IV for $(Z, XZ)$.

\paragraph{Estimation under Example \ref{example::g-estimation-snm}}\label{problem::snm-estimation-example}

We can estimate the $\beta$'s by solving the empirical version of the estimating equations in Example \ref{example::g-estimation-snm}. We first estimate the two propensity scores and obtain the centered treatment 
$$
\check{Z}_{1i} = Z_{1i} - \hat{e}(1,X_{0i})
$$
 at time point 1 and 
 $$
 \check{Z}_{2i} = Z_{2i}  - \hat{e}(1, Z_{1i}, X_{1i}, X_{0i})
 $$
 at time point 2.

Show that we can estimate $\beta_2$ and $\beta_3$ by running TSLS of $Y_i$ on $(Z_{2i}, Z_{1i} Z_{2i} )$ with $( \check{Z}_{2i} ,  Z_{1i}\check{Z}_{2i} )$ as the IV for $(Z_{2i}, Z_{1i} Z_{2i} )$, and then we can estimate $\beta_1$ by running TSLS of $Y_i - (\hat{\beta}_2+\hat{\beta}_3 Z_{1i}) Z_{2i}$ on $Z_{1i}$ with $\check{Z}_{1i} $ as the IV for $Z_{1i}$.

\paragraph{g-formula with a treatment at multiple time points}\label{problem::g-formula-multiple-time}

Extend the discussion to the setting with $K$ time points. The temporal ordering (but not the causal diagram) of the variables is
$$
X_0 \rightarrow Z_1 \rightarrow X_1 \rightarrow Z_2 \rightarrow \cdots  \rightarrow  X_{K-1}  \rightarrow Z_K .
$$
Introduce the notation $\overline{Z}_k = (Z_1, \ldots, Z_k)$ and $\overline{X}_k = (X_0, X_1, \ldots, X_k)$ with lower case $\overline{z}_k $ and $\overline{x}_k$  denoting the corresponding  realized values. With $k=0$, we have $\overline{X}_0 = X_0$ and $\overline{Z}_0$ is empty. 
Each unit has $2^K$ potential outcomes:
$$
Y(\overline{z}_K) \text{ for all } z_1,\ldots, z_K = 0,1.
$$
Assume sequential ignorability below.

\begin{assumption}
[sequential ignorability at multiple time points]\label{assume::si-multiple}
We have
$$
Z_k \ind Y(\overline{z}_K) \mid (\overline{Z}_{k-1}, \overline{X}_{k-1})
$$
for all $k=1,\ldots, K$ and all $z_1,\ldots, z_K = 0,1.$
\end{assumption}

Prove Theorem \ref{thm::g-formula-multiple} below.

\begin{theorem}
[g-formula with multiple time points]
\label{thm::g-formula-multiple}
Under Assumption \ref{assume::si-multiple}, 
$$
E\{ Y(\overline{z}_K)  \} = E\left[ \cdots    
E\{   E(Y\mid \overline{z}_K, \overline{X}_{K-1}) \mid \overline{z}_{K-1}, \overline{X}_{K-2} \}
\cdots \mid z_1,  X_0 \right] .
$$
\end{theorem}

Remark: 
In Theorem \ref{thm::g-formula-multiple}, I use the simplified notation ``$\overline{z}_k$'' for  ``$\overline{Z}_k = \overline{z}_k $.''
With discrete $X$, Theorem \ref{thm::g-formula-multiple} reduces to
\begin{eqnarray*}
 E\{ Y(\overline{z}_K)  \}  
&=& 
\sum_{x_0} \sum_{x_1} \cdots   \sum_{x_{K-1}} 
 E(Y\mid \overline{z}_K, \overline{x}_{K-1})   \\
 && 
\qquad \qquad \cdot  \pr(x_{K-1}\mid \overline{z}_{K-1}, \overline{x}_{K-2}) 
 \cdots
 \pr(x_1\mid z_{1}, x_0)
 \pr(x_0) ; 
\end{eqnarray*} 
with continuous $X$, Theorem \ref{thm::g-formula-multiple} reduces to
\begin{eqnarray*}
  E\{ Y(\overline{z}_K)  \}  
&=& 
\int 
 E(Y\mid \overline{z}_K, \overline{x}_{K-1})   \\
 && \cdot 
 f(x_{K-1}\mid \overline{z}_{K-1}, \overline{x}_{K-2}) 
 \cdots
 f(x_1\mid z_{1}, x_0)
 f(x_0)
 \d \overline{x}_{K-1}   . 
\end{eqnarray*}

 \paragraph{IPW with treatments at multiple time points}\label{problem::ipw-multiple-time}

Inherit the setting of Problem \ref{problem::g-formula-multiple-time}. Define the propensity score at $K$ time points as
\begin{eqnarray*}
e(z_1, X_0) &=& \pr(Z_1 = z_1\mid X_0) ,\\
\vdots \\
e(z_k, \overline{Z}_{k-1},  \overline{X}_{k-1}) &=&  \pr(Z_k = z_k \mid \overline{Z}_{k-1},  \overline{X}_{k-1}) ,\\
\vdots \\
e(z_K, \overline{Z}_{K-1},  \overline{X}_{K-1}) &=&  \pr(Z_K = z_K \mid \overline{Z}_{K-1},  \overline{X}_{K-1}) .
\end{eqnarray*}

Prove Theorem \ref{thm::ipw-multiple} below assuming overlap implicitly.

\begin{theorem}
[IPW with multiple time points]\label{thm::ipw-multiple}
Under Assumption \ref{assume::si-multiple},
$$
  E\{ Y(\overline{z}_K)  \}   = E\left\{  
  \frac{1(Z_1 = z_1) \cdots  1( Z_K = z_K) Y}{    e(z_1, X_0)  \cdots  e(z_K, \overline{Z}_{K-1},  \overline{X}_{K-1})   }
  \right\}.
$$
\end{theorem}

Based on Theorem \ref{thm::ipw-multiple}, construct the Horvitz--Thompson and Hajek estimators.

 \paragraph{MSM with treatments at multiple time points}\label{problem::msm-multiple-time}

The number of potential outcomes grows exponentially with $K$. The formulas in Problems \ref{problem::g-formula-multiple-time} and \ref{problem::ipw-multiple-time} are not directly applicable in finite samples. We can impose the following structural assumptions on the potential outcomes.

\begin{definition}
[MSM with multiple time points]\label{definition::msm-multiple}
Assume 
$$
  E\{ Y(\overline{z}_K)  \mid X_0\}  = f(\overline{z}_K, X_0 ; \beta  ).
$$
\end{definition}

Two leading examples of Definition \ref{definition::msm-multiple} are 
$$
  E\{ Y(\overline{z}_K)  \mid X_0\}   = \beta_0 + \beta_1 \sum_{k=1}^K z_k + \beta_2\tran X_0 
  $$
and
$$ 
 E\{ Y(\overline{z}_K)  \mid X_0\}   = \beta_0 +  \sum_{k=1}^K \beta_k z_k + \beta_{K+1}\tran X_0 .
 $$

If we know all the potential outcomes, we can solve $\beta$ from the following minimization problem:
$$
\beta = \arg \min_b  \sum_{  \overline{z}_K  }  E\{   Y(\overline{z}_K) -  f(\overline{z}_K, X_0 ; \beta  )\}^2.
$$
Theorem \ref{thm::ipw-multiple} below shows that under Assumption \ref{assume::si-multiple}, we can solve $\beta$ from a minimization problem that only involves observables.

\begin{theorem}
[IPW for MSM with multiple time points]\label{thm::ipw-multiple}
Under Assumption \ref{assume::si-multiple},
$$
\beta = \arg \min_b  \sum_{  \overline{z}_K  }    E\left[ 
  \frac{1(Z_1 = z_1) \cdots 1( Z_K = z_K) }{    e(z_1, X_0)  \cdots  e(z_K, \overline{Z}_{K-1},  \overline{X}_{K-1})    }
  \{ Y -  f(\overline{z}_K, X_0 ; \beta  )\} ^2
  \right] .
$$
\end{theorem}

   \paragraph{Structural nested model with treatments at multiple time points}\label{problem::snm-multiple-time}

Inherit the setting from Problem \ref{problem::g-formula-multiple-time} and the notation from Problem \ref{problem::ipw-multiple-time}. This problem presents a general structural nested model. 

\begin{definition}
[structural nested model with multiple time points]
\label{def::snm-multiple}
The conditional effect at time $k$ is 
$$
E\{  Y(\overline{z}_k,0) - Y( \overline{z}_{k-1},0) \mid \overline{z}_k , \overline{X}_{k-1} \} = g_k( \overline{z}_k, \overline{X}_{k-1}; \beta) 
$$
for all $\overline{z}_k$ and all $k=1,\ldots, K$. 
\end{definition}

In Definition \ref{def::snm-multiple},  a logical restriction is
$$
g_k(0,  \overline{z}_{k-1}, \overline{X}_{k-1}; \beta)    = 0
$$ 
for all $\overline{z}_{k-1}$ and all $k=1,\ldots, K$.

Define 
$$
U_k(\beta) = Y - \sum_{s=1}^k  g_s( \overline{Z}_s, \overline{X}_{s-1}; \beta) 
$$
for all $k=1,\ldots, K$. Theorem  \ref{thm::snm-estimating-equation-multiple} below extends Theorem \ref{thm::estimating-equation-snm}.

\begin{theorem}
\label{thm::snm-estimating-equation-multiple}
Under Assumption \ref{assume::si-multiple} and Definition \ref{def::snm-multiple},  we have 
$$
E\left[    h_k( \overline{Z}_{k-1}, \overline{X}_{k-1} ) \{  Z_k - e(1, \overline{Z}_{k-1}, \overline{X}_{k-1}) \}  U_k(\beta) 
\right] = 0
$$
for any functions $h_k$ $(k=1,\ldots, K)$, provided that   the moment exists. 
\end{theorem}

Remark: Choosing appropriate $h_k$'s, we can estimate $\beta$ by solving the empirical version of Theorem \ref{thm::snm-estimating-equation-multiple}.

  \paragraph{Recommended reading} 

\citet{robins2000marginal} reviewed the MSM. \citet{naimi2017introduction} reviewed the g-methods.

\appendix 
\part{Appendices}

\renewcommand{\theparagraph}{\Alph{chapter}.\arabic{paragraph}}

\chapter{Probability and Statistics}
 \label{chapter::basic-prob-append}

This book assumes that the readers have basic knowledge of probability theory and statistical inference. Therefore, this chapter is not a comprehensive review of probability and statistics. For easy reference, I review the key concepts that are crucial for the main text.

\section{Probability}

\subsection{Pearson correlation coefficient and squared multiple correlation coefficient}
\label{sec::measure-pearson-r2}

For two random variables $Y$ and $X$, define the Pearson correlation coefficient as
$$
\rho_{YX} = \frac{  \cov(Y, X)  }{  \sqrt{  \var(Y)   \var(X) }   }
$$
which measures the linear dependence of $Y$ on $X$. The definition is symmetric in $Y$ and $X$ in that
$$
\rho_{YX}  = \rho_{XY} .
$$

With a random variable $Y$ and a random vector $X$, define the squared multiple correlation coefficient as
$$
R^2_{YX} = \textup{corr}^2(Y, X )
= \frac{   \cov(Y, X) \cov(X)^{-1} \cov(X, Y)  }{  \var(Y) }
$$
where $\cov(Y, X) $ is a row vector and $\cov(X, Y)$ is a column vector. It also measures the linear dependence of $Y$ on $X$. But this definition is not symmetric in $Y$ and $X$.

\subsection{Multivariate Normal random vector}
\label{sec::multivariate-normal}
 
A multivariate Normal random vector $\textsc{N}(\mu, \Sigma)$ is determined by its mean vector $\mu$ and covariance matrix $\Sigma$. Partition it into two parts: 
\[
\left(\begin{array}{c}
Y_{1}\\
Y_{2}
\end{array}\right)\sim\textsc{N}\left(\left(\begin{array}{c}
\mu_{1}\\
\mu_{2}
\end{array}\right),\left(\begin{array}{cc}
\Sigma_{11} & \Sigma_{12}\\
\Sigma_{21} & \Sigma_{22}
\end{array}\right)\right).
\]

First, the marginal distributions are Normal:
$$
Y_{1}  \sim\textsc{N}\left(\mu_{1},\Sigma_{11}\right),\quad 
Y_{2}  \sim\textsc{N}\left(\mu_{2},\Sigma_{22}\right).
$$
Second, if $\Sigma_{22}$ is positive definite, then the conditional distribution is also Normal:
\[
Y_{1}\mid Y_{2}=y_{2}\sim\textsc{N}\left(\mu_{1}+\Sigma_{12}\Sigma_{22}^{-1}(y_{2}-\mu_{2}),\Sigma_{11}-\Sigma_{12}\Sigma_{22}^{-1}\Sigma_{21}\right).
\]

 \subsection{$\chi^2$ and $t$ distributions}
\label{subsection::chi2-t-distribution}

Assume $X_1,\ldots, X_n$ are IID $\N01$. Then 
$$
\sumn X_i^2 
$$
follows a $\chi^2_n$ distribution with degrees of freedom $n$. The $\chi^2_n$ distribution has mean $n$ and variance $2n$.

Assume $X\sim \N01, Q_n\sim \chi^2_n$ and $X\ind Q_n$. Then
$$
\frac{X}{  \sqrt{Q_n / n} }
$$
follows a $t_n$ distribution with degrees of freedom $n$. The $t_n$ distribution has mean 0 if $n>1$. When $n=1$, the $t_1$ distribution is also called the Cauchy distribution, which does not have a finite mean.

\subsection{Cauchy--Schwarz inequality}\label{sec::cauchy-schwarz-inequality}

The  Cauchy--Schwarz inequality has many forms. With two random variables $A$ and $B$, we have
$$
 | E(AB)  | \leq \sqrt{   E(A^2) E(B^2)  }
$$
with equality holding when $B = \beta A$ for some $\beta$. Centering  $A$ and $B$ to have mean 0, we have
$$
| \cov(A, B)  | \leq \sqrt{   \var(A) \var(B)  } 
$$
with equality holding when $B = \alpha + \beta A$ for some $\alpha$ and  $\beta$.

When $A$ and $B$ are uniform random variables over finite sets $\{ a_1, \ldots, a_n \}$ and $\{ b_1, \ldots, b_n \}$ respectively, we have
$$
 \left|  \sumn a_i b_i  \right| \leq \sqrt{  \sumn a_i^2   \sumn b_i^2  }
$$
with equality holding if there exists   $\beta$ such that $b_i =  \beta a_i$ for all $i$'s.

\subsection{Tower property and variance decomposition}\label{appendix::tower-property}

Given random variables or vectors $A, B, C$, we have
$$
E(A) = E\{   E(A \mid B) \} 
$$
and
$$
E(A\mid C) = E\{   E(A \mid B, C) \mid C \}. 
$$

Given a random variable $A$ and random variables or vectors $B, C$, we have
$$
\var(A) = E\{   \var(A \mid B) \}  +  \var\{   E(A \mid B) \} 
$$
and
$$
\var(A\mid C) = E\{   \var(A \mid B, C)  \mid C \}  +  \var\{   E(A \mid B, C)\mid C \}  .
$$
When I was in graduate school, my professors Carl Morris and Joe Blitzstein called this formula the {\it Eve's Law} due to the ``EVVE'' form of the formula. They then went back to call the first tower property the {\it Adam's Law.}

Similarly, we can decompose the covariance as
$$
\cov(A_1, A_2) = E\left\{ \cov(A_1, A_2\mid B)   \right\} + \cov\{ E(A_1\mid B), E(A_2\mid B)  \}
$$
and
$$
\cov(A_1, A_2\mid C) = E\left\{ \cov(A_1, A_2\mid B, C)  \mid C \right\} + \cov\{ E(A_1\mid B, C), E(A_2\mid B, C) \mid C  \}.
$$

\subsection{Limiting theorems}

\begin{definition}
[convergence in probability]\label{def::converge-in-prob}
A sequence of random variables $(X_n)_{n\geq 1}$ converges to $X$ in probability, if for every $\varepsilon > 0$, we have
$$
\pr (   |X_n - X| > \varepsilon  ) \rightarrow 0
$$
as $n\rightarrow \infty$. 
\end{definition}

\begin{definition}
[convergence in distribution]\label{def::converge-in-distribution}
A sequence of random variables $(X_n)_{n\geq 1}$ converges to $X$ in distribution, if 
$$
\pr(X_n \leq x) \rightarrow \pr(X\leq x)
$$
for all continuity point $x$ of $ \pr(X\leq x) $, as $n\rightarrow \infty$. 
\end{definition}

Convergence in probability is stronger than convergence in distribution. Definitions \ref{def::converge-in-prob} and \ref{def::converge-in-distribution} are useful for stating the following two fundamental theorems on the sample average of independent and identically distributed (IID) random variables.

\begin{theorem}[law of large numbers]
\label{thm::lln}
If $X_1,\ldots, X_n \iidsim X$ with $E| X| <\infty$, then $\bar{X} = n^{-1} \sumn X_i \rightarrow E(X)$ in probability. 
\end{theorem}

The law of large numbers in Theorem \ref{thm::lln} states that the sample average is close to the population mean in the limit. 

\begin{theorem}[central limit theorem (CLT)]
\label{thm::clt}
If $X_1,\ldots, X_n \iidsim X$ with $\var (X) <\infty$, then 
$$ 
\frac{   \bar{X} - E(X)   }
{ \sqrt{ \var (X) /n    } } \rightarrow \N01
$$  
in distribution. 
\end{theorem}

The CLT in Theorem \ref{thm::clt} states that the standardized sample average is close to a standard Normal random variable in the limit.

Theorems \ref{thm::lln} and \ref{thm::clt} assume IID random variables for convenience. 
There are also many laws of large numbers and CLTs for the sample mean of independent or weakly dependent random variable \citep[e.g.,][]{durrett2019probability}.

\subsection{Delta method}\label{section::delta-method}

The delta method is a power tool to derive the asymptotic Normality of nonlinear functions of an asymptotically Normal random vector. I review a special case of the delta method below.

\begin{theorem}
[delta method]\label{thm::delta-method}
Assume $\sqrt{n} (X_n - \mu) \rightarrow \textsc{N}(0, \Sigma)$ in distribution and the function $g(x)$ has non-zero derivative $ \nabla g(\mu)   $ at $\mu$. Then
$$
\sqrt{n} \{  g(X_n) - g(\mu) \}  \rightarrow \textsc{N} (0,   (\nabla g(\mu) ) \tran  \Sigma \nabla g(\mu)  )
$$
in distribution.
\end{theorem}  

I will omit the proof of Theorem \ref{thm::delta-method}. It is intuitive based on the first-order Taylor expansion:
$$
g(X_n) - g(\mu) \approx  (\nabla g(\mu) ) \tran  (X_n - \mu) .
$$
So $\sqrt{n} \{  g(X_n) - g(\mu) \} $ is close to the linear transformation of $\textsc{N}(0, \Sigma)$, which is $\textsc{N} (0,   (\nabla g(\mu) ) \tran  \Sigma \nabla g(\mu)  )$.

As illustrations, we can use the delta method to obtain the asymptotic Normality of the ratio and product. 

\begin{example}
[asymptotic Normality for the ratio]\label{example::asymptotic-normal-ratio}
Assume 
\begin{equation}\label{eq::normal-ratio}
\sqrt{n} \begin{pmatrix}
Y_n - \mu_Y \\
X_n - \mu_X
\end{pmatrix}
\rightarrow 
\textsc{N}\left(
\begin{pmatrix}
0\\
0
\end{pmatrix},
\begin{pmatrix}
\sigma_Y^2 & \sigma_{YX} \\
 \sigma_{YX} &  \sigma_X^2 
\end{pmatrix}
\right)
\end{equation}
in distribution with $\mu_X \neq 0.$ Apply Theorem \ref{thm::delta-method} to obtain that
\begin{equation}\label{eq::ratio-normality}
\sqrt{n} \left(   \frac{Y_n}{X_n}  -  \frac{\mu_Y}{\mu_X} \right)
\rightarrow
\textsc{N}\left(0,    \frac{  \sigma_Y^2 }{ \mu_X^2 }  + \frac{ \mu_Y^2 \sigma_X^2}{ \mu_X^4}  - \frac{2\mu_Y \sigma_{YX}}{\mu_X^3} \right)
\end{equation}
in distribution. 
In the special case that $X_n$ and $Y_n$ are asymptotically independent with $\sigma_{YX} = 0$, the asymptotic variance of $Y_n/X_n$ simplifies to $\sigma_Y^2 / \mu_X^2 + \mu_Y^2 \sigma_X^2 / \mu_X^4$.
I leave the details to Problem \ref{hw::asymptotic-normality-ratio}.
\end{example}

The asymptotic variance in Example \ref{example::asymptotic-normal-ratio} is a little cumbersome. An easier way to memorize it is based on the following approximation:
\begin{equation}\label{eq::ratio-slutsky}
\frac{Y_n}{X_n}  -  \frac{\mu_Y}{\mu_X}  = \frac{Y_n -  \mu_Y/\mu_X \cdot X_n  }{X_n} \approx   \frac{Y_n -  \mu_Y/\mu_X \cdot X_n  }{\mu_X}, 
\end{equation}
so the asymptotic variance of the ratio equals the asymptotic variance of 
$$
 \frac{Y_n -  \mu_Y/\mu_X \cdot X_n  }{\mu_X},
 $$
which is a linear combination of $Y_n$ and $X_n$. Slutsky's theorem can make the approximation in \eqref{eq::ratio-slutsky} rigorous but it is beyond this book.

\begin{example}
[asymptotic Normality for the product]\label{example::asymptotic-normal-product}
Assume \eqref{eq::normal-ratio}. Apply Theorem \ref{thm::delta-method} to obtain that
\begin{equation}\label{eq::product-normality}
\sqrt{n} \left(   X_n  Y_n  -  \mu_X \mu_Y \right)
\rightarrow
\textsc{N}\left(0,     \mu_Y^2 \sigma_X^2 + \mu_X^2 \sigma_Y^2 + 2 \mu_X \mu_Y \sigma_{XY} \right)
\end{equation}
in distribution. In the special case that $X_n$ and $Y_n$ are asymptotically independent with $\sigma_{YX} = 0$, the asymptotic variance of $X_n Y_n$ simplifies to $\mu_Y^2 \sigma_X^2 + \mu_X^2 \sigma_Y^2$.
I leave the details to Problem \ref{hw::asymptotic-normality-product}.
\end{example}

\section{Statistical inference}
\label{sec::statistical-inference}

\subsection{Point estimation}

Assume that $\theta$ is the parameter of interest. Oftentimes, the problem also contains other parameters not of interest, denoted by $\eta$. Statisticians call $\eta$ the {\it nuisance parameter}. Based on the data, we can compute an estimator $\hat\theta$. Throughout this book, we take the frequentist's perspective by assuming that $\theta$ is a fixed number and $\hat\theta$ is random due to the randomness of data. Two basic requirements for an estimator are below. 

\begin{definition}
[unbiasedness]\label{def::unbiasedness}
The estimator $\hat\theta$ is {\it unbiased} for $\theta$ if 
$$
E(\hat\theta) = \theta
$$
for all possible values of $\theta$ and $\eta$. 
\end{definition}

\begin{definition}
[consistency]\label{def::consistency}
The estimator $\hat\theta$ is {\it consistent} for $\theta$ if 
$$
 \hat\theta  \rightarrow \theta
$$
in probability as the sample size approaches to infinity, for all possible values of $\theta$ and $\eta$. 
\end{definition}

Unbiasedness requires that the mean of the estimator is identical to the parameter of interest. Consistency requires that the estimator is close to the true parameter in the limit. Unbiasedness does not imply consistency, and consistency does not imply unbiasedness either. Unbiasedness can be restrictive because it is impossible even in some simple statistics problems. Consistency is often the basic requirement in most statistics problems.

\subsection{Confidence interval}

A point estimator $\hat\theta$ is a random variable that differs from the true parameter $\theta$. Statisticians are often interested in finding an interval that covers the true parameter with a certain given probability. This interval is computed based on the data, and it is random. 

\begin{definition}
[confidence interval]\label{def::confidence-interval}
A data-dependent interval $[  \hat\theta_\textsc{l} , \hat\theta_\textsc{u} ] $ is a confidence interval for $\theta$ with coverage probability at least $1-\alpha$ if
$$
\pr(    \hat\theta_\textsc{l}    \leq \theta \leq  \hat\theta_\textsc{u}  ) \geq 1-\alpha 
$$
for all possible values of $\theta$ and $\eta$.  
\end{definition}

\begin{definition}
[asymptotic confidence interval]\label{def::asymptotic-confidence-interval}
A data-dependent interval $[  \hat\theta_\textsc{l} , \hat\theta_\textsc{u} ] $ is an asymptotic confidence interval for $\theta$ with coverage probability at least $1-\alpha$ if
$$
\pr(    \hat\theta_\textsc{l}   \leq \theta \leq  \hat\theta_\textsc{u}  )   \rightarrow    1-\alpha'   ,\qquad \text{ as } n\rightarrow \infty
$$
with $\alpha '  \leq \alpha$, for all possible values of $\theta$ and $\eta$.     
\end{definition}

A standard choice is $\alpha = 0.05$. 
In Definitions \ref{def::confidence-interval} and \ref{def::asymptotic-confidence-interval}, the coverage probabilities can be larger than the nominal level $1-\alpha$. That is, the definitions allow for over-coverage but do not allow for under-coverage. With over-coverage, we say that the confidence interval is conservative. Of course, we hope the confidence interval to be as narrow as possible. Otherwise, the definition of the confidence interval can be arbitrary.

\subsection{Hypothesis testing}

  Many applied problems can be formulated as testing a hypothesis:
  $$
  H_0: \theta = 0 .
  $$
The decision rule $\phi$ is a binary function of the data: $\phi = 1$ if we reject $H_0$; $\phi = 0$ if we fail to reject   $H_0$. The type one error rate of the test is the probability of rejection if the null hypothesis holds. I review the definition below.

\begin{definition}[type one error rate]
\label{def::size-test}
When $H_0$ holds, define the type one error rate of the test $\phi$ as the maximum value of the probability
$$
\pr(\phi = 1)
$$
over all possible values of $\theta$ and $\eta$.     
\end{definition}

A standard choice is to make sure that the type one error rate is below $\alpha = 0.05$. The type two error rate of the test is the probability of no rejection if the null hypothesis does not hold. I review the definition below.

\begin{definition}[type two error rate]
\label{def::type2-error-test}
When $H_0$ does not hold, define the type two error rate of the test $\phi$ as the maximum value of the  probability
$$
\pr(\phi = 0)
$$
over all possible values of $\theta$ and $\eta$.     
\end{definition}

Given the control of the type one error rate under $H_0$, we hope the type two error rate is as low as possible when $H_0$ does not hold.

\subsection{Wald-type confidence interval and test}\label{section::wald-type-ci-test}

Many statistics problems have the following structure. The parameter of interest is $\theta$. We first find a consistent estimator $\hat\theta$ that converges in probability to $\theta$, and show that it is asymptotically Normal with mean $\theta$ and variance $v$ which may depend on $\theta$ as well as the nuisance parameter $\eta$. We then find a consistent estimator $\hat v$ for $v$, based on analytic formulas or the bootstrap reviewed in Chapter \ref{sec::bootstrap} later. 
The square root of $\hat v$ is called the {\it standard error}. 
We finally construct the Wald-type confidence interval for $\theta$ as
$$
\hat{\theta} \pm z_{1-\alpha/2} \sqrt{ \hat{v} }
$$  
where $z_{1-\alpha/2}$ is the $1-\alpha/2$ upper quantile of the standard Normal random variable.
It covers $\theta$ with probability approximately $1-\alpha$. When this interval excludes a particular $c$, for example, $c = 0$, we reject the null hypothesis $H_0(c): \theta = c$, which is called the Wald test.

 \subsection{Duality between constructing confidence sets and testing null hypotheses}
 \label{section::duality-ci-testing}

Consider the statistical inference problem for a scalar parameter $\theta$. A fundamental result in statistics is that constructing confidence sets for $\theta$ is equivalent to testing null hypotheses about $\theta$. This is often called the duality between constructing confidence sets and testing null hypotheses. 

Section \ref{section::wald-type-ci-test} has reviewed the duality based on the  Wald-type confidence interval and test. The duality also holds in general. Assume that $\hat\Theta$ is a $(1-\alpha)$-level confidence set for $\theta$:
$$
\pr(  \theta \in \hat\Theta  ) \geq  1-\alpha.
$$
Then we can reject the null hypothesis $H_0(c): \theta = c$ if $c$ is not in the set $\hat\Theta$. This is a valid test because when $\theta$ indeed equals $c$, we have the correct type one error rate $\pr(  \theta \not  \in \hat\Theta ) \leq  \alpha$. 
Conversely, if we test a sequence of null hypotheses $H_0(c): \theta = c$, we can obtain the corresponding $p$-values, $p(c)$, as a function of $c$. Then the values of $c$ that we fail to reject at level $\alpha$ form a confidence set for $\theta$:
$$
 \hat\Theta = 
\{  c:  p(c) \geq \alpha \} = \{  c: \text{ fail to reject } H_0(c) \text{ at level } \alpha \} . 
$$
It is a valid confidence set because 
$$
\pr( \theta \in  \hat\Theta ) = 
\pr\{  \text{fail to reject } H_0(\theta) \text{ at level } \alpha \} \geq   1-\alpha.
$$

Here I use ``confidence set'' instead of ``confidence interval'' because $ \hat\Theta$ based on inverting tests may not be an interval. 
See the use of the duality in Chapters \ref{sec::fc-problem} and \ref{sec::frt-invert-interval}.

\section{Inference with two-by-two tables}

\subsection{Fisher's exact test}
\label{sec::fisher-exact-test}

Fisher  proposed an exact test for $H_0: p_1 = p_0$ under the statistical model: 
$$
n_{11} \sim \text{Binomial}(n_1, p_1),\quad
n_{01} \sim \text{Binomial}(n_0, p_0),\quad 
n_{11} \ind n_{01}.
$$
The table below summarizes the data.

\begin{center}
\begin{tabular}{cccc}
\hline 
 & 1 & 0 & row sum \\
 \hline 
sample 1 & $n_{11}$ & $n_{10}$ & $n_1$ \\
sample 0 & $n_{01}$ & $n_{00}$ & $n_0$ \\
column sum & $n_{\cdot 1}$ & $n_{\cdot 0}$ & $n$ \\
\hline 
\end{tabular}
\end{center}

He argued that the sums $n_{11}  + n_{01}  =  n_{\cdot 1}$ and $n_{10}  + n_{00}  =  n_{\cdot 0}$ contain little information about the difference between $p_1$ and $p_0$, and conditional on them, $n_{11}$  follows a  Hypergeometric distribution that does not depend on the unknown parameter $p_1=p_0$ under $H_0$:
$$
\pr(n_{11} = k )  = \frac{  \binom{ n_{\cdot 1} }{  k }  \binom{  n -  n_{\cdot 1} }{  n_1 - k   }  }{  \binom{  n }{ n_1 }  }.
$$
In \ri{R}, the function \ri{fisher.test} implements this test.

\subsection{Estimation with two-by-two tables}\label{section::asymptotic-inference-2x2}

Based on the model in Section \ref{sec::fisher-exact-test}, we can estimate the parameters $p_1$ and $p_0$ by sample frequencies:
$$
\hat p_1 = \frac{n_{11}}{n_1} ,\quad \hat p_0 = \frac{n_{01}}{n_0}.
$$
Therefore, we can estimate the risk difference, log risk ratio, and log odds ratio 
\begin{eqnarray*}
\RD &=& p_1 - p_0,\\ 
\log \RR &=&\log \frac{ p_1}{p_0},\\ 
\log \OR &=& \log \frac{ p_1/(1-p_1)}{p_0/(1-p_0)}
\end{eqnarray*}
by the sample analogs
\begin{eqnarray*}
\hat \RD &=& \hat  p_1 -\hat  p_0,\\ 
\log \hat \RR &=& \log\frac{ \hat p_1}{\hat p_0},\\ 
\log \hat \OR &=&\log \frac{ \hat p_1/(1-\hat p_1)}{\hat p_0/(1-\hat p_0)}  = \log\frac{n_{11}n_{00}}{n_{10}n_{01}}.
\end{eqnarray*}
Based on the asymptotic approximation (see Problem \ref{hw::variance-estimators-2x2}), the estimated variance for the above parameters are 
\begin{eqnarray*}
&&\frac{ \hat  p_1 (1-\hat  p_1) }{n_1} + \frac{ \hat  p_0 (1-\hat  p_0) }{n_0},\\
&& \frac{   1-\hat  p_1 }{n_1 \hat  p_1} + \frac{   1-\hat  p_0 }{n_0 \hat  p_0 },\\
&& \frac{ 1 }{n_1  \hat  p_1 (1-\hat  p_1)} + \frac{ 1 }{n_0 \hat  p_0 (1-\hat  p_0)},
\end{eqnarray*}
respectively.
The log transformation above yields better Normal approximations because the risk ratio and odds ratio are always positive.

\section{Two famous problems in statistics}

\subsection{Behrens--Fisher problem}\label{sec::bf-problem}

Consider the two-sample problem with $n_1$ units under the treatment and $n_0$ units under the control, respectively. Assume the outcomes under the treatment $\{  Y_i: Z_i = 1 \}$ are IID from $\textsc{N} (\mu_1, \sigma_1^2)$ and the outcomes under the control  $\{  Y_i: Z_i = 0 \}$ are IID from $\textsc{N} (\mu_0, \sigma_0^2)$, respectively. The goal is to test $H_0: \mu_1 = \mu_0$.

Start with the easier case with $\sigma_1^2  = \sigma_0^2.$ Coherent with Chapter \ref{ch::frt-cre}, let 
$$
\hat{\bar{Y}}(1) = n_1^{-1} \sum_{Z_i =1}  Y_i ,\quad
\hat{\bar{Y}}(0) = n_0^{-1} \sum_{Z_i =0}  Y_i
$$
denote the sample means of the outcomes under the treatment and control, respectively. 
A standard result is that 
$$
t_\textup{equal} = 
\frac{  \hat{\bar{Y}}(1) - \hat{\bar{Y}}(0)    }{  \sqrt{\frac{n}{n_1 n_0(n-2)}  \left[  \sum_{Z_i=1}  \{Y_i - \hat{\bar{Y}}(1)\}^2
+ \sum_{Z_i=0}  \{Y_i - \hat{\bar{Y}}(0)\}^2
    \right]  }    }   \sim t_{n-2}.
$$     
Based on $t_\textup{equal} $, we can construct a test for $H_0.$

Now consider the more difficult case with possibly different $\sigma_1^2  $ and $ \sigma_0^2.$ The distribution of $t_\textup{equal}$ is no longer $t_{n-2}.$ Estimating the variances separately, we can also define
$$
t_\textup{unequal} = \frac{    \hat{\bar{Y}}(1)  -   \hat{\bar{Y}}(0)    }{  \sqrt{  \frac{  \hat{S}^2(1)  }{n_1}  + \frac{\hat{S}^2(0)}{n_0}  }   },
$$
where 
$$
\hat{S}^2(1) = (n_1 - 1)^{-1} \sum_{Z_i =1} \{ Y_i  -\hat{\bar{Y}}(1)   \}^2,\quad
\hat{S}^2(0) = (n_0 - 1)^{-1} \sum_{Z_i =0} \{ Y_i  -\hat{\bar{Y}}(0)   \}^2 
$$
are the sample variances of the outcomes under the treatment and control, respectively.  Unfortunately, the exact distribution of $t_\textup{unequal} $ depends on the known variances. Testing $H_0$ without assuming equal variances is the famous Behrens--Fisher problem.  With large sample sizes $n_1$ and $n_0$, the CLT ensures that $t_\textup{unequal} $ is approximately $\N01$. So we can construct an approximate test for $H_0.$ By duality, a large-sample Wald-type confidence interval for $\mu_1 - \mu_0$ is
$$
 \hat{\bar{Y}}(1)  -   \hat{\bar{Y}}(0)   \pm z_{1-\alpha/2}    \sqrt{  \frac{  \hat{S}^2(1)  }{n_1}  + \frac{\hat{S}^2(0)}{n_0}  }     
$$
where $z_{1-\alpha/2}$ is the $1-\alpha/2$ upper quantile of the standard Normal random variable.

\subsection{Fieller--Creasy problem}\label{sec::fc-problem}

Consider the two-sample problem with $n_1$ units under the treatment and $n_0$ units under the control, respectively. Assume the outcomes under the treatment $\{  Y_i: Z_i = 1 \}$ are IID from $\textsc{N} (\mu_1, 1)$ and the outcomes under the control $\{  Y_i: Z_i = 0 \}$ are IID from $\textsc{N} (\mu_0, 1)$, respectively. The goal is to estimate $\gamma = \mu_1 / \mu_0$. We can use $\hat\gamma = \hat{\bar{Y}}(1)  / \hat{\bar{Y}}(0) $ to estimate $\gamma$. However, the point estimator has a complicated distribution, which does not yield a simple procedure to construct the confidence interval for $\gamma$.

Fieller's confidence interval can be formulated as inverting tests for a sequence of null hypotheses: $H_0(c): \gamma = c$. Under $H_0(c)$, we have 
$$
\frac{  \hat{\bar{Y}}(1)  - c \hat{\bar{Y}}(0) }{  \sqrt{ 1/n_1 + c^2/n_0  } } \sim \N01 
$$
which motivates the confidence interval
$$
\left\{
c: \Big | \frac{  \hat{\bar{Y}}(1)  - c \hat{\bar{Y}}(0) }{  \sqrt{ 1/n_1 + c^2/n_0  } } \Big | \leq z_{1-\alpha/2}
\right\}
$$
where $z_{1-\alpha/2}$ is the $1-\alpha/2$ upper quantile of the standard Normal random variable.

\section{Monte Carlo method in statistics}\label{sec::monte-carlo-method}

The Monte Carlo method is a powerful tool in statistics. I will review its basic use in approximating expectations or averages, which is fundamental in understanding the idea of FRT  introduced in Chapter \ref{ch::frt-cre}.

If our goal is to calculate $\theta = E\{  g(Y)\}$ with $Y$ being a random variable, we can simply draw IID samples $Y_1,\ldots, Y_n$ from the distribution of $Y$ and obtain the moment estimator for $\theta$:
$$
\hat\theta = n^{-1} \sumn g(Y_i).
$$

As a special case, $Y$ is a uniform distribution over $\{ y_1, \ldots, y_N \}$ and $\theta = N^{-1} \sum_{i=1}^N g(y_i)$. We can draw $n$ IID samples $\{ Y_1, \ldots, Y_n \}$ from $\{ y_1, \ldots, y_N \}$ to obtain the moment estimator $\hat\theta $ defined above. This is called {\it sampling with replacement}, which is different from {\it sampling without replacement} reviewed in Chapter \ref{chapter::SRS-appendix}.

\section{Bootstrap}
\label{sec::bootstrap}

It is often very tedious to derive the variance formulas for complex estimators. \citet{Efron_1979} proposed the bootstrap as a general tool for variance estimation.  There are many versions of the bootstrap \citep{davison1997bootstrap}. In this book, we only need the most basic one: the nonparametric bootstrap, which will be simply called the bootstrap.

Consider the generic setting with 
$$
Y_1,\ldots, Y_n \iidsim Y,
$$
where $Y_i$ can be a general random element denoting the observed data for unit $i$. An estimator $\hat\theta$ is a function of the observed data: $\hat\theta = T(Y_1, \ldots, Y_n)$. When $T$ is a complex function, it may not be easy to obtain the variance or asymptotic variance of $\hat\theta$. 

The uncertainty of $\hat\theta$ is driven by the IID sampling of $Y_1,\ldots, Y_n$ from the true distribution. Although the true distribution is unknown, it can be well approximated by its empirical version 
$$
\hat{F}_n(y) = n^{-1} \sumn I(Y_i \leq y),
$$
when the sample size $n$ is large. If we believe this approximation, we can simulate $\hat\theta$ by sampling 
$$
(Y_1^*,\ldots, Y_n^* ) \iidsim \hat{F}_n(y).
$$ 
Because $\hat{F}_n(y) $ is a discrete distribution with mass $1/n$ on each observed data point, the simulation of $\hat\theta$ reduces to the following procedure: 
\begin{enumerate}
\item
sample $(Y_1^*,\ldots, Y_n^* ) $ from $\{Y_1,\ldots, Y_n \}$ with replacement;
\item
compute $\hat{\theta}^* = T(Y_1^*, \ldots, Y_n^*)$;
\item
repeat the above two steps $B$ times to obtain the bootstrap replicates $\{  \hat{\theta}^*_1, \ldots, \hat{\theta}^*_B \}$.
\end{enumerate}

We can then approximate the (asymptotic) variance of $\hat\theta$ by the sample variance of the bootstrap replicates:
$$
\hat{V}_\text{boot} = (B-1)^{-1} \sum_{b=1}^B  (\hat{\theta}^*_b - \bar{\theta}^*)^2,
$$
where $\bar{\theta}^* = B^{-1}\sum_{b=1}^B \hat{\theta}^*_b $. The bootstrap confidence interval based on the Normal approximation is then
$$
\hat\theta \pm  z_{1-\alpha/2} \sqrt{\hat{V}_\text{boot} },
$$
where $z_{1-\alpha/2}$ is the $1-\alpha/2$ upper quantile of the standard Normal random variable.

I use the following simple example to illustrate the idea of the bootstrap. With $n=100$ IID observations $Y_i$'s from $\textsc{N}(1,1)$, the sample mean should be close to $1$ with variance $1/100 = 0.01$. Over $500$ simulations, the classic variance estimator and the bootstrap variance estimator with $B=200$ both have average values close to $0.01$. 

\begin{lstlisting}
> ## sample size
> n  = 100
> ## number of Monte Carlo simulations
> MC = 500
> ## number of bootstrap replicates B
> n.boot = 200
> simulation = replicate(MC, {
+   Y = rnorm(n, 1, 1)
+   boot.mu.hat = replicate(n.boot, {
+     index = sample(1:n, n, replace = TRUE)
+     mean(Y[index])
+   })
+   c(mean(Y), var(Y)/n, var(boot.mu.hat))
+ })
> ## summarize the results
> apply(simulation, 1, mean)
[1] 0.997602961 0.010006303 0.009921895
\end{lstlisting}

\section{Homework problems}

\paragraph{Independent but not IID data}
\label{hw::inid}

Assume that the $Y_i$'s are independent with mean $\mu_i$ and variances $\sigma_i^2$ for $i=1,\ldots, n$. The parameter of interest is 
$
\mu = n^{-1} \sum_{i=1}^n \mu_i.
$
Show that 
$
\hat\mu = n^{-1} \sum_{i=1}^n Y_i
$
is unbiased for $\mu$ and find its variance. Show that the usual variance estimator for IID data
$$
\hat{v} = \{  n(n-1) \}^{-1} \sum_{i=1}^n (Y_i -  \hat{\mu})^2
$$
is a conservative estimator for the variance of $\hat\mu$ in the sense that
$$
E(\hat{v}) - \var( \hat\mu ) = \{  n(n-1) \}^{-1} \sum_{i=1}^n (\mu_i - \mu)^2 \geq 0. 
$$

Remark: Consider a simpler case with $\mu_i = \mu$ and $\sigma_i^2 = \sigma^2$ for all $i=1,\ldots, n$. The sample mean is unbiased for $\mu$ with variance $\sigma^2/n$. Moreover, an unbiased estimator for the variance $\sigma^2/n$ is $\hat \sigma^2/n = \hat v$, where $\hat\sigma^2 = (n-1)^{-1} \sum_{i=1}^n (Y_i -  \hat{\mu})^2$. 
This problem states a more general result for the case with independent but not identically distributed observations. The result has an important implication for Theorem \ref{thm::varianceest-mp} for the matched-pairs experiment discussed in Chapter \ref{chapter::mpe}.

\paragraph{Asymptotic Normality of ratio}\label{hw::asymptotic-normality-ratio}

Prove \eqref{eq::ratio-normality}.

\paragraph{Asymptotic Normality of product}\label{hw::asymptotic-normality-product}

Prove \eqref{eq::product-normality}.

\paragraph{Product of two independent Normals}\label{hw-product-Normals}

Assume $X\sim \textsc{N}(\mu_X, \sigma_X^2), Y\sim \textsc{N}(\mu_Y, \sigma_Y^2)$ and $X\ind Y$. Show that
$$
\var(XY) = \sigma_X^2 \sigma_Y^2 +\mu_X^2 \sigma_Y^2 + \mu_Y^2 \sigma_X^2 .
$$

Remark: This problem gives a non-asymptotic form of Example \ref{example::asymptotic-normal-product}.

\paragraph{Variance estimators in two-by-two tables}\label{hw::variance-estimators-2x2}

Use the delta method to show the variance estimators in Section \ref{section::asymptotic-inference-2x2}.

\paragraph{Fisher weighting}\label{hw::fisher-weighting}

Assume that we have $p$ independent unbiased estimators $\hat{\theta}_1, \ldots, \hat{\theta}_p$ for a common parameter $\theta$:
$$
E(\hat{\theta}_j) = \theta,\quad (j=1,\ldots, p).
$$
The estimator  $\hat{\theta}_j$ has variance $v_j$ $(j=1,\ldots, p)$. 

Construct a new estimator
$$
\hat{\theta} = \sum_{j=1}^p w_j \hat{\theta}_j
$$
with nonrandom constants $w_j$'s. 
Show that $\hat{\theta} $ is unbiased for estimating $\theta$ if $\sum_{j=1}^p w_j = 1$. Show that the optimal $w_j$'s such that $\hat{\theta} $ has the smallest variance under the constraint $\sum_{j=1}^p w_j = 1$ are
$$
w_j^* =  \frac{1/v_j}{ \sum_{j'=1}^p   1/v_{j'} }, \quad (j=1,\ldots, p).
$$
Show that the resulting estimator $\hat{\theta}^* = \sum_{j=1}^p w_j^* \hat{\theta}_j$ has variance
$$
\var(\hat{\theta}^*) =  \frac{1 }{ \sum_{j=1}^p   1/v_{j} }. 
$$

Remark: This is called Fisher weighting. The optimal weights are proportional to the inverses of the variances.

\chapter{Linear and Logistic Regressions}
\label{appendix::basic-linear-regression}
 
 \section{Population ordinary least squares}\label{sec::populationOLS}

Assume that $(x_{i},y_{i})_{i=1}^{n} \iidsim (x,y)$, where $x $ is a $p$-dimensional random scalar or vector and $y $ is a random scalar. 
Below I will use $(x,y)$ to denote a general observation, dropping the subscript $i$ for simplicity. 
Define the population ordinary least squares (OLS) coefficient as
\[
\beta=\arg\min_{b } E\left\{ (y-x \tran b)^{2}\right\} .
\]
The objective function is quadratic in $b$, so we can show that the minimizer is 
$$
\beta=\left\{ E\left(xx \tran \right)\right\} ^{-1}E\left(xy\right)
$$
if the moments exist and $E(xx \tran )$ is invertible. 

With $\beta$, we can define $x \tran \beta$ as the {\it linear projection} of $y$ on $x$, and define 
\begin{equation}
\varepsilon=y-x \tran \beta\label{eq:populationresidualols}
\end{equation}
as the {\it population residual}. 
By the definition of $\beta$, we can verify that
$$
E(x\varepsilon)=E\left\{ x(y-x \tran \beta)\right\} =E(xy)-E(xx \tran )\beta=0. 
$$

\begin{example}
[population OLS with the intercept]
If we include $1$ as a component of $x$, then 
$$
E(\varepsilon)
=E(y-x \tran \beta)
=0
$$
which further implies that $\cov(x,\varepsilon)=0$.
So with an intercept in $\beta$, the mean of the population residual
must be zero, and it is uncorrelated with other covariates by construction. 
\end{example}

\begin{example}
[univariate population OLS with the intercept]\label{eg::ols1-intercept}
An important special case is that for scalars $x$ and $y$, we can define
$$
(\alpha, \beta)  = \arg\min_{a,b} E \{  (y - a - bx)^2 \}
$$
which have explicit formulas
$$
\beta = \frac{  \cov(x,y) }{\var(x)},\quad 
\alpha = E(y)  - \beta E(x).
$$
\end{example}

 \begin{example}
[univariate population OLS without an intercept]
Without intercept, we can define 
$$
\gamma = \arg\min_{c} E \{  (y  - c x)^2 \}
$$
which equals
$$
\gamma = \frac{  E(xy) }{E(x^2)}.
$$
When $x$ has mean zero, $ \gamma$ equals the $\beta $ in Example \ref{eg::ols1-intercept}. 
\end{example}

We can also rewrite (\ref{eq:populationresidualols}) as
\begin{equation}
y=x \tran \beta+\varepsilon,\label{eq:population-ols-decomposition}
\end{equation}
which holds by the definition of the population OLS coefficient and residual without any modeling
assumption. We call (\ref{eq:population-ols-decomposition}) the population
OLS decomposition.

 \section{Sample ordinary least squares}\label{sec::sampleOLS}

Based on data $(x_{i},y_{i})_{i=1}^{n} \iidsim (x,y)$, we can easily
obtain the moment estimator for the population OLS coefficient
\[
\hat{\beta}=\left(n^{-1}\sumn x_{i}x_{i} \tran \right)^{-1}\left(n^{-1}\sumn x_{i}y_{i}\right),
\]
and the residuals $ \hat{\varepsilon}_i = y_i - x_i \tran  \hat{\beta} . $ 
This is called the sample OLS or simply the OLS. The OLS coefficient 
$\hat{\beta}$ minimizes the residual sum of squares
\[
\hat \beta=\arg\min_{b } n^{-1} \sumn  (y_i-x_i \tran b)^{2} ,
\]
so it must satisfy 
$$
\sumn x_i (y_i-x_i \tran \hat{\beta}) = 0,
$$
which is sometimes called the {\it Normal equation.}
 The fitted values, also called the {\it linear projections} of $y_i$ on $x_i$, equal
 $$
 \hat{y}_i = x_i\tran \hat\beta \quad (i=1,\ldots, n).
 $$

 Using the matrix notation
 $$
 X = \begin{pmatrix}
 x_1\tran \\
 \vdots \\
 x_n\tran 
 \end{pmatrix},\quad
 Y = \begin{pmatrix}
 y_1 \\
 \vdots \\
 y_n
 \end{pmatrix} ,
 $$
 we can write the OLS coefficient as
$$
\hat{\beta} = (X \tran X)^{-1} X\tran Y
$$ 
and the fitted vector as
 $$
 \hat{Y} = X\hat{\beta} =  X (X \tran X)^{-1} X\tran Y.
 $$
Define the hat matrix as
$$
H =  X (X \tran X)^{-1} X\tran.
$$
Then we also have $ \hat{Y}  = HY$, justifying the name ``hat matrix.'' The diagonal elements of $H $, $h_{ii}$'s, are often called the {\it  leverage scores}.

Assuming finite fourth moments of $(x,y)$, we can use the law of large numbers and the CLT to show that 
$$
\sqrt{n}(\hat{\beta}-\beta)\rightarrow\text{N} ( 0, V )
$$
in distribution with $V =  B^{-1} M B^{-1}$,  where $B =   E(xx \tran ) $ and $M= E(\varepsilon^{2}xx \tran )$. 
So a moment estimator for the asymptotic variance of $\hat{\beta}$ is  
\begin{equation}\label{eq::v-hat-ehw}
\hat{V}_\textsc{ehw} =  
n^{-1}
\left(n^{-1}\sumn x_{i}x_{i} \tran \right)^{-1}
\left(n^{-1}\sumn  \hat{\varepsilon}_i^2  x_{i}x_{i} \tran \right) 
\left(n^{-1}\sumn x_{i}x_{i} \tran \right)^{-1} ,
\end{equation}
which is called the Eicker--Huber--White (EHW) robust covariance
estimator \citep{eicker1967limit, huber::1967, white::1980}. We can show that $n \hat{V}_\textsc{ehw}  \rightarrow  V$ in probability. Based on $\hat{\beta}$ and $\hat{V}_\textsc{ehw} $, we can make inference about the population OLS coefficient $\beta$.

There are many variants of the EHW robust covariance estimator based on the leverage scores \citep{long2000using}. In particular, the HC1 variant modifies $\hat{\varepsilon}_i^2$ to $\hat{\varepsilon}_i^2/(n-p)$,
the HC2 variant modifies $\hat{\varepsilon}_i^2$ to $\hat{\varepsilon}_i^2 /(1 - h_{ii}) $, 
and
the HC3 variant modifies $\hat{\varepsilon}_i^2$ to $\hat{\varepsilon}_i^2/(1 - h_{ii})^2$, in the definition of $\hat{V}_\textsc{ehw}$.

\section{Frisch--Waugh--Lovell Theorem}
\label{appendix::fwl-theorem}

The Frisch--Waugh--Lovell (FWL) theorem has two versions: one at the population level and the other at the sample level. It reduces multivariate OLS to univariate OLS and therefore facilitates the understanding and calculation of the OLS coefficients. Below I will present special cases of the FWL theorem which are enough for this book. 

\begin{theorem}
[population FWL]\label{thm::population-FWL}
The coefficient of $x_1$ in the OLS fit of $y$ on $(x_1, x_2, \ldots, x_p)$ equals the coefficient of $\tilde{x}_1$ in the OLS fit of $y$ or $\tilde{y}$ on $\tilde{x}_1$, where $\tilde{y}$ is the residual from the OLS fit of $y$ on $(x_2, \ldots, x_p)$ and $\tilde{x}_1$ is the residual from the OLS fit of $x_1$ on $(x_2, \ldots, x_p)$.
\end{theorem}

In Theorem \ref{thm::population-FWL},  residualizing $x_1$ is crucial but residualizing $y$ is not.

\begin{theorem}
[sample FWL]\label{thm::sample-FWL}
With data $(Y, X_1, X_2, \ldots, X_p)$ containing column vectors, the coefficient of $X_1$ in the OLS fit of $Y$ on $(X_1, X_2, \ldots, X_p)$  equals the coefficient of $\tilde{X}_1$ in the OLS fit of $Y$ or $\tilde{Y}$ on $\tilde{X}_1$, where $\tilde{Y}$ is the residual vector from the OLS fit of $Y$ on $(X_2, \ldots, X_p)$ and $\tilde{X}_1$ is the residual vector  from the OLS fit of $X_1$ on $(X_2, \ldots, X_p)$. 
\end{theorem}

Again, in Theorem \ref{thm::sample-FWL}, residualizing $X_1$ is crucial but residualizing $Y$ is not. 
\citet{ding2021frisch} gives more numerical properties related to the sample FWL theorem.

\section{Linear model}\label{section::linear-model-var}

Sometimes, we impose a stronger model assumption which requires the conditional mean of $y$ given $x$ is linear:
$$
E(y\mid x)=x \tran \beta
$$
or, equivalently, 
\[
y=x \tran \beta+\varepsilon \qquad  \text{with} \qquad  E(\varepsilon\mid x)=0,
\]
which is called the restricted mean model. Under this model,  the population OLS coefficient is the true parameter of interest:
\begin{align*}
\left\{ E(xx \tran )\right\} ^{-1}E(xy) & =\left\{ E(xx \tran )\right\} ^{-1}E\left\{ xE(y\mid x)\right\} \\
 & =\left\{ E(xx \tran )\right\} ^{-1}E(xx \tran \beta)\\
 & =\beta.
\end{align*}
Moreover, the population OLS coefficient does not depend on the distribution
of $x$. The asymptotic inference in Section \ref{sec::populationOLS} applies to this model too.

In the special case with $\var(\varepsilon \mid x) = \sigma^2$, 
the asymptotic variance of the OLS coefficient reduces to
$$
V = \sigma^2 \{ E(xx \tran ) \}^{-1}
$$
so a simpler moment estimator for the asymptotic variance of $\hat{\beta}$ is 
\begin{equation}
\hat{V}_\textsc{ols} = \hat\sigma^2 \left( \sumn x_{i}x_{i} \tran \right)^{-1}
\label{eq::v-hat-ols}
\end{equation}
where $\hat\sigma^2 =  (n-p)^{-1}  \sum_{i=1}^n \hat{\varepsilon}_i^2$ is an unbiased estimator for $\sigma^2$, recalling that $n$ denotes the sample size and $p$ denotes the dimension of $x$. This is the standard covariance estimator from the \ri{lm} function.

Based on the \ri{BostonHousing} data, we first obtain the standard output from the \ri{lm} function. 

\begin{lstlisting}
> library("mlbench")
> data(BostonHousing)
> ols.fit = lm(medv ~ ., data = BostonHousing)
> summary(ols.fit)

Call:
lm(formula = medv ~ ., data = BostonHousing)

Residuals:
    Min      1Q  Median      3Q     Max 
-15.595  -2.730  -0.518   1.777  26.199 

Coefficients:
              Estimate Std. Error t value Pr(>|t|)    
(Intercept)  3.646e+01  5.103e+00   7.144 3.28e-12 ***
crim        -1.080e-01  3.286e-02  -3.287 0.001087 ** 
zn           4.642e-02  1.373e-02   3.382 0.000778 ***
indus        2.056e-02  6.150e-02   0.334 0.738288    
chas1        2.687e+00  8.616e-01   3.118 0.001925 ** 
nox         -1.777e+01  3.820e+00  -4.651 4.25e-06 ***
rm           3.810e+00  4.179e-01   9.116  < 2e-16 ***
age          6.922e-04  1.321e-02   0.052 0.958229    
dis         -1.476e+00  1.995e-01  -7.398 6.01e-13 ***
rad          3.060e-01  6.635e-02   4.613 5.07e-06 ***
tax         -1.233e-02  3.760e-03  -3.280 0.001112 ** 
ptratio     -9.527e-01  1.308e-01  -7.283 1.31e-12 ***
b            9.312e-03  2.686e-03   3.467 0.000573 ***
lstat       -5.248e-01  5.072e-02 -10.347  < 2e-16 ***
\end{lstlisting}

In \ri{R}, the \ri{lm} function can compute $\hat{\beta}$, and the \ri{hccm} function in the package \ri{car} can compute $\hat{V}_\textsc{ehw} $ as well as its variants. Below we compare the $t$-statistics based on different choices of the standard errors. In this example, the EHW standard errors differ a lot for some regression coefficients. 

\begin{lstlisting}
> library("car")
> ols.fit.hc0 = sqrt(diag(hccm(ols.fit, type = "hc0")))
> ols.fit.hc1 = sqrt(diag(hccm(ols.fit, type = "hc1")))
> ols.fit.hc2 = sqrt(diag(hccm(ols.fit, type = "hc2")))
> ols.fit.hc3 = sqrt(diag(hccm(ols.fit, type = "hc3")))
> tvalues = summary(ols.fit)$coef[,1]/
+   cbind(summary(ols.fit)$coef[,2], 
+         ols.fit.hc0, 
+         ols.fit.hc1, 
+         ols.fit.hc2, 
+         ols.fit.hc3)
> colnames(tvalues) = c("ols", "hc0", "hc1", "hc2", "hc3")
> round(tvalues, 2)
               ols   hc0   hc1   hc2   hc3
(Intercept)   7.14  4.62  4.56  4.48  4.33
crim         -3.29 -3.78 -3.73 -3.48 -3.17
zn            3.38  3.42  3.37  3.35  3.27
indus         0.33  0.41  0.41  0.41  0.40
chas1         3.12  2.11  2.08  2.05  2.00
nox          -4.65 -4.76 -4.69 -4.64 -4.53
rm            9.12  4.57  4.51  4.43  4.28
age           0.05  0.04  0.04  0.04  0.04
dis          -7.40 -6.97 -6.87 -6.81 -6.66
rad           4.61  5.05  4.98  4.91  4.76
tax          -3.28 -4.65 -4.58 -4.54 -4.43
ptratio      -7.28 -8.23 -8.11 -8.06 -7.89
b             3.47  3.53  3.48  3.44  3.34
lstat       -10.35 -5.34 -5.27 -5.18 -5.01
\end{lstlisting}

\section{Weighted least squares}
\label{sec::appendix-wls}

Assuming that $(w_i, x_i, y_i) \iidsim (w, x, y)$ with $w\neq 0$. At the population level, we can define the weighted least squares (WLS) coefficient as
$$
\beta_w = \arg\min_b  E\{  w (y-x \tran b)^{2}    \},
$$
which satisfies
$$
E  \{  w x (y-x \tran \beta_w ) \} = 0
$$
and thus equals
$$
\beta_w  = \{  E  ( w xx\tran ) \}^{-1} E(  w xy )
$$
if $E  ( w xx\tran )$ is invertible. 

At the sample level, we can define the WLS coefficient as
$$
\hat{\beta}_w = \arg\min_b \sumn w_i (y_i-x_i \tran b)^{2},
$$
which satisfies
$$
\sumn w_i x_i (y_i -x_i \tran \hat\beta_w ) = 0
$$
and thus equals
$$
\hat \beta_w  =  \left(  n^{-1} \sumn   w_i x_ix_i\tran \right)^{-1} \left(   n^{-1} \sumn   w_i x_i y_i \right)
$$
if $\sumn   w_i x_ix_i\tran$ is invertible. 

In \ri{R}, we can specify \ri{weights} in the \ri{lm} function to implement WLS.

\section{Logistic regression}\label{sec::logistic-regression}

 \subsection{Model}
Technically, we can apply the OLS procedure even if the outcome $y$ is binary. However, it is a little awkward to have predicted probabilities outside the range of $[0,1]$. This motivates us to consider the
following model:
\[
\pr(y_{i}=1\mid x_{i})=g(x_{i} \tran \beta),
\]
where $g(\cdot):\mathbb{R}\rightarrow[0,1]$ is a monotone function,
and its inverse is often called the {\it  link function}. The $g(\cdot)$ function can be any distribution function of a random variable, but we will focus on the logistic form:
$$
g(z) =  \frac{ e^z }{1 + e^z} = (1 + e^{-z})^{-1}. 
$$
We can also write the logistic model
as
$$
\pr(y_{i}=1\mid x_{i}) =\frac{e^{x_{i} \tran \beta}}{1+e^{x_{i} \tran \beta}},
$$
or, equivalently, with the definition $\text{logit}(z) = \log\frac{z}{1-z}$, we have 
\[
\text{logit}\left\{ \pr(y_{i}=1\mid x_{i})\right\}  =x_{i} \tran \beta.
\]

Assume that $x_{i1}$ is binary. 
Under the logistic model, we have
\begin{eqnarray*}
\beta_1 &=& \text{logit}\left\{ \pr(y_{i}=1\mid x_{i1} = 1,  \ldots )\right\}  - \text{logit}\left\{ \pr(y_{i}=1\mid x_{i1} = 0,\ldots)\right\}   \\
&=& \log \frac{  \pr(y_{i}=1\mid x_{i1} = 1, \ldots)/\pr(y_{i}=0\mid x_{i1} = 1, \ldots)     }
{  \pr(y_{i}=1\mid x_{i1} = 0, \ldots)/\pr(y_{i}=0\mid x_{i1} = 0, \ldots)    } ,
\end{eqnarray*}
where $\ldots$ contains all other regressor $ x_{i2}, \ldots, x_{ip} $.
Therefore, the coefficient $\beta_1$ equals the log of the odds ratio of $x_{i1}$ on $y_i$ conditional on other regressors.

\subsection{Maximum likelihood estimate}\label{sec::mle-logistic}

Let $\pr(y_{i}=1\mid x_{i}) = \pi(x_{i},\beta)$. 
To estimate the parameter $\beta$, we can maximize the following likelihood function:
\begin{align*}
L(\beta)  
&=\prod_{i=1}^{n}\left\{ \pi(x_{i},\beta)\right\} ^{y_{i}}\left\{ 1-\pi(x_{i},\beta)\right\} ^{1-y_{i}} \\
& =\prod_{i=1}^{n}\left\{ \frac{\pi(x_{i},\beta)}{1-\pi(x_{i},\beta)}\right\} ^{y_{i}}\left\{ 1-\pi(x_{i},\beta)\right\} \\
 & =\prod_{i=1}^{n}\left(e^{x_{i} \tran \beta}\right)^{y_{i}}\frac{1}{1+e^{x_{i} \tran \beta}}\\
 & =\prod_{i=1}^{n}\frac{e^{y_{i}x_{i} \tran \beta}}{1+e^{x_{i} \tran \beta}}.
\end{align*}
Let $\hat{\beta}$ denote the maximizer, which is called the maximum likelihood estimate (MLE). Taking the log of $L(\beta)  $ and differentiating it with respect to $\beta$, we can show that the MLE must satisfy the first-order condition:
$$
\sumn  x_i  \{ y_i - \pi(x_{i}, \hat\beta)\} = 0.
$$
If $x_i$ contains the intercept, the MLE must satisfy 
$$
\sumn    \{ y_i - \pi(x_{i}, \hat\beta)\} = 0,
$$ 
that is, the average of the observed $y_i$'s must be identical to the average of the fitted probabilities $ \pi(x_{i}, \hat\beta)$'s.

Using the general theory for the MLE, we can show that it is consistent for the true parameter $\beta$ and is asymptotically Normal:
$$
\sqrt{n}( \hat{\beta} -  \beta) \rightarrow \text{N}(0, V)
$$
in distribution, where 
$
V = E [  \pi(x_{i},\beta) \{1-  \pi(x_{i},\beta)\} x x\tran   ].
$
So we can approximate the covariance matrix of $\hat{\beta} $ by 
$$
n^{-1} \sumn \pi(x_{i}, \hat\beta) \{1-  \pi(x_{i},\hat\beta)\}   x_i x_i\tran.
$$
In \ri{R}, the \ri{glm} function can find the MLE and report the estimated covariance matrix. We use the \ri{lalonde} data to illustrate the logistic regression with the binary outcome indicating positive real earnings in 1978.

\begin{lstlisting}
> library(Matching)
> data(lalonde)
> logit.re78 = glm(I(re78>0) ~ ., family = binomial, 
+                  data = lalonde)
> summary(logit.re78)

Call:
glm(formula = I(re78 > 0) ~ ., family = binomial, data = lalonde)

Deviance Residuals: 
    Min       1Q   Median       3Q      Max  
-2.1789  -1.3170   0.7568   0.9413   1.0882  

Coefficients:
              Estimate Std. Error z value Pr(>|z|)  
(Intercept)  1.910e+00  1.241e+00   1.539   0.1238  
age         -2.812e-03  1.533e-02  -0.183   0.8545  
educ        -2.179e-02  7.831e-02  -0.278   0.7808  
black       -1.060e+00  5.041e-01  -2.103   0.0354 *
hisp         2.741e-01  6.967e-01   0.393   0.6940  
married      7.577e-02  3.057e-01   0.248   0.8042  
nodegr      -1.984e-01  3.460e-01  -0.573   0.5664  
re74         7.857e-06  3.173e-05   0.248   0.8044  
re75         4.016e-05  6.058e-05   0.663   0.5074  
u74         -6.177e-02  4.095e-01  -0.151   0.8801  
u75          1.505e-02  3.518e-01   0.043   0.9659  
treat        5.412e-01  2.222e-01   2.435   0.0149 *
\end{lstlisting}

\subsection{Extension to the case-control study}\label{sec::case-control}

In case-control studies, sampling is conditional on the binary outcome, that is, units with outcomes $y_i=1$ and $y_i=0$ are sampled with different probabilities. Let $s_i$ be the sampling indicator. In case-control studies, we have
$$
\pr(s_i =1\mid x_i, y_i) = \pr(s_i =1\mid  y_i) 
$$
as a function of $y_i$, and we only observe units with $s_i = 1$.

Based on the model and the sampling mechanism of the case-control study, it does not seem obvious whether or not we can still use logistic regression to estimate the coefficients. Nevertheless, \citet{prentice1979logistic} proved a positive results. In case-control studies, logistic regression can still consistently estimate all the coefficients except for the intercept.

\subsection{Logistic regression with weights}

Sometimes, unit $i$ has weight $w_i$. Then we can fit a weighted logistic regression by solving
$$
\sumn  w_i x_i  \{ y_i - \pi(x_{i}, \hat\beta)\} = 0.
$$
In \ri{R}, we can specify \ri{weights} in the \ri{glm} function to implement weighted logistic regression.

\section{Homework problems}

\paragraph{Sample WLS with the intercept}\label{hw::sample-ols-intercept}

Assume the regressor $x_i$ contains the intercept. Show that 
\begin{equation}\label{eq::OLS-mean-data}
\bar{y}_w = \bar{x}_w\tran \hat{\beta}_w
\end{equation}
where $ \bar{x}_w =  \sumn w_i  x_i / \sumn w_i $ and $ \bar{y}_w =  \sumn w_i  y_i / \sumn w_i $ are the weighted averages of the $x_i$'s and $y_i$'s.

\paragraph{Population OLS with a binary regressor}\label{problem::pols-binary}

Assume $x$ is binary. Define the population OLS: 
$$
(\alpha, \beta)  = \arg\min_{(a,b)} E \{  (y - a - bx)^2 \} .
$$
Show that $\beta = E(y\mid x=1) - E(y\mid x=0)$ and $\alpha = E(y\mid x=0)$.

\paragraph{Univariate WLS}\label{hw::wls-uni-binary}

As a special case of WLS, define 
$$
(\hat \alpha_w, \hat \beta_w  ) = \arg\min_{(a,b)} \sumn w_i(y_i - a - b x_i)^2
$$
where $w_i\geq 0$. Show that
\begin{eqnarray}\label{eq::wls-beta}
 \hat \beta_w  = \frac{  \sumn w_i (x_i -  \bar{x}_w) (y_i - \bar{y}_w) }{  \sumn w_i (x_i - \bar{x}_w)^2 }
\end{eqnarray}
and 
\begin{eqnarray}\label{eq::wls-alpha}
\hat \alpha_w = \bar{y}_w -  \hat \beta_w \bar{x}_w,
\end{eqnarray}
where $ \bar{x}_w =  \sumn w_i  x_i / \sumn w_i $ and $ \bar{y}_w =  \sumn w_i  y_i / \sumn w_i $ are the weighted averages of the $x_i$'s and $y_i$'s.

Further assume that the $x_i$'s are binary. Show that  
\begin{eqnarray}\label{eq::wls-10}
\hat \beta_w  = \frac{ \sumn  w_i  x_i  y_i  }{ \sumn w_i  x_i  } - \frac{ \sumn w_i  (1-x_i)  y_i  }{ \sumn w_i  (1-x_i)  }.
\end{eqnarray}
That is, if the regressor is binary in the univariate WLS, the coefficient of the regressor equals the difference in the weighted means.

Remark: To prove \eqref{eq::wls-10}, use an appropriate reparametrization of the WLS problem. Otherwise, the derivation can be tedious.

\paragraph{OLS with orthogonal regressors}\label{hw::ols-orthogonal}

Consider sample OLS fit of an $n$-vector $Y$ on an $n\times p$ matrix $X $, with coefficient $\hat{\beta}$. Partition $X$ into $X = (X_1, X_2)$, where $X_1$ is an $n\times k$ matrix and $X_2$ is an $n\times l$ matrix, with $p=k+l$. Correspondingly, partition $\hat{\beta}$ into 
$$
\hat{\beta} = \begin{pmatrix}
\hat{\beta}_1\\
\hat{\beta}_2
\end{pmatrix} . 
$$

Assume $X_1$ and $X_2$ are orthogonal, that is, $X_1\tran X_2 = 0$. Show that $\hat{\beta}_1$ equals the coefficient from OLS of $Y$ on $X_1$ and $\hat{\beta}_2$ equals the coefficient from OLS of $Y$ on $X_2$, respectively.

\paragraph{OLS with a non-degenerate transformation of the regressors}\label{hw::ols-transform-regressors}

Define $\hat{\beta}$ as the coefficient from the sample OLS fit of an $n$-vector $Y$ on an $n\times p$ matrix $X $. 
Let $\Gamma$ be a $p\times p$ non-degenerate matrix, and define $X' = X\Gamma$. 
Define $\hat{\beta} ' $ as the coefficient from the sample OLS fit of  $Y$ on $X' $. 

Show that 
$$
 \hat{\beta}  =  \Gamma\hat{\beta} ' .
$$

\paragraph{Variances of the OLS estimator}\label{hw::variance-ols-estimator}

Assume $(x_{i},y_{i})_{i=1}^{n} \iidsim (x,y)$ and 
$$
E(y\mid x) = x\tran \beta,\quad \var(y\mid x) = \sigma^2(x).
$$
Show that the OLS estimator $\hat{\beta}$ has conditional mean
$$
E( \hat{\beta} \mid x_1,\ldots, x_n ) = \beta 
$$
and conditional variance
$$
\var( \hat{\beta} \mid x_1,\ldots, x_n )
= \left(\sumn x_{i}x_{i} \tran \right)^{-1}
\left(\sumn \sigma^2(x_i) x_{i}x_{i} \tran \right)
\left(\sumn x_{i}x_{i} \tran \right)^{-1}.
$$

Remark: This problem may provide some intuition for the variance estimators
\eqref{eq::v-hat-ehw} and \eqref{eq::v-hat-ols}.

\chapter{Some Useful Lemmas for Simple Random Sampling From a Finite Population}
\chaptermark{Simple Random Sampling} 
 \label{chapter::SRS-appendix}

\section{Lemmas}
 
Simple random sampling is a basic topic in standard survey sampling textbooks \citep[e.g.,][]{cochran1953}. Below I review some results for simple random sampling that are useful for design-based inference in the CRE in Chapters \ref{ch::frt-cre} and \ref{chapter::neyman-cr}. I define simple random sampling based on the distribution of the inclusion indicators.

\begin{definition}
[simple random sampling]\label{definition-srs}
A simple random sample of size $n_1$ consists of a subset from a finite population of $n$ units indexed by $i=1,\ldots, n$. 
Let $\bm{Z} = (Z_1, \ldots, Z_n)$ be the inclusion indicators of the $n$ units with $Z_i=1$ if unit $i$ is sampled and $Z_i=0$ otherwise.
The vector $\bm{Z}$ can take $\binom{n}{n_1}$ possible permutations of a vector of $n_1$ 1's and $n_0$ 0's, and each has equal probability.
\end{definition}

By Definition \ref{definition-srs}, simple random sampling is a special form of {\it sampling without replacement} because it does not allow for repeatedly sampling the same unit. Lemma \ref{lemma::randomization} below summarizes the first two moments of the inclusion indicators.

\begin{lemma}
\label{lemma::randomization}
Under simple random sampling, we have 
$$
E ( Z_i )  = \frac{n_1}{ n}, \quad 
\var(Z_i) = \frac{  n_1 n_0 }{n^2}, \quad 
\cov(Z_i, Z_j) = - \frac{n_1 n_0 } { n^2 (n-1)  }.
$$ 
In more compact forms, we have 
$$
E(\bm{Z}) = \frac{n_1}{n} \bm{1}_n,\quad
\cov(\bm{Z}) = \frac{n_1n_0}{n(n-1)} \bm{P}_n,
$$
where $ \bm{1}_n$ is a $n$-dimensional vector of $1$'s, and $\bm{P}_n = \bm{I}_n  - n^{-1} \bm{1}_n \bm{1}_n\tran $ is the $n\times n$ projection matrix orthogonal to $ \bm{1}_n.$
\end{lemma}

Let $\{ c_1, \ldots, c_n\} $ be a finite population with mean $ \bar{c}  = \sumn c_i/n$ and variance
$$
S_c^2 = (n-1)^{-1} \sumn (c_i -  \bar{c} )^2;
$$
let $\{  d_1, \ldots, d_n \} $ be another finite population with mean $ \bar{d}   = \sumn d_i/n$ and variance
$$
S_d^2 =  (n-1)^{-1}   \sumn (d_i -  \bar{d}  )^2;
$$
their covariance is
$$
S_{cd} =  (n-1)^{-1}   \sumn (c_i -  \bar{c} ) (d_i -  \bar{d}  ).
$$
Under simple random sampling, the sample means are
$$
 \hat{\bar{c}}  =  n_1^{-1} \sumn Z_i c_i,\quad 
 \hat{\bar{d}}   = n_1^{-1} \sumn Z_i d_i;
$$
sample variances are
$$
 \hat{S}_c^2   = (n_1 - 1)^{-1} \sumn Z_i (c_i -  \hat{\bar{c}} )^2,\quad 
 \hat{S}_d^2   = (n_1 - 1)^{-1}   \sumn Z_i (d_i -  \hat{\bar{d}}  )^2;
$$
the sample covariance is
$$
 \hat{S}_{cd}   = (n_1 - 1)^{-1}  \sumn Z_i (c_i -  \hat{\bar{c}} ) (d_i -  \hat{\bar{d}}  ). 
$$

Lemma \ref{lemma::samplemean} below gives the moments of the sample means $ \hat{\bar{c}} $ and $ \hat{\bar{d}}  $. 

\begin{lemma}
\label{lemma::samplemean}
Under simple random sampling, the sample means are unbiased for the population means:
$$
E( \hat{\bar{c}} ) =  \bar{c} ,\quad
E( \hat{\bar{d}}  ) =  \bar{d}  .
$$
Their variances and covariance are
\begin{eqnarray*}
\var\left(   \hat{\bar{c}} \right) = \frac{ n_0}{n n_1  }  S_c^2 ,\quad 
\var\left(   \hat{\bar{d}}  \right) = \frac{ n_0}{n n_1  }  S_d^2, \quad 
\cov\left(  \hat{\bar{c}}  ,   \hat{\bar{d}}  \right) = \frac{ n_0}{n n_1 } S_{cd}. 
\end{eqnarray*}
\end{lemma}

In the variance formulas in Lemma \ref{lemma::samplemean},  the coefficient $  n_0 / (n n_1 )  = 1/n_1 \times (1-n_1/n)$ in Lemma \ref{lemma::samplemean} is different from $1/n_1$ under IID sampling. The additional term $1-n_1/n = n_0/n$ is called the {\it finite population correction} factor.

Lemma \ref{lemma::variance-unbiased} below gives the unbiasedness of the sample variances and covariance for estimating the population analogs. 
 
\begin{lemma}\label{lemma::variance-unbiased}
Under simple random sampling, the sample variances and covariance are unbiased for their population versions:
$$
E( \hat{S}_c^2  ) = S_c^2,\quad 
E( \hat{S}_d^2  ) = S_d^2,\quad
E( \hat{S}_{cd} ) = S_{cd}  .
$$
\end{lemma}

An important practical question is to make inference about $ \bar{c} $ based on the simple random sample. This requires a more precise characterization of the distribution of its unbiased estimator $ \hat{\bar{c}} $. The finite-sample exact distribution of $ \hat{\bar{c}} $ depends on the whole finite population $\{  c_1, \ldots, c_n \}$, which is unknown. The following finite population CLT characterizes the asymptotic distribution of $ \hat{\bar{c}} $ based on its first two moments. 

\begin{lemma}[finite population CLT]
\label{lemma::finite-population-clt}
Under Under simple random sampling, as $n\rightarrow \infty$, if
$$
  \frac{ \max_{1\leq i \leq n}  (c_i -  \bar{c}  )^2 }{ \min(n_1, n_0)  S_c^2  } \rightarrow 0,
$$
then 
$$
\frac{  \hat{\bar{c}}  -  \bar{c}  }{   \sqrt{  \frac{ n_0}{n n_1  }  S_c^2}    } \rightarrow \textsc{N}(0,1)
$$
in distribution, and
$
 \hat{S}_c^2   / S_c^2   \rightarrow 1
$
in probability. 
\end{lemma}

Lemma \ref{lemma::finite-population-clt} justifies the Wald-type $1-\alpha$ confidence interval for $ \bar{c} $: 
$$
 \hat{\bar{c}}  \pm z_{1-\alpha/2}  \sqrt{  \frac{ n_0}{n n_1  }   \hat{S}_c^2  } 
$$
where $z_{1-\alpha/2} $ is the $1-\alpha/2$ upper quantile of the standard Normal random variable. 

\section{Proofs}

\begin{myproof}{Lemma}{\ref{lemma::randomization}}
By symmetry, the $Z_i$'s have the same mean, so
$$
n_1 = \sumn Z_i = E\left (  \sumn Z_i  \right  )  = n E(Z_i)
\Longrightarrow  E(Z_i) = n_1/n .
$$
Because $Z_i$ is a Bernoulli random variable, its variance is 
$$
\var(Z_i) = \frac{n_1}{ n} \left(1 - \frac{n_1}{n} \right) = \frac{n_1 n_0}{n^2}.
$$
By symmetry again, the $Z_i$'s have the same variance and the pairs $(Z_i, Z_j)$'s have the same covariance, so 
$$
0 = \var\left(   \sumn Z_i \right)   = n   \var(Z_i) + n(n-1) \cov(Z_i, Z_j)
$$
which implies that  
$$
\cov(Z_i, Z_j) = - \frac{n_1 n_0 } { n^2 (n-1)  }  \quad (i\neq j).
$$
\end{myproof}

\begin{myproof}{Lemma}{\ref{lemma::samplemean}}
The unbiasedness of the sample mean follows from linearity. For example,
$$
E( \hat{\bar{c}} ) = E\left(   \frac{1}{n_1} \sumn Z_i c_i \right) =   \frac{1}{n_1} \sumn E(Z_i) c_i =  \bar{c} .
$$

The covariance of the sample means is
\begin{eqnarray*}
&&\cov( \hat{\bar{c}} ,  \hat{\bar{d}}  ) \\
&=& \cov\left\{      \frac{1}{n_1} \sumn Z_i (c_i -  \bar{c}  ) ,   \frac{1}{n_1} \sumn Z_i (d_i -  \bar{d}   )\right\} \\
&=& \frac{1}{n_1^2} \left[ \sumn \var(Z_i) (c_i -  \bar{c}  ) (d_i -  \bar{d}   ) + 
\sum_{i\neq j} \cov(Z_i, Z_j) (c_i -  \bar{c}  )(d_j -  \bar{d}   ) 
\right] \\
&=& \frac{1}{n_1^2} \left[ \frac{n_1 n_0}{n^2} \sumn   (c_i -  \bar{c}  )(d_i -  \bar{d}   )
-\frac{n_1 n_0 } { n^2 (n-1)  }
\sum_{i\neq j}  (c_i -  \bar{c}  )(d_j -  \bar{d}   ) 
\right] .
\end{eqnarray*}
Because 
$$
0 =   \sumn   (c_i -  \bar{c}  )  \sumn   (d_i -  \bar{d}   )  
= 
\sumn   (c_i -  \bar{c}  )(d_i -  \bar{d}   )   + 
\sum_{i\neq j}  (c_i -  \bar{c}  )(d_j -  \bar{d}   ) ,
$$
the covariance of the sample means reduces to
\begin{eqnarray*}
&&\cov( \hat{\bar{c}} ,  \hat{\bar{d}} )  \\
&=& \frac{1}{n_1^2} \left[ \frac{n_1 n_0}{n^2} \sumn   (c_i -  \bar{c}  ) (d_i -  \bar{d}   ) 
+\frac{n_1 n_0 } { n^2 (n-1)  }
\sumn   (c_i -  \bar{c}  ) (d_i -  \bar{d}  ) 
\right]  \\
&=& \frac{n_0}{nn_1} S_{cd}.
\end{eqnarray*}
The variance formulas are just special cases with $ \hat{\bar{c}}  =  \hat{\bar{d}}  $. 
\end{myproof}

\begin{myproof}{Lemma}{\ref{lemma::variance-unbiased}} 
We prove only the sample covariance term because the formulas for sample variances are special cases.
We have the following decomposition:
\begin{eqnarray*}
(n_1 - 1)  \hat{S}_{cd}  &=& \sumn Z_i  (c_i -  \hat{\bar{c}} ) (d_i -  \hat{\bar{d}}  ) \\
&=& \sumn Z_i \{    (c_i -  \bar{c}  ) - (  \hat{\bar{c}}   -  \bar{c} )  \}  \{    (d_i -  \bar{d}   ) - (  \hat{\bar{d}}    -  \bar{d}  )  \}\\
&=&\sumn Z_i    (c_i -  \bar{c}  ) (d_i -  \bar{d}   )  -  n_1(  \hat{\bar{c}}   -  \bar{c} )   (  \hat{\bar{d}}    -  \bar{d}  ).
\end{eqnarray*}
Taking expectations on both sides, we have
\begin{eqnarray*}
E\{ (n_1 - 1)  \hat{S}_{cd}  \} &=&   \sumn E(Z_i)    (c_i -  \bar{c}  )  (d_i -  \bar{d}   )   -  n_1 E\{ (  \hat{\bar{c}}   -  \bar{c} )   (  \hat{\bar{d}}    -  \bar{d}  ) \} \\
&=& \frac{n_1}{n}  \sumn     (c_i -  \bar{c}  ) (d_i -  \bar{d}   ) - n_1 \frac{n_0}{nn_1} S_{cd} \\
&=& S_{cd} \left\{   \frac{n_1(n-1)}{n} - \frac{n_0}{n} \right\} \\
&=& (n_1 - 1) S_{cd},
\end{eqnarray*}
and the conclusion follows by dividing both sides by $n_1-1$. 
\end{myproof}

\begin{myproof}{Lemma}{\ref{lemma::finite-population-clt}}
\citet{hajek1960limiting} gave a proof of the CLT for simple random sampling, and \citet{lehmann2006nonparametrics} gave a more accessible version of the proof. \citet{li2017general} modified the CLT as presented in Lemma \ref{lemma::finite-population-clt}, and proved the consistency of the sample variance. Due to the technical complexities, I omit the proof. 
\end{myproof}

\section{Comments on the literature}

Survey sampling and experimental design have been deeply connected ever since \citet{neyman1934two, neyman1935statistical}'s seminal work. \citet{li2017general} and \citet{mukerjee2018using} made many theoretical ties between these two areas.

\section{Homework Problems}

\paragraph{Sampling without replacement and the Hypergeometric distribution}
\label{problem::srs-hypergeometric}

Consider a special case of Lemma \ref{lemma::samplemean} with $c_i$'s being binary. Assume the total number of 1's equals $T$ so the total number of 0's equals $n-T$. Let $t = \sumn Z_i c_i$ denote the total number of 1's in the sample of size $n_1$. 

Find the distribution, mean, and variance of $t$. 

Remark: $t$ follows a Hypergeometric distribution.

\paragraph{Vector form of Lemma \ref{lemma::samplemean}}
\label{para::srs-vector-form}

Assume the $c_i$'s are vectors and define 
$$
S_c^2 = (n-1)^{-1} \sumn (c_i  - \bar{c})(c_i -  \bar{c}) \tran ,\quad
\hat S_c^2 = (n_1-1)^{-1} \sumn Z_i (c_i  - \hat{\bar{c}})(c_i - \hat{\bar{c}}) \tran .
$$
Show that 
$$
E(\hat{\bar{c}}) = \bar c,\quad
\cov( \hat{\bar{c}} ) = \frac{n_0}{ nn_1} S_c^2,\quad
E(\hat S_c^2 ) = S_c^2 .
$$

\bibliographystyle{apalike}
\bibliography{causal}

\printindex

\end{document}